\documentclass[11pt,letterpaper,onecolumn]{article}

\usepackage[T1]{fontenc}
\usepackage{lmodern}
\usepackage[letterpaper,margin=1in]{geometry}
\usepackage{microtype}
\usepackage{mathtools}
\usepackage{amssymb}
\usepackage{amsthm}
\usepackage{graphicx}
\usepackage[font=normalsize,labelfont=bf,skip=6pt]{caption}
\usepackage{needspace}
\usepackage{colortbl}
\usepackage{tikz}
\usetikzlibrary{arrows.meta,positioning,shapes.geometric,calc}
\usepackage{aliascnt}
\usepackage[hidelinks]{hyperref}
\usepackage[capitalise,nameinlink,noabbrev]{cleveref}
\usepackage{authblk}

\allowdisplaybreaks[2]

\theoremstyle{plain}
\newtheorem{theorem}{Theorem}[section]

\newaliascnt{lemma}{theorem}
\newtheorem{lemma}[lemma]{Lemma}
\aliascntresetthe{lemma}

\newaliascnt{proposition}{theorem}
\newtheorem{proposition}[proposition]{Proposition}
\aliascntresetthe{proposition}

\newaliascnt{corollary}{theorem}
\newtheorem{corollary}[corollary]{Corollary}
\aliascntresetthe{corollary}

\newaliascnt{claim}{theorem}

\aliascntresetthe{claim}

\theoremstyle{definition}
\newaliascnt{definition}{theorem}
\newtheorem{definition}[definition]{Definition}
\aliascntresetthe{definition}

\newaliascnt{procedure}{theorem}

\aliascntresetthe{procedure}

\theoremstyle{remark}
\newaliascnt{remark}{theorem}
\newtheorem{remark}[remark]{Remark}
\aliascntresetthe{remark}

\crefname{theorem}{Theorem}{Theorems}
\Crefname{theorem}{Theorem}{Theorems}
\crefname{lemma}{Lemma}{Lemmas}
\Crefname{lemma}{Lemma}{Lemmas}
\crefname{proposition}{Proposition}{Propositions}
\Crefname{proposition}{Proposition}{Propositions}
\crefname{corollary}{Corollary}{Corollaries}
\Crefname{corollary}{Corollary}{Corollaries}
\crefname{claim}{Claim}{Claims}
\Crefname{claim}{Claim}{Claims}
\crefname{definition}{Definition}{Definitions}
\Crefname{definition}{Definition}{Definitions}
\crefname{procedure}{Procedure}{Procedures}
\Crefname{procedure}{Procedure}{Procedures}
\crefname{remark}{Remark}{Remarks}
\Crefname{remark}{Remark}{Remarks}

\usepackage{mathrsfs,booktabs,tabularx,array,enumitem}
\definecolor{repositorylink}{HTML}{1F5A94}
\IfFileExists{fontawesome5.sty}{\usepackage{fontawesome5}
\newcommand{\GitHubIcon}{\mbox{\faGithub}\nobreakspace}}{\newcommand{\GitHubIcon}{}}
\setlist[itemize]{leftmargin=1.6em,itemsep=0.25em,topsep=0.3em}
\setlist[enumerate]{leftmargin=*,itemsep=0.3em,topsep=0.3em}
\newenvironment{proofstrategy}{\par\noindent\textit{Proof strategy.}\ }{\par}
\hypersetup{pdftitle={A Dichotomy for Complex Boolean Holant with Binary Disequality},
pdfauthor={Chenghua Liu and Boning Meng}}
\newcommand{\AlgNums}{\overline{\mathbb Q}}
\newcommand{\Ftwo}{\mathbb F_2}
\newcommand{\cF}{\mathcal F}
\newcommand{\cG}{\mathcal G}
\newcommand{\cS}{\mathcal S}
\newcommand{\cA}{\mathscr A}
\newcommand{\cP}{\mathscr P}
\newcommand{\cL}{\mathscr L}
\newcommand{\cM}{\mathscr M}
\newcommand{\cT}{\mathscr T}
\newcommand{\cC}{\mathscr C}

\newcommand{\angles}[1]{\left\langle #1\right\rangle}
\newcommand{\Holant}{\operatorname{Holant}}
\newcommand{\KHolant}{\operatorname{K\text{-}Holant}}
\newcommand{\EOTract}{\mathsf{EOTract}}
\newcommand{\Tract}{\mathsf{Tract}}
\newcommand{\ReTract}{\mathsf{ReTract}}
\newcommand{\DihZ}[1]{{D_{#1}^{(Z)}}}
\newcommand{\DihY}[1]{{D_{#1}^{(Y)}}}
\newcommand{\DihXzero}[1]{{D_{#1}^{(X,0)}}}
\newcommand{\DihXone}[1]{{D_{#1}^{(X,1)}}}
\newcommand{\CycX}[1]{{C_{#1}^{(X)}}}
\newcommand{\KleinStd}{{V_4^{(\mathrm{std})}}}
\newcommand{\KleinEx}{{V_4^{(\mathrm{ex})}}}
\newcommand{\TetInt}{{\mathcal T^{(\mathrm{int})}}}
\newcommand{\TetExt}{{\mathcal T^{(\mathrm{ext})}}}
\newcommand{\OctDD}{{\mathcal O^{(\mathrm{dd})}}}
\newcommand{\OctTT}{{\mathcal O^{(\mathrm{tt})}}}
\newcommand{\OctDT}{{\mathcal O^{(\mathrm{dt})}}}
\newcommand{\OctTD}{{\mathcal O^{(\mathrm{td})}}}
\newcommand{\OctForm}[1]{{\mathcal O^{(#1)}}}
\newcommand{\IcoZero}{{\mathcal I^{(0)}}}
\newcommand{\IcoOne}{{\mathcal I^{(1)}}}

\newcommand{\PresK}{\mathsf{Pres}_{K}}
\newcommand{\Stable}{\mathsf{Stable}}
\newcommand{\Even}{\mathsf{Even}}
\newcommand{\Close}{\mathsf{Close}}

\newcommand{\FLag}{\mathsf{FLag}}
\newcommand{\FP}{\mathsf{FP}}
\newcommand{\SharpP}{\#\mathsf P}
\newcommand{\PGL}{\operatorname{PGL}}
\newcommand{\GL}{\operatorname{GL}}
\newcommand{\Mat}{\operatorname{Mat}}
\newcommand{\diag}{\operatorname{diag}}
\newcommand{\Span}{\operatorname{span}}
\newcommand{\rank}{\operatorname{rank}}
\newcommand{\wt}{\operatorname{wt}}
\newcommand{\supp}{\operatorname{supp}}
\newcommand{\arity}{\operatorname{arity}}
\newcommand{\EO}{\mathrm{EO}}
\newcommand{\Eq}{=}
\newcommand{\Neq}{\neq}
\newcommand{\eps}{\varepsilon}
\newcommand{\one}{\mathbf 1}
\newcommand{\zero}{\mathbf 0}
\newcommand{\DeltaZero}{\Delta_0}
\newcommand{\DeltaOne}{\Delta_1}
\newcommand{\RMcore}{\mathrm{RM}_8}
\newcommand{\Hcore}{\mathrm{H}_6}
\newcommand{\leT}{\leq_{\mathrm T}}
\newcommand{\equivT}{\equiv_{\mathrm T}}

\newcolumntype{L}[1]{>{\raggedright\arraybackslash}p{#1}}

\newcommand{\DEQedgenumber}[1]{%
  \ifnum#1<10\relax#1\else
    \begingroup
    \setbox0=\hbox{0}%
    \hbox to \wd0{\hss\vphantom{0}\raisebox{.14\ht0}{\scalebox{.72}{#1}}\hss}%
    \endgroup
  \fi}

\definecolor{pinkfoldmarkcolor}{HTML}{B74B7D}
\newcommand{\pinkfoldmark}{%
  \raisebox{-0.15em}{%
    \begin{tikzpicture}[x=0.0375em,y=-0.0375em,baseline=(current bounding box.south)]
      \path[use as bounding box] (0,0) rectangle (24,24);
      \draw[draw=pinkfoldmarkcolor,line width=0.06375em,line join=miter,line cap=butt]
        (12,2) -- (20.660,7) -- (20.660,17) -- (12,22) -- (3.340,17) -- (3.340,7) -- cycle
        (12,12) -- (20.660,7)
        (12,12) -- (3.340,7)
        (12,12) -- (12,22);
    \end{tikzpicture}%
  }%
}

\begin{document}
\title{A Dichotomy for Complex Boolean Holant\\[0.15em]
with Binary Disequality}

\author[1]{Chenghua Liu}
\author[2]{Boning Meng}

\affil[1]{Institute of Software, Chinese Academy of Sciences, Beijing, China}
\affil[2]{University of Regensburg, Regensburg, Germany}
\affil[ ]{\texttt{liuch.russell@gmail.com}, \texttt{mengboning2013@gmail.com}}
\date{}

\hypersetup{pageanchor=false}
\begin{titlepage}
\maketitle
\thispagestyle{empty}

\begin{abstract}
\normalsize
We classify Boolean Holant problems with a supplied binary disequality signature and an arbitrary fixed finite set of algebraic complex-valued signatures. Each problem is polynomial-time computable or \(\SharpP\)-hard, and the tractability criterion is decidable. The proof combines arity reduction with the growth of the available binary transfer group and uses stable pairs to recover the structure of higher-arity signatures. We also prove a dichotomy for the disequality-edge model when the signature set contains an endpoint-nondegenerate irreducible eight-vertex signature, resolving the corresponding arity-four cases in the main proof.
\end{abstract}
\end{titlepage}

\clearpage
\pagenumbering{arabic}
\hypersetup{pageanchor=true}

\section{Introduction}
\label{s:sec:p1-introduction}

Many counting problems ask for a global sum determined by local
constraints: each assignment receives the product of its local weights,
and these products are summed over all assignments.  A complexity
dichotomy identifies the choices of local signatures that admit
polynomial-time evaluation and proves \(\SharpP\)-hardness for every
remaining choice.  On the Boolean domain, this program has a long history
in counting constraint satisfaction problems (\(\#\mathrm{CSP}\)).
Creignou and Hermann classified the unweighted case
\cite{CreignouHermann1996}.  Later dichotomies treated nonnegative rational
weights \cite{DyerGoldbergJerrum2009}, rational weights of mixed signs
\cite{BulatovDyerGoldbergJalseniusRicherby2009}, and complex weights
\cite{CaiLuXia2014}.  Complex weights permit cancellations that yield
tractable cases absent from the nonnegative setting
\cite{CaiLuXia2014}, making the algebra of the weights part of the
classification.
Over arbitrary finite domains, counting-CSP dichotomies have likewise
been established for unweighted constraints
\cite{Bulatov2013,DyerRicherby2013}, nonnegative rational weights
\cite{BulatovDyerGoldbergJalseniusJerrumRicherby2012}, nonnegative algebraic
weights \cite{CaiChenLu2016}, and algebraic complex weights
\cite{CaiChen2017}.

The Holant framework brings the availability of local constraints into
this classification problem \cite{CaiLuXia2009}.  A counting-CSP instance
can be represented by a bipartite graph whose variable vertices carry
equality constraints, enforcing agreement among all occurrences of a
variable.  Holant allows the available vertex signatures to be prescribed
without supplying these equalities.  This accommodates problems such as
counting perfect matchings, where each vertex requires exactly one
incident edge to be selected.  It also makes the construction of useful
local signatures a part of the classification proof.  Holographic
transformations, originating in Valiant's algorithms \cite{Valiant2008},
give another source of tractability: a simultaneous change of basis can
expose structure in the local signatures while preserving the value of the
total sum.

Boolean Holant dichotomies have addressed both restrictions on the
signatures and assumptions about auxiliary signatures.  With no prescribed
auxiliary signatures, classifications were obtained for symmetric
real-valued signatures \cite{HuangLu2016}, symmetric complex-valued
signatures \cite{CaiGuoWilliams2016}, arbitrary nonnegative signatures
\cite{LinWang2018}, and arbitrary real-valued signatures
\cite{ShaoCai2020}.  Here symmetry means that a signature depends only on
the number of ones in its input.  Another line of work supplies unary
signatures: \(\Holant^*\) supplies all of them
\cite{CaiLuXia2011HolantStar}, \(\Holant^+\) supplies four distinguished
ones \cite{Backens2017}, and \(\Holant^c\) supplies the two pinnings
\(\DeltaZero=[1,0]^{\mathsf T}\) and
\(\DeltaOne=[0,1]^{\mathsf T}\).
The last model was classified for symmetric complex signatures
\cite{CaiHuangLu2012Holantc}, arbitrary real signatures
\cite{CaiLuXia2018RealHolantc}, and algebraic complex signatures
\cite{Backens2021,Backens2025}.  The real-valued classification introduced
the local-affine tractable family \(\cL\), which also occurs here.

Meng, Wang, Xia, and Zheng proved a tensor-factor extraction theorem and
classified sets containing a nonzero odd-arity signature. Their condition
(PC) lists seven cases \cite[conference version, Theorem~30]{MengWangXiaZheng2025}.
For two cases involving weighted Eulerian orientations
\cite{MengWangXia2025EO}, the original bound was \(\FP^{\mathsf{NP}}\).
Guan, Shao, and Shi gave polynomial-time algorithms for these cases
\cite{GuanShaoShi2026}. We write \(\Tract\) for the resulting seven-case
criterion, the most general tractability condition currently known for
algebraic complex-valued Boolean Holant.
Even-arity signatures present a different resource constraint.  Every
gadget built entirely from them has an even number of external inputs,
because each internal edge consumes two inputs.  Such gadgets cannot
supply a unary signature, so binary signatures are the smallest
positive-arity objects available for further constructions.  We study
the case in which binary disequality, requiring two bits to differ, is
supplied.

There is also a precedent for prescribing disequality in counting CSP.
In \(\#\mathrm{CSP}_d\), each variable occurs a number of times divisible
by \(d\).  Cai, Fu, and Shao classified
\(\#\mathrm{CSP}_d(\Neq_2,\cF)\) for complex signatures
\cite[full version, Section~5.1 and Theorem~5.3]{CaiFuShao2020Entanglement}.
Shao and Cai used its hardness conclusion in their real Holant dichotomy
\cite[full version, Theorem~2.24]{ShaoCai2020}.

Prescribing binary disequality is useful for three further reasons.
First, ordinary edges enforce equality, and the supplied binary
signature enforces disequality. Together they restrict the binary groups
in Xia's framework \cite[Section~4 and Table~2]{Xia2026FrameworkV1} to
the marked forms analyzed here. Second, inserting binary vertices lets
us use disequality edges with equality supplied as a vertex signature.
We can thus choose the edge model suited to a construction while keeping
the original signatures in their original coordinates. Third, the classification applies to any Holant problem for which
a reduction obtains disequality. For the complementary problems, the
inability to obtain it supplies a structural restriction for further
classification.

\paragraph{The problem and main results.}
A Boolean signature of arity \(n\) is a function
\(f:\{0,1\}^n\to\mathbb C\).  In a Holant instance, a graph vertex
of degree \(n\) carries such a signature, with its incident edges ordered.
For a fixed finite set \(\cF\) of signatures, the problem
\(\Holant(\cF)\) evaluates
\[
  \sum_{\sigma:E\to\{0,1\}}
  \prod_{v\in V}f_v\bigl(\sigma|_{E(v)}\bigr),
  \qquad f_v\in\cF.
\]
Here \(E(v)\) is the ordered list of edge incidences at \(v\).
\Cref{subsec:holant-definitions} gives the conventions for loops and
multiple edges.  The graph and its vertex labels form the input. The set
\(\cF\) is fixed.  Write
\[
 X=(\Neq_2)=\begin{pmatrix}0&1\\1&0\end{pmatrix},
 \qquad I=(\Eq_2)=\begin{pmatrix}1&0\\0&1\end{pmatrix}.
\]
Our problem is \(\Holant(\cF,X):=\Holant(\cF\cup\{X\})\).
Thus any number of degree-two vertices may enforce opposite values on
their two incident edges.

With \(X\) supplied, three of the seven cases of \(\Tract\) are
excluded (\cref{lem:tract-X-anchor}). Write \(\ReTract\)
for the restriction to the other four. In one case every signature is a
tensor product of signatures of arity at most two. In each of the other
three, a common change of basis puts the edge and all vertex signatures
in the affine class \(\cA\), the product class \(\cP\), or the local-affine
class \(\cL\) (Section~\ref{s:short:tools}). The prefix \emph{Re} records this
restriction and the correspondence with the tractable cases of the
real-valued Holant dichotomy of Shao and Cai~\cite{ShaoCai2020}.

\begin{theorem}[Binary-disequality Holant dichotomy]
\label{s:thm:p1-deq2-dichotomy}
For every finite set of algebraic complex-valued Boolean signatures \(\cF\),
the problem \(\Holant(\cF,X)\) is polynomial-time computable when
\(\ReTract(\cF\cup\{X\})\) holds and is \(\SharpP\)-hard otherwise.
The tractability criterion is decidable from exact algebraic input.
\end{theorem}

The proof also classifies sets containing a particular kind of
arity-four signature, without a supplied equality signature.  Following Meng, Wang, Xia, and Zheng
\cite{MengWangXiaZheng2025}, write
\(\KHolant(\cF):=\Holant(\Neq_2\mid\cF)\) for the bipartite Holant
problem with disequality on the left and signatures from \(\cF\) on the
right.  Equivalently, each connection between signature inputs enforces
disequality.

A quaternary signature \(q\) is
\emph{endpoint-nondegenerate} when
\[
 q({\mathbf{x}})=0\quad\text{for odd }\wt({\mathbf{x}}),\qquad q(0000)q(1111)\ne0.
\]
Here \(\wt(\mathbf{x})\) counts the ones in \(\mathbf{x}\). Only the
eight even-weight inputs can have nonzero values, giving the eight-vertex
form used below.
A nonzero positive-arity signature is \emph{irreducible} if it has no factorization
\(a\otimes b\) on two nonempty blocks of inputs, allowing a permutation of inputs and a nonzero scalar.

\begin{theorem}[Quaternary \(\KHolant\) dichotomy]
\label{s:thm:p1-quaternary-kholant-dichotomy}
Let \(\cF\) be a finite set of algebraic complex-valued Boolean signatures.
Suppose that \(\cF\) contains an endpoint-nondegenerate irreducible
quaternary signature.  Then \(\KHolant(\cF)\) is polynomial-time computable
if \(\ReTract(\cF)\) holds and is \(\SharpP\)-hard otherwise.  The
tractability criterion is decidable from exact algebraic input.
\end{theorem}

The dichotomy for the eight-vertex model
\cite[arXiv version~2, Theorem~3.1]{CaiFu2023} classifies a single
eight-vertex signature \(q\) with nonzero endpoints. Even if \(q\) alone
is tractable, using it with other signatures can make the problem hard.
Theorem~\ref{s:thm:p1-quaternary-kholant-dichotomy} settles the complexity
of the whole set whenever a construction yields an endpoint-nondegenerate
irreducible quaternary signature.

\paragraph{From existing tools to stable pairs.}
The odd-arity classification and tensor-factor extraction of
\cite{MengWangXiaZheng2025} provide two starting points. Obtaining a
nonzero odd-arity signature invokes a known dichotomy. In the cases
considered here, their factor-extraction theorem allows us to simulate
irreducible factors and retain them for subsequent constructions. Two existing approaches suggest
how to use these tools. Shao and Cai's real-valued proof uses arity
descent \cite[Sections~3--4]{ShaoCai2020}, but its orthogonality and norm
arguments do not directly control cancellation for arbitrary complex
signatures. Xia's finite-group approach
\cite[Sections~4--5]{Xia2026FrameworkV1} organizes the available binary
constructions. The remaining task is to determine how these constructions
constrain higher-arity signatures. We couple these approaches by tracking a nonbinary
irreducible signature \(f\) of minimum arity together with a group \(G\).
After the preliminary reductions, \(G\) consists of the transfer classes
\([BX]\) of all ordered nonzero binaries \(B\) directly realizable by
gadgets over the current set. Brackets identify matrices up to nonzero scalar.

Close a pair of inputs of \(f\) with an available binary and retain
the irreducible factors of the output by reduction. An unresolved
nonbinary factor has smaller arity. A binary factor whose transfer class
lies outside \(G\) gives a different form of progress. Once retained,
it can be used directly in gadgets over the enlarged set. The complete
group strictly enlarges, so we can test the same \(f\) with new
contractions even when its arity has not decreased.
Figure~\ref{s:fig:complete-group-growth} shows the possible strict
inclusions. Bounded arity descent, a uniform bound on group orders,
and finite processing at each stage ensure termination.

If neither form of progress occurs and the complexity remains unresolved,
every prescribed one-copy contraction, including those obtained by attaching realizable
binaries to selected inputs, gives zero or a product of retained binaries on a perfect
matching. This is the stable condition. We call the outputs cards.
Fix the pair of inputs being closed and compare its nonzero cards. To
recover a common factor, we first seek a matching compatible with the
cards. Even when the pairings agree, the binary factors on an edge from
which we want to extract a common factor must be proportional across
the cards.

Gadgets test these requirements by preserving a proposed disagreement down to
bounded arity, where exhaustive classifications apply. When the
contractions recover the slices of \(f\), a common factor of the cards
lifts to \(f\), contradicting irreducibility. Other outcomes determine
the complexity or enlarge the binary group. In the cyclic case the
surviving arity-eight core, the indicator of the Reed--Muller code
\(RM(1,3)\) up to the equivalences in Section~\ref{s:short:groups}, permits
group enlargement. In the standard Klein-four
and internal tetrahedral cases the whole set can admit a common real
or affine basis, respectively.

The short account below gives the classification argument and selected
proof mechanisms. The full paper follows the shared references. Numbers prefixed by F refer to its statements and proofs. Recognition takes the signature
set as input, whereas the signature set is fixed in each evaluation problem.

\section{Classification results and reductions}
\label{s:short:tools}
In the disequality-edge model, the equality constraint of an ordinary edge is realized
by two \(X\)-edges joined through an \(I\)-vertex, since \(XIX=I\).
Conversely, an \(X\)-edge becomes an \(X\)-vertex in the ordinary model,
and an \(I\)-vertex can be suppressed. Inserting and suppressing these
binary vertices, using \(X^2=I\), gives
\[
 \Holant(\cF,X)\equivT\KHolant(\cF\cup\{I\}).
\]
Thus proving Theorem~\ref{s:thm:p1-deq2-dichotomy} amounts to classifying
disequality-edge problems whose signature sets contain \(I\).
For counting problems \(A,B\), write \(A\leT B\) if \(A\) has a
polynomial-time Turing reduction to \(B\), and \(A\equivT B\) if both
reductions hold. A finite set \(\cS\) is \emph{simulated} by
\(\cF\) if \(\KHolant(\cF\cup\cS)\leT\KHolant(\cF)\).
We use \emph{realized} for a signature obtained by a gadget.

The second model equivalence is a holographic transformation
\cite{Valiant2008}. Set
\[
 K=\frac1{\sqrt2}\begin{pmatrix}1&1\\i&-i\end{pmatrix},
 \qquad K^{\mathsf T}K=X,
 \qquad K\cF=\{K^{\otimes\arity(f)}f:f\in\cF\}.
\]
Transforming every signature by \(K\) makes contraction along an ordinary
edge equal to the original contraction against \(X\). The partition
function is unchanged, so
\[
 \KHolant(\cF)\equivT\Holant(K\cF).
\]

The tractable classes used here are the affine class \(\cA\)
\cite{CaiLuXia2009,CaiLuXia2014}, the product class \(\cP\)
\cite{DyerGoldbergJerrum2009,CaiLuXia2014}, and the local-affine class
\(\cL\) \cite{CaiLuXia2018RealHolantc}. We use their standard
definitions from these papers. The fourth case is the tensor closure
\(\angles{\cT}\), where \(\cT\) consists of signatures of arity at most
two. Splitting these tensor products reduces evaluation to paths and cycles.

Holographic transformations explain the common-transformability
condition. For edge signature \(E\in\{I,X\}\) and
\(\cC\in\{\cA,\cP,\cL\}\), it requires a single invertible matrix \(M\)
such that
\[
 M^{\mathsf T}EM\in\cC,\qquad M^{-1}\cF\subseteq\cC.
\]
Here \(M^{-1}\) acts on every input of each signature. Transforming all
vertices by \(M^{-1}\) and all edges by \(M^{\mathsf T}EM\) preserves
the partition function. Both conditions put the entire network in the
same tractable class, using one common basis.
Write \(\ReTract_E\) for these three possibilities together with
\(\cF\subseteq\angles{\cT}\). Theorem~\ref{s:thm:p1-deq2-dichotomy}
uses \(E=I\) and Theorem~\ref{s:thm:p1-quaternary-kholant-dichotomy}
uses \(E=X\). Elsewhere \(\ReTract=\ReTract_I\).
The corresponding criteria in the two models satisfy
\[
\begin{aligned}
 \ReTract_X(\cF)&\ \Longleftrightarrow\ \ReTract(K\cF),\\
 \ReTract(\cF\cup\{X\})&\ \Longleftrightarrow\ \ReTract(K(\cF\cup\{I\})).
\end{aligned}
\]
The proofs are in \cref{subsec:tract-predicate} and
the discussion following \cref{lem:p1-equality-wire-equivalence}.

Both predicates use the zero and nullary preprocessing of \cref{subsec:tract-predicate}. In Theorem~\ref{s:thm:p1-deq2-dichotomy},
\(X\in\cF\cup\{X\}\) excludes the three additional cases of \(\Tract\).
In the equivalent disequality-edge model the set contains \(I\), so its
image under \(K\) contains \(K^{\otimes2}I=\diag(1,-1)\), which excludes
the same cases (\cref{lem:tract-X-anchor}).

A nonzero odd-arity signature, including a unary, invokes the known
classification and algorithms \cite{MengWangXiaZheng2025,GuanShaoShi2026}.
A nonzero singular binary has unary tensor factors. For a nonsingular binary \(B\), the transfer \(BX\)
has infinite projective order if no positive power is a scalar matrix.
Such a binary yields a rank-one binary by interpolation. For the signature sets containing \(I\), factor
extraction and the unary reductions therefore resolve these cases
(\cref{lem:binary-interpolation} and
\cref{lem:common-unary-exit}).

The factor-extraction theorem of \cite{MengWangXiaZheng2025} lets us
simulate the factors of an available tensor product unless all supports
lie on the same side of the middle Hamming layer (including that layer).
Weighted Eulerian-orientation results classify this exception
\cite{MengWangXia2025EO,GuanShaoShi2026}. Since \(I(00)=I(11)=1\),
a set containing \(I\) has support on both sides. We retain all
previous signatures when passing to the set containing the simulated
factors. The problems remain Turing-equivalent, hardness transfers to
the preceding problem, and the same common basis establishes the
tractable condition for the preceding set.
We call the set factor-closed once all its nonzero positive-arity
irreducible tensor factors have been retained.

Two further reductions will be used in the proof of Theorem~\ref{s:thm:p1-quaternary-kholant-dichotomy}.
A generalized equality is supported exactly on the all-zero and all-one
strings, with both values nonzero. At arity at least four it permits
the unary reduction. An equality anchor is a realized quaternary for
which one change of basis turns the edge into binary equality and the
quaternary into four-ary equality, up to nonzero scalars. It likewise
permits a nonzero unary to be simulated. These reductions classify the
whole set (\cref{lem:pure-ge} and
\cref{lem:equality-anchor}).

\section{From an eight-vertex signature to a signature-set dichotomy}
\label{s:short:eight}
The dichotomy for the eight-vertex model classifies \(\Holant(X\mid q)\)
for a single endpoint-nondegenerate eight-vertex signature \(q\)
\cite[arXiv version~2, Theorem~3.1]{CaiFu2023}. Its hard case gives
hardness for any set realizing \(q\). In the tractable cases we must
determine how \(q\) interacts with the rest of the set. We allow \(q\)
to be reducible here because later gadgets can produce such a signature
and supply its factors.

\begin{proposition}[Eight-vertex reduction]
\label{s:short:eight-reduction}
Let \(\cF\) be a finite factor-closed retained set directly realizing an
endpoint-nondegenerate eight-vertex signature \(q\). Then
\(\KHolant(\cF)\) is \(\SharpP\)-hard, or \(\Tract(K\cF)\) holds,
or the original \(q\) factors across two pairs of inputs. In the last
case both binary factors can be adjoined by reduction, and one is a
nonsingular weighted equality \(\diag(p,q')\), with \(pq'\ne0\).
\end{proposition}

To prove the proposition, we examine the three exceptions of the
single-signature theorem. Product transformability
leaves three structural possibilities: a generalized equality, a product
of two weighted equalities, or a dense even-parity form that realizes an
equality anchor. The generalized equality and the anchor yield a simulated unary as
explained in Section~\ref{s:short:tools}, and therefore classify the whole set. The product of two binaries
is the factorization outcome of the proposition. The classification of product forms uses the
support and the ranks of the three pair flattenings, so it identifies
factors of \(q\) itself.

The endpoint conditions reduce local-affine transformability to affine
transformability. In the affine case, a common diagonal change of basis
preserves the disequality edge up to scalar and puts \(q\) into a normal
form with support of size two, four, or eight. The diagonal change
preserves the two nonzero endpoints of \(q\), so factor extraction
remains available throughout this analysis.

Support two gives a generalized equality, so the same unary reduction
applies. For support four, constant assignments on the two input pairs
define logical bits with
coefficient matrix \(\left(\begin{smallmatrix}A&B\\C&D\end{smallmatrix}\right)\),
where \(AD=\pm BC\). If \(AD=-BC\), a two-copy gadget gives a generalized
equality. If \(AD=BC\), the normal form factors into two weighted
equalities. Undoing the diagonal change gives the required factorization
of the original \(q\).

On full even-parity support, the constructions give an equality anchor,
a nonzero rank-one binary, or a tensor product of nonsingular weighted equalities.
In the last case, extracting a factor and normalizing it to equality
allows a single-input flip, producing a quaternary with full odd-parity
support. A finite analysis of its contractions gives an equality anchor,
a nonzero rank-one binary, a binary of infinite projective transfer order,
or an endpoint-nondegenerate eight-vertex signature with support of size six.
The anchor invokes the equality-anchor reduction. A rank-one binary
supplies a unary factor, while an infinite-order binary first yields a
rank-one binary by interpolation. The support-six output lies outside
the three transformable classes and is hard by the dichotomy for the
eight-vertex model \cite[arXiv version~2, Theorem~3.1]{CaiFu2023}.
Thus each full-support case classifies the entire retained set.

The reductions transfer these conclusions back to the original set. The normal forms and finite gadget analysis,
including the exact computations and coverage arguments, are given in
\cref{thm:cert-interface-affine-q4-normal-form},
\cref{thm:cert-interface-affine-q4-gadget-exits}, and
\cref{thm:cert-interface-affine-q4-odd}.

In Theorem~\ref{s:thm:p1-quaternary-kholant-dichotomy}, irreducibility
excludes the factorization of \(q\). Its two nonzero endpoints give
support on both sides of the middle layer. \Cref{lem:tract-X-anchor} therefore reduces \(\Tract(K\cF)\) to
\(\ReTract(K\cF)\). An irreducible \(q\) of arity four also excludes
the tensor-product case. By the criterion correspondence in
Section~\ref{s:short:tools}, the three common-transformability cases give
the tractable side of the theorem (\cref{cor:p1-eight-vertex-reduced-interface}).

\section{Stabilizing a signature and its binary group}
\label{s:short:deck}
We return to Theorem~\ref{s:thm:p1-deq2-dichotomy}. By
Section~\ref{s:short:tools} we may work with disequality edges and
supplied \(I\). After factor extraction and the binary and odd-arity
reductions, choose a nonbinary irreducible \(f\) of minimum
arity \(n=\nu(\cF)\) in the retained factor-closed set \(\cF\).
Its arity is even. We follow the pair \((f,G(\cF))\), recording
\(\cF\) as part of the state because it determines the available gadgets.

For an ordered binary \(B\), define its transfer by \(T_B=BX\).
Series composition realizes \(BXC\), with \(T_{BXC}=T_BT_C\).
After the binary reductions, every nonzero realizable binary is
nonsingular with finite projective transfer order. We call such a binary
\emph{safe}. Put
\[
 G(\cF)=\{[BX]:B\text{ is a safe ordered binary directly realizable over }\cF\}.
\]
Here \([BX]\) is the transfer class of \(B\), identifying matrices
up to nonzero scalar. This \emph{complete transfer group} includes the
classes of all realizable binary gadgets, of arbitrary size.
It is finite. In a fixed coefficient field \(L\), an eigenvalue ratio
of order \(r\) satisfies \(\varphi(r)\le2[L:\mathbb Q]\), bounding
the possible orders. Powers supply inverses. The faithful adjoint
representation and Burnside's finite-exponent theorem give finiteness
\cite{Burnside1905}. The finite subgroups of \(\PGL_2(\mathbb C)\)
then give a uniform order bound in this field (\cref{lem:bounded-torsion}).

Fix realizable ordered representatives \(B_g\) for \(g\in G\).
A \(G\)-matching signature is
\[
 c\bigotimes_{e\in M}B_{g_e}^{(e)},\qquad c\ne0,
\]
where \(M\) is an oriented perfect matching of the inputs and the
superscript places the binary on that ordered pair. The scalar absorbs
the choice of representatives. To test \(f\), attach a representative,
in either orientation, to any subset of its inputs through \(X\)-edges.
Write \(f^\delta\) for the resulting dressed signature. Close one ordered pair \((i,j)\)
with another representative \(B_h^\omega\), where
\(B_h^{\mathrm{id}}=B_h\) and \(B_h^{\mathrm{rev}}=B_h^{\mathsf T}\).
The \emph{matching deck} is the finite family
\[
 \mathfrak D_G(f)=\{0\}\ \cup\
 \left\{\rho\!\left(\partial_{(i,j)}^{XB_h^\omega X}f^\delta\right)
 :\delta,i\ne j,h,\omega,\rho\right\},
\]
which collects the outputs obtained by applying the allowed binary
dressings to a single copy of \(f\) and then closing one self-loop.
Here \(\partial_{(i,j)}^Q u\) sums the two inputs against the kernel
\(Q=XB_h^\omega X\), with one \(X\)-edge at each end of the closing binary. The permutation \(\rho\) ranges over residual input orders. Exactly one pair
is closed and at least two inputs remain. Every nonzero card is a direct
gadget output.

\begin{definition}[Stable state]
\label{s:short:stable}
Let \(\cF\) be finite, factor-closed, and contain \(I\). Let
\(f\in\cF\) be a minimum-arity nonbinary irreducible signature, and
let \(G=G(\cF)\) be finite with fixed realizable binary representatives.
The state \((\cF,G,f)\) is \emph{deck-stable} if every nonzero card
of \(\mathfrak D_G(f)\) is a \(G\)-matching signature and all its
factors are retained in \(\cF\).
\end{definition}

Thus stability says that every allowed self-loop, including the specified
input dressings, gives zero or a tensor product of already available
binaries. Including input dressings makes this condition apply to the changes of
binary action used in later comparisons.

\begin{lemma}[Stabilization, \cref{lem:minimum-core-deck}]
\label{s:short:stabilization}
Let \(\cF_0\ni I\) be a retained factor-closed finite set, and let
\(f\in\cF_0\) be an even nonbinary irreducible signature of minimum arity.
The reductions either establish that \(\KHolant(\cF_0)\) is
polynomial-time computable or \(\SharpP\)-hard, or yield a finite retained
factor-closed extension \(\cF'\supseteq\cF_0\) with
\(\KHolant(\cF')\equivT\KHolant(\cF_0)\) such that either its minimum
nonbinary arity is smaller or \((\cF',G(\cF'),f)\) is deck-stable.
The procedure can strictly enlarge the complete binary group before
reaching either extension.
\end{lemma}

Process each card and retain its irreducible factors using factor
extraction. A nonbinary factor has arity at most \(n-2\).
Odd-arity factors and unsafe binaries invoke the established complexity
classifications. Otherwise there are two distinct forms of progress:
\begin{enumerate}[label=\textnormal{(\roman*)}]
\item \emph{Lowering the arity.} An even nonbinary factor gives an
extension with smaller minimum nonbinary arity. Choose a minimum-arity
irreducible in this extension as the new core signature and restart
stabilization.
\item \emph{Enlarging the binary group.} Retain the new binary factors.
If a safe binary \(B\) has \([BX]\notin G\), the complete transfer group
strictly enlarges. Keep the same core \(f\) and regenerate its deck
using the larger group, which supplies additional contraction tests.
\end{enumerate}
Each extension retains the previous signatures, including \(I\), and
preserves Turing equivalence. Binary factors whose transfer classes
already lie in \(G\) do not by themselves give strict group growth.

Iterating these steps must eventually reach a stable pair unless the
complexity is determined first. The two forms of progress are finite
for different reasons.
Each replacement of the core strictly decreases its arity. Since arity
is an integer bounded below, only finitely many such replacements occur.
Each group enlargement is strict, and all previous binary resources
remain available even when the core changes. In the original
coordinates, gadget evaluation and normalized factor extraction stay
in one fixed number field. The complete groups therefore form an
increasing chain with the uniform order bound established above, so
only finitely many strict enlargements occur.

For a fixed core and group, the deck and its factorization trees are
finite, with each distinct card of this deck processed once. After the last core replacement
and group enlargement, this finite processing either determines the
complexity or leaves every nonzero card as a tensor product of retained
binaries from the current group. This is the stable condition, relative
to the final retained set and its complete group.

\section{Marked groups and their possible growth}
\label{s:short:marked}
The complete binary group determines the contractions available in the
matching argument. Its abstract type is cyclic, dihedral, or one of
the Platonic groups \(A_4,S_4,A_5\), by the classification of finite
subgroups of \(\PGL_2(\mathbb C)\). The supplied \(I\) puts \(x=[X]\)
in \(G\). Set \(Z=\diag(1,-1)\), \(Y=ZX\), and \(j=[Z]\).
Figure~\ref{s:fig:complete-group-growth} shows the possible strict
inclusions \(G<G^+\). Its superscripts distinguish groups of the
same abstract type that supply different contractions in our coordinates.

\begin{figure}[p]
\centering
\begin{tikzpicture}[
  x=1cm,y=1cm,
  state/.style={
    draw,rounded corners=2pt,align=center,text width=3.05cm,
    minimum height=1.05cm,inner sep=2pt,font=\normalsize
  },
  strictedge/.style={
    -{Stealth[length=2.2mm]},semithick,rounded corners=4pt,
    preaction={draw=white,line width=3.5pt,solid,-}
  },
  reroute/.style={
    -{Stealth[length=2.2mm]},semithick,rounded corners=4pt,
    preaction={draw=white,line width=3.5pt,solid,-}
  },
  edgetag/.style={
    circle,draw,fill=white,minimum size=3.5mm,
    inner sep=.35pt,font=\bfseries
  }
]
  \node[state] (c2) at (0,0)
    {$\CycX{2}=\langle x\rangle$};

  \node[state] (vex) at (-5.4,-2.8)
    {exotic Klein-four\\$\KleinEx=\DihXone{4}$};
  \node[state] (cyc) at (0,-2.8)
    {proper cyclic\\$\CycX{N}$, $N>2$ even};
  \node[state] (vp) at (5.4,-2.8)
    {standard Pauli\\$\KleinStd=\DihZ{4}$\\$=\DihY{4}$};

  \node[state] (aa) at (-3.5,-5.8)
    {$\DihXone{2m}$\\$m\ge4$ even};
  \node[state] (mm) at (3.5,-5.8)
    {$\DihZ{2m}$\\$m\ge3$};

  \node[state] (te) at (-5.4,-9.0)
    {$\TetExt$\\external $A_4$};
  \node[state] (bb) at (0,-9.0)
    {$\DihY{2m}$ or\\$\DihXzero{2m}$\\$m\ge3$};
  \node[state] (tt) at (5.4,-9.0)
    {$\TetInt$\\internal $A_4$};

  \node[state] (op) at (-5.4,-12.2)
    {$\OctTT$ or $\OctDT$\\$S_4$};
  \node[state] (oo) at (1.8,-12.2)
    {$\OctDD$ or $\OctTD$\\$S_4$};
  \node[state] (ii) at (5.8,-12.2)
    {$A_5$\\$\IcoZero,\IcoOne$};

  \draw[strictedge] (c2.south) -- (cyc.north);
  \draw[strictedge] (c2.south west) -- (vex.north east);
  \draw[strictedge] (c2.south east) -- (vp.north west);

  \draw[strictedge]
    (c2.south west)
    -- (-2.15,-1.05)
    -- (-2.15,-3.75)
    -- (-1.25,-3.75)
    -- (-1.25,-8.05)
    -- ([xshift=-8mm]bb.north);

  \draw[strictedge]
    (c2.east)
    -- (7.55,0)
    -- (7.55,-5.8)
    -- (mm.east);

  \draw[strictedge]
    (cyc.south west) -- (aa.north east);

  \draw[reroute]
    (cyc.south)
    -- (0,-8.05)
    -- ([xshift=-3mm]bb.north);

  \draw[strictedge]
    (vex.south east) -- (aa.north west);

  \draw[reroute]
    (vex.south east)
    -- (-3.0,-4.20)
    -- (-1.60,-4.20)
    -- (-1.60,-8.05)
    -- ([xshift=-12mm]bb.north);

  \draw[strictedge]
    (vp.south west)
    -- (3.0,-4.45)
    -- (1.60,-4.45)
    -- (1.60,-8.05)
    -- ([xshift=10mm]bb.north);

  \draw[strictedge]
    (vp.south west) -- (mm.north east);

  \draw[reroute]
    (aa.south)
    -- (-3.5,-7.75)
    -- (bb.north west);

  \draw[strictedge] (vex.south) -- (te.north);
  \draw[strictedge] (vp.south) -- (tt.north);

  \draw[strictedge]
    (bb.south west) -- (op.north east);

  \draw[strictedge]
    (bb.south east) -- (oo.north west);

  \draw[strictedge]
    (mm.south) -- ([xshift=6mm]oo.north);

  \draw[reroute]
    (te.south) -- (op.north);

  \draw[strictedge]
    (tt.south west) -- ([xshift=12mm]oo.north);

  \draw[strictedge]
    (tt.south east) -- (ii.north west);

  \node[edgetag] at (.38,-1.38) {\DEQedgenumber{1}};
  \node[edgetag] at (-3.05,-1.20) {\DEQedgenumber{2}};
  \node[edgetag] at (2.75,-.95) {\DEQedgenumber{3}};
  \node[edgetag] at (-2.55,-2.65) {\DEQedgenumber{4}};
  \node[edgetag] at (7.93,-3.25) {\DEQedgenumber{5}};
  \node[edgetag] at (-2.20,-4.70) {\DEQedgenumber{6}};
  \node[edgetag] at (.38,-5.55) {\DEQedgenumber{7}};
  \node[edgetag] at (-4.10,-4.45) {\DEQedgenumber{8}};
  \node[edgetag] at (-1.95,-6.75) {\DEQedgenumber{9}};
  \node[edgetag] at (1.95,-6.75) {\DEQedgenumber{10}};
  \node[edgetag] at (4.10,-4.35) {\DEQedgenumber{11}};
  \node[edgetag] at (-3.90,-7.00) {\DEQedgenumber{12}};
  \node[edgetag] at (-5.78,-6.95) {\DEQedgenumber{13}};
  \node[edgetag] at (5.78,-6.95) {\DEQedgenumber{14}};
  \node[edgetag] at (-2.80,-10.18) {\DEQedgenumber{15}};
  \node[edgetag] at (.72,-10.20) {\DEQedgenumber{16}};
  \node[edgetag] at (2.75,-7.55) {\DEQedgenumber{17}};
  \node[edgetag] at (-5.78,-10.62) {\DEQedgenumber{18}};
  \node[edgetag] at (3.82,-10.63) {\DEQedgenumber{19}};
  \node[edgetag] at (6.55,-10.55) {\DEQedgenumber{20}};
\end{tikzpicture}

\vspace{.15em}
\begingroup
\normalsize
\setlength{\tabcolsep}{2.5pt}
\renewcommand{\arraystretch}{1.08}

\def\gref#1{%
  \tikz[baseline=(n.base)]
  \node[
    draw,circle,minimum size=3.5mm,
    inner sep=.35pt,font=\bfseries
  ] (n) {\DEQedgenumber{#1}};%
}

\begin{tabularx}{\linewidth}{
  @{}r>{\raggedright\arraybackslash}X@{\hspace{7pt}}
  r>{\raggedright\arraybackslash}X@{}
}
\gref{1}&$\CycX{2}\to \CycX{N}$
&\gref{11}&$\KleinStd\to\DihZ{2m}$, $2\mid m$.\\

\gref{2}&$\CycX{2}\to \KleinEx$
&\gref{12}&$\DihXone{2m}\to\DihXzero{4m}$,
  $j$ enters $G^+$.\\

\gref{3}&$\CycX{2}\to \KleinStd$
&\gref{13}&$\KleinEx\to\TetExt$.\\

\gref{4}&$\CycX{2}\to\DihY{2m}$, $m$ odd.
&\gref{14}&$\KleinStd\to\TetInt$.\\

\gref{5}&$\CycX{2}\to\DihZ{2m}$, $m$ odd.
&\gref{15}&$\DihY{8}\to\OctTT$ or
  $\DihXzero{8}\to\OctDT$.\\

\gref{6}&$\CycX{N}\to\DihXone{2m}$, $N\mid m$.
&\gref{16}&$\DihY{8}\to\OctDD$ or
  $\DihXzero{8}\to\OctDD$.\\

\gref{7}&$\CycX{N}\to\DihXzero{2m}$,
  $N\mid m$, $j$ enters $G^+$.
&\gref{17}&$\DihZ{8}\to\OctDD$ or $\OctTD$.\\

\gref{8}&$\KleinEx\to\DihXone{2m}$,
  $m\equiv2\pmod4$.
&\gref{18}&$\TetExt\to\OctDT$,
  $j$ enters $G^+$.\\

\gref{9}&$\KleinEx\to\DihXzero{2m}$,
  $4\mid m$, $j$ enters $G^+$.
&\gref{19}&$\TetInt\to\OctDD$.\\

\gref{10}&$\KleinStd\to\DihY{2m}$ or $\DihXzero{2m}$, $2\mid m$.
&\gref{20}&$\TetInt\to\IcoZero$ or $\IcoOne$.\\
\end{tabularx}

\medskip
\noindent Here \(j\) enters \(G^+\) when \(j\notin G\) and \(j\in G^+\).
The parameters satisfy strict inclusion, with \(m\ge3\) for dihedral
targets and even \(m\ge4\) for \(\DihXone{2m},\DihXzero{2m}\).
\endgroup

\caption{Possible strict growth of the complete transfer group with the
supplied equality fixed. Numbered arrows indicate subgroup inclusions
between marked forms, with their parameter conditions listed below the
graph. Divisibility steps within a family are suppressed. Grouped nodes
collect distinct marked forms at the same level of the diagram.}
\label{s:fig:complete-group-growth}
\end{figure}
\clearpage

Reversing the inputs of a binary replaces
its transfer \(T\) by \(XT^{\mathsf T}X\), with
\([XT^{\mathsf T}X]=j[T]^{-1}j\). Since reversal preserves
realizability and \(G\) is closed under inverses, \(jGj=G\). Thus
\[
 \widehat G=\langle G,j\rangle,\qquad [\widehat G:G]\le2,
 \qquad \langle x,j\rangle\cong V_4.
\]
The positions of \(x\) and \(j\) in \(\widehat G\), including whether \(j\in G\),
distinguish the marked forms.

\paragraph{Cyclic and dihedral groups.}
Write \(C_N\) for the abstract cyclic group of order \(N\) and
\(D_{2m}\) for the abstract dihedral group of order \(2m\).
Let \(\mu_m\) be the \(m\)-th roots of unity and put
\(D(t)=\diag(1,t)\), \(C(t)=(1+t)I+i(1-t)Y\), and
\(A(t)=(1+t)I+(1-t)X\). In the original coordinates the cyclic
transfer group is \(\CycX{N}=\{[A(t)]:t\in\mu_N\}\), with \(N\)
even. Its smallest form is \(\CycX{2}=\langle x\rangle\). For \(N>2\),
\(j\notin G\) and \(\widehat G\cong D_{2N}\). The dihedral forms are
\[
\begin{aligned}
 \DihZ{2m}&=\{[D(t)],[XD(t)]:t\in\mu_m\},&
 \DihY{2m}&=\{[C(t)],[XC(t)]:t\in\mu_m\},\\
 \DihXzero{2m}&=\{[A(t)],[ZA(t)]:t\in\mu_m\},&
 \DihXone{2m}&=\{[A(t)],[ZA(\eta t)]:t\in\mu_m\}.
\end{aligned}
\]
Here \(m\ge3\), with even \(m\) in the last two forms and
\(\eta\in\mu_{2m}\setminus\mu_m\). The subscript gives the group order.
The superscripts \(Z,Y,X\) specify the rotation axis, with rotation
matrices in \(\Span\{I,Z\}\), \(\Span\{I,Y\}\), or \(\Span\{I,X\}\).
The labels \((X,0)\) and \((X,1)\) distinguish the unshifted and shifted
reflection cosets. In the former \(j\in G\), whereas in the latter
\(j\notin G\). At the Klein-four boundary the two groups are
isomorphic to \(V_4\). In our coordinates the standard Pauli subgroup is
\(\KleinStd=\DihZ{4}=\DihY{4}\) and contains \(j\). The exotic subgroup
is \(\KleinEx=\DihXone{4}\) and does not contain \(j\).

\paragraph{Platonic groups.}
The tetrahedral forms are \(\TetInt\), with \(G=\widehat G\cong A_4\), and
\(\TetExt\), with \(G\cong A_4\) normal in \(\widehat G\cong S_4\).
Their superscripts record whether \(j\) is internal or external to \(G\).
For the four octahedral forms \(\OctDD,\OctTT,\OctDT,\OctTD\),
one has \(G=\widehat G\cong S_4\). Each superscript gives the cycle types of
\((x,j)\), where \(d\) denotes a double transposition and \(t\) a
transposition. Finally \(\IcoZero,\IcoOne\) are two coordinate forms
of the same abstract marked icosahedral type, with \(G=\widehat G\cong A_5\).
Explicit matrices and the complete marked classification appear in
\cref{thm:marked-finite-group-routing}.

If a retained extension gives \(G<G^+\), a path in
Figure~\ref{s:fig:complete-group-growth} leads from the marked type of
\(G\) to that of \(G^+\), with divisibility steps within a family
suppressed (\cref{prop:finite-marked-successor-routing}).
The binaries realizable over the enlarged set determine which inclusions
occur and which additional contractions can be used.

\section{Reconstruction from stable contractions}
\label{s:short:mechanism}
Fix a stable pair with core \(f\) and retained set \(\cF\). Every allowed one-copy self-loop on
\(f\) gives zero or a tensor product of retained binaries. We use these
contractions to recover enough of \(f\) to determine the complexity of \(\KHolant(\cF)\) or
to contradict the irreducibility of \(f\). The difficulty is that each
nonzero card comes with its own pairing of the remaining inputs and its
own binary factors on those pairs.

\paragraph{Comparing contractions at a fixed pair.}
Fix a pair \(p\) of inputs of \(f\). Closing \(p\) with different binary
kernels gives cards on the same remaining inputs. We can therefore
compare their values at each assignment. Write
\(\Phi_p^f(Q)=\partial_p^Q f\) for the card obtained with kernel \(Q\).
For example, suppose three available kernels satisfy \(Q_3=aQ_1+bQ_2\),
and write \(C_i=\Phi_p^f(Q_i)\). Linearity of contraction gives
\[
 C_3(z)=aC_1(z)+bC_2(z)
\]
for every assignment \(z\) to the remaining inputs. Wherever \(C_1\)
and \(C_2\) both vanish, \(C_3\) must vanish as well. A proposed choice
of matchings and factors that violates this identity at even one
assignment is ruled out.

\paragraph{Testing a disagreement at low arity.}
Suppose two cards pair one input with different partners. To test
whether these pairings are compatible, we keep the inputs that exhibit
the difference and close pairs elsewhere. On disjoint pairs these
closures commute with the contraction at \(p\). The two smaller cards
can thus be compared as contractions of the same smaller signature.
Each closure sums coefficients, so a choice of kernel can erase the
very entries that distinguish the pairings. The group-specific argument
must exhibit available kernels that keep the relevant entries nonzero.
If the difference survives down to bounded arity, the low-arity
classification can rule it out or yield a further reduction.

Once the matching is fixed, extracting a common binary on a chosen
edge requires the factors there to be proportional across the nonzero
cards. To test this, fix the inputs outside that edge and list the four
values obtained by varying its two inputs. If two such vectors, from
different cards or different assignments, were not proportional, no
single binary could factor out on that edge. A gadget testing this
possibility must preserve entries that witness the failure of
proportionality.

\paragraph{Lifting a common factor to \(f\).}
Suppose the available kernels span \(\Mat_2(\mathbb C)\). Choose four
that form a basis and close the same ordered pair \(p\) with each.
Write \(f_{ab}(z)=f(a,b,z)\), ordering the inputs with \(p\) first
and letting the remaining assignment \(z\) vary. The coordinate kernel
selecting \(ab\) is a linear combination of the four basis kernels,
so \(f_{ab}\) is the same linear combination of their cards. If the nonzero
cards all share a factor \(P\) on the same nonempty block of remaining
inputs, each slice has the form \(P\otimes r_{ab}\), allowing
\(r_{ab}=0\). Putting the slices back together gives \(f=P\otimes r\)
up to input order, where the four \(r_{ab}\) are the slices of \(r\)
obtained by fixing \(p\) to \(ab\). Since \(r\) contains the inputs of
\(p\), this factorization contradicts irreducibility. If the kernels
recover only combinations of slices, the group analysis must determine
what those combinations imply.

Some comparisons require additional gadgets, including gadgets built
from two copies of \(f\). Stability applies to the prescribed one-copy
cards. For each additional gadget we first extract the factors of its
output and apply the reductions of Section~\ref{s:short:deck}
(\cref{lem:matching-synthesis-measure} and
\cref{lem:token-coverage}).

\section{How the group cases use the common mechanism}
\label{s:short:groups}
\paragraph{Dihedral groups.}
For \(G=\DihZ{2m}\), \(m\ge3\), the allowed kernels at a fixed pair
give two families of cards, up to nonzero scalars,
\[
 E(t)=f_{00}+t f_{11},\qquad O(t)=f_{10}+t f_{01},
 \qquad t\in\mu_m.
\]
Within each family, the support constraints force one matching.
On that matching the coefficient minors have degree at most two in
\(t\). They vanish at three distinct phases, forcing all but at most
one binary factor to be fixed up to scalar (\cref{lem:app-nd-sector-ruling}). For arity at least eight, suppose
two cards have different matchings or differ on two edges of the same
matching. Monomial closures outside the deleted pair preserve this
disagreement in an arity-six contraction of \(f\). Its two inherited
cards are incompatible with a matching product. Yet its first closure
is a stable card, and further monomial closures preserve the matching
product whenever the result is nonzero. This rules out the disagreement (\cref{lem:app-nd-localization} and \cref{lem:app-nd-bounded-consumer}).
If two cards differ on edge \(e\), a third differing from the first
on another edge would differ from the second on both. Thus the
nonzero cards share a product on all but one common edge (any edge
suffices if all cards are proportional).
Two phase evaluations in each family recover the four slices, so this
product factors out of \(f\) by Section~\ref{s:short:mechanism}.

At arity six a direct support and pencil analysis forces factorization
(\cref{lem:nd-six-prime-base}). At arity four invertible
dressings give an eight-vertex signature to which
Section~\ref{s:short:eight} applies. The forms \(\DihY{2m}\) and
\(\DihXzero{2m}\) reduce to the standard analysis by transformations of
the whole network, including its edges. For \(\DihXone{2m}\), once the
support matching is fixed, weighted two-copy gadgets compare the ratios
of coefficients obtained by changing the assignment on one edge.
After the lower-arity outputs have been processed, the resulting
identities force these ratios to be independent of the other edge
assignments. This yields the required factorization. \Cref{sec:normalized-dihedral} gives the transformations and
the low-arity classifications.

\paragraph{Cyclic groups.}
In the original coordinates \(G=\CycX{N}\) and its transfer span is
\(\Span\{I,X\}\). Write \(P=\left(\begin{smallmatrix}1&1\\1&-1\end{smallmatrix}\right)\).
Since \(P^{-1}A(t)P=2D(t)\), the group in Walsh coordinates is
\(P^{-1}GP=C_N^{\rm std}=\{[D(t)]:t\in\mu_N\}\).
For \(\widetilde f=(P^{-1})^{\otimes n}f\), the closures test the
sum and difference of two slices. The nonzero cards are matching
products whose overlaps constrain both support and coefficients.
After the binary and arity-four reductions, a surviving core forces
\(N=2\), excluding the \(N>2\) cyclic node of
Figure~\ref{s:fig:complete-group-growth}. In the original coordinates
\[
 f\sim\RMcore:=\one[RM(1,3)],\qquad G=\CycX{2}=\{[I],[X]\}.
\]
Its Walsh-coordinate group is \(C_2^{\rm std}=\{[I],[Z]\}\).
The code \(RM(1,3)\) consists of the eight-bit evaluation vectors of
affine linear functions on \(\Ftwo^3\). Here \(\sim\) allows a nonzero
scalar, input permutation, and realizable input \(X\)-dressings.
To identify it, the matching slices reconstruct a doubly-even self-dual
support, with a unique weight-four word through each triple of
coordinates. This forces the Reed--Muller support. Comparisons on its
affine planes then fix the coefficients up to a sign character, removed
by the allowed dressings. Applying the coefficient constraint after
every available single-input dressing also forces the complete group
to be \(\CycX{2}\) in the original coordinates (\cref{lem:rm-support-classification}
and \cref{lem:rm-safe-group-collapse}).

This core supplies a reduction that strictly enlarges the binary group. Once \(\RMcore\) is available and the binary and arity-four
cases have been resolved, the limited-appearance reduction of Shao and
Cai \cite[full version, Section~8.2, Lemmas~8.9--8.10]{ShaoCai2020}
permits adjoining the Bell signatures. In the original coordinates the binaries
\(Z,Y\) supply transfer classes \([ZX]=[Y]\) and \([YX]=[Z]\).
Their matrix representatives lie outside \(\Span\{I,X\}\). The complete group therefore strictly enlarges
and stabilization resumes. This is the shortcut stated in \cref{rem:rm-binary-enlargement-shortcut}, which also describes
the stronger internal structural classification retained in the full proof.

\paragraph{Klein-four groups.}
The standard Pauli subgroup \(\KleinStd\) supplies the four Bell kernels \(I,X,Z,Y\).
For two signatures of arity at least five, suppose their corresponding
cards are linearly dependent for every input pair and each of the four
kernels. The relation may depend on the pair and the kernel, and zero
cards are allowed. Then the signatures themselves are linearly dependent
(\cref{lem:v4-pauli-projective-determination}). Apply
this to a retained signature and its complex conjugate. If the
signature is not proportional to a real signature, some pair card
must have the same obstruction. Descent reaches arity four, where
classification resolves the set or enlarges the group, unless an
earlier reduction already gives progress. If all retained signatures
are proportional to real signatures in the same coordinates, the real
Holant dichotomy \cite{ShaoCai2020} applies.

For \(\KleinEx\), the common matching method
forces a proper factor. The arity-six classification supplies the
base case. At higher arity the four cards at a fixed deleted pair share
one residual matching. If an edge has a fixed binary factor across
these cards, reconstruction extracts it from \(f\). Otherwise the
four labels on each edge vary by a permutation. Crossing two edges
compares these permutations and creates a support incompatible with
a matching product. The arity-eight boundary needs two such contraction
choices. From arity ten one suffices. This reduces the branch to the
arity-four analysis. \Cref{thm:standard-v4-relative}
and \cref{thm:exotic-v4-relative-rigidity} give the two conclusions.

\paragraph{Platonic groups.}
Fix the deleted pair \(p\) and regard each available kernel as a
projective point. For three collinear points, a linear relation among
their kernel representatives gives the same relation among their cards
under \(\Phi_p^f\). This constrains the possible matching products as in
Section~\ref{s:short:mechanism}. A common factor can lift to \(f\).
In the octahedral and icosahedral cases a surviving disagreement can
be localized to arity six. Finite classifications then give a quaternary
signature covered by Section~\ref{s:short:eight} or a binary whose transfer
class enlarges the group (\cref{sec:platonic-decks}).
For \(\TetExt\), the gadgets determine the complexity, strictly enlarge
the group, or expose a proper factor of \(f\), contradicting its
irreducibility in the last case (\cref{thm:external-a4-deck-rigidity}).

For \(\TetInt\) the whole retained set may admit a common affine basis.
Work in the coordinates of the internal tetrahedral form. If a remaining
irreducible is nonaffine, the shortening argument exposes a factor
or produces a smaller signature that is still nonaffine in those coordinates
(\cref{lem:a4rel-physical-code-shortening}). Iteration
would reach the already excluded cases of arity at most eight. Applying this argument
in the same basis to every remaining irreducible gives the common affine
presentation.
\section{Completing the dichotomy}
For a set containing \(I\), stabilization brings each unresolved case
to a stable pair. The group arguments compare its contractions to
reconstruct a proper factor, contradicting irreducibility, or reduce a
disagreement to low arity. Together with the structural exceptions
and the reductions that enlarge the binary group, these arguments
classify stable pairs (\cref{thm:matching-deck-classification}).
Factor extraction and stabilization then give the dichotomy for sets
containing \(I\) (\cref{thm:p1-eq2-kholant-dichotomy}).

The model equivalence and criterion correspondence in
Section~\ref{s:short:tools} transfer this dichotomy to \(\Holant(\cF,X)\),
giving Theorem~\ref{s:thm:p1-deq2-dichotomy}. The recognition procedure
of \cref{thm:predicate-algorithms} tests the four cases
of \(\ReTract(\cF\cup\{X\})\) on the exact algebraic input, so the
criterion is decidable.

\section*{AI methodology statement}
\addcontentsline{toc}{section}{AI methodology statement}
We used OpenAI's GPT-5.6 Sol to assist with mathematical exploration and proof development for the arguments in this paper. In Sections~\ref{sec:normalized-dihedral}--\ref{sec:platonic-decks}, we used AI to carry out most of the analysis within each group case, including developing proofs and writing code for the computations used in those proofs. All companion verification code was generated by OpenAI Codex and subsequently reviewed by the authors.

We used OpenAI's GPT-6 Astra to improve the writing and presentation, and support the \href{https://github.com/liuchliuch/holant-binary-disequality-lean}{\pinkfoldmark\nobreak\hspace{0.2em}Lean formalization}. We formalized the proof in Lean and incorporated two corrections to the original arXiv version (\href{https://arxiv.org/abs/2609.00219v1}{arXiv:2609.00219v1}). We corrected the signs in the two mixed-mark identities in Equation~(132) and restored a common nonzero scalar factor in Lemma~9.19 and its application in the proof of Lemma~9.20. The corrected statements were verified in Lean. The authors independently checked the mathematical claims, proofs, and computational results and take full responsibility for the correctness and originality of the paper.
\label{s:end:main}

\clearpage
\phantomsection
\addcontentsline{toc}{section}{References}

\newcommand{\etalchar}[1]{$^{#1}$}

\clearpage
\setcounter{section}{0}
\setcounter{equation}{0}
\setcounter{figure}{0}
\setcounter{table}{0}
\setcounter{footnote}{0}
\setcounter{page}{1}
\renewcommand{\thepage}{F-\arabic{page}}
\renewcommand{\thesection}{F.\arabic{section}}
\renewcommand{\theHsection}{full.\arabic{section}}
\renewcommand{\theequation}{F.\arabic{equation}}
\renewcommand{\theHequation}{full.\arabic{equation}}
\renewcommand{\thefigure}{F.\arabic{figure}}
\renewcommand{\theHfigure}{full.\arabic{figure}}
\renewcommand{\thetable}{F.\arabic{table}}
\renewcommand{\theHtable}{full.\arabic{table}}
\pdfbookmark[0]{Full paper}{deq.fullpaper}
\section*{Full paper}

\section{Introduction}
\label{sec:p1-introduction}

Many counting problems ask for a global sum determined by local
constraints: each assignment receives the product of its local weights,
and these products are summed over all assignments.  A complexity
dichotomy identifies the choices of local signatures that admit
polynomial-time evaluation and proves \(\SharpP\)-hardness for every
remaining choice.  On the Boolean domain, this program has a long history
in counting constraint satisfaction problems (\(\#\mathrm{CSP}\)).
Creignou and Hermann classified the unweighted case
\cite{CreignouHermann1996}.  Later dichotomies treated nonnegative rational
weights \cite{DyerGoldbergJerrum2009}, rational weights of mixed signs
\cite{BulatovDyerGoldbergJalseniusRicherby2009}, and complex weights
\cite{CaiLuXia2014}.  Complex weights permit cancellations that yield
tractable cases absent from the nonnegative setting
\cite{CaiLuXia2014}, making the algebra of the weights part of the
classification.
Over arbitrary finite domains, counting-CSP dichotomies have likewise
been established for unweighted constraints
\cite{Bulatov2013,DyerRicherby2013}, nonnegative rational weights
\cite{BulatovDyerGoldbergJalseniusJerrumRicherby2012}, nonnegative algebraic
weights \cite{CaiChenLu2016}, and algebraic complex weights
\cite{CaiChen2017}.

The Holant framework brings the availability of local constraints into
this classification problem \cite{CaiLuXia2009}.  A counting-CSP instance
can be represented by a bipartite graph whose variable vertices carry
equality constraints, enforcing agreement among all occurrences of a
variable.  Holant allows the available vertex signatures to be prescribed
without supplying these equalities.  This accommodates problems such as
counting perfect matchings, where each vertex requires exactly one
incident edge to be selected.  It also makes the construction of useful
local signatures a part of the classification proof.  Holographic
transformations, originating in Valiant's algorithms \cite{Valiant2008},
give another source of tractability: a simultaneous change of basis can
expose structure in the local signatures while preserving the value of the
total sum.

Boolean Holant dichotomies have addressed both restrictions on the
signatures and assumptions about auxiliary signatures.  With no prescribed
auxiliary signatures, classifications were obtained for symmetric
real-valued signatures \cite{HuangLu2016}, symmetric complex-valued
signatures \cite{CaiGuoWilliams2016}, arbitrary nonnegative signatures
\cite{LinWang2018}, and arbitrary real-valued signatures
\cite{ShaoCai2020}.  Here symmetry means that a signature depends only on
the number of ones in its input.  Another line of work supplies unary
signatures: \(\Holant^*\) supplies all of them
\cite{CaiLuXia2011HolantStar}, \(\Holant^+\) supplies four distinguished
ones \cite{Backens2017}, and \(\Holant^c\) supplies the two pinnings
\(\DeltaZero=[1,0]^{\mathsf T}\) and
\(\DeltaOne=[0,1]^{\mathsf T}\).
The last model was classified for symmetric complex signatures
\cite{CaiHuangLu2012Holantc}, arbitrary real signatures
\cite{CaiLuXia2018RealHolantc}, and algebraic complex signatures
\cite{Backens2021,Backens2025}.  The real-valued classification introduced
the local-affine tractable family \(\cL\), which also occurs here.

Meng, Wang, Xia, and Zheng proved a tensor-factor extraction theorem and
classified complex-valued Boolean Holant when a nonzero odd-arity signature
is present \cite[Theorems~29 and~30]{MengWangXiaZheng2025}. They collected
seven cases in condition~(PC). For two cases defined by support, the
weighted Eulerian-orientation results of \cite{MengWangXia2025EO} gave
an \(\mathrm{FP}^{\mathrm{NP}}\) upper bound. Guan, Shao, and Shi subsequently
gave polynomial-time algorithms for these cases \cite{GuanShaoShi2026}.
Taken together, the seven cases give the most general known sufficient condition
for tractability of algebraic complex-valued Boolean Holant. We denote
their disjunction by \(\Tract\).
Even-arity signatures present a different resource constraint.  Every
gadget built entirely from them has an even number of external inputs,
because each internal edge consumes two inputs.  Such gadgets cannot
supply a unary signature, so binary signatures are the smallest
positive-arity objects available for further constructions.  We study
the case in which binary disequality, requiring two bits to differ, is
supplied.

There is also a precedent for prescribing disequality in counting CSP.
In \(\#\mathrm{CSP}_d\), each variable occurs a number of times divisible
by \(d\).  Cai, Fu, and Shao classified
\(\#\mathrm{CSP}_d(\Neq_2,\cF)\) for complex signatures
\cite[full version, Section~5.1 and Theorem~5.3]{CaiFuShao2020Entanglement}.
Shao and Cai used its hardness conclusion in their real Holant dichotomy
\cite[full version, Theorem~2.24]{ShaoCai2020}.

Prescribing binary disequality is useful for three further reasons.
First, ordinary edges enforce equality and the supplied binary signature
enforces disequality. Equivalently, we may use disequality edges and supply
binary equality as a vertex signature (\cref{lem:p1-equality-wire-equivalence}).
Together these constraints restrict the realizable binaries in Xia's finite-group
framework \cite[Section~4 and Table~2]{Xia2026FrameworkV1} to the cyclic,
dihedral, Klein-four, and Platonic forms in \cref{thm:marked-finite-group-routing}.
Second, switching between the two formulations only requires inserting or
removing binary vertices. Every signature in the original set keeps its
entries, so we can use whichever edge model suits a construction or
tractable case. Third, the resulting
classification can be applied to other Holant problems whenever a
reduction makes binary disequality available.
When disequality cannot be obtained, that failure gives a structural
restriction on the remaining problems.

\paragraph{The problem and main results.}
A Boolean signature of arity \(n\) is a function
\(f:\{0,1\}^n\to\mathbb C\).  In a Holant instance, a graph vertex
of degree \(n\) carries such a signature, with its incident edges ordered.
For a fixed finite set \(\cF\) of signatures, the problem
\(\Holant(\cF)\) evaluates
\[
  \sum_{\sigma:E\to\{0,1\}}
  \prod_{v\in V}f_v\bigl(\sigma|_{E(v)}\bigr),
  \qquad f_v\in\cF.
\]
Here \(E(v)\) is the ordered list of edge incidences at \(v\).
\Cref{subsec:holant-definitions} gives the conventions for loops and
multiple edges.  The graph and its vertex labels form the input. The set
\(\cF\) is fixed.  Write
\[
 X=(\Neq_2)=\begin{pmatrix}0&1\\1&0\end{pmatrix},
 \qquad I=(\Eq_2)=\begin{pmatrix}1&0\\0&1\end{pmatrix}.
\]
Our problem is \(\Holant(\cF,X):=\Holant(\cF\cup\{X\})\).
Thus any number of degree-two vertices may enforce opposite values on
their two incident edges.

With \(X\) supplied, \cref{lem:tract-X-anchor} rules out three of the
seven cases of \(\Tract\). We write \(\ReTract\) for the disjunction of the
remaining four. The signature set lies in the tensor closure \(\angles{\cT}\)
of signatures of arity at most two, or it is transformable into the affine
\(\cA\), product \(\cP\), or local-affine \(\cL\) class. In each transformability case,
one change of basis must place the edge signature of the model and
all the vertex signatures in that class.
The classes and common transformability
are defined in \cref{subsec:standard-classes,def:common-transformability},
and the four cases are collected in \cref{subsec:tract-predicate}.
Here ``Re'' refers both
to the restriction of the general tractability criterion \(\Tract\)
to these four tractable cases and to their correspondence with the
tractable cases in the real-valued Holant dichotomy of Shao and
Cai~\cite{ShaoCai2020}.

\begin{theorem}[Binary-disequality Holant dichotomy]
\label{thm:p1-deq2-dichotomy}
For every finite set of algebraic complex-valued Boolean signatures \(\cF\),
the problem \(\Holant(\cF,X)\) is polynomial-time computable when
\(\ReTract(\cF\cup\{X\})\) holds and is \(\SharpP\)-hard otherwise.
The tractability criterion is decidable from exact algebraic input.
\end{theorem}

With disequality edges, the proof also classifies sets containing an
endpoint-nondegenerate irreducible signature of arity four, without
requiring a supplied equality signature. Following Meng, Wang, Xia, and Zheng
\cite{MengWangXiaZheng2025}, write
\(\KHolant(\cF):=\Holant(\Neq_2\mid\cF)\) for the bipartite Holant
problem with disequality on the left and signatures from \(\cF\) on the
right. Each connection between signature inputs thus enforces disequality.
In the following theorem, transformability in \(\ReTract\) uses this edge \(X\).

A quaternary signature \(q\) is
\emph{endpoint-nondegenerate} when
\[
 q({\mathbf{x}})=0\quad\text{for odd }\wt({\mathbf{x}}),\qquad q(0000)q(1111)\ne0.
\]
Here \(\wt({\mathbf{x}})\) counts the ones in \({\mathbf{x}}\). The eight
even-weight inputs are the allowed configurations of an eight-vertex
signature. The two endpoint entries are required to be nonzero.
A nonzero positive-arity signature is \emph{irreducible} if it has no factorization
\(a\otimes b\) on two nonempty blocks of inputs, allowing a permutation of inputs and a nonzero scalar.

\begin{theorem}[Quaternary \(\KHolant\) dichotomy]
\label{thm:p1-quaternary-kholant-dichotomy}
Let \(\cF\) be a finite set of algebraic complex-valued Boolean signatures.
Suppose that \(\cF\) contains an endpoint-nondegenerate irreducible
quaternary signature.  Then \(\KHolant(\cF)\) is polynomial-time computable
if \(\ReTract(\cF)\) holds and is \(\SharpP\)-hard otherwise.  The
tractability criterion is decidable from exact algebraic input.
\end{theorem}

Cai and Fu's eight-vertex classification determines the complexity for
a single eight-vertex signature \(q\) with nonzero endpoints
\cite[arXiv version~2, Theorem~3.1]{CaiFu2023}. If \(q\) alone is tractable,
its interaction with the remaining signatures can still make the problem hard.
\Cref{sec:p1-eight-vertex} treats these interactions. Consequently, whenever
a later reduction realizes an endpoint-nondegenerate irreducible eight-vertex
signature, \cref{thm:p1-quaternary-kholant-dichotomy} classifies the entire available set.

\subsection{Proof strategy}
\label{subsec:intro-proof-strategy}

Two approaches to Boolean Holant dichotomies motivate our proof.
Shao and Cai use arity descent in the real-valued setting, reducing a
signature outside the relevant tensor-product class to a smaller such
signature or a low-arity case \cite[Sections~3--4]{ShaoCai2020}.
Their orthogonality and norm arguments rely on real-valuedness, or
the corresponding conjugation symmetry after a holographic
transformation.  Arbitrary complex signatures lack this guarantee,
so the descent needs another way to control cancellations.
Xia's approach organizes the remaining problems by their finite
binary groups \cite[Sections~4--5]{Xia2026FrameworkV1}.
This specifies the binary constructions available in each case, but
leaves open what they force about higher-arity signatures.

We combine the two approaches in the disequality-edge formulation, where
binary equality is available by \cref{lem:p1-equality-wire-equivalence}.
After the preliminary reductions in \cref{sec:p1-shared-terminals}
(see also \cite{CaiFuXia2018SixVertex,CaiFuShao2020ARS}), choose a nonbinary
irreducible signature \(f\) of minimum arity in the current set \(\cF\).
The complete transfer group \(G\) of \(\cF\) consists of the projective
classes \([BX]\) as \(B\) ranges over all nonsingular ordered binary
signatures directly realizable over \(\cF\), with transfer matrices identified
up to nonzero scalar. Close a pair of inputs
of \(f\) using such a binary. By \cref{lem:factor-saturation}, we can retain the resulting signature's
irreducible factors by reduction. An unresolved nonbinary factor has smaller
arity than \(f\) and continues the descent.

A contraction that produces only binary factors can also advance the proof.
If a retained factor has transfer class outside \(G\), the complete group
of the enlarged set strictly contains \(G\). The factor is available by
reduction, so gadgets over the enlarged set can use binaries that were not
directly realizable over \(\cF\). We can now test the same \(f\) with
additional contractions. \Cref{fig:complete-group-growth} shows the
possible strict inclusions. Which one occurs depends on the binaries
realizable over the enlarged set.

If neither the minimum arity nor the group changes and the reductions have
not settled the complexity, processing the prescribed contractions leaves
every nonzero output as a product of retained binaries on a perfect matching
of the remaining inputs. We call the contraction outputs \emph{cards}.
Together with zero, the cards obtained from one copy of \(f\) by attaching
available binaries to selected inputs and closing one pair form its
\emph{matching deck}. When every nonzero card has such a decomposition
with its factors retained, the state is \emph{deck-stable}
(\cref{def:matching-deck,def:deck-stable-state}).
The group-order bound, arity descent, and finite processing within each
stage give the descent measure of \cref{lem:matching-synthesis-measure}.
\Cref{lem:token-coverage} verifies descent in all four group branches.
Since each extension retains the preceding
set, hardness transfers back along the reductions, and a common tractable
basis restricts to that set (\cref{lem:downward-inheritance}).

\subsection{Structural arguments in the group cases}
\label{subsec:intro-group-arguments}

The availability of both \(\Eq_2\) and \(\Neq_2\) restricts the
forms of the finite binary groups in the given basis.
In the disequality-edge formulation, equality supplies a fixed
involution in the group, while reversal of binary inputs imposes
an additional symmetry.  Together these constraints refine the
abstract group classification into the specific forms of
\cref{thm:marked-finite-group-routing}.  These forms give explicit
restrictions on the supports and coefficients of the cards.

Fix a pair of inputs of \(f\). Each nonzero card obtained by closing this
pair has a matching decomposition, but two cards may pair the same remaining
input with different partners. Even when their matchings agree, extracting
a common binary factor on an edge requires the factors there to be proportional
across the cards being compared. The proportionality scalar can be absorbed
into the rest of each product. The group arguments use further contractions
to test a disagreement at lower arity or use compatible factors to recover
a factor of \(f\), as explained in \cref{subsec:shared-matching-methods}.

\paragraph{Dihedral groups.}
The group binaries constrain the supports and coefficient ratios
of the cards, but these constraints must agree across contractions.
Superposing two matchings gives alternating cycles. \Cref{sec:normalized-dihedral}
uses these cycles to localize matching disagreements to six inputs. In the standard case, excluding these
disagreements forces a common factor, which two phase evaluations
recover in the original signature. This contradicts irreducibility.
The weighted dihedral case requires additional support and
coefficient reconstruction, including Laurent interpolation, to obtain
the same reduction to arity four.

\paragraph{Cyclic groups.}
Fewer independent contractions are available in this case.
After the arity-four cases are resolved or yield group enlargement,
\cref{sec:proper-deck} identifies the surviving minimum-arity signature,
up to input permutations, bit flips and a nonzero scalar, with the
indicator of the length-eight Reed--Muller code \(RM(1,3)\). The proof
reconstructs its support and coefficients by comparing contractions across
input pairs. This signature then realizes sums over affine fibres
(\cref{lem:rm-radon-accessible}).  These sums test the supports and
coefficients of the rest of the set.  When the tests produce no
further reduction, they place the whole set and the edge in the
affine class in one basis.

\paragraph{Klein-four groups.}
In the standard Pauli form, compare two signatures of the same arity
above four by contracting each input pair with each of the four binary
Bell signatures. If the corresponding cards are linearly dependent in
every comparison, the signatures are linearly dependent. \Cref{sec:v4-deck}
applies this test to each signature and its complex conjugate to detect
whether the signature can be made real by rescaling. A failure descends to a quaternary signature with the
same obstruction, whose analysis resolves the complexity or produces
further progress.  If every signature can be made real, the real
Holant dichotomy applies to the whole set.  The other Klein-four
form requires a direct factorization argument.  Its six-input
classification forces agreement among residual factors, which the
crossed contractions then use to obtain factors at higher arities.
These factors exclude a surviving nonbinary irreducible signature.

\paragraph{Platonic groups.}
Fix a pair of inputs and regard the available contraction kernels as
projective points. Linear relations among their representatives pass to
the corresponding cards and constrain their matchings. \Cref{sec:platonic-decks}
uses these relations to obtain a common factor or localize differing
matchings to six inputs. A common factor contradicts irreducibility,
while localized disagreements give small signatures to which the
results for arity-four signatures or binary-group analysis can be applied.  The
octahedral and icosahedral analyses
then produce quaternary signatures or new binaries. The internal
tetrahedral case requires testing whether the whole set is affine
in one basis.  Stabilizer-code constructions shorten a non-affine
signature while preserving its failure of affine membership, unless
a complexity conclusion, factor, or larger group is obtained first.
Once the descent reaches at most eight inputs, the low-arity
classifications rule out a surviving non-affine obstruction.  This
completes the induction to a common affine basis for the remaining
set and its edge.

The low-arity classifications that require exact finite enumeration,
together with their completeness arguments, are stated in
Appendix~\ref{sec:exact-finite-certificates}. \Cref{proofsec:matching-deck-synthesis}
assembles the four branches, and \cref{sec:p1-synthesis} transfers their
conclusions to the original problem. The recognition procedure tests the criterion
on the input set itself (\cref{lem:tract-X-anchor,thm:predicate-algorithms}).
\section{Preliminaries}
\label{sec:preliminaries}

We fix the common algebraic, tensor, and reduction conventions, define the
signature classes, and state the imported dichotomies used across the proof.
The constructive recognition and retained-reduction arguments are in
\cref{sec:p1-shared-terminals}.  Specialized tools are introduced where
their applications begin.

\subsection{Foundational algebraic conventions}
\label{subsec:foundational-algebra}

\subsubsection{Sets, maps, and exact fields}
\label{subsubsec:set-field-exact}

We use \([n]=\{1,\ldots,n\}\), \([0]=\varnothing\), the indicator
\(\one[\mathsf P]\), cardinality \(|S|\), complement \(S^c\) in the
stated ambient set, disjoint union \(\sqcup\), restriction
\(f|_S\), and fibres \(\pi^{-1}(y)=\{x:\pi(x)=y\}\).
Tuples and binary strings are ordered lexicographically.

For a field \(L\), write \(L^\times=L\setminus\{0\}\),
\(\Mat_{m\times n}(L)\) for the \(m\)-by-\(n\) matrices over \(L\),
and \(\Mat_n(L)=\Mat_{n\times n}(L)\).
We use \(\mathbb Z_m=\mathbb Z/m\mathbb Z\) and the binary field
\(\Ftwo=\mathbb Z_2\).  The function \(\varphi(m)\) counts the residue
classes modulo \(m\) coprime to \(m\).

We write \(\AlgNums\subseteq\mathbb C\) for the field of algebraic numbers.
We write \(\overline L\) for an algebraic closure, \([L':L]\) for a
finite extension degree, and \(L(\alpha)\) for the generated field.
An \(L\)-conjugate of \(\alpha\) is its image under an embedding
\(L(\alpha)\hookrightarrow\overline L\) fixing \(L\). \(\bar z\) denotes
complex conjugation.  Exact algebraic values are represented by a minimal
polynomial and a rational isolating region selecting the intended complex
root \cite{Cohen1993}.

\subsubsection{Linear, polynomial, and tensor conventions}
\label{subsubsec:linear-polynomial-tensor}

All vector spaces are finite-dimensional unless stated otherwise.  We use
\(\ker T\), \(\operatorname{im}T\), \(\rank T\), \(\Span_L S\),
\(U\oplus V\), \(W^\vee=\operatorname{Hom}_L(W,L)\), and \(W/U\)
with their usual meanings.
An affine subspace is a coset \({\mathbf{a}}+U\), with translation space \(U\).
For matrices, \(A^{\mathsf T}\) denotes transpose, \(\operatorname{tr}A\)
trace, \(I_n\) the identity, and \(\GL_n(L)\) the invertible matrices.
\(\GL(V)\) denotes the invertible linear maps on \(V\).
A pivot is a selected nonzero entry used in elimination or normalization.
An \(L\)-matrix algebra is an \(L\)-subspace of \(\Mat_n(L)\) closed
under multiplication. It is unital if it contains \(I_n\) and
transpose-stable if it is closed under transpose.  Matrix units \(E_{ij}\)
use the stated index order. Boolean indices give \(E_{00},E_{01},E_{10},E_{11}\).
We use the polynomial ring \(L[x_1,\ldots,x_n]\) and monomials
\({\mathbf{x}}^{\mathbf{u}}=\prod_jx_j^{u_j}\), \({\mathbf{u}}\in\mathbb Z_{\ge0}^n\).

With fixed bases, tensor products have coordinates
\[
 ({\mathbf{v}}_1\otimes\cdots\otimes {\mathbf{v}}_r)_{i_1,\ldots,i_r}
   =({\mathbf{v}}_1)_{i_1}\cdots({\mathbf{v}}_r)_{i_r}.
\]
Such a product is a simple tensor.  The map \(A^{\otimes r}\) applies
\(A\) to every factor.  A bipartite flattening groups selected tensor
coordinates into a row index and the complementary coordinates into a
column index. Its flattening rank is the resulting matrix rank.

\subsubsection{Groups and projective spaces}
\label{subsubsec:group-projective}

We use \(H\le G\) for subgroups, \(H\lhd G\) for normal subgroups,
\([G:H]\) for index, \(G/N\) for a quotient, and \(\langle S\rangle\)
for the subgroup generated by \(S\).  A group presentation is written
\(\langle S:\mathcal R\rangle\).  The order of an element is its least
positive identity power, or infinity. An involution has order two.
The exponent of a finite group is the least common multiple of its element
orders.  Bounded torsion means that one integer \(m\ge1\) satisfies
\(g^m=1\) for every element.  We use the center \(Z(G)\) and automorphism group
\(\operatorname{Aut}(G)\).
For a group action, \(Gx\) and \(G_x\) denote the orbit and stabilizer.
A linear representation is faithful when its homomorphism to \(\GL(V)\)
is injective.  An action of \(H\) by automorphisms of \(N\) gives
\(N\rtimes H\), with product
\((n,h)(n',h')=(n(h\cdot n'),hh')\).  The normalizer and centralizer are
\[
 N_G(H)=\{g:gHg^{-1}=H\},\qquad
 C_G(H)=\{g:gh=hg\text{ for every }h\in H\}.
\]
Write \(S_m,A_m,C_m,D_{2m}\), and \(V_4\) for the symmetric,
alternating, cyclic, dihedral, and Klein four groups.  Our convention is
\(D_{2m}=\langle r,s:r^m=s^2=1,\ srs=r^{-1}\rangle\), of order \(2m\):
\(\langle r\rangle\) consists of rotations and its other coset of
reflections.  For a finite set \(V\), \(\operatorname{Sym}(V)\) is its
permutation group. A double transposition is a product of two disjoint
transpositions.

For a finite-dimensional \(L\)-space \(V\), its projective space is
\(\mathbb P_L(V):=(V\setminus\{0\})/L^\times\). When \(L=\mathbb C\) or is
clear we write \(\mathbb P(V)\).  The class of \({\mathbf{v}}\ne0\) is \([{\mathbf{v}}]\), and
the line through distinct \([{\mathbf{u}}],[{\mathbf{v}}]\) is
\(\mathbb P(\Span_L\{{\mathbf{u}},{\mathbf{v}}\})\).  A projective subspace \(\mathbb P(U)\) has
dimension \(\dim U-1\).  An invertible linear map induces a projective
automorphism, and scalar multiples induce the same one.  We put
\[
 \PGL(V):=\GL(V)/(L^\times I_V),\qquad
 \PGL_n(L):=\GL_n(L)/(L^\times I_n).
\]
Here \(I_V\) is the identity linear map on \(V\).
The diagonal torus in \(\PGL_n(L)\) is the subgroup of projective classes of
nonsingular diagonal matrices.
For a nonsingular \(2\)-by-\(2\) matrix \(T\), the \emph{projective order}
of \([T]\in\PGL_2(\AlgNums)\) is the least \(m\ge1\) for which
\(T^m=\lambda I\) for some \(\lambda\in\AlgNums^\times\), and is infinite
if no such \(m\) exists.

\subsection{Complexity conventions}
\label{subsec:complexity-conventions}

We use polynomial-time Turing reductions \(\leT\), allowing adaptive exact
oracle queries. \(\equivT\) means reduction in both directions.  The
classes \(\FP\), \(\SharpP\), and \(\FP^{\mathsf{NP}}\) have their usual
meanings.  For algebraic-valued problems, \(\SharpP\)-hardness means that
every integer-valued \(\SharpP\) function reduces to the exact algebraic
oracle, including polynomial-time postprocessing of its answers.

Signature sets, their coefficient fields, and fixed gadgets are fixed
independently of the input network.  Thus polynomial time measures the
network's encoding length. Exact answers use the chosen effective number
field.  Recognition of a tractable signature set is a separate task whose
input consists of finite exact signature tables.  We prove decidability
for that task without an encoding-size complexity bound.

\subsection{Signatures, tensors, and Holant instances}
\label{subsec:holant-definitions}

We use the fields \(\AlgNums\) and \(\Ftwo\) fixed above.  For \(r\ge0\), an
arity-\(r\) Boolean signature is a function
\(f:\{0,1\}^r\to\AlgNums\), identified with its lexicographically ordered
column tensor.  Its \(r\) input positions are ordered, and this order is
part of the data.  A \emph{relation} is a \(\{0,1\}\)-valued
signature, identified with its support.  Arity zero, one, two, three, and four are called
nullary, unary, binary, ternary, and quaternary.  We write \(\arity(f)=r\) and
\(\supp(f)=\{{\mathbf{x}}:f({\mathbf{x}})\ne0\}\).  For a Boolean
string \({\mathbf{x}}\in\{0,1\}^r\), we also write
\(\supp({\mathbf{x}}):=\{j\in[r]:x_j=1\}\). The argument type distinguishes string
support from signature support.
Whole vectors are written in bold, while their scalar coordinates and
input-position labels are not. All binary-vector arithmetic is over \(\Ftwo\), with \(\oplus\) used when
bitwise addition should be explicit.  The symbols \(\zero^r\) and
\(\one^r\) denote the all-zero and all-one strings, respectively. At \(r=0\),
both denote the unique empty string.

For \(T\subseteq[r]\), \(\one_T\in\{0,1\}^r\) denotes the incidence
vector of \(T\). In particular \(\one_\varnothing=\zero^r\).  We write
\({\mathbf{e}}_j=\one_{\{j\}}\) for the \(j\)-th standard basis vector whenever the
ambient arity is clear.

We write \(\diag(a,b)\) for the diagonal matrix with entries \(a,b\).
For \(s\in\mathbb C\), put \(D(s):=\diag(1,s)\). This matrix is
invertible exactly when \(s\ne0\), and has algebraic entries when \(s\) is algebraic.

Following \cite[Section~2.3.1]{MengWangXiaZheng2025}, the signature
matrix \(M_f=M_{x_1\cdots x_\ell,x_{\ell+1}\cdots x_r}(f)\) has
rows indexed by the first \(\ell\) bits and columns by the remaining
bits, both in lexicographic order.  Its entries are the corresponding
values of \(f\). The subscripts may also specify any other ordered partition of the input positions. For a binary signature,
\(M_f=(f(a,b))_{a,b\in\{0,1\}}\). We also use \(f\) for this matrix
when the input order is fixed.

For \(\tau\in S_r\), define the reordering of inputs
\((\tau f)({\mathbf{x}})=f(x_{\tau^{-1}(1)},\ldots,x_{\tau^{-1}(r)})\).
With this convention, \(\tau(\sigma f)=(\sigma\circ\tau)f\).
On disjoint ordered inputs, \((f\otimes g)({\mathbf{x}},{\mathbf{y}})=f({\mathbf{x}})g({\mathbf{y}})\).
A nonzero positive-arity signature is \emph{irreducible} if, up to permutation of inputs and nonzero scalar, it cannot
be written as \(a\otimes b\) with both \(a\) and \(b\) of positive arity.
A nonzero positive-arity signature that is not irreducible is
\emph{reducible}. We use \emph{irreducible} as in
\cite[Section~2.3.4]{MengWangXiaZheng2025}. The positive-arity convention
excludes scalar factors.  Prime factorization into irreducible signatures is
unique up to these operations and factor order
\cite[Lemma~16]{MengWangXiaZheng2025}.  For any class \(\cC\) of
signatures, its tensor closure \(\angles{\cC}\) consists of tensor products
of members of \(\cC\), up to permutation of inputs and nonzero scalar.  By
convention, it also contains zero and the arity-zero empty product.
Thus \(\angles{\cC}\) denotes tensor closure, whereas \(\langle S\rangle\)
denotes the subgroup generated by \(S\).
A \emph{tensor factorization} partitions the ordered inputs into blocks and
writes the tensor, up to a nonzero scalar and permutation of inputs, as
a tensor product on those blocks.  For example, a \(2+2\) factorization of a quaternary has two binary
factors.

For \({\mathbf{a}}\in\{0,1\}^r\), define \(\delta_{\mathbf{a}}:\{0,1\}^r\to\AlgNums\) by
\[
  \delta_{\mathbf{a}}({\mathbf{x}})=\one[{\mathbf{x}}={\mathbf{a}}].
\]
We use the unary pinnings
\[
  \DeltaZero=[1,0]^{\mathsf T},
  \qquad
  \DeltaOne=[0,1]^{\mathsf T},
\]
which are point masses forcing the incident bit to \(0\) and \(1\),
respectively.  

For \(r\ge1\), a \emph{complementary-pair signature} (CP) has the form
\begin{equation*}
  c_0\delta_{\mathbf{a}}+c_1\delta_{\bar {\mathbf{a}}},
  \qquad c_0,c_1\in\AlgNums,\quad c_0c_1\ne0,
\end{equation*}
where \(\bar {\mathbf{a}}\) is the bitwise complement.  It is a
\emph{generalized equality} (GE) when \({\mathbf{a}}=\zero^r\): its support is
\(\{\zero^r,\one^r\}\), with the two nonzero weights still arbitrary.
Replacing each argument bit \(x_j\) by \(x_j\oplus a_j\)
puts a CP signature in GE form as a tensor identity.

Define the nullary equality/unit by \((\Eq_0)(\varnothing):=1\).  More generally,
for \(r\ge1\), define the arity-\(r\) equality signature by
\[
  (\Eq_r)(x_1,\ldots,x_r)=\one[x_1=\cdots=x_r].
\]

The binary equality and disequality tensors are
\[
  I=(\Eq_2)=
  \begin{pmatrix}1&0\\0&1\end{pmatrix},
  \qquad
  X=(\Neq_2)=
  \begin{pmatrix}0&1\\1&0\end{pmatrix}.
\]

A \emph{weighted equality} is a binary signature with matrix \(\diag(p,q)\),
where \(p,q\in\AlgNums\). It is nonsingular when \(pq\ne0\).
Binary equality is the case \(p=q=1\).

Graphs are finite multigraphs with ordered incidences at each vertex.
Loops and parallel edges are allowed, and a loop contributes two incidences
to the degree. A matching is a family of disjoint two-element subsets,
and is perfect when it covers its ground set.

A closed signature network \(\Omega\) has ordinary tensor vertices
\(v\), each assigned a signature \(f_v\) and a linearly ordered set
\(H_v\) of \(\arity(f_v)\) half-edges.  The sets \(H_v\) are
disjoint.  Edges pair these half-edges in a perfect matching.  Loops join
two half-edges of one vertex, and parallel edges are allowed.  Each half-edge
has its own Boolean variable.  An edge \(e=\{h,h'\}\) contributes
\(W_e(\tau_h,\tau_{h'})\), and a vertex contributes \(f_v(\tau|_{H_v})\).
The partition function sums their product over all half-edge assignments.
The ordinary model uses \(W_e=I\). The disequality-edge model uses \(W_e=X\).

Write \(Z_{\mathbf W}(\Omega)\) for this partition function with edge kernels
\(\mathbf W=(W_e)_e\), and abbreviate it to \(Z_W(\Omega)\) when every
edge has kernel \(W\).
An instance uses either the ordinary model \(Z_I\) or the
disequality-edge model \(Z_X\),
unless a displayed \(I\) or \(X\) is supplied as an explicit degree-two
signature/gadget.  The problems \(\Holant(\cF)\) and \(\KHolant(\cF)\) ask for
\(Z_I(\Omega)\) and \(Z_X(\Omega)\), respectively, on closed networks
over \(\cF\). When all edge kernels are \(I\), summing the
half-edge variables forces the two endpoints of each edge to agree and
recovers the usual one-bit-per-edge multigraph definition (including loops).
For signature sets \(\cG\) and \(\cF\), the bipartite problem
\(\Holant(\cG\mid\cF)\) is the ordinary (\(I\)-edge) half-edge problem on a
bipartite network: vertices in the left part are assigned signatures from
\(\cG\), and vertices in the right part are assigned signatures from
\(\cF\).  For \(X\)-edges no one-bit compression is made.  Subdivision inserts degree-two vertices on an edge.
Subdividing every \(X\)-edge by a degree-two \(X\)-vertex, with ordinary equality edges on its
two sides, gives the equivalent bipartite formulation
\begin{equation*}
  \KHolant(\cF)\equivT\Holant(X\mid\cF).
\end{equation*}
Throughout, comma-separated signature or signature-set arguments mean
adding their union. For example, \(\Holant(\cF,a,b)\) abbreviates
\(\Holant(\cF\cup\{a,b\})\).

We write
\[
  \Holant(\cF,X)=\Holant(\cF\cup\{X\})
\]
for the same ordinary equality-edge problem with the additional binary
signature \(X\) available as a vertex.
Subdividing every ordinary edge by an \(\Eq_2\)-vertex gives
\[
  \Holant(\cF)=\Holant(\Eq_2\mid\cF).
\]

An \emph{open gadget} is a finite signature network with ordered, labelled
dangling inputs. Fixing their bits and summing over the internal variables
gives its signature \(T_\Omega\). Let \(\partial\Omega\) be the ordered list
of unpaired vertex half-edges and \(E_{\rm int}(\Omega)\) the internal edges.
For this boundary convention,
\[
 T_\Omega({\mathbf z})=
 \sum_{\substack{\tau:\bigsqcup_v H_v\to\{0,1\}\\
                 \tau|_{\partial\Omega}={\mathbf z}}}
 \prod_v f_v(\tau|_{H_v})
 \prod_{\{h,h'\}\in E_{\rm int}(\Omega)}W_{\{h,h'\}}(\tau_h,\tau_{h'}).
\]
We also allow an isolated edge \(\mathsf E_W\) as a binary gadget with
signature \(W\). A supplied \(I\)- or \(X\)-signature instead labels a
tensor vertex. A closed component contributes its partition function,
and the empty network contributes the nullary unit \(1\).
Joining inputs contracts their bits against the edge kernel, as specified
in \cref{subsec:reduction-conventions}.
Appendix~\ref{app:general-boundaries} gives the equivalent general
representation with free edge endpoints and its boundary-coordinate
correction. Both representations use the same replacement signature
\(T_\Omega\).

\subsection{Zero and nullary preprocessing}
\label{subsec:zero-nullary}

An instance containing an identically zero signature has value zero.
A nullary signature contributes a scalar: delete its vertex and multiply
that scalar back into the final value. The empty residual instance has
value \(1\). After this preprocessing, we continue to write \(\cF\) for
the input signature set and assume it contains neither zero nor nullary
signatures. The tractable predicates are evaluated after this preprocessing.
Intermediate constructions may still produce zero or nullary signatures,
which are handled as specified in each argument.

\subsection{Holographic transformations and the
  \texorpdfstring{\(K\)}{K}-basis}
\label{subsec:holographic}

Let \(i\in\mathbb C\) satisfy \(i^2=-1\), and fix the \(K\)-coordinate matrix
\hypertarget{eq:K-matrix}{}
\begin{equation*}
  K=\frac1{\sqrt2}
  \begin{pmatrix}1&1\\ i&-i\end{pmatrix},
  \qquad
  K^{-1}=\frac1{\sqrt2}
  \begin{pmatrix}1&-i\\1&i\end{pmatrix}.
\end{equation*}

The transformations below rewrite every tensor and its incident edge
consistently, giving exact identities between partition functions.  Their
fixed algebraic coefficients lie in a finite extension of the input
coefficient field.  We follow the exact evaluation model of
\cref{subsec:complexity-conventions}.

For tensors or matrices, write \(A\doteq B\) when
\(A=\lambda B\) for some \(\lambda\in\AlgNums^\times\).
Throughout the complexity arguments, we ignore fixed nonzero scalar
factors of signatures.  Rescaling a signature multiplies an instance value
by a computable nonzero factor, so it does not change the complexity.
This convention does not discard relative coefficients in sums or
assignment-dependent weights.  

In \(\Holant(\cG\mid\cF)\), write each left-side signature \(g\)
of arity \(r\) as a row vector and each right-side signature \(f\) of
arity \(s\) as a column vector, both in lexicographic order. For
\(M\in\GL_2(\AlgNums)\), replace them by
\[
  gM^{\otimes r},\qquad (M^{-1})^{\otimes s}f,
\]
respectively. On each edge the two matrices cancel by \(MM^{-1}=I\),
so the transformed instance has the same value. Write
\(gM:=gM^{\otimes r}\), \(Mf:=M^{\otimes s}f\), and
\[
  \cG M:=\{gM:g\in\cG\},\qquad M\cF:=\{Mf:f\in\cF\}.
\]
Applying the transformation in both directions gives
\[
  \Holant(\cG\mid\cF)
  \equivT\Holant(\cG M\mid M^{-1}\cF).
\]
For binary \(f\), the right tensor action has signature matrix
\(M_{Mf}=M M_f M^{\mathsf T}\). For \cref{thm:p1-deq2-dichotomy},
\cref{lem:p1-equality-wire-equivalence} gives an equivalent disequality-edge
problem with \(I\) supplied. For a signature in ordinary
coordinates, write \(\widehat f=K^{-1}f\) and
\(\widehat{\cS}=K^{-1}\cS\).

For ordinary Holant, the left signature \((\Eq_2)\) has row vector
\((1,0,0,1)\). After right multiplication by \(M^{\otimes2}\), its
\((a,b)\)-entry is \(\sum_{x\in\Ftwo}M_{xa}M_{xb}=(M^{\mathsf T}M)_{ab}\).
Consequently,
\begin{equation*}
  \Holant(\Eq_2\mid\cF)
  =\Holant((\Eq_2)M\mid M^{-1}\cF),
  \qquad (\Eq_2)M=M^{\mathsf T}M.
\end{equation*}

Write
\begin{equation*}
  GO(X)
  =
  \{D\in\GL_2(\AlgNums):D^{\mathsf T}XD=\lambda X
    \text{ for some }\lambda\in\AlgNums^\times\}
\end{equation*}
for legal global transformations preserving the disequality edge
\(X\) projectively.  For a closed network, if \(D\Omega\) applies \(D\) at
every ordinary vertex and
\(D^{\mathsf T}XD=\lambda X\), then
\begin{equation}
  Z_X(D\Omega)=\lambda^{|E_{\rm int}(\Omega)|}Z_X(\Omega).
  \label{eq:GOX-global-scalar}
\end{equation}
Thus the two problems differ only by the known edge-count scalar. For \(m\ge1\), define
the group of \(m\)-th roots of unity by
\[
  \mu_m=\{z\in\AlgNums:z^m=1\}.
\]
An element \(\omega\in\mu_m\) is a \emph{primitive} \(m\)-th root of
unity when \(m\) is the least positive integer satisfying \(\omega^m=1\).
Since \(K^{\mathsf T}K=X\), putting
\(\widehat{\cF}=K^{-1}\cF\) gives
\begin{equation}
  \Holant(\cF)
  \equivT
  \Holant(X\mid\widehat{\cF})
  \equivT
  \KHolant(\widehat{\cF}).
  \label{eq:K-equivalence}
\end{equation}
For original-coordinate signatures and signature sets, the hats in
\cref{eq:K-equivalence} record passage to \(K\)-coordinates.  Once we work
with a signature set already in \(K\)-coordinates, we suppress
these hats and denote its member signatures simply by \(f\).  Unless stated
otherwise, signatures and signature sets appearing in a \(\KHolant\) problem
are understood to be in \(K\)-coordinates, with edge signature \(X\).

\subsection{Gadgets, interpolation, and factor reductions}
\label{subsec:reduction-conventions}

A \emph{direct gadget} over \(\cF\) is a finite \(X\)-edge network of
\(\cF\)-vertices, possibly using the edge \(\mathsf E_X\), with an
ordered exposed boundary. If its signature, defined in
\cref{subsec:holant-definitions}, is \(T_\Omega=\rho h\), where \(\rho\in\AlgNums^\times\) is a nonzero scalar, then
\begin{equation}
  \KHolant(\cF,h)
  \leT
  \KHolant(\cF).
  \label{eq:direct-gadget-reduction}
\end{equation}
A fixed gadget has constant size.  A gadget realizing \(\rho h\),
\(\rho\ne0\), implements the same Turing reduction after an instance with
\(N\) occurrences is rescaled by \(\rho^{-N}\).
We call such an \(h\) \emph{realizable} over \(\cF\).  More
generally, a finite set \(S\) is \emph{available by reduction} over
\(\cF\) if
\[
  \KHolant(\cF,S)\leT\KHolant(\cF).
\]
Joining two boundary inputs contracts their signature coordinates against
the default edge kernel \(W\), in the inherited input order. On an ordinary
boundary, this inserts a \(W\)-edge. An isolated \(\mathsf E_W\) splices
into one edge, and closing its two ends contributes
\(\sum_{a,b}W(a,b)^2=2\). Vertex replacement gives
\cref{eq:direct-gadget-reduction}.
The same rules for general free endpoints, and their compatibility with
\(GO(X)\)-transport, are proved in Appendix~\ref{app:general-boundaries}.
Every realizable signature is available by reduction.  Tensor factorization
and interpolation may also give reductions without realizing the resulting
signature by a gadget.  A gadget using such a signature therefore gives a
reduction to the preceding set, but need not be realizable over that set.
Algebraic operations on tensors may establish identities or select
contractions. Making their outputs available still requires a gadget or
a reduction. A \emph{slice} fixes selected argument values. Implementing
it by pinnings requires the corresponding signatures. In particular,
bit flips at individual inputs are used only when available or inside a
stated global transform. Interpolation recovers finitely many unknown
coefficients from oracle evaluations of a parameterized gadget family by
solving an invertible linear system. It supplies an oracle reduction.

Let \(f\) have arity \(r\).  For an ordered pair of distinct inputs
\(p=(i,j)\), a binary contraction kernel
\(C:\Ftwo^2\to\AlgNums\), and an assignment to the remaining inputs \({\mathbf{z}}\) indexed by
\([r]\setminus\{i,j\}\) in increasing input order, let
\({\mathbf{z}}^{i\leftarrow a,j\leftarrow b}\) be the length-\(r\) string obtained by
inserting \(a,b\) at inputs \(i,j\).  Define the pair contraction by
\begin{equation*}
  (\partial_p^C f)({\mathbf{z}})
  =
  \sum_{a,b\in\Ftwo}C(a,b)
  f\bigl({\mathbf{z}}^{i\leftarrow a,j\leftarrow b}\bigr).
\end{equation*}
In the notation of Shao--Cai
\cite[full version, Section~2.4]{ShaoCai2020}, their
\(\widehat\partial_{ij}\widehat f\) and \(\partial_{ij}f\) are,
respectively, our \(\partial_{(i,j)}^X\widehat f\) and
\(\partial_{(i,j)}^I f\).  Input order fixes the ordered pair \((i,j)\).
A \emph{pair card} of \(f\) is one such contraction
\(\partial_p^C f\). We shorten this to \emph{card} when the contraction
is specified.  A card is \emph{proper} if its arity is strictly smaller
than its parent's.  Iterated Pauli and exotic cards are defined where
contractions on lists of disjoint pairs are used.

For binary matrices
\(U,V\), the Frobenius pairing is bilinear:
\begin{equation*}
  \langle U,V\rangle_{\mathrm{Frob}}
  =\operatorname{tr}(U^{\mathsf T}V)
  =\sum_{a,b\in\Ftwo}U(a,b)V(a,b),
\end{equation*}
without conjugation. A \emph{binary chain} is a serial
composition of ordered binary gadgets joined by \(X\)-edges.  A
\emph{dressed pair card} \(\partial_p^Cf\) uses a closing binary
chain over the given signature set, attached through the two edge kernels.  Suppose
the closing subnetwork has ordered boundary tensor \(B\).  Its two
attachment \(X\)-edges complement both boundary bits, so
\begin{equation*}
  C(a,b)
  =\sum_{u,v\in\Ftwo}X(a,u)B(u,v)X(v,b)
  =(XBX)(a,b)
  =B(1-a,1-b).
\end{equation*}
Exchanging the two ordered boundary inputs replaces \(B\) by
\(B^{\mathsf T}\), and the same attachment therefore gives
\(XB^{\mathsf T}X\).  These formulas fix the input order used
below. Every dressed pair card so constructed is a gadget. Linear
combinations of cards are only analytic comparisons.

For every ordered binary signature \(B\), its transfer matrix is
\(T_B:=BX\).  If \(B\) is nonsingular, its transfer class is
\([T_B]=[BX]\in\PGL_2(\AlgNums)\).  A nonzero singular matrix has a
class only in the projective matrix space, and the zero matrix has no
projective class.

\subsection{Charge and Hamming layers}
\label{subsec:charge}

All charge terminology in this subsection is in \(K\)-coordinates.
For \({\mathbf{x}}\in\{0,1\}^r\), define its Hamming weight by
\(\wt({\mathbf{x}}):=|\{j:x_j=1\}|\) and its doubled charge by
\[
  q_r({\mathbf{x}})=2\wt({\mathbf{x}})-r.
\]
A \emph{Hamming layer} is the set of strings of one fixed Hamming weight.  A
string has even or odd parity according as its weight is even or odd, and a
signature has \emph{even-parity support} when it vanishes on every odd-parity string.
We distinguish \(\supp f\subseteq M\) from \(\supp f=M\). Coefficients
in a prescribed support set may vanish unless equality is required.
Define the Hamming half-spaces
\[
  HW_r^{\ge}=\{{\mathbf{x}}:q_r({\mathbf{x}})\ge0\},
  \quad
  HW_r^{\le}=\{{\mathbf{x}}:q_r({\mathbf{x}})\le0\},
  \quad
  HW_r^{=}=\{{\mathbf{x}}:q_r({\mathbf{x}})=0\},
\]
and define \(HW_r^{>},HW_r^{<}\) analogously.  A signature is
\(HW^{\ge}\) (respectively, \(HW^{\le}\)) when its support lies in
\(HW_r^{\ge}\) (respectively, \(HW_r^{\le}\)).  A signature set is \emph{common
one-sided} when all members are \(HW^{\ge}\) or all are \(HW^{\le}\).
Otherwise it is \emph{two-sided}, meaning that both charge signs occur
across the supports of its members. A signature is \emph{two-sided} when its own
support contains strings of both charge signs. Common one-sided support
is the exception to tensor-factor extraction in \cref{thm:external-factor}.
An arity-\(2d\) \emph{EO signature} is supported in \(HW_{2d}^=\).

In a closed \(X\)-edge instance, each contributing assignment has one
\(0\)-bit and one \(1\)-bit on every structural edge. Hence
\(\sum_v q_{\arity(f_v)}(\tau|_{H_v})=0\).
For an open gadget, the same cancellation equates the total vertex charge
with the boundary charge. The general-boundary calculation, including
the free-endpoint correction, is given in
Appendix~\ref{app:boundary-charge}.

\begin{remark}[Notation relative to prior work]
\label{rem:mwxz-notation}
We follow \cite[Sections~2.1--2.3]{MengWangXiaZheng2025} in writing
\(\widehat{\cF}=K^{-1}\cF\), \(HW^{\ge}\), \(HW^{\le}\), \(HW^=\),
and \emph{EO signature}.  We also use their equality and disequality notation \(\Eq_r\) and
\(\Neq_2\), with matrix aliases \(I=(\Eq_2)\) and \(X=(\Neq_2)\).
Subscripts in \(HW_r^\bullet\) record arity.

\end{remark}

\subsection{Binary codes and affine geometry}
\label{subsec:binary-affine-conventions}

The support reconstructions in
\cref{sec:normalized-dihedral,sec:proper-deck,sec:v4-deck} use binary
affine spaces and coefficient ratios to describe supports and phases.
We fix their shared Boolean and affine conventions here. The
\hyperref[subsubsec:symmetric-binary-spaces]{symmetric-space tools}
are introduced with the flat-Lagrangian networks.

\label{subsubsec:boolean-affine-tools}

For a finite coordinate set \(S\), put
\(\Ftwo^S:=\{{\mathbf{x}}:S\to\Ftwo\}\).  A \emph{binary code} of length \(n\) is a
subset of \(\Ftwo^n\). Its elements are \emph{codewords}.  It is linear when
it is a subspace and affine when it is a coset of a linear code.  The truth
table of a Boolean function lists its values in the fixed lexicographic input
order.  For a commutative ring \(R\), a Boolean polynomial over \(R\) is an
element of the quotient ring of \(R[x_1,\ldots,x_n]\) by the ideal generated
by the polynomials \(x_j^2-x_j\), \(j\in[n]\), and hence has a unique
multilinear representative.  Its degree is the largest number of
distinct variables in a nonzero monomial, and a mixed term contains at least
two distinct variables.  An affine-linear form is a constant plus a linear
form.  A \emph{quadratic phase} is \(i^{Q({\mathbf{x}})}\), where
\(Q:\Ftwo^n\to\mathbb Z_4\) has multilinear degree at most two and even
coefficients on mixed terms. We call \(Q\) its quadratic phase polynomial. A character of an additive group is a group
homomorphism into the multiplicative group of the coefficient field.  These
Boolean normal forms are instances of Boolean-lattice Möbius inversion
\cite[Chapters~1--2]{Carlet2021}.

For an affine subspace \({\mathbf{a}}+U\), \(U\) is its \emph{translation space}.
An affine map \({\mathbf{a}}+U\to {\mathbf{b}}+V\) has the form
\({\mathbf{a}}+{\mathbf{u}}\mapsto {\mathbf{b}}+T({\mathbf{u}})\) for a linear map \(T:U\to V\).  If an affine surjection
\(\pi:{\mathbf{a}}+U\twoheadrightarrow D\) onto an affine space \(D\) is fixed, an
\emph{affine section} is an
affine map \(\sigma:D\to {\mathbf{a}}+U\) with
\(\pi\circ\sigma=\mathrm{id}_D\).  For \(S\subseteq[n]\), in increasing
coordinate order, define coordinate projection and zero extension by
\[
 \pi_S:\Ftwo^n\to\Ftwo^S,\quad \pi_S({\mathbf{x}})=(x_i)_{i\in S},
 \qquad
 \iota_S:\Ftwo^S\to\Ftwo^n,\quad
 (\iota_S{\mathbf{y}})_i=\begin{cases}y_i,&i\in S,\\0,&i\notin S.\end{cases}
\]

For \(Q:\Ftwo^d\to R\), with \(R\) a commutative ring, its
\emph{Boolean Möbius expansion} is the unique multilinear expression
\[
 Q({\mathbf{x}})=\sum_{S\subseteq[d]}\widehat Q_S\prod_{j\in S}x_j,
 \qquad
 \widehat Q_S
 =\sum_{T\subseteq S}(-1)^{|S|-|T|}Q(\one_T).
\]
The coefficient table \((\widehat Q_S)_S\) is the Boolean Möbius transform.
The formula is interpreted in the stated coefficient ring, including
\(\Ftwo\) and \(\mathbb Z_4\).  Direct substitution is the Möbius-inversion
proof of uniqueness.  This additive transform is distinct from the
multiplicative Möbius inversion used later.

\subsection{Bell signatures}
\label{subsec:pauli-notation}

The Bell tensors form a common basis for binary contractions.
We fix their ordered signs.

We also fix the sign and skew matrices
\[
  Z=\begin{pmatrix}1&0\\0&-1\end{pmatrix},
  \qquad
  Y=ZX=\begin{pmatrix}0&1\\-1&0\end{pmatrix}.
\]

For \(d,\ell,u,v\in\Ftwo\), define the ordered Bell signature
\begin{equation*}
  B_{d,\ell}(u,v)
  =
  \one[u\oplus v=d](-1)^{\ell u}.
\end{equation*}
Write \(B_{\boldsymbol{\gamma}}=B_{d,\ell}\) for \({\boldsymbol{\gamma}}=(d,\ell)^{\mathsf T}\).  Input order matters
for \(B_{1,1}\), which is negated under reversal.

With truth-table order \(00,01,10,11\), these four signatures are
\begin{equation}
 \begin{aligned}
  B_{0,0}&=(1,0,0,1)=I,&
  B_{0,1}&=(1,0,0,-1)=Z,\\
  B_{1,0}&=(0,1,1,0)=X,&
  B_{1,1}&=(0,1,-1,0)=Y.
 \end{aligned}
 \label{eq:bell-matrix-dictionary}
\end{equation}

A \emph{Pauli binary} is a nonzero scalar multiple of one of the four
ordered Bell tensors \(B_{\boldsymbol{\gamma}}\), retaining its input order.
It is \emph{realizable} over a signature set when a direct gadget over that
set realizes it up to a nonzero scalar, as in
\cref{subsec:reduction-conventions}.

\begin{remark}[The Bell-signature dictionary of Shao--Cai]
The tuple in \cref{eq:bell-matrix-dictionary} is Shao--Cai's
\((=^+_2,=^-_2,\neq^+_2,\neq^-_2)\) \cite[Section~2]{ShaoCai2020}.
\(Y^{\mathsf T}=-Y\).  Their holographic matrix named \(Z\) is our \(K\),
whereas our \(Z\) is only the diagonal sign matrix.
\end{remark}

The Bell tables satisfy the
exact bilinear orthogonality identity
\[
 \langle B_{\boldsymbol{\gamma}},B_{{\boldsymbol{\gamma}}'}\rangle_{\mathrm{Frob}}
 =2\,\one[{\boldsymbol{\gamma}}={\boldsymbol{\gamma}}'].
\]

For an ordered binary tensor \(A\) and an ordered pair of inputs \((p,q)\),
the notation \(A^{(p,q)}\) places the first and second inputs of \(A\) on
inputs \(p\) and \(q\), respectively. For an ordered pair \(e=(p,q)\), we also write \(A^{(e)}\). For an even finite input set \(R\), a
\emph{Bell product} on \(R\) is a nonzero scalar multiple of
\[
 \bigotimes_{e=(p_e,q_e)\in M}B_{{\boldsymbol{\gamma}}_e}^{(p_e,q_e)},
 \qquad {\boldsymbol{\gamma}}_e\in\Ftwo^2,
\]
where \(M\) is a perfect matching of \(R\) and each pair of \(M\) has been
given the displayed orientation.  Reversing a pair carrying \(B_{1,1}\)
only changes the overall scalar sign, so this class is independent of the
chosen orientations.  For \(R=\varnothing\), a Bell product is a nonzero
nullary signature.

\subsection{Tractable signature sets and imported dichotomies}
\label{sec:predicate}

\subsubsection{The standard signature classes}
\label{subsec:standard-classes}

We include the zero signature in every class below.  The affine class
\(\cA\) consists of signatures
\begin{equation}
  f({\mathbf{x}})=\lambda\,\one[A{\mathbf{x}}={\mathbf{b}}]\,i^{Q({\mathbf{x}})},
  \label{eq:affine-class}
\end{equation}
where \(\lambda\in\AlgNums\setminus\{0\}\) and, for
\(r=\arity(f)\), there is an \(m\ge0\) with
\(A\in\Ftwo^{m\times r}\) and \({\mathbf{b}}\in\Ftwo^m\).  The function
\(Q:\{0,1\}^r\to\mathbb Z_4\), where
\(\mathbb Z_4=\mathbb Z/4\mathbb Z\), has a multilinear representative
\[
 Q({\mathbf{x}})=q_0+\sum_{j=1}^r q_jx_j
      +\sum_{1\le j<\ell\le r}q_{j\ell}x_jx_\ell\pmod4,
 \qquad q_{j\ell}\in\{0,2\},
\]
so it has degree at most two and even mixed coefficients.
Equivalently, the phase is a product of fourth-root phases of affine linear
forms interpreted as \(\{0,1\}\)-valued integers.
Membership in \(\cA\) always refers to the displayed computational
coordinates.

The product class \(\cP\) consists of zero and all signatures of the form
\(\lambda\prod_jh_j\), where \(0\ne\lambda\in\AlgNums\) and every factor is
either a nonzero unary signature \(u(x_i)\), a binary equality
\(I(x_i,x_j)\), or a binary disequality \(X(x_i,x_j)\) on designated
signature variables.  Empty products and repeated variables are allowed.
An inconsistent product denotes the zero signature.

Define the primitive eighth root and the phase matrix by
\[
 \zeta_8:=e^{\pi i/4}=\frac{1+i}{\sqrt2},
 \qquad T_2:=D(\zeta_8).
\]
The local-affine class \(\cL\) consists of the signatures \(f\) such that
\begin{equation*}
 \left(\bigotimes_{j=1}^{\arity(f)}T_2^{\alpha_j}\right)f\in\cA
 \qquad\text{for every }{\boldsymbol{\alpha}}\in\supp(f).
\end{equation*}

Let \(\cT\) be the set of all unary and binary signatures, and define the
signature class \(\cM\) by
\begin{equation*}
  \cM=\{f:f({\mathbf{x}})=0\text{ whenever }\wt({\mathbf{x}})>1\}.
\end{equation*}

The Reed--Muller arguments also use the following subclass of \(\cA\),
whose supports are affine translates of self-dual binary codes.
Let \(L\) be a linear subspace of \(\Ftwo^{2k}\).  Its orthogonal complement
under the standard dot product
\({\mathbf{x}}\mathbin{\cdot}{\mathbf{y}}:=\sum_{j=1}^{2k}x_jy_j\in\Ftwo\) is
\[
 L^\perp
 =
 \{{\mathbf{y}}\in\Ftwo^{2k}:{\mathbf{x}}\mathbin{\cdot}{\mathbf{y}}=0
     \text{ for every }{\mathbf{x}}\in L\}.
\]
Call \(L\) \emph{self-dual for the dot product} when \(L=L^\perp\), and an arity-\(2k\)
nonzero signature \emph{flat-Lagrangian} when it has the form
\begin{equation*}
  f({\mathbf{x}})=c\,\one[{\mathbf{x}}\in {\mathbf{a}}+L],\qquad
  0\ne c\in\AlgNums,\quad {\mathbf{a}}\in\Ftwo^{2k},\quad L=L^\perp.
\end{equation*}
Let \(\FLag\) denote the zero signature together with all flat-Lagrangian
signatures of every even arity.
\begin{definition}[Common transformability and \(K\)-presentation]
\label{def:common-transformability}
\label{def:common-K-presentation}
Let \(\cC\) be a signature class closed under nonzero scalars and permutations of inputs.  A
finite signature set \(\cF\) is \(\cC\)-transformable if one
\(M\in\GL_2(\AlgNums)\) satisfies
\begin{equation*}
 M^{\mathsf T}M\in\cC,\qquad M^{-1}\cF\subseteq\cC.
\end{equation*}
Equivalently, for \(\widehat{\cF}=K^{-1}\cF\), one matrix
\(U=K^{-1}M\) gives the common \(K\)-presentation
\begin{equation*}
 U^{\mathsf T}XU\in\cC,\qquad U^{-1}\widehat{\cF}\subseteq\cC.
\end{equation*}
We abbreviate this condition by
\begin{equation}
 \PresK(\cF;\cC)
 \quad\Longleftrightarrow\quad
 \exists U\in\GL_2(\AlgNums):
 \quad U^{\mathsf T}XU\in\cC,
 \qquad U^{-1}(K^{-1}\cF)\subseteq\cC.
 \label{eq:common-K-presentation-predicate}
\end{equation}
If \(\cF\) is already in \(K\)-coordinates, its common
\(\cC\) \(K\)-presentation is precisely
\[
 \PresK(K\cF;\cC)\quad\Longleftrightarrow\quad
 \exists U\in\GL_2(\AlgNums):\quad
 U^{\mathsf T}XU\in\cC,\qquad U^{-1}\cF\subseteq\cC.
\]
These conditions include the edge and require one basis
for the entire signature set, not one basis per signature.  We use this definition
for \(\cC\in\{\cA,\cP,\cL\}\) and for the flat-Lagrangian subclass.
The character-flat subclass is introduced where it is used in
\cref{proofsubsec:rm-pauli-tools}.
\end{definition}

Since \(\FLag\subseteq\cA\), a common flat-Lagrangian
\(K\)-presentation is a common affine \(K\)-presentation.

\subsubsection{The Eulerian-orientation predicate}
\label{subsec:eo-predicate}

Two cases of \(\Tract\) use the same Eulerian-orientation criterion.
In the common one-sided case, we restrict each signature to its central
Hamming layer. In the single-layer case with both charge signs represented
across the set, we append pins to each signature to obtain central support. For an arity-\(r\) signature \(g\), its central
restriction is
\begin{equation*}
  g|_{\EO}({\mathbf{x}})=g({\mathbf{x}})\one[q_r({\mathbf{x}})=0].
\end{equation*}
It is zero for odd \(r\). We omit zero restrictions from the signature
set and define
\begin{equation*}
  \cF|_{\EO}=\{g|_{\EO}:g\in\cF,\ g|_{\EO}\not\equiv0\}.
\end{equation*}
This omission is a convention for the set. In a closed instance over a
common-one-sided set, an occurrence whose central restriction is zero
makes the instance value zero.

Let \(g\ne0\) be a signature of arity \(k\) supported on the single Hamming
layer of weight \(d\).  Its \emph{EO completion} \cite[Section~6.2]{MengWangXia2025EO} is the ordered tensor
\begin{equation}
  g_{\to\EO}
  =
  \begin{cases}
    g\otimes\DeltaZero^{\otimes(2d-k)},
      & 2d\ge k,\\
    g\otimes\DeltaOne^{\otimes(k-2d)},
      & 2d<k.
  \end{cases}
  \label{eq:eo-completion}
\end{equation}
The original inputs precede the added unaries. A zeroth power is the nullary
unit.  The completed signature has even arity and central support, hence is
EO\@.  For a signature set of nonzero single-layer signatures, define its
EO-completion signature set by
\begin{equation*}
  \cF_{\to\EO}=\{g_{\to\EO}:g\in\cF\}.
\end{equation*}

For an arity-\(2d\) EO signature \(g\), an oriented perfect pairing
\(\pi=((a_1,b_1),\ldots,(a_d,b_d))\) has unordered pairs partitioning
\([2d]\).  For \({\mathbf{z}}\in\Ftwo^d\), let \({\mathbf{y}}\in\Ftwo^{2d}\) be the unique string
with \(y_{a_j}=z_j\) and \(y_{b_j}=1-z_j\). The pairing induces
\begin{equation*}
  g^\pi(z_1,\ldots,z_d)=g({\mathbf{y}}).
\end{equation*}

\begin{definition}[The predicate \(\EOTract\)]
\label{def:eo-tract}
For a finite EO signature set \(\cF\), \(\EOTract(\cF)\) holds if there exist a
single sign \(\varepsilon\in\{+1,-1\}\) and a single class
\(\cC\in\{\cA,\cP\}\), both shared by the entire signature set, such that:
\begin{enumerate}
\item for every \(g\in\cF\) and all
      \({\boldsymbol{\alpha}},{\boldsymbol{\beta}},{\boldsymbol{\gamma}}\in\supp(g)\), if
      \({\boldsymbol{\delta}}={\boldsymbol{\alpha}}\oplus{\boldsymbol{\beta}}\oplus{\boldsymbol{\gamma}}\), then
      \({\boldsymbol{\delta}}\in\supp(g)\) or
      \(\varepsilon q_{\arity(g)}({\boldsymbol{\delta}})>0\);
\item for every \(g\in\cF\) and every oriented perfect pairing \(\pi\)
      of its inputs, \(g^\pi\in\cC\).
\end{enumerate}
In particular, \(\EOTract(\varnothing)\) holds.
\end{definition}

The common-sign condition follows \cite[Theorem~26]{MengWangXia2025EO}.
The opposite-pair restrictions defining the affine and product EO classes
appear in \cite[Definition~22]{MengWangXia2024TypicalEO}.
Our compressed form substitutes \(y_{b_j}=1-y_{a_j}\).
After ordering the inputs as \((a_1,\ldots,a_d,b_1,\ldots,b_d)\), the
full zero-extended restriction is
\[
 R_\pi g({\mathbf z},{\mathbf w})
 =g({\mathbf z},{\mathbf w})\prod_{j=1}^dX(z_j,w_j)
 =g^\pi({\mathbf z})\prod_{j=1}^dX(z_j,w_j).
\]
For \(\cC\in\{\cA,\cP\}\), adjoining these disequality factors preserves
\(\cC\), so \(g^\pi\in\cC\) implies \(R_\pi g\in\cC\).
Conversely, each \(z_j\) determines exactly one \(w_j\), and
\[
 \sum_{{\mathbf w}\in\Ftwo^d}R_\pi g({\mathbf z},{\mathbf w})=g^\pi({\mathbf z}).
\]
This variable elimination preserves affine and product normal forms:
in the affine case it is affine-variable summation, and in the product
case substituting \(w_j=1-z_j\) preserves unary and equality/disequality
factors. Hence \(R_\pi g\in\cC\) if and only if \(g^\pi\in\cC\).
These are identities of normal forms. Gadget availability follows the
conventions of \cref{subsec:reduction-conventions}.

The source papers use unoriented pairs.  Reversing one pair applies the
bit permutation \(z_j\mapsto1-z_j\), and reordering pairs is a permutation of inputs. Both preserve \(\cA\) and \(\cP\).  Hence the oriented
predicate and the source predicate are equivalent.  For a fixed signature
set, the number of pairings and orientations is a constant independent of
the network input.  If the signature set itself is input, these are finite
exact checks within the decidability procedure of
\cref{subsec:complexity-conventions}. No polynomial bound in that input's
encoding length is asserted.

\subsubsection{The tractability predicates}
\label{subsec:tract-predicate}

The reductions use all seven cases of \(\Tract\) for intermediate
signature sets. \Cref{lem:tract-X-anchor} identifies when these reduce
to the four cases of \(\ReTract\).

When either predicate below is applied to an arbitrary finite input set, its
argument is first reduced to the nonzero positive-arity part according to
\cref{subsec:zero-nullary}.  The definitions may therefore be stated for
preprocessed signature sets without carrying a separate superscript in the
public criteria.

For a preprocessed signature set \(\cF\), define its \(K\)-coordinate image
\(\widehat{\cF}=K^{-1}\cF\).  Define \(\Tract(\cF)\) to be the disjunction
of the following seven cases.

\begin{enumerate}[label=\textnormal{(\arabic*)}]
\item \label{case:one-sided}
      The signature set \(\widehat{\cF}\) is common one-sided, in the explicit
      sense that
      \[
       [\,\forall\widehat f\in\widehat{\cF},\
       \supp(\widehat f)\subseteq HW_{\arity(\widehat f)}^{\ge}\,]
       \quad\text{or}\quad
       [\,\forall\widehat f\in\widehat{\cF},\
       \supp(\widehat f)\subseteq HW_{\arity(\widehat f)}^{\le}\,],
      \]
      and \(\EOTract(\widehat{\cF}|_{\EO})\) holds.

\item \label{case:single-weight}
      Every \(\widehat f\in\widehat{\cF}\) is supported on one Hamming
      weight \(d_f\).  Across the signature set, some string in a signature support has negative
      charge and some (possibly other) string in a signature support has positive charge, and
      \(\EOTract(\widehat{\cF}_{\to\EO})\) holds, with the completion
      operation defined in \cref{eq:eo-completion}.

\item \label{case:arity-two}
      \(\cF\subseteq\angles{\cT}\).

\item \label{case:matching}
      \(\cF\subseteq\angles{K\cM}\), or
      \(\cF\subseteq\angles{KX\cM}\).

\item \label{case:affine}
      \(\cF\) is \(\cA\)-transformable.

\item \label{case:product}
      \(\cF\) is \(\cP\)-transformable.

\item \label{case:local-affine}
      \(\cF\) is \(\cL\)-transformable.

\end{enumerate}

\begin{remark}
The matrix symbols in \cref{case:matching} act tensorwise at the arity of
each signature.  In particular,
\[
\begin{aligned}
 f\in\angles{K\cM}
 &\Longleftrightarrow
 \text{ every irreducible factor of }K^{-1}f\text{ lies in }\cM,\\
 f\in\angles{KX\cM}
 &\Longleftrightarrow
 \text{ every irreducible factor of }XK^{-1}f\text{ lies in }\cM.
\end{aligned}
\]
\end{remark}

The four cases indexed by \(\cT,\cA,\cP,\cL\) form the reduced
tractability predicate used in the main theorems.

\begin{definition}[The reduced tractability predicate \(\ReTract\)]
\label{def:tract-X}
For a preprocessed signature set \(\cF\), define
\hypertarget{eq:tract-X}{}
\begin{equation*}
\begin{split}
 \ReTract(\cF)\quad\Longleftrightarrow\quad
 &\cF\subseteq\angles{\cT},
 \\[-0.2em]
 &\text{or \(\cF\) is \(\cA\)-transformable,}
 \\[-0.2em]
 &\text{or \(\cF\) is \(\cP\)-transformable,}
 \\[-0.2em]
 &\text{or \(\cF\) is \(\cL\)-transformable.}
\end{split}
\end{equation*}
Thus \(\ReTract\) is exactly the disjunction of
\cref{case:arity-two,case:affine,case:product,case:local-affine}.
\end{definition}

To make the edge condition explicit, for \(E\in\{I,X\}\) write
\(\ReTract_E(\cF)\) for the same four cases, with transformability
requiring \(M^{\mathsf T}EM\in\cC\) and
\(M^{-1}\cF\subseteq\cC\) for one common \(M\).
The introduction suppresses \(E\), using \(I\) in
\cref{thm:p1-deq2-dichotomy} and \(X\) in
\cref{thm:p1-quaternary-kholant-dichotomy}. Elsewhere the unadorned
\(\ReTract\) is \(\ReTract_I\), as in \cref{def:tract-X}.
Since \(K^{\mathsf T}K=X\),
\hypertarget{eq:retract-K-coordinates}{}
\[
 \ReTract_X(\cF)\quad\Longleftrightarrow\quad\ReTract(K\cF).
\]
Indeed, substituting \(M=KU\) identifies the edge condition
\(M^{\mathsf T}M\in\cC\) with \(U^{\mathsf T}XU\in\cC\) and gives
\(M^{-1}K\cF=U^{-1}\cF\). Invertible single-input maps also preserve
\(\angles{\cT}\). Thus the \(K\)-change expresses the same four
tractable cases in ordinary-Holant coordinates.

\begin{lemma}[Reduction to \(\ReTract\)]
\label{lem:tract-X-anchor}
Let \(\cF\) be a preprocessed signature set. If \(K^{-1}\cF\) contains a
two-sided irreducible signature of arity at least two, then
\[
 \Tract(\cF)\quad\Longleftrightarrow\quad\ReTract(\cF).
\]
In particular, this equivalence holds if \(X\in\cF\) or
\(Z=\diag(1,-1)\in\cF\).
\end{lemma}

\begin{proof}
Let \(g\in K^{-1}\cF\) be two-sided and irreducible, with
\(n:=\arity(g)\ge2\). Its support contains strings of both charge signs
and hence of different Hamming weights. Thus
\cref{case:one-sided,case:single-weight} are impossible.

By irreducibility and the characterization preceding \cref{def:tract-X},
\cref{case:matching} would require \(g\in\cM\) or
\(X^{\otimes n}g\in\cM\). A signature in \(\cM\) is supported at
weights at most one, whose doubled charges are at most \(2-n\le0\).
Complementing all inputs reverses the charge sign. Either possibility
therefore makes \(g\) one-sided, a contradiction. Only the four cases
of \(\ReTract\) remain. The reverse implication holds by definition.

For the binary special cases, \(K^{-1}X=-iZ\) and \(K^{-1}Z=I\).
Both images have matrix rank two and nonzero values at \(00\) and \(11\),
so they are irreducible and two-sided.
\end{proof}

In \cref{thm:p1-deq2-dichotomy}, \(\cF\cup\{X\}\) contains \(X\), so the
first binary special case applies directly. In \cref{thm:p1-eq2-kholant-dichotomy},
\(I\in\cF\) gives \(Z=K^{\otimes2}I\in K\cF\), so the second applies.
For \cref{thm:p1-quaternary-kholant-dichotomy}, the nonzero values at
\(0000\) and \(1111\) make the irreducible quaternary two-sided.
Its arity also excludes \cref{case:arity-two}, since invertible maps at
individual inputs preserve tensor irreducibility. Thus only the three
transformability cases of \(\ReTract\) remain under that hypothesis.

\subsubsection{Imported dichotomies and decomposition}
\label{subsec:imported-results}

Unless an ordinary \(\Holant\) symbol or an explicit source edge model
\(\Holant(X\mid\cdot)\) is displayed, the imported \(\KHolant\) statements
below use \(K\)-coordinates (so the edge tensor is \(X\)). Applying
the fixed \(K\)-change gives the corresponding ordinary statement.  All
signature sets in these statements are fixed and preprocessed.
In particular, every later use of \cref{thm:external-odd} on a signature set
\(\cF\) means the holographic equivalence
\[
 \KHolant(\cF)\equivT\Holant(K\cF),
 \qquad
 K\cF=\{K^{\otimes\arity(g)}g:g\in\cF\},
\]
followed by the ordinary theorem on \(K\cF\) and the inverse holographic equivalence.
The known nonzero edge and gadget scalars are restored by the record.

\begin{theorem}[Weighted Eulerian orientations
{\cite[Theorem~26]{MengWangXia2025EO}; \cite[Theorem~1.1]{GuanShaoShi2026}}]
\label{thm:external-eo}
For each fixed finite preprocessed algebraic EO signature set \(\cF\), if \(\EOTract(\cF)\) holds,
then \(\KHolant(\cF)\in\FP\). Otherwise
\(\KHolant(\cF)\) is \(\SharpP\)-hard.
\end{theorem}

\begin{theorem}[Odd-arity Holant
{\cite[full version, Theorem~33; conference version, Theorem~30]{MengWangXiaZheng2025}}]
\label{thm:external-odd}
Let \(\cF\) be a finite preprocessed algebraic signature set containing a
nonzero odd-arity signature.  Then \(\Holant(\cF)\in\FP\) if
\(\Tract(\cF)\) holds, and is \(\SharpP\)-hard otherwise.
\end{theorem}

Its first two cases were originally in \(\FP^{\mathsf{NP}}\). The
stated \(\FP\) conclusion follows from
\cite[Theorem~1.1 and Corollary~1.2]{GuanShaoShi2026}.
Thus a set containing a nonzero odd-arity signature is covered by this
classification. The remaining structural analysis concerns even-arity
signatures. The next theorem allows us to separate their tensor factors.

\begin{theorem}[Tensor-factor extraction
{\cite[full version, Theorem~32 and Remark~39; conference version,
Theorem~29]{MengWangXiaZheng2025}}]
\label{thm:external-factor}
Let \(\cF\) be a finite preprocessed algebraic signature set in \(K\)-coordinates, and let \(h=a\otimes b\) be a nonzero tensor product with
algebraic factors \(a,b\) of positive arity (equivalently, apply the source theorem
to the ordinary set \(K(\cF\cup\{h\})\)).  Any nullary scalar factor is
absorbed during preprocessing.  Unless
\(\cF\cup\{h\}\) is common one-sided,
\[
  \KHolant(\cF,a,b)\equivT\KHolant(\cF,h).
\]
\end{theorem}

In the common-one-sided exception, \(\KHolant(\cF,h)\) is in \(\FP\)
when \(\EOTract((\cF\cup\{h\})|_{\EO})\) holds and is
\(\SharpP\)-hard otherwise, by the one-sided clause of the cited theorem
and \cref{thm:external-eo}.

In the remaining case, the displayed equivalence permits factor extraction.
We next justify
iterating this reduction and handle any odd-arity factors that it produces.

\section{Shared Reductions and Effective Recognition}
\label{sec:p1-shared-terminals}

The structural classification needs both algorithms for its tractable cases
and reductions that apply across several branches. We first establish
recognition and evaluation, then show how adding factors and changing basis
preserve the connection to the original problem. The remaining tools use
binary interpolation, single-occurrence replication, and higher equalities
to obtain nonzero unaries. The odd-arity dichotomy then classifies the whole
signature set. Direct realization and Turing access follow
\cref{subsec:reduction-conventions}.

\subsection{Recognition and evaluation of tractable cases}
\label{subsec:shared-recognition}

Recognition of the predicates in \cref{sec:predicate} combines finite
support and factor tests with common-basis tests for the transformable
cases. Polynomial constraints decide the latter and produce an exact
witness for the edge and all signatures. For each fixed signature set,
the resulting representations give polynomial-time instance evaluation.

\subsubsection{Effective algebraic geometry}
\label{subsubsec:effective-algebraic-geometry}

The transformability test in \cref{lem:algebraic-transform-witness} asks
for one matrix satisfying polynomial conditions for the edge and every
signature. Work over the exact coefficient field \(L\). For an ideal
\(I\subseteq L[x_1,\ldots,x_n]\), write \(V_{\overline L}(I)\) for its
common zero set. Appendix~\ref{app:exact-algebraic-tools} records the
algebraic tools used to decide whether this set is empty and to recover
an exact witness when it is nonempty.

\begin{lemma}[Effective algebraic transformability witnesses]
\label{lem:algebraic-transform-witness}
Let \(E\in\Mat_2(\AlgNums)\), let \(\cF\) be a finite algebraic
signature set, and let \(\cC\in\{\cA,\cP,\cL\}\).  Suppose there is a matrix
\(U\in\GL_2(\mathbb C)\) such that
\[
  U^{\mathsf T}EU\in\cC,
  \qquad
  U^{-1}\cF\subseteq\cC,
\]
where membership in \(\cC\) allows complex values for the free scalar and
unary parameters in its normal forms.  Then such a matrix exists in
\(\GL_2(\AlgNums)\).  Moreover, for exact algebraic input, the existence of a
matrix is decidable and an algebraic matrix can be produced.
\end{lemma}

\begin{proof}
Let \(L\) be a number field containing the entries of \(E\) and \(\cF\),
as well as \(i\) and \(\zeta_8=(1+i)/\sqrt2\) (equivalently \(i\) and
\(\sqrt2\)).  There are only finitely many combinatorial normal-form
types compatible with the arities: affine supports and phase coefficients,
product partitions with equality or disequality signs and unary weights,
and, for \(\cL\), the finitely many prescribed
\(T_2\)-twisted affine conditions.  Enumerate one shared type choice for the transformed edge and
all transformed signatures.

For \(\cP\), repeated factors cause no infinitude: at fixed arity we
canonically enumerate the support set, tensor partition, factor
multiplicities, equality/disequality signs, unary phase type, and permutation of inputs.  On a fixed pair of variables, repeated copies of the same
\(I\)- or \(X\)-factor are idempotent and are retained as a single factor.
The simultaneous presence of an \(I\)- and an \(X\)-factor makes the whole
product zero.  A loop \(I(x,x)\) is the scalar one and a loop \(X(x,x)\)
is zero.  Unary factors on one variable are multiplied and combined into
one unary (or zero). These simplifications give a finite canonical enumeration.
For \(\cL\), enumerate the prospective transformed support set and affine
phase data. Requiring entries to be zero outside this support and nonzero
on it, together with the twist conditions, gives finitely many polynomial
equalities and inequations.  Thus every candidate, including one whose
transformed support is not known in advance, is one of finitely many
constructible systems.

Fix one such branch, and let \(\mathbf x\) be the tuple of all its scalar
variables: the four entries of \(U\), the scalar and unary parameters in the
chosen normal forms, and the auxiliary variables introduced below.  Introduce
an auxiliary scalar variable \(z\) and impose
\[
 z\det U=1.
\]
Here \(\operatorname{adj}(U)\) denotes the classical adjugate, characterized
by \(U\operatorname{adj}(U)=(\det U)I\). Hence on this solution set
\(U^{-1}=z\operatorname{adj}(U)\).  After this substitution, equality of
every transformed tensor entry with its proposed normal form is a polynomial
equation over \(L\).  If the branch also has the nonvanishing polynomial
constraints \(p_1,\ldots,p_s\ne0\), introduce one more auxiliary scalar
variable \(t\) and replace all these inequations by the single equation
\[
 t\prod_{\ell=1}^s p_\ell=1.
\]
Thus the branch is exactly the zero set of a finitely generated ideal
\(I\subseteq L[\mathbf x]\).

Compute a Gr\"obner basis of \(I\) over the exact number field \(L\).
For the algebraic closure \(\overline L\subseteq\mathbb C\) of this
embedded field, ideal membership and the weak Nullstellensatz give
\[
 V_{\overline L}(I)=\varnothing
 \quad\Longleftrightarrow\quad
 1\in I,
\]
and the latter is equivalent to the reduced Gr\"obner basis being
\(\{1\}\).  Hence emptiness of each branch is decidable.  Moreover, a
complex solution makes \(I\) proper, so the weak Nullstellensatz supplies
a point over \(\overline L=\AlgNums\). Conversely every such point is a
complex solution.  In a zero-dimensional branch, standard elimination
followed by triangular decomposition produces exact minimal polynomials and
isolating data for one point.  A nonempty positive-dimensional branch is
also effective: enumerate algebraic tuples fairly, first by total exact
encoding height and then lexicographically, and test every displayed
equation exactly.  The weak Nullstellensatz guarantees that this search
eventually reaches a point of the branch. Reading the \(U\)-coordinates
gives the required algebraic witness. These effective procedures follow
from Gr\"obner-basis theory \cite{Cohen1993,CoxLittleOShea2015}, without
an encoding-size or running-time bound.
\end{proof}

\subsubsection{Evaluation for signatures in \texorpdfstring{\(\cM\)}{M}}
\label{subsubsec:one-hot-evaluation}

For signatures in \(\cM\), each vertex accepts at most one incident
bit equal to one. This reduces evaluation to trees and unicyclic components,
as the following lemma shows. Here a factor graph records signature vertices
and their incidences with edge kernels. A unicyclic component has exactly
one cycle, including a loop or a two-cycle formed by parallel edges.
Cycles are identified under reversal and cyclic changes of starting point.
Suppression removes subdivision vertices.

\begin{lemma}[Evaluation for \(\cM\)]
\label{lem:one-hot-matching-evaluation}
Fix a finite algebraic signature set, and let \(\Gamma\) be a closed
factor-expanded finite \(X\)-edge network over it (each
occurrence has first been replaced by its fixed irreducible factors, with
the induced input connections and scalar).  If a vertex signature is
zero, return zero. Otherwise, for every vertex \(w\), let \(h_w\) denote its
nonzero signature, define its arity by \(d_w:=\arity(h_w)\),
and suppose
\(\supp(h_w)\subseteq\{\zero^{d_w},{\mathbf{e}}_1,\ldots,{\mathbf{e}}_{d_w}\}\).  Define its
weight at the all-zero string \(a_w:=h_w(\zero^{d_w})\) and weight-one input coefficients
\(b_{w,p}:=h_w({\mathbf{e}}_p)\).  Then \(Z_X(\Gamma)\) is computable using
\(O(|V(\Gamma)|+|E_{\rm int}(\Gamma)|)\) arithmetic operations in the fixed
coefficient field, and hence in polynomial bit time. \(V(\Gamma)\) is the
vertex set.
\end{lemma}

\begin{proof}
For an \(X\)-edge exactly one of its two half-edges receives bit \(1\).
Consequently assignments are in bijection with choices of one endpoint of
each edge, subject to at most one chosen endpoint at each vertex.  For such a
feasible endpoint-choice assignment \(\omega\), write
\(\operatorname{occ}_\omega(w)\in\{0,1\}\) for the number of selected
endpoints at \(w\), and let \(s_w(\omega)\) be the selected input when it
exists.  The weight of \(\omega\) is
\[
  \prod_{\operatorname{occ}_\omega(w)=0}a_w
  \prod_{\operatorname{occ}_\omega(w)=1}b_{w,s_w(\omega)}.
\]
The product of the fixed nonzero scalar factors recorded during the
factor expansion is multiplied into the final component product (and a
zero expansion scalar was already handled by returning zero), so it does
not change the recurrence below.
For a connected component, let \(e\) and \(v\) denote its edge and vertex
counts.  A component with \(e>v\) has value zero.  Every surviving component
has \(e=v-1\) (a tree) or \(e=v\) (unicyclic).
On a rooted tree, let \(p_w\) be the parent-edge input at \(w\),
and let \(p_{wc}\) be the input at \(w\) of the child edge \(wc\).  If
\(F_w(1)\) (resp. \(F_w(0)\)) denotes the subtree sum when the parent edge
selects \(w\) (resp.\ its parent), then
\[
\begin{aligned}
 F_w(1)&=b_{w,p_w}\prod_cF_c(1),\\
 F_w(0)&=a_w\prod_cF_c(1)+
 \sum_c b_{w,p_{wc}}F_c(0)\prod_{d\ne c}F_d(1).
\end{aligned}
\]
At the root use the second line without a parent term.  Order the children as
\(c_1,\ldots,c_m\).  Products and sums are evaluated without division by
the prefix products
\(P_j:=\prod_{\ell<j}F_{c_\ell}(1)\) and suffix products
\(S_j:=\prod_{\ell>j}F_{c_\ell}(1)\), for which
\(P_jS_j=\prod_{\ell\ne j}F_{c_\ell}(1)\) even
when some factors vanish.  For a unicyclic component delete one cycle
edge \(e\), with endpoint inputs \(p_u\) at \(u\) and \(p_v\) at \(v\).
Enumerate the two legal choices \(p\in\{p_u,p_v\}\), and let \(v(p)\)
be the vertex incident to the chosen input.  The inputs are distinct even
when \(e\) is a loop.  For each choice, root the resulting tree at
\(v(p)\) and impose that \(p\) is already occupied through \(e\):
at that root use
\[
 b_{v(p),p}\prod_cF_c(1).
\]
This forces every remaining incident edge at that root to select its child.
Use the ordinary
tree recurrence on all proper subtrees and perform this same forced-root
calculation separately for each \(p\). Sum the two values.
The prefix/suffix evaluations use linearly many field operations.  Every
intermediate entry is a sum of at most exponentially many products of
\(O(|V|+|E_{\rm int}|)\) fixed coefficients.  In a fixed rational basis of
their number field, a common denominator and the numerator sizes therefore
have \(O(|V|+|E_{\rm int}|)\) bits, giving polynomial bit time.
\end{proof}

\subsubsection{Decidability and polynomial-time evaluation}
\label{subsec:predicate-algorithms}

\begin{theorem}[Recognition and evaluation]
\label{thm:predicate-algorithms}
For every finite algebraic signature set \(\cF\), the predicate \(\Tract(\cF)\),
with the preprocessing convention of \cref{subsec:zero-nullary}, is
decidable.  If it holds, then \(\Holant(\cF)\in\FP\).
\end{theorem}

\begin{proof}
The two EO cases are decided by finitely many exact tests of support,
layers, affine subspaces, and phases. Enumerate all oriented pairings and the
four shared \((\varepsilon,\cC)\)-choices.  A signature's tensor
decomposition can be found by enumerating all nontrivial bipartitions
\(S\mid S^c\), where
\(\varnothing\ne S\subsetneq[\arity(f)]\) and
\(S^c=[\arity(f)]\setminus S\), and testing whether the corresponding
matrix flattening has rank one. Recursively applying this test to the
factors decides \cref{case:arity-two,case:matching}.  Meanwhile,
\cref{lem:algebraic-transform-witness} with \(E=I\) decides
\cref{case:affine,case:product,case:local-affine} and returns the common
witness.

For the two EO cases, evaluation uses value-preserving transformations
of the input instance. Charge conservation reduces \cref{case:one-sided} to
\cref{thm:external-eo}: in a closed contributing assignment all local
charges have the common sign and sum to zero, so every local charge is zero
and only the central restrictions contribute. Deleting the off-central
entries therefore preserves the value. In \cref{case:single-weight}, an
occurrence of arity \(k_f\) and weight \(d_f\) has charge
\(c_f=2d_f-k_f\). Each connected component \(C\) must satisfy
\begin{equation*}
  \sum_{v\in V(C)}c_{f_v}=0.
\end{equation*}
Failure gives zero. Otherwise construct an auxiliary EO instance:
replace each occurrence \(f_v\) by its fixed completion
\((f_v)_{\to\EO}\), and in each component pair the newly appended
\(\DeltaZero\)-inputs (from \(c_{f_v}>0\)) with the appended
\(\DeltaOne\)-inputs (from \(c_{f_v}<0\)) by \(X\)-edges.  The charge
balance makes the two numbers equal, and every original assignment has a
unique extension to this auxiliary instance (and conversely), so its
partition function is unchanged
\cite[Section~6.2, proof of Corollary~68]{MengWangXia2025EO}.
The finite completed signature set is EO, so \cref{thm:external-eo}
supplies the polynomial-time evaluator.
After expansion into irreducible factors, the arity-two branch is a graph of maximum
degree two and is evaluated by transfer matrices on paths and cycles.
For the matching cases, first apply the common \(K\)-coordinate
equivalence, the possible global \(X\)-toggle, and expansion into irreducible factors.
The toggle is a change preserving the disequality edge because \(X^{\mathsf T}XX=X\).
After this change, every counted irreducible factor \(h\) satisfies
\(h({\mathbf{x}})=0\) for \(\wt({\mathbf{x}})>1\).  Only structural equality kernels, or degree-two equality
representations introduced solely to state the source theorem, are
suppressed before the factor graph is built. An explicit arity-two signature
occurrence remains a factor vertex unless it has first been eliminated by the
path/cycle preprocessing of \cref{lem:p1-equality-wire-equivalence}, with its
scalar recorded.  Thus
\cref{lem:one-hot-matching-evaluation} gives a polynomial-time evaluator
using linearly many field operations. The
expansion has constant size per occurrence because the signature set is
fixed.  Finally, a common change-of-basis matrix \(M\) gives
\[
  \Holant((\Eq_2)M\mid M^{-1}\cF),
  \qquad(\Eq_2)M=M^{\mathsf T}M,
\]
with both sides in the same class
\(\cC\in\{\cA,\cP,\cL\}\).  Forgetting the
bipartition yields an ordinary \(\Holant(\cC)\) instance, computable in
\(\FP\): affine instances are quadratic Gauss sums, product instances split
into unary/equality/disequality components, and local-affine instances are a
restriction of the corrected complex-valued \(\Holant^c\) evaluator of
\cite{Backens2021,Backens2025} (the real-valued presentation
\cite[Corollary~3.1]{CaiLuXia2018RealHolantc} gives the same local routine).
\end{proof}

\subsection{Retained reductions and normalization}
\label{sec:retained-normalization}

Fix the preprocessed signature set \(\cF_0\) at the beginning of a
structural branch.  We form finite signature sets from \(\cF_0\) by
adding realizable signatures, adding factors through the decomposition
theorem, or, within the eight-vertex reduction, applying a global \(GO(X)\)
change of basis.  We call a set obtained in this way a \emph{retained set}.
Each addition is justified by its gadget or factor reduction, and each
change of basis acts on the whole set. Interpolation is used only to
resolve the current case.

We record the current coordinates and signature set as \((A,\cF)\), with
\[
 A\in GO(X),\qquad A\cF_0\subseteq\cF,
 \qquad \KHolant(\cF)\equivT\KHolant(\cF_0).
\]
Adding signatures in the current basis leaves \(A\) unchanged.
The normalization data in \cref{lem:normalization-ledger} records the
coefficient field and pulled-back group. The eight-vertex interface
returns its outputs in the original coordinates, as specified in
\cref{cor:p1-eight-vertex-reduced-interface,sec:p1-binary-phase}.

For each signature, specify its gadget or reduction, input order, and
changes of basis.  Inverse transformations express it in any earlier basis.
Nonzero scalar factors follow the convention in \cref{subsec:holographic}.

\subsubsection{Finite induction and termination}
\label{subsec:worklist-termination}

The reductions in \cref{subsec:safe-matching-cards} retain tensor factors
and sometimes enlarge the finite group of binary transfers.  The orders
below distinguish finite processing of a fixed family from a restart with
a smaller arity or larger group.

\label{subsubsec:worklist-termination}

A strict order is irreflexive and transitive.  It is \emph{well founded}
when it has no infinite descending chain.  A finite multiset over a set
\(E\) records elements of \(E\) with multiplicity. The multiset sum \(\uplus\) adds multiplicities.  Given a strict order \(<\) on \(E\) and finite multisets \(A,B\) over
\(E\), its strict multiset extension is defined by \(A<_{\rm mul}B\) when
\[
 B=Z\uplus X,\qquad
 A=Z\uplus Y,\qquad X\ne\varnothing,
\]
for finite multisets \(X,Y,Z\) over \(E\), and every \(y\in Y\) satisfies
\(y<x\) for some \(x\in X\).  The multiset
extension of a well-founded strict order is well founded
\cite{DershowitzManna1979}.  Lexicographic order compares the first
coordinate at which two tuples differ. Finite lexicographic products of
well-founded orders are well founded by induction on the number of
coordinates.  Well-founded induction proves a property at \(x\) assuming it
at every strict predecessor of \(x\).

We index unfinished signature occurrences by a finite multiset.  Each
occurrence retains its realizing gadget or reduction and its ordered inputs.
To identify duplicate signatures, divide by the first nonzero coefficient
in the fixed input order.  Repeated
occurrences are still counted with multiplicity in the induction measures.
The reductions test their listed cases in order. The first applicable case
determines the next step.

\subsubsection{Factor closure}
\label{subsec:factor-saturation}

Record each recursive tensor factorization as a tree, retaining the ordered
blocks of inputs, permutation of inputs, and nonzero scalar at every split.  The leaves
are the irreducible factors.

For a preprocessed finite signature set \(\cF\), choose an exact
irreducible tensor factorization of every member and let \(\Pi(\cF)\) be the
joint signature set of all irreducible factors, with ordered blocks of inputs retained.
We call \(\cF\) \emph{factor-closed} when
\(\Pi(\cF)\subseteq\cF\).

\begin{lemma}[Factor closure]
\label{lem:factor-saturation}
If \(\cF\) is not common one-sided, then
\begin{equation}
  \KHolant(\cF)
  \equivT
  \KHolant(\cF,\Pi(\cF)).
  \label{eq:factor-saturation}
\end{equation}
Moreover, every invocation of factor closure is finite. The factorization
procedure and its strictly decreasing potential are given in the proof.
\end{lemma}

\begin{proof}
Start with one root occurrence for each member of the current
finite retained set and add a factor only through the decomposition theorem.  Let \(h_u\) denote the signature carried by a nonzero node \(u\).
Such a node, carrying its ordered block of inputs, is either
proved irreducible, or is replaced by one split
\[
 h_u\doteq a_u\otimes b_u,
 \qquad 1\leq\arity(a_u),\arity(b_u)<\arity(h_u),
\]
with the nonzero scalar and the permutation of inputs recorded with the factorization.
Record and discard nullary scalars, and remove irreducible unary nodes
before forming the unresolved queue. Normalize each node by its first
nonzero entry and retain its ordered input block. Identical normalized
signatures may share one retained signature, while their occurrences remain
distinct work items. Let \(\mathcal W\) be the multiset of unresolved
occurrences, all of arity at least two. The potential
\begin{equation}
  \Phi(\mathcal W)
  =
  \sum_{u\in\mathcal W}(\arity(h_u)-1)
  \label{eq:factor-potential}
\end{equation}
is a nonnegative integer. Splitting an item of arity \(r+s\) removes
\(r+s-1\) and adds at most \((r-1)+(s-1)=r+s-2\). Already pending or
processed normalized children only lower this amount further, and identifying
an irreducible signature removes its summand. Thus every step strictly
decreases \(\Phi\), and \(\Pi(\cF)\) is the finite set of canonical
irreducible leaves. Its independence of processing order follows from
uniqueness of irreducible tensor factorization in \cref{subsec:holant-definitions}.

Apply the decomposition theorem at each queued split
\(h_u=a_u\otimes b_u\) to the current signature set
\cite[full version, Theorem~32 and Remark~39]{MengWangXiaZheng2025}.
The original non-one-sided signature set remains a subset, so the
exceptional case never reappears.  Iteration adds every factor, while
set inclusion gives the reverse reduction. These two directions give
\cref{eq:factor-saturation}. All reduction scalars and ordered blocks
are available for later gadgets.
\end{proof}

For the initial reduction, each original signature can also be reconstructed
as the tensor product of its irreducible factors, with its input permutation
and nonzero scalar restored. Consequently, outside the common-one-sided
case,
\[
 \KHolant(\cF)\equivT\KHolant(\Pi(\cF)).
\]
We may therefore take the initial signature set to consist of irreducible
signatures and continue to denote it by \(\cF\). If a nonzero odd-arity
factor occurs, \cref{thm:external-odd} applies after the fixed
\(K\)-change. If all factors have arity at most two, the evaluation in
\cref{thm:predicate-algorithms} applies. The remaining analysis starts with
even-arity irreducible signatures and at least one of arity at least four.
Later constructions can have reducible outputs. At those steps we use the
extension \(\cF\cup\Pi(\cF)\) in \cref{eq:factor-saturation}, retaining
the outputs as well as their factors for the subsequent group comparison.

For a factor-closed signature set \(\cF\) containing an irreducible
signature of arity at least three, define its minimum arity of a nonbinary irreducible signature by
\begin{equation*}
 \nu(\cF)
 =\min\{\arity(h):h\in\cF
          \text{ is irreducible and }\arity(h)\ge3\}.
\end{equation*}
This parameter is used only after the branch in which every irreducible factor has
arity at most two has been removed.

\subsubsection{Preservation under taking subsets and global normalization}

To recover the tractability criterion for the original set, we must both
restrict to a subset and undo any global normalization. The next two
lemmas justify these steps.

\begin{lemma}[Preservation under taking subsets]
\label{lem:downward-inheritance}
If \(\cF\subseteq\cF'\) are preprocessed signature sets and
\(\Tract(K\cF')\) holds, then \(\Tract(K\cF)\) holds.
\end{lemma}

\begin{proof}
All cases except the single-weight one are hereditary by definition.
In the latter case a signature subset either retains both charge signs and the
same EO-completion witness, or becomes one-sided.  In the latter event, its
central members are unchanged by both restriction and EO completion, whereas
all off-central members restrict to zero.  Hence its nonzero central
restrictions form a subset of the original completion signature set, and the
same shared EO witness applies.
\end{proof}

\begin{lemma}[\(GO(X)\)-invariance of transported tractability]
\label{lem:GOX-tract-invariance}
Let \(\cF\) be a finite preprocessed signature set and let
\(D\in GO(X)\).  Then
\begin{equation*}
  \Tract(K\cF)
  \quad\Longleftrightarrow\quad
  \Tract\bigl(K(D\cF)\bigr).
\end{equation*}
\end{lemma}

\begin{proof}
Write the matrix \(D=\begin{psmallmatrix}a&b\\ c&d\end{psmallmatrix}\), and
let \(\lambda\in\AlgNums^\times\) be its conformal multiplier, defined by
\(D^{\mathsf T}XD=\lambda X\).  Comparing entries gives
\(ac=bd=0\) and \(ad+bc=\lambda\ne0\).  Thus \(D\) is diagonal or
anti-diagonal.  The former preserves every string in the support. The latter
complements it. On a fixed layer its multiplier is constant.  Hence the two
EO cases are preserved (with the support side and shared sign
possibly exchanged), irreducible factor arities are unchanged, and
\(\angles{\cM}\) and \(\angles{X\cM}\) are preserved or exchanged.

Define the conjugate matrix \(A=KDK^{-1}\), so
\(A^{\mathsf T}A=\lambda I\).  If the matrix \(W\) is a common
\(\cC\)-transformability witness, define the matrix
\(W':=AW\).  It satisfies
\[
  (W')^{-1}A(K\cF)=W^{-1}(K\cF)\subseteq\cC,
  \qquad
  (W')^{\mathsf T}W'
  =\lambda W^{\mathsf T}W\in\cC
\]
for \(\cC\in\{\cA,\cP,\cL\}\).  Applying the same argument to \(D^{-1}\)
proves the converse.
\end{proof}

The following lemma records the data needed to transport reductions,
group enlargements, and tractable presentations back to the original
signature set. Its continuing-state use is confined to the eight-vertex
reduction.

\begin{lemma}[Normalization data]
\label{lem:normalization-ledger}
Let \(\cF_0\) be the signature set at the beginning of a structural
branch in which coordinate transport is permitted.  Let \(L_0\) be a fixed
number field containing all entries of
\(\cF_0\), their complex conjugates, and all algebraic constants already fixed by
the coordinate conventions.  Thus \(L_0\) is the fixed initial coefficient field, not a
field enlarged after each normalization.  Every continuing normalization
inside that branch is an explicitly recorded \(GO(X)\) change of coordinates, and tensor
splits are performed after pulling back by the current \(A\), using a nonzero
pivot and row--column ratios.  Thus the pulled-back data (rather than the
displayed current coordinates) stay over \(L_0\), and the pulled-back group
remains inside \(\PGL_2(L_0)\).  The
structural reductions maintain a current signature set \(\cF\),
a matrix \(A\in GO(X)\), and its nonzero conformal factor
\(\mu_A\in\AlgNums^\times\), satisfying
\begin{equation}
  A\cF_0\subseteq\cF,\qquad
  A^{\mathsf T}XA=\mu_A X,\qquad
  \KHolant(\cF)\equivT\KHolant(\cF_0).
  \label{eq:normalization-ledger}
\end{equation}
Every positive-arity \(h\in\cF\), up to a nonzero scalar, also has the
field-compatible form
\begin{equation*}
 h=A^{\otimes\arity(h)}h_0
 \quad\text{for a tensor }h_0\text{ over }L_0.
\end{equation*}
At a residual node where every directly realizable nonsingular transfer has
finite projective order, these transfers form a group. Denote by
\(G(\cF)\) the group of their classes \([BX]\), where \(B\) ranges over
directly realizable nonsingular ordered binaries over \(\cF\).  Then the
pulled-back group
\begin{equation*}
 \overline G(A,\cF)
 :=[A]^{-1}G(\cF)[A]
 \le \PGL_2(L_0)
\end{equation*}
is unchanged by a global normalization and is monotone when signatures are added.
Adding realizable signatures and factor closure preserve \(A\).
If \(D\in GO(X)\) is applied globally to the current signature set, the record
updates by
\hypertarget{eq:normalization-ledger-update}{}
\begin{equation*}
  (A,\cF)\longmapsto(DA,D\cF).
\end{equation*}
The new reduction state again satisfies \cref{eq:normalization-ledger}, up to the
known edge scalar from \cref{eq:GOX-global-scalar}.  Before a continuing
branch leaves \cref{sec:p1-eight-vertex}, apply the inverse of the accumulated
matrix \(A\).  The returned reduction state therefore has \(A=I\) in the fixed original coordinates and retains any supplied equality signature.
Generic interpolation is used only as a terminal transition and is not a
record-preserving update.  If an interpolation output were ever allowed to
re-enter a continuing finite-group state, the effective coefficient field
and the pulled-back group would first have to be enlarged explicitly.

If \(\Tract(K\cF)\) holds, then
\(\Tract(K\cF_0)\) holds.  Separately, if
\(V\) is a common \(\cC\) \(K\)-presentation of \(\cF\), where
\(\cC\) is closed under nonzero scalar multiplication, then the pullback
matrix
\begin{equation*}
  W:=A^{-1}V
\end{equation*}
is a common \(\cC\) \(K\)-presentation of \(\cF_0\).  In particular, this
applies to \(\cA,\cP,\cL\) and to the flat-Lagrangian subclass.
\end{lemma}

\begin{proof}
The assertions hold initially for \((I,\cF_0)\).  Adding realizable signatures and
factor closure preserve them by
\cref{eq:direct-gadget-reduction,eq:factor-saturation}.  A global
\(D\in GO(X)\), with nonzero conformal factor
\(\mu_D\in\AlgNums^\times\) defined by \(D^{\mathsf T}XD=\mu_DX\), replaces \(A\) by
\(DA\). Then
\((DA)^{\mathsf T}X(DA)=\mu_D\mu_AX\), and
\cref{eq:GOX-global-scalar} supplies the stated equivalence.  Conversely,
the transported inclusion \(A\cF_0\subseteq\cF\), followed by the
inverse global transformation, reduces the initial problem to the current
one. 

Inductively pull every construction through \(A\).  Gadget contraction
stays over \(L_0\). At a tensor split, normalizing one nonzero entry and
taking row--column ratios chooses the factors over \(L_0\).  Moreover
\(A^{\mathsf T}X\doteq XA^{-1}\).  Thus, if a current binary \(B\) has the
record form \(B=A^{\otimes2}B_0\) for a binary tensor \(B_0\) over \(L_0\),
its projective transfer pulls back by
conjugation:
\[
 [BX]=[A(B_0X)A^{-1}].
\]
The identity transfer is realized by the disequality edge, and chaining realizes
products.  At such a finite-projective-order residual node, the inverse of each transfer is a
positive power of that transfer, so the realizable classes form a group.
The displayed conjugation shows that global normalization leaves
\(\overline G(A,\cF)\) unchanged, while adding signatures can only add realizable transfers.  This proves the field and group assertions.
Finally, if \(\Tract(K\cF)\) holds, preservation under taking subsets first gives
\(\Tract(K(A\cF_0))\), and \cref{lem:GOX-tract-invariance} then gives
\(\Tract(K\cF_0)\).  If \(V\) is an explicit common \(K\)-presentation, then
\(W=A^{-1}V\) satisfies
\[
 W^{-1}\cF_0\subseteq V^{-1}\cF\subseteq\cC,
 \qquad
 W^{\mathsf T}XW=\mu_A^{-1}V^{\mathsf T}XV\in\cC,
\]
which proves the last assertion.
\end{proof}
\subsection{Binary Interpolation}
\label{proofsubsec:central-descent}

A singular nonzero binary gives unary factors. For an infinite-order
transfer, interpolation first gives a rank-one binary. The following lemma
establishes both reductions, including the size and query bounds needed
for polynomial-time interpolation.

\begin{lemma}[Binary interpolation]
\label{lem:binary-interpolation}
Let \(\cF\) be a finite algebraic, retained, two-sided signature set,
let \(B\ne0\) be an ordered binary directly realizable over \(\cF\),
and define its transfer matrix \(A=BX\).
\begin{enumerate}
\item If \(B\) has the rank-one factorization \(B=u\otimes v\), where
      \(u\) and \(v\) are nonzero algebraic unary signatures, then
      \[
        \KHolant(\cF,u,v)\leT\KHolant(\cF).
      \]
\item If \(A\) is nonsingular and has infinite projective order, there is
      a nonzero rank-one binary \(H\) such that
      \[
        \KHolant(\cF,H)\leT\KHolant(\cF).
      \]
      For \(m\ge1\) designated occurrences of \(H\), the interpolation uses
      \(m+1\) oracle queries, \(O(m^2)\) added vertices per query, and hence
      \(O(m^3)\) added vertices over all queries.
\end{enumerate}
\end{lemma}

\begin{proof}
For the singular case, choose a nonzero algebraic pivot of the rank-one matrix
and take the corresponding algebraic row and column factors (absorbing the
pivot into one factor). Then \(B=u\otimes v\) with \(u,v\) algebraic, and
\cref{thm:external-factor} applies.  Two-sidedness excludes its one-sided
exception.  For the other case, a length-\(t\) chain realizes the ordered
binary signature \(B_t:=A^tX\).  If \(A\) is diagonalizable, choose an eigenbasis matrix
\(V\in\GL_2(\AlgNums)\) and nonzero eigenvalues
\(\lambda_0,\lambda_1\in\AlgNums^\times\) such that
\(A=V\diag(\lambda_0,\lambda_1)V^{-1}\).  For \(j\in\{0,1\}\), let
\(E_{jj}\) be the \(2\times2\) matrix with \((a,b)\)-entry
\(\one[a=b=j]\), for \(a,b\in\{0,1\}\), and define the rank-one binary
matrix \(H_j=VE_{jj}V^{-1}X\).  Then
\[
 B_t=\lambda_0^tH_0+\lambda_1^tH_1,
\]
with \(H_0,H_1\) rank one.  Replacing the \(m\) designated \(H_0\)-vertices by \(B_t\) and
dividing by \(\lambda_1^{tm}\) yields evaluations of a polynomial \(P(z)\)
of degree at most \(m\) at the distinct points
\(z=(\lambda_0/\lambda_1)^t\), \(1\le t\le m+1\). The coefficient of
\(z^m\) is the all-\(H_0\) value.  These points are distinct because
\([A]\) has infinite projective order, so \(\lambda_0/\lambda_1\) is not a
root of unity.

In the remaining infinite-order Jordan case, write \(A=\lambda I+N\), where
\(\lambda\in\AlgNums^\times\) is the eigenvalue and \(N\) is a nonzero
nilpotent \(2\times2\) matrix satisfying \(N^2=0\).  Then
\(B_t=\lambda^t(X+t\lambda^{-1}NX)\).  After normalization the oracle value
is a polynomial in \(t\) of degree at most \(m\), whose coefficient of
\(t^m\) is the value with the rank-one binary \(\lambda^{-1}NX\) at every
designated vertex.
Indeed, \(N\ne0\) and \(N^2=0\) in dimension two imply
\(\rank(N)=1\), and right multiplication by \(X\) preserves
rank.
Both cases use \(m+1\) queries. \(t\le m+1\) gives \(O(m^2)\) replacement
size for each query and \(O(m^3)\) in total.  The eigenvalues and basis
matrices are fixed algebraic numbers.  Their powers and the interpolation
systems have polynomial encoding length, so exact interpolation is
polynomial-time postprocessing.
\end{proof}

Both reductions give access to a nonzero unary after factor extraction.
The dichotomy in \cref{lem:common-unary-exit} below then applies to the
whole current set. Thus the later unresolved binary analysis concerns
nonsingular transfers of finite projective order.

\subsection{Single-occurrence reductions and nonzero unaries}
\label{subsec:single-occurrence-carriers}

The next reductions start with one occurrence of an auxiliary signature.
A fixed gadget passes that signature to the next copy and supplies one
additional output. Repeating the construction supplies as many outputs
as needed while using the auxiliary signature only once.
The single-occurrence replication result in
\cref{lem:single-occurrence-replication} originates in an unpublished
expanded version of \cite{MengWangXiaZheng2025}, which is not publicly
available. We include the statement and proof here, with the ordered
connections explicit.

Define the ordered carrier signature
\[
  \Delta=\DeltaZero\otimes\DeltaOne,
\]
with the displayed order of its two inputs.  For a signature \(g\), the
superscripts \(g^{=1}\) and \(g^{\leq1}\) mean that an oracle instance contains,
respectively, exactly one or at most one \(g\)-vertex.  

\begin{lemma}[Universal one-pair reduction]
\label{lem:universal-directed-pair}
For every fixed finite algebraic signature set \(\cF\),
\[
  \KHolant(\cF,\Delta^{\leq1})
  \leT
  \KHolant(\cF).
\]

\end{lemma}

\begin{proof}
The zero-occurrence case is already an instance over \(\cF\).  Consider an
instance with one \(\Delta\)-vertex.  Delete that vertex, retaining its two
incident \(X\)-attachments as edges to ordered free boundary inputs.
With the boundary bits read at these free endpoints, the raw signature is
\[
  H=\begin{pmatrix}h_{00}&h_{01}\\h_{10}&h_{11}\end{pmatrix},
  \qquad\text{the desired one-\(\Delta\) value is }h_{01}.
\]
Each free endpoint carries an \(X\)-edge, so the signature for vertex
replacement is \(XHX\) by \cref{eq:half-edge-boundary-tensor}.
Closing the two exposed inputs by an \(X\)-edge gives an oracle value
\[
  S=h_{01}+h_{10}.
\]
Let \(\mathcal B_\cF\) be the set of ordered binary tensors realized by
finite \(\cF\)-networks.  If every \(B\in\mathcal B_\cF\) satisfies
  \(B_{01}=B_{10}\), then \(XHX\in\mathcal B_\cF\) is symmetric, so
  \(H\) is symmetric and the desired value is
  \(S/2\).

Otherwise fix once and for all a realized ordered binary
\[
  B=\begin{pmatrix}a&b\\c&d\end{pmatrix},
  \qquad b\ne c.
\]
Insert its fixed network at the deleted location in the two boundary
orientations.  The two oracle values are
\[
  T_+=ah_{00}+bh_{01}+ch_{10}+dh_{11},
  \qquad
  T_-=ah_{00}+ch_{01}+bh_{10}+dh_{11}.
\]
Thus
\[
  h_{01}
  =\frac12\left(S+\frac{T_+-T_-}{b-c}\right).
\]
The ordered inputs are essential in the subtraction.  The witness network for
\(B\) is fixed with \(\cF\), hence has constant size, and the reduction
uses at most three oracle queries.
\end{proof}

The preceding reduction allows one carrier occurrence. To obtain
unrestricted access to a new signature, we use a gadget that preserves
the carrier while producing another output. The next lemma states the
required replication and nonzero-closure conditions for ordinary Holant
and \(\KHolant\), with equality and disequality edges, respectively.

\begin{lemma}[Single-occurrence replication]
\label{lem:single-occurrence-replication}
Let \(\cF\) be a fixed signature set and let \(g,h\) be nonzero signatures.  For ordinary Holant, all connections use \(I\). Suppose that
\begin{enumerate}
\item \(\Holant(\cF,g^{=1})\leT\Holant(\cF)\);
\item a fixed gadget \(C\) over \(\cF\) has three disjoint ordered
      boundary blocks \(A,P,H\), where
      \(|A|=|P|=\arity(g)\) and \(|H|=\arity(h)\).
      Connecting one \(g\)-vertex coordinatewise to \(A\) realizes
      \(\rho\,g(P)\otimes h(H)\), for a nonzero scalar
      \(\rho\in\AlgNums^\times\). Between consecutive copies
      \(C_\ell,C_{\ell+1}\), connect \(P_\ell\) coordinatewise to
      \(A_{\ell+1}\), with no other input identifications.
      For every \(k\ge1\), this chain begins with exactly one
      \(g\)-vertex and realizes
      \(\rho^k g(P_k)\otimes\bigotimes_{\ell=1}^k h(H_\ell)\), with
      pairwise disjoint output blocks \(P_k,H_1,\ldots,H_k\); and
\item some fixed \(\Holant(\cF,g^{=1})\)-instance \(J\) has
      \(Z_I(J)\ne0\).
\end{enumerate}
Then
\[
  \Holant(\cF,h)\leT\Holant(\cF).
\]
For K-Holant, all connections instead use \(X\), assumption~1 is
\(\KHolant(\cF,g^{=1})\leT\KHolant(\cF)\), and assumption~3
requires a \(\KHolant(\cF,g^{=1})\)-instance \(J\) with \(Z_X(J)\ne0\).
With item~2 imposed using these \(X\)-edges, the conclusion is
\[
  \KHolant(\cF,h)\leT\KHolant(\cF).
\]
\end{lemma}

\begin{proof}
By the ordered chain composition in item~2, iterating the gadget
produces \(g\otimes h^{\otimes k}\) with one \(g\)-vertex, size \(O(k)\),
and a nonzero scalar.  Hence an instance containing one \(g\) and \(k\) designated
\(h\)-vertices reduces to an instance containing exactly one \(g\). Divide
out the scalar.  For an instance \(\Omega\) over \(\cF\cup\{h\}\), take the
disjoint union \(\Omega\sqcup J\).  Its value is
\(Z_I(\Omega)Z_I(J)\), so division by
the fixed nonzero number \(Z_I(J)\) inserts the one carrier required by the
first reduction.  The context and \(J\) are fixed, so the construction has
size linear in the number of \(h\)-occurrences.  Composing with item~1
proves the ordinary-Holant claim. For K-Holant, the same construction
uses \(X\)-edges and the disjoint-union value is
\(Z_X(\Omega)Z_X(J)\). Division by the stipulated nonzero \(Z_X(J)\)
and the K-Holant version of item~1 give its stated reduction.
\end{proof}

\begin{center}
\begin{tikzpicture}[font=\normalsize,box/.style={draw,minimum width=9mm,minimum height=7mm}]
\node[draw,circle] (g) at (0,0) {\(g\)};
\node[box] (c1) at (1.6,0) {\(C_1\)};
\node (dots) at (3.1,0) {\(\cdots\)};
\node[box] (ck) at (4.6,0) {\(C_k\)};
\draw (g)--node[above]{\(A_1\)}(c1);
\draw (c1)--node[above]{\(P_1\)}(dots);
\draw (dots)--node[above]{\(A_k\)}(ck);
\draw (ck)--node[above]{\(P_k\)}(6,0);
\draw (c1.south)--node[right]{\(H_1\)}(1.6,-.8);
\draw (ck.south)--node[right]{\(H_k\)}(4.6,-.8);
\end{tikzpicture}
\par\normalsize The chain in item~2 of \cref{lem:single-occurrence-replication}.
Each connection represents the corresponding ordered block of edges.
\end{center}

Once these reductions provide a nonzero unary, the imported odd-arity
dichotomy applies to the extended set. Its conclusion transfers to the
original set as follows.

\begin{lemma}[Dichotomy with a nonzero unary]
\label{lem:common-unary-exit}
Let \(\cF\) be a signature set and let \(u\ne0\) be unary.
In ordinary Holant,
if \(\Holant(\cF,u)\leT\Holant(\cF)\), then
\(\Holant(\cF)\) is \(\SharpP\)-hard unless \(\Tract(\cF)\) holds. In the disequality-edge model,
if \(\KHolant(\cF,u)\leT\KHolant(\cF)\), then
\(\KHolant(\cF)\) is \(\SharpP\)-hard unless \(\Tract(K\cF)\) holds.
\end{lemma}

\begin{proof}
For the first assertion, apply \cref{thm:external-odd} to
\(\cF\cup\{u\}\).  A tractable presentation of that set restricts to one for
\(\cF\).  For the second, transform to ordinary coordinates:
\[
  \KHolant(\cF,u)
  \equivT
  \Holant(K\cF,Ku).
\]
The signature \(Ku\) is a nonzero unary, so the same argument applies.
\end{proof}

For unary columns \(g,u\), their contraction is the bilinear scalar
\[
  g^{\mathsf T}u:=\sum_{x\in\{0,1\}}g(x)u(x),
\]
with no complex conjugation.
For the higher-equality construction below, the carrier must have a
nonzero closure. The next lemma supplies a companion unary with this
property and a reduction allowing one occurrence of the pair.

\begin{lemma}[Unary companion]
\label{lem:unary-companion}
Let \(\cF\) be a fixed finite algebraic signature set and let \(g\ne0\) be
unary.  There is an algebraic unary \(u\) such that
\[
  g^{\mathsf T}u\ne0,
  \qquad
  \Holant(\cF,(g\otimes u)^{=1})\leT\Holant(\cF).
\]
\end{lemma}

\begin{proof}
We first prove the claim for \(g=\DeltaZero\).  Let \(\mathcal B_\cF\) be
the set of ordered binary tensors realized by finite \(\cF\)-networks with
ordinary equality edges.  Deleting one occurrence of
\(\DeltaZero\otimes u\), where \(u=[s,t]^{\mathsf T}\), leaves a binary
context \(H=(h_{ij})\), and the desired value is
\(sh_{00}+th_{01}\).  Equality-closing the exposed inputs gives the oracle value
\[
  S:=h_{00}+h_{11}.
\]
We choose \(u\) by the following exhaustive alternatives.

If every \(H\in\mathcal B_\cF\) satisfies \(h_{00}=h_{11}\), take
\(u=[2,0]^{\mathsf T}\). Then the desired value is \(S\).  Otherwise fix
\[
  G_1=\begin{pmatrix}a_1&b_1\\c_1&d_1\end{pmatrix}
  \in\mathcal B_\cF,
  \qquad a_1\ne d_1.
\]
If every realized binary has equal off-diagonal entries, then
\(b_1=c_1=b\).  For \(b=0\), take \(u=[1,0]^{\mathsf T}\).  Inserting
\(G_1\) gives the oracle value \(T:=a_1h_{00}+d_1h_{11}\), whence
\[
  h_{00}=\frac{T-d_1S}{a_1-d_1}.
\]
For \(b\ne0\), take \(u=[a_1-d_1,2b]^{\mathsf T}\).  The same insertion,
now using \(h_{01}=h_{10}\), gives
\[
  T-d_1S=(a_1-d_1)h_{00}+2bh_{01}.
\]

It remains that some
\[
  G_2=\begin{pmatrix}a_2&b_2\\c_2&d_2\end{pmatrix}
  \in\mathcal B_\cF
  \quad\text{satisfies}\quad b_2\ne c_2.
\]
Take \(u=[a_1-d_1,b_1+c_1]^{\mathsf T}\).  Let \(T_1\) be the value obtained
by inserting \(G_1\), and let \(T_2,T_3\) use the two orientations of
\(G_2\).  Then
\[
  \frac{T_2-T_3}{b_2-c_2}=h_{01}-h_{10},
\]
and therefore
\[
  T_1-d_1S+
  \frac{c_1}{b_2-c_2}(T_2-T_3)
  =(a_1-d_1)h_{00}+(b_1+c_1)h_{01}.
\]
In every case the first entry of \(u\) is nonzero, so
\(\DeltaZero^{\mathsf T}u\ne0\).

Now write the unary column vector \(g=[a,b]^{\mathsf T}\).  If
\(a^2+b^2=0\), then projectively
\(g\doteq[1,i]^{\mathsf T}\) or \(g\doteq[1,-i]^{\mathsf T}\).  In the first
case take \(u=[1,-i]^{\mathsf T}\), and in the second take
\(u=[1,i]^{\mathsf T}\).  Under the ordinary-to-disequality-edge
\(K^{-1}\)-transformation the pair is, up to
a nonzero scalar and input order,
\(\DeltaZero\otimes\DeltaOne\).  The reduction follows from
\cref{lem:universal-directed-pair}, and \(g^{\mathsf T}u\ne0\).

If \(a^2+b^2\ne0\), choose the scalar \(r\in\AlgNums\) with
\(r^2=a^2+b^2\) and define the matrix
\[
  Q=\frac1r\begin{pmatrix}a&b\\b&-a\end{pmatrix}.
\]
Then \(Q^{\mathsf T}Q=I\) and \(Qg=r\DeltaZero\).  Apply the point-mass
case to \(Q\cF\), obtaining \(u'=[s,t]^{\mathsf T}\) with \(s\ne0\), and
define the unary vector \(u=Q^{-1}u'\).  Because \(Q^{\mathsf T}Q=I\), the
holographic identity leaves every ordinary equality edge unchanged and pulls
the reduction back to \(\cF\), while
\[
  g^{\mathsf T}u=(Qg)^{\mathsf T}u'=rs\ne0.\qedhere
\]
\end{proof}

\begin{corollary}[Higher equality exposes unaries]
\label{cor:higher-equality-unary}
Let \(d\ge4\) and let \(\cF\) be a finite algebraic signature set.
In ordinary Holant, the reduction
\(\Holant(\cF\cup\{\Eq_d\})\leT\Holant(\cF)\) implies
\[
 \Holant(\cF\cup\{\DeltaZero,\DeltaOne\})\leT\Holant(\cF).
\]
In the disequality-edge model, the reduction
\(\KHolant(\cF\cup\{\Eq_d\})\leT\KHolant(\cF)\) implies
\[
 \KHolant(\cF\cup\{\DeltaZero,\DeltaOne\})\leT\KHolant(\cF).
\]
\end{corollary}

\begin{proof}
First work with ordinary equality edges and add the accessible \(\Eq_d\).
For \(j\in\{0,1\}\), apply \cref{lem:unary-companion} to
\({\mathbf{e}}_j=\delta_j\), obtaining \(u_j\) with \(u_j(j)\ne0\) and the
one-occurrence carrier \(v_j={\mathbf{e}}_j\otimes u_j\).  Attach the \({\mathbf{e}}_j\)-input of
that carrier to one input of \(\Eq_d\).  Up to input order the output is
\[
  v_j\otimes {\mathbf{e}}_j^{\otimes(d-2)}.
\]
Closing the carrier through equality has value \(u_j(j)\ne0\), so
\cref{lem:single-occurrence-replication} gives access to
\({\mathbf{e}}_j^{\otimes(d-2)}\).  Passing temporarily to \(K\)-coordinates, the unary
\(K^{-1}{\mathbf{e}}_j\) has two nonzero entries. Hence this tensor power is not in the
common-one-sided exception of \cref{thm:external-factor}.  Factor extraction
therefore exposes \({\mathbf{e}}_j\) after transforming back.

Now work in \(\KHolant\) and use the carrier \(\Delta\).  Attach its two factors,
through \(X\)-edges and in the two orientations, to two copies of
\(\Eq_d\).  The output is
\[
  \Delta^{\otimes(d-1)}
  =\Delta\otimes\Delta^{\otimes(d-2)}
\]
up to input order.  The \(X\)-contraction of \(\Delta\) has value one, so
\cref{lem:universal-directed-pair,lem:single-occurrence-replication} give
access to \(\Delta^{\otimes(d-2)}\).  The available \(\Eq_d\) contains both
endpoint strings \(\zero^d,\one^d\), of doubled charges \(-d,+d\), and hence
excludes the common-one-sided exception. Applying
\cref{thm:external-factor} exposes \(\DeltaZero\) and \(\DeltaOne\).  Finally
compose with the assumed reduction that removes \(\Eq_d\).
\end{proof}

Either point mass is a nonzero unary signature, so
\cref{lem:common-unary-exit} determines the complexity of the original
problem. Thus a reduction obtaining \(\Eq_d\), for \(d\ge4\), ends the
current structural case.

\subsection{Equality and generalized-equality anchors}
\label{proofsubsec:equality-anchor}

We now apply the higher-equality reduction after a common change of
basis, and then adapt the carrier construction to arbitrary nonzero
endpoint weights. The first application requires the same basis change to turn
the edge into equality and the quaternary signature into a multiple of
four-ary equality. We call this configuration an equality anchor.

\begin{definition}[Equality anchor]
\label{def:equality-anchor}
A signature set \(\cF\) has an \emph{equality anchor} if it directly
realizes an ordered quaternary signature \(g\) for which there exist a matrix
\(U\in\GL_2(\AlgNums)\) and nonzero algebraic scalars
\(\kappa,\lambda\in\AlgNums^\times\) satisfying
\[
  (U^{-1})^{\mathsf T}XU^{-1}=\kappa\,(\Eq_2),
  \qquad
  U^{\otimes4}g=\lambda\,(\Eq_4).
\]
\end{definition}

\begin{lemma}[Dichotomy from a transformed quaternary equality]
\label{lem:equality-anchor}
Let \(\cF\) be a finite algebraic signature set with an equality anchor
\(g\), and fix witnesses \(U,\kappa,\lambda\) as in
\cref{def:equality-anchor}.  Then either \(\KHolant(\cF)\) is
\(\SharpP\)-hard or \(\Tract(K\cF)\) holds.
\end{lemma}

\begin{proof}
Apply the disequality-edge holographic identity with \(M=U^{-1}\) to the whole
problem.  It sends
each right signature \(f\) to \(Uf\) and the disequality edge to
\((U^{-1})^{\mathsf T}XU^{-1}=\kappa(\Eq_2)\). For a queried instance
\(\Omega\), the edge factor \(\kappa^{|E_{\rm int}(\Omega)|}\) is known.
Thus the direct anchor becomes
\(Ug=\lambda(\Eq_4)\), so \(\Eq_4\) is accessible in the ordinary-Holant
problem over \(U\cF\).  By
\cref{cor:higher-equality-unary}, a nonzero unary \(u\) is accessible there.
Pulling the reduction back exposes the nonzero unary \(U^{-1}u\) over
\(\cF\).  The conclusion is now \cref{lem:common-unary-exit}.
\end{proof}

\begin{lemma}[Dichotomy from a generalized equality]
\label{lem:pure-ge}
Suppose a retained finite algebraic signature set \(\cF\) directly realizes
the arity-\(q\) generalized-equality signature
\[
  G_q:=c_0\delta_{\zero^q}+c_1\delta_{\one^q},
  \qquad c_0,c_1\in\AlgNums^\times,\qquad q\ge4.
\]
Then either \(\KHolant(\cF)\) is \(\SharpP\)-hard or
\(\Tract(K\cF)\) holds.
\end{lemma}

\begin{proof}
Direct realizability gives
\(\KHolant(\cF\cup\{G_q\})\leT\KHolant(\cF)\).
We use this extension in the construction below.
Use the ordered carrier \(\Delta\) from
\cref{subsec:single-occurrence-carriers}.  Attach its
two factors through \(X\)-edges, in opposite orientations, to one input
of each of two copies of \(G_q\).  The two copies are forced onto opposite
endpoint strings, and the external tensor is
\[
  c_0c_1\,\Delta^{\otimes(q-1)}
  =
  c_0c_1\,\Delta\otimes\Delta^{\otimes(q-2)}
\]
up to input order.  The \(X\)-contraction of the carrier has value one.
Thus \cref{lem:universal-directed-pair,lem:single-occurrence-replication}
give access to \(\Delta^{\otimes(q-2)}\).  Since \(G_q\) contains
both endpoint strings, of doubled charges \(-q,+q\),
\cref{thm:external-factor} has no common-one-sided
exception and exposes \(\DeltaZero\) and \(\DeltaOne\).  Apply
\cref{lem:common-unary-exit} over \(\cF\cup\{G_q\}\).  A hard conclusion composes
back to \(\cF\), while
\(\Tract(K(\cF\cup\{G_q\}))\) implies \(\Tract(K\cF)\) by
\cref{lem:downward-inheritance}.
\end{proof}
\section{Eight-Vertex Signatures with Nonzero Endpoints}
\label{sec:p1-eight-vertex}

The eight-vertex dichotomy classifies the problem with a single
quaternary signature. When that signature is realized over a larger set,
its hard case transfers immediately, but its tractable cases do not
classify the other signatures. We analyze those product, affine, and
local-affine cases using the unary and equality reductions of
\cref{sec:p1-shared-terminals}. They either classify the whole retained
set or factor the original quaternary into weighted equalities. Its
irreducibility in \cref{thm:p1-quaternary-kholant-dichotomy} rules out
this last outcome. The group analysis also calls this reduction on
possibly reducible quaternary signatures and can retain their factors.

\begin{theorem}[Quaternary eight-vertex boundary
{\cite[arXiv version~2, Theorem~3.1]{CaiFu2023}}]
\label{thm:external-eight-vertex}
Let \(M\) be an algebraic quaternary signature satisfying
\[
 \supp(M)\subseteq\{{\mathbf{x}}\in\{0,1\}^4:\wt({\mathbf{x}})=2\}\cup\{0000,1111\},\qquad
 M(0000)M(1111)\ne0.
\]
Then \(\Holant(X\mid M)\) is \(\SharpP\)-hard unless there exist
\(\cC\in\{\cA,\cP,\cL\}\) and \(U\in\GL_2(\AlgNums)\) such that
\[
  U^{\mathsf T}XU\in\cC,\qquad U^{-1}M\in\cC.
\]
Every exception is in \(\FP\).  The source permits \(U\) over \(\mathbb C\).
\Cref{lem:algebraic-transform-witness} makes it algebraic.
\end{theorem}

\begin{definition}[Endpoint-nondegenerate eight-vertex signature]
\label{def:p1-endpoint-eight-vertex}
An \emph{endpoint-nondegenerate eight-vertex signature} is a quaternary
signature \(q\) satisfying
\[
 q({\mathbf{x}})=0\quad\text{when \(\wt({\mathbf{x}})\) is odd},
 \qquad q(0000)q(1111)\ne0.
\]
Thus its support is contained in the eight even-parity strings, while both
endpoint values are nonzero.
\end{definition}
We use the two entries \(q(0000),q(1111)\), the six weight-two entries, and
the three flattenings with row--column splits
\(12\mid34\), \(13\mid24\), and \(14\mid23\), respectively. For example,
\(12\mid34\) uses \((x_1,x_2)\) as row index and \((x_3,x_4)\) as column
index.  We list their ranks in this order.
A complementary pair of two-element input blocks is \(J,J^c\subseteq[4]\) with
\(|J|=2\).

We first handle product transformability and reduce local-affine
transformability to affine transformability. The affine case then uses
the finite interfaces in
\cref{thm:cert-interface-affine-q4-normal-form,thm:cert-interface-affine-q4-gadget-exits,thm:cert-interface-affine-q4-odd}.
For an irreducible quaternary, the continuing factorization is impossible,
giving \cref{thm:p1-quaternary-kholant-dichotomy}.

Throughout this section, weighted equality and nonsingularity have the
meaning fixed in \cref{subsec:holant-definitions}.  For a signature class
\(\cC\), the statement that one signature \(M\) is \(\cC\)-transformable
uses the singleton convention: the set \(\{M\}\) and the transformed disequality edge have one common transformability witness, as in
\cref{def:common-transformability}.

\subsection{Product and local-affine exceptions}

For a product-transformable signature, we distinguish the forms that
yield an equality reduction from the form that splits into two binaries.

\begin{lemma}[Product-transformable eight-vertex signatures]
\label{lem:p1-eight-vertex-product-transform}\leavevmode\newline
Let \(M\) be an endpoint-nondegenerate eight-vertex signature.  If \(M\) is
\(\cP\)-transformable in the sense of
\cref{thm:external-eight-vertex}, then exactly one of the following
structural outcomes applies (up to a permutation of inputs):
\begin{enumerate}
\item \(M\) is a generalized equality;
\item \(M\) is a tensor product of two nonsingular weighted equalities;
\item \(M\) has full even-parity support, has the dense form
      \cref{eq:p1-eight-vertex-product-dense-form}, and every retained set
      directly realizing \(M\) also directly realizes an equality anchor.
\end{enumerate}
In particular, every
\(\cP\)-transformable endpoint-nondegenerate eight-vertex signature has support size \(2\), \(4\), or
\(8\).
\end{lemma}

\begin{proof}
Choose an algebraic witness matrix
\(T=(\begin{smallmatrix}a&b\\c&d\end{smallmatrix})\) by
\cref{lem:algebraic-transform-witness}, and define the transformed signature
\(H=(T^{-1})^{\otimes4}M\in\cP\).  The transformed edge is
\[
 T^{\mathsf T}XT=
 \begin{pmatrix}2ac&ad+bc\\ad+bc&2bd\end{pmatrix}.
\]
Because this transformed edge is nonsingular and belongs to \(\cP\), its
support is either equality or disequality.  Disequality support makes
\(T\) monomial.  Equality support gives \(ad+bc=0\). Define the nonzero
scalar ratio \(\lambda:=b/a\).  Then
\[
 T=\diag(a,c)\begin{pmatrix}1&\lambda\\1&-\lambda\end{pmatrix},
 \qquad ac\lambda\ne0.
\]

If \(H\) is irreducible, product type makes it a CP signature
for some \({\mathbf{s}}\in\Ftwo^4\), with nonzero coefficients
\(A,B\in\AlgNums^\times\):
\[
 H=A\delta_{\mathbf{s}}+B\delta_{\bar {\mathbf{s}}},\qquad AB\ne0.
\]
In the monomial case the two nonzero endpoint values force
\(\{{\mathbf{s}},\bar {\mathbf{s}}\}=\{\zero^4,\one^4\}\), giving outcome~1.  Otherwise, for
\(k=\wt({\mathbf{s}})\) and \(w=\wt({\mathbf{x}})\),
\[
 M({\mathbf{x}})=a^{4-w}c^w(-1)^{{\mathbf{s}}\cdot {\mathbf{x}}}
 \bigl(A\lambda^k+B\lambda^{4-k}(-1)^w\bigr).
\]
The odd-weight zeros are equivalent to \(B/A=\lambda^{2k-4}\).  Hence all
even entries are nonzero.  Define the nonzero overall scale
\(\rho:=2A\lambda^k a^4\) and weight-ratio scalar \(r:=c/a\). The entries
then have the form
\begin{equation}
 M({\mathbf{x}})=\rho r^{\wt({\mathbf{x}})}(-1)^{{\mathbf{s}}\cdot {\mathbf{x}}}
 \one[\wt({\mathbf{x}})\text{ is even}],\qquad \rho r\ne0.
 \label{eq:p1-eight-vertex-product-dense-form}
\end{equation}

For this dense case choose nonzero algebraic scalars
\(q_0,q_1\in\AlgNums^\times\) satisfying the first two equations below,
define the diagonal matrix \(Q:=\diag(q_0,q_1)\), and define the
normalization matrix \(R:=QT^{-1}\):
\[
 q_0^2=2ac,\qquad q_1/q_0=i\lambda,\qquad R=QT^{-1}.
\]
Then, since
\(\lambda^{2k-4}(i\lambda)^{4-2k}=(-1)^k\), there is a nonzero scalar
\(\eta\in\AlgNums^\times\) such that
\begin{equation}
 (R^{-1})^{\mathsf T}XR^{-1}=I,\qquad
 R^{\otimes4}M=\eta\bigl(\delta_{\mathbf{s}}+(-1)^k\delta_{\bar {\mathbf{s}}}\bigr).
 \label{eq:p1-eight-vertex-product-anchor-normal-form}
\end{equation}
Define the signature
\(F_{\mathbf{s}}=\delta_{\mathbf{s}}+(-1)^k\delta_{\bar {\mathbf{s}}}\).  In these coordinates the
disequality edge is equality by \cref{eq:p1-eight-vertex-product-anchor-normal-form}.  A nonzero
multiple of \(\Eq_4\) uses at most two copies of \(F_{\mathbf{s}}\): one for
\(k=0,4\). For \(k=2\), expose the two
\(s=0\) inputs of the first and the two \(s=1\) inputs of the second, pairing
the remaining inputs. For \(k=1\), expose the three majority-zero inputs of
the first and the minority-one input of the second, join the first's
minority input to one majority input of the second, and equality-close the
second's last two majority inputs.  The construction for \(k=3\) complements all bits.
In each two-copy construction,
an all-equal external string has exactly one compatible internal assignment,
while every other external string has zero contribution.  The resulting
signature is \(\Eq_4\) for \(k=2\) and \(-(\Eq_4)\) for \(k=1,3\).
Pulling it back through \(R\) gives a
direct gadget \(g\) over \(M\) with \(R^{\otimes4}g=\lambda'(\Eq_4)\),
\(\lambda'\ne0\). Together with \cref{eq:p1-eight-vertex-product-anchor-normal-form}
this is outcome~3.

If \(H\) is not irreducible, neither is \(M\).  Its nonzero endpoints make
every factor nonzero at both endpoints.  Even parity excludes an odd-sized
block, so the only proper tensor-factor partition is
\([4]=S\sqcup S^c\) with \(|S|=|S^c|=2\), where
\(S^c=[4]\setminus S\). The odd-weight zeros then force both binary factors
to be diagonal.  Thus they are nonsingular weighted equalities and
\[
 M(\zero^4)M(\one^4)=M(\one_S)M(\one_{S^c}).
\]
This is outcome~2 and completes the classification.
\end{proof}

\begin{corollary}[Reductions from product-transformable eight-vertex signatures]
\label{cor:p1-eight-vertex-product-operational}
Let \(\cF\) be a retained, factor-closed finite algebraic signature
set that directly realizes a product-transformable endpoint-nondegenerate eight-vertex signature
\(M\).  Then at least one of the following holds:
\begin{enumerate}[label=\textnormal{(\roman*)}]
\item \(\KHolant(\cF)\) is \(\SharpP\)-hard;
\item \(\Tract(K\cF)\) holds;
\item both factors in the two-binary factorization of \(M\), including
      a nonsingular weighted-equality factor, can be added to \(\cF\)
      by the factor-closure reduction. The extension is finite, retained,
      and factor-closed, and its \(\KHolant\) problem reduces to
      \(\KHolant(\cF)\).
\end{enumerate}
\end{corollary}

\begin{proof}
Apply \cref{lem:p1-eight-vertex-product-transform}.  Its generalized-equality
form is closed by \cref{lem:pure-ge}, and its dense form is closed by
\cref{lem:equality-anchor}.  These give alternatives~(i)--(ii), with
hardness composing to \(\cF\) and tractability descending to
\(\cF\) by \cref{lem:downward-inheritance}.

In the remaining form, add the directly realized \(M\) and retain the
complete factor batch from its decomposition across two pairs of inputs.  The
endpoint-nondegenerate eight-vertex signature is nonzero on both endpoint
strings \(\zero^4\) and \(\one^4\), whose doubled charges are \(-4\) and
\(+4\), respectively, so the common-one-sided exception in
\cref{lem:factor-saturation} is impossible.  The resulting finite
factor-closed signature set reduces to \(\cF\) and
contains both nonsingular weighted-equality factors.  This is
alternative~(iii).
\end{proof}

It remains to treat the affine and local-affine cases. The next lemma
shows that even parity lets us replace a local-affine witness by an
affine witness, so one affine analysis will cover both.

\begin{lemma}[Local-affine reduction to affine]
\label{lem:p1-eight-vertex-local-affine-collapse}
Let \(M\) be a nonzero quaternary even-parity signature.  If one matrix
\(T\in\GL_2(\mathbb C)\) satisfies
\[
  T^{\mathsf T}XT\in\cL,
  \qquad (T^{-1})^{\otimes4}M\in\cL,
\]
then there is a matrix \(T_0\in\GL_2(\mathbb C)\) with
\[
  T_0^{\mathsf T}XT_0\in\cA,
  \qquad (T_0^{-1})^{\otimes4}M\in\cA.
\]
If \(T\in\GL_2(\AlgNums)\), then \(T_0\) is algebraic.
\end{lemma}

\begin{proof}
For any nonempty affine set \(S\subseteq\Ftwo^n\), choose a base point
\({\mathbf{s}}\in S\) and write \(S={\mathbf{s}}+C\), where \(C\leq\Ftwo^n\) is a linear subspace.  The subspace

\[
  C=\{{\mathbf{x}}\oplus {\mathbf{y}}:{\mathbf{x}},{\mathbf{y}}\in S\},
\]

is independent of \({\mathbf{s}}\).  It is self-orthogonal if
\({\mathbf{u}}\mathbin{\cdot}{\mathbf{v}}=0\) over \(\Ftwo\) for all \({\mathbf{u}},{\mathbf{v}}\in C\).

Use the phase scalar \(\zeta_8\) from \cref{subsec:standard-classes}.  For an arity-\(n\) signature
\(F\) and \({\mathbf{z}},{\mathbf{y}}\in\Ftwo^n\), define the local twist selected by \({\mathbf{z}}\) as
\[
  \Theta_{\mathbf{z}}:=\bigotimes_{j=1}^nT_2^{z_j},
  \qquad
  (\Theta_{\mathbf{z}}F)({\mathbf{y}})
  =\zeta_8^{\langle {\mathbf{z}},{\mathbf{y}}\rangle_{\mathbb Z}}F({\mathbf{y}}),
  \qquad
  \langle {\mathbf{z}},{\mathbf{y}}\rangle_{\mathbb Z}:=\sum_{j=1}^n z_jy_j.
\]
By the local-affine criterion
\cite[Definition~3.1 and Theorem~3.2]{CaiLuXia2018RealHolantc}, the
translation space of every support is self-orthogonal, and
\(\Theta_{\mathbf{z}}F\in\cA\) for every \({\mathbf{z}}\in\supp(F)\) when \(F\in\cL\).  Apply
this first to
the symmetric nonsingular binary \(E=T^{\mathsf T}XT\).  Self-orthogonality
rules
out full support, and nonsingularity rules out singleton support. Hence
\(\supp(E)\) is equality or disequality.  Disequality support would give
\[
  E=pX\notin\cL,
\]
since \(01\in\supp(E)\) and the twist required by this string in the support has
relative phase \(\zeta_8\notin\mu_4\), so it is not affine.  Thus
\(E=\diag(p,q)\), where \(p,q\in\AlgNums^\times\) are its nonzero diagonal
entries (their algebraicity follows from \(E\in\cL\)). The \(00\)-twist is
the identity, so
\(q/p\in\mu_4\).  Writing
\(T=(\begin{smallmatrix}a&b\\c&d\end{smallmatrix})\), the off-diagonal
equation yields \(q/p=-(b/a)^2\).  Hence, with the scalar parameters
\(\rho:=b/a\in\mu_8\) and \(t:=c/a\in\mathbb C^\times\), using the diagonal family defined in
\cref{subsec:holant-definitions},
\[
  T=aD(t)PD(\rho),
  \quad at\ne0,\quad \rho^8=1.
\]
Remove the right factor \(D(\rho)\) by defining \(T_0:=aD(t)P\).
To prove that this is an affine witness, we examine the resulting
signature through the sum
\[
  W_t({\mathbf{y}})=
  \sum_{\substack{{\mathbf{x}}\in\Ftwo^4\\\wt({\mathbf{x}})\equiv0\pmod 2}}
  t^{-\wt({\mathbf{x}})}M({\mathbf{x}})(-1)^{{\mathbf{x}}\cdot {\mathbf{y}}}.
\]
Let \(N:=(T^{-1})^{\otimes4}M\) denote the transformed local-affine
signature.  Since \(P^{-1}=P/2\), define
\[
  \kappa_0:=(2a)^{-4}\ne0,
  \qquad
  W'_t:=\kappa_0W_t.
\]
\begin{equation}
  (T^{-1})^{\otimes4}M({\mathbf{y}})=\rho^{-\wt({\mathbf{y}})}W'_t({\mathbf{y}}),
  \qquad
  (T_0^{-1})^{\otimes4}M({\mathbf{y}})=W'_t({\mathbf{y}}),
  \label{eq:p1-eight-vertex-local-affine-delete}
\end{equation}
Since \(N\in\cL\) and \(\rho\in\mu_8\),
\(W'_t({\mathbf{y}})=\rho^{\wt({\mathbf{y}})}N({\mathbf{y}})\) is algebraic for every \({\mathbf{y}}\).
Moreover,
\[
  T_0^{\mathsf T}XT_0
  =2a^2t\diag(1,-1)
  =p\diag(1,-1)\in\cA,
\]
and parity gives \(W'_t(\bar {\mathbf{y}})=W'_t({\mathbf{y}})\).

The support of \(N\), which is also the support of
\(W'_t\), is an affine coset \({\mathbf{s}}+C\).  Self-orthogonality and the
identity \(W'_t(\bar {\mathbf{y}})=W'_t({\mathbf{y}})\), which makes this support
closed under bitwise complement, give
\(\one^4\in C\) and \(1\le\dim C\le2\).  For dimension one, the support is
a complementary pair on which \(W'_t\) has equal values, so
\(W'_t\in\cA\).
For dimension two, after permuting inputs,
\[
  C=\Span_{\Ftwo}\{\one^4,{\mathbf{w}}\},
  \qquad \wt({\mathbf{w}})=2.
\]
For \(u,v\in\Ftwo\), index the four strings in the support by
\({\mathbf{x}}_{uv}:={\mathbf{s}}\oplus u\one^4\oplus v{\mathbf{w}}\).  For a signature \(F\) that is nonzero on
these four strings, define its cross-ratio functional by
\(\operatorname{cr}(F):=F({\mathbf{x}}_{00})F({\mathbf{x}}_{11})/
(F({\mathbf{x}}_{10})F({\mathbf{x}}_{01}))\).  Complement symmetry gives
\(\operatorname{cr}(W'_t)=1\), while the nowhere-zero restriction of
any member of \(\cA\) with this support has cross ratio \(\pm1\).  The
quotient \(\Ftwo^4/C\) has exactly three nonzero classes.  Up to a permutation of inputs and \({\mathbf{w}}\leftrightarrow\bar {\mathbf{w}}\), each of the two odd cosets has
representative \(({\mathbf{s}},{\mathbf{w}})=(0001,0011)\), and the unique even coset has
representative \(({\mathbf{s}},{\mathbf{w}})=(0101,0011)\).  Thus the displayed cases exhaust
all three nonzero quotient classes.  In both representatives
the \(\rho^{-\wt({\mathbf{y}})}\) factor has cross ratio one and the direct four-point
check gives
\[
 \langle {\mathbf{s}},{\mathbf{x}}_{00}\rangle_{\mathbb Z}
 +\langle {\mathbf{s}},{\mathbf{x}}_{11}\rangle_{\mathbb Z}
 -\langle {\mathbf{s}},{\mathbf{x}}_{10}\rangle_{\mathbb Z}
 -\langle {\mathbf{s}},{\mathbf{x}}_{01}\rangle_{\mathbb Z}=2.
\]
The local twist selected by \({\mathbf{s}}\) is \(N_{\mathbf{s}}:=\Theta_{\mathbf{s}}N\), so
\(N_{\mathbf{s}}({\mathbf{y}})=\zeta_8^{\langle {\mathbf{s}},{\mathbf{y}}\rangle_{\mathbb Z}}N({\mathbf{y}})\).  It would therefore
have cross ratio \(\zeta_8^2=i\), contradicting \(N_{\mathbf{s}}\in\cA\).  Hence
\({\mathbf{s}}+C=C\).

If the complement-symmetric values of \(W'_t\) are \(A\) on the endpoints
and \(B\) on the central pair, then \(N\) has values
\((A,\rho^{-4}A,\rho^{-2}B,\rho^{-2}B)\).  Since \(0\in\supp(N)\), its
selected local twist is the identity, so \(N\in\cA\) and
\(\rho^{-2}B/A\in\mu_4\).  As \(\rho^2\in\mu_4\), also \(B/A\in\mu_4\),
whence \(W'_t\in\cA\).  Now
\cref{eq:p1-eight-vertex-local-affine-delete} proves the
claim.  If the original witness is algebraic, so are \(t\) and \(T_0\).
\end{proof}

\subsection{The finite affine quaternary certificate}
\label{proofsubsec:affine-certificate}

The affine witness may initially use an arbitrary invertible matrix.
We first reduce it to a diagonal change of basis, which preserves the
disequality edge up to scalar. The proof derives a finite list of matrix
forms and applies the classification in
Appendix~\ref{appsec:cert-eight-vertex}, checked by
\path{certificates/verify_affine_q4.py}. The corollary then uses the
resulting signatures in reductions for the whole retained set.

\begin{proposition}[Affine quaternary certificate]
\label{prop:p1-affine-eight-vertex-certificate}
Let \(M\) be an algebraic affine-transformable
endpoint-nondegenerate eight-vertex signature.  Then \(M\) admits an
algebraic invertible diagonal matrix
\(D_{\mathrm{aff}}=\diag(d_0,d_1)\), with
\(d_0,d_1\in\AlgNums^\times\), such that
\[
  (D_{\mathrm{aff}}^{-1})^{\otimes4}M\in\cA,
  \qquad D_{\mathrm{aff}}^{\mathsf T}XD_{\mathrm{aff}}\doteq X.
\]
For some \(J\subseteq[4]\) with \(|J|=2\), write
\(J^c:=[4]\setminus J\).  In these coordinates its support is exactly one
of
\[
  \{\zero^4,\one^4\},
  \qquad
  \{\zero^4,\one^4,\one_J,\one_{J^c}\},
  \qquad
  \{{\mathbf{x}}:\wt({\mathbf{x}})\text{ is even}\}.
\]
The first support is a generalized equality.
\end{proposition}

\begin{proof}
By \cref{lem:algebraic-transform-witness}, choose an algebraic affine change-of-basis matrix
\(U=(\begin{smallmatrix}a&b\\c&d\end{smallmatrix})\), and define the
transformed binary edge \(E:=U^{\mathsf T}XU\).  A nonsingular symmetric affine binary has
disequality support, equality support, or full support.  In the first
case,
\[
  U=D_0K_0,\qquad K_0\in\{I,X\},
\]
for an invertible diagonal \(D_0\).
In the second, \(ad+bc=0\), so
\[
 U=D_0\begin{pmatrix}1&r\\1&-r\end{pmatrix},
 \qquad r\in\mu_8,
\]
for an invertible diagonal \(D_0\).  In the full-support case, after
normalizing its \(00\)-entry, \(E\) has the affine form
\[
 \begin{pmatrix}1&u\\u&-u^2\end{pmatrix},
 \qquad u\in\mu_4.
\]
Writing \(U=D_0(\begin{smallmatrix}1&r\\1&s\end{smallmatrix})\) gives
\(r+s=2u\) and \(rs=-u^2\). Hence
\(\{r,s\}=\{u(1+\sqrt2),u(1-\sqrt2)\}\).
Denote the resulting eighteen-element set of representative matrices by
\(\mathscr K_{\mathrm{aff}}\). Its explicit list appears in
\cref{thm:cert-interface-affine-q4-normal-form}.  These three calculations give precisely
\[
 U=D_0K_0,\qquad K_0\in\mathscr K_{\mathrm{aff}},
\]
which proves completeness of the list.

Now define the affine signature \(f=(U^{-1})^{\otimes4}M\in\cA\).  Since
\(M\) is an endpoint-nondegenerate eight-vertex signature and \(D_0\)
is diagonal, \(K_0^{\otimes4}f\) satisfies the hypotheses of
\cref{thm:cert-interface-affine-q4-normal-form}.  The certificate is projective, so the
normalization factor for \(f\) can be absorbed into a global scalar.
The same classification theorem therefore yields
\[
  \gamma,t\in\AlgNums^\times
\]
and a normalized affine signature \(h\in\cA\), with
\(h(\zero^4)=1\), such that
\[
 K_0^{\otimes4}f
 =\gamma D(t)^{\otimes4}h.
\]
Moreover, \(\supp(h)\) is one of the three supports displayed above.
Consequently the
single algebraic diagonal matrix \(D_\star=D_0D(t)\) satisfies
\((D_\star^{-1})^{\otimes4}M\doteq h\), and
\(D_\star^{\mathsf T}XD_\star\doteq X\).  Taking
\(D_{\mathrm{aff}}=D_\star\) proves the asserted diagonal
normalization.
\end{proof}

\begin{corollary}[Reductions from affine quaternaries]
\label{cor:p1-affine-eight-vertex-operational}
Let \(\cF\) be a retained, factor-closed finite algebraic signature
set that directly realizes an affine-transformable endpoint-nondegenerate eight-vertex signature
\(M\).
Then at least one of the following holds:
\begin{enumerate}[label=\textnormal{(\roman*)}]
\item \(\KHolant(\cF)\) is \(\SharpP\)-hard;
\item \(\Tract(K\cF)\) holds;
\item the complete factor batch from the gadget construction can be
      added to \(\cF\) by reduction, giving a finite retained factor-closed
      extension whose \(\KHolant\) problem reduces to \(\KHolant(\cF)\). The original
      signature \(M\) has, up to a permutation of inputs and a nonzero
      scalar, a genuine factorization across two pairs of inputs.
      One of its proper binary factors is a nonsingular weighted equality
      \(E:=\diag(p,q)\), where \(p,q\in\AlgNums^\times\), retained in
      this extension through the factor-closure reduction.
\end{enumerate}
\end{corollary}

\begin{proof}
Choose \(D_\star=D_{\mathrm{aff}}\) from
\cref{prop:p1-affine-eight-vertex-certificate}.  Apply the global change
\(D_\star^{-1}\), and normalize the resulting quaternary to an affine
signature \(h\) with \(h(\zero^4)=1\).  Thus
\[
 M\doteq D_\star^{\otimes4}h.
\]
Keep every direct output over the current signature set, and add proper
tensor factors only through factor closure.  Record these
constructions and the chosen normal form in the normalization data of
\cref{lem:normalization-ledger}. The normalized signature \(h\) has
nonzero entries at both endpoints, so every resulting set is two-sided and
\cref{lem:factor-saturation} applies.

The normal form has three possible supports. The support-two form is a
generalized equality. For support four, the distinction between an
equality construction and a factorization is determined by the following
coefficient matrix. Let \(J\sqcup J^c=[4]\) be the two pairs of inputs.  For \(u,v\in\Ftwo\), put
\({\mathbf x}_{uv}:=u\one_J\oplus v\one_{J^c}\), and define the logical coefficient matrix by
\[
 F:=\bigl(h({\mathbf x}_{uv})\bigr)_{u,v\in\Ftwo}
   =\begin{pmatrix}A&B\\ C&D\end{pmatrix},\qquad ABCD\ne0.
\]
Here \(u\) is the row bit and \(v\) is the column bit.
For an ordered binary tensor \(E\), write \(E^{(J)}\) for its placement on
the two inputs of \(J\) in their inherited order, and define
\(E^{(J^c)}\) analogously.
If \(AD=-BC\), the two-copy \(X\)-edge gadget in
\cref{thm:cert-interface-affine-q4-gadget-exits}\textnormal{(ii)} directly realizes a nonzero pure
generalized equality.  If \(AD=BC\), then, on these two blocks,
\[
  h=E_1^{(J)}\otimes E_2^{(J^c)}
\]
for nonsingular weighted equalities \(E_1,E_2\).  Since \(D_\star\) is
diagonal,
\[
  M\doteq
  (D_\star E_1D_\star^{\mathsf T})^{(J)}\otimes
  (D_\star E_2D_\star^{\mathsf T})^{(J^c)}.
\]
Thus the original endpoint-nondegenerate eight-vertex signature itself has a
genuine factorization across two pairs of inputs into nonsingular weighted
equalities.  Add \(M\) and apply factor closure to
obtain both factors by reduction.

On full even-parity support,
\cref{thm:cert-interface-affine-q4-gadget-exits}\textnormal{(iii)}, together with
\cref{lem:p1-eight-vertex-product-transform}, gives, over the current signature set,
either a directly realized equality anchor, a directly realized nonzero
rank-one binary, or a directly realized product
\[
  cD(r)\otimes D(r),
  \qquad c\in\AlgNums^\times,
  \qquad r\in\mu_4,
\]
where \(D(r)\) is a nonsingular weighted-equality matrix.
In the last case, \cref{lem:factor-saturation} makes \(D(r)\) available
by reduction. We use it to construct an odd-parity signature and continue
the full-support analysis. Choose
the scalar \(s\in\mu_8\) with \(s^2=r^{-1}\), and use \(D(s)\) for the equality normalization.  Then
\[
  (D(s))^{\mathsf T}XD(s)=sX,
  \qquad
  D(s)D(r)(D(s))^{\mathsf T}=I.
\]
Apply this normalization and define
the transformed signature \(h_r=(D(s))^{\otimes4}h\). Attach one input of
\(h_r\) by a \(X\)-edge
to one input of the retained equality \(I\), leaving the other equality input
external.  This is a gadget over the current signature set,
and it complements that input of \(h_r\), producing, up to a nonzero scalar, an odd-parity quaternary \(p\) that is nonzero on all eight
odd-weight strings and is used in
\cref{thm:cert-interface-affine-q4-odd}.  Because \(s\in\mu_8\), its normalized
nonzero entries are fourth roots of unity.

The twelve disequality-loop and equality-mediated binary gadgets of
\cref{thm:cert-interface-affine-q4-odd} are therefore realizable at this node.  If a test gives a nonzero rank-one binary or a nonsingular binary of
infinite projective order, the unary reductions below apply. Otherwise
the same theorem classifies the remaining signatures, including the
simultaneous-zero cases of its loop tests. Its exceptional family directly
realizes the certified endpoint-nondegenerate eight-vertex signature with
support of cardinality six.  For every other remaining signature, choose a certified
\(\delta\in\mathbb Z_4\), choose \(\tau_\delta\in\mu_8\) with
\(\tau_\delta^2=i^\delta\), and apply the further legal global-normalization
matrix \(D(\tau_\delta)\).  Let \(p_\delta\) denote the certified
phase-shifted quaternary signature with odd-parity support from
\cref{thm:cert-interface-affine-q4-odd}.  Since \(p\) is supported only at
Hamming weights one and three,
\[
  D(\tau_\delta)^{\otimes4}p\doteq p_\delta.
\]
\Cref{thm:cert-interface-affine-q4-odd} now gives, through its disequality self-loop
and two-copy constructions, either a nonzero rank-one binary or a
product-transformable endpoint-nondegenerate eight-vertex quaternary in the
dense form \cref{eq:p1-eight-vertex-product-dense-form}. The latter directly
yields an equality anchor by
\cref{lem:p1-eight-vertex-product-transform}.

For the eight-vertex signature with support of size six,
\cref{lem:p1-eight-vertex-product-transform} rules out
\(\cP\)-transformability. The diagonal normalization in
\cref{prop:p1-affine-eight-vertex-certificate} preserves support and allows
only sizes \(2,4,8\), ruling out \(\cA\)-transformability.
\Cref{lem:p1-eight-vertex-local-affine-collapse} then rules out
\(\cL\)-transformability, so \cref{thm:external-eight-vertex} gives hardness.

The support-two case and the support-four case \(AD=-BC\) are resolved by
\cref{lem:pure-ge}. Equality anchors are resolved by
\cref{lem:equality-anchor}.  A nonzero rank-one binary exposes a unary by
\cref{lem:binary-interpolation}, and an infinite-order binary first
interpolates such a binary.  In both cases
\cref{lem:common-unary-exit} applies.  An exposed odd-arity signature is
resolved by \cref{thm:external-odd} after the holographic \(K\)-coordinate equivalence.
Every case with full even-parity support above therefore gives hardness or tractability.

The sole continuing case, the support-four form \(AD=BC\), gives the
displayed factorization of the original \(M\) into two nonsingular
weighted-equality factors after undoing the diagonal normalization.
By \cref{lem:normalization-ledger}, hardness reductions compose back
to \(\cF\), tractable conclusions descend to \(K\cF\), and this
continuing state returns to the original coordinates.
\end{proof}

\subsection{Consequences for the Full Signature Set}

We now combine the three transformability cases. The conclusion either
classifies the entire signature set or supplies proper binary factors of
the original quaternary for the later group analysis.

\begin{corollary}[Reductions from eight-vertex signatures]
\label{cor:p1-eight-vertex-reduced-interface}
Let \(\cF\) be a retained, factor-closed finite algebraic
signature set containing a directly realizable endpoint-nondegenerate eight-vertex signature \(M\).
Then at least one of the following holds:
\begin{enumerate}[label=\textnormal{(\roman*)}]
\item \(\KHolant(\cF)\) is \(\SharpP\)-hard;
\item \(\Tract(K\cF)\) holds;
\item the complete factor batch from a genuine factorization of the
      original \(M\) across two pairs of inputs can be added to
      \(\cF\) by reduction, giving a finite retained factor-closed
      extension whose \(\KHolant\) problem reduces to \(\KHolant(\cF)\). This extension
      contains a proper tensor factor of \(M\) of the form
      \[
        E:=\diag(p,q),\qquad p,q\in\AlgNums^\times.
      \]
\end{enumerate}
In item~(iii), the factor \(E\) is available by reduction from the original \(M\).
\end{corollary}

\begin{proof}
If all six weight-two entries vanish, the signature is a generalized
equality and is closed by
\cref{lem:pure-ge}.  Otherwise apply the Cai--Fu quaternary boundary
\cref{thm:external-eight-vertex}.  Its hard branch is item~(i).  Its
product-transformable branch is handled by
\cref{cor:p1-eight-vertex-product-operational}.  A local-affine branch first
collapses to affine by \cref{lem:p1-eight-vertex-local-affine-collapse} and then
uses \cref{cor:p1-affine-eight-vertex-operational}. An affine branch uses that corollary directly.

These two corollaries give the three stated outcomes.
Composing the reductions gives hardness for \(\cF\), and restricting the
tractability criterion gives the same conclusion for \(K\cF\).  Apply
\cref{lem:normalization-ledger} to return to the original basis.
The sole continuing output is the retained nonsingular weighted-equality
factor in item~(iii).
\end{proof}

The continuing outcome factors the original quaternary signature, so the
irreducibility hypothesis in \cref{thm:p1-quaternary-kholant-dichotomy} rules it out.

\begin{proof}[Proof of \cref{thm:p1-quaternary-kholant-dichotomy}]
Choose an endpoint-nondegenerate irreducible quaternary
\(q\in\cF\).  Apply \cref{subsec:zero-nullary}, retaining the name \(\cF\).
The signature \(q\) survives, and the tractability criterion is unchanged.
By the \hyperlink{eq:retract-K-coordinates}{coordinate identity in
Section~\ref{subsec:tract-predicate}}, the criterion in the theorem is
\(\ReTract(K\cF)\) in ordinary-Holant coordinates.
Since \(q\) is two-sided, \cref{lem:tract-X-anchor} gives
\[
 \Tract(K\cF)\quad\Longleftrightarrow\quad\ReTract(K\cF).
\]
If \(\ReTract(K\cF)\) holds, then
\cref{thm:predicate-algorithms,eq:K-equivalence} gives
\(\KHolant(\cF)\in\FP\), and the root evaluator restores the deleted
nullary scalars. Now suppose that \(\ReTract(K\cF)\) fails, so
\(\Tract(K\cF)\) fails as well. The two-sided signature \(q\)
excludes the common-one-sided exception. By
\cref{lem:factor-saturation}, adding \(\Pi(\cF)\) gives a finite
factor-closed extension equivalent to the original problem. By
\cref{lem:downward-inheritance}, its tractability criterion also fails.
Continue to denote this extension by \(\cF\). It still contains the
original irreducible signature \(q\).

Apply \cref{cor:p1-eight-vertex-reduced-interface} to \(q\) over
\(\cF\).  Its tractable case is excluded by the preceding paragraph.
Its continuing outcome gives a genuine \(2+2\) factorization of the original
\(q\), contradicting irreducibility. Hence \(\KHolant(\cF)\) is
\(\SharpP\)-hard.
The factor-closure equivalence and preprocessing transfer this hardness
to the original input problem. Finally, the same predicate equivalence and
\cref{thm:predicate-algorithms} give decidability.
\end{proof}
\section{Equality-Accessible Matching Decks}
\label{sec:weighted-equality}
\label{sec:p1-binary-phase}

This section organizes the structural proof of the binary-disequality
dichotomy. By the equivalence below, we work with disequality edges and
the supplied binary equality \(I=(\Eq_2)\). The central task is to combine
the classification of binary signatures with reductions of higher-arity
irreducible signatures. Binary transfers determine the available
contractions, while factors of their outputs can supply new transfers
and change the group under analysis.

After the unary reductions have settled the singular and infinite-order
binary cases, the remaining transfers form a finite group. The supplied
\(I\) gives the transfer class \([X]\), and reversal makes \([Z]\)
normalize the group. These two distinguished classes restrict its form.
We then choose a minimum-arity nonbinary irreducible signature and
contract it using the available binaries. Retaining the outputs and their
factors can settle the complexity, lower the minimum arity, or
enlarge the complete transfer group. When the remaining process stabilizes, the
contraction outputs are products of binaries on perfect matchings.
This is the structure analyzed in the four group branches.

The progress mechanism gives those branches a common task: determine the
complexity of the current set or produce a reduction that advances the
structural analysis. A fixed number field bounds strict group enlargement,
and finite processing controls the steps that leave both the group and
minimum arity unchanged. A further induction treats the unresolved
irreducible occurrences in the whole set. These ingredients connect the
local constructions in \cref{sec:normalized-dihedral,sec:proper-deck,sec:v4-deck,sec:platonic-decks}
to the final dichotomy proof in \cref{sec:p1-synthesis}.

Throughout, the supplied \(I\) stays in the original coordinates.
The eight-vertex reductions return to these coordinates, and any additional
weighted equality is retained as a binary signature. Continuing changes
of basis preserve both \(X\) and \(I\) projectively. Other basis changes
are used within reductions that settle the complexity. A \emph{resolved
complexity leaf}, or \emph{resolved leaf}, means hardness,
\(\Tract\) for the entire retained set,
or a common tractable \(K\)-presentation as in
\cref{def:common-K-presentation}.

\begin{lemma}[Equality-edge-model equivalence]
\label{lem:p1-equality-wire-equivalence}
For every signature set \(\cF\),
\begin{equation}
  \KHolant(\cF,I)
  \equivT
  \Holant(\cF,X),
  \label{eq:intro-wire-bridge}
\end{equation}
where the right side uses ordinary equality edges and has \(X\) as an
available binary signature.
\end{lemma}

\begin{proof}
Forward, subdivide each \(X\)-edge by an \(X\)-vertex and suppress
degree-two \(I\)-vertices as equality edges.  Conversely, splice the two
incident inputs of the ordinary \(X\)-vertices after first preprocessing the
subgraph induced by those vertices.  Every such component is a path or a
cycle, because \(X\) has arity two.  A path with two incidences at
non-\(X\) vertices collapses to \(I\) or \(X\) according to the parity of
the number of \(X\)-vertices. A cycle with no external incidence contributes
zero for odd length and the scalar \(2\) for even length.  (A self-loop is
the length-one case.)  These are exact identities obtained by summing the
internal equality-edge bits, and they also cover adjacent \(X\)-vertices and
parallel edges.  For each resulting \(X\) kernel use one structural
\(\mathsf E_X\) edge (no additional tensor vertex). For each resulting
\(I\) kernel, and for each ordinary equality edge whose endpoints are not
in an \(X\)-component, use
\[
  \text{input}-X-I-X-\text{input}.
\]
Indeed, \(\sum_{u,v}X(a,u)I(u,v)X(v,b)=I(a,b)\).  Multiply the final
answer by the known product of the even-cycle scalars (or return zero for
an odd cycle).  The preprocessing and gadget replacements are linear-size and
preserve ordered incidences.
\end{proof}

The signatures in \(\cF\) keep their entries under this gadget replacement.
The four tractable cases correspond as well:
\hypertarget{eq:retract-edge-bridge}{}
\[
 \ReTract(\cF\cup\{X\})
 \quad\Longleftrightarrow\quad
 \ReTract_X(\cF\cup\{I\})
 \quad\Longleftrightarrow\quad
 \ReTract(K(\cF\cup\{I\})).
\]
The last equivalence is the \hyperlink{eq:retract-K-coordinates}{coordinate
identity in Section~\ref{subsec:tract-predicate}}. The arity-two case is
unchanged by adjoining either binary. For
\(\cC\in\{\cA,\cP,\cL\}\), nonsingular symmetric binaries in
\(\cC\) are closed under inversion and under \(EBE\).
The latter is \cref{lem:symmetric-binary-pullback}. Inversion preserves
the binary affine and product normal forms, and the local-affine forms
\(a\diag(1,i^k)\) identified in its proof.
For an ordinary-Holant witness \(M\), put
\(E=M^{\mathsf T}M\) and \(B=M^{-1}XM^{-\mathsf T}\).
Then the disequality edge becomes \(M^{\mathsf T}XM=EBE\), and the
supplied \(I\) becomes \(E^{-1}\), both in \(\cC\).
Conversely, for a disequality-edge witness \(U\), put
\(E=U^{\mathsf T}XU\) and \(B=U^{-1}IU^{-\mathsf T}\).
The ordinary edge becomes \(U^{\mathsf T}U=EBE\), and the supplied
\(X\) becomes \(U^{-1}XU^{-\mathsf T}=BEB\), both in \(\cC\).
The same witness transforms every member of \(\cF\) in each direction.

Consequently, the following statement is equivalent to
\cref{thm:p1-deq2-dichotomy} and is the target of the structural analysis
beginning here.  Its proof is completed in \cref{sec:p1-synthesis}.

\begin{theorem}[Equality-accessible \(\KHolant\) dichotomy]
\label{thm:p1-eq2-kholant-dichotomy}
For every finite set \(\cF\) of algebraic complex-valued Boolean signatures
with \(I=(\Eq_2)\in\cF\), the problem \(\KHolant(\cF)\) is polynomial-time
computable when \(\ReTract(K\cF)\) holds and is \(\SharpP\)-hard otherwise.
The tractability criterion is decidable from exact algebraic input.
\end{theorem}

For a set containing \(I\), the transformed set contains
\(Z=K^{\otimes2}I=\diag(1,-1)\).  Thus \cref{lem:tract-X-anchor}
reduces the full criterion \(\Tract\) to the four cases of
\(\ReTract\).

Since \(I(00)=I(11)=1\), the supplied \(I\) excludes the common one-sided support exception, so
\cref{lem:factor-saturation} allows the initial reduction to irreducible
signatures while retaining \(I\). Odd-arity factors are covered by
\cref{thm:external-odd}, and a set of binary factors is covered by the
arity-two case. We therefore begin the remaining analysis with a finite
set \(\cF\) of nonzero even-arity irreducible signatures, containing
\(I\) and at least one signature of arity at least four. Such a set is
factor-closed. Later outputs and their factors are retained as they are
added. The final proof in \cref{sec:p1-synthesis} transfers the resulting
classification to the original set. We first control the binary transfers.

\subsection{Binary transfers and complete groups}
\label{subsec:binary-transfer-control}
Binary chains turn composition into matrix multiplication.  Bounding the
group they generate gives a finite set of contraction kernels for the
matching deck.

For an ordered binary signature \(B\), recall its transfer
\[
  T_B=BX.
\]
Joining \(B\) to another ordered binary signature \(C\) in series through
one \(X\)-edge realizes \(BXC\),
and therefore
\[
  T_{BXC}=T_BT_C.
\]
For a transfer matrix \(T\), define its reversal by the anti-involution
\begin{equation}
  T^\#:=XT^{\mathsf T}X;
  \qquad
  T^\#=(\det T)ZT^{-1}Z\quad\text{if }T\text{ is nonsingular}.
  \label{eq:transfer-reversal}
\end{equation}
Here ``anti-involution'' means explicitly that
\[
  (ST)^\#=T^\#S^\#,
  \qquad
  (T^\#)^\#=T
\]
for all transfer matrices \(S,T\).

\begin{lemma}[Bounded torsion]
\label{lem:bounded-torsion}
Fix a number field \(L\), and let \(G\) be the set of projective
transfers of all directly realizable nonzero nonsingular ordered
binary gadgets over a fixed signature set with entries in \(L\).
If \(G\ne\varnothing\) and every element of \(G\) has finite
projective order, then \(G\) is a finite subgroup of
\(\PGL_2(L)\).
\end{lemma}

\begin{proof}
Define the field degree \(D=[L:\mathbb Q]\).  Let \([T]\in G\) have
projective order
\(r\).  By definition,
\[
  T^r=\eta I
\]
for some \(\eta\in L^\times\), and \(r\) is the least positive
integer with this property.

The minimal polynomial \(m_T\) divides \(x^r-\eta\), which is square-free
because \(\gcd(x^r-\eta,rx^{r-1})=1\) in characteristic zero.
Thus \(T\) is diagonalizable over the algebraic closure.  Choose
an eigenbasis matrix \(V\in\GL_2(\AlgNums)\) and eigenvalues
\(\lambda_1,\lambda_2\in\AlgNums^\times\) such that
\[
  V^{-1}TV=\diag(\lambda_1,\lambda_2).
\]
Define the eigenvalue ratio
\[
  \zeta:=\frac{\lambda_1}{\lambda_2}.
\]
The equality \(T^r=\eta I\) gives \(\zeta^r=1\), and the minimality of
\(r\) shows that \(\zeta\) is a primitive \(r\)-th root of unity.
Moreover,
\[
  \zeta+\zeta^{-1}
  =
  \frac{(\operatorname{tr}T)^2}{\det T}-2
  \in L.
\]
For \(r>2\),
\[
  [\mathbb Q(\zeta+\zeta^{-1}):\mathbb Q]
  =\frac{\varphi(r)}2,
\]
so \(\varphi(r)\le2D\), also valid for \(r=1,2\).  The integers with this
bound form an effectively finite set: each prime power \(p^a\mid r\)
satisfies \(p^{a-1}(p-1)\le2D\), bounding both \(p\) and \(a\).
Consequently all projective orders divide the computable exponent
\[
  e_D=\operatorname{lcm}\{r:\varphi(r)\le2D\}.
\]

Series composition gives product closure.  A positive power of every
element is the identity, so the identity and inverses also belong to
\(G\).  Hence \(G\) is a subgroup.  Put
\[
 \mathfrak{sl}_2(L)
 :=\{A\in\Mat_2(L):\operatorname{tr}A=0\}.
\]
The adjoint representation is
\[
  \operatorname{Ad}:\PGL_2(L)\hookrightarrow
  \GL(\mathfrak{sl}_2(L))\cong\GL_3(L),
  \qquad
  \operatorname{Ad}([T])(A)=TAT^{-1}.
\]
The conjugation homomorphism has trivial kernel: a matrix commuting with all trace-zero matrices
is scalar.  Burnside's finite-exponent theorem states that a subgroup
\(H\le\GL_n(\mathbb C)\) satisfying \(h^e=I\) for one fixed \(e\) and
every \(h\in H\) is finite \cite{Burnside1905}.  Apply it to
\(\operatorname{Ad}(G)\le\GL_3(\mathbb C)\) with \(e=e_D\).
Thus all binary gadgets, despite their unbounded sizes, have only finitely
many projective transfer classes.
\end{proof}

For a finite algebraic retained, factor-closed, two-sided signature set
\(\cF\), call a nonzero directly realizable ordered binary \emph{safe} if it
is nonsingular and its transfer has finite projective order.  Call it
\emph{unsafe} if it is singular or its transfer has infinite projective
order.  Call the set \emph{binary-residual} when every nonzero realizable ordered
binary is safe.  We work in this case. Otherwise
\cref{lem:binary-interpolation,lem:common-unary-exit} resolves the branch.  Series composition and
\cref{lem:bounded-torsion} then give a finite group: take \(L\) to be the
number field generated by the entries of \(\cF\), which also contains the
entries of every direct gadget transfer.
The group is
\hypertarget{eq:complete-safe-group}{}
\begin{equation*}
  G(\cF)
  =
  \{[BX]:B\text{ is a safe ordered binary directly realizable over }\cF\}.
\end{equation*}
The definition uses all directly realizable binaries.  Below we distinguish
cases according to this group for each signature set.  The decision algorithm for the resulting tractability
cases is given by \cref{thm:predicate-algorithms}. It does not
enumerate gadgets of unbounded size.  We choose realizable representatives as follows.

\paragraph{Realizable binary representatives.}
\label{par:binary-representatives}

For each class \(g\) of a finite realized projective group
\(G\le\PGL_2(\AlgNums)\), choose an ordered binary gadget \(B_g\)
with transfer \(T_{B_g}=B_gX\) and \([T_{B_g}]=g\). For
\(1=[I]\), take \(B_1:=\mathsf E_X=X\), so \(T_{B_1}=I\).
The empty network is the nullary unit. Every realizable ordered binary
with transfer class \(g\) has the form \(B=\rho B_g\) for a unique
\(\rho\in\AlgNums^\times\). Retain the multiplication and reversal
scalars for these representatives, with the input orders and reduction
data of \cref{sec:retained-normalization}. This fixes the scalar factors
when projective identities are used in gadgets.

Whenever the proof says that an external input is X-translated, it means
attaching the binary chain in the current set whose transfer class is \([X]\) and
recording its nonzero scalar in the input data.  The displayed chain realizes this action.

\begin{lemma}[Relative completeness test for realizable binaries]
\label{lem:actual-binary-completeness}
Let \(\cF_0\subseteq\cF\) be finite algebraic retained,
factor-closed signature sets with \(I\in\cF_0\), and suppose that
every nonzero binary directly realizable over \(\cF_0\) is safe.  Put
\(H=G(\cF_0)\).  The following exhaustive ordered case distinction applies:
\begin{enumerate}
\item if \(\cF\) directly realizes a nonzero singular ordered binary,
      its nonzero unary factors are available by the factor reduction;
\item otherwise, if \(\cF\) directly realizes a nonsingular ordered
      binary of infinite projective transfer order, a nonzero rank-one
      binary is available by interpolation;
\item otherwise every nonzero binary directly realizable over \(\cF\) is safe, \(H\le G(\cF)\), and
      every nonsingular realizable binary \(B\) over \(\cF\) with
      \([BX]\notin H\) witnesses the strict inclusion \(H<G(\cF)\).
\end{enumerate}
In the third case, if \(G(\cF)=H\), every nonzero realizable binary over
\(\cF\) is projectively one of the fixed representatives \(B_h\),
\(h\in H\).
\end{lemma}

\begin{proof}
The supplied \(I\) has both charge signs, so both signature sets are two-sided.
A singular binary is handled by
\cref{lem:binary-interpolation,lem:common-unary-exit}, and an infinite-order
binary by \cref{lem:binary-interpolation}.  If neither exists, \(\cF\)
is binary-residual, so \cref{lem:bounded-torsion} defines \(G(\cF)\).
Every gadget over \(\cF_0\) remains a gadget over \(\cF\), hence
\(H\le G(\cF)\).  The remaining assertions follow directly from the
definitions and the fixed representative choice.
\end{proof}

In the first two cases, \cref{lem:common-unary-exit} determines the
complexity after factor extraction. In the third, a new transfer class
gives strict group enlargement, while equality of the groups permits
reuse of the fixed binary representatives.

\begin{remark}[Connection with binary-signature classifications]
The singular, nonsingular infinite-projective-order, and nonsingular
finite-projective-order split is the \(\KHolant\) form of the binary split
used by Cai--Lu--Xia \cite{CaiLuXia2009,CaiLuXia2018RealHolantc}.  The four
ordered Bell signatures
\(\mathcal B=\{B_{0,0},B_{0,1},B_{1,0},B_{1,1}\}\) represent the projective
Bell classes of Shao--Cai \cite{ShaoCai2020}. Beyond the binary classification, we need the
complete transfer group and uniform control of its higher-arity cards.
\end{remark}

\paragraph{Finite projective groups and marked matrices.}
The classification of finite projective groups gives the uniform order
bound needed for stabilization. We record its concrete models here and
analyze the distinguished classes \([X]\) and \([Z]\) in
\cref{sec:matching-decks}.

We use the quaternion
matrix basis
\begin{equation}
  Q_0=\begin{pmatrix}1&0\\0&1\end{pmatrix},\quad
  Q_1=\begin{pmatrix}0&i\\i&0\end{pmatrix},\quad
  Q_2=\begin{pmatrix}0&1\\-1&0\end{pmatrix},\quad
  Q_3=\begin{pmatrix}i&0\\0&-i\end{pmatrix}.
  \label{eq:quaternion-basis}
\end{equation}
It is a basis of \(\Mat_2(\mathbb C)\). Moreover, \(Q_0=I\),
\(Q_j^2=-I\) for \(1\le j\le3\), and
\(Q_jQ_k=-Q_kQ_j\) for distinct \(j,k\in\{1,2,3\}\).  These identities are
what the adjective \emph{quaternion} records.

For the concrete Platonic models below, \(\mathrm{int}\) records that
\([Z]\) belongs to the tetrahedral group, \(dd\) records that \([X],[Z]\)
are double transpositions in the octahedral group, and \(0\) labels the
base icosahedral display.

\begin{proposition}[Finite subgroups of \(\PGL_2(\mathbb C)\)]
\label{prop:finite-pgl2-classification}
Every finite subgroup of \(\PGL_2(\mathbb C)\) has exactly one of the following
isomorphism types, and subgroups of the same type are projectively conjugate:
\[
  C_m,\qquad D_{2m},\qquad A_4,\qquad S_4,\qquad A_5,
\]
where \(m\ge1\) for \(C_m\), \(m\ge2\) for \(D_{2m}\),
\(|C_m|=m\), and \(|D_{2m}|=2m\).  If \(\omega\) is a primitive
\(m\)-th root of unity, explicit cyclic and dihedral models are
\[
 C_m^{\rm std}
 =
 \left\{
   \left[\begin{pmatrix}1&0\\0&\omega^k\end{pmatrix}\right]:
   0\le k<m
 \right\},
\]
and
\[
 D_{2m}^{\rm std}
 =
 C_m^{\rm std}
 \cup
 \left\{
   \left[
     X\begin{pmatrix}1&0\\0&\omega^k\end{pmatrix}
   \right]:
   0\le k<m
 \right\}.
\]
The diagonal classes are the rotations, and the anti-diagonal coset
consists of the reflections.

In the basis \cref{eq:quaternion-basis}, take the tetrahedral model
\begin{equation}
\begin{aligned}
 \TetInt
 ={}&
 \{[Q_0],[Q_1],[Q_2],[Q_3]\}\\
 &{}\cup
 \{[Q_0+\epsilon_1Q_1+\epsilon_2Q_2+\epsilon_3Q_3]:
       \epsilon_j\in\{1,-1\}\},
\end{aligned}
\label{eq:tetrahedral-domain}
\end{equation}
and the octahedral model
\begin{equation}
 \OctDD
 =
 \TetInt
 \cup
 \{[Q_\mu+\epsilon Q_\nu]:
       0\le\mu<\nu\le3,\ \epsilon\in\{1,-1\}\}.
 \label{eq:octahedral-domain}
\end{equation}
Finally, define the golden-ratio scalar and its reciprocal by
\[
  \phi:=\frac{1+\sqrt5}{2},
  \qquad
  \tau:=\phi^{-1},
\]
and define the icosahedral model by
\begin{equation}
\begin{aligned}
 \IcoZero
 =\TetInt\ \cup\
 \bigl\{[v_0Q_0+v_1Q_1+v_2Q_2+v_3Q_3]:{}&
 (v_0,v_1,v_2,v_3)\\
 &{}=\pi(0,\epsilon_1,\epsilon_2\phi,\epsilon_3\tau),\
 \ \pi\in A_4,\ \epsilon_j\in\{1,-1\}\bigr\},
\end{aligned}
\label{eq:icosahedral-domain}
\end{equation}
where \(A_4\) acts by even coordinate permutations and projective
duplicates are removed.  The additional part of \(\IcoZero\) has
\(48\) projective classes.  Thus
\[
  \TetInt\cong A_4,\qquad
  \OctDD\cong S_4,\qquad
  \IcoZero\cong A_5,
\]
of orders \(12,24,60\), respectively.
\end{proposition}

\begin{proof}
The conjugacy classification is classical
\cite[Section~5.1]{Beardon1983}.  The first two displays are the standard
rotation and reflection models.  Direct matrix multiplication verifies
closure, cardinalities, and element-order distributions of the three
remaining displayed sets, identifying them respectively with
\(A_4,S_4,A_5\).
\end{proof}

For the concrete models in \cref{prop:finite-pgl2-classification}, the
finite matrix checks in its proof consist of reducing products modulo
nonzero scalar, testing closure of the displayed projective sets, and
counting their distinct classes and element orders.  The scalar factors
remain separate from the realizable-gadget records above.

We now choose a minimum-arity nonbinary irreducible signature and retain the factors
of all its dressed contractions.

\subsection{Stable matching decks and restart control}
\label{subsec:safe-matching-cards}

Fix a retained, factor-closed set \(\cF\ni I\) with complete transfer
group \(G=G(\cF)\). Attach available binaries to a minimum-arity
nonbinary irreducible signature and close one pair of inputs. A nonbinary
factor of the output has smaller arity. If only binary factors remain,
they either settle the complexity, enlarge the group, or form a product
on a perfect matching. The \emph{matching deck} collects these contractions,
using the ordered representatives from \cref{par:binary-representatives}.

We first stabilize this deck for the selected signature. We then give
the descent conditions for subsequent constructions in the group branches.
The final proof also uses an outer induction over unresolved irreducible
occurrences to pass from the selected signature to the whole set.

For an even finite input set \(R\), a \emph{perfect matching} is a partition
of \(R\) into two-element subsets.  An \emph{oriented perfect matching}
also chooses an order within each pair, and a matching product places one
ordered binary factor on every oriented pair.

\begin{definition}[Matching deck]
\label{def:matching-deck}
Let \(\cF\ni I\) be factor-closed and let \(G=G(\cF)\).  On a fixed
ordered input set, a \emph{\(G\)-matching signature} of arity \(2m\) is
\begin{equation*}
  c\bigotimes_{e\in M}B_{g_e}^{(e)},
  \qquad
  c\ne0,\quad g_e\in G,
\end{equation*}
where \(M\) is an oriented perfect matching in the preceding sense and
\(B_{g_e}^{(e)}\) places the chosen ordered representative \(B_{g_e}\) on
the oriented edge \(e\).  We allow a simultaneous permutation of inputs. The
scalar \(c\) absorbs all representative scalars, so the notion depends only
on \(G\).

For \(\omega\in\{\mathrm{id},\mathrm{rev}\}\), put
\(B_g^{\mathrm{id}}=B_g\) and
\(B_g^{\mathrm{rev}}=B_g^{\mathsf T}\).  Let \(\delta\) specify a subset of the inputs and, at each selected input, a representative
\(B_g\) and one of these two orientations.  Attach its recorded binary through
one \(X\)-edge and retain the other binary input in the same position.
Write \(f^\delta\) for the resulting realizable dressed signature.
Here the dressing acts on the specified inputs of \(f\), before the pair
contraction defining a deck card below.

For each \(\delta\), choose distinct ordered inputs \((i,j)\), a closing
representative \(B_h\) with orientation \(\omega\), and a residual input order \(\rho\), leaving at least two inputs.  The resulting card is
\[
 \rho\!\left(\partial_{(i,j)}^{XB_h^\omega X}f^\delta\right).
\]
The \emph{matching deck} \(\mathfrak D_G(f)\) contains zero and all such
one-copy proper pair cards.  Thus one copy of \(f\) is used and exactly one
ordered input pair is closed. The binary dressings and closing chain use the
fixed representatives.  Every nonzero card is a direct gadget output.
\end{definition}

\begin{definition}[Deck-stable state]
\label{def:deck-stable-state}
A retained reduction state \((\cF,G,f)\) is \emph{deck-stable} if \(\cF\) is
finite, factor-closed, and contains \(I\); \(f\in\cF\) is a
minimum-arity irreducible nonbinary signature; \(G=G(\cF)\) is the
complete transfer group, which is finite, with fixed safe binary representatives;
and the deck for the resulting set, with all card factors retained, consists of zero
and \(G\)-matching signatures.  Realizability is always relative to \(\cF\).
Added factors are only available by reduction over their predecessor.
\end{definition}

Adding signatures available by reduction gives a Turing-equivalent problem:
the defining reduction removes the added signatures, and inclusion gives
the reverse reduction.  Direct outputs use
\cref{eq:direct-gadget-reduction}. Their factors use
\cref{lem:factor-saturation}.

To prove stabilization, fix the finite family of contractions that can
be processed while the group and minimum arity stay unchanged. We record
this family and its factorization trees as follows. Fix \((A,\cF,f)\) and
\(H:=\overline G(A,\cF)\), the complete transfer group in the original
coordinates. Throughout this stabilization and descent argument, \(H\)
has this meaning. While \(H\) and \(\nu(\cF)\) remain
unchanged, fix one representative of each group class. A one-copy proper
card is specified by the input dressings, the ordered deleted pair,
the closing representative \(B_h\) and its orientation \(\omega\)
(giving the kernel \(XB_h^\omega X\)), and the residual input order.
Let \(\mathscr U(A,f,H)\) consist of zero and the signatures obtained by
pulling these cards back by \(A\) and dividing each nonzero result by its
first nonzero entry, with the residual inputs kept in order.

Each nonzero member of \(\mathscr U(A,f,H)\) indexes one \emph{token} in
the local queue, even if several constructions yield that member. Store
one such construction, its discarded scalar, and the factorization tree of
the selected card from \cref{lem:factor-saturation}. Equal factors at
different positions in the tree remain distinct occurrences.
Write \(\mathscr U^{\mathrm{fac}}(A,f,H)\) for the token set with this
attached factor work, and write
\[
 Q_{\rm done}\subseteq\mathscr U^{\mathrm{fac}}(A,f,H),
 \qquad
 Q=\mathscr U^{\mathrm{fac}}(A,f,H)\setminus Q_{\rm done}
\]
for the completed and unfinished tokens, respectively. Processing a factor
occurrence takes place within its token's tree, so the queue stays within
this finite set while \(H\) and \(\nu\) are fixed. A lower \(\nu\), a
strict enlargement of the pulled-back group, or a complexity conclusion
stops the local construction. The binary equality \(I\) and the original
coordinates remain fixed throughout.

The next lemma either produces a deck-stable state, resolves the complexity
of the current set, or exposes a smaller nonbinary irreducible signature.

\begin{lemma}[Deck stabilization under factor access]
\label{lem:minimum-core-deck}
Let \(\cF_0\) be a retained, factor-closed finite algebraic
signature set containing the supplied equality \(I\), and let
\(f\in\cF_0\) be an even-arity irreducible nonbinary signature of
arity \(n=\nu(\cF_0)\).

Then an established reduction determines the complexity, or
\(\Tract(K\cF_0)\) holds, or one obtains one of the following
retained extensions:
\begin{enumerate}
\item a finite retained factor-closed set
      \(\cF_{<}\supseteq\cF_0\) such that
      \[
        \KHolant(\cF_{<})\equivT\KHolant(\cF_0),
        \qquad \nu(\cF_{<})<n;
      \]
      every nonbinary irreducible signature retained in this successor has even arity;
\item a finite retained factor-closed set
      \(\cF_\star\supseteq\cF_0\) such that
\begin{equation*}
  \KHolant(\cF_\star)
  \equivT
  \KHolant(\cF_0),
\end{equation*}
and, for \(G_\star=G(\cF_\star)\), every member of
\(\mathfrak D_{G_\star}(f)\) is zero or a
\(G_\star\)-matching signature.
\end{enumerate}

The stable successor retains the chosen representatives, the regenerated
complete deck, and every card factor.  Such factors are available by reduction over,
but need not be directly realizable from, their predecessor.
\end{lemma}

\begin{proof}
Construct states \((A_i,\cF_i,H_i,Q_i)\), where
\(H_i=\overline G(A_i,\cF_i)\le\PGL_2(L_0)\) is in the fixed
pulled-back coordinates and \(Q_i\) is the unfinished queue defined above.
At stage \(i\), apply the reductions for unsafe binaries,
consider the complete group of the resulting set \(G_i=G(\cF_i)\), retain one chosen representative of every class, and generate each normalized card in
\(\mathscr U(A_i,f,H_i)\) at most once.  Preprocess and jointly
close the resulting finite batch under irreducible factors to obtain \(\cF_{i+1}\).
The supplied \(I\) excludes the one-sided exception.  Inclusion supplies the
reverse reduction at every exposure step, so every \(\cF_i\) is
Turing-equivalent to \(\cF_0\).

Every nonbinary card factor has arity at most \(n-2\).  Odd factors are
handled by \cref{thm:external-odd}. An even one gives successor~1.
Otherwise only binaries remain.
  No global normalization is made in this fixed-\(I\) procedure.  Completeness
  therefore gives \(H_i\le H_{i+1}\). Only a strict inclusion
  \(H_i\subsetneq H_{i+1}\) is a successor with a strictly larger complete transfer group, and it alone triggers
  regeneration of the complete deck.

Use the fixed pulled-back field \(L_0\) from
\cref{lem:normalization-ledger}.  Gadget evaluation and pivot-normalized
rank-one splitting are performed in pulled-back coordinates, so
\(H_i\le\PGL_2(L_0)\).  Let \(d_{L_0}:=[L_0:\mathbb Q]\) be the field
degree, and define the torsion exponent and group-order bound by
\begin{equation}
  e_{L_0}=\operatorname{lcm}\{r\ge1:\varphi(r)\le2d_{L_0}\},
  \qquad
  M_{L_0}=\max\{2e_{L_0},60\},
  \label{eq:deck-group-bound}
\end{equation}
where \(\varphi\) denotes Euler's totient function.
\Cref{prop:finite-pgl2-classification,lem:bounded-torsion} give
\(|H_i|\le M_{L_0}\).  Hence the strict subgroup chain has at most
\(M_{L_0}-1\) changes. Between changes, stabilization processes only the
tokens indexed by the nonzero cards in \(\mathscr U(A_i,f,H_i)\). Each
token is removed once, and its factorization tree is finite by the decreasing
potential \(\Phi\) in \cref{eq:factor-potential}. A lower-arity factor,
new group class, or complexity conclusion stops the current stage.
Consequently there is a first stage at which both \(H_{i+1}=H_i\) and
the queue is empty. Call this the stabilized stage.
At that stage define the stable retained set and its complete group by
\[
  \cF_\star=\cF_{i+1},
  \qquad G_\star=G_{i+1}.
\]
Every \(0\ne h\in\mathfrak D_{G_\star}(f)\) then has retained irreducible binary
factors \(B_1,\ldots,B_m\) satisfying
\[
  [B_jX]\in G(\cF_\star)=G_\star.
\]
Writing \(B_j=\beta_jB_{g_j}\), their blocks form a perfect matching, so
\[
  h=c\bigotimes_{j=1}^m B_{g_j}
\]
for \(c\ne0\), as required.
\end{proof}

\smallskip
\noindent\textit{Remark (The role of stabilization).}
The contraction and factor arguments interact through the complete
transfer group. Direct gadgets use the current set, whereas a factor
may become available only after a reduction. Once retained, that factor
can participate in further gadgets and enlarge the group, changing both
the available contractions and the applicable group case. The procedure
therefore retains every factor, compares the complete groups, and
regenerates the deck after strict enlargement. This comparison includes
multi-copy binary gadgets beyond the finite batch that produced the factor.

The fixed number field bounds the number of strict enlargements. Between
them, the finite card indices and their factorization trees control the
remaining work. Thus stabilization supplies a common starting point for
the group-specific arguments: each can establish rigidity of the selected
signature or obtain progress that is handled by the same descent proof.
The conditions below make this requirement precise for constructions
beyond the one-copy deck.

Later constructions can lower \(\nu\) or enlarge the complete group.
We call either change an \emph{outer deck successor}. Processing a fixed
family while both remain unchanged is a \emph{within-node closure}.

\begin{definition}[Processing a finite family of signatures]
\label{def:relative-stratum-protocol}
For deck-stable \((\cF,G,f)\), take a finite family of signatures
retained in or directly realized over \(\cF\), with their input orders
and realization scalars.  Add them, preprocess, jointly close the enlarged set under irreducible factors,
apply the reductions for odd and unsafe factors, and consider the group of the resulting set.
After a strict enlargement, regenerate the complete deck. We refer to
these steps as processing the family. Each added signature has its
realization or reduction, as stipulated above. In the fixed partition
\(\mathscr U^{\mathrm{fac}}=Q_{\rm done}\sqcup Q\), each token moves from
\(Q\) to \(Q_{\rm done}\) once its card and all factor occurrences in its
tree have been processed. A completed token stays in \(Q_{\rm done}\).
For families indexed by constructions, completion is recorded at the
specified index even when two constructions give the same signature.
The descent conditions for continuations with the same group and
minimum arity are stated in \cref{def:frozen-token-coverage} and verified
in the group-specific arguments.

Applications of the equality-accessible arity-four boundary also satisfy
the requirements stated after \cref{thm:equality-q4-boundary}.

The current case is complete once finitely many such families resolve every
branch and \(Q=\varnothing\).  Under the conditions of \cref{def:frozen-token-coverage}, a transition leaving the pulled-back
group and \(\nu\) unchanged completes a pending index, decreases its stored
factor potential, or certifies that none remains, and hence is a within-node
closure. Otherwise, \(\nu\) decreases or the pulled-back group enlarges.
\end{definition}

\paragraph{Standing predicates for relative cases.}
To keep the group-specific statements short, write
\hypertarget{eq:stable-predicate}{}
\hypertarget{eq:even-prime-predicate}{}
\begin{align*}
 \Stable(\cF,G,f)
 &\Longleftrightarrow
   (\cF,G,f)\text{ is deck-stable in the sense of
   \cref{def:deck-stable-state}},
   \\
 \Even(\cF)
 &\Longleftrightarrow
   \forall g\in\Pi(\cF),\quad
   \arity(g)\ge3\Longrightarrow \arity(g)\in2\mathbb Z.
\end{align*}

Each later group branch must show that its constructions satisfy the
following conditions. In particular, a step that preserves both \(H\)
and \(\nu\) must consume work in the already fixed finite family.

\begin{definition}[Conditions for descent]
\label{def:frozen-token-coverage}
Fix a deck-stable tuple \((\cF,G,f)\), a coordinate matrix \(A\), the
pulled-back group \(H=\overline G(A,\cF)\), and a fixed partition
\[
 \mathscr U^{\mathrm{fac}}(A,f,H)=Q_{\rm done}\sqcup Q,
\]
where \(Q\) is the unfinished queue.  Let \(\mathfrak B\) be a collection of admissible families.  Processing any
member must give, on each branch, one of:
\begin{enumerate}
\item a resolved complexity leaf;
\item a successor with smaller \(\nu\);
\item a strict enlargement of the pulled-back complete group;
\item unchanged \((H,\nu)\), in which case every continuing card or factor occurrence retained for processing is indexed by a token of \(Q\), and its processing is
      confined to that token's stored card, reduction, and attached factorization tree.  The processing procedure either moves at least one such token to
      \(Q_{\rm done}\), or, while processing one fixed unfinished token,
      strictly decreases the factor potential of its stored worklist.
\end{enumerate}
 A token is marked done only after its full factorization tree is exhausted.
\end{definition}

For a family \(\mathfrak C\) of tractable signature classes,
\begin{equation}
 \Close_{\mathfrak C}(\cF,G,f)
 \label{eq:relative-close-predicate}
\end{equation}
means that finitely many applications of \cref{def:relative-stratum-protocol}
resolve all branches for \((\cF,G,f)\), with \(Q=\varnothing\), and every surviving
tractable case with a common presentation satisfies
\(\PresK(K\cF;\cC)\) for some \(\cC\in\mathfrak C\).
Established complexity conclusions and outer deck successors are also
part of completion. The class restriction applies to cases certified by
a common presentation.
\(\mathfrak C=\varnothing\) means that no tractable case with a common presentation
remains.  Thus \(\Close\) is a completion predicate for this reduction procedure,
whereas tensor closure refers to the class \(\angles{\cC}\).  The predicates
retain all procedure hypotheses and transition requirements.

The verification is necessarily branch-by-branch and uses the reductions
proved in
\cref{sec:normalized-dihedral,sec:proper-deck,sec:v4-deck,sec:platonic-decks}.
It is therefore deferred to \cref{sec:p1-synthesis}, after those four
arguments are available.

The following lexicographic measure strictly decreases on every continuing
step satisfying these conditions.

\begin{lemma}[Deck-transition dichotomy and termination]
\label{lem:matching-synthesis-measure}
Fix the initial number field \(L_0\).  At a continuing all-even,
binary-residual deck-stable node, carry the reduction data \((A,\cF,f)\), a
finite token queue \(Q\), and the supplied equality \(I\).  Let
\(\mathfrak B\) satisfy \cref{def:frozen-token-coverage}, let
\(\cF^+\) be the result of processing one batch in \(\mathfrak B\) over
\(\cF\), and define
the pulled-back group by
\[
  \overline G(A,\cF)=[A]^{-1}G(\cF)[A].
\]
Then
\begin{equation}
  \KHolant(\cF^+)\equivT\KHolant(\cF).
  \label{eq:deck-transition-equivalence}
\end{equation}
Batch outputs are direct over \(\cF\), while their factors need only be
available by reduction there.

After considering odd-arity signatures, unsafe binaries, and unary factors over the whole set, the only
continuing transitions are:
\begin{enumerate}
\item \(\overline G(A,\cF^+)=\overline G(A,\cF)\) and
      \(\nu(\cF^+)=\nu(\cF)\): process a previously unfinished
      token of \(Q\), complete it or strictly decrease its factor potential,
      or certify that \(Q\) is empty.
\item \(\overline G(A,\cF^+)=\overline G(A,\cF)\) and
      \(\nu(\cF^+)<\nu(\cF)\): choose a new minimum-arity nonbinary irreducible signature and reapply
      \cref{lem:minimum-core-deck}.
\item \(\overline G(A,\cF)\subsetneq\overline G(A,\cF^+)\): choose
      a minimum-arity nonbinary irreducible signature in the resulting set, regenerate its deck, and restart marked
      routing.
\end{enumerate}

At every continuing node define the pair by
\begin{equation*}
 \Xi(A,\cF)
 =\bigl(M_{L_0}-|\overline G(A,\cF)|,\nu(\cF)\bigr).
\end{equation*}
Refine it to the tuple
\begin{equation}
  \Xi_{\mathrm{full}}(A,\cF,f,Q)
  =\bigl(M_{L_0}-|\overline G(A,\cF)|,\nu(\cF),
        |Q|,\,
        \Phi_Q\bigr).
  \label{eq:deck-closure-measure}
\end{equation}
For each \(u\in Q\), let \(\mathcal W_u\) be the multiset of still-open
occurrences in its attached factor worklist, and define
\[
 \Phi_Q:=\sum_{u\in Q}\Phi(\mathcal W_u),
\]
using the potential in \cref{eq:factor-potential}.
The displayed tuple is ordered lexicographically.  Items~2--3 strictly
decrease its first two coordinates.  In item~1, completing a token strictly
decreases the third coordinate until the queue is empty.  If a token remains
unfinished during a genuine factor split, the fourth coordinate \(\Phi_Q\)
strictly decreases.  If an exposed factor changes the pulled-back group, the
operation is instead item~3 and is charged to the strict subgroup chain.
\end{lemma}

\begin{proof}
Gadget replacement and factor closure prove
\cref{eq:deck-transition-equivalence}. \(I\) excludes the one-sided
exception, and inclusion reverses the reduction.  Adding signatures cannot increase \(\nu\).  The continuing state remains in fixed-\(I\)
coordinates.  New classes can arise only from an exposed factor. Such a
class is precisely a strict successor in the pulled-back group.  By
  \cref{lem:normalization-ledger,eq:deck-group-bound}, every continuing group
lies in \(\PGL_2(L_0)\) and has order at most \(M_{L_0}\).  Therefore a lower
  \(\nu\) decreases the second coordinate of \(\Xi\), strict subgroup
  inclusion decreases the first.  A fixed-\((H,\nu)\) step either decreases the
  finite queue counter in \(\Xi_{\mathrm{full}}\) or, within its fixed token, decreases
\(\Phi\). By \cref{def:frozen-token-coverage,def:relative-stratum-protocol},
item~1 processes only a previously fixed token of \(Q\), so these
decreases exhaust every fixed-\((H,\nu)\) continuation.
\end{proof}

\paragraph{Reading the transition measure.}
The table matches each outcome of \cref{lem:matching-synthesis-measure}
with the coordinate that decreases. The field \(L_0\), original
coordinates and \(I\) remain fixed.  Write
\(H^+=\overline G(A,\cF^+)\) and \(\nu^+=\nu(\cF^+)\).
\begin{center}
\begin{tabularx}{\linewidth}{@{}L{.23\linewidth}>{\raggedright\arraybackslash}XL{.20\linewidth}@{}}
\toprule
Outcome & Next operation & Descent or exit\\
\midrule
Resolved complexity leaf & Return the established whole-set conclusion. & Terminal\\[3pt]
\(H\subsetneq H^+\) & Choose a new minimum-arity nonbinary irreducible signature, regenerate its deck,
and restart marked routing. & First coordinate of \(\Xi_{\mathrm{full}}\)\\[3pt]
\(H^+=H,\ \nu^+<\nu\) & Choose a new minimum-arity nonbinary irreducible signature and reapply
\cref{lem:minimum-core-deck}. & Second coordinate\\[3pt]
Same \((H,\nu)\), completed token & Move that token from \(Q\) to
\(Q_{\rm done}\) after its full factor tree is processed. & Third coordinate\\[3pt]
Same \((H,\nu)\), unfinished token & Continue inside its fixed factor
tree. A genuine factor split decreases \(\Phi_Q\). & Fourth coordinate\\
\bottomrule
\end{tabularx}
\end{center}
When \(Q\) is empty, the local family has been processed.

\begin{definition}[The conclusion required of each construction]
\label{def:protocol-discharged}
Write
\[
 \Sigma=(A,\cF,H,f,Q),
 \qquad H=\overline G(A,\cF),
\]
with the record, deck-stability, fixed queue, and fixed-\(I\) data of
\cref{def:relative-stratum-protocol,def:frozen-token-coverage}.  Write
\[
 \Xi_{\mathrm{full}}(\Sigma)
 :=\Xi_{\mathrm{full}}(A,\cF,f,Q)
 =\bigl(M_{L_0}-|H|,\nu(\cF),|Q|,\Phi_Q\bigr)
\]
for the lexicographic measure in \cref{eq:deck-closure-measure}.

For each concrete construction at \(\Sigma\), we prove that processing its
outputs by \cref{def:relative-stratum-protocol} gives either
\begin{enumerate}
\item a resolved complexity leaf: hardness, a tractable case satisfying \(\Tract\), or
      a tractable case with a common \(K\)-presentation; or
\item a deck-stable tuple \(\Sigma'\), with the same required realization and
      induction data, satisfying
      \[
        \Xi_{\mathrm{full}}(\Sigma')<_{\rm lex}\Xi_{\mathrm{full}}(\Sigma).
      \]
\end{enumerate}
We call the tuple in the second outcome a \emph{strict stable successor}.
This outcome includes a decrease of \(\nu\), a strict enlargement of
the pulled-back complete group, completion of a fixed token, or a strict
decrease of that token's factor potential.  Every
same-\((H,\nu)\) continuation must be indexed by the already fixed queue
and processes only its stored cards, reductions, and factorization trees. A
continuing successor includes this queue data and the strict descent
asserted in item~2.

Item~1 resolves the complexity question. Item~2 establishes strict
progress and continues the induction at \(\Sigma'\). A \emph{resolved
construction}, in the terminology of the relative branch statements,
is one for which either item has been proved, including realizability,
factor access, and descent. A \emph{resolved complexity leaf} always
means item~1.
The assertion \(\Close_{\mathfrak C}(\cF,G,f)\) additionally covers
every case for the current \((H,\nu)\) and restricts the surviving common
presentations to \(\mathfrak C\), as in \cref{eq:relative-close-predicate}.
\end{definition}

For a selected stable tuple \(\Sigma\), the continuing outcome of
\cref{def:protocol-discharged} decreases \(\Xi_{\mathrm{full}}\). To classify the whole
set, the final proof in \cref{sec:p1-synthesis} also uses an outer worklist
induction over unresolved irreducible occurrences. Completing the current
occurrence and choosing another of the same arity advances that outer
induction. This separates progress for the selected signature from
completion of the entire set.

\subsection{The main stable-state theorem and four-way routing}
\label{sec:matching-decks}

The preceding reductions leave a deck-stable tuple
\((\cF,G,f)\).  The next theorem is proved in
\cref{sec:p1-synthesis} from the four branches routed below.  Together with
stabilization, it gives the equality-accessible main theorem
\cref{thm:p1-eq2-kholant-dichotomy}.

\begin{theorem}[Relative matching-deck classification]
\label{thm:matching-deck-classification}
Assume
\[
 \Stable(\cF,G,f),\qquad \arity(f)=\nu(\cF).
\]
In particular, the deck-stable state contains the supplied
equality \(I\).  If \(\Tract(K\cF)\) holds, then
\(\KHolant(\cF)\in\FP\). Otherwise \(\KHolant(\cF)\) is
\(\SharpP\)-hard.  Equivalently, by \cref{lem:tract-X-anchor}, the
tractable condition is \(\ReTract(K\cF)\).
\end{theorem}

In each tractable case reached after the hardness reductions, the proof
supplies one common product, affine, or flat-Lagrangian
\(K\)-presentation. By \cref{eq:common-K-presentation-predicate}, the same
basis also places the transformed edge in that class.

The supplied equality gives \(x=[X]\in G\). Reversal then makes
\(j=[Z]\) normalize \(G\). Let \(\widehat G:=\langle G,j\rangle\) be the
extension obtained by adjoining the mark. The reversal identity gives
\(G\triangleleft \widehat G\) and \([\widehat G:G]\le2\), while the distinct commuting
involutions \(x,j\) generate a Klein-four subgroup of \(\widehat G\).
Their positions refine the abstract finite-group classification into the
four branches used below. We record the group facts and matrix notation
needed to make this refinement explicit.
\label{sec:marked-finite-groups}

\begin{lemma}[Finite projective-group facts]
\label{lem:finite-pgl2-facts}
The following properties will be used below.
\begin{enumerate}
\item A cyclic group has at most one involution, and it has one exactly when
      its order is even.
\item For
      \[
        D_{2m}=\langle r,s:r^m=s^2=1,\ srs=r^{-1}\rangle,
      \]
      the rotations \(\langle r\rangle\cong C_m\) form a normal index-two
      subgroup, and every element outside it is a reflection.  If \(m\) is
      odd, the rotation group is the unique index-two subgroup and the
      center is trivial.  If \(m\) is even and \(m\ge4\), the center is
      \(\{1,r^{m/2}\}\). For \(m=2\), \(D_4\cong V_4\) and all three
      nonidentity elements are central.  Every subgroup of a cyclic or
      dihedral group is cyclic or dihedral.
\item The three involutions of \(A_4\), together with the identity, form
      its unique normal \(V_4\).  The group has no index-two subgroup, and
      the centralizer of an involution is \(V_4\).
\item The unique index-two subgroup of \(S_4\) is \(A_4\).  Its
      involutions are six transpositions and three double transpositions.
      The centralizer of a transposition has order \(4\), whereas the
      centralizer of a double transposition is \(D_8\).  The three double
      transpositions are the nonidentity elements of the unique normal
      \(V_4\).  Conjugation is transitive both on ordered pairs of distinct
      double transpositions and on ordered pairs of disjoint transpositions.
\item The group \(A_5\) is simple and has no index-two subgroup.  Its
      fifteen involutions form one conjugacy class, and the centralizer of
      each is \(V_4\).
\item The element-order distributions are
      \[
      \begin{array}{c|rrrr}
          &1&2&3&\text{order }4\text{ or }5\\ \hline
        A_4&1&3&8&0\\
        S_4&1&9&8&6\text{ of order }4\\
        A_5&1&15&20&24\text{ of order }5.
      \end{array}
      \]
\item For the displayed models,
      \[
        N_{\PGL_2(\mathbb C)}(\TetInt)=\OctDD,
        \qquad
        N_{\PGL_2(\mathbb C)}(\OctDD)=\OctDD,
        \qquad
        N_{\PGL_2(\mathbb C)}(\IcoZero)=\IcoZero.
      \]
\item The projective centralizers of the distinguished involutions are
      \[
      C_{\PGL_2(\mathbb C)}([Z])
      =
      \left\{
      \left[\begin{pmatrix}a&0\\0&b\end{pmatrix}\right]:ab\ne0
      \right\}
      \sqcup
      \left\{
      \left[\begin{pmatrix}0&a\\b&0\end{pmatrix}\right]:ab\ne0
      \right\}.
      \]
      Thus \([Z]\) is the only nonidentity diagonal involution, while
      every nonsingular anti-diagonal class is an involution.  Moreover,
      \[
        C_{\PGL_2(\mathbb C)}\bigl(\langle[X],[Z]\rangle\bigr)
        =\langle[X],[Z]\rangle\cong V_4.
      \]
      Every finite subgroup of the displayed diagonal torus is cyclic.
\end{enumerate}
\end{lemma}

\begin{proof}
Items~1--6 follow directly from the cyclic and dihedral presentations and
the usual permutation models of \(A_4,S_4,A_5\)
\cite{DixonMortimer1996}.  For item~7, conjugation
embeds each normalizer modulo its projective centralizer into the
corresponding automorphism group.  The projective centralizers are trivial.
\(\operatorname{Aut}(A_4)\cong S_4\), the displayed \(\OctDD\)
normalizes \(\TetInt\), and every automorphism of \(S_4\) is inner.
Since \(\operatorname{Aut}(A_5)\cong S_5\), an additional outer normalizer
of \(\IcoZero\) would give a finite projective group \(H\) of order \(120\)
containing \(A_5\).  The classification forces \(H\cong C_{120}\) or
\(D_{120}\). Every subgroup of a cyclic or dihedral group is cyclic or
dihedral, so \(H\) cannot contain \(A_5\).  Item~8 follows by solving
\(UZ\doteq ZU\), and then imposing the same identity with \(X\).  The ratio
map \([\diag(a,b)]\mapsto a/b\) embeds the diagonal torus into
\(\mathbb C^\times\). Every finite subgroup of \(\mathbb C^\times\) is a cyclic group
of roots of unity.
\end{proof}

The abstract classification is up to projective conjugacy. Our group
cases also retain the ordered marks \([X],[Z]\) in the original
coordinates. A global change of basis or a binary gadget is used only
with its corresponding holographic identity or realization.

\paragraph{Matrix and marked-pair dictionary.}
Combining \cref{eq:quaternion-basis,eq:bell-matrix-dictionary} gives the
matrix identities
\hypertarget{eq:global-matrix-aliases}{}
\begin{equation*}
 I=Q_0=B_{0,0},\qquad
 iX=Q_1=iB_{1,0},\qquad
 Y=Q_2=B_{1,1},\qquad
 iZ=Q_3=iB_{0,1}.
\end{equation*}
Consequently the marked projective notation used throughout is
\hypertarget{eq:global-mark-aliases}{}
\begin{equation*}
 \begin{aligned}
  1&=[I]=[Q_0]=[B_{0,0}],&
  x&=[X]=[Q_1]=[B_{1,0}],\\
  j&=[Z]=[Q_3]=[B_{0,1}],&
  xj&=[Q_2]=[B_{1,1}].
 \end{aligned}
\end{equation*}
Here the last identity is projective: \(XZ=-Y\), whereas \(ZX=Y\).  For
the transfer convention \(T_B=BX\), the four Bell binaries give
\begin{equation*}
\begin{array}{c|c|c|c}
 B&\text{Bell name}&[B]&[T_B]=[BX]\\ \hline
 I&B_{0,0}&1&x\\
 X&B_{1,0}&x&1\\
 Z&B_{0,1}&j&xj\\
 Y&B_{1,1}&xj&j
\end{array}
\end{equation*}
The subscripted \(B_{d,\ell}\) is a fixed ordered Bell signature, whereas
\(B_g\) denotes a chosen realizable binary representative of a general
transfer class \(g\).  Reversal fixes \(I,X,Z\) and sends
\(Y=B_{1,1}\) to \(-Y\), without changing its projective class.

Use the diagonal family \(D(t)\) defined in
\cref{subsec:holant-definitions}.
For the dihedral forms, the subscript gives the group order and the
superscript records the direction of the rotation subgroup: \(Z\), \(Y\),
or \(X\). The labels \((X,0)\) and \((X,1)\) distinguish the unshifted
and shifted reflection cosets, respectively.
For \(m\ge2\), define the matrix family \(C(t)\) and the first two marked
dihedral families
\begin{align*}
 \DihZ{2m}
 &=D_{2m}^{\rm std}=\{[D(t)],[XD(t)]:t\in\mu_m\},
 \\
 C(t)&=(1+t)I+i(1-t)Y,\\
 \DihY{2m}
 &=\{[C(t)],[XC(t)]:t\in\mu_m\}.
\end{align*}
For even \(N\ge2\), define the cyclic group
\[
 A(t)=(1+t)I+(1-t)X,\qquad \CycX{N}:=\{[A(t)]:t\in\mu_N\}.
\]
Its unique involution is \(x=[X]=[A(-1)]\).
For even \(m\), fix \(\eta\in\mu_{2m}\setminus\mu_m\) and define
\begin{equation*}
 \DihXone{2m}
 =
 \CycX{m}
 \cup
 \{[ZA(\eta t)]:t\in\mu_m\}.
\end{equation*}
In the same even parameter range, define the unshifted marked form
\begin{equation*}
 \DihXzero{2m}
 =
 \{[A(t)],[ZA(t)]:t\in\mu_m\}.
\end{equation*}
These are projective displays.  Any suppressed factor of two belongs to
the coordinate model and does not normalize a chosen representative.

The symbols \(\TetInt,\TetExt\),
\(\OctDD,\OctTT,\OctDT,\OctTD\), and
\(\IcoZero,\IcoOne\) denote the marked tetrahedral, octahedral,
and icosahedral forms specified in the theorem. Here \(\mathrm{int}\)
and \(\mathrm{ext}\) record whether \(j\in G\) or \(j\notin G\).
For octahedral forms, \(d\) denotes a double transposition and \(t\) a
transposition, in the order \((x,j)\). The labels \(0,1\) distinguish
the two icosahedral coordinate displays of the same abstract marked type.
\Cref{sec:platonic-decks} gives the explicit fixed-equality representatives
and coefficient correspondences.

\begin{theorem}[Marked finite-group routing]
\label{thm:marked-finite-group-routing}
Let \(\cF\) be a retained factor-closed set containing the
supplied equality \(I\), and let \(G=G(\cF)\) be its complete transfer group (finite, with safe binary representatives).  Recall the ordered marked pair and its generated
group:
\[
 (x,j):=([X],[Z]),\qquad x\in G,\qquad \widehat G:=\langle G,j\rangle.
\]
In the fixed coordinates containing the supplied \(I\), exactly
one of the following alternatives holds:
\begin{enumerate}
\item \(G\cong D_{2m}\), \(m\ge3\), and the marked configuration
      \((G;x,j)\) has one of the forms denoted by
      \(\DihZ{2m},\DihY{2m},\DihXone{2m},\DihXzero{2m}\).  The last
      two families occur only for even \(m\).
\item \(G=\CycX{N}\cong C_N\), where \(N\) is even and \(j\notin G\).  This is
      the cyclic branch.
\item \(G\cong V_4\), and \(G\) is either the standard Pauli group
      \(\KleinStd=\DihZ{4}=\DihY{4}\) or the exotic group
      \(\KleinEx=\DihXone{4}\).
\item \(G\) is one of the two marked tetrahedral forms
      \(\TetInt,\TetExt\), the four fixed-equality marked
      octahedral forms
      \(\OctDD,\OctTT,\OctDT,\OctTD\), or one of the two icosahedral displays
      \(\IcoZero,\IcoOne\).
\end{enumerate}

\smallskip
\noindent\emph{Positions of the marks.}
All conjugacy statements in this paragraph are taken in \(\widehat G\).  In every
form, \(x\) and \(j\) are distinct commuting involutions and
\(\langle x,j\rangle=\{1,x,j,xj\}\cong V_4\).  In the four dihedral forms the
marked positions are
\begin{align*}
 \DihZ{2m}:&\quad
 x=[XD(1)],\quad j=[D(-1)],\quad xj=[XD(-1)],\qquad
 \widehat G=\begin{cases}\DihZ{2m},&2\mid m,\\ \DihZ{4m},&2\nmid m;\end{cases}\\
 \DihY{2m}:&\quad
 x=[XC(1)],\quad j=[XC(-1)],\quad xj=[C(-1)],\qquad
 \widehat G=\begin{cases}\DihY{2m},&2\mid m,\\ \DihY{4m},&2\nmid m;\end{cases}\\
 \DihXone{2m}:&\quad
 x=[A(-1)]\in G,\quad j=[ZA(1)]\in \widehat G\setminus G,\quad
 xj=[ZA(-1)]\in \widehat G\setminus G;\\[-2pt]
 &\quad
 \widehat G=\{[A(t)]:t\in\mu_{2m}\}\cup
   \{[ZA(t)]:t\in\mu_{2m}\}\cong D_{4m};\\
 \DihXzero{2m}:&\quad
 x=[A(-1)],\quad j=[ZA(1)],\quad xj=[ZA(-1)],\qquad
 \widehat G=G=\DihXzero{2m}.
\end{align*}
For \(G=\DihZ{2m}\), the central half-turn of \(\widehat G\) is \(j\) and the other two marks are reflections.
For \(G=\DihY{2m}\), it is \(xj\) and the reflections are \(x,j\).
In both cases \(j\in G\) exactly when \(m\) is even. The two marked reflections are conjugate in \(\widehat G\) exactly when \(4\mid m\). In
\(\DihXone{2m}\), where \(m\) is even, \(x\) is the central half-turn and
\(j,xj\) are conjugate reflections in \(\widehat G\setminus G\).
In \(\DihXzero{2m}\), \(x\) is again the central half-turn, while
\(j,xj\in G\) are reflections. They are conjugate in \(G\) exactly when
\(4\mid m\).
When \(m\) is odd in either \(\DihZ{2m}\) or \(\DihY{2m}\), one has
\(j,xj\in \widehat G\setminus G\). When \(m\) is even, all three marks lie in \(G\).

In the cyclic branch \(G=\CycX{N}\), the order \(N\) is even,
\(j,xj\in \widehat G\setminus G\), \(\widehat G\cong D_{2N}\), \(x\) is the central half-turn, and
\(j,xj\) are reflections, conjugate exactly when \(4\mid N\).

At \(m=2\), the standard Pauli form has \(\widehat G=G\), with \(x,j,xj\) the three
nonidentity elements (three singleton conjugacy classes) of \(\KleinStd\). The
exotic form has \(j,xj\in \widehat G\setminus G\), \(\widehat G\cong D_8\), central half-turn
\(x\), and conjugate outside reflections \(j,xj\).  In \(\TetInt\),
\(\widehat G=G\cong A_4\) and \(x=[Q_1],j=[Q_3],xj=[Q_2]\) are the three double
transpositions, hence one conjugacy class.  In \(\TetExt\),
\(G\cong A_4\), \(G\triangleleft \widehat G\), and \(\widehat G\cong S_4\).
\(x=[Q_1]\in G\) is a double transposition,
and \(j=[Q_3],xj=[Q_2]\in \widehat G\setminus G\) are disjoint transpositions in one
  class.  In \(\OctDD\), \(\widehat G=G\cong S_4\) and \(x,j,xj\) are the three
  double transpositions.  In \(\OctTT\), \(x,j\) are disjoint
  transpositions and \(xj\) is a double transposition.  In
  \(\OctDT\), \(x\) is a double transposition and \(j,xj\) are
  transpositions. In \(\OctTD\), \(j\) is a double transposition
  and \(x,xj\) are transpositions.  In all four cases \(\widehat G=G\cong S_4\).
  Finally, in
each icosahedral coordinate display \(\widehat G=G\cong A_5\), and
\(x=[Q_1],j=[Q_3],xj=[Q_2]\) lie in its unique involution class.
\end{theorem}

\begin{proof}
The reversal identity
\([T]\mapsto j[T]^{-1}j=[ZT^{-1}Z]\) (with \(j=[Z]\)) and completeness give
\[
 jGj=G,\qquad G\triangleleft \widehat G,\qquad [\widehat G:G]\le2.
\]
The two distinct projective involutions \(x,j\) commute, so
\(\KleinStd=\langle x,j\rangle\le\widehat G\).  By
\cref{prop:finite-pgl2-classification}, this leaves an even dihedral group
or one of \(A_4,S_4,A_5\).

Suppose first that \(G\cong D_{2m}\) with \(m\ge3\).  Then \(\widehat G\) is also
dihedral.  Indeed, if \(\widehat G\) were Platonic and \([\widehat G:G]=2\), then \(G\)
would be an index-two normal subgroup.  By
\cref{lem:finite-pgl2-facts}, the only such subgroup among the Platonic
groups is \(A_4\triangleleft S_4\), which is not dihedral.  The case
\(\widehat G=G\) is immediate.  Choose a rotation generator \(r\) and a reflection
generator \(s\) so that
\(\widehat G=\langle r,s:r^n=s^2=1,\ srs=r^{-1}\rangle\).
Since \(\widehat G\) contains \(\langle x,j\rangle\cong V_4\), \(n\) is even.
Exactly one of \(x,j,xj\) is the central half-turn \(r^{n/2}\), and
the other two are reflections.  If the central mark is \(j\) or \(xj\),
we obtain \(\DihZ{2m}\) or \(\DihY{2m}\), respectively.
If it is \(x\), membership \(j\in G\) gives \(\DihXzero{2m}\),
while \(j\notin G\) gives \(\DihXone{2m}\).
For \(j\in G\), one has \(\widehat G=G\), so \(n=m\) is even.
For \(j\notin G\), one has \(n=2m\) and
\(G=\langle r^2,r^ks\rangle\) for some \(k\in\{0,1\}\).
The central half-turn \(r^m\) of \(\widehat G\) belongs to \(G\) exactly when \(m\) is even.
Thus \(m\) is odd in the first two outside-mark cases and even in
\(\DihXone{2m}\), where \(x\in G\).  This exhausts the four forms
and their parameter ranges.

In \(\DihXzero{2m}\), the mark \(j=[Z]\) belongs to the complete group.
Its fixed chosen representative \(B_j\) satisfies
\([B_jX]=j\), hence \(B_j\doteq Y\).  This realizable binary and the already
available \(I\) give the signature-preserving edge switch used in
\cref{thm:marked-A0-reduction}. The distinguished equality is never
replaced or re-anchored.

The marked eigenline equations give the displayed matrices, with
\[
 ZC(t)Z=tC(t^{-1}),\qquad
 x=[XC(1)],
\]
while \(x=[A(-1)]\) is central in both \(\DihXone{2m}\) and
\(\DihXzero{2m}\). Hence the four marked configurations are legally distinct.

Next suppose that \(G\) is cyclic.  By
\cref{lem:finite-pgl2-facts}, its unique involution is \(x\).  A class
commuting with \([X]\) lies in \(\Span\{I,X\}\) or
\(\Span\{Z,Y\}\), and every nonsingular member of the latter is another
involution.  Thus all transfer classes have representatives in
\(\Span\{I,X\}\). Their eigenvalue ratios form \(\mu_N\), where \(N=|G|\).
Thus \(G=\CycX{N}\), \(N\) is even, and \(j\notin G\). Conjugation by \(j\) inverts the eigenvalue ratio of every
class in this algebra, so
\[
 \widehat G=G\rtimes\langle j\rangle\cong D_{2N}.
\]
Consequently, \(x\) is the central half-turn and \(j,xj\) are the two
marked reflections.

It remains to consider \(G\cong V_4\), corresponding to \(m=2\) in the
dihedral notation.  Here
\[
 \KleinStd=\DihZ{4}=\DihY{4}=\{[I],[X],[Z],[Y]\}.
\]
For \(\eta=i\),
\[
 \frac{ZA(i)}{1-i}=Y+iZ,\qquad
 \frac{ZA(-i)}{1+i}=Y-iZ,
\]
so \(\KleinEx=\{[I],x,[Y+iZ],[Y-iZ]\}\). Conversely, every subgroup \(G\cong V_4\) containing \(x\) has the form
\[
 \{[I],x,[aZ+bY],[bZ+aY]\},\qquad a^2\ne b^2,
\]
and \(jGj=G\) gives either \(ab=0\) (the Pauli form) or
\(a^2=-b^2\) (the exotic form \(\KleinEx\)).

Thus the Pauli form has \(\widehat G=G\), with \(x,j,xj\) its three nonidentity
elements.  In the exotic form, \(j,xj\notin G\) and
\(\widehat G=\langle G,j\rangle\cong D_8\). Relative to this \(\widehat G\), \(x\) is the central
half-turn and \(j,xj\) are conjugate reflections in the outside coset.

It remains to route the Platonic groups while retaining the ordered marked
pair.  Let \(G\cong A_4\).  If \(j\in G\), then \(\widehat G=G\), and the three
nonidentity elements \(x,j,xj\) of \(\KleinStd\) are precisely the
double transpositions.  This is \(\TetInt\).  If \(j\notin G\), then
\(|\widehat G|=24\).  By
\cref{prop:finite-pgl2-classification,lem:finite-pgl2-facts}, the
finite-\(\PGL_2\) classification and the subgroup structure of cyclic and
dihedral groups force
\(\widehat G\cong S_4\) with \(G\triangleleft\widehat G\) and \(G\cong A_4\).  Thus \(x\in G\) is a
double transposition, whereas the odd involutions \(j,xj\in \widehat G\setminus G\)
are transpositions.  Since they are distinct and commute, they are disjoint.
This is \(\TetExt\).

Now let \(G\cong S_4\). If \(j\notin G\), then
\(\widehat G=\langle G,j\rangle\) has order \(48\) and is cyclic or dihedral by
\cref{prop:finite-pgl2-classification}, contrary to containing \(S_4\)
by \cref{lem:finite-pgl2-facts}. Hence \(j\in G\).

  The two commuting involutions may independently lie in the transposition
  or double-transposition class, subject to the class of their product.  The
  four resulting ordered class patterns are
  \[
    (dd;d),\qquad (tt;d),\qquad (dt;t),\qquad (td;t),
  \]
  where the entries record the classes of \((x,j;xj)\).  Conjugation in
  \(S_4\) is transitive on ordered pairs in each pattern by
  \cref{lem:finite-pgl2-facts}.  They are therefore exactly the four
  fixed-equality marked types
  \[
    \OctDD,\qquad \OctTT,\qquad
    \OctDT,\qquad \OctTD.
  \]
  Their ordered marks distinguish the four types in the fixed coordinates.
  Concrete representatives and direct finite exits are recorded in
  \cref{sec:platonic-decks} and
  Appendix~\ref{appsec:cert-platonic}.

For \(G\cong A_5\), if \(j\notin G\), then \(\widehat G\) has order \(120\)
and is cyclic or dihedral by \cref{prop:finite-pgl2-classification},
contrary to containing \(A_5\) by \cref{lem:finite-pgl2-facts}.
Thus \(\widehat G=G\). Finally, \cref{lem:finite-pgl2-facts} shows that \(A_5\) has one
involution class, with \(x,j,xj\) in that class.  Once an ordered pair
\((x,j)\) is fixed, two conjugators differ by
\[
 C_{\PGL_2(\mathbb C)}(\langle x,j\rangle)=\langle x,j\rangle.
\]
Each of these four projective classes has a representative in \(GO(X)\).
  Hence the abstract marked type is unique.  Fixed-equality coordinates give
  the two displays \(\IcoZero,\IcoOne\). \Cref{sec:platonic-decks} verifies their
  explicit representatives and a direct coefficient-and-card correspondence
  that leaves the supplied \(I\) untouched.  This deferred coordinate calculation is not used
to classify the abstract group, so the four-way list is exhaustive.
\end{proof}

Each of the four group-specific sections,
\cref{sec:normalized-dihedral,sec:proper-deck,sec:v4-deck,sec:platonic-decks},
proves a relative theorem for its corresponding stable tuple.  It either
determines the complexity at that tuple, or
produces a reduction step whose processing yields a strictly smaller
active tuple under \(\Xi_{\mathrm{full}}\).  In the second case, the deck is stabilized
again and the new tuple is routed anew.  Movement between distinct
same-arity irreducible signatures in a proof for the whole signature set remains governed by the
separate outer worklist induction of \cref{def:protocol-discharged}.

Coefficient and open-network correspondences in the branch proofs
leave the retained set and supplied \(I\) fixed. If factor access
enlarges \(G\), reapply \cref{lem:minimum-core-deck} and the marked
classification to the new complete group.

\begin{remark}[Connection with Xia's binary-group taxonomy]
\label{rem:xia-binary-taxonomy}
Version~1 of Xia's framework
\cite[Sections~1 and~4]{Xia2026FrameworkV1} organizes the finite projective
binary residue into nine \emph{unmarked} rotation-group types.  Here the
group consists of transfer classes \([T_B]=[BX]\), rather than projective
classes \([B]\) of the underlying binary matrices, and the supplied binary equality fixes the
marks
\[
  x=[X]\in G,
  \qquad
  j=[Z],
  \qquad
  jGj=G.
\]
Consequently, our routing refines Xia's taxonomy by the ordered marked pair
\((x,j)\).  The table compares the unmarked types with the cases compatible with
the supplied \(I\) in this section.
\begin{center}
\normalsize
\setlength{\tabcolsep}{4pt}
\renewcommand{\arraystretch}{0.92}
\begin{tabularx}{0.96\linewidth}{@{}L{0.21\linewidth}X@{}}
\toprule
Xia's unmarked type & Corresponding case in this proof\\
\midrule
\(C_1\) &
absent at a continuing node, where a nonidentity transfer is already retained\\
\(C_2\) &
the order-two cyclic case \(\CycX{2}\), including the Reed--Muller case of \cref{sec:proper-deck}\\
\(C_n\), \(n\ge3\) &
  the fixed-\(I\) cyclic case \(\CycX{N}\ni[X]\), with even \(N\ge4\)\\
\(D_{2m}\), \(m\) odd &
\(\DihZ{2m},\DihY{2m}\), with odd \(m\ge3\)\\
\(V_4=D_4\) &
\(\KleinStd,\KleinEx\)\\
\(D_{2m}\), even \(m\ge4\) &
\(\DihZ{2m},\DihY{2m},\DihXzero{2m},\DihXone{2m}\)\\
\(A_4\) & \(\TetInt,\TetExt\)\\
\(S_4\) & \(\OctDD,\OctTT,\OctDT,\OctTD\)\\
\(A_5\) & the two fixed-equality coordinate displays\\
\bottomrule
\end{tabularx}
\end{center}
The marked classification and the corresponding signature-set
reductions are proved here.
\end{remark}

\subsection{Strict growth of the complete transfer group}
\label{subsec:complete-group-growth}

The preceding classification identifies the branch for a fixed complete
group. A later contraction or factor reduction may enlarge that group and
return the proof to a different branch. \Cref{fig:complete-group-growth}
shows how these branches fit together, retaining the marked distinctions
that determine which later argument applies.

Work in the fixed coordinates containing the supplied \(I\).  Let
\(\cF\subseteq\cF^+\) be a retained factor-closed set,
put
\[
 G=G(\cF),\qquad G^+=G(\cF^+),
\]
and suppose that the unsafe-binary case distinction has not already applied.  If a
newly exposed ordered binary \(B\) is safe and \([BX]\notin G\), then
\cref{lem:actual-binary-completeness} gives the strict inclusion
\(G<G^+\).  If \([BX]\in G\), its transfer class is already present. We still
consider the complete group of the resulting set, which may realize
additional multi-copy binaries beyond the batch that exposed \(B\).
Accordingly, \(G^+\) is the complete group of realizable transfer classes, and need not equal
the subgroup generated by \(G\) and the first new class.  Whenever
\(G<G^+\), regenerate the deck and reroute before invoking a
group-specific theorem.

\begin{figure}[p]
\centering
\begin{tikzpicture}[
  x=1cm,y=1cm,
  state/.style={
    draw,rounded corners=2pt,align=center,text width=3.05cm,
    minimum height=1.05cm,inner sep=2pt,font=\normalsize
  },
  strictedge/.style={
    -{Stealth[length=2.2mm]},semithick,rounded corners=4pt,
    preaction={draw=white,line width=3.5pt,solid,-}
  },
  reroute/.style={
    -{Stealth[length=2.2mm]},semithick,rounded corners=4pt,
    preaction={draw=white,line width=3.5pt,solid,-}
  },
  edgetag/.style={
    circle,draw,fill=white,minimum size=3.5mm,
    inner sep=.35pt,font=\bfseries
  }
]
  \node[state] (c2) at (0,0)
    {$\CycX{2}=\langle x\rangle$};

  \node[state] (vex) at (-5.4,-2.8)
    {exotic Klein-four\\$\KleinEx=\DihXone{4}$};
  \node[state] (cyc) at (0,-2.8)
    {proper cyclic\\$\CycX{N}$, $N>2$ even};
  \node[state] (vp) at (5.4,-2.8)
    {standard Pauli\\$\KleinStd=\DihZ{4}$\\$=\DihY{4}$};

  \node[state] (aa) at (-3.5,-5.8)
    {$\DihXone{2m}$\\$m\ge4$ even};
  \node[state] (mm) at (3.5,-5.8)
    {$\DihZ{2m}$\\$m\ge3$};

  \node[state] (te) at (-5.4,-9.0)
    {$\TetExt$\\external $A_4$};
  \node[state] (bb) at (0,-9.0)
    {$\DihY{2m}$ or\\$\DihXzero{2m}$\\$m\ge3$};
  \node[state] (tt) at (5.4,-9.0)
    {$\TetInt$\\internal $A_4$};

  \node[state] (op) at (-5.4,-12.2)
    {$\OctTT$ or $\OctDT$\\$S_4$};
  \node[state] (oo) at (1.8,-12.2)
    {$\OctDD$ or $\OctTD$\\$S_4$};
  \node[state] (ii) at (5.8,-12.2)
    {$A_5$\\$\IcoZero,\IcoOne$};

  \draw[strictedge] (c2.south) -- (cyc.north);
  \draw[strictedge] (c2.south west) -- (vex.north east);
  \draw[strictedge] (c2.south east) -- (vp.north west);

  \draw[strictedge]
    (c2.south west)
    -- (-2.15,-1.05)
    -- (-2.15,-3.75)
    -- (-1.25,-3.75)
    -- (-1.25,-8.05)
    -- ([xshift=-8mm]bb.north);

  \draw[strictedge]
    (c2.east)
    -- (7.55,0)
    -- (7.55,-5.8)
    -- (mm.east);

  \draw[strictedge]
    (cyc.south west) -- (aa.north east);

  \draw[reroute]
    (cyc.south)
    -- (0,-8.05)
    -- ([xshift=-3mm]bb.north);

  \draw[strictedge]
    (vex.south east) -- (aa.north west);

  \draw[reroute]
    (vex.south east)
    -- (-3.0,-4.20)
    -- (-1.60,-4.20)
    -- (-1.60,-8.05)
    -- ([xshift=-12mm]bb.north);

  \draw[strictedge]
    (vp.south west)
    -- (3.0,-4.45)
    -- (1.60,-4.45)
    -- (1.60,-8.05)
    -- ([xshift=10mm]bb.north);

  \draw[strictedge]
    (vp.south west) -- (mm.north east);

  \draw[reroute]
    (aa.south)
    -- (-3.5,-7.75)
    -- (bb.north west);

  \draw[strictedge] (vex.south) -- (te.north);
  \draw[strictedge] (vp.south) -- (tt.north);

  \draw[strictedge]
    (bb.south west) -- (op.north east);

  \draw[strictedge]
    (bb.south east) -- (oo.north west);

  \draw[strictedge]
    (mm.south) -- ([xshift=6mm]oo.north);

  \draw[reroute]
    (te.south) -- (op.north);

  \draw[strictedge]
    (tt.south west) -- ([xshift=12mm]oo.north);

  \draw[strictedge]
    (tt.south east) -- (ii.north west);

  \node[edgetag] at (.38,-1.38) {\DEQedgenumber{1}};
  \node[edgetag] at (-3.05,-1.20) {\DEQedgenumber{2}};
  \node[edgetag] at (2.75,-.95) {\DEQedgenumber{3}};
  \node[edgetag] at (-2.55,-2.65) {\DEQedgenumber{4}};
  \node[edgetag] at (7.93,-3.25) {\DEQedgenumber{5}};
  \node[edgetag] at (-2.20,-4.70) {\DEQedgenumber{6}};
  \node[edgetag] at (.38,-5.55) {\DEQedgenumber{7}};
  \node[edgetag] at (-4.10,-4.45) {\DEQedgenumber{8}};
  \node[edgetag] at (-1.95,-6.75) {\DEQedgenumber{9}};
  \node[edgetag] at (1.95,-6.75) {\DEQedgenumber{10}};
  \node[edgetag] at (4.10,-4.35) {\DEQedgenumber{11}};
  \node[edgetag] at (-3.90,-7.00) {\DEQedgenumber{12}};
  \node[edgetag] at (-5.78,-6.95) {\DEQedgenumber{13}};
  \node[edgetag] at (5.78,-6.95) {\DEQedgenumber{14}};
  \node[edgetag] at (-2.80,-10.18) {\DEQedgenumber{15}};
  \node[edgetag] at (.72,-10.20) {\DEQedgenumber{16}};
  \node[edgetag] at (2.75,-7.55) {\DEQedgenumber{17}};
  \node[edgetag] at (-5.78,-10.62) {\DEQedgenumber{18}};
  \node[edgetag] at (3.82,-10.63) {\DEQedgenumber{19}};
  \node[edgetag] at (6.55,-10.55) {\DEQedgenumber{20}};
\end{tikzpicture}

\vspace{.15em}
\begingroup
\normalsize
\setlength{\tabcolsep}{2.5pt}
\renewcommand{\arraystretch}{1.08}

\def\gref#1{%
  \tikz[baseline=(n.base)]
  \node[
    draw,circle,minimum size=3.5mm,
    inner sep=.35pt,font=\bfseries
  ] (n) {\DEQedgenumber{#1}};%
}

\begin{tabularx}{\linewidth}{
  @{}r>{\raggedright\arraybackslash}X@{\hspace{7pt}}
  r>{\raggedright\arraybackslash}X@{}
}
\gref{1}&$\CycX{2}\to \CycX{N}$
&\gref{11}&$\KleinStd\to\DihZ{2m}$, $2\mid m$.\\

\gref{2}&$\CycX{2}\to \KleinEx$
&\gref{12}&$\DihXone{2m}\to\DihXzero{4m}$,
  $j$ enters $G^+$.\\

\gref{3}&$\CycX{2}\to \KleinStd$
&\gref{13}&$\KleinEx\to\TetExt$.\\

\gref{4}&$\CycX{2}\to\DihY{2m}$, $m$ odd.
&\gref{14}&$\KleinStd\to\TetInt$.\\

\gref{5}&$\CycX{2}\to\DihZ{2m}$, $m$ odd.
&\gref{15}&$\DihY{8}\to\OctTT$ or
  $\DihXzero{8}\to\OctDT$.\\

\gref{6}&$\CycX{N}\to\DihXone{2m}$, $N\mid m$.
&\gref{16}&$\DihY{8}\to\OctDD$ or
  $\DihXzero{8}\to\OctDD$.\\

\gref{7}&$\CycX{N}\to\DihXzero{2m}$;
  $N\mid m$, $j$ enters $G^+$.
&\gref{17}&$\DihZ{8}\to\OctDD$ or $\OctTD$.\\

\gref{8}&$\KleinEx\to\DihXone{2m}$;
  $m\equiv2\pmod4$.
&\gref{18}&$\TetExt\to\OctDT$,
  $j$ enters $G^+$.\\

\gref{9}&$\KleinEx\to\DihXzero{2m}$;
  $4\mid m$, $j$ enters $G^+$.
&\gref{19}&$\TetInt\to\OctDD$.\\

\gref{10}&$\KleinStd\to\DihY{2m}$ or $\DihXzero{2m}$, $2\mid m$.
&\gref{20}&$\TetInt\to\IcoZero$ or $\IcoOne$.\\
\end{tabularx}

\medskip
\noindent Here \(j\) enters \(G^+\) when \(j\notin G\) and \(j\in G^+\).
The parameters satisfy strict inclusion, with \(m\ge3\) for dihedral
targets and even \(m\ge4\) for \(\DihXone{2m},\DihXzero{2m}\).
\endgroup

\caption{Possible strict growth of the complete transfer group with the
supplied equality fixed. Numbered arrows indicate subgroup inclusions
between marked forms, with their parameter conditions listed below the
graph. Divisibility steps within a family are suppressed. Grouped nodes
collect distinct marked forms at the same level of the diagram.}
\label{fig:complete-group-growth}
\end{figure}

Recall the marked extension \(\widehat G=\langle G,j\rangle\) from \cref{thm:marked-finite-group-routing}.
The diagram tracks inclusions of \(G\). An outside mark \(j\) contributes
to growth only when it enters the complete group of the enlarged set.
For example, adjoining \(j\) to \(\DihXone{2m}\) gives
\(\DihXzero{4m}\). The standard Klein-four node is
\(\KleinStd=\DihZ{4}=\DihY{4}\), and the exotic node is
\(\KleinEx=\DihXone{4}\).

\begin{proposition}[Finite marked-successor routing]
\label{prop:finite-marked-successor-routing}
Let \(\cF\subseteq\cF^+\) be retained factor-closed sets containing
the supplied equality \(I\), both in the binary-residual regime, and suppose
\[
 G=G(\cF)<G^+=G(\cF^+).
\]
After recomputing the complete group and applying the marked classification
\cref{thm:marked-finite-group-routing}, the marked type of \(G^+\) is
reachable from the marked type of \(G\) by a directed
path in \cref{fig:complete-group-growth}, allowing the suppressed
same-family divisibility steps.  The paths describe the group inclusions compatible with the fixed marks.
Their occurrence for a particular signature set depends on its realizable
binaries.
\end{proposition}

\begin{proof}
The fixed marks \(x=[X]\) and \(j=[Z]\), together with reversal closure, give
\[
 x\in G<G^+,
 \qquad jGj=G,
 \qquad jG^+j=G^+.
\]
Hence \(\widehat G\le\langle G^+,j\rangle\), and if \(j\in G^+\) then
\(\widehat G\le G^+\).  Apply
\cref{prop:finite-pgl2-classification,lem:finite-pgl2-facts} and retain this
marked-pair constraint throughout.

Suppose first that \(G=\CycX{N}\), where \(N\) is even.  For \(N>2\), every
embedding in a dihedral overgroup places \(\CycX{N}\) in the rotation subgroup;
therefore its unique involution \(x\) is the central half-turn.  If
  \(j\notin G^+\), the routed form is \(\DihXone{2m}\); if \(j\in G^+\),
  the routed form is \(\DihXzero{2m}\).  Enlarging first inside the cyclic family accounts for
the displayed condition \(N\mid m\).  The group \(\CycX{2}=\langle x\rangle\)
is exceptional because \(x\) may instead be a reflection, giving the
odd-parameter entrances to \(\DihZ{2m}\) and \(\DihY{2m}\); the
even-parameter entrances factor through one of the displayed Klein forms.

For the Klein boundary, \(\KleinStd\) already contains \(j\), so a
dihedral overgroup has all three marked involutions and routes to
\(\DihZ{2m}\), \(\DihY{2m}\), or \(\DihXzero{2m}\) with even \(m\).
These three cases are distinguished by which of \(j,xj,x\), respectively,
is the central half-turn of the dihedral overgroup.  In the exotic
form \(\KleinEx=\DihXone{4}\), the mark \(j\) lies outside.
The reflection parameters in \(\DihXone{2m}\) form \(\mu_{2m}\setminus\mu_m\).
Since \(\mu_4\setminus\mu_2=\{i,-i\}\), for even \(m>2\) we have
\[
 \DihXone{4}<\DihXone{2m}
 \quad\Longleftrightarrow\quad
 i^m=-1
 \quad\Longleftrightarrow\quad
 m\equiv2\pmod4.
\]
When \(4\mid m\), the inclusion \(\KleinEx<\DihXzero{2m}\) factors
through the subgroup \(\langle\KleinEx,j\rangle=\DihXzero{8}\).
More generally,
\[
  \langle\DihXone{2m},j\rangle=\DihXzero{4m}.
\]
The remaining dihedral-over-dihedral possibilities are precisely the
routine divisibility inclusions suppressed from the drawing.  In the
  \((X,1)\)-family, an odd index preserves the shifted reflection coset;
  an even index enters an unshifted \(\DihXzero{2n}\) form.  Thus all cyclic, Klein, and dihedral
successors follow the displayed paths.

It remains to attach the Platonic groups.  The unique normal Klein four in
\(A_4\), together with whether \(j\) belongs to the current group, gives
\[
 \KleinStd<\TetInt,
 \qquad
 \KleinEx<\TetExt.
\]
The displayed models in
\cref{eq:tetrahedral-domain,eq:octahedral-domain,eq:icosahedral-domain}
give
\[
 \TetInt<\OctDD,
 \qquad
 \TetInt<\IcoZero\cap\IcoOne.
\]
For the external tetrahedral form, adjoining \(j\) gives the mixed fixed-\(I\)
form \(\OctDT\).

The only marked dihedral groups of parameter at least three that can enter
a Platonic successor have parameter four, with fixed-mark inclusions
\[
 \DihZ{8}<\OctDD\text{ or }\OctTD,
\qquad
 \DihY{8}<\OctDD\text{ or }\OctTT,
\qquad
 \DihXzero{8}<\OctDD\text{ or }\OctDT.
\]
Indeed, an \(S_4\) dihedral subgroup of parameter at least three has
parameter \(3\) or \(4\), by the element orders in
\cref{lem:finite-pgl2-facts}.  At parameter \(3\), the marked extension
\(\widehat G\) has a rotation of order \(6\), which \(S_4\) does not.
At parameter \(4\), the \(\DihZ{8}\), \(\DihY{8}\), and
\(\DihXzero{8}\) marked positions give the displayed inclusions, while
\(\widehat{\DihXone{8}}\cong D_{16}\) cannot occur. The domain and
second-form matrices in \cref{eq:octahedral-domain,eq:octahedral-second-form},
together with the conjugations and mark checks in
\cref{eq:octahedral-mixed-conjugators,eq:octahedral-mixed-forms,eq:octahedral-mixed-mark-check},
verify the three inclusions. Here the second-form similarity identifies
subgroup classes. The same argument for \(A_5\)
leaves only parameters \(3\) and \(5\), but their marked extensions contain
rotations of order \(6\) and \(10\), absent from \(A_5\).  No
parameter-at-least-three dihedral group lies in \(A_4\): the only possible
parameter is \(3\), and an order-six subgroup would have index two,
contrary to \cref{lem:finite-pgl2-facts}.

Finally, \(\TetExt\) cannot grow to a group isomorphic to \(A_5\), because the
target contains \(j\) and would therefore contain
\(\langle\TetExt,j\rangle\cong S_4\), whereas
\(24\nmid60\).  The groups \(S_4\) and \(A_5\) have no strict finite
overgroup in \(\PGL_2(\mathbb C)\): the finite-group classification and
orders exclude every remaining type.  This proves exhaustiveness.
\end{proof}

Every strict inclusion increases \(|G|\), and therefore decreases the
group coordinate of \(\Xi_{\mathrm{full}}\) in \cref{lem:matching-synthesis-measure}.
The bounded-field argument bounds the possible orders along a run,
including the same-family divisibility steps suppressed in the figure.
Thus the diagram complements the arity descent: when a construction
produces new binary transfers instead of a smaller irreducible signature,
the proof continues in a strictly larger group.

\subsection{Shared methods for the four branch proofs}
\label{subsec:shared-matching-methods}
\label{sec:equality-q4-boundary}
\label{proofsec:finite-certificates}

The group branches use two common tools. Binary contractions determine
either a full two-input slice or only part of it, which separates the
cyclic analysis from the other cases. At arity four, a shared gadget
construction connects the stable deck to the eight-vertex reductions.

\subsubsection{Two-input slices and contractions}
\label{subsubsec:two-port-fibres}

For two binary chains \(L,R\), order each matrix with the free input
as its row and the joined input as its column. Joining their second
inputs through \(X\) gives the external tensor \(LXR^{\mathsf T}\).
This fixes the transpose convention for the contraction kernels below.

For an arity-\(r\) tensor \(f\), ordered inputs \(p=(i,j)\), and an assignment to the remaining inputs \({\mathbf z}\), the \emph{two-input slice} is the binary
matrix
\[
 F^f_{p,{\mathbf z}}(a,b):=
 f\bigl({\mathbf z}^{i\leftarrow a,j\leftarrow b}\bigr),
 \qquad a,b\in\Ftwo.
\]
If \(\mathscr K\le\Mat_2(\mathbb C)\) is a space of contraction matrices, its
Frobenius-orthogonal space is
\[
 \mathscr K^{\perp_{\mathrm{Frob}}}:=
 \{F\in\Mat_2(\mathbb C):
   \langle C,F\rangle_{\mathrm{Frob}}=0\text{ for every }C\in\mathscr K\}.
\]
Thus contractions against \(\mathscr K\) determine the slice modulo
\(\mathscr K^{\perp_{\mathrm{Frob}}}\).

We determine the transfer span and use its fixed identification with
contraction kernels to recover two-input slices.

\begin{definition}[The transfer span \(\mathscr K_G\)]
\label{def:effective-kernel-algebra}
For a retained set in the fixed-\(I,X\) coordinates, with
complete transfer group \(G\), let
\(\mathscr K_G\) be the complex span of the transfer matrices
\(T_B=BX\) of all finite ordered two-input chains \(B\) made from the fixed representatives \(B_g\), \(g\in G\), the explicit \(I\)- or
\(X\)-edge links, and their input reversals.  This is equivalently the span
of independently dressed \(I/X\) kernels after the fixed right
multiplication by \(X\), which is invertible and therefore only relabels
the corresponding slice coordinates. For a closing binary \(B\), its
contraction kernel is \(C=XBX=XT_B\). Thus
\(\langle C,F\rangle_{\mathrm{Frob}}=\langle T_B,XF\rangle_{\mathrm{Frob}}\).
We use this fixed identification when reconstructing slices from
\(\mathscr K_G\).
\end{definition}

The information determined by binary contractions depends on
\(\mathscr K_G\).  The following theorem gives its two possible values.

Define the unnormalized Walsh matrix \(P\) and, for a signature \(f\)
already in \(K\)-coordinates, its Walsh-coordinate tensor \(\widetilde f\) by
\begin{equation}
  P:=\begin{pmatrix}1&1\\1&-1\end{pmatrix},
  \qquad
  \widetilde f:=(P^{-1})^{\otimes\arity(f)}f.
  \label{eq:Walsh-coordinates}
\end{equation}

The normalized Walsh matrix used in the proof of
\cref{thm:dressed-kernel-observability} is
\(S=P/\sqrt2\). Thus \(SXS=Z\), whereas \(P^{\mathsf T}XP=2Z\).
Since \(P^{-1}A(t)P=2D(t)\), the cyclic group in Walsh coordinates is
\(P^{-1}\CycX{N}P=C_N^{\rm std}\).

\begin{theorem}[The span of realizable binary transfers]
\label{thm:dressed-kernel-observability}
Let \(\cF\) be a retained set with normalized \(I,X\), and let
\(G=G(\cF)\).  With \(\mathscr K_G\) as in
\cref{def:effective-kernel-algebra},
\[
  \mathscr K_G\in
  \bigl\{\Span\{I,X\},\Mat_2(\mathbb C)\bigr\}.
\]
Moreover,
\[
 \Span\{I,X\}^{\perp_{\mathrm{Frob}}}
 =
 \left\{
  \begin{pmatrix}u&v\\-v&-u\end{pmatrix}:u,v\in\mathbb C
 \right\}.
\]
\end{theorem}

If \(\mathscr K_G=\Mat_2(\mathbb C)\), four realizable kernel evaluations
determine each two-input slice.  If \(\mathscr K_G=\Span\{I,X\}\), they
determine it modulo the displayed orthogonal complement.

\begin{proof}
Because \(B_1=X\), the span contains \(I\), and the explicit \(I\)-link
gives \(X\).  Series composition, with the second chain taken in its
inward (reversed) input order, has tensor in the original basis \(BXC\) and satisfies
\(T_{BXC}=T_BT_C\), so the span is an algebra.  Reversal gives
\(T^\#=XT^{\mathsf T}X\). Since \(X\) is in the span,
\(T^{\mathsf T}=XT^\#X\) is also in it.  Thus \(\mathscr K_G\) is unital
and transpose-stable.

Conjugate by \(S=P/\sqrt2\), so \(SXS=Z\).  For \(A\) in the conjugated algebra, define its diagonal
and off-diagonal parts by \(D:=(A+ZAZ)/2\) and \(O:=(A-ZAZ)/2\).  If some
\(O\ne0\), write
\(O=\left(\begin{smallmatrix}0&b\\c&0\end{smallmatrix}\right)\).  Since \(Z\)
and \(O\) lie in the algebra, so does
\(ZO=\left(\begin{smallmatrix}0&b\\-c&0\end{smallmatrix}\right)\).  If
\(bc\ne0\), \(O+ZO=2bE_{01}\) and \(O-ZO=2cE_{10}\), where \(E_{01}\)
and \(E_{10}\) use Boolean row and column indices. If one entry is zero,
transpose-stability supplies the missing unit.  Products supply the diagonal
units, hence the algebra is \(\Mat_2(\mathbb C)\).  If every \(O=0\), it is diagonal and contains \(I,Z\), so it is
\(\Span\{I,Z\}\), which conjugates back to \(\Span\{I,X\}\).

In the full case, four realizable transfer matrices can be greedily chosen as a
basis. Their corresponding kernel evaluations reconstruct every two-input slice
under the fixed identification above.  In the proper case, orthogonality to
\(I,X\) is \(a+d=b+c=0\), giving exactly
\(\left(\begin{smallmatrix}u&v\\-v&-u\end{smallmatrix}\right)\).
\end{proof}

For the finite groups routed above, observability identifies the routing split
intrinsically:
\[
 G\text{ is cyclic}
 \quad\Longleftrightarrow\quad
 \mathscr K_G=\Span\{I,X\},
\]
\[
 G\text{ is dihedral, Klein four, or Platonic}
 \quad\Longleftrightarrow\quad
 \mathscr K_G=\Mat_2(\mathbb C).
\]
Indeed, the cyclic transfer classes lie in \(\Span\{I,X\}\), whereas a
finite projective subgroup of the units of this proper algebra is cyclic.
Thus \cref{sec:proper-deck} treats \(\mathscr K_G=\Span\{I,X\}\).
The other three branches have \(\mathscr K_G=\Mat_2(\mathbb C)\).

To see what the kernel span tells us about a stable \(f\), fix an ordered
pair \(p\) of its inputs and vary the closing kernel, keeping the other
inputs in the same order. If three available kernels satisfy
\(C_3=\alpha C_1+\beta C_2\), their cards satisfy
\[
  \partial_p^{C_3}f=\alpha\,\partial_p^{C_1}f+\beta\,\partial_p^{C_2}f .
\]
At an assignment where the first two cards vanish, the third must vanish
as well. The proposed matchings and binary factors must give values
consistent with this identity at every assignment.

When the kernels span \(\Mat_2(\mathbb C)\), choose four that form a basis.
Fixing \(p\) to \(a,b\in\{0,1\}\) gives a residual signature \(f_{ab}\)
with \(f_{ab}({\mathbf z})=F^f_{p,{\mathbf z}}(a,b)\).
Thus each \(f_{ab}\) is a linear combination of the four cards. Suppose the nonzero cards among them all have
a common factor \(h\) on the same nonempty block of remaining inputs. Then
\(f_{ab}=h\otimes r_{ab}\) for each \(a,b\), allowing \(r_{ab}=0\).
Define \(r\) by \(r(a,b,\mathord\cdot)=r_{ab}\) in the input order
with \(p\) first. Then \(f=h\otimes r\) up to input order. Since \(r\)
contains the two inputs in \(p\), this contradicts irreducibility.
In the cyclic case the contractions recover only combinations of the four residual signatures.
\Cref{sec:proper-deck} compares these combinations across different input pairs.

\subsubsection{The arity-four reductions}

We first process the binary contractions of a quaternary signature. This
local step checks which complete group applies before the common gadget
construction is used.

For a quaternary tensor \(q\) with ordered inputs \(1,2,3,4\), let
\(M_{x_ix_j,x_kx_\ell}(q)\) be its \(4\times4\) signature matrix, with row index
\((x_i,x_j)\) and column index \((x_k,x_\ell)\).  For a finite
complete transfer group \(G\) and its fixed representatives
\(B_g\), \(g\in G\), define
\[
 \mathcal M_4(G)=
 \left\{
 c\,B_g(x_i,x_j)B_h(x_k,x_\ell):
 \begin{array}{l}
 c\in\AlgNums^\times,\ ((i,j),(k,\ell))\text{ is an oriented}\\[-2pt]
 \text{perfect matching of }[4],\quad g,h\in G
 \end{array}
 \right\}.
\]
Thus \(\mathcal M_4(G)\) is the arity-four \(G\)-matching notion of \cref{def:matching-deck}.  Define the
complete oriented one-copy binary batch as follows.  For an ordered binary
card \(C\), set \(C^{\mathrm{id}}=C\) and
\(C^{\mathrm{rev}}=C^{\mathsf T}\), and put
\[
 \mathscr D_G^{(1)}(q)
 =
 \{0\}\cup
 \left\{
   \bigl(\partial_{(i,j)}^{XB_gX}q\bigr)^\epsilon:
   i,j\in[4],\ i\ne j,\ g\in G,\
   \epsilon\in\{\mathrm{id},\mathrm{rev}\}
 \right\}.
\]
The order \((i,j)\) records the orientation of the deleted pair, and the
choice of \(\epsilon\) records either orientation of the residual pair.
Every displayed member is a direct gadget: \(XB_gX\) is the realizable kernel
obtained by attaching
\(B_g\) through its two \(X\)-edges, and residual reversal is a permutation of inputs.  In this local case analysis, a
\emph{nonbinary irreducible signature} means an irreducible signature of arity at least
three.

\begin{lemma}[Binary contractions of a quaternary signature]
\label{lem:finite-q4-dispatch}
Let \(\cF\) be a finite algebraic retained, factor-closed,
  binary-residual signature set containing the supplied equality
\(I\), suppose every retained nonbinary irreducible signature has even arity, and put
\(G=G(\cF)\).  Let \(q\) be a realizable signature over \(\cF\), with
its ordered inputs, realization, and nonzero scalar recorded.  Assume that
\(q\) is endpoint-nondegenerate, irreducible, and
\[
 q\notin\mathcal M_4(G).
\]
The three flattenings satisfy
\[
 \rank M_{x_1x_2,x_3x_4}(q)\ge2,\qquad
 \rank M_{x_1x_3,x_2x_4}(q)\ge2,\qquad
 \rank M_{x_1x_4,x_2x_3}(q)\ge2.
\]
These inequalities follow from irreducibility and give the rank
conditions used in the later arity-four arguments.

Let \(\cF_q\) be the preprocessed signature set obtained by
adding \(q\) and all nonzero tensor cards in
\(\mathscr D_G^{(1)}(q)\), followed by factor closure of this
finite batch.  Then
\[
 \KHolant(\cF_q)\equivT\KHolant(\cF),
\]
and one of the following holds:
\begin{enumerate}[label=\textnormal{(\roman*)}]
\item an established complexity conclusion applies, including one obtained from an
      odd-arity or unary signature, a nonzero singular binary, or
      interpolation from a realizable infinite-order binary. Alternatively,
      a retained lower-arity nonbinary irreducible signature gives an
      outer successor with smaller minimum arity;
\item \(\cF_q\) is binary-residual and
      \[
        G<G_q:=G(\cF_q),
      \]
      with the strict inclusion witnessed by an ordered binary \(B\) realizable
      over \(\cF_q\) such that \([BX]\in G_q\setminus G\);
\item \(\cF_q\) is binary-residual, \(G_q=G\), \(q\) is a
      minimum-arity nonbinary irreducible signature in \(\cF_q\), and every member
      of \(\mathscr D_G^{(1)}(q)\) is zero or projectively equal to one of the
      fixed \(B_g\), \(g\in G\).  All nonzero card factors are retained. On
      this continuing branch every nonzero card has a trivial factorization tree.
\end{enumerate}
Moreover, on branch~(iii), let \(r\) be an auxiliary realizable quaternary
signature over \(\cF_q\), assume \(r\notin\mathcal M_4(G)\), and suppose
that \(r\) has a genuine \(2+2\) factorization.  Let
\(\cF_{q,r}\) be the preprocessed retained, factor-closed signature set
obtained by adding \(r\) and its nonzero factor batch.  Then either
an established reduction determines the complexity or
\[
  G<G(\cF_{q,r}).
\]
Thus there is no same-group continuing branch.  The case distinction is finite.
\end{lemma}

\begin{proof}
The batch \(\mathscr D_G^{(1)}(q)\) is finite: there are six unordered
deleted pairs, two orders for the closing kernel, two residual orders, and
finitely many closing-representative choices \(B_g\).  Its tensor cards are
realizable over \(\cF\), since \(q\) and every \(B_g\) have direct realizations.  Adding them therefore gives one Turing reduction by
\cref{eq:direct-gadget-reduction}, while inclusion gives the reverse one.
Because every resulting set contains \(I\) and is two-sided,
\cref{lem:factor-saturation} may retain the finite factorization tree without
changing the problem up to Turing equivalence.

Dispatch every odd-arity or unary factor and every singular binary factor.
If \(\cF_q\) directly realizes any nonsingular binary of infinite
projective transfer order, apply \cref{lem:binary-interpolation}. This is
item~(i).  Otherwise \(\cF_q\) is binary-residual, so
\cref{lem:bounded-torsion} defines the finite group
\(G_q=G(\cF_q)\).  Retained inclusion gives \(G\le G_q\).  If the
inclusion is strict, the definition of \(G_q\) supplies a realizable binary
representative of a class in \(G_q\setminus G\), proving item~(ii).
The group-enlargement witness is a realizable binary gadget over \(\cF_q\), possibly
using several copies of \(q\).

It remains that \(G_q=G\).  Every nonzero card in
\(\mathscr D_G^{(1)}(q)\) is then nonsingular and has transfer class in
\(G_q=G\), hence is projectively one of the fixed \(B_g\).  Such a binary
has no proper positive-arity tensor factorization.  The tensor \(q\) remains
irreducible under tensor factorization.  Any nonbinary irreducible signature of arity below four would have
arity three and would already have triggered the odd-arity reduction, so
\(q\) has minimum arity of a nonbinary irreducible signature.  This proves item~(iii).

Finally, write \(r\doteq C_1\otimes C_2\) for the auxiliary factorization.
Factor closure makes \(C_1,C_2\) signatures of
\(\cF_{q,r}\). They are not asserted to have been realizable over
\(\cF_q\).  If either is singular, or if the resulting set realizes an infinite-order binary, an established reduction determines the complexity.  Otherwise
\(\cF_{q,r}\) is binary-residual and
\(G\le G(\cF_{q,r})\).  If equality held, both factors would be
nonsingular and would represent classes in \(G\), whence
\(r\in\mathcal M_4(G)\), contradicting the stated hypothesis.  The inclusion
is therefore strict.
\end{proof}

Only item~(iii) of \cref{lem:finite-q4-dispatch} is passed to the
following arity-four construction. The preceding proof uses the finite
local batch and the complete-group comparison, before any appeal to
the global descent measure. The complete stabilized deck permits the
following gadget construction.  Its output is a realizable nonmatching eight-vertex
signature to which \cref{cor:p1-eight-vertex-reduced-interface} applies.

\begin{theorem}[Quaternary signatures with stable matching decks]
\label{thm:equality-q4-boundary}
Assume
\[
 \Stable(\cF,G,q),\qquad \arity(q)=\nu(\cF)=4,
\]
and that \(G\) is one of
\begin{enumerate}
\item \(G=\{1,[X]\}\cong C_2\);
\item \(G=\{[D(t)],[XD(t)]:t\in\mu_m\}\), \(m\ge3\);
\item the standard Pauli \(\KleinStd\);
\item the exotic Klein group
  \begin{equation*}
   \KleinEx
   =
   \left\{[I],[X],
   \left[\begin{pmatrix}i&1\\-1&-i\end{pmatrix}\right],
   \left[\begin{pmatrix}-i&1\\-1&i\end{pmatrix}\right]\right\}.
  \end{equation*}
\end{enumerate}
Then a gadget using at most two copies of \(q\) produces an
endpoint-nondegenerate eight-vertex signature that is not \(G\)-matching.
\end{theorem}

\begin{proof}
The gadget in \cref{thm:cert-interface-equality-q4} uses at most two
copies of the selected signature and gives the asserted nonmatching
endpoint-nondegenerate eight-vertex output. Its directly realized signature
\(r\) satisfies
\[
  \KHolant(\cF,r)
  \leT\KHolant(\cF).\qedhere
\]
\end{proof}

To apply \cref{thm:equality-q4-boundary}, first regenerate the current
quaternary card and its complete oriented one-copy deck, close under
factors, and process them as in \cref{lem:finite-q4-dispatch}.  Auxiliary cards used inside the finite
certificate are local to that proof and are not appended to \(Q\).  Only an added factor batch returned by the eight-vertex reduction can be passed to an
outer closure.

The quantifiers include zero cards, both input orientations for all six
contracted pairs, every current-group representative assignment on the
remaining matched edges, and the fixed representative scalars. Apply
\cref{cor:p1-eight-vertex-reduced-interface} to the output over the whole
retained set. In items~2--4 and the one-copy case of item~1, invertible
dressings preserve irreducibility, so only a hard or tractable case remains.

In the exceptional \(\CycX{2}\) subcase, write
\(a_S=\widetilde q(\one_S)\).  Call it \emph{mixed-sign} when
\(a_\varnothing=a_{[4]}=0\) and two nonzero complementary weight-two
blocks \(\{S,\bar S\}\) have different ratios
\(a_{\bar S}/a_S\in\{1,-1\}\), as in
\cref{thm:cert-interface-equality-q4}\textnormal{(i)}. Only this mixed-sign
\(\CycX{2}\) subcase can reach the continuing factor outcome after the two-copy
construction. Retain its complete added factor batch. The last assertion of
\cref{lem:finite-q4-dispatch} resolves the complexity or gives a strictly
larger complete transfer group, since the output is nonmatching. Every added
factor remains available by reduction over the predecessor, and only this
outer successor is passed back to stabilization.

Every bounded enumeration appears as an exact proposition in
Appendix~\ref{sec:exact-finite-certificates}. The main text supplies the reduction for the whole signature set and its direction.  The coverage map
\path{certificates/theorem-to-script-map.md} records the manuscript anchors
and finite claims checked by each entry point, and
\path{certificates/README.md} gives the replay instructions and the
package-wide computational scope.  The next four sections use the available contraction kernels to constrain
higher-arity signatures in their respective group cases. Each supplies
a complexity conclusion or proves the strict progress required by
\cref{def:protocol-discharged}. \Cref{sec:p1-synthesis} combines these
arguments with stabilization and the outer induction over irreducible
occurrences to classify the whole signature set. We begin with the
marked dihedral case.
\section{Marked Dihedral Matching Decks}
\label{sec:normalized-dihedral}

This section proves the marked-dihedral case of the deck-stable
classification. A deck-stable signature has only zero or \(G\)-matching
proper cards. The difficulty is that different cards may use different
matchings or different binary factors. We show how these local choices
force a proper factor of a higher-arity signature, contradicting its
irreducibility. At arity four, realizable eight-vertex signatures give
the complexity conclusion for the entire retained set.

For the standard group \(\DihZ{2m}\), monomial kernels let us shorten
alternating cycles and preserve selected differences between two cards.
This reduces a disagreement to six inputs, where the pencil
classification and a finite support analysis apply. When the cards
agree outside one binary factor, two phase evaluations recover the four
residual signatures obtained by fixing the deleted pair, giving a common
tensor factor of the original signature.

The other marked groups require control of the edge matrix as well as
the binary transfers. Whole-network equivalences carry
\(\DihY{2m}\) and \(\DihXzero{2m}\) to the standard case.
For \(\DihXone{2m}\), explicit weighted kernels preserve the local
comparisons. The support and coefficient reconstruction developed below
supplies the weighted-kernel argument, including its six-input case.
Each branch gives the conclusion required in
\cref{def:protocol-discharged}; their combination proves the following theorem.

\begin{theorem}[Marked-dihedral deck rigidity]
\label{thm:marked-dihedral}
Assume
\[
 \Stable(\cF,G,f),\qquad \Even(\cF),
\]
and suppose that the complete transfer group has one of the following marked forms:
\[
 G\in
 \{\DihZ{2m}:m\ge3\}
 \cup
 \{\DihY{2m}:m\ge3\}
 \cup
 \{\DihXzero{2m}:m\ge4,\ 2\mid m\}
 \cup
 \{\DihXone{2m}:m\ge4,\ 2\mid m\}.
\]
Then \(\Close_{\{\cP\}}(\cF,G,f)\).
\end{theorem}

The four cases are proved in
\cref{thm:normalized-dihedral,thm:marked-B-reduction,%
thm:marked-A0-reduction,thm:marked-A-rigidity}.

\paragraph{Contractions in the standard case.}
Until \cref{sec:marked-dihedral-sectors}, fix \(G=\DihZ{2m}\),
\(m\ge3\). Let \(n:=\arity(f)\).  We use the conventions of
\cref{subsec:reduction-conventions,def:relative-stratum-protocol,%
def:protocol-discharged} throughout.

If \(n=4\), first regenerate the complete deck of \(f\) and close it under
irreducible factors, apply \cref{thm:equality-q4-boundary}, and apply the eight-vertex reduction to its realizable
endpoint-nondegenerate output using
\cref{cor:p1-eight-vertex-reduced-interface}.  The added factor batch is
closed by \cref{lem:actual-binary-completeness,lem:factor-saturation}. It
gives a resolved complexity leaf or a strict stable successor.  Thus the arity-four
base is resolved, and throughout the remaining standard \(\DihZ{2m}\)
proof we may assume
\[
 n=2k+2\ge6.
\]

For every \(t\in\mu_m\), the representative convention of
\cref{subsec:reduction-conventions} fixes realizable ordered binary outputs
\begin{equation*}
  B_{[D(t)]}=\sigma_t^{\mathrm{diag}}D(t)X,
  \qquad
  B_{[XD(t)]}=\sigma_t^{\mathrm{anti}}XD(t)X,
  \qquad
  \sigma_t^{\mathrm{diag}},\sigma_t^{\mathrm{anti}}\ne0.
\end{equation*}
For a deleted set \(p=\{i,j\}\) equipped with the input order \((i,j)\),
write the four residual signatures obtained by fixing that pair as
\begin{equation*}
  A:=f(00,\mathord\cdot),\quad
  B:=f(01,\mathord\cdot),\quad
  C:=f(10,\mathord\cdot),\quad
  D_{11}:=f(11,\mathord\cdot),
\end{equation*}
and define the associated \((n-2)\)-ary \emph{sector pencils} by
\hypertarget{eq:nd-sector-pencils}{}
\begin{equation*}
 E(t):=A+tD_{11},
 \qquad
 O(t):=C+tB.
\end{equation*}
Here \(E(t)\) is the equal/even sector associated with the deleted-pair
values \(00,11\), while \(O(t)\) is the unequal/odd sector associated with
\(10,01\).  A pencil means a one-parameter linear family \(U+tV\).
The corresponding contraction-kernel spaces
\[
 \Span\{E_{00},E_{11}\}
 \quad\text{and}\quad
 \Span\{E_{01},E_{10}\}
\]
are the even and odd \emph{parity planes}, with the displayed matrix
units as coordinate axes.
For \(t\in\mu_m\), the \(X\)-edges give the exact realizable outputs
\(\sigma_t^{\mathrm{diag}}O(t)\) and \(\sigma_t^{\mathrm{anti}}E(t)\), since
\[
  XB_{[D(t)]}X
    =\sigma_t^{\mathrm{diag}}XD(t),
  \qquad
  XB_{[XD(t)]}X
    =\sigma_t^{\mathrm{anti}}D(t).
\]
For \(t\in\mu_m\), these outputs are deck cards with the displayed
nonzero scalars. The four slices fix the inputs of the deleted pair and
leave all other inputs free. They are the coefficient tensors of the two
sector pencils and will be called their \emph{pencil endpoints}.
Two phase evaluations recover these tensors by linear inversion.
Later \(K_E(t),K_O(t)\in\Mat_2(\mathbb C)\) denote the varying
binary factors of the two sector pencils.

We use \emph{deck card} for a member of the matching deck in
\cref{def:matching-deck}. A sequence of contractions gives a
\emph{contraction output}; constructions using two copies give
\emph{two-copy outputs}.

Write a nonsingular monomial binary, up to scalar, as
\begin{equation*}
  M_\alpha(a)(x,y)
  =\one[x\oplus y=\alpha]a^x,
  \qquad
  \alpha\in\Ftwo,\quad a\ne0.
\end{equation*}
Here \(\alpha\) is its support parity and \(M_\alpha(a)=D(a)X^\alpha\). In particular, \(M_0(a)=D(a)\), while \(XD(t)=tM_1(t^{-1})\) for \(t\ne0\). Reversal changes only the weight and a nonzero scalar.

\subsection{Trace covers and alternating-cycle localization}
\label{proofsubsec:nd-trace-cover}

Consider two nonzero sampled cards of the same signature.  Their matchings and
binary factor classes can disagree.  We will contract common parts while
preserving a disagreement, leaving only four or six inputs to analyze.

\paragraph{Projective tensor geometry.}
\phantomsection\label{subsec:projective-incidence-conventions}

A matching product is a simple tensor after its inputs are grouped into
pairs. Segre geometry expresses this factorization by rank equations.

\label{subsubsec:projective-tensor-geometry}

We use the projective-space conventions of
\cref{subsubsec:group-projective}.  Whenever a projective object occurs in a
tensor equation, a displayed or previously fixed nonzero
representative is understood. We set \(\mathbb P(0):=\varnothing\).

A nonzero tensor \({\mathbf{v}}_1\otimes\cdots\otimes {\mathbf{v}}_s\) is \emph{simple}.  For finite-dimensional spaces \(V_1,\ldots,V_s\), the
Segre map is
\[
 \mathbb P(V_1)\times\cdots\times\mathbb P(V_s)
 \longrightarrow \mathbb P(V_1\otimes\cdots\otimes V_s),
 \qquad ([{\mathbf{v}}_1],\ldots,[{\mathbf{v}}_s])\longmapsto[{\mathbf{v}}_1\otimes\cdots\otimes {\mathbf{v}}_s].
\]
Its image is the \emph{Segre variety}. Its affine cone consists of zero and
the simple tensors.  The Segre ideal is
generated by the \(2\)-by-\(2\) minors of the bipartite flattenings.  A
\emph{one-factor Segre ruling} is a projective linear family in which all
factors but one are fixed, and every projective line contained in a Segre
variety belongs to such a ruling \cite[Chapter~2]{Landsberg2012}.

For two perfect matchings \(M_0,M_1\), an \emph{alternating
\(M_0/M_1\)-cycle} is an even cycle in \(M_0\mathbin\triangle M_1\) whose
edges alternate between the two matchings.  We call a sequence of the
contractions below a \emph{trace cover}. It closes common edges and
shortens alternating cycles while retaining a chosen discrepancy.  Its ordered contraction graph records the connections, edge orientations,
and external-input order.

\begin{definition}[Disagreements between matching products]
\label{def:nd-live-card-disagreement}
The \emph{factor class} of an ordered nonzero binary \(B\) is its
projective class \([B]\). Normalize the sampled deck cards by their
nonzero realization scalars. Two nonzero cards
\emph{disagree} if their residual matchings differ, or if their ordered
binary factor classes differ on at least two edges of the same
matching.  A parity change changes the factor class.  Variation on one designated edge, including its parity, is allowed.

A one-factor Segre ruling, as defined in
\hyperref[subsubsec:projective-tensor-geometry]{the projective tensor conventions}, allows one binary factor to
vary and fixes the other factors on the matching. Thus products on one
such ruling differ in at most that factor class.
\end{definition}

Consider two realizable nonzero sector cards
\begin{equation}
  P_0=\bigotimes_{e\in M_0}B_{0,e},
  \qquad
  P_1=\bigotimes_{e\in M_1}B_{1,e},
  \label{eq:app-nd-two-products}
\end{equation}
where every \(B_{\epsilon,e}\) is a nonsingular monomial binary from the
normalized group.

We close common edges whose ordered binary factors agree projectively.
Connecting one input on each of two residual edges by a realizable
binary fuses those edges in each product. The following identities show
how to choose these contractions while preserving selected parity
differences or nonzero determinants.

\begin{definition}[Parity and coefficient discrepancies]
\label{def:app-nd-tc-tag}
After aligning the common edges of two sector products, a residual matched
factor is a \emph{parity discrepancy} if its support parity changes between the two products.  It is a \emph{coefficient discrepancy} if the parities agree but the two
ordered nonzero-entry vectors are not projectively proportional. In that
case the associated nonzero \(2\times2\) determinant records the coefficient discrepancy.

\end{definition}

\begin{lemma}[Compatible contractions and cycle shortening]
\label{lem:app-nd-compatible-closure}
Common factors with the same projective class in \cref{eq:app-nd-two-products} can be closed by
realizable group binaries while keeping both products nonzero.  An alternating
\(M_0/M_1\)-cycle can be shortened by two inputs without cancellation, and
a selected coefficient determinant is only multiplied by a nonzero scalar.
Two edges with discrepancies can be fused by a realizable connection while preserving
any distinguished remaining discrepancy.
\end{lemma}

\begin{proof}
For ordered monomial factors one has, up to the fixed orientation scalar,
\begin{align*}
 \sum_{x,y}M_\alpha(a)(x,y)M_\alpha(t)(x,y)&=1+at,\\
 \sum_zM_\alpha(a)(u,z)M_\beta(b)(z,v)
  &=b^\alpha M_{\alpha+\beta}
       \bigl(ab^{1-2\alpha}\bigr)(u,v).
\end{align*}
Thus on a common edge the two product traces are nonzero affine functions
of \(t\in\mu_m\). Together they exclude at most two phases.  Since
\(m\ge3\), one allowed phase keeps both nonzero.  Closing an edge then
splices the two incident edges in the other matching.  The displayed fusion has a unique internal lift: each supported
external assignment extends to exactly one supported internal assignment.
Thus there is no cancellation, and every selected minor is multiplied
by nonzero trace and fusion factors.

For an ordered edge with a discrepancy \(q=(u_q,v_q)\), let
\(\alpha_{\xi,q}\in\Ftwo\) denote its support parity in product
\(\xi\in\Ftwo\).  Put
\[
 \delta_q:=\alpha_{0,q}+\alpha_{1,q}\in\Ftwo
 \qquad(q\in\{r,s\}),\qquad
 \alpha_{\xi,q}=\alpha_{0,q}+\xi\delta_q.
\]
One ordered connection \(M_\gamma(t)\) fuses the two factors in each
product.  Applying the displayed fusion identity twice gives
\[
 M_\alpha(a)M_\gamma(t)M_\beta(b)
 =t^\alpha b^{\alpha\oplus\gamma}
 M_{\alpha\oplus\gamma\oplus\beta}
 \!\left(a\,t^{1-2\alpha}b^{1-2(\alpha\oplus\gamma)}\right).
\]
For each supported exposed string the internal lift is unique, and the
fusion scalar is nonzero.  The fused parity difference is
\(\delta_r+\delta_s\).  A mixed pair therefore leaves a parity
discrepancy. Equal types leave at most one coefficient discrepancy, which
may also cancel.  A distinguished discrepancy outside the two spliced
edges is preserved: its determinant is multiplied only by the nonzero
orientation and fusion scalars of the two products.  Each connection
removes two inputs, whether or not the two spliced discrepancies cancel.
All operations use only the \(X\)-edge, the explicit normalized
\(I\)-link,
and fixed realizable group representatives.
\end{proof}

\begin{lemma}[Binary factor classes and a quaternary contraction]
\label{lem:app-nd-coefficient-line-factor}
For distinct \(t_0,t_1\in\mu_m\), suppose the four nonzero realizable cards
represented by \(E(t_0),E(t_1),O(t_0),O(t_1)\) at a deleted pair \(p\)
have a common residual matching edge \(e\), with the same support parity.
If all four ordered factors on \(e\) are proportional to one binary \(B_e\), the signature \(f\) has the exact tensor factorization
\[
  f=B_e\otimes h
\]
with a nonsingular monomial binary \(B_e\).  Otherwise choose two cards
whose ordered factors on \(e\) are nonproportional.  If their other
matching edges and support parities agree, a contraction sequence retaining
\(p\cup e\) has a nonzero \(2\times2\) flattening minor.  Without this
parity compatibility no simultaneous trace is asserted.  The factor
identity is analytic. Its factors are exposed only by factor closure.
\end{lemma}

\begin{proof}
For the nonproportional pair with the stated compatibility, close every
other common matching edge successively.  At each step
the two traces exclude at most two roots, so
\cref{lem:app-nd-compatible-closure} keeps both products nonzero.  The two
surviving \(e\)-vectors remain nonproportional.  They are contractions of
the resulting realizable quaternary by two kernels on \(p\). Hence that quaternary has a
nonzero flattening minor.  The two kernels are linearly independent: two
distinct phases are independent inside one parity plane, while an even and
an odd kernel have different support parities.  Thus nonproportional
retained \(e\)-vectors really give flattening rank at least two.

For the factor alternative all four ordered \(e\)-factors are proportional. Choose their common representative
\[
 B_e(x_u,x_v)=\one[x_u\oplus x_v=\alpha]b_{x_u},
 \qquad b_0b_1\ne0,
\]
and absorb each card's nonzero scalar into its complementary tensor.  The
two-card inversions
\[
 A=\frac{t_1E(t_0)-t_0E(t_1)}{t_1-t_0},\quad
 D_{11}=\frac{E(t_1)-E(t_0)}{t_1-t_0},\qquad
 C=\frac{t_1O(t_0)-t_0O(t_1)}{t_1-t_0},\quad
 B=\frac{O(t_1)-O(t_0)}{t_1-t_0}
\]
put every slice in \(\mathbb C B_e\otimes V\), where \(V\) is the tensor
space on the residual inputs outside \(e\).  Reassembling the four
slices gives the displayed tensor identity.
\end{proof}

\begin{lemma}[Two-card localization to six inputs]
\label{lem:app-nd-localization}
Fix the deleted pair \(p\), and let two nonzero realizable sampled
\(p\)-cards of the same signature disagree.  There is a realizable
one-copy contraction \(h\), retaining \(p\) and four further inputs, whose
two inherited \(p\)-cards remain nonzero and either have different
residual matchings or differ on both edges of their common residual
matching.  The contraction closes only inputs outside \(p\) and uses
\(O(\arity(f))\) fixed realizable group representatives.  In particular,
\(h\) is not a \(G\)-matching signature.
\end{lemma}

\begin{proof}
For each selected pair of cards, choose contractions that keep both
outputs nonzero and preserve their disagreement.

First suppose the two residual matchings differ.  Choose one alternating
cycle as distinguished.  On every other alternating cycle, close an edge
of the first matching by a kernel of the same support parity.  The trace
in the first product excludes at most one phase, whereas the other
product has a unique monomial lift.  Thus both products remain nonzero.
Repeat until that nondistinguished cycle becomes one common edge.

Every common edge outside the distinguished cycle can now be absorbed
into the cycle without any parity compatibility assumption.  Write that
edge as \(\{u,w\}\), select a cycle vertex \(v\), and connect
\(u,v\) by any realizable monomial group kernel.  If the two cycle neighbours
of \(v\) are \(v_0,v_1\), the new matching edges are respectively
\(\{w,v_0\}\) and \(\{w,v_1\}\).  Thus the alternating cycle is
unchanged except that \(w\) replaces \(v\).  Both internal lifts are
unique and have nonzero weights, including when the old common edge had
different support parities in the two products.  Absorb every remaining
common edge in this way.  Finally shorten the distinguished cycle, using
the same trace-and-fusion operation, until it is a four-cycle.  Stop
before closing an edge of that four-cycle.  The two residual four-input
matchings are therefore still different.

Now suppose the residual matching agrees and at least two factor classes
differ.  Select two such edges whose discrepancies will be retained.  Every other edge whose two support
parities agree can be traced directly: the two traces exclude at most
two phases, so \(m\ge3\) leaves an allowed phase keeping both products nonzero.  The two selected edges are untouched by these traces.

An edge outside the selected pair with different support parities is instead
fused with one selected edge by a realizable binary connection.  The other selected edge remains
untouched.  If the selected edge used in the fusion has equal parities but different
factor classes, the fused edge has different support parities and
retains a parity discrepancy.  If both fused edges have different support parities, the
new support parities agree, but the ratio of their normalized fused
weights is
\[
  \kappa t^{\pm2},\qquad \kappa\ne0.
\]
This follows directly from the three-factor fusion identity in
\cref{lem:app-nd-compatible-closure}.  At most two phases make this ratio
equal to one.  Choose another allowed phase. The fused edge then carries a
coefficient discrepancy.  In both cases the operation reduces the number of edges
by one and leaves two edges whose factor classes differ between the products.  Continue until
exactly these two edges remain.

Perform the selected contractions on the original signature, leaving the
inputs of \(p\) open.  Tensor contraction commutes with either of the two
original \(p\)-kernels, so its inherited cards are precisely the two
nonzero products just constructed, with their nonzero scalars.
Each operation removes two inputs, giving the stated size bound.

If \(h\) were a product on three matching edges, deleting \(p\) would
either trace out one of its matching edges, leaving the other two factors
fixed, or fuse its two incident matching edges, leaving the third factor
fixed.  Whenever the two resulting cards are nonzero, their residual
matching is therefore the same and at most the fused factor can vary.
This contradicts either of the two constructed configurations.
\end{proof}

\begin{lemma}[Reduction to four inputs or a tensor factor]
\label{lem:app-nd-bounded-consumer}
Every arity-six signature produced in \cref{lem:app-nd-localization} has a complete
reduction through its proper factor and deck batches.  It cannot
close as a \(G\)-matching signature.  A quaternary child is processed
by its own complete factor and deck batch. A crossed factorization into safe binaries
completes that child only.
\end{lemma}

\begin{proof}
Let \(h\) be the realizable six-input output, with its inherited nonzero cards,
input order, realization scalar, and originating fixed token retained.
By \cref{lem:app-nd-localization}, \(h\) is not a three-factor matching
product.  Add \(h\), jointly close the enlarged set under irreducible factors, and close every proper
factor and every proper oriented deck card.  Odd or unary factors and
unsafe binaries or safe binaries whose transfer classes lie outside the current group follow their
established exits.  Any proper nonbinary irreducible signature has arity at most four and
is processed by the lower-arity induction, including the arity-four
quaternary deck classification stated in the setup.  Thus this processing does
not use the six-input case being proved.

If no resolved leaf or strict successor occurs, all surviving proper
factors are \(G\)-matching signatures.  A proper factorization of
\(h\) would therefore make \(h\) itself a matching product, a
contradiction.  Otherwise \(h\) is irreducible and every nonzero proper
pair card is a \(G\)-matching signature.  \Cref{lem:nd-six-prime-base} excludes this case by the sector-ruling and
binary-pencil lemmas together with the Boolean support classification.

For a realizable quaternary child, first close its complete factor batch.
If it factors into safe current-group binaries, this completes the child
card index.  Otherwise use the established binary exits or retain its irreducible
quaternary occurrence, regenerate the complete oriented deck, and apply
\cref{thm:equality-q4-boundary,cor:p1-eight-vertex-reduced-interface}.
Every returned factor batch is closed by
\cref{lem:actual-binary-completeness,lem:factor-saturation}.
A factorization completes the corresponding child output. The selected
signature remains under analysis.

In particular, for an active deck-stable signature of arity at least eight there
is an immediate contradiction: the first proper pair contraction outside
\(p\) is zero or a matching product, and further monomial contractions
preserve that property.  Yet localization keeps the two disagreeing
inherited cards nonzero.  The same argument applies after the proper
lower-arity batches of a later occurrence have been closed.
\end{proof}

\paragraph{Common-factor consequence.}
\phantomsection\label{par:nd-common-factor-consequence}
Let \(\mathscr S\) be a nonempty family of nonzero matching products on
one fixed, nonempty oriented matching \(M\).  Suppose that, for every
\(S,T\in\mathscr S\), their binary factor classes differ on at most one
edge of \(M\).  Then there are an edge \(e\in M\) and a nonzero product
\(P\) on \(M\setminus\{e\}\) such that
\[
  \mathscr S\subseteq\mathbb C P\otimes V_e,
  \qquad V_e:=(\mathbb C^2)^{\otimes e}.
\]
Here \((\mathbb C^2)^{\otimes e}:=\bigotimes_{j\in e}\mathbb C^2\),
with factors in the fixed input order, and proportional
factors are absorbed into the factor on \(e\).  If \(S,T\) differ on
\(e\), a third sample differing from \(S\) on another edge would differ
from \(T\) on two edges.  If all samples are proportional, choose any
edge.  This proves the assertion, including a matching with only one edge,
where \(P\) is a nonzero scalar.

Apply the preceding localization and reduction to each pair of nonzero
samples at a fixed deleted pair.  The closures need preserve only that
pair and may differ between comparisons.  On the continuing branch they
give the hypothesis above for the union of the two sectors, so the same
\(P\) works in both.  Each nonzero pencil has at least two nonzero samples
at distinct phases because \(m\ge3\).  Inverting those two evaluations
puts both endpoint slices in \(\mathbb C P\otimes V_e\). An identically
zero pencil has two zero endpoint slices.  Thus all four slices share
\(P\), which gives the factorization of \(f\) used below.

\begin{lemma}[Sector-ruling rigidity]
\label{lem:app-nd-sector-ruling}
Let \(U+tV\), \(t\in\mu_m\), \(m\ge3\), be a tensor pencil on an
even number of inputs whose samples are zero or current
\(\DihZ{2m}\)-matching signatures.  Its nonzero samples have one fixed
matching and all but at most one binary factor class fixed projectively.
In particular this holds for each of \(E(t)\) and \(O(t)\).
\end{lemma}

\begin{proof}
We use only the sampled matching hypothesis.  A proportional pencil has
one projective value.  Otherwise
its \(m\) samples are nonzero and projectively distinct.  Write
\(N=2^s\) for the support size of an \(s\)-edge matching product.
Two different matching supports \(S_0,S_1\) are affine cosets of the
same dimension, so \(|S_0\cap S_1|\le N/2\).  Every nontrivial linear
combination of the corresponding tensors has at least
\(2N-2|S_0\cap S_1|\ge N\) supported strings outside the intersection.
For a third sample to have exactly \(N\) strings requires equality and
cancellation on the whole intersection.  A single common string fixes that
linear combination projectively.  There is therefore at most one such
third matching point, contradicting \(m\ge4\).

For \(m=3\), every ratio of two nonzero coefficients within a matching
sample lies in \(\mu_3\).  Use a string of \(S_0\cap S_1\) to express
the required cancellation, and choose one string in each of
\(S_0\setminus S_1\) and \(S_1\setminus S_0\).  Their coefficient
ratio in the third sample lies in \(-\mu_3\), disjoint from \(\mu_3\).
This is again impossible.  Thus all sample supports coincide.

The weight-two strings of the common support translation space determine its
matching uniquely, including the parity of each support coset.  In that
fixed tensor factorization every \(2\times2\) Segre flattening
minor is a polynomial in \(t\) of degree at most two.  It vanishes at
all \(m\ge3\) sampled phases and hence identically.  The whole
projective line consists of simple tensors.  The Segre-line criterion
in \hyperref[subsubsec:projective-tensor-geometry]{the projective tensor conventions} therefore permits variation
in only one factor, proving the assertion.
\end{proof}

\subsection{Binary pencils and endpoint compatibility}
\label{proofsubsec:nd-endpoints}

By \cref{lem:app-nd-sector-ruling}, each sector has one
fixed residual matching and at most one varying factor class.  This
subsection classifies those binary factor pencils and proves the
six-input base for the standard case. Further endpoint comparisons
will be used in the weighted reconstruction below.

\subsubsection{Möbius geometry}
\label{subsubsec:mobius-geometry}

These M\"obius conventions control the
ratio of the two nonzero entries in a binary factor.  That ratio is a
fractional-linear function of the contraction phase.  Requiring it to
remain in the same finite phase group forces the two endpoint matrices
onto complementary coordinate axes.

The Riemann sphere is
\(\widehat{\mathbb C}:=\mathbb C\cup\{\infty\}\cong\mathbb P(\mathbb C^2)\),
with affine chart \(z=[z:1]\) and \(\infty=[1:0]\).  A Möbius transformation
is a projective automorphism of this sphere. In the affine chart it is
\[
 z\longmapsto\frac{az+b}{cz+d},
 \qquad ad-bc\ne0,
\]
where \(cz+d=0\) is sent to \(\infty\) and \(\infty\) is interpreted
projectively.  A \emph{generalized circle} is a Euclidean circle
\(\{z:|z-z_0|=r\}\), \(r>0\), or a real affine line together with
\(\infty\). Such a line has the form
\(\{z:\operatorname{Re}(\overline\alpha z)=\beta\}\) for
\(\alpha\in\mathbb C^\times\) and \(\beta\in\mathbb R\).
Möbius transformations carry generalized circles to
generalized circles \cite[Chapter~3]{Beardon1983}.  The unit circle is
\(S^1:=\{z\in\mathbb C:|z|=1\}\). Its counterclockwise cyclic order is the
order of arguments modulo \(2\pi\).

\begin{lemma}[M\"obius stabilizer of a root polygon]
\label{lem:nd-root-mobius}
Let \(m\ge3\), and let the M\"obius map
\[
 \phi:\widehat{\mathbb C}\longrightarrow\widehat{\mathbb C},
 \qquad
 \phi:z\longmapsto\frac{c+dz}{a+bz},
 \qquad ad-bc\ne0,
\]
have no pole on \(\mu_m\) and permute \(\mu_m\).  Then
\[
 \phi(z)=\alpha z
 \qquad\text{or}\qquad
 \phi(z)=\alpha/z,
 \qquad \alpha\in\mu_m.
\]
\end{lemma}

\begin{proof}
A M\"obius map sends generalized circles to generalized circles.  Three
distinct roots and their images force \(\phi\) to preserve the unit circle.
Its action on the regular \(m\)-gon preserves or reverses cyclic order, so it
is a rotation or a rotation followed by reflection.  The corresponding
M\"obius maps agree with \(\phi\) on at least three points and hence agree
identically.
\end{proof}

Use the even and odd parity planes and their ordered coordinate axes
defined above.  For a nonsingular
monomial matrix, its \emph{ordered nonzero-entry ratio} is the coefficient
of the second displayed axis divided by that of the first, in the fixed input
orientation.

\begin{lemma}[Classification of binary pencils]
\label{lem:nd-root-pencil}
Let \(A,B\in\Mat_2(\mathbb C)\) be binary signatures, and define the binary
matrix pencil
\[
 K(t):=A+tB,
 \qquad t\in\mu_m,\qquad m\ge3.
\]
Suppose every sampled binary signature \(K(t)\) is zero or a nonsingular
monomial binary whose ordered nonzero-entry ratio lies in \(\mu_m\).  Then
either \(A,B\) are proportional, or
there are a fixed parity plane with coordinate axes \(E_0,E_1\),
\(\lambda\ne0\), and \(\alpha\in\mu_m\) such that
\[
 K(t)=\lambda(E_0+\alpha tE_1)
 \qquad\text{or}\qquad
 K(t)=\lambda(tE_0+\alpha E_1).
\]
Here \(K(1)=A+B\) is a sampled value, whereas \(A=K(0)\) is an analytic
endpoint because \(0\notin\mu_m\).  In the application below, \(K(t)\) is
the one varying binary factor of a higher-arity realizable matching card. Its coefficients are recovered from the sector pencil.
\end{lemma}

\begin{proof}
In the nonproportional case no sample is zero and the projective samples are
distinct.  Two of the at least three samples lie in one of the two monomial
parity planes. They span the pencil, so all samples lie in that plane.  In
these axes, write the matrix pencil with coefficient scalars \(a,b,c,d\)
and projective coordinate ratio \(\phi\) as
\[
 K(t)=(a+tb)E_0+(c+td)E_1,
 \qquad \phi(t)=\frac{c+td}{a+tb}.
\]
Nonproportionality gives \(ad-bc\ne0\), and nonsingularity excludes zeros
and poles on \(\mu_m\).  Thus \(\phi\) permutes \(\mu_m\).  By
\cref{lem:nd-root-mobius}, it equals \(\alpha t\) or \(\alpha/t\).
Cross-multiplication gives a degree-two identity valid on at least three
points.  Its coefficients yield respectively
\(b=c=0,d=\alpha a\), or \(a=d=0,c=\alpha b\), which are the two displayed
forms.  A zero sample already makes the endpoints proportional.
\end{proof}

Use the following names for the three binary-pencil outcomes of
\cref{lem:nd-root-pencil}:
\[
 \begin{array}{ccl}
 \mathsf Z&:&\text{identically zero},\\
 \mathsf P&:&\text{nonzero proportional},\\
 \mathsf C&:&\text{nonproportional complementary-axis}.
 \end{array}
\]

For a deleted pair \(p\), factor the even-sector card into the common matching
factor \(P_E\) and endpoint matrix pencil \(K_E(t)\), after ordering the
\(2k=\arity(f)-2\) residual inputs of a nonconstant even ruling:
\begin{equation*}
  E(t)=P_E\otimes K_E(t),
  \qquad
  K_E(t):=K_{E,0}+tK_{E,1},
  \qquad K_E(t),K_{E,0},K_{E,1}\in\Mat_2(\mathbb C),
\end{equation*}
with \(k-1\) fixed factors. Define \(O(t)\) and its binary factor pencil
\(K_O(t)\) analogously.  To obtain one tensor identity with a fixed scalar, take
two distinct nonzero samples, use their linear inversion to reconstruct the
two analytic endpoint tensors, normalize the common factor \(P_E\) by a
fixed linear functional \(\eta_E(P_E)=1\), and define
\[
 K_E(t)=(\eta_E\otimes\operatorname{id})(E(t)).
\]
The normalization fixes the common factor once for the whole pencil.
The whole sector inherits the type of this exact residual binary pencil.
An identically zero whole pencil is type \(\mathsf{Z}\), and a nonzero proportional
whole pencil is type \(\mathsf{P}\) before any varying factor is selected.  In a
\(\mathsf{C}/\mathsf{C}\) branch the variable endpoints are complementary rank-one matrices.

\subsubsection{The six-input base}

\begin{lemma}[Direct six-input base case]
\label{lem:nd-six-prime-base}
Let \(h\ne0\) be a retained arity-six signature in the standard
\(\DihZ{2m}\) sector, where \(m\ge3\).  Suppose every realizable oriented
pair card of \(h\), in both kernel parities and at every phase in
\(\mu_m\), is zero or a product of two nonsingular current-group
monomial binaries.  Then, after a permutation of inputs, \(h\) has an exact
nontrivial factorization of type \(1+5\) or \(2+4\).  In particular,
\(h\) is not irreducible.
\end{lemma}

\begin{proof}
Apply \cref{lem:app-nd-sector-ruling,lem:nd-root-pencil} at each
deleted pair \(p\).  A type-\(\mathsf P\)
sector is a scalar pencil times one fixed nonsingular matching product
on the four residual inputs.  A type-\(\mathsf C\) sector is one fixed
nonsingular binary factor tensored with a complementary-axis pencil.
Its two endpoint supports are complementary singletons on the varying
edge.  A type-\(\mathsf Z\) sector vanishes identically.

If a pair has type \(\mathsf{Z}/\mathsf{P}\), then \(h=U_p\otimes P\) for a nonzero
binary \(U_p\) and a nonzero residual matching product \(P\).  If it
has type \(\mathsf{Z}/\mathsf{C}\), the fixed binary factor of the \(\mathsf C\) ruling factors
from \(h\).  The reversed cases are identical, and \(\mathsf{Z}/\mathsf{Z}\) would give
\(h=0\).  Each remaining case containing \(\mathsf Z\) therefore proves the
assertion.

Suppose a pair has type \(\mathsf{P}/\mathsf{P}\).  Write
\[
 h=U_p\otimes P_0+V_p\otimes P_1,
\]
where \(U_p,V_p\ne0\) are supported on the two opposite parities of
\(p\), and \(P_0,P_1\) are nonsingular matching products on the four
residual inputs.  Choose a residual pair \(e\) belonging to neither
matching.  Such a pair exists because the two matchings contain at most
four of the six residual pairs.  Contract \(e\) by either realizable
parity kernel at phase \(1\).  This crosses the two factors of each
\(P_j\).  For every supported \(p\)-string, the output has precisely
two complementary residual strings, each with one internal lift and a
nonzero coefficient.  Different \(p\)-strings cannot cancel.  The
resulting realizable quaternary therefore has
\[
 2\bigl(\lvert\supp U_p\rvert+\lvert\supp V_p\rvert\bigr)
\]
nonzero entries.  Six or eight entries contradict the card hypothesis.
Otherwise both supports are singletons.  Their opposite-parity
\(p\)-strings differ in exactly one bit, so the other input of
\(p\) is fixed.  Its point mass is a unary tensor factor of \(h\).

For type \(\mathsf{P}/\mathsf{C}\), choose a residual pair \(e\) belonging to neither
the \(\mathsf P\) matching nor the fixed/varying matching of the \(\mathsf C\)
ruling.  Apply the same realizable parity contraction.  Each supported
\(\mathsf P\)-sector \(p\)-string leaves two residual strings.  Each of the two
\(\mathsf C\)-sector \(p\)-strings leaves exactly one: the singleton varying
endpoint and the kernel parity force the bit on the fixed factor, and
then force both remaining residual bits.  These lifts are unique and
all coefficients are nonzero.  If the \(\mathsf P\) sector has two
\(p\)-strings, the card has six nonzero entries.  If it has one, the
card has four nonzero entries but its projection onto \(p\) has
three strings.  A product of two nonsingular monomial binaries projects
onto any two inputs in exactly two strings when those inputs are matched,
and in four strings otherwise.  Both alternatives contradict the card
hypothesis.  Type \(\mathsf{C}/\mathsf{P}\) is identical.

Unless one of the stated factorizations has already occurred, every
deleted pair must consequently have type \(\mathsf{C}/\mathsf{C}\).  Fix one such pair.
Its fixed binary factors and complementary-axis endpoints give exactly
the eight-string support in
\cref{eq:cert-nd-q6-four-flat-support}.  The support classification in
\cref{prop:cert-nd-q6-support-lowering} either gives an injective
\(I/X\)-contraction with six or eight projected strings, or identifies
a three-edge matching coset.  In the first case, injectivity gives one
nonzero summand per output string, so this realizable quaternary contradicts
the card hypothesis.  In the second case, one parity sector at any
edge of that matching is identically zero, contradicting the conclusion
that every deleted pair is \(\mathsf{C}/\mathsf{C}\).  This completes the proof.

All contractions in this proof are applied directly to \(h\).
The factors are exact tensor factors, and are exposed only through
\cref{thm:external-factor,lem:factor-saturation} when the reduction
procedure uses them.
\end{proof}

\subsection{Completion of the standard dihedral case}

\begin{theorem}[Standard \(\DihZ{2m}\)-sector deck rigidity]
\label{thm:normalized-dihedral}
Assume
\[
 \Stable(\cF,\DihZ{2m},f),\qquad \Even(\cF),\qquad
 \arity(f)=\nu(\cF)=2k+2,\qquad m\ge3,
\]
where
\[
 \DihZ{2m}
 =\{[D(t)],[XD(t)]:t\in\mu_m\}.
\]
Then \(\Close_{\{\cP\}}(\cF,\DihZ{2m},f)\).
\end{theorem}

\begin{proof}[Completion of \cref{thm:normalized-dihedral}]
If \(\arity(f)=4\), regenerate and close its complete deck.
In the standard \(\DihZ{2m}\) case the output of
\cref{thm:equality-q4-boundary} is an invertible dressing of the irreducible
quaternary.  It is therefore a realizable endpoint-nondegenerate irreducible
eight-vertex signature.  Apply
\cref{thm:p1-quaternary-kholant-dichotomy} to its signature set.
Direct realization and preservation under taking subsets give a hard or
tractable case satisfying \(\ReTract(K\cF)\) for the entire retained set \(\cF\).

For \(\arity(f)=6\), \cref{lem:nd-six-prime-base} contradicts the
irreducibility of the distinguished irreducible signature.  Now let \(n:=\arity(f)\ge8\).
At any deleted pair \(p\), two disagreeing nonzero sampled cards would
give the six-input output in \cref{lem:app-nd-localization}.  The first
proper contraction of the distinguished irreducible signature is zero or a matching product, and
further monomial contractions preserve that property.  This contradicts
the two inherited disagreeing nonzero cards.  Hence all samples at \(p\)
share one matching and all factor classes outside one designated edge \(e\).
The \hyperref[par:nd-common-factor-consequence]{common-factor consequence}
following \cref{lem:app-nd-bounded-consumer}, applied to both sectors,
therefore gives \(f=P\otimes q_{p,e}\). Here \(P\) is a nonzero product
containing at least one binary factor, contradicting irreducibility.

Consequently no higher-arity deck-stable signature branch remains.  The quaternary
case already resolves the whole signature set, and completes this branch. These are the allowed outcomes of
\(\Close_{\{\cP\}}(\cF,\DihZ{2m},f)\).
\end{proof}

The restriction \(m\ge3\) is essential here.  For \(m=2\), two forbidden
trace roots can exhaust \(\mu_2\), and evaluation on \(\mu_2\) does not
determine a polynomial of coordinate degree two.  That case belongs to the
separate normalized \(\KleinStd\) analysis.

\subsection{Marked coordinates and whole-network reductions}
\label{sec:marked-dihedral-sectors}

For the nonstandard marked groups, the same binary transfers acquire a
different edge matrix.  We first compute that matrix, then determine
which parts of the standard proof survive.  The matrices \(U_\tau\)
below change analytic coordinates.  The \(\DihY{2m}\) and
\(\DihXzero{2m}\) proofs additionally give whole-network
equivalences. The \(\DihXone{2m}\) proof instead keeps the original
network and tracks the resulting weights.

\begin{lemma}[Marked-sector coordinates and edge matrices]
\label{lem:marked-sector-dictionary}\leavevmode\newline
Use the form labels \(Z,Y,(X,0),(X,1)\) as sector indices.
For \(\tau=Z,Y\), let \(m\ge3\) and set
\begin{equation*}
 U_{Z}=I,\qquad
 U_{Y}=I-iX.
\end{equation*}
For \(\tau=(X,1)\), let \(m\ge4\) be even, fix
\(\eta\in\mu_{2m}\setminus\mu_m\), choose \(s_0\in\AlgNums\) with
\(s_0^2=\eta^{-1}\), and set
\begin{equation*}
 U_{X,1}=
 \begin{pmatrix}1&1\\1&-1\end{pmatrix}\diag(1,s_0),
 \quad s_0^2=\eta^{-1}.
\end{equation*}
Conjugation by \(U_\tau\) sends \(\DihZ{2m}\) projectively onto the
corresponding marked group.  Define the transformed edge matrix \(S_\tau\) by
\[
 S_\tau=U_\tau^{\mathsf T}XU_\tau.
\]
Then, projectively,
\begin{equation*}
 S_{Z}=X,\qquad
 S_{Y}=I,\qquad
 S_{X,1}=\diag(1,-\eta^{-1}).
\end{equation*}
For \(T\in\DihZ{2m}\), choose a fixed realizable marked-sector binary
\(B_T\) whose sector-coordinate transfer is \([T]\), and set
\(B'_T:=U_\tau^{-1}B_TU_\tau^{-\mathsf T}\).  Then, for
\(T,U\in\DihZ{2m}\),
\begin{equation}
 B'_T\doteq T S_\tau^{-1},
 \qquad
 (TS_\tau^{-1})S_\tau(US_\tau^{-1})
 =TU S_\tau^{-1}.
 \label{eq:marked-sector-fusion}
\end{equation}
Thus multiplication and reversal are preserved, and every monomial fusion
has the same unique internal lift. Consequently such a fusion introduces no
cancellation.
\end{lemma}

\begin{proof}
\[
 U_{Y}D(t)U_{Y}^{-1}\doteq C(t),\qquad
 U_{X,1}D(t)U_{X,1}^{-1}\doteq A(t),
 \qquad
 U_{X,1}XD(t)U_{X,1}^{-1}
 \doteq ZA(\eta t).
\]
\[
 U_{Y}^{\mathsf T}XU_{Y}=-2iI,
 \qquad
 U_{X,1}^{\mathsf T}XU_{X,1}
 =2\diag(1,-\eta^{-1}).
\]
Congruence gives \(B'_T\doteq TS_\tau^{-1}\), and the fusion
identity is matrix multiplication. All omitted scalars are fixed nonzero.
\end{proof}

\paragraph{Sector coordinates and available kernels.}
For an \(r\)-ary current-coordinate signature \(h\), write
\[
 h^{(\tau)}=(U_\tau^{-1})^{\otimes r}h
\]
for its analytic sector-coordinate tensor.  A \emph{sector-\(X\) dressing} is input dressing by the fixed realizable binary \(B_X\). By
\cref{eq:marked-sector-fusion}, its transported transfer fuses as \(X\).
Put
\[
 \beta_{Y}:=1,
 \qquad
 \delta:=-\eta^{-1},\quad
 \beta_{X,1}:=\delta^{-1},
 \qquad
 \mathscr R_\tau:=\beta_\tau\mu_m,
\]
where the quantities involving \(\eta\) and \(\delta\) apply only in the
\((X,1)\)-sector.  For a quotient in \(\mathscr R_\tau\), division
by \(\beta_\tau\) gives a point of \(\mu_m\).  This division acts on coefficient ratios.  In one monomial parity plane, a
\emph{complementary-axis pencil} has its two coefficient matrices on
the two coordinate axes.  A sample with exactly one vanishing coordinate
is a singular binary.

\begin{lemma}[Symmetric binary pullback]
\label{lem:symmetric-binary-pullback}
Let \(\cC\in\{\cA,\cP,\cL\}\).  If
\(E,F\in\cC\) are nonsingular symmetric binaries, then
\(EFE\in\cC\).
\end{lemma}

\begin{proof}
Closure under contraction proves the affine case, and the
equality/disequality normal forms prove the product case.  For the
local-affine case, write
\(E=\left(\begin{smallmatrix}a&b\\b&c\end{smallmatrix}\right)\).
The diagonal twists in the definition of \(\cL\) preserve support, which
must be affine.  Nonsingularity excludes support size one, and affine
support excludes size three.  Full support is also impossible: the
\(\alpha=00\) condition requires \(b/a\in\mu_4\), whereas the
\(\alpha=01\) twist requires \(\zeta_8 b/a\in\mu_4\).
On anti-diagonal support symmetry gives \(E=bX\), and the same
\(\alpha=01\) twist again has a nonzero-entry ratio outside \(\mu_4\).
Consequently \(E=a\diag(1,i^k)\), with \(a\ne0\) and
\(k\in\mathbb Z/4\mathbb Z\).  Conversely these four diagonal forms
satisfy both supported twists, \(\alpha=00,11\), and belong to \(\cL\).
Writing \(F=d\diag(1,i^\ell)\) in the same way gives
\(EFE=a^2d\diag(1,i^{2k+\ell})\in\cL\), as required.
\end{proof}

\begin{theorem}[Signature-set-wide reduction of the \(\DihY{2m}\) sector]
\label{thm:marked-B-reduction}
If
\[
 \Stable(\cF,\DihY{2m},f),\qquad
 \Even(\cF),\qquad m\ge3,
\]
then \(\Close_{\{\cP\}}(\cF,\DihY{2m},f)\).
\end{theorem}

The tractable cases here include
\(\PresK(K\cF;\cA)\) and \(\PresK(K\cF;\cL)\). The product case is
the only remaining tractable case.

\begin{proof}
Define the transformed signature set
\(\Gamma:=U_{Y}^{-1}\cF\) and the transformed active irreducible signature
\[
 f^{(Y)}:=(U_{Y}^{-1})^{\otimes\arity(f)}f\in\Gamma.
\]
This whole-problem change gives
\begin{equation*}
 U_{Y}^{\mathsf T}XU_{Y}=-2iI,
 \qquad
 U_{Y}^{-1}IU_{Y}^{-\mathsf T}=\frac{i}{2}X,
 \qquad
 U_{Y}^{-1}XU_{Y}^{-\mathsf T}=\frac{i}{2}I.
\end{equation*}
Thus edge and equality exchange, and \cref{lem:p1-equality-wire-equivalence}
gives
\begin{equation*}
 \KHolant(\cF)
 \equivT
 \Holant(\Gamma,X)
 \equivT
 \KHolant(\Gamma,I).
\end{equation*}
Define the signature set
\(\cF^\sharp:=\Gamma\cup\{I\}\).  Here the \emph{open-network
bijection} is the vertexwise change by \(U_{Y}^{-1}\), together
with the displayed edge/equality exchange, applied also to the ordered open
boundary legs. The inverse change restores the original network.  It
preserves arity, factorization, minimum arity of a nonbinary irreducible signature, and parity of
the arity of every irreducible signature. Hence \(\Even(\cF^\sharp)\).

For a realizable nonsingular \(b\) over \(\cF^\sharp\), its pullback
\(B\) satisfies
\begin{equation*}
 B\doteq U_{Y}bU_{Y}^{\mathsf T},
 \qquad
 [BX]=[U_{Y}bU_{Y}^{-1}],
\end{equation*}
using \(U_{Y}^{\mathsf T}X=-2iU_{Y}^{-1}\).  Infinite
order pulls back to the interpolation reduction of
\cref{lem:binary-interpolation}. Otherwise
\cref{lem:actual-binary-completeness} applies to the pullback transfer
\([BX]\).  Conjugating by \(U_{Y}^{-1}\) first puts
\([b]\) in \(\DihZ{2m}\).  Since \([X]\in\DihZ{2m}\), closure under
right multiplication then puts the realizable transformed transfer
\([bX]\) in \(\DihZ{2m}\).

Conversely, for the chosen representative \(B_T\), define its
nonzero representative scalar \(\beta_T\) by
\(B_TX=\beta_TU_{Y}TU_{Y}^{-1}\).  Then
\begin{equation*}
 B'_T
 =U_{Y}^{-1}B_TU_{Y}^{-\mathsf T}
 =\frac{i\beta_T}{2}T.
\end{equation*}
Since \([D(t)X]=[XD(t^{-1})]\) and
\([XD(t)X]=[D(t^{-1})]\), every class occurs and
\begin{equation*}
 G(\cF^\sharp)=\DihZ{2m}.
\end{equation*}
The open-network bijection also transports the complete deck: absorbing an
equality edge replaces a label by \(XT,TX\), or \(XTX\), which permute
\(\DihZ{2m}\).  Hence zeros, matching products, factor closure, and
minimum arity among irreducible signatures correspond, so the transformed state is deck-stable.
In particular,
\[
 \Stable(\cF^\sharp,\DihZ{2m},f^{(Y)}),
 \qquad
 \arity(f^{(Y)})=\nu(\cF^\sharp)=\arity(f).
\]

Apply \cref{thm:normalized-dihedral}.  Complete closure of each transported
concrete batch is resolved.  On the remaining branch there are
\(\cC\in\{\cA,\cP,\cL\}\) and one \(V\in\GL_2(\AlgNums)\) such that
\[
 V^{-1}(\Gamma\cup\{I\})\subseteq\cC,
 \qquad V^{\mathsf T}XV\in\cC.
\]

Define the common change-of-basis matrix \(W:=U_{Y}V\). Then
\(W^{-1}\cF\subseteq\cC\).  With
\[
 E=V^{\mathsf T}XV,
 \qquad F=V^{-1}IV^{-\mathsf T},
\]
both \(E,F\) are nonsingular symmetric members of \(\cC\), and
\cref{lem:symmetric-binary-pullback} gives
\[
 W^{\mathsf T}XW
 =-2iV^{\mathsf T}IV
 =-2iEFE\in\cC.
\]
Thus \(W\) is common.  The \(\cA,\cL\) cases are resolved
tractable cases satisfying \(\Tract(K\cF)\). \(\cP\) is the remaining tractable case.
The reductions above use \(U_{Y}\) for the whole network.
\(W=U_{Y}V\) records its contribution to the final witness.
\end{proof}

\begin{theorem}[Reduction of the
  \texorpdfstring{\(\DihXzero{2m}\)}{D2m (X,0)} sector]
\label{thm:marked-A0-reduction}
Assume
\[
 \Stable(\cF,\DihXzero{2m},f),\qquad
 \Even(\cF),\qquad m\ge4,\quad 2\mid m.
\]
Then \(\Close_{\{\cP\}}(\cF,\DihXzero{2m},f)\).
\end{theorem}

\begin{proof}
Use the Walsh matrix \(P\) from \cref{eq:Walsh-coordinates}, with
\(P^{-1}=P/2\).
Direct multiplication gives, projectively,
\begin{equation}
 PD(t)P^{-1}\doteq A(t),
 \qquad
 PXD(t)P^{-1}\doteq ZA(t),
 \label{eq:marked-A0-conjugacy}
\end{equation}
so \(P^{-1}\DihXzero{2m}P=\DihZ{2m}\).  More importantly,
\begin{equation}
 P^{\mathsf T}XP=2Z,
 \qquad
 P^{-1}IP^{-\mathsf T}=\frac12I.
 \label{eq:marked-A0-fixed-I}
\end{equation}
Thus this whole-problem change preserves the available binary equality.

Make available by reduction the fixed representative \(B_j\) of the mark
\(j=[Z]\), retaining its realization scalar and input orientation.  Since
\([B_jX]=j\), one has \(B_j\doteq Y\).  Put
\[
 \cF^+=\cF\cup\{B_j\},
 \qquad
 \Gamma=P^{-1}\cF^+,
 \qquad
 f^{(X,0)}=(P^{-1})^{\otimes\arity(f)}f.
\]
After recording the two nonzero representative scalars, use \(I,Y\in\Gamma\). Indeed
\[
 P^{-1}IP^{-\mathsf T}=\frac12I,
 \qquad
 P^{-1}YP^{-\mathsf T}=-\frac12Y.
\]
Write \(\Holant^Z(\Gamma)\) for the homogeneous \(Z\)-edge half-edge
problem.  The whole-problem change in
\cref{eq:marked-A0-fixed-I} gives
\[
 \KHolant(\cF^+)\equivT\Holant^Z(\Gamma),
\]
where the factor \(2\) per edge is known.  The exact local identities
\begin{equation}
 Z=XIXYX,
 \qquad
 X=-ZIZYZ
 \label{eq:marked-A0-edge-switch}
\end{equation}
give a signature-preserving open-network equivalence
\begin{equation}
 \Holant^Z(\Gamma)\equivT\KHolant(\Gamma).
 \label{eq:marked-A0-open-equivalence}
\end{equation}
For the forward reduction, replace a \(Z\)-edge by three \(X\)-edges with the available ordered vertices \(I,Y\). For the reverse reduction, replace an \(X\)-edge by three \(Z\)-edges with the same two
vertices.  Reversing the ordered \(Y\)-vertex changes only its recorded
nonzero sign.  Define
\[
 \Gamma^\sharp:=\Gamma\cup\{Z\}.
\]
The first identity in \cref{eq:marked-A0-edge-switch} directly realizes
\(Z\) in the \(X\)-edge model, so adding it changes neither the
problem nor the set of directly realizable signatures.

The two edge presentations therefore have the same projective set
\(\mathscr B\) of directly realizable nonsingular ordered binary tensors.
In the \(Z\)-edge model, open-network transport and
\cref{eq:marked-A0-conjugacy} give
\[
 \{[bZ]:b\in\mathscr B\}
 =P^{-1}\DihXzero{2m}P
 =\DihZ{2m}.
\]
Since \(m\) is even,
\([Y]=[XD(-1)]\in\DihZ{2m}\).  Hence, in the \(X\)-edge
presentation,
\hypertarget{eq:marked-A0-complete-group}{}
\begin{equation*}
 G(\Gamma^\sharp)
 =\{[bX]:b\in\mathscr B\}
 =\{[bZ][ZX]:b\in\mathscr B\}
 =\DihZ{2m}[Y]
 =\DihZ{2m}.
\end{equation*}

The complete deck transports as well.  Under \(P^{-1}\), the original
\(X\)-edge deck is the corresponding \(Z\)-edge deck.  Translating any
target dressing or closing edge by the second identity in
\cref{eq:marked-A0-edge-switch}, and then absorbing the inserted \(I,Y\)
binary chain into the adjacent fixed chosen representative, merely permutes
the \(\DihZ{2m}\) labels.  The same argument with the first identity gives
the inverse correspondence.  Thus zeros, matching products, factorization trees,
minimum arity of a nonbinary irreducible signature, and parity of the irreducible factors' arities correspond.  Therefore
\[
 \Stable(\Gamma^\sharp,\DihZ{2m},f^{(X,0)}),
 \qquad
 \arity(f^{(X,0)})=\nu(\Gamma^\sharp)=\arity(f),
 \qquad
 \Even(\Gamma^\sharp).
\]

Apply \cref{thm:normalized-dihedral}.  Every resolved complexity leaf, concrete factor
batch, and strict successor transports back through the two open-network
equivalences.  On a surviving tractable case with a common presentation, there are
\(\cC\in\{\cA,\cP,\cL\}\) and \(V\in\GL_2(\AlgNums)\) such that
\[
 V^{-1}\Gamma^\sharp\subseteq\cC,
 \qquad
 E:=V^{\mathsf T}XV\in\cC.
\]
Set \(W:=PV\).  Then \(W^{-1}\cF\subseteq\cC\).  Because
\(Z\in\Gamma^\sharp\), also
\[
 F:=V^{-1}ZV^{-\mathsf T}\in\cC.
\]
The binaries \(E,F\) are nonsingular and symmetric, so
\cref{lem:symmetric-binary-pullback} and \(XZX=-Z\) give
\[
 EFE=V^{\mathsf T}XZXV=-V^{\mathsf T}ZV,
 \qquad
 W^{\mathsf T}XW
 =2V^{\mathsf T}ZV
 =-2EFE\in\cC.
\]
Thus \(W\) is a common change-of-basis matrix for the original problem.  The
\(\cA,\cL\) cases are resolved tractable cases satisfying \(\Tract(K\cF)\) and \(\cP\) is
the remaining tractable case.  The witness \(W=PV\) combines the
whole-network equivalence with the transformed common presentation.
\(I\) remains available throughout.
\end{proof}

\subsection{The weighted \texorpdfstring{\(\DihXone{2m}\)}{D2m (X,1)} case}
\label{subsec:weighted-dihedral-case}

In the shifted dihedral case, a matching describes the support, but
factorization also requires the coefficients to split across its edges.
We test this by joining two copies. The resulting identities force the
ratio for changing one edge assignment to be independent of the others.
We develop the test in standard coordinates and then show how to realize
the required contractions with the weighted kernels of \(\DihXone{2m}\).

\begin{theorem}[\(\DihXone{2m}\)-sector deck rigidity]
\label{thm:marked-A-rigidity}
Assume
\[
 \Stable(\cF,\DihXone{2m},f),\qquad \Even(\cF),
 \qquad m\ge4,\quad 2\mid m.
\]
Then \(\Close_{\{\cP\}}(\cF,\DihXone{2m},f)\).
\end{theorem}

\subsubsection{Relative stability and endpoint compatibility}
\label{subsubsec:nd-relative-endpoints}

We work in the standard \(\DihZ{2m}\) coordinates up to and including
\cref{lem:nd-relative-reentry}, using the
\hyperlink{eq:nd-sector-pencils}{standard kernels and sector pencils}
with \(m\ge3\) and even arity \(n=2k+2\ge6\). The support and coefficient
reconstruction applies to the selected occurrences under the deck hypotheses
stated below. \Cref{lem:marked-A-gauge-transport,lem:marked-A-primitive-transport}
then transfer these results to the weighted \(\DihXone{2m}\) kernels.
We begin by specifying relative deck stability.

\begin{definition}[Occurrence-level relative deck stability]
\label{def:occurrence-relative-deck-stability}
For a retained occurrence \(h\) at current reduction data \((\cF,G)\), write
\[
 \operatorname{RelStable}(\cF,G;h)
\]
when its complete oriented deck has been regenerated,
every already obtained proper card and its complete factorization tree
have been processed, the complete group remains \(G\), and, on the continuing branch,
every surviving nonzero proper card is a \(G\)-matching signature, while every
unresolved lower-arity nonbinary occurrence has been placed earlier in the
occurrence induction.  An active deck-stable minimum occurrence is
automatically relatively deck-stable.
\end{definition}

Under this hypothesis, the proper outputs needed below have already
been processed. \Cref{lem:nd-relative-reentry} checks the use of the
reconstruction results with this hypothesis, and
\cref{proofsubsec:nd-completion} records their occurrence-level descent.

For a fixed deleted pair \(p=\{i,j\}\), a \emph{residual matching} of a
nonzero card
\[
 P=\bigotimes_{\{u,v\}\in\mathcal M_p}B_{uv}
\]
is the perfect matching \(\mathcal M_p\) of the residual inputs
\([n]\setminus p\).  Its matching translation space is
\[
 M(\mathcal M_p)=
 \Span\{{\mathbf{e}}_u+{\mathbf{e}}_v:\{u,v\}\in\mathcal M_p\}.
\]
The even and odd families have the same residual matching at \(p\)
when their unordered edge sets agree, equivalently when their matching
translation spaces agree.
Support-parity and ordered factor-class disagreements remain separate
discrepancies.  For example, on residual inputs \(3,4,5,6\), the pairings
\(\{\{3,4\},\{5,6\}\}\) in both sectors give the same matching, whereas that
pairing and \(\{\{3,5\},\{4,6\}\}\) form a matching disagreement whose
union contains an alternating four-cycle.  The support integration below relates these local matchings for
different deleted pairs.

For the following support comparison suppose that both sector
pencils have type \(\mathsf{C}\). All ordered pairs containing \(\mathsf Z\) or \(\mathsf P\) are
handled separately in \cref{lem:app-nd-endpoint-type} below.

Compare one realizable nonzero even card with one realizable nonzero odd card at the fixed
deleted pair \(p\).  If their residual matchings differ, or their ordered factor classes differ
on at least two edges (including parity changes), this is a cross-sector
disagreement and is consumed
by \cref{lem:app-nd-localization,lem:app-nd-bounded-consumer}.  Hence the
continuing branch has one common residual matching with all nondesignated
factor classes aligned.

If the two sector rulings have the same residual matching but designate
different varying edges, close only their common fixed factors.  Up to affine
translations, the remaining support, in the distinguished bits \(x,y\) and
the two varying-edge bits \(u,v\), is
\begin{equation*}
  \begin{aligned}
    x=y&\ \Longrightarrow\ u=x+\alpha,\quad v\text{ free},\\
    x\ne y&\ \Longrightarrow\ v=x+\beta,\quad u\text{ free}.
  \end{aligned}
\end{equation*}
A realizable parity kernel imposing \(u=x+\alpha\) has a unique lift and leaves
a six-string quaternary signature, hence not a two-monomial product.  Its
support contains a complementary pair. Realizable logical \(X\)-dressings
may move that pair to the two endpoints.  Add this quaternary signature and
its irreducible factors.  A retained nonbinary factor has arity at
most four: for the active minimum-arity nonbinary irreducible signature it is a strict lower-\(\nu\)
successor, and for a later occurrence it is assigned to the lower-arity
occurrence induction.  Stabilize the state of that quaternary occurrence, regenerate its complete
oriented deck, and only then invoke the equality-accessible quaternary
boundary and eight-vertex reduction.  Thus the different-edge branch is closed.
We may therefore assume below that both sectors use one common varying
edge.

To assemble the four slices into one affine support, their offsets must
fit the same affine translation space. After aligning the common factors,
write the four pencil-endpoint supports as
\begin{equation*}
  \supp h_\xi={\mathbf{a}}_\xi+L,
  \qquad
  h_{00}=A,\ h_{01}=B,\ h_{10}=C,\ h_{11}=D_{11},
\end{equation*}
where
\begin{equation*}
  L=\Span\{{\mathbf{e}}_{u_r}+{\mathbf{e}}_{v_r}:1\le r<k\},
  \qquad
  \dim L=k-1.
\end{equation*}
Define the endpoint displacement vectors
\begin{equation*}
  {\mathbf{d}}_E={\mathbf{a}}_{00}+{\mathbf{a}}_{11},
  \qquad
  {\mathbf{d}}_O={\mathbf{a}}_{01}+{\mathbf{a}}_{10}.
\end{equation*}
Then the realizable even/odd card support translation spaces are
\begin{equation*}
  L_E=L+\Span_{\Ftwo}\{{\mathbf{d}}_E\},
  \qquad
  L_O=L+\Span_{\Ftwo}\{{\mathbf{d}}_O\},
  \qquad {\mathbf{d}}_E,{\mathbf{d}}_O\notin L.
\end{equation*}
The condition we need is
\begin{equation}
 {\mathbf{a}}_{00}+{\mathbf{a}}_{01}+{\mathbf{a}}_{10}+{\mathbf{a}}_{11}\in L
 \label{eq:app-nd-parallelogram}
\end{equation}
Equivalently,
\[
 {\mathbf{d}}_E+{\mathbf{d}}_O\in L,
 \qquad
 [{\mathbf{a}}_{00}]+[{\mathbf{a}}_{11}]=[{\mathbf{a}}_{01}]+[{\mathbf{a}}_{10}]
       \quad\text{in }\Ftwo^{n-2}/L,
 \qquad
 L+\Span_{\Ftwo}\{{\mathbf{d}}_E\}=L+\Span_{\Ftwo}\{{\mathbf{d}}_O\}.
\]
This condition is independent of the representatives
\({\mathbf{a}}_\xi\), since changing one representative changes the four-term sum by
an element of \(L\).

\begin{lemma}[Realizable-card endpoint test]
\label{lem:app-nd-parallelogram}
In the complementary-axis \(\mathsf{C}/\mathsf{C}\) branch with one common varying edge, the
residual matchings of
the realizable nonzero cards represented by \(E(t)\) and \(O(s)\) agree if
and only if
\cref{eq:app-nd-parallelogram} holds.
Failure is localized using the realizable cards, not the endpoint slices.
\end{lemma}

\begin{proof}
A monomial matching is determined by its weight-two strings, so the two
matchings agree exactly when their translation spaces agree.  Since
\({\mathbf{d}}_E,{\mathbf{d}}_O\notin L\),
\[
  L_E=L_O
  \quad\Longleftrightarrow\quad
  {\mathbf{d}}_E+{\mathbf{d}}_O\in L
  \quad\Longleftrightarrow\quad
  {\mathbf{a}}_{00}+{\mathbf{a}}_{01}+{\mathbf{a}}_{10}+{\mathbf{a}}_{11}\in L.
\]
Failure therefore invokes
\cref{lem:app-nd-localization,lem:app-nd-bounded-consumer} on the realizable
cards and is resolved.
\end{proof}

\begin{lemma}[Zero and proportional endpoint pencils]
\label{lem:app-nd-exceptional-endpoints}
Every endpoint pair in which at least one sector has type \(\mathsf Z\) or \(\mathsf P\)
yields at least one of the following concrete outcomes:
\begin{enumerate}
\item the branch is impossible because it would give \(f=0\);
\item all four slices have one common residual tensor, giving an exact
      binary--residual factorization of \(f\);
\item all but one residual matching factor are common, giving an exact
      factorization \(f=P\otimes q\), where \(q\) is a nonzero quaternary
      factor on the deleted pair and the exceptional residual edge;
\item realizable nonzero cards have a disagreement, which is localized and
      consumed by \cref{lem:app-nd-localization,lem:app-nd-bounded-consumer}.
\end{enumerate}
In item~3 the quaternary has matching-flattening rank two whenever one of
the two pencils has type \(\mathsf{C}\).  The factor \(q\) is available by the factor theorem and is a strict lower-\(\nu\) successor.
\end{lemma}

\begin{proof}
For a type-\(\mathsf Z\) even pencil, two-card inversion gives \(A=D_{11}=0\), hence the
support condition \(x_i\oplus x_j=1\).  If the odd pencil also has type
\(\mathsf Z\), then \(f=0\), contrary to the active nonzero irreducible signature.  If it has type
\(\mathsf P\), the same inversion gives
\begin{equation*}
 f=G_p\otimes H_O,
  \qquad
  G_p=\begin{pmatrix}0&b\\ c&0\end{pmatrix},
\end{equation*}
where \(G_p\) is the displayed binary signature on \(p\) and \(H_O\) is the
complementary \((n-2)\)-ary signature.  This is an exact factorization of \(f\), and its factors
are exposed only by \cref{thm:external-factor}.  If the odd pencil has type
\(\mathsf C\), write its exact ruling as \(O(t)=P_O\otimes K_O(t)\), where \(P_O\)
is the product of the \(k-1\) fixed residual factors and \(K_O(t)\) is the
nonproportional binary pencil on its varying edge \(e\).  Two-card inversion
puts both \(B\) and \(C\) in \(P_O\otimes\Mat_2(\mathbb C)\). The zero
slices \(A,D_{11}\) lie there as well.  Hence
\[
 f=P_O\otimes q_{p,e}.
\]
The complementary-axis endpoints of \(K_O\) are nonproportional, so the
\(p\mid e\) logical flattening of \(q_{p,e}\) has rank two.  Apply
    \cref{thm:external-factor,lem:factor-saturation}. The quaternary available by reduction is a
deck-stable lower-\(\nu\) successor, while every binary factor is sent through
\cref{lem:actual-binary-completeness}.  Exchanging the sectors covers every
type-\(\mathsf Z\) case.  A single zero sample in a type-\(\mathsf P\) pencil is included in
the exact proportional inversion and is never used by itself to infer a
factor.

For a proportional even pencil, inversion gives
\begin{equation*}
  A=aH_E,
  \qquad
  D_{11}=dH_E.
\end{equation*}
Choose a nonzero realizable even card as the representative of the line
\(\mathbb C H_E\), and compare it with two nonzero realizable odd cards.  Any
matching change or variation on at least two residual factor classes,
including parity changes, is a disagreement and is consumed by
\cref{lem:app-nd-localization,lem:app-nd-bounded-consumer}.  If no bounded
output occurs, the two sectors have one common residual matching and differ
on at most one designated edge \(e\).  Let \(P\) be the product of all fixed
factors outside \(e\).  Four-card inversion then puts
\(A,B,C,D_{11}\in P\otimes\Mat_2(\mathbb C)\), and therefore proves the exact
identity
\[
 f=P\otimes q_{p,e}.
\]
If no edge differs, absorb the last common factor into \(P\). This is the
stronger identity \(f=G_p\otimes H\).  If one edge differs, the remaining
quaternary is nonzero. In a \(\mathsf{P}/\mathsf{C}\) or \(\mathsf{C}/\mathsf{P}\) case its \(p\mid e\) flattening has
rank two because the \(\mathsf C\)-endpoints are nonproportional.  In the \(\mathsf{P}/\mathsf{P}\)
case different projective factor classes give the same rank-two conclusion, while
equal classes give the stronger binary--residual factorization.  Only after
this exact identity has been proved are \(P\) and \(q_{p,e}\) exposed by
\cref{thm:external-factor,lem:factor-saturation}.  This covers \(\mathsf{P}/\mathsf{P}\),
\(\mathsf{P}/\mathsf{C}\), \(\mathsf{C}/\mathsf{P}\), and their possible zero samples.
\end{proof}

By \cref{lem:app-nd-exceptional-endpoints}, it remains to analyze two
nonproportional sector pencils.

We now describe the six-input contraction output obtained by retaining
the deleted pair and two residual edges. Label these inputs by
\(p=\{1,2\}\) and \(R=\{3,4,5,6\}\).  For each sector
\(\epsilon=x_1\oplus x_2\), let
\(M_\epsilon=\{e_\epsilon,e'_\epsilon\}\) be its residual matching, where
\(e_\epsilon=\{u_\epsilon,v_\epsilon\}\) is the fixed edge and
\(e'_\epsilon=(r_\epsilon,s_\epsilon)\) is the ordered varying edge.  There
are parameters \(a_\epsilon,b_\epsilon,o_\epsilon\in\Ftwo\) such that its
six-input support is parametrized by
\begin{equation}
 \begin{aligned}
 (x_1,x_2)&=(z,z+\epsilon),\\
 x_{u_\epsilon}\oplus x_{v_\epsilon}&=a_\epsilon,\\
 (x_{r_\epsilon},x_{s_\epsilon})
   &=(o_\epsilon+z,o_\epsilon+z+b_\epsilon),
 \end{aligned}
 \qquad \epsilon,z\in\Ftwo.
 \label{eq:app-nd-four-flat-support}
\end{equation}
For fixed \((\epsilon,z)\), the parity equation on the fixed edge has two
solutions and the varying edge is determined.  Hence each of the four
endpoint slices indexed by \((\epsilon,z)\) is a two-string affine flat, each
sector has four strings, and the two sectors have the eight-string support used
in \cref{eq:app-nd-four-flat-support}.  The varying-edge assignments for
\(z=0,1\) are bitwise complements.  The coefficients on these slices need not be equal.

For a perfect matching \(M\) on \(2r\) inputs and parities
\((c_e)_{e\in M}\), the associated \emph{matching coset} is
\[
 \{{\mathbf{x}}\in\Ftwo^{2r}:x_u+x_v=c_{\{u,v\}}
       \text{ for every }\{u,v\}\in M\}.
\]

\begin{lemma}[A six-to-four contraction]
\label{lem:app-nd-six-to-four}
Suppose a standard \(\DihZ{2m}\) localization leaves exactly the distinguished
pair \(p\) and two residual monomial pairs, with support as in \cref{eq:app-nd-four-flat-support}.  Then either the six-input support is a single three-edge
matching coset, or a realizable \(I/X\)-contraction realizes a nonzero
quaternary of support size six or eight.  In the latter case, the output is
not a product of current-group monomial binaries and the procedure does not
re-enter a six-input state.
\end{lemma}

\begin{proof}
The explicit support in \cref{eq:app-nd-four-flat-support} is the Boolean
family exhaustively covered by
\cref{thm:cert-interface-nd-q6}.  That classification either identifies
the global matching coset or supplies a pair and parity \(\gamma\) with
injective projection.  Contracting it by the realizable parity-\(\gamma\)
kernel---\(I\) for \(\gamma=0\) and \(X\) for \(\gamma=1\)---gives one
nonzero internal extension per string,
so there is no cancellation and the output signature has the certified
six- or eight-string support.  Add this direct output and close its complete
factor batch.  A singular/unsafe factor, unary or odd factor, or a smaller
irreducible signature is handled by the stabilized procedure. A safe binary whose transfer class lies outside the
current group gives a successor with a strictly larger complete transfer group.  If the output itself
is irreducible, retain it as a lower-arity quaternary occurrence, stabilize its state, and
regenerate its complete oriented deck, and only then invoke the
equality-accessible quaternary boundary and eight-vertex reduction.  Since a product of two
nonsingular current-group monomial binaries has support size four, the
support-six/eight output cannot close as an unrecorded matching product.
This completes the reduction of the arity-six tensor through its
four-input output and factor batch.
\end{proof}

\begin{lemma}[Reduction of the endpoint cases]
\label{lem:app-nd-endpoint-type}
For every deleted pair \(p\), all endpoint combinations containing type
\(\mathsf Z\) or \(\mathsf P\) are resolved.  On the remaining branch, the
endpoint pair is \(\mathsf{C}/\mathsf{C}\), \cref{eq:app-nd-parallelogram} holds, and there is one
common residual matching.
\end{lemma}

\begin{proof}
The \(\mathsf{Z}/\mathsf{Z}\) entry gives \(f=0\).  Every \(\mathsf{Z}/\mathsf{P}\), \(\mathsf{Z}/\mathsf{C}\), \(\mathsf{P}/\mathsf{P}\),
\(\mathsf{P}/\mathsf{C}\), and transposed entry is covered by
\cref{lem:app-nd-exceptional-endpoints}.

It remains to take both types equal to \(\mathsf C\).  A matching change or two
different ordered factor classes invokes the trace cover
\cref{lem:app-nd-localization,lem:app-nd-bounded-consumer}.  After closing
common factors with the same projective class, two different variable edges give the six-string quaternary constructed above. Apply the lower-arity stabilized-quaternary route
given there.  For one common variable edge, failure of
\cref{eq:app-nd-parallelogram} is again a realizable-card disagreement.  If the
parallelogram holds, then, because \(n\ge6\), retain the varying edge and one
fixed residual edge together with \(p\). This gives the arity-six tensor
of \cref{eq:app-nd-four-flat-support}. By \cref{lem:app-nd-six-to-four},
any nonmatching quaternary output is processed through its full factor batch.
The sole support-level remaining case is a three-edge matching coset.
These alternatives exhaust \(\mathsf{C}/\mathsf{C}\).
\end{proof}

\subsubsection{Global support integration and the Reed--Muller alternative}
\label{proofsubsec:nd-global-support}

The \(\mathsf Z/\mathsf P/\mathsf C\) endpoint classification has resolved every case
containing \(\mathsf Z\) or \(\mathsf P\) and every
\(\mathsf{C}/\mathsf{C}\) family with a disagreement. The remaining case satisfies
\cref{eq:app-nd-parallelogram}, and the two sectors have the same residual
matching at every deleted pair.  These local translation spaces determine a self-dual code. A weight-two
codeword identifies a matched pair and then the global matching coset.
If there is no such codeword, the weight-four supports determine
\(RM(1,3)\). A realizable contraction with a minor proportional to
\(1-t^2\) excludes that remaining case.

\begin{lemma}[Endpoint-support saturation]
\label{lem:app-nd-endpoint-support-saturation}
Assume the standard \(\DihZ{2m}\) analysis is on the remaining endpoint-
type branch.  Then, for each deleted pair
\(p=\{i,j\}\), there is a fixed-edge matching translation space \(L_p\) on the
residual coordinates and endpoint differences \({\mathbf{d}}_{E,p},{\mathbf{d}}_{O,p}\notin L_p\)
such that the four endpoint slices have full support on affine cosets
\({\mathbf{a}}_{ab}+L_p\), with every coefficient on each such coset nonzero.  Moreover
\[
 M_p:=L_p+\Span_{\Ftwo}\{{\mathbf{d}}_{E,p}\}
     =L_p+\Span_{\Ftwo}\{{\mathbf{d}}_{O,p}\}
\]
is the support translation space of the equal- and unequal-\(p\) realizable cards. It is the
translation space \(M(\mathcal M_p)\) of the residual matching coset,
abbreviated \(M_p\), and hence \(M_p=M_p^\perp\).
\end{lemma}

\begin{proof}
On the remaining \(\mathsf{C}/\mathsf{C}\) branch,
\cref{lem:app-nd-sector-ruling} gives one fixed
nonzero monomial factor on every residual matching edge except the designated
variable edge.  Two samples with nonzero output and \cref{eq:app-nd-parallelogram} reconstruct
each slice by the four-card inversion used in
\cref{lem:app-nd-coefficient-line-factor}. Hence every string in the resulting
affine matching coset has a nonzero coefficient.  More explicitly, the two
complementary-axis endpoint matrices have disjoint monomial supports, so each
assignment to the remaining inputs is supplied by exactly one endpoint. Its coefficient is a
product of nonzero fixed monomial entries and a nonzero endpoint entry, with
no cancellation between endpoints. For
\({\mathbf{d}}_{E,p}={\mathbf{a}}_{00}+{\mathbf{a}}_{11}\) and
\({\mathbf{d}}_{O,p}={\mathbf{a}}_{01}+{\mathbf{a}}_{10}\), the parallelogram equation says
\({\mathbf{d}}_{E,p}+{\mathbf{d}}_{O,p}\in L_p\), which identifies the two displayed descriptions
of \(M_p\) and the four offsets.  Matchings belonging to different deleted
pairs are integrated into one common code by
\cref{lem:app-nd-support-integration}, rather than compared directly here.
\end{proof}

\phantomsection
\label{par:reed-muller-code}

For \(d\ge1\),
\[
 \operatorname{AGL}(d,2):=\Ftwo^d\rtimes\GL_d(\Ftwo)
\]
acts by \({\mathbf{x}}\mapsto A{\mathbf{x}}+{\mathbf{b}}\).  An \emph{affine relabelling} of inputs indexed by
\(\Ftwo^d\) is the corresponding permutation of those inputs.  Indexing the
eight coordinates by \(\Ftwo^3\) in lexicographic order, define the
first-order Reed--Muller code by
\[
 RM(1,3):=
 \{(\ell({\mathbf{t}}))_{{\mathbf{t}}\in\Ftwo^3}:
   \ell({\mathbf{t}})=a+u_1t_1+u_2t_2+u_3t_3,\ 
   a,u_1,u_2,u_3\in\Ftwo\}.
\]
Thus \(RM(1,3)\) consists of the sixteen truth tables of affine-linear
functions on \(\Ftwo^3\).  The natural action satisfies
\(\operatorname{AGL}(2,2)\cong S_4\).
\(\operatorname{AGL}(3,2)\) is two-transitive and preserves \(RM(1,3)\),
because affine precomposition preserves affine truth tables
\cite[Section~4.1]{Carlet2021}.

We now regard the support as a set of binary strings, using the projections
and affine operations of \cref{subsubsec:boolean-affine-tools}.  The
local matching equations will determine a self-dual binary code.  A
weight-two codeword identifies a global matched pair. If no such codeword
exists, the weight-four codewords identify the eight-coordinate Reed--Muller
code.  This reconstructs support only. Coefficients are treated afterward.

For a function
\(F:\Ftwo^d\to R\) into an abelian group, its additive derivative is
\[
 (\Delta_{\mathbf{a}}F)({\mathbf{x}}):=F({\mathbf{x}}+{\mathbf{a}})-F({\mathbf{x}}).
\]
The identities \(\Delta_{\mathbf{a}}\Delta_{\mathbf{b}}F=0\) for all \({\mathbf{a}},{\mathbf{b}}\) hold exactly when
\(F({\mathbf{x}})=F(0)+\ell({\mathbf{x}})\) for a group homomorphism
\(\ell:(\Ftwo^d,+)\to(R,+)\): the identities make every
\(\Delta_{\mathbf{a}}F\) constant, and the resulting
\(\ell({\mathbf{x}}):=F({\mathbf{x}})-F(0)\) is additive.

\begin{lemma}[Self-dual Steiner reconstruction]
\label{lem:self-dual-steiner-reconstruction}
Let \(D\le\Ftwo^n\) be self-dual, where \(n>4\), and suppose that \(D\)
has no codeword of weight two.  If every three distinct coordinates lie in the
support of a unique weight-four codeword of \(D\), then \(n=8\). After
identifying the coordinates with \(\Ftwo^3\), one has \(D=RM(1,3)\).
\end{lemma}

\begin{proof}
The weight-four supports form a \emph{Steiner quadruple system}: a family of
four-element blocks in which every three coordinates belong to exactly one
block.  Self-orthogonality
of \(D\) makes two blocks meet evenly. Uniqueness through triples then makes
the intersection of two distinct blocks either empty or a pair.

Fix a coordinate \(0\).  Define a binary operation \(+\) on the coordinates
by \(x+x=0\), \(x+0=x\), and, for distinct nonzero \(x,y\), let \(x+y\) be
the fourth point in the block through
\(0,x,y\).  This operation is commutative and every element has order two.
For the associativity check, define the auxiliary points \(u:=x+y\),
\(v:=y+z\), and \(w:=u+z\).  The symmetric
difference of the blocks \(\{0,x,y,u\}\) and \(\{0,y,z,v\}\) is the
weight-four codeword \(\{x,u,z,v\}\).  Symmetric difference with
\(\{0,u,z,w\}\) gives the block \(\{0,x,v,w\}\).  Uniqueness of the
block through \(0,x,v\) gives
\[
 (x+y)+z=w=x+v=x+(y+z).
\]
If an auxiliary point coincides---for example \(z=x+y\)---the same identity
reduces immediately to \(x+x=0\). Thus the displayed distinct-block
calculation and these collision cases cover all inputs.  The coordinates form an
elementary abelian group \(V\cong\Ftwo^r\), and \(n=2^r\).  If
\(\{a,b,c,d\}\) is any block, symmetric difference with the block through
\(0,a,b\) shows \(d=a+b+c\). Hence the blocks are exactly the affine
planes of \(V\).

Let \(B\) be the binary span of the block indicator vectors.  An affine plane is a translate of a two-dimensional
subspace of \(V\).  Orthogonality to every such plane is equivalent to
vanishing of every second additive derivative
\(\Delta_{\mathbf{a}}\Delta_{\mathbf{b}}F({\mathbf{x}})\). Hence \(B^\perp\) is exactly the space of truth
tables of affine-linear functions \(F:V\to\Ftwo\).  Thus
\(\dim B=2^r-r-1\).  Since \(B\subseteq D\) and
\(\dim D=2^{r-1}\), the inequality \(2^r-r-1\le2^{r-1}\), with \(r>2\),
forces \(r=3\). Equality gives \(D=B=RM(1,3)\).
\end{proof}

\begin{lemma}[Support integration]
\label{lem:app-nd-support-integration}
Assume the remaining endpoint-type case, so that
\cref{eq:app-nd-parallelogram} holds for every deleted pair.  Then
\(\supp(f)={\mathbf{a}}+D\) for a self-dual binary linear code \(D\le\Ftwo^n\).
Exactly one of the following support alternatives remains:
\begin{enumerate}
\item \(D\) is the matching translation space of one perfect matching on
      all \(n\) inputs, so \({\mathbf{a}}+D\) is one \emph{global matching coset};
\item \(n=8\) and, after relabelling the inputs, \(D=RM(1,3)\).
\end{enumerate}
The second alternative is incompatible with the standard \(\DihZ{2m}\) deck for \(m\ge3\).  Consequently the sole
remaining support is one global matching coset.
\end{lemma}

\begin{proof}
Apply \cref{lem:app-nd-endpoint-support-saturation}.  Translation by the
\(00\)-slice gives, for any fixed deleted pair \(p\),
\begin{equation*}
  {\mathbf{a}}+D,\qquad
  D=(\{00\}\times L_p)
    +\Span\{(01,{\mathbf{a}}_{00}+{\mathbf{a}}_{01}),
             (10,{\mathbf{a}}_{00}+{\mathbf{a}}_{10})\}.
\end{equation*}
The two displayed generators are independent modulo
\(\{00\}\times L_p\). Hence \(\dim D=n/2\).  For an arbitrary deleted
pair \(p=\{i,j\}\), define the equal-coordinate slice
\(H_p:=\{{\mathbf{x}}\in D:x_i=x_j\}\).  The endpoint saturation and
the parallelogram give
\[
  \pi_{\{i,j\}}:D\longrightarrow\Ftwo^2\ \text{surjective},\qquad
  \pi_{[n]\setminus\{i,j\}}(H_{\{i,j\}})=M_{\{i,j\}},
\]
where \(M_p=M_p^\perp\) is the residual matching translation space from
\cref{lem:app-nd-endpoint-support-saturation}.  To see
self-orthogonality, take arbitrary \({\mathbf{x}},{\mathbf{y}}\in D\).  Among the \(n>4\)
coordinate columns
\((x_\ell,y_\ell)\in\Ftwo^2\), two are equal. Relabel them \(i,j\).  Then
\({\mathbf{x}},{\mathbf{y}}\in H_p\), so their residual projections lie in \(M_p\) and have zero
mutual dot product.  Their contribution on the deleted pair is
\(x_i y_i+x_j y_j=0\).  Hence \({\mathbf{x}}\cdot {\mathbf{y}}=0\), proving
\(D\subseteq D^\perp\).  Since \(\dim D=n/2\), equality follows:
\(D=D^\perp\).

Suppose first that \({\mathbf{e}}_u+{\mathbf{e}}_v\in D\).  Self-duality implies
\(x_u=x_v\) for every \({\mathbf{x}}\in D\), so \(H_{\{u,v\}}=D\).  Put
\(E=[n]\setminus\{u,v\}\).  Endpoint saturation gives
\[
 \pi_E(D)=M_{\{u,v\}},
\]
the translation space of the residual perfect matching.  For
\({\mathbf{y}}\in M_{\{u,v\}}\), choose a lift \({\mathbf{x}}\in D\).  Subtracting
\(x_u({\mathbf{e}}_u+{\mathbf{e}}_v)\) gives the zero extension \(\iota_E({\mathbf{y}})\in D\).  Hence
\[
 D=\Span_{\Ftwo}\{{\mathbf{e}}_u+{\mathbf{e}}_v\}
   \oplus\iota_E(M_{\{u,v\}}),
\]
which is exactly the translation space of the global perfect matching
\(\{\{u,v\}\}\cup\mathcal M_{\{u,v\}}\).  This is a support identity
only: it does not yet assert a tensor factorization of the coefficients.

It remains that \(D\) has no weight-two codeword.  Fix \(i\ne j\) and put
\(E=[n]\setminus\{i,j\}\).  The projection
\(H_{\{i,j\}}\to\Ftwo^E\) is injective, because its only possible nonzero nullspace
is \({\mathbf{e}}_i+{\mathbf{e}}_j\).  The functional \({\mathbf{x}}\mapsto x_i+x_j\) is nonzero on \(D\),
since otherwise \({\mathbf{e}}_i+{\mathbf{e}}_j\in D^\perp=D\).  Thus
\(\dim H_{\{i,j\}}=n/2-1\), and endpoint saturation identifies its image with
\(M_{\{i,j\}}\).  For a third coordinate \(\ell\), the residual matching
pairs \(\ell\) with a unique \(q\).  The unique lift of
\({\mathbf{e}}_\ell+{\mathbf{e}}_q\in M_{\{i,j\}}\) cannot have \(00\) on \(i,j\), since that would
be a weight-two codeword.  It is therefore
\({\mathbf{e}}_i+{\mathbf{e}}_j+{\mathbf{e}}_\ell+{\mathbf{e}}_q\in D\).  Hence every three coordinates lie in a unique
weight-four support, and
\cref{lem:self-dual-steiner-reconstruction} gives \(n=8\) and
\(D=RM(1,3)\).

To exclude this residue, identify the eight coordinate positions with
\(\Ftwo^3\) and, after an affine relabelling, take the deleted pair to be
\(\{000,001\}\).  On its equal-coordinate slice the residual matching is
\[
 \{010,011\},\qquad \{100,101\},\qquad \{110,111\}.
\]
If \(y_1,y_2,y_3\) are the three logical matching bits, the common bit on
the deleted pair is
\[
 b=y_1+y_2+y_3.
\]
Contract the deleted pair by the realizable sector phase \(t\).  The projection
of the equal slice is injective, so every string in the residual support has one
internal lift and its coefficient is multiplied by \(t^b\).  Fixing
\(y_3\) and inspecting the \(2\times2\) face indexed by \(y_1,y_2\), the
phase exponents are \((0,1,1,0)\), or their complement.  Thus the face has
one of the raw forms
\[
 \begin{pmatrix}A&tB\\tC&D\end{pmatrix},
 \qquad
 \begin{pmatrix}tA&B\\C&tD\end{pmatrix},
 \qquad A,B,C,D\ne0.
\]
The realizable \(t=1\) card lies in the fixed Segre ruling, so \(AD=BC\).
Nonzero row and column scalings therefore give, for some
\(\varepsilon\in\Ftwo\), the normalized block in the original basis
\hypertarget{eq:app-nd-RM-minor}{}
\begin{equation*}
 \begin{aligned}
  H_t&=\kappa\diag(r_0,r_1)C_\varepsilon(t)\diag(s_0,s_1),\\
  C_0(t)&=\begin{pmatrix}1&t\\t&1\end{pmatrix},
  &C_1(t)&=\begin{pmatrix}t&1\\1&t\end{pmatrix},\\
  \det H_t&=(-1)^\varepsilon
      \kappa^2r_0r_1s_0s_1(1-t^2).
 \end{aligned}
\end{equation*}
with all scalar factors nonzero.  If an orientation
produces \(t^{-1}\), clear its nonzero monomial factor.  Since \(m\ge3\),
some realizable \(t\in\mu_m\setminus\{\pm1\}\) exists, and the displayed
nonzero minor contradicts the fixed Segre ruling.
\end{proof}

The support is therefore one global matching coset. To obtain a tensor
factorization, we still need to separate its coefficients. We first
handle the degenerate six-input coefficient tensors, then use two-copy
outputs to prove the coefficient product formula.

\subsubsection{The six-input coefficient calculation}
\label{proofsubsec:nd-six-port}

We now return to the matching-supported six-input coefficient cases retained
by the support analysis.  Trace-cover six-input disagreements have already been
reduced to quaternary and resolved. The only remaining case is the fixed
three-edge-matching coefficient pattern.  This part first handles
all degenerate coefficient cases and then analyzes the coefficient tensor with all eight entries nonzero.

For a quaternary contraction whose support is contained in two fixed residual
matching edges, write its logical coefficient matrix as
\[
 Q=\begin{pmatrix}a&b\\c&d\end{pmatrix}.
\]
We include the cases \(0<|\supp Q|<4\).  A \emph{support-matching disagreement} means that its support in the original variables uses a residual pairing different from the fixed pairing. This is
the combinatorial disagreement of the trace-cover reduction.  By contrast,
\emph{not \(G\)-matching} means that the signature has no tensor
factorization into current-group binaries.  These two notions are not
interchangeable.

For each directly realized quaternary, first add it with its input order, scalar, and reduction. Jointly add all its irreducible factors. Test all three \(2+2\)
flattenings and consider every exposed factor. Make a realizable endpoint
dressing. Invoke \cref{lem:finite-q4-dispatch} only after tensor
irreducibility, endpoint nondegeneracy, non-\(G\)-matching status, and both
crossed rank hypotheses have been verified.

\begin{lemma}[Reductions of six-input coefficient tensors]
\label{lem:nd-six-coefficient-dispatch}
Let \(h\) be a directly realized six-input signature whose support is exactly the
eight-string coset of a fixed three-edge matching.  Before a quotient of its logical coefficients is
formed, close every already constructed four- or six-input batch as follows:
\begin{enumerate}
\item a zero quaternary is converted into an exact factorization of its parent arity-six tensor by
      the proportional-slice identity below;
\item a support-matching disagreement is sent to
      \cref{lem:app-nd-localization,lem:app-nd-bounded-consumer};
\item for a quaternary on the two retained matching edges, test \(Q\): if
      \(\det Q=0\), factor it exactly. If \(\det Q\ne0\), use the
      quaternary reduction just described;
\item jointly close all outputs under irreducible factors and consider every unary,
      singular, unsafe, infinite-order, or safe binary factor whose transfer class lies outside \(G\);
\item if no preceding concrete output applies, retain \(h\) only when all
      eight logical coefficients are nonzero.
\end{enumerate}
A quaternary with \(0<|\supp Q|<4\) enters item~3.
Every noncurrent factor or nonmatching quaternary outcome gives a resolved leaf or
strict successor.  If a nonzero child quaternary factors into two current-group
binaries, complete that output, record its rank-one equation, and continue the
finite analysis of the same originating arity-six tensor.
\end{lemma}

\begin{proof}
Let \(u_0,u_1\) be the two standard-basis binary signatures supported on
the two allowed states of a selected matching edge.  Write
\[
 h=u_0\otimes H_0+u_1\otimes H_1.
\]
An allowed phase kernel has two nonzero coefficients
\(\alpha_t,\beta_t\), and its quaternary contraction is
\[
 q_t=\alpha_tH_0+\beta_tH_1,
 \qquad \alpha_t\beta_t\ne0.
\]
If \(q_t=0\), first suppose \(H_0=0\). Then \(H_1=0\), so \(h=0\) and its token is completed directly.  Otherwise \(H_0\ne0\) and
\[
 H_1=-\frac{\alpha_t}{\beta_t}H_0,
 \qquad
 h=\left(u_0-\frac{\alpha_t}{\beta_t}u_1\right)\otimes H_0.
\]
The zero quaternary output therefore gives an exact
binary--quaternary factorization of \(h\).  Close the complete
factor batch by \cref{thm:external-factor,lem:factor-saturation}.  A
nonbinary factor gives a quaternary or lower-\(\nu\) successor. Unary factors and unsafe binaries enter their established
reductions. A safe binary \(B\) with \([BX]\notin G\) strictly enlarges the complete group.  If
all irreducible factors are already safe current-group binaries, the parent is a
\(G\)-matching signature and its fixed token is completed.

Now let \(q_t\ne0\) have logical coefficient matrix \(Q\) on the two retained matching
edges.  If \(\det Q=0\), then \(Q\) has rank one and the quaternary signature has an
exact \(2+2\) factorization.  When \(0<|\supp Q|<4\), at least one
logical factor has a zero entry, so one exposed binary is singular.
If both exposed factors are safe current-group binaries, complete this quaternary output, retain the equation \(\det Q=0\), and resume the coefficient analysis
of the same parent arity-six tensor. The child factorization alone does not complete the
parent token.
If \(\det Q\ne0\), its fixed-matching flattening has rank two.  Add the
realizable card with its input order, scalar, and reduction.  If \(Q\) is
supported on exactly two entries, nonzero determinant forces
them to be opposite logical entries.  Realizable logical \(X\)-translations
then make the quaternary a nonzero generalized equality, and
\cref{lem:pure-ge} applies.  If exactly three entries are nonzero, realizable
logical flips put a nonzero opposite pair at the two endpoints, so the card is
endpoint-nondegenerate.  After factor closure and removal of every
crossed rank-one factorization, it is irreducible, non-\(G\)-matching, and
has both crossed ranks at least two.  The case with four nonzero logical coefficients is
endpoint-nondegenerate after the same endpoint choice.  Thus the
quaternary reduction just described is applicable. Invoke
\cref{lem:finite-q4-dispatch,thm:equality-q4-boundary,%
cor:p1-eight-vertex-reduced-interface} only after their hypotheses have been
verified.  A support-matching disagreement instead enters the typed
trace-cover reduction.  These routes also process every exposed factor by
\cref{lem:actual-binary-completeness}.

For the matching-supported arity-six tensor retained from
\cref{lem:app-nd-support-integration}, the support is already the complete
eight-string global-matching coset, so its eight logical coefficients are
nonzero. The case analysis processes each phase contraction produced below
and each output of a later mixed two-copy batch within the originating
fixed local queue.  We may therefore write its coefficients as \(c_{abc}\),
\(a,b,c\in\Ftwo\), and normalize \(c_{000}=1\).
\end{proof}

\begin{lemma}[One-coordinate six-input minor localization]
\label{lem:nd-q6-minor-localization}
Let \(q=(q_{ab\xi})_{a,b,\xi\in\Ftwo}\) be the logical coefficient tensor of
an arity-six tensor whose support is contained in a fixed three-edge matching.
Close every quaternary and factor batch produced below by the
quaternary reduction just described and the binary case analysis.  On any branch
with no resolved leaf or strict stable successor, fully close every factor
batch.  Each resulting arity-six tensor is then either zero, a \(G\)-matching signature, or has rank at most one in each of
\[
 a\mid(b,\xi),\qquad b\mid(a,\xi),\qquad \xi\mid(a,b)
\]
The first two alternatives also have all three ranks at most one.  Their
rank equations are recorded before their child tokens are completed, so
the conclusion holds at every allowed phase assignment in any later
two-copy batch to which this lemma is applied.
\end{lemma}

\begin{proof}
Consider \(a\mid(b,\xi)\).  Close the \(\xi\)-edge by every allowed phase,
obtaining the logical matrix pencil
\[
 A+tB
\]
in the \(a,b\) coordinates.  If \(\det(A+tB)\ne0\) at an allowed phase, the resulting quaternary signature enters the preceding nonzero-minor reduction.  If the
determinant vanishes for every \(t\in\mu_m\), it is a degree-two polynomial
with at least three roots and is therefore identically zero.

The two-dimensional rank-one-line classification is elementary.  Write
nonzero rank-one endpoints as \(A={\mathbf{u}}{\mathbf{v}}^{\mathsf T}\) and
\(B={\mathbf{x}}{\mathbf{y}}^{\mathsf T}\).  The mixed coefficient of
\(\det(A+tB)\) is, up to sign,
\(\det[{\mathbf{u}},{\mathbf{x}}]\det[{\mathbf{v}},{\mathbf{y}}]\).  Hence either \({\mathbf{u}},{\mathbf{x}}\) are proportional or
\({\mathbf{v}},{\mathbf{y}}\) are proportional. Zero or proportional endpoints satisfy the same
conclusion directly.  In the first case \(A,B\) have one common
\(a\)-factor, so the \(a\mid(b,\xi)\) flattening of \(q\) has rank at most
one.  In the second case they have one common \(b\)-factor, giving an exact
binary factor of \(q\). Take factor closure and apply the factorization reduction.
If closing this factorization gives a resolved leaf or strict stable
successor, the parent branch ends.  Otherwise all exposed factors are
current-group matching factors, so \(q\) itself is a \(G\)-matching signature and all three logical flattenings have rank at most one. Record
these equations before completing its child token.  Thus every branch
continuing the parent analysis has the stated \(a\)-rank equation.
Repeating the same argument after permuting \(a,b,\xi\) proves the other two
rank bounds.
\end{proof}

After \cref{lem:nd-six-coefficient-dispatch}, all eight logical
coefficients on the continuing branch are nonzero.  Retain their notation and
normalize the remaining global-matching six-input support to
\(X^{(1,2)}\otimes X^{(3,4)}\otimes X^{(5,6)}\), where the superscript records
the ordered pair of inputs on which the binary disequality tensor \(X\) acts.
For each selected pair of matching edges, use both realizable parity-zero and
parity-one kernels.  The parity-zero contraction on the first two edges
gives
\[
 c_{000}c_{111}=c_{001}c_{110}.
\]
The parity-one contractions on the three choices of edge pair give the
remaining equalities among
\[
 c_{001}c_{110},\qquad
 c_{010}c_{101},\qquad
 c_{011}c_{100}.
\]
Every nonzero determinant is handled by the quaternary reduction just described.  Hence, on the continuing branch, all four products are equal:
\begin{equation}
  c_{000}c_{111}
  =c_{001}c_{110}
  =c_{010}c_{101}
  =c_{011}c_{100}.
  \label{eq:app-nd-six-relations}
\end{equation}
Define
\[
  \alpha=\frac{c_{100}}{c_{000}},\qquad
  \beta=\frac{c_{010}}{c_{000}},\qquad
  \gamma=\frac{c_{001}}{c_{000}}.
\]
These three quantities are exposed by explicit factor cards.
For \(\alpha\), join inputs \(3\) and \(5\), the first inputs of the second
and third matching edges, by the realizable parity-zero kernel
\(M_0(1)=I\).  The remaining logical coefficient matrix, with the first matching edge
as its row edge, is
\[
 \begin{pmatrix}
 c_{000}&c_{011}\\
 c_{100}&c_{111}
 \end{pmatrix}.
\]
Its determinant vanishes by \cref{eq:app-nd-six-relations}, so exact factor
extraction exposes on the first matching edge a binary with ordered ratio
\(\alpha^{\pm1}\).  Cycling the three matching edges exposes
\(\beta^{\pm1}\) and \(\gamma^{\pm1}\) in the same way.  Process every exposed binary by \cref{lem:actual-binary-completeness}. Hence on the continuing
branch
\[
 \alpha,\beta,\gamma\in\mu_m.
\]

Fuse the fixed representatives with transfers
\(D(\alpha^{-1})\),
\(D(\beta^{-1})\), and
\(D(\gamma^{-1})\) onto the three oriented matching edges, and
divide by the recorded common scalar \(c_{000}\).  Equation
\cref{eq:app-nd-six-relations} then gives
\[
 N_r({\mathbf z})=
 \begin{cases}
  1,&\wt({\mathbf z})\le1,\\
  r,&\wt({\mathbf z})\ge2,
 \end{cases}
 \qquad {\mathbf z}\in\Ftwo^3,\qquad
 r:=\frac{c_{000}c_{110}}{c_{100}c_{010}}.
\]
It remains to prove that \(r\) is also a group phase.  On this normalized six-input contraction output, contract inputs \(1\) and \(3\), the first inputs
of the first and second matching edges, by the realizable parity-one kernel
\(M_1(u)=M_{\alpha=1}(a=u)\), \(u\in\mu_m\). In the displayed input
orientation this is
\[
 M_1(u)=\begin{pmatrix}0&1\\ u&0\end{pmatrix}.
\]
Write the bits on the six inputs as
\(x_1,\ldots,x_6\). The support matching means
\(x_2=1-x_1\), \(x_4=1-x_3\), and \(x_6=1-x_5\).
The kernel joining inputs \(1\) and \(3\) therefore imposes
\(x_1\oplus x_3=1\) and contributes the factor \(u^{x_1}\).
After this contraction, use \(x_2\) (with \(x_4=1-x_2\)) as the row bit
and \(x_5\) (with \(x_6=1-x_5\)) as the column bit.  Ordering the rows and
columns changes only the nonzero representative scalar or reverses
an input.
The resulting quaternary signature on inputs
\(2,4,5,6\), up to reversal and a nonzero representative scalar, is
\[
 \begin{pmatrix}1&r\\ u&ur\end{pmatrix}
 =\begin{pmatrix}1\\u\end{pmatrix}
  \begin{pmatrix}1&r\end{pmatrix}.
\]
This is an exact tensor factorization, and its two binary factors have
ordered ratios \(u^{\pm1}\) and \(r^{\pm1}\).  Expose them through
\cref{thm:external-factor,lem:factor-saturation} before testing the complete transfer group.  Process both factors by \cref{lem:actual-binary-completeness} within
the selected occurrence's fixed local queue.  Thus the coefficient
calculation continues only when both factors are safe with transfer classes in
\(\DihZ{2m}\).  On that branch,
\(r^{\pm1}\in\mu_m\), and hence \(r\in\mu_m\).  \hypertarget{eq:app-nd-Nr}{}
Thus the parameter in the displayed coefficient tensor \(N_r\) belongs to
\(\mu_m\).
Closing the third matching pair gives the quaternary whose logical
coefficient matrix is
\hypertarget{eq:app-nd-Qr}{}
\begin{equation*}
  Q_r=
  \begin{pmatrix}
    2&1+r\\
    1+r&2r
  \end{pmatrix},
  \qquad
  \det Q_r=-(r-1)^2.
\end{equation*}
Attach the supplied \(I\) through an \(X\)-edge at the second input of
each residual matching pair. These two realizable bit flips give a
quaternary supported on \(\{0000,0011,1100,1111\}\), with the same
logical coefficient matrix \(Q_r\). Its endpoint entries \(2,2r\) are
nonzero. For \(r\ne-1\), all four logical coefficients are nonzero, and
their cross-ratio is
\[
 \operatorname{cr}(Q_r)=\frac{(Q_r)_{00}(Q_r)_{11}}
                 {(Q_r)_{01}(Q_r)_{10}}
          =\frac{4r}{(1+r)^2}.
\]
If this support-four quaternary is affine- or local-affine-transformable,
\cref{lem:p1-eight-vertex-local-affine-collapse} reduces the latter case
to affine transformability. The diagonal normal form of
\cref{prop:p1-affine-eight-vertex-certificate} then gives an affine
reference signature on the same four-point support. On the two equal-input
pairs, a diagonal change of basis rescales the rows and columns of
\(Q_r\), preserving its cross-ratio. Clause~\textnormal{(ii)} of
\cref{thm:cert-interface-affine-q4-gadget-exits} therefore gives
\(\operatorname{cr}(Q_r)\in\{1,-1\}\).
The value \(+1\) is equivalent here to
\(\det Q_r=0\), hence to \(r=1\), which is the product case removed below.
Therefore a nonproduct affine signature has \(\operatorname{cr}(Q_r)=-1\), giving
\(r+r^{-1}=-6\).  Every other outcome is handled by the quaternary reduction.
This is a direct quaternary signature with its inherited input order and nonzero
representative scalar.  Add it and its complete batch of irreducible factors
before applying the quaternary reduction.  If \(r=1\), then
\(N_1\equiv1\), so the parent arity-six tensor itself is the product of its
three matching binaries. Complete the originating fixed token after the
binary case distinction.  If \(r\ne1\), the displayed fixed-matching minor is
nonzero, and the quaternary reduction of
\cref{lem:app-nd-bounded-consumer,lem:nd-six-coefficient-dispatch} applies.
For the exceptional value \(r=-1\), the equality exit is the
following explicit two-copy construction.  Define the arity-six signature
\[
 F_r(z_1,\ldots,z_6)=
 \prod_{j=1}^3\one[z_{2j-1}\oplus z_{2j}=1]\,N_r((z_1,z_3,z_5)),
\]
where the three displayed logical bits are the first inputs of the three
paired-disequality edges.  Take two copies, leave inputs \(1,2\) of each copy
external, and join corresponding inputs \(3,4,5,6\) by four realizable equality
links.  The resulting quaternary tensor is
\hypertarget{eq:app-nd-r-minus-one-wiring}{}
\begin{equation*}
 W_r(a,b,c,d)=\sum_{u,v,w,z\in\Ftwo}
 F_r(a,b,u,v,w,z)F_r(c,d,u,v,w,z).
\end{equation*}
Direct substitution into \(N_{-1}\) gives
\[
 W_{-1}(a,b,c,d)=4\,\one[(a,b,c,d)\in\{0101,1010\}].
\]
Applying a realizable \(X\)-translation on the second external input of each
copy sends these two strings to \(0000,1111\), so the output is
\(4(\Eq_4)\) in the displayed normalization.  This terminates by
\cref{cor:higher-equality-unary,lem:common-unary-exit}.  Otherwise the
affine cross-ratio test would force
\(r+r^{-1}=-6\), impossible for a root of unity.  Hence every nonproduct
case not already terminated enters
\cref{cor:p1-eight-vertex-reduced-interface}.

\subsubsection{Laurent polynomials}
\label{subsubsec:laurent-polynomials}
\label{subsec:laurent-gauge-conventions}

The minors below are polynomials in the phase variables. Reversals and
weighted kernels can also introduce negative exponents, so we use the
following Laurent-polynomial convention. For a commutative ring \(R\),
a Laurent polynomial is a finite sum
\[
 P({\boldsymbol\lambda})=\sum_{{\mathbf{u}}\in\mathbb Z^r}c_{\mathbf{u}}{\boldsymbol\lambda}^{\mathbf{u}},
 \qquad
 {\boldsymbol\lambda}^{\mathbf{u}}:=\prod_{j=1}^r\lambda_j^{u_j},
\]
with \(c_{\mathbf{u}}\in R\) and all but finitely many \(c_{\mathbf{u}}\) zero. Equivalently,
\(P\in R[\lambda_1^{\pm1},\ldots,\lambda_r^{\pm1}]\).  Coefficient
extraction is \([{\boldsymbol\lambda}^{\mathbf{u}}]P:=c_{\mathbf{u}}\).  When a
Boolean vector is used as an exponent, its entries are lifted from
\(\{0,1\}\) to \(\mathbb Z\), so exponent arithmetic is never modulo two.
Multiplication by a monomial with sufficiently large nonnegative
exponents clears all negative exponents. The coordinate degree in
\(\lambda_j\) of a nonzero \(P\) is
\[
 \deg_{\lambda_j}P
 :=\max\{u_j:c_{\mathbf{u}}\ne0\}-\min\{u_j:c_{\mathbf{u}}\ne0\}.
\]
By convention, \(\deg_{\lambda_j}0:=-\infty\). Consequently, every finite
upper degree bound used below includes the zero Laurent polynomial.

\begin{lemma}[Grid interpolation for bounded Laurent degree]
\label{lem:nd-grid-interpolation}
Let \(m\ge3\), let \(c_1,\ldots,c_r\in\mathbb C^\times\), and let
\(P(\lambda_1,\ldots,\lambda_r)\) be a Laurent polynomial such that,
after multiplying by a monomial, its degree in each variable is at most
two.  If \(P\) vanishes on the product of cosets
\(\prod_{j=1}^r c_j\mu_m\), then \(P\) is identically zero.
\end{lemma}

\begin{proof}
After the monomial shift, substitute \(\lambda_j=c_jt_j\).  Fix all
variables except the last one.  The result is a univariate polynomial of
degree at most two with at least three distinct roots \(t_j\in\mu_m\),
hence is zero.  Induction on the number of variables proves that every
coefficient polynomial vanishes.  The nonzero monomial used to clear the
Laurent exponents never affects a value on the stated cosets.
\end{proof}

\subsubsection{Polarized coefficient integration}
\label{proofsubsec:nd-polarization}

The remaining tensor has a fixed matching support and nonzero logical
coefficients.  We test their factorization by the following two-copy
construction, called \emph{polarization}.  Each phase assignment gives a realizable arity-six tensor.  Its rank-one equations give polynomial identities in the phases.
One set of minors makes the coefficient ratio constant within each of
two layers. A second minor equates the ratios between the layers.
Interpolation then gives a ratio independent of all other logical inputs,
which yields the coefficient product formula.

Let \(w:\Ftwo^s\to\mathbb C^\times\), \(s\ge3\), be the logical coefficient
tensor on the global matching, with every matching edge in its fixed
orientation.  Fix distinct logical coordinates \(i,k\) and take two equally
oriented copies.  Leave the entire \(i\)-edge of each copy external.  On the
two \(k\)-edges, join one pair of corresponding inputs by a realizable
equality link and leave the other two inputs external. After one recorded
\(X\)-dressing they form the third external matching edge and carry one
common logical bit \(\xi\).  For every \(\ell\notin\{i,k\}\), close the two
\(\ell\)-edges by one equality link and one phase-dressed equality link.
Unique monomial lifts give, after division by the nonzero representative scalar, the logical coefficient tensor of the direct six-input output
\begin{equation}
  q^{\,i,k}_{ab\xi}({\boldsymbol\lambda})
  =\sum_{\mathbf z}{\boldsymbol\lambda}^{\mathbf z} w_i(a,\xi,{\mathbf z})w_i(b,\xi,{\mathbf z}),
  \qquad
  {\boldsymbol\lambda}^{\mathbf z}=\prod_{\ell\notin\{i,k\}}\lambda_\ell^{z_\ell}.
  \label{eq:app-nd-polarized-card}
\end{equation}
Here \(w_i(a,\xi,{\mathbf z})\) means that coordinates \(i,k\) of \(w\) are
\(a,\xi\), respectively, and the other coordinates are ordered as \({\mathbf z}\).
The fixed realization scalars may replace \(\mu_m\) in each variable by a
nonzero coset \(c_\ell\mu_m\). \Cref{lem:nd-grid-interpolation} applies to these cosets.  Their Cartesian product is the \emph{polarization grid}.
Every grid coordinate is nonzero. \Cref{lem:nd-grid-interpolation} turns the
allowed phase evaluations into identities among the original coefficients.

\begin{lemma}[Polarized determinant]
\label{lem:app-nd-polarized-determinant}
Assume the global matching support has been fixed by
\cref{lem:app-nd-support-integration}, and use
\cref{lem:nd-q6-minor-localization} to retain every one-coordinate rank
equation across completed current-group factor batches.  Form every realizable two-copy output
\(q^{\,i,k}({\boldsymbol\lambda})\) in \cref{eq:app-nd-polarized-card}.  Suppose that,
after the concrete six-input case analysis has removed every violating batch,
each retained two-copy output has rank at most one in each of its three
one-logical-coordinate flattenings
\[
 a\mid(b,\xi),\qquad b\mid(a,\xi),\qquad \xi\mid(a,b).
\]
Any two-copy output violating this hypothesis is resolved before
this lemma is invoked
by \cref{lem:app-nd-bounded-consumer,lem:nd-six-coefficient-dispatch,%
lem:actual-binary-completeness}.  Then
\begin{equation}
  w({\mathbf z})=w(\zero^s)\prod_{i=1}^s\rho_i^{z_i}
  \label{eq:app-nd-coefficient-product}
\end{equation}
for nonzero scalars \(\rho_i\).  Consequently the signature is a tensor product
of binary signatures on the global perfect matching.  This is an exact
tensor identity. The individual factors are made available only through
\cref{thm:external-factor,lem:factor-saturation} and are then sent through
the binary case analysis.
\end{lemma}

\begin{proof}
Fix \(i\ne k\) and \(\xi\in\Ftwo\), and abbreviate
\[
 u_{\mathbf z}:=w_i(0,\xi,{\mathbf z}),\qquad v_{\mathbf z}:=w_i(1,\xi,{\mathbf z}).
\]
Rank at most one of the \(a\mid b\) face, including a zero output, gives the
determinant polynomial
\begin{equation*}
  \Delta_\xi({\boldsymbol\lambda})
  :=
  \left(\sum_{\mathbf z}{\boldsymbol\lambda}^{\mathbf z}u_{\mathbf z}^2\right)
  \left(\sum_{\mathbf z}{\boldsymbol\lambda}^{\mathbf z}v_{\mathbf z}^2\right)
  -
  \left(\sum_{\mathbf z}{\boldsymbol\lambda}^{\mathbf z}u_{\mathbf z}v_{\mathbf z}\right)^2.
\end{equation*}
This polynomial vanishes on the full product of phase cosets.  Its coordinate
degree is at most two, so \cref{lem:nd-grid-interpolation} makes it the zero
polynomial.  If \({\mathbf z},{\mathbf z}'\) are adjacent in the remaining \((s-2)\)-cube, the
coefficient of \({\boldsymbol\lambda}^{{\mathbf z}+{\mathbf z}'}\) is
\begin{equation*}
  (u_{\mathbf z}v_{{\mathbf z}'}-u_{{\mathbf z}'}v_{\mathbf z})^2.
\end{equation*}
All values of \(w\) are nonzero. Hence
\(w_i(1,\xi,{\mathbf z})/w_i(0,\xi,{\mathbf z})\) is independent of \({\mathbf z}\) for fixed \(\xi\).

To equate the ratios between the two \(\xi\)-layers, use an
\(a\mid\xi\) face of the same arity-six output.  Rank at most one of
\(a\mid(b,\xi)\) gives the genuine face minor
\begin{equation*}
  \Gamma({\boldsymbol\lambda})
  =q^{\,i,k}_{000}({\boldsymbol\lambda})q^{\,i,k}_{101}({\boldsymbol\lambda})
   -q^{\,i,k}_{001}({\boldsymbol\lambda})q^{\,i,k}_{100}({\boldsymbol\lambda})=0.
\end{equation*}
The polynomial \(\Gamma\) vanishes on the full product of phase cosets and
has coordinate degree at most two.  Hence
\cref{lem:nd-grid-interpolation} makes it identically zero.
At \({\boldsymbol\lambda}^{2{\mathbf z}}\), only the ordered pair \(({\mathbf z},{\mathbf z})\) contributes, and its
coefficient is
\begin{equation*}
  w_i(0,0,{\mathbf z})w_i(0,1,{\mathbf z})
  \bigl(w_i(0,0,{\mathbf z})w_i(1,1,{\mathbf z})
        -w_i(1,0,{\mathbf z})w_i(0,1,{\mathbf z})\bigr).
\end{equation*}
It vanishes, so the edge ratio in coordinate \(i\) is the same for
\(\xi=0\) and \(\xi=1\).  Combining this equality with the preceding
constancy over the remaining \((s-2)\)-coordinate cube shows that
\[
 \rho_i:=\frac{w(1,{\mathbf x}_{-i})}{w(0,{\mathbf x}_{-i})}
\]
is independent of every coordinate of \({\mathbf x}_{-i}\), where \({\mathbf x}_{-i}\) is the
ordered tuple of all coordinates except \(i\), and \(w(b,{\mathbf x}_{-i})\) inserts
\(b\) in coordinate \(i\).  Repeat for each \(i\).
Starting at \(\zero^s\) and flipping the nonzero coordinates of \({\mathbf z}\) in any order
then multiplies by the corresponding constant edge ratios, giving
\cref{eq:app-nd-coefficient-product}.  The vanishing two-face minors make
the result independent of the order of the flips.
\end{proof}

Together, the two minors make every logical-coordinate ratio constant
on the coefficient cube. The interpolation uses the full phase grid
specified above.

\subsubsection{Composition of reductions and occurrence-level descent}
\label{proofsubsec:nd-completion}

The reconstruction results supply an exact factorization when all
local comparisons continue. We now record how their gadget outputs and
factor reductions compose, and how processing further occurrences
decreases the occurrence measure.

Every contraction above uses fixed representatives. The only
analytic objects are slices, pencil endpoints, inverse evaluation formulas,
and coefficient comparisons.  The two-root choices are legal for
\(m\ge3\).  Unique-lift monomial splices are cancellation-free. Phase
traces and two-copy sums may cancel and are handled by their explicit zero-output analysis or by grid interpolation.
Realizable four- or six-input outputs therefore have the direct-gadget reduction direction.  In
contrast, \cref{lem:app-nd-coefficient-line-factor,%
lem:app-nd-support-integration,lem:app-nd-polarized-determinant} use analytic
reconstruction only to prove tensor identities. Their proper factors are
added through \cref{thm:external-factor,lem:factor-saturation} with the Turing reduction supplied by the factor theorem recorded.  All bounded
outputs have size \(O(\arity(f))\).  Their
complete concrete batches are resolved by
\cref{lem:app-nd-bounded-consumer,lem:nd-six-coefficient-dispatch,%
def:relative-stratum-protocol,lem:matching-synthesis-measure}. In particular,
no nonmatching arity-six tensor is recursively re-entered.

For relative worklist invocations, let \(\mathcal W\) be the finite
multiset of unresolved irreducible occurrences in the retained factorization trees. Distinct occurrences of the same normalized signature are counted separately.  For
each unresolved occurrence \(u\), with underlying irreducible
signature \(h_u\), let \(A\) denote the current coordinate matrix and put
\(H:=\overline G(A,\cF)\).  Fix the
occurrence's own finite card/factor universe
\[
 \mathscr U^{\mathrm{fac}}(A,h_u,H)
 =Q_{u,\mathrm{done}}\sqcup Q_u
\]
when that occurrence first enters the worklist.  Let \(\Phi_u\) be its
stored factor potential.  Define
\[
 \Theta_{\mathrm{MD}}(\mathcal W)
 =
 \{\!\{
   (\arity(h_u),|Q_u|,\Phi_u):u\in\mathcal W
 \}\!\},
\]
using strict lexicographic order on each triple and its well-founded multiset
extension, with the conventions of
\cref{subsubsec:worklist-termination}.

Certifying an occurrence removes its triple.  If the selected signature
itself factors, its triple is replaced by triples having strictly smaller
first coordinate.  If processing one child-card index exposes lower-arity
occurrences while the parent remains unresolved, that index is first completed
or its factor potential strictly decreases. Hence the old parent triple is
replaced by a smaller parent triple together with child triples of smaller
arity.  This also decreases the multiset measure.  The bounded four- or six-input reductions
create no fresh unresolved same-arity occurrence: their complete factorization trees are processed inside the originating \(Q_u\).

\begin{lemma}[Relative further applications]
\label{lem:nd-relative-reentry}
Let \(h\) be an irreducible occurrence of even arity in the current
reduction data \((\cF,\DihZ{2m})\) in the standard \(\DihZ{2m}\) coordinates, and suppose
\(\operatorname{RelStable}(\cF,\DihZ{2m};h)\), under the standing
assumption \(m\ge3\).  Every invocation of deck stability of a minimum-arity irreducible signature
in the preceding active-signature argument remains valid with \(h\) as the
selected occurrence and with the other local hypotheses unchanged.
In particular, this applies to
\cref{lem:app-nd-sector-ruling,lem:app-nd-endpoint-type,%
lem:app-nd-support-integration,lem:nd-six-coefficient-dispatch,%
lem:nd-q6-minor-localization,lem:app-nd-polarized-determinant}.
Specifically, every
surviving nonzero proper card is a \(\DihZ{2m}\)-matching signature, and
every lower-arity nonbinary occurrence exposed during its processing is
handled earlier in the occurrence induction.  The chosen representatives,
current group and fixed local queue are those of \(h\).  These are the
replacements for the minimum-arity argument in the preceding active-signature
proof. All its other hypotheses and reduction directions are retained.
\end{lemma}

\begin{proof}
The sector-ruling, endpoint, support, six-input, and polarization arguments
use minimum arity only to exclude a surviving proper nonmatching card or
nonbinary factor.  Under the present hypothesis each such object has
already produced a resolved leaf or strict stable successor, or has been
certified as a \(\DihZ{2m}\)-matching signature.  All other inputs---the
fixed representatives, normalized group, even arity, and tensor
irreducibility of \(h\)---are unchanged.  Thus every local construction and
reduction used in the active-signature proof remains applicable.
\end{proof}

\subsubsection{Weighted kernels and arity-four reductions}

We now return to the marked coordinates of
\cref{lem:marked-sector-dictionary}. The following support calculation
applies to both displayed marked forms. Its \(\DihXone{2m}\) case
supplies the arity-four step for the weighted argument.

\begin{lemma}[Supports of marked quaternaries]
\label{lem:marked-q4-support-atlas}
Let \(q\ne0\) be a retained quaternary occurrence in a current
\(\DihY{2m}\)-state (\(m\ge3\)) or \(\DihXone{2m}\)-state (even
\(m\ge4\)).  Suppose its complete oriented deck has been
regenerated and every member of that deck is zero or a safe current-group
binary.  Let
\(\tau=Y\) or \((X,1)\), respectively, and define the
sector-coordinate signature
\[
  q^{(\tau)}:=(U_\tau^{-1})^{\otimes4}q.
\]
After realizable sector-\(X\) dressings, for some
\(J\subseteq\{1,2,3\}\), nonzero \(a,h\), and nonzero
\(b_\ell,d_\ell\) (\(\ell\in J\)),
\begin{equation}
 q^{(\tau)}
 =a\delta_{\zero^4}+h\delta_{\one^4}
  +\sum_{\ell\in J}
   \bigl(b_\ell\delta_{{\mathbf p}_\ell}
        +d_\ell\delta_{\overline{\mathbf p}_\ell}\bigr),
 \label{eq:marked-q4-complement-form}
\end{equation}
where \({\mathbf p}_1,{\mathbf p}_2,{\mathbf p}_3\) represent the three complementary pairs of
weight-two strings.  The assertion includes all six contracted pairs, both
orientations, all zero-card patterns, and the fixed representative scalars.
\end{lemma}

\begin{proof}
For an ordered input pair with residual slices \((A,B,C,D_{11})\), the complete
card pencils, up to their nonzero representative scalars, are
\begin{equation*}
\begin{array}{ccc}
\toprule
 &E(t)&O(t)\\ \midrule
 \DihY{2m}&A+tD_{11}&tB+C\\
 \DihXone{2m}&A+\delta tD_{11}&tB+\delta C,
 \quad \delta=-\eta^{-1},\\
\bottomrule
\end{array}
\qquad t\in\mu_m.
\end{equation*}
Their effective kernels are \(D(t),XD(t)\) for \(\DihY{2m}\), and
\[
 D(\delta)D(t)=D(\delta t),
 \qquad
 D(\delta)XD(t)=\begin{pmatrix}0&t\\ \delta&0\end{pmatrix},
\]
for \(\DihXone{2m}\).  The diagonal kernels are symmetric.  Define the
off-diagonal kernel
\(K^{\rm off}(t):=\left(\begin{smallmatrix}0&t\\\delta&0\end{smallmatrix}\right)\).
It satisfies
\begin{equation}
 K^{\rm off}(t)^{\mathsf T}
 =\frac{t}{\delta}K^{\rm off}(\delta^2/t),
 \qquad \delta^2/t\in\mu_m;
 \label{eq:marked-A-card-reversal}
\end{equation}
so reversal permutes the labels and restores a nonzero scalar. The
\(\DihY{2m}\) formula is \(\delta=1\).

With \(\beta_\tau\) fixed above, three
phases imply that
each two-coordinate pencil is zero, lies on a fixed line with ordered
ratio in \(\mathscr R_\tau\), or has its endpoints on the two
complementary coordinate axes.  Indeed, a zero makes the endpoints
proportional.  Otherwise the quotient \(\phi(t)\) of the two independent
affine coordinates is injective on \(\mu_m\) and takes its values in the
fixed coset \(\beta_\tau\mu_m\).  Hence
\(\psi(t)=\beta_\tau^{-1}\phi(t)\) permutes \(\mu_m\), and
\cref{lem:nd-root-mobius} gives
\(\psi(t)=\lambda t\) or \(\lambda/t\).  Cross-multiplication leaves the
opposite coordinate axes.  Multiplication by \(\beta_\tau^{-1}\) is only an
analytic scaling of one endpoint coordinate.   Formula \cref{eq:marked-A-card-reversal} covers the opposite
orientation with its nonzero scalar.

Apply this three-type pencil classification to both card columns for
each of the six deleted pairs.  After analytic quotient normalization, the
Boolean support constraints are exactly those enumerated in
\cref{thm:cert-interface-equality-q4}\textnormal{(ii)}: every nonempty solution consists of
one total parity together with a nonempty subset of its four complementary
string pairs.  Quotient normalization changes no zero pattern, so the same
exact enumeration applies in marked coordinates.  Choose a nonzero
complementary pair and use realizable sector-\(X\) dressings, justified by
\cref{eq:marked-sector-fusion}, to send it to \(0000,1111\).  The other
pairs of nonzero entries are then among the three complementary weight-two blocks, giving
\cref{eq:marked-q4-complement-form}.  This includes both orientations and
every zero-card pattern. The resulting support description is in sector coordinates.
\end{proof}

In \cref{eq:marked-q4-complement-form}, an included complementary
weight-two block has \(b_\ell d_\ell\ne0\). All other blocks are zero.  Ratios below use the
displayed order of the two nonzero binary entries.

\begin{lemma}[The \(\DihXone{2m}\) quaternary coset]
\label{lem:marked-A-q4}
Let \(m\ge4\) be even, let \(G(\cF)=\DihXone{2m}\), and let
\(0\ne q\) be a retained quaternary occurrence satisfying
\[
 \operatorname{RelStable}(\cF,\DihXone{2m};q).
\]
Suppose every member of the complete oriented dressed deck of \(q\) is
zero or a safe \(\DihXone{2m}\)-binary.  Then the quaternary analysis yields an established
resolved complexity leaf or a successor with a strictly larger complete transfer group, a genuine factorization of
\(q\) into two binary signatures whose complete factor batch is retained,
or a realizable endpoint-nondegenerate eight-vertex signature in the current coordinates. In this last alternative the original
sector-coordinate signature also satisfies
\[
 (U_{X,1}^{-1})^{\otimes4}q\in\cP.
\]
\end{lemma}

\begin{proof}
Apply \cref{lem:marked-q4-support-atlas} and fix the resulting realizable
sector-\(X\) dressings.  In sector coordinates those dressings are
invertible monomial transformations, so they preserve membership in
\(\cP\), reducibility, all flattening ranks, and exact factorization.  We may therefore work with
\cref{eq:marked-q4-complement-form} and undo the dressings afterward.

Recall the phase scalar \(\delta\) and ratio coset \(\mathscr R_{X,1}\):
\[
 \delta=-\eta^{-1},\qquad
 \mathscr R_{X,1}=\delta^{-1}\mu_m=\eta\mu_m.
\]
Since \(\eta\notin\mu_m\) and \(m\) is even,
\begin{equation*}
 \mathscr R_{X,1}\cap\mu_m=\varnothing,
 \qquad \mathscr R_{X,1}^{-1}=\mathscr R_{X,1}.
\end{equation*}
By \cref{lem:marked-sector-dictionary}, every safe sector binary has ratio
in \(\mathscr R_{X,1}\).  In \cref{eq:marked-q4-complement-form}, denote the endpoint
pair by \((a,h)\) and an active central pair by \((b_i,d_i)\).

For one active block, the corresponding diagonal pencil is
\begin{equation*}
 \diag(a+\delta t d_i,\ b_i+\delta t h),
 \qquad t\in\mu_m.
\end{equation*}
The root-pencil type classification in
\cref{lem:marked-q4-support-atlas} excludes a
sample with exactly one vanishing coordinate and forces
\begin{equation*}
 (b_i,d_i)=(a\rho_i,h/\rho_i),
 \qquad \rho_i\in\mathscr R_{X,1},
 \qquad b_i d_i=ah,
\end{equation*}
whereas an inactive block gives the diagonal card pencil
\[
 \diag(a,\delta t h),\qquad t\in\mu_m.
\]
Its ordered ratio \(\delta t h/a\) belongs to
\(\mathscr R_{X,1}=\delta^{-1}\mu_m\).  Since
\(\delta^2=\eta^{-2}\in\mu_m\), this is equivalent to
\[
 h/a\in\delta^{-2}\mu_m=\mu_m.
\]

If two blocks \(j,k\) are active, the crossing deleted pair and the realizable
reflections
\[
 \begin{pmatrix}0&t\\ \delta&0\end{pmatrix},
 \qquad t\in\mu_m
\]
give, up to common endpoint factors,
\[
 t a\rho_j+\delta h/\rho_k,
 \qquad
 t a\rho_k+\delta h/\rho_j.
\]
Writing
\[
 L(t)=ta+\frac{\delta h}{\rho_j\rho_k},
\]
these entries are \(\rho_jL(t)\) and \(\rho_kL(t)\).  Every nonzero sample
therefore has ratio
\(\rho_j/\rho_k\in\mu_m\), because both \(\rho_j,\rho_k\) lie in the same
coset \(\mathscr R_{X,1}\).  This ratio is outside the coset
\(\mathscr R_{X,1}\).  The nonzero affine function \(L(t)\) vanishes for at most
one phase, so another allowed phase gives a nonsingular binary whose transfer class lies outside the current group.
Complete binary closure by \cref{lem:actual-binary-completeness} gives a
resolved complexity leaf or a successor with a strictly larger complete transfer group, contradicting a remaining
deck-stable branch.

Thus at most one block is active.  If exactly one block \(i\) is active,
then \(ah-b_id_i=0\), giving a genuine \(2+2\) factorization exposed by
\cref{thm:external-factor,lem:factor-saturation}.  Its two binaries are
then considered individually: a safe binary \(B\) with \([BX]\notin G\) gives a successor with a strictly larger complete transfer group,
and unsafe binary factors reach their established
complexity conclusions.

With no active block, let \(\lambda=h/a\in\mu_m\).  Before any further
rotation, the sector-coordinate tensor is already
\[
 a\delta_{0000}+h\delta_{1111},
\]
a generalized equality and hence a member of \(\cP\).  Undoing the fixed
sector-\(X\) dressings preserves \(\cP\), proving
\[
 (U_{X,1}^{-1})^{\otimes4}q\in\cP.
\]
Moreover \(\eta^2/\lambda\in\mu_m\), so it is the product of four available
phase parameters.  Realizable rotations with this product of input phases give the
sector tensor
\[
 a\bigl(\delta_{0000}+\eta^2\delta_{1111}\bigr).
\]
Since \(s_0^4=\eta^{-2}\),
\begin{align*}
 U_{X,1}^{\otimes4}
 a\bigl(\delta_{0000}+\eta^2\delta_{1111}\bigr)
 &=
 a\bigl((1,1)^{\otimes4}+(1,-1)^{\otimes4}\bigr)\\
 &=2a\,\one[\wt({\mathbf{x}})\text{ is even}].
\end{align*}
This is a current-coordinate endpoint-nondegenerate eight-vertex signature.  The rotations are
realizable by \cref{eq:marked-sector-fusion}. \(U_{X,1}\) only verifies
the identity.
\end{proof}

To apply \cref{lem:marked-A-q4} to the selected quaternary occurrence
or a quaternary output of a higher-arity construction, first add it with its input order,
scalar, and reduction, jointly add all its irreducible factors, consider the complete transfer group of the resulting set, and regenerate its complete oriented one-copy deck.  Test all
three \(2+2\) flattenings.  Every rank-one flattening is retained as
an exact factor identity.  Unary factors and unsafe binaries
reach their established reductions, and a safe binary whose transfer class lies outside
\(\DihXone{2m}\) gives a successor with a strictly larger complete transfer group.  If all exposed binary factors
are safe current-group binaries, the child quaternary is a \(\DihXone{2m}\)-matching signature and its originating fixed token is completed. If the quaternary itself is
the selected irreducible occurrence, such a factorization contradicts its
irreducibility.

On the remaining branch the quaternary is irreducible, every member of its complete deck
is zero or a safe binary whose transfer class lies in \(\DihXone{2m}\), and the closure just performed gives
\[
 \operatorname{RelStable}(\cF,\DihXone{2m};q)
\]
in the sense of \cref{def:occurrence-relative-deck-stability}.  For a later
occurrence, processing of its fixed local queue and prior processing of every
lower-arity occurrence give this certificate directly.  Apply \cref{lem:marked-A-q4}, which starts from the support in sector
coordinates given by \cref{lem:marked-q4-support-atlas} and constructs
the required eight-vertex signature in current coordinates.

In the no-active-block branch the output signature is
\(2a\,\one[\wt({\mathbf{x}})\text{ is even}]\), with \(a\ne0\).
It is irreducible: the GE signature used to construct
it has rank two across every nontrivial partition, and the invertible
transformations at individual inputs preserve these ranks.  The output is realizable,
has nonzero endpoints, and is an eight-vertex signature.  Apply the already
proved \cref{thm:p1-quaternary-kholant-dichotomy} to the signature set. This gives a hard or tractable case satisfying \(\ReTract(K\cF)\) for the
entire current signature set, with the reduction supplied by realization. The procedure is the
\emph{marked-coordinate reduction}.
\phantomsection\label{par:marked-coordinate-reduction}

\subsubsection{Gauges and product weights}
\label{subsubsec:gauge-product-weights}

An \emph{input-wise gauge} is an invertible diagonal rescaling on specified
tensor inputs. On a binary flattening it gives invertible diagonal row or
column scaling.  Its use as a gadget requires a separate realization.  A genuine character of
\((\Ftwo^r,+)\) has the form
\({\mathbf z}\mapsto\prod_j\epsilon_j^{z_j}\) with
\(\epsilon_j\in\{1,-1\}\).  More generally,
\[
 {\mathbf z}\longmapsto\prod_j\gamma_j^{z_j},
 \qquad \gamma_j\in\mathbb C^\times,
\]
is a \emph{product weight}. It is a
character exactly when every \(\gamma_j^2=1\).

For the remainder of the \(\DihXone{2m}\)-sector, fix
\[
 \delta:=-\eta^{-1},
\]
with the input-wise gauge defined above. In particular,
\((D(\delta)^{(i)}h)({\mathbf{x}})=\delta^{x_i}h({\mathbf{x}})\).

To compare coefficients across several outputs, we need the kernel
weights to act on the same parent inputs. Call a finite family of
marked-sector contraction outputs a \emph{gauge-coherent comparison batch} if its members use the same ordered contraction
multigraph, the same orientations on all contracted edges, and the same
external-input order, and differ only in the chosen phase or product at
designated kernels.  For a set \(\mathscr S\) of parent inputs write
\[
 D(\delta)^{(\mathscr S)}:=\prod_{i\in\mathscr S}D(\delta)^{(i)},
 \qquad
 (D(\delta)^{(\mathscr S)}h)({\mathbf{x}})
   =\delta^{\sum_{i\in\mathscr S}x_i}h({\mathbf{x}}).
\]
The embedded single-input operators commute.  The set \(\mathscr S\), and
hence the gauged parent \(D(\delta)^{(\mathscr S)}h\), is fixed throughout
one comparison batch.  Different ordered contraction graphs may yield
different sets.

\begin{lemma}[Explicit weighted-gauge transport]
\label{lem:marked-A-gauge-transport}
Let \(m\ge4\) be even and use the fixed
\(\delta=-\eta^{-1}\), where \(\eta\in\mu_{2m}\setminus\mu_m\), and the
diagonal gauge matrix \(D(\delta)\) above.  In the analytic \(\DihXone{2m}\)-sector the
effective diagonal and reflection kernels are
\begin{equation*}
 K^0(t)=D(\delta) D(t)=\diag(1,\delta t),
 \qquad
 K^1(t)=D(\delta) XD(t)=\begin{pmatrix}0&t\\ \delta&0\end{pmatrix},
 \qquad t\in \mu_m.
\end{equation*}
For every binary kernel \(C\) and every two-input slice \(g\) on ordered
inputs \((i,j)\), pointwise in all remaining variables, every occurrence of
\(D(\delta)\) can be moved from the kernel to the corresponding endpoint of
the parent tensor:
\begin{equation}
 \langle D(\delta) C,g\rangle_{\mathrm{Frob}}
   =\langle C,D(\delta)^{(i)}g\rangle_{\mathrm{Frob}},
 \qquad
 \langle C D(\delta),g\rangle_{\mathrm{Frob}}
   =\langle C,D(\delta)^{(j)}g\rangle_{\mathrm{Frob}},
 \label{eq:marked-A-gauge-push}
\end{equation}
where \(i\) and \(j\) are the first and second ordered inputs and
\(D(\delta)^{(i)}\) means multiplication by \(\delta^{x_i}\).  Consequently
for every gauge-coherent comparison batch there is one fixed set
\(\mathscr S\),
independent of the phase and product choices, such that the entire batch
is exactly the corresponding standard \(\DihZ{2m}\) batch on the one
locally gauged parent \(D(\delta)^{(\mathscr S)}h\), followed by invertible
row/column gauges and the substitutions
\begin{equation}
 \widehat t=c\,t^{\varepsilon},\qquad
 c\in\{1,\delta,\delta^{-1}\},\quad \varepsilon\in\{1,-1\}.
 \label{eq:marked-A-gauge-phase-substitution}
\end{equation}
In particular:
\begin{enumerate}
\item support, zero or incomplete-support status, matching translation spaces, all flattening ranks,
      and exact factorization are unchanged by the gauges;
\item whenever both product coefficients on a common edge are nonzero and their
      trace is not identically zero, its trace has the form
      \(\kappa(1+\theta t^{\varepsilon})\) with
      \(\kappa\theta\ne0\) and \(\varepsilon\in\{-1,1\}\). A one-term
      trace has no forbidden phase, while an identically-zero product is a
      case of the zero-output or incomplete-support analysis, according to the other product.  Whenever
      the corresponding selected minor is
      defined, its transported value is a nonzero monomial in \(\delta,t\)
      times the standard selected determinant;
\item reversal sends \(t\) to \(\delta^2/t\in \mu_m\), so both orientations
      use the same finite phase set, and the ordered ratios of all safe sector binaries lie in the
      single coset \(\delta^{-1}\mu_m\).
\item the four samples
      \(E(t_0),E(t_1),O(t_0),O(t_1)\) at one fixed ordered deleted pair,
      the two products of one fixed trace-cover graph, the phase family
      producing a quaternary from an arity-six tensor, and the two-copy outputs on the complete polarization grid
      are gauge-coherent.  Hence every joint multi-card comparison in the
      transported proof is made on one common locally gauged parent.
\end{enumerate}
\end{lemma}

\begin{proof}
The two displayed kernel identities are direct multiplication.  For the
first identity in \cref{eq:marked-A-gauge-push}, expand the Frobenius pairing:
\(\sum_{a,b}\delta^aK(a,b)g(a,b)\). The second is identical with the
factor \(\delta^b\).  Transposing or reversing an ordered kernel merely
switches \(i\) and \(j\) and contributes the nonzero scalar in
  \cref{eq:marked-A-card-reversal}.  Induction over the contraction edges
  therefore pushes all \(D(\delta)\)'s to vertex inputs.  For a fixed
  ordered contraction graph, the endpoint at which each \(D(\delta)\) lands is
  determined only by the ordered edge, not by its phase or by the choice of
  diagonal versus reflection product.  Hence every member of a
  gauge-coherent batch has the same parent-input set \(\mathscr S\).  Reversed cards
  are first put in the common orientation using
  \cref{eq:marked-A-card-reversal}, which only changes the scalar
  and replaces the phase by \(\delta^2/t\).  On a matching product the
  remaining external gauges are row and column gauges of its binary
  factors, so they preserve support, nonsingularity, ranks, and exact
  tensor-factor identities.  Unrelated ordered contraction graphs may induce
  different sets \(\mathscr S\), but their raw coefficients are never compared:
  support assertions are gauge-invariant, and coefficient identities pull
  back with their explicit nonzero product-weight factors.

Pushing the entire gauge to the parent leaves the phase unchanged, so
\(c=1\).  Alternatively, keeping the raw parent uses
\(K^0(t)=D(\delta t)\) and \(K^1(t)=\delta XD(t/\delta)\), giving
\(c=\delta\) and \(c=\delta^{-1}\), respectively.  These are the
substitutions in \cref{eq:marked-A-gauge-phase-substitution}. A reversed reflection has
\(\delta^2/t\), and \(\delta^2=\eta^{-2}\in \mu_m\).  A trace of two monomial
factors is a sum of two nonzero monomials.  If the two terms have the same
exponent and cancel, that product is identically zero (the full card may
consequently be zero or partially supported). Otherwise, after
extracting one term it is \(\kappa(1+\theta t)\) with
\(\kappa\theta\ne0\) (or a one-term trace).  A unique-lift
fusion multiplies a selected minor by the nonzero entry and orientation
scalars. A row/column gauge multiplies a determinant by the product of its
diagonal entries.  This proves the second item and also shows that a zero
or partially supported card is zero or partially supported before and after transport.
Finally, \(\delta^{-1}\mu_m=\eta \mu_m\) is disjoint from \(\mu_m\) and is invariant
under inversion because \(\eta^{-2}\in \mu_m\).  This proves the ratio and
orientation assertions.
\end{proof}

\paragraph{The six-input argument with weighted kernels.}
\label{par:marked-A-six-base}
The weighted kernels also preserve the support argument of
\cref{lem:nd-six-prime-base}. We record this observation to explain the
support part of the transport below. The proof of
\cref{thm:marked-A-rigidity} uses the full reconstruction in
\cref{lem:marked-A-primitive-transport}.  Work analytically with
\(h^{(X,1)}=(U_{X,1}^{-1})^{\otimes6}h\).
The exact sector-ruling argument uses only monomial supports and
Segre factorization.  In its individual binary pencils, the ratios of safe sector binaries
lie in \(\delta^{-1}\mu_m\). Dividing that one ratio by
\(\delta^{-1}\) gives the same \(\mathsf Z/\mathsf P/\mathsf C\) support classification.
This is a separate analytic calculation for each pencil, not a common
input gauge or an available binary.  In the proof of \cref{lem:nd-six-prime-base}, replace the
parity kernels \(I,X\) by the realizable effective kernels
\[
 D(\delta) D(1)=\begin{pmatrix}1&0\\0&\delta\end{pmatrix},
 \qquad
 D(\delta) XD(1)=\begin{pmatrix}0&1\\\delta&0\end{pmatrix}.
\]
They have the same respective supports and nonzero entries.  Every
cross-pair lift in the \(\mathsf{P}/\mathsf{P}\) and \(\mathsf{P}/\mathsf{C}\) arguments and every
injective lift in the support classification of
\cref{prop:cert-nd-q6-support-lowering} therefore acquires only a nonzero
weight. All support counts and projection obstructions are unchanged.
A matching coset still has an identically zero opposite-parity sector.
The resulting exact factorization of \(h^{(X,1)}\) pulls back by
the tensor powers of \(U_{X,1}\) to an exact
factorization of the signature \(h\) in the original basis, with its nonzero scalar.  

\begin{lemma}[Transport of contractions and reconstruction in the \(\DihXone{2m}\) sector]
\label{lem:marked-A-primitive-transport}
Let
\[
 \Stable(\cF,\DihXone{2m},f_0),\qquad \Even(\cF),
 \qquad m\ge4\text{ even}.
\]
Use the phase group \(\mu_m\) and recall the fixed scalar, gauge matrix,
and ratio coset
\[
 \delta=-\eta^{-1},\qquad
 D(\delta)=\diag(1,\delta),\qquad \mathscr R_{X,1}=\delta^{-1}\mu_m.
\]
Let \(h\in\Pi(\cF)\) be a selected unresolved nonbinary irreducible signature
occurrence of even arity at least six.  Regenerate and close its complete
realizable oriented deck over the current signature set.  Suppose that this
occurrence is relatively deck-stable and has produced neither a resolved
complexity leaf nor a strict stable successor. In particular, all lower-arity
irreducible occurrences exposed by its proper cards or factors have already been
processed, and every surviving binary factor has its transfer class in the current
complete group.  Then every concrete local comparison batch in the
standard \(\DihZ{2m}\) proof has a gauge-coherent counterpart in the original basis.  Its
required invariants are the following:
\begin{enumerate}
\item The realizable kernels are monomial, their traces remain affine in
      \(t\in \mu_m\), reversal permutes \(\mu_m\), and the ordered ratio of every safe sector binary lies
      in the fixed coset \(\mathscr R_{X,1}\).  Analytically dividing that ratio by
      \(\delta^{-1}\) gives the pencil normal form.  Hence all support,
      ruling, trace-cover, coefficient-minor, endpoint-type, and
      bounded-reduction arguments transport with their zero cases.
      Unique-lift monomial fusions remain cancellation-free.
\item For a logical coefficient tensor
      \(c=(c_{\mathbf z})_{{\mathbf z}\in\Ftwo^d}\) on an oriented matching and orientation
      signs \(\epsilon_j\in\{1,-1\}\), the product-weight change
      \(\widetilde c_{\mathbf z}=\delta^{\sum_j\epsilon_jz_j}c_{\mathbf z}\) sends the logical ratios of safe binary factors from \(\mathscr R_{X,1}\) to \(\mu_m\) while preserving
      \cref{eq:app-nd-six-relations} and the cross ratio \(r\).  The final
      quaternary signature is handled by the marked-coordinate reduction.
\item Fixed \(\delta\)-weights do not alter support integration.  They
      multiply the Reed--Muller obstruction by a nonzero scalar.  In the
      polarized determinant, they merely reparameterize the full grid and
      multiply each tested coefficient by a nonzero product-weight value.
\end{enumerate}
Consequently every branch either
\begin{enumerate}
\item reaches a resolved complexity leaf or a strict stable successor;
\item returns an exact proper-factor or child-card factor batch to the
      lower-arity occurrence induction; or
\item certifies
      \[
       (U_{X,1}^{-1})^{\otimes\arity(h)}h\in\cP.
      \]
\end{enumerate}
When \(h=f_0\), the first two alternatives are resolved in the
sense of \cref{def:protocol-discharged}.  For a later occurrence they are
charged to the occurrence-level induction described in
\cref{proofsubsec:nd-completion}. No decrease of the active tuple's
\(\Xi_{\mathrm{full}}\) is asserted.
The bounded outputs are realizable quaternary or arity-six tensors in the
current coordinates.  Factor alternatives are exact tensor identities
pulled back to those coordinates and exposed through
\cref{thm:external-factor,lem:factor-saturation}.  Their Turing reductions
are retained in the worklist.  The conclusion concerns the selected
occurrence \(h\).
\end{lemma}

\begin{proof}
By \cref{lem:marked-sector-dictionary}, a realizable sector binary is a
nonzero multiple of \(TD(\delta)^{-1}\), and its effective kernel is
\(D(\delta) T\).
The gauge identity \cref{lem:marked-A-gauge-transport} moves each kernel
weight to a specified parent input.  It gives the effective kernels
\hypertarget{eq:marked-A-effective-kernels}{}
\begin{equation*}
 D(\delta) D(t)=\diag(1,\delta t),
 \qquad
 D(\delta) XD(t)=\begin{pmatrix}0&t\\ \delta&0\end{pmatrix},
\end{equation*}
so the two sector pencils are \(A+\delta tD_{11}\) and \(tB+\delta C\).  For a
nonproportional pencil, multiply the ratio of a safe sector binary by \(\delta\) and use the
  coset-normalized pencil argument. The resulting ratio is in
\(\mu_m\), so \cref{lem:nd-root-mobius,lem:nd-root-pencil} applies.  This is
an analytic rescaling only.  Reversal sends \(t\) to
\(\delta^2/t\in \mu_m\), and the gauge lemma shows that a common-edge trace
has at most one forbidden phase in each product.  Hence the two-product
trace cover still has an admissible phase because \(m\ge4\), while every
zero or partially supported card is preserved exactly by the invertible gauges.
By item~4 of \cref{lem:marked-A-gauge-transport}, each comparison of
samples, products, or a full phase grid uses one common locally gauged
parent. Its determinant identity gives, with an explicitly nonzero
scalar \(\kappa\),
\[
 \det_{X,1}(t)
   =\kappa\,\delta^e t^u
      \det_{Z}(c\,t^{\varepsilon}),
\qquad e,u\in\mathbb Z,\quad c\ne0,\quad\varepsilon\in\{-1,1\},
\]
where \(\det_{Z}\) denotes the same selected minor for the
locally gauged tensor in the standard sector,
so selected discrepancies, the six-to-four Laurent coefficient, and the
endpoint minor are nonzero exactly in the corresponding standard
calculation.  Therefore
\cref{lem:app-nd-sector-ruling,lem:app-nd-compatible-closure,%
lem:app-nd-coefficient-line-factor,lem:app-nd-localization,%
lem:app-nd-bounded-consumer,lem:app-nd-parallelogram,%
lem:app-nd-exceptional-endpoints,lem:app-nd-endpoint-type,%
lem:app-nd-six-to-four}
transport with the same realizable closures.  Every bounded quaternary output
is handled by the marked-coordinate reduction,
and every exact factor identity is pulled
back by the corresponding tensor power of \(U_{X,1}\) and
then handled with a recorded Turing reduction.  For the active minimum
occurrence these transported concrete batches are resolved.  For a later
occurrence, the role of minimum arity is replaced by the closed proper
batches and the earlier lower-arity occurrences, exactly as in the
hypothesis check of \cref{lem:nd-relative-reentry}. The changes of kernel
are justified separately by \cref{lem:marked-A-gauge-transport}.
Each batch is completely closed inside that occurrence's fixed local
queue and charged to the outer occurrence induction.  On the branch
surviving the corresponding closure, the standard support, ruling, and
coefficient invariants remain available.

Use the product-weight convention of
\cref{subsubsec:gauge-product-weights}. For this logical
coefficient cube put
\[
 \chi({\mathbf z}):=\prod_j\gamma_j^{z_j}\qquad(\gamma_j\in\mathbb C^\times).
\]
Multiplication by \(\chi\) preserves supports and all two-face
parallelogram identities.  It also preserves every complementary-pair
product up to one common nonzero scalar.  For oriented matched pairs, the logical ratio of a safe binary factor is
\(\delta^{-\epsilon_j}h_j\), \(h_j\in \mu_m\),
\(\epsilon_j\in\{1,-1\}\).  The product-weight change
\[
  \widetilde c_{\mathbf z}
  =\delta^{\sum_j\epsilon_jz_j}c_{\mathbf z}
\]
puts all ratios in \(\mu_m\), multiplies both sides of every equality in
\cref{eq:app-nd-six-relations} equally, and leaves
\[
  r=\frac{c_{000}c_{110}}{c_{100}c_{010}}
\]
unchanged.  Here the product-weight change is analytic.  Since
\(\delta^2\in \mu_m\), absorb orientation changes into the phases and write
the ratios of the three initial safe binary factors as \(\delta^{-1}h_j\), \(h_j\in \mu_m\).
The crossing factor cards used in the standard six-input calculation
expose these raw ratios: the extra \(\delta t\) multiplies a column of
the card and does not change the ratio on its unclosed matching edge.
Realizable rotations \(D(h_j^{-1})\) therefore leave the signature in the current coordinates
\[
 c_{\mathbf z}=\delta^{-\wt({\mathbf z})}N_r({\mathbf z}),
 \qquad N_r({\mathbf z})=\begin{cases}1,&\wt({\mathbf z})\le1,\\r,&\wt({\mathbf z})\ge2,\end{cases}
\]
after its nonzero realization scalar is recorded.  To check \(r\in \mu_m\)
without realizing the analytic weight change, use the standard
\(X_{12}X_{34}X_{56}\) support and close inputs \((1,3)\) by \(K^1(t)\).
The resulting quaternary signature, with logical row \(x_2\) and column
\(x_5\), has coefficient matrix
\[
 \begin{pmatrix}1&r/\delta\\t/\delta&tr/\delta^2\end{pmatrix}
 =\binom{1}{t/\delta}(1,r/\delta).
\]
Expose its exact binary factors through the factor theorem.  Unless a
binary exit occurs, \(r/\delta\in\delta^{-1}\mu_m\), proving \(r\in \mu_m\).
Closing the third matching pair of the signature in the current coordinates by \(K^1(t)\) gives
\[
 Q(t)=\diag(1,\delta^{-1})
 \begin{pmatrix}1+t&r+t\\r+t&r(1+t)\end{pmatrix}
 \diag(1,\delta^{-1}),\qquad
 \det Q(t)=\delta^{-2}(r-1)(t^2-r).
\]
If \(r=1\), the displayed arity-six signature already factors.  Otherwise
at most two phases satisfy \(t^2=r\). Choose another phase in \(\mu_m\).
The resulting nonzero quaternary signature enters the marked-coordinate
reduction with the displayed nonzero minor.

Support integration is Boolean and unchanged.  On its sole Reed--Muller
residue the raw face, in one orientation, is
\[
 \begin{pmatrix}A&\delta tB\\\delta tC&D\end{pmatrix},
 \qquad A,B,C,D\ne0.
\]
The realizable \(t=1\) matching card gives \(AD=\delta^2BC\), so its
determinant is \(\delta^2BC(1-t^2)\).  The complementary exponent pattern
gives the same obstruction up to sign and a nonzero scalar.  It is nonzero
for some \(t\in \mu_m\setminus\{\pm1\}\).  In polarization,
each grid variable is merely reparameterized as
\(\lambda_j=c_jt_j\), \(c_j\in\mathbb C^\times\), and the remaining weights form
a nonzero product weight \(\chi\).  The grid is a product of cosets
\(c_j\mu_m\), so \cref{lem:nd-grid-interpolation} applies without requiring
\(c_j\in \mu_m\).  Hence the adjacent coefficient becomes
\[
 [\lambda^{{\mathbf z}+{\mathbf z}'}]\Delta_\xi
 =\chi({\mathbf z})\chi({\mathbf z}')(u_{\mathbf z}v_{{\mathbf z}'}-u_{{\mathbf z}'}v_{\mathbf z})^2
 \qquad\text{for adjacent }{\mathbf z},{\mathbf z}'.
\]
The cross-slice coefficient is similarly multiplied by \(\chi({\mathbf z})^2\).
Thus the full grid and all vanishing tests in
\cref{lem:app-nd-support-integration,lem:nd-q6-minor-localization,%
lem:app-nd-polarized-determinant}
remain valid.

The transported support integration therefore gives one global matching
support for the selected occurrence, and the transported polarization
identities integrate its nonzero coefficients into a product of binary
factors on that matching.  This is precisely
\((U_{X,1}^{-1})^{\otimes\arity(h)}h\in\cP\).
If an intermediate identity factors the selected signature \(h\) itself,
replace its occurrence by the irreducible occurrences of its proper
factors.  If instead only a constructed child card factors, process that
child's complete factorization tree inside the selected occurrence's fixed
local queue. The token for \(h\) remains unresolved until its analysis
resumes.  Together with the
product certificate above, this proves the occurrence-level alternatives.
\end{proof}

\subsubsection{Completion of the weighted dihedral case}

\begin{proof}[Proof of \cref{thm:marked-A-rigidity}]
Fix \(U_{X,1}\) from \cref{lem:marked-sector-dictionary} and
let \(n=\arity(f)\) be the active minimum arity.  If \(n=4\), the
\hyperref[par:marked-coordinate-reduction]{marked-coordinate reduction},
using \cref{lem:marked-A-q4} after its prescribed factor and deck batches,
gives a resolved complexity leaf for the whole signature set or an established strict successor.
For \(n=6\), apply \cref{lem:marked-A-primitive-transport} to the active
minimum occurrence.  It is relatively deck-stable by
\cref{def:occurrence-relative-deck-stability}. On a branch without a
resolved leaf or strict successor, its lower-arity batches have already
been processed.  The weighted support and polarization calculation in
that lemma gives a proper factorization, contradicting irreducibility.

Suppose \(n\ge8\), and work analytically with
\(f^{(X,1)}=(U_{X,1}^{-1})^{\otimes n}f\).
The two-card localization proof transports directly to the realizable
weighted kernels: common-edge traces exclude at most two phases. Joining
two parity discrepancies gives a fused weight ratio
\(\kappa t^{\pm2}\), \(\kappa\ne0\). Cycle shortening or vertex
replacement has the same unique-lift matching support.  These are the
explicit kernel and fusion identities in
\cref{lem:marked-A-gauge-transport,lem:app-nd-localization}.
As \(m\ge4\), each required allowed phase exists.  A disagreeing pair
of sampled cards would therefore localize to a nonmatching arity-six tensor.  But the first proper gadget contraction of the distinguished irreducible signature is
zero or a product of safe current-group binaries, and subsequent
contractions preserve that property by the group fusion law.  The
inherited nonzero disagreeing cards give a contradiction.

Thus at a fixed deleted pair all nonzero samples have one matching and
one common varying edge.  The
\hyperref[par:nd-common-factor-consequence]{common-factor consequence}
fixes the other factor classes across both sectors. Its hypothesis concerns
projective factor classes on the common matching and is unchanged by the
coherent invertible gauges.
For each weighted pencil \(A+\delta tD_{11}\) and \(tB+\delta C\),
evaluation at two distinct nonzero phases gives an invertible linear
system for its two slices, since \(\delta\ne0\). The four recovered
slices have one common nonzero binary product factor, proving an exact
proper factorization of \(f^{(X,1)}\).  Pulling it back by the
invertible matrices at individual inputs \(U_{X,1}\) gives a proper
factorization of the signature \(f\) in the current coordinates, again impossible.

No higher-arity deck-stable signature branch survives, and the quaternary route
already resolves the entire retained set.  This proves \(\Close_{\{\cP\}}(\cF,\DihXone{2m},f)\).
\end{proof}

\subsection{Completion of the marked-dihedral classification}

\begin{proof}[Proof of \cref{thm:marked-dihedral}]
If \(G=\DihZ{2m}\), then deck stability and \(\Even(\cF)\) give
\(\arity(f)=\nu(\cF)=2k+2\) for some \(k\), so apply
\cref{thm:normalized-dihedral}.  If \(G=\DihY{2m}\), apply
\cref{thm:marked-B-reduction}.  If \(G=\DihXzero{2m}\), apply
\cref{thm:marked-A0-reduction}.  If \(G=\DihXone{2m}\), apply
\cref{thm:marked-A-rigidity}.  These are exactly the four cases in the
statement.
\end{proof}
\section{Cyclic Matching Decks and Reed--Muller Rigidity}
\label{sec:proper-deck}

In this case the transfer span is \(\mathscr K_G=\Span\{I,X\}\).
In Walsh coordinates, the corresponding contractions impose equality
of two bits with a possible phase on their common value. In the remaining
higher-arity case, the matching cards determine the Reed--Muller signature
\(\RMcore\) up to the equivalences in \cref{thm:reed-muller-rigidity}.
If the core remains in this case under every realizable single-input
dressing, \cref{lem:rm-safe-group-collapse} forces the current complete
group to be \(\CycX{2}\). This excludes every larger cyclic group in the
surviving case. After the binary and arity-four cases have been resolved,
the reduction in \cref{rem:rm-binary-enlargement-shortcut} can adjoin
Bell signatures and continue the complexity classification with a strictly
larger group. For the synthesis in \cref{sec:p1-synthesis}, we prove the
stronger conclusion that the entire retained
set and the edge admit one common flat-Lagrangian presentation in the surviving cyclic branch.

To obtain this presentation, we use \(\RMcore\) to test the other retained
signatures. Networks built from the core realize flat-Lagrangian signatures.
They also realize Radon contractions of a signature \(g\). On a chosen
hyperplane, the new signature takes the value \(g(\mathbf x)+g(\mathbf x+\mathbf v)\)
on each fibre \(\{\mathbf x,\mathbf x+\mathbf v\}\). It has two
fewer inputs, allowing a failure of affine structure to be localized at
lower arity. An induction, followed by affine patching and recognition, yields one common
flat-Lagrangian presentation for the whole set.
Appendix~\ref{proofsubsec:rm-pauli-tools} extends the network results
to all four Bell signatures.

Use the lexicographic indexing of the eight inputs by \(\Ftwo^3\) fixed in
\cref{par:reed-muller-code}.
Define the eight-ary Reed--Muller indicator signature
\(\RMcore:\Ftwo^8\to\{0,1\}\) by
\begin{equation}
  \RMcore({\mathbf{x}}):=\one[{\mathbf{x}}\in RM(1,3)].
  \label{eq:RM-core}
\end{equation}

For \(S\subseteq\Ftwo^r\), write \(\one[S]\) for the indicator
signature \({\mathbf{x}}\mapsto\one[{\mathbf{x}}\in S]\).

\begin{theorem}[Reed--Muller deck rigidity]
\label{thm:reed-muller-rigidity}
Assume
\[
 \Stable(\cF,G,f),\qquad
 \arity(f)=\nu(\cF),\qquad
 \mathscr K_G=\Span\{I,X\}.
\]
Then \(\Close_{\{\FLag\}}(\cF,G,f)\).  In the remaining tractable case with a common flat-Lagrangian presentation,
\[
 G=\{[I],[X]\},\qquad
 f\sim\RMcore=\one[RM(1,3)],
\]
where \(\sim\) means equality up to a nonzero scalar, a permutation of
inputs, and realizable single-input \(X\)-dressings.
\end{theorem}

For \(s\in\AlgNums\), the Walsh coordinates satisfy
\begin{equation*}
  P^{-1}\!\left[\frac12
  \begin{pmatrix}1+s&1-s\\1-s&1+s\end{pmatrix}\right]P
  =D(s),
  \qquad
  P^{\mathsf T}XP=2Z
\end{equation*}

\subsection{Cyclic phase pencils and binary attachments}
\label{subsec:proper-cyclic-phase-pencils}

The cyclic group supplies a finite family of phase contractions.  The next
three lemmas describe which binary pencils they permit and which changes
to a quaternary signature their realizable input dressings can realize.

A \emph{sampled phase pencil} evaluates a pencil \(U+tV\) on a finite
root-of-unity group, here \(\mu_N\), retaining zero members and the
projective classes of nonzero members.

For each integer \(N\ge3\), use the diagonal phase matrices
\(D(s)\), \(s\in\mu_N\), defined in \cref{subsec:holant-definitions}.
Realizable networks use \(X\)-edges. \Cref{eq:Walsh-coordinates}
gives their analytic matrix \(P^{\mathsf T}XP=2Z\) in the tables below.

For \(N\ge3\) and \(a,b,c,d\in\AlgNums\), define the condition
\(\mathcal R_N(a,b,c,d)\) by declaring that, for some \(\alpha\in\mu_N\), at least
one of the following systems holds:
\begin{align}
  c&=-\alpha a,&d&=-\alpha b,
  &&\text{(proportional)},
  \label{eq:odd-dih-R-proportional}\\
  b&=c=0,&d&=\alpha a,
  &&\text{(diagonal monomial)},
  \label{eq:odd-dih-R-diagonal}\\
  a&=d=0,&c&=\alpha b,
  &&\text{(anti-diagonal monomial)}.
  \label{eq:odd-dih-R-antidiagonal}
\end{align}
The alternatives may overlap and include the all-zero quadruple.

For a quaternary signature \(q\), write \(\widetilde q\) for its
Walsh coordinates from \cref{eq:Walsh-coordinates}. When it is
supported on even Hamming weight, define its endpoint and central Walsh
coordinates by
\begin{equation}
  p=\widetilde q_{0000},\qquad
  t=\widetilde q_{1111},\qquad
  x_{ab}=\widetilde q_{\one_{\{a,b\}}},\qquad 1\le a<b\le4.
  \label{eq:odd-dih-q4-coordinates}
\end{equation}
Here \(\one_{\{a,b\}}\in\Ftwo^4\) has ones precisely at inputs
\(a,b\), so \(x_{ab}\) indexes one of the six weight-two strings.

\begin{lemma}[Cyclic phase-pencil rigidity, including zeros]
\label{lem:odd-dih-RN}
For every integer \(N\ge3\), the condition
\(\mathcal R_N(a,b,c,d)\) is equivalent to the following
condition: for every \(s\in\mu_N\),
\[
  \diag(a-sb,c-sd)Z
  \quad\text{is zero or is projectively }D(\eta)
  \text{ for some }\eta\in\mu_N.
\]
\end{lemma}

\begin{proof}
If \(a=b=c=d=0\), both conditions hold, so assume otherwise.  Set
\[
  u(s)=-c+sd,\qquad v(s)=a-sb.
\]
The phase-card condition says that \(u(s)\) and \(v(s)\) are either both zero
or both nonzero, and in the latter case
\(u(s)/v(s)\in\mu_N\).  If the two affine linear forms are dependent, the
possibility that exactly one is identically zero is excluded by this
condition.  Hence \(u=\alpha v\) for some \(\alpha\in\mu_N\), giving
\cref{eq:odd-dih-R-proportional}.  If they are independent, neither form
vanishes on \(\mu_N\), and
\(\psi(s):=u(s)/v(s)\) is a nonconstant M\"obius
transformation.  The card condition gives
\(\psi(\mu_N)\subseteq\mu_N\). Injectivity and equality of the finite
cardinalities give \(\psi(\mu_N)=\mu_N\).  Thus
\cref{lem:nd-root-mobius} applies and gives
\(\psi(s)=\alpha s\) or \(\alpha/s\), with
\(\alpha\in\mu_N\).  Cross multiplication yields
\cref{eq:odd-dih-R-diagonal,eq:odd-dih-R-antidiagonal}.  Direct substitution
proves the converse, including the all-zero case.
\end{proof}

For the two cyclic lemmas below, fix a finite algebraic retained set \(\cF\) in \(K\)-coordinates, with edge signature \(X\), and
fix an integer \(N\ge3\).  For every \(s\in\mu_N\), fix a realizable ordered binary gadget
\(C_s\) over \(\cF\), with a specified orientation, whose analytic Walsh
matrix is
\[
  \widetilde C_s=\kappa_sD(s)Z,
  \qquad \kappa_s\in\AlgNums^\times.
\]
The nonzero scalar \(\kappa_s\) is recorded with the gadget.  Throughout
these lemmas, ``realizable'' and ``directly realizable'' are relative to this same
\(\cF\), and every displayed phase-card matrix is divided by its nonzero scalar. In the cyclic case,
\(P^{-1}G(\cF)P=C_N^{\rm std}=\{[D(s)]:s\in\mu_N\}\). The Walsh transfer class of \(C_s\)
is \([\widetilde C_sZ]=[D(s)]\), so these binaries represent all classes
in that form.

\begin{lemma}[Cyclic quaternary phase family]
\label{lem:odd-cyc-q4-family}
Let \(q\) be a realizable quaternary signature over \(\cF\), and suppose
\(\widetilde q=(P^{-1})^{\otimes4}q\) has even-parity support.  Define \(p,t,x_{ab}\) by \cref{eq:odd-dih-q4-coordinates}.  Assume
that, for every complementary pair, both orders of its two residual inputs,
and every \(s\in\mu_N\), contraction with the fixed chosen representative
\(C_s\) gives either the zero card or a nonsingular binary card whose Walsh
transfer class belongs to
\(C_N^{\rm std}=\{[D(\eta)]:\eta\in\mu_N\}\).  Equivalently, define
\[
  \mathfrak I=\bigl\{\{1,2\},\{1,3\},\{1,4\}\bigr\},
  \qquad I^c=[4]\setminus I,
\]
and write \(x_I:=x_{ab}\) for \(I=\{a,b\}\).  For every
\(I\in\mathfrak I\), \(s\in\mu_N\), and
\(\varepsilon\in\{+,-\}\), the corresponding normalized realizable Walsh card
has matrix
\begin{align*}
 B_{I,+}(s)&=\diag(p-sx_I,x_{I^c}-st),\\
 B_{I,-}(s)&=\diag(p-sx_{I^c},x_I-st).
\end{align*}
Here \(+\) records deletion of \(I\) and \(-\) deletion of \(I^c\). Both
orders of the remaining two inputs are part of the assumption.  Each displayed
card is zero or satisfies
\[
  [B_{I,\varepsilon}(s)Z]\in C_N^{\rm std}.
\]
The right factor \(Z\) comes from the Walsh transfer convention.  Organize the eight coordinates into the endpoint block
\(\mathbf v_0=(p,t)\) and the three central complement blocks
\(\mathbf v_I=(x_I,x_{I^c})\), \(I\in\mathfrak I\).  Then exactly one of the
following holds:
\begin{enumerate}
\item If \(p,t\ne0\), then \(\lambda=t/p\in\mu_N\), and each central
      block \(\mathbf v_I\) is zero or
      \hypertarget{eq:odd-cyc-q4-nonzero-endpoints}{}
\begin{equation*}
        \mathbf v_I=(-p\lambda\alpha_I^{-1},-p\alpha_I),
        \qquad \alpha_I\in\mu_N.
      \end{equation*}
\item If \(p=t=0\), each central block \(\mathbf v_I\) is zero or
      \[
        \mathbf v_I=\rho_I(1,\alpha_I),
        \qquad \rho_I\in\AlgNums^\times,
        \quad \alpha_I\in\mu_N.
      \]
\end{enumerate}
In particular, exactly one of \(p,t\) cannot be nonzero.
Conversely, every even Walsh coefficient tensor in these alternatives satisfies the stated
phase-card condition under every phase contraction.
\end{lemma}

\begin{proof}
For a complementary block \(I,I^c\), the phase card is
\[
  \diag(p-sx_I,x_{I^c}-st),
  \qquad s\in\mu_N.
\]
Apply \cref{lem:odd-dih-RN} to
\((a,b,c,d)=(p,x_I,x_{I^c},t)\), and then interchange \(I,I^c\).
If \(p,t\ne0\), the anti-diagonal system is impossible.  If one of the
two complementary-deletion systems is diagonal, then \(x_I=x_{I^c}=0\) and
\(t/p\in\mu_N\). The other system is then diagonal as well.
In the remaining case both systems are proportional.  Taking their
parameters to be \(\alpha,\beta\in\mu_N\) gives
\[
 x_{I^c}=-\alpha p,\qquad x_I=-\beta p,\qquad
 t=\alpha\beta p.
\]
Thus \(t/p=\alpha\beta\in\mu_N\), and choosing
\(\alpha_I=\alpha\) gives the displayed parametrization.  If exactly one
endpoint is nonzero, neither monomial system is possible and the two
proportional systems force respectively \(x_I=0,x_{I^c}\ne0\) and
\(x_{I^c}=0,x_I\ne0\), a contradiction.  If \(p=t=0\), proportional and
diagonal systems give only the zero block. Every nonzero block is therefore
anti-diagonal, \(\mathbf v_I=\rho_I(1,\alpha_I)\).  Direct substitution checks
both complementary deletions and every card arising from a zero complementary
block, proving the converse.
\end{proof}
\begin{lemma}[Realizable quaternary cyclic dressing]
\label{lem:odd-cyc-four-port-dressing}
Let \(q\) be a realizable quaternary signature over \(\cF\).  For
\(r_1,r_2,r_3,r_4\in\mu_N\), attach the fixed representative \(C_{r_i}\), in
its specified orientation, to input \(i\) of \(q\) through one \(X\)-edge, leaving the other input of \(C_{r_i}\) exposed.  Denote the
normalized output signature by \(h\).  After division of the direct-gadget
boundary tensor by the nonzero scalar
\(2^4\prod_i\kappa_{r_i}\), its Walsh coordinates are
\begin{equation*}
  \widetilde h_{\mathbf{y}}
  =\widetilde q_{\mathbf{y}}\prod_{i=1}^4r_i^{y_i}.
\end{equation*}
Define the map \(F_N:\mu_N^4\to\mu_N^4\) by
\begin{equation*}
  F_N(r_1,r_2,r_3,r_4)
  =(R,r_{12},r_{13},r_{14}),
  \qquad
  R=\prod_{i=1}^4r_i,
  \quad r_I=\prod_{i\in I}r_i,
\end{equation*}
and put \((\mu_N)^2:=\{z^2:z\in\mu_N\}\).  A target
\((R,u,v,w)\in\mu_N^4\) lies in the image of \(F_N\) exactly when
\begin{equation}
  \frac{uvw}{R}\in(\mu_N)^2.
  \label{eq:cyclic-dressing-image}
\end{equation}
For odd \(N\), \(F_N\) is bijective.
\end{lemma}

\begin{proof}
At input \(i\), the fixed Walsh matrix
\(\kappa_{r_i}D(r_i)Z\), followed by the analytically transformed disequality edge \(P^{\mathsf T}XP=2Z\), acts as
\(2\kappa_{r_i}D(r_i)\).  The four attachments therefore give the stated
realizable dressing and scalar.  Necessity is
\(uvw/R=r_1^2\). For sufficiency choose
\[
  r_1^2=uvw/R,
  \qquad
  r_2=u/r_1,
  \qquad
  r_3=v/r_1,
  \qquad
  r_4=w/r_1.
\]
These choices satisfy
\(F_N(r_1,r_2,r_3,r_4)=(R,u,v,w)\). For odd \(N\), squaring on
\(\mu_N\) is bijective.
\end{proof}

\subsection{The arity-four case}

The cyclic groups in \cref{rem:xia-binary-taxonomy} have transfer span
\(\mathscr K_G=\Span\{I,X\}\). We treat arity four before reconstructing
the higher-arity Reed--Muller signature.

\begin{lemma}[The cyclic arity-four case]
\label{lem:proper-cyclic-q4}
Let \((\cF,G,q)\) be a deck-stable retained state with
the supplied equality \(I\) available,
\(\arity(q)=\nu(\cF)=4\), transfer span \(\mathscr K_G=\Span\{I,X\}\), and finite cyclic \(G\ni[X]\).  Then either an
unsafe realizable binary is exposed, factor access strictly enlarges the
complete transfer group of the resulting set, or a gadget using at most two copies of
\(q\) realizes an irreducible endpoint-nondegenerate eight-vertex signature.  Thus one obtains a resolved complexity leaf, an outer deck successor
under \cref{lem:matching-synthesis-measure}, or the tractable case
\(\Tract(K\cF)\).
\end{lemma}

\begin{proof}
In Walsh coordinates the transfer span is diagonal. Therefore
\begin{equation*}
  P^{-1}GP=C_N^{\rm std}=\{[D(s)]:s\in\mu_N\},
  \qquad 2\mid N.
\end{equation*}
Here \([X]\) becomes \([Z]=[D(-1)]\), and chosen representatives are
proportional to \(D(s)Z\).  To justify the parity reduction, fix an ordered
deleted pair and an odd-weight string \({\mathbf{y}}\) on the two residual inputs.  Up to
a fixed nonzero representative scalar and a possible sign, the corresponding
off-diagonal phase-card entry is \(u+sv\), where \(u\) and \(v\) are the
contributions of the two Walsh coefficients whose deleted-pair assignments
are \(00\) and \(11\), respectively, and whose assignment to the remaining inputs is \({\mathbf{y}}\).  A zero or
cyclic card is diagonal, so this entry vanishes.  At the two allowed phases \(s=1,-1\) we obtain \(u+v=u-v=0\), hence \(u=v=0\).  Running through
all six deleted pairs and both input orders (equivalently, the complete
ordered deck) forces every odd-weight Walsh coefficient to vanish.  Thus
\(\widetilde q\) is supported on even Walsh parity, as required by
\cref{lem:odd-cyc-q4-family}.  Item~1 of
\cref{thm:equality-q4-boundary} handles \(N=2\). Hence
assume \(N\ge4\) and use the endpoint and complement-block coordinates of
\cref{eq:odd-dih-q4-coordinates}, indexed by
\[
  I_1=\{1,2\},\qquad I_2=\{1,3\},\qquad I_3=\{1,4\}.
\]
The cyclic quaternary-family result in \cref{lem:odd-cyc-q4-family} leaves two
endpoint regimes.  We treat the zero-endpoint regime first. In the nonzero
regime, the dressing-image criterion further separates successful dressing
from the residual square obstruction.
Define the block phase parameters \(\alpha_i:=\alpha_{I_i}\) and, for
nonzero endpoints, the endpoint ratio \(\lambda:=t/p\).
Call the complementary block indexed by \(I_i\) \emph{active} when
\((x_{I_i},x_{I_i^c})\ne(0,0)\), and \emph{inactive} otherwise.

For each two-copy output below, let \(\cF^+\) be the signature set after preprocessing, complete factor closure, and group
recomputation.  Zero, nullary, unary, odd, singular, and infinite-order
factors are handled by the corresponding reductions.  The remaining calculations concern
the all-even finite-group branch.  Proper factors become available over
\(\cF^+\) through the factor theorem, with its Turing reduction and
nonzero normalization scalars recorded.

\emph{Zero endpoints.}
Write each complementary block with scale \(\rho_i\) as
\((x_{I_i},x_{I_i^c})=\rho_i(1,\alpha_i)\), taking \(\rho_i=0\)
for an inactive block and then choosing any \(\alpha_i\in\mu_N\), and define the auxiliary scalar
\(X_i:=\alpha_i\rho_i^2\).  For each \(i\), choose the remaining
indices \(j<k\), so \(\{i,j,k\}=\{1,2,3\}\), and order
\(I_i=(1,i+1)\).  Connect the first input
of one copy to the first input of the other by the chosen representative
\(D(\alpha_j\alpha_k)Z\), and connect the second inputs by \(I\) (with the
same orientation on both copies).  If \(Q_i\) is the \(I_i\mid I_i^c\)
flattening, write the normalized output in Walsh coordinates as \(\widetilde g_i\), where
\[
 \widetilde g_i=Q_i^{\mathsf T}K_iQ_i,\qquad
 K_i=(D(\alpha_j\alpha_k)Z)\otimes I
     =\diag(1,1,-\alpha_j\alpha_k,-\alpha_j\alpha_k).
\]
Order the residual inputs as the two inputs of \(I_i^c\) in the first copy
followed by those in the second copy, each in inherited order.  The only
potentially nonzero complementary pairs of \(\widetilde g_i\) are
\[
 E=(\widetilde g_i(0000),\widetilde g_i(1111)),
 \qquad
 D=(\widetilde g_i(0101),\widetilde g_i(1010)).
\]
Their entries are
\begin{equation*}
 \begin{aligned}
  (E_0,E_1)
    &=(-\alpha_j\alpha_k\rho_i^2,\ \alpha_i^2\rho_i^2),\\
  (D_0,D_1)
    &=(\alpha_j(X_j-X_k),\ -\alpha_k(X_j-X_k)),
 \end{aligned}
\end{equation*}
and, whenever both displayed pairs \(E\) and \(D\) are nonzero,
\[
 \frac{D_1/D_0}{E_1/E_0}=\left(\frac{\alpha_k}{\alpha_i}\right)^2.
\]
Set
\[
 d_i:=(X_j-X_k)^2-X_i^2.
\]
We now verify endpoint nondegeneracy, including the case of an inactive
\(i\)-block.  If \(\rho_i\ne0\), the binary-gadget attachment criterion
\cref{lem:odd-cyc-four-port-dressing,eq:cyclic-dressing-image} is applied
in the displayed residual-input order.  Put
\((u,v,w)=(r_{12},r_{13},r_{14})\) for these four output positions.
The entries at \(0101\) and \(1010\) are multiplied by \(R/v\)
and \(v\), respectively.  Take
\[
 R=(E_1/E_0)^{-1},
 \qquad u=1,
 \qquad v=-\alpha_j/\alpha_i,
 \qquad w=R/v.
\]
All four parameters lie in \(\mu_N\), since \(N\) is even, and
\(uvw/R=1\).  Moreover
\[
 E_1R=E_0,\qquad
 D_0R/v=D_1v=\frac{\alpha_j\alpha_k}{\alpha_i}(X_j-X_k).
\]
Thus the two entries in each displayed block are equal.  Since the Walsh
support is even, the two endpoint values in the original basis coincide. Up to one common
nonzero scalar their common value is
\[
 (X_j-X_k)-X_i.
\]
It is nonzero whenever \(d_i\ne0\).  If instead \(\rho_i=0\),
then \(E_0=E_1=0\), while \(d_i\ne0\) implies
\(D_0D_1\ne0\).  Set
\[
 u=v=1,
 \qquad R=D_1/D_0=-\alpha_k/\alpha_j,
 \qquad w=R.
\]
Here \(uvw/R=1\), so the same criterion applies and makes the two dressed
entries of the \(D\)-block equal and nonzero.  Inverse Walsh transformation again gives two
nonzero endpoint values in the original basis.  Thus every index with \(d_i\ne0\),
active or inactive, yields an endpoint-nondegenerate eight-vertex
  signature.  In the displayed four-string matching-code support, the two
  crossed \(2+2\) flattenings have rank at least two. The remaining logical
  flattening has determinant, up to a nonzero scalar,
\begin{equation*}
  \alpha_j\alpha_kd_i.
\end{equation*}
If some \(d_i\ne0\), the corresponding realizable output has no binary
factor.  The dressing is a product of nonsingular local maps, so it preserves
tensor factorizations.  Even parity and its complementary endpoints exclude a unary factor,
so the dressed endpoint-nondegenerate eight-vertex signature is irreducible.

It remains to exclude simultaneous vanishing of the three minors.  If
\(d_1=d_2=d_3=0\), choose signs
\(\epsilon_i\in\{1,-1\}\) such that
\hypertarget{eq:proper-cyclic-simultaneous-zero}{}
\begin{equation*}
 \begin{aligned}
  X_2-X_3&=\epsilon_1X_1,\\
  X_3-X_1&=\epsilon_2X_2,\\
  X_1-X_2&=\epsilon_3X_3.
 \end{aligned}
\end{equation*}
The coefficient matrix of this system is
\[
 M_\epsilon=
 \begin{pmatrix}
  -\epsilon_1&1&-1\\
  -1&-\epsilon_2&1\\
  1&-1&-\epsilon_3
 \end{pmatrix},
 \qquad
 \det M_\epsilon
 =-(\epsilon_1\epsilon_2\epsilon_3+
     \epsilon_1+\epsilon_2+\epsilon_3).
\]
For the two constant sign triples this determinant is nonzero.  For each of
the other six, direct substitution gives, up to permutation, precisely the
kernel lines
\(\Span_{\mathbb C}\{(1,1,0)\}\),
\(\Span_{\mathbb C}\{(1,0,1)\}\), and
\(\Span_{\mathbb C}\{(0,1,1)\}\).  Hence every nonzero solution is a permutation of
  \((\xi,\xi,0)\), with \(\xi\ne0\).  Since \(q\ne0\), not all three
  \(X_i\) vanish.

Relabel so that \((X_1,X_2,X_3)=(\xi,\xi,0)\).  Then
\(\rho_1\rho_2\ne0\) and \(\rho_3=0\).  In the
\(I_3\mid I_3^c\) flattening of \(\widetilde q\), order the odd assignments
on either side as \(10,01\).  Its only nonzero block is
\begin{equation*}
 \begin{pmatrix}
  \rho_1&\rho_2\\
  \alpha_2\rho_2&\alpha_1\rho_1
 \end{pmatrix},
 \qquad
 \det=\alpha_1\rho_1^2-\alpha_2\rho_2^2=X_1-X_2=0.
\end{equation*}
All four entries are nonzero, so the block has rank one and
\(\widetilde q\) has a genuine binary-by-binary tensor factorization across
\(I_3\mid I_3^c\).  The Walsh change is a product of invertible local maps,
so \(q\) has the same factorization, contradicting its irreducibility.
Therefore some \(d_i\ne0\), as required.

\emph{Nonzero endpoints and successful dressing.}
Define the ratio scalar \(R:=\lambda^{-1}\) and the dressing scalar
\(r_I:=\epsilon_I\alpha_I\lambda^{-1}\) on each active block, where
\(\epsilon_I\in\{\pm1\}\). Abbreviate
\(\epsilon_i:=\epsilon_{I_i}\).  An inactive block removes the square
constraint: its otherwise irrelevant dressing parameter may be chosen so
that the criterion's product equation \(uvw/R=1\) holds.
With all three active,
\cref{lem:odd-cyc-four-port-dressing,eq:cyclic-dressing-image} applies
exactly when
\(\epsilon_1\epsilon_2\epsilon_3
 \alpha_1\alpha_2\alpha_3\) is a square.  If
\(N\equiv2\pmod4\), signs remove the obstruction. When \(4\mid N\), this
works precisely if \(\alpha_1\alpha_2\alpha_3\) is a square.  The dressed
endpoint block is \((p,p)\). Every active complementary block becomes
\((-\epsilon_Ip,-\epsilon_Ip)\), while every inactive block remains
\((0,0)\).  Let \(h\) be
the resulting signature in the original basis.  Equality of complementary Walsh blocks
makes \(h\) even-parity, while its even Walsh support makes it
  complement-invariant, meaning
  \(h({\mathbf{y}})=h({\mathbf{y}}+\one^4)\) for every \({\mathbf{y}}\in\Ftwo^4\). It is nonzero because
  \(p\ne0\).  Choose an even
string \({\mathbf{x}}\) with \(h({\mathbf{x}})\ne0\), and attach realizable \(X\)-dressings on
exactly the coordinates where \({\mathbf{x}}\) is one.  This even translation preserves
parity, and the translated signature \(h'\) satisfies
\[
 h'(0000)=h({\mathbf{x}})=h({\mathbf{x}}+\one^4)=h'(1111)\ne0.
\]
Thus \(h'\) is an endpoint-nondegenerate eight-vertex signature.  Every
dressing used here is an invertible single-input map, so it also preserves the
irreducibility of \(q\).

\emph{Square obstruction.}
The only remaining case is
\begin{equation*}
 4\mid N,\qquad
 p,t\ne0,\qquad
 \text{all three blocks active},\qquad
 \alpha_1\alpha_2\alpha_3\notin(\mu_N)^2.
\end{equation*}
  Normalize \(p=1\).  For a permutation \((i,j,k)\) of \((1,2,3)\),
  abbreviate
  \[
   (a,b,c):=(\alpha_i,\alpha_j,\alpha_k),
  \]
  and order
the \(I_i\mid I_i^c\) flattening as \(00,01,10,11\):
\begin{equation*}
 Q_i=
 \begin{pmatrix}
  1&0&0&-a\\
  0&-b&-c&0\\
  0&-\lambda c^{-1}&-\lambda b^{-1}&0\\
  -\lambda a^{-1}&0&0&\lambda
 \end{pmatrix}.
\end{equation*}
Let \(D:=Z\otimes Z\) be the diagonal matrix on the four two-bit assignments, and let \(S\) be
the permutation matrix that swaps \(01,10\).  The realizable joins are
\begin{align*}
 \widetilde G_i^{\parallel}&=4Q_iDQ_i^{\mathsf T},
 \\
 \widetilde G_i^{\times}&=4Q_iDSQ_i^{\mathsf T}.
\end{align*}
The crossed one is
\begin{equation*}
 \frac14\widetilde G_i^{\times}
 =
 \begin{pmatrix}
  1+a^2&0&0&-\lambda(a+a^{-1})\\
  0&-2bc&-2\lambda&0\\
  0&-2\lambda&-2\lambda^2/(bc)&0\\
  -\lambda(a+a^{-1})&0&0&\lambda^2(1+a^{-2})
 \end{pmatrix}.
\end{equation*}
By \cref{lem:actual-binary-completeness}, an unsafe phase card is already a
binary and resolves the complexity. Otherwise the cyclic pencil
of the displayed central block forces, when \(1+a^2\ne0\),
\[
 \frac{2\lambda}{1+a^2}\in\mu_N.
\]
Since \(a\) is a root of unity, \(|1+a^2|=2\) gives \(a^2=1\), whereas
\(1+a^2=0\) gives \(a^2=-1\).  Hence
\begin{equation*}
 \alpha_i^2\in\{1,-1\}\qquad(i=1,2,3).
\end{equation*}

If \(8\mid N\), all fourth roots are squares, contradicting the
  obstruction.  Thus \(N\equiv4\pmod8\), and an odd number of the three
  squares equals \(-1\).  Choose the permutation \((i,j,k)\) so that
  \(\alpha_i^2=a^2=-1\), define the ratio scalar
  \(r:=\lambda/(bc)\), and evaluate the straight join.  Its sole nonzero
block has rank one and factors through the nonsingular anti-diagonal matrix
\(M_1(r)=D(r)X\), giving
\begin{equation*}
  \frac14\widetilde G_i^\parallel
  =\pm2 M_1(r)\otimes M_1(r),
  \qquad
  M_1(r)=\begin{pmatrix}0&1\\ r&0\end{pmatrix},
  \qquad r\ne0,
\end{equation*}
on the two residual copies of \(I_i\).  \Cref{lem:factor-saturation} exposes
the nonsingular anti-diagonal \(M_1(r)\). Its transfer obeys
\[
  (M_1(r)Z)^2=-rI,
  \qquad
  (M_1(r)Z)D(s)(M_1(r)Z)^{-1}=sD(s^{-1})\quad(s\in\mu_N).
\]
The class \([M_1(r)Z]\) is an involution outside \(C_N^{\rm std}\) and acts on it by inversion.
Together they generate a subgroup isomorphic to \(D_{2N}\).
After factor closure, the complete transfer group \(G(\cF^+)\) in the original coordinates contains
\(G(\cF)\) and the additional class \([P M_1(r)Z P^{-1}]\), so \(G(\cF)<G(\cF^+)\). The first two cases
pass a realizable endpoint-nondegenerate eight-vertex signature to
\cref{cor:p1-eight-vertex-reduced-interface}, while the third strictly enlarges the
complete group.  This proves the lemma.
\end{proof}

The arity-four case therefore gives a resolved complexity leaf or an
outer deck successor. We next reconstruct the higher-arity core needed
for \cref{thm:reed-muller-rigidity}.

\subsection{Reconstruction of the minimum Reed--Muller core}

Let \(f\) be a minimum-arity nonbinary irreducible signature of the retained set,
and let its arity be \(n:=\arity(f)\).  If \(n\) is odd, apply
\cref{thm:external-odd} to the ordinary-coordinate set \(K\cF\).  Its
hardness conclusion pulls back through the holographic \(K\)-coordinate
equivalence, while its tractable conclusion is exactly \(\Tract(K\cF)\).
Thus, on the continuing branch, \(n\) is even.  Every
proper card is zero or a nonsingular \(G\)-matching signature. In Walsh
coordinates the latter is a weighted-equality product.  Define its analytic
Walsh coefficients by
\[
  a_S:=\widetilde f(\one_S),
  \qquad S\subseteq[n].
\]
For distinct \(i,j\), label and normalize the two realizable cards obtained
from the \(I/X\)-closures as \(C_{ij}^{+}\) and \(C_{ij}^{-}\) so that,
for every \(T\subseteq[n]\setminus\{i,j\}\),
\begin{equation}
  \widetilde C_{ij}^{+}(\one_T)=2(a_T+a_{Tij}),
  \qquad
  \widetilde C_{ij}^{-}(\one_T)=2(a_T-a_{Tij}),
  \label{eq:rm-card-pencil}
\end{equation}
where \(Tij=T\cup\{i,j\}\).  A nonzero card has the form
\begin{equation}
  c\prod_{\{p,q\}\in M}
  \one[z_p=z_q]r_{pq}^{z_p},
  \qquad c\prod_{\{p,q\}\in M}r_{pq}\ne0,
  \label{eq:rm-weighted-matching}
\end{equation}
for a perfect matching \(M\). It is nonzero at zero, and its supported
weight-two strings are exactly the edges of \(M\).  Arity four is
\cref{lem:proper-cyclic-q4}, so assume \(n\ge6\).

Define the zero-card graph \(Z_f\) on \([n]\), writing \(ij\) for the
unordered pair \(\{i,j\}\), by
\[
  ij\in E(Z_f)
  \quad\Longleftrightarrow\quad
  C_{ij}^{+}=C_{ij}^{-}=0.
\]
By \cref{eq:rm-card-pencil},
\begin{equation*}
  ij\in E(Z_f)
  \quad\Longleftrightarrow\quad
  \supp\widetilde f\subseteq\{{\mathbf z}:z_i+z_j=1\}.
\end{equation*}

\begin{lemma}[Elimination of the zero-card graph]
\label{lem:rm-zero-graph}
\label{lem:rm-no-zero-edge}
Let \(f\ne0\) be the fixed minimum-arity nonbinary irreducible signature of arity \(n\).  Every connected
component of \(Z_f\) is complete bipartite.  If \(n>4\), then
\(E(Z_f)=\varnothing\).
\end{lemma}

\begin{proof}
Zero-edge equations alternate along paths, so every nontrivial connected
component is complete bipartite.  Suppose that \(uv\in E(Z_f)\).  Since
\(n>4\), the set \([n]\setminus\{u,v\}\) has at least three vertices.  The
graph induced there is a disjoint union of bipartite graphs and therefore
is not complete. Choose distinct \(i,j\notin\{u,v\}\) with
\(ij\notin E(Z_f)\).  At least one of \(C_{ij}^{+},C_{ij}^{-}\) is then a
nonzero realizable card.  Because \(u,v\) remain external and
\(\supp\widetilde f\subseteq\{{\mathbf z}:z_u+z_v=1\}\), that card vanishes at the
all-zero boundary string.  This contradicts \cref{eq:rm-weighted-matching},
under which every nonzero card is nonzero at zero.  Hence
\(E(Z_f)=\varnothing\).
\end{proof}

Thus, for the remaining arities, no coordinate pair is invisible to both
realizable contractions: every pair supports at least one nonzero matching card.
We may therefore compare the local card matchings through the common pencil
\cref{eq:rm-card-pencil}.  The first global datum extracted from that
comparison is the following weight-two graph.

\paragraph{The weight-two graph.}

Assume henceforth that \(n\ge6\) and \(E(Z_f)=\varnothing\).  Define the
weight-two edge set
\begin{equation*}
  W_f:=\{ij:1\le i<j\le n,\ a_{ij}\ne0\}.
\end{equation*}
Let \(\mathcal W_f:=([n],W_f)\) be the corresponding simple graph.  For
\(S\subseteq[n]\), write
\[
 W_f[S]:=\{pq\in W_f:p,q\in S\},
 \qquad
 W_f[S]-k:=W_f[S\setminus\{k\}]
\]
for the restricted edge sets.

\begin{lemma}[Matching consistency]
\label{lem:rm-matching-consistency}
Let \(f\) be the fixed minimum-arity nonbinary irreducible signature of arity \(n\ge6\), and assume
\(E(Z_f)=\varnothing\).  Then the following statements hold.
\begin{enumerate}
\item \(a_\varnothing\ne0\).
\item \(W_f\) is a matching.
\item If both \(C_{ij}^{+}\) and \(C_{ij}^{-}\) are nonzero, their
      support matchings are equal.
\item Every edge \(pq\in W_f\) is a global Walsh equality:
\[
  \supp\widetilde f\subseteq\{{\mathbf z}:z_p=z_q\}.
\]
\end{enumerate}
\end{lemma}

\begin{proof}
If \(a_\varnothing=0\), then
\(\widetilde C_{ij}^{\pm}(\zero)=\pm2a_{ij}\).  Since
\(E(Z_f)=\varnothing\)
and matching cards are nonzero at zero, every \(a_{ij}\ne0\).  The two
\(ij\)-cards would then cover every edge on the remaining at least four
inputs, impossible for two matchings.  Hence \(a_\varnothing\ne0\).

For a deleted pair \(i,j\), let \(P,Q\) be the support matchings when both
cards survive, and write \(P\mathbin\triangle Q\) for their symmetric
difference.  Their weight-two entries imply
\begin{equation*}
  W_f[[n]\setminus\{i,j\}]\subseteq P\cup Q,
  \qquad
  P\mathbin\triangle Q
  \subseteq W_f[[n]\setminus\{i,j\}].
\end{equation*}
Indeed, for \(pq\in W_f[[n]\setminus\{i,j\}]\) the two pencil values cannot
both vanish. If one whole card vanishes, the other has value
\(\pm4a_{pq}\) at \(pq\).  Thus, with one nonzero card,
\(W_f[[n]\setminus\{i,j\}]\) lies in its perfect
matching.  We use the following elementary graph argument.  If a vertex of
\(\mathcal W_f\) had degree at least three, delete two vertices outside it and its
three neighbours (possible since \(n\ge6\)). The induced graph has degree at
least three, whereas the union of two perfect matchings has maximum degree
two (and one matching has maximum degree one), a contradiction.  Hence every
component of \(\mathcal W_f\) is a path or a cycle.

Suppose a path component has at least two edges.  If two vertices lie outside
that component, delete them. Otherwise (then the component has at least
\(n-1\ge5\) vertices) delete two consecutive vertices at one end, leaving a
path with at least two edges.  For this deleted pair, one nonzero card is
impossible.  If both cards are nonzero, the two edges at the middle of the
retained path must lie one in
\(P\) and one in \(Q\), hence in \(P\triangle Q\).  Alternating along the
path forces an edge of \(P\triangle Q\) at an endpoint, but
\(P\triangle Q\) is a disjoint union of cycles and
\(P\triangle Q\subseteq W_f[[n]\setminus\{i,j\}]\), whose endpoint has no
second edge.  This is impossible.  A cycle is handled similarly: a triangle
is retained by deleting vertices outside it and cannot alternate. For a
4-cycle delete one cycle vertex and one outside vertex, and for a cycle of
length at least five delete two adjacent cycle vertices, leaving a path with
at least two edges.  Thus every component has at most one edge, so \(W_f\)
is a matching.  Finally, for any pair with both cards nonzero,
\(P\triangle Q\) is a union of cycles contained in the matching \(W_f\).
Therefore \(P\triangle Q=\varnothing\) and \(P=Q\).

For \(pq\in W_f\), suppose \({\mathbf{z}}\in\supp\widetilde f\) and
\(z_p\ne z_q\).  Choose distinct \(i,j\in[n]\setminus\{p,q\}\) with
\(z_i=z_j\).  Both \(ij\)-cards vanish on this restriction because every
nonzero one contains \(pq\). \Cref{eq:rm-card-pencil} then kills both
lifts, contradicting \({\mathbf{z}}\in\supp\widetilde f\).  Hence \(z_p=z_q\) on the
support.
\end{proof}

When \(W_f\ne\varnothing\), \cref{lem:rm-matching-consistency} supplies
global equality constraints.  The remaining issue is whether the
coefficients separate across these pairs.  The following lemma recovers
that product structure from diagonal restrictions.

\begin{lemma}[Diagonal-slice integration]
\label{lem:diagonal-slice-integration}
Let \(r\ge3\), and let \(h:\Ftwo^r\to\mathbb C^\times\) satisfy
\(h(\zero)=1\).  Suppose that, for every \(i\ne j\), the restriction
\(h|_{t_i=t_j}\), viewed as a signature on the common bit and the other
\(r-2\) bits, is a product of unary signatures (equivalently, is a simple
tensor).  If \(r\ge4\), then
\[
  h({\mathbf t})=\prod_{i=1}^r\rho_i^{t_i}
  \qquad(\rho_i\in\mathbb C^\times).
\]
If \(r=3\), there are
\(\rho_1,\rho_2,\rho_3,\kappa\in\mathbb C^\times\) such that, for distinct
\(i,j\),
\begin{equation}
  h({\mathbf{e}}_i)=\rho_i,\qquad
  h({\mathbf{e}}_i+{\mathbf{e}}_j)=\kappa\rho_i\rho_j,\qquad
  h(\one^3)=\kappa\rho_1\rho_2\rho_3.
  \label{eq:diagonal-slice-q3-table}
\end{equation}
\end{lemma}

\begin{proof}
Use the multiplicative subset expansion
\begin{equation}
  h({\mathbf t})=\prod_{\varnothing\ne S\subseteq[r]}
       \theta_S^{\prod_{i\in S}t_i},
   \qquad \theta_S\in\mathbb C^\times.
  \label{eq:multiplicative-moebius-expansion}
\end{equation}
Multiplicative M\"obius inversion determines its parameters:
\[
  \theta_S
  =\prod_{T\subseteq S}h(\one_T)^{(-1)^{|S|-|T|}}.
\]
The parameter \(\theta_S\) measures the interaction among the coordinates
in \(S\). Unary factors account exactly for the parameters with
\(|S|=1\).  Thus the coefficient tensor is simple if and only if
\(\theta_S=1\) for every \(|S|\ge2\).  On setting \(t_i=t_j\), a parameter indexed by a
nonempty \(U\subseteq[r]\setminus\{i,j\}\) and the common bit is
\begin{equation}
  \theta_{U\cup\{i\}}
  \theta_{U\cup\{j\}}
  \theta_{U\cup\{i,j\}}=1.
  \label{eq:diagonal-slice-moebius-relation}
\end{equation}
Parameters \(\theta_S\) with \(S\cap\{i,j\}=\varnothing\) and
\(|S|\ge2\) must also equal one.

Assume \(r\ge4\).  Choosing \(i,j\) outside \(S\) first gives
\(\theta_S=1\) for \(2\le|S|\le r-2\).  For \(|S|=r-1\), choose distinct
\(i,j\in S\) and define the index set \(U:=S\setminus\{i,j\}\) in
\cref{eq:diagonal-slice-moebius-relation}. The other two factors have
index-set cardinality \(r-2\), so \(\theta_S=1\).  Taking
\(U=[r]\setminus\{i,j\}\) then gives \(\theta_{[r]}=1\).  Only the unary
parameters remain in \cref{eq:multiplicative-moebius-expansion}, proving
the product formula.

Let \(r=3\).  The three instances of
\cref{eq:diagonal-slice-moebius-relation} imply
\[
  \theta_{12}=\theta_{13}=\theta_{23}=:\kappa,
  \qquad \theta_{123}=\kappa^{-2}.
\]
Substitution in \cref{eq:multiplicative-moebius-expansion}, with
\(\rho_i=\theta_{\{i\}}\), gives
\cref{eq:diagonal-slice-q3-table}.
\end{proof}

\begin{lemma}[Integration of a nonempty weight-two graph]
\label{lem:rm-weight-two-integration}
Let \(f\) be the fixed minimum-arity nonbinary irreducible signature of arity \(n\ge6\), with
\(E(Z_f)=\varnothing\).  If \(W_f\ne\varnothing\), then either \(n=6\) and a
constant-size two-copy gadget realizes the even-parity signature of arity
four, or \(f\) admits a nontrivial tensor factorization.  In particular, a
minimum-arity nonbinary irreducible signature of arity
at least eight has \(W_f=\varnothing\).
\end{lemma}

\begin{proof}
Fix \(E_1=\{u,v\}\in W_f\) and define the residual input set
\(R:=[n]\setminus E_1\).  For \(k\in R\), let
\(A_k:=C_{uk}^{+}\).  Since
\(\widetilde A_k(\zero)=2a_\varnothing\ne0\), this is a nonzero matching card.
The global equality on \(E_1\) gives the unique lift
\begin{align*}
  \widetilde A_k(z_v,{\mathbf z}_{R\setminus\{k\}})
  &=2\widetilde f(z_u=z_v,z_k=z_v,{\mathbf z}_{R\setminus\{k\}}),\\
  \widetilde A_k(\zero)&=2a_\varnothing\ne0,
  \\
  \widetilde A_k({\mathbf{e}}_p+{\mathbf{e}}_q)&=2a_{pq}
  \quad(p,q\in R\setminus\{k\}),\\
  \widetilde A_k({\mathbf{e}}_v+{\mathbf{e}}_q)&=2a_{\{u,v,k,q\}}
  \quad(q\in R\setminus\{k\}).
\end{align*}
Its matching therefore consists of \(W_f[R]-k\) and at most one edge from
\(v\).  To justify the last assertion, let \(U\) be the set of vertices not
covered by \(W_f\). It excludes \(u,v\).  For every \(k\), all vertices of
\(U\setminus\{k\}\) remain on the card boundary and have no available
weight-two edge, except that at most one can be paired through \(v\).
If \(W_f=\{E_1\}\), then \(|U|=n-2\ge4\), and the preceding restriction
already gives a contradiction.  Otherwise a covered vertex of \(W_f\) lies
in \(R\), so choose \(k\in R\setminus U\). Hence \(|U|\le1\).
Since \(|U|=n-2|W_f|\) is even, \(U=\varnothing\).  Thus \(W_f\) is
perfect. Write
its edges \(E_1,\ldots,E_s\), where \(s=n/2\).

By the global equality conclusion of
\cref{lem:rm-matching-consistency}, for every string \({\mathbf z}\) constant on each
\(E_r\), let \(t_r({\mathbf z})\) denote that common edge bit.  There is a logical
coefficient tensor \(h:\Ftwo^s\to\mathbb C\) such that
\begin{equation*}
  \widetilde f({\mathbf z})=
  \begin{cases}
    h(t_1({\mathbf z}),\ldots,t_s({\mathbf z})),
      &{\mathbf z}|_{E_r}\text{ is constant for every }r,\\
    0,&\text{otherwise}.
  \end{cases}
\end{equation*}
For two distinct \(W_f\)-edges \(E_i,E_j\), close one endpoint of \(E_i\)
to one endpoint of \(E_j\), forming separately the two cards with the
realizable \(+\) and \(-\) kernels.  Since
cross coefficients between different \(W_f\)-edges vanish and
\(a_\varnothing\ne0\), both cards are nonzero.  Both matching supports
impose \(t_i=t_j\). The \(+/-\) choice changes only the sign character in
that common logical bit.  Since \(s=n/2\ge3\),
every \({\mathbf t}\in\Ftwo^s\) has a pair \(i\ne j\) with \(t_i=t_j\). Use the
plus card for that pair.  The boundary string obtained from \({\mathbf t}\)
by retaining the mates of \(E_i,E_j\) and the two inputs of every other
\(E_r\) lies in this card's matching support.  The closure equation has a
unique compatible lift in the global equality code, namely the string with
edge bits \({\mathbf t}\), so the plus-card value is a fixed nonzero scalar times \(h({\mathbf t})\)
(there is no second lift with which it could cancel).  A nonzero matching
product is nonzero at every supported boundary string. Hence \(h({\mathbf t})\ne0\) for
every \({\mathbf t}\), and \(h\) has full support.
Rescale it by \(a_\varnothing^{-1}\) (absorbing
this nonzero scalar into \(f\)), so that \(h(\zero)=1\).  Every cross-pair
diagonal slice \(t_i=t_j\) is nonzero and rank one.  By
\cref{lem:diagonal-slice-integration},
\(s\ge4\) would make \(f\) a product of \(s\) nonsingular binaries, which
is the factorization alternative of the lemma.  Hence, on the remaining
branch, \(s=3\). Use the parameters
\(\rho_1,\rho_2,\rho_3,\kappa\ne0\) of
\cref{eq:diagonal-slice-q3-table}.  Let \(H_i^\pm\) be the \(2\times2\)
logical coefficient matrix obtained by closing \(E_i\) with the \(\pm\)
kernel.  For \(\{i,j,k\}=\{1,2,3\}\), one has
\begin{equation*}
  \det H_i^{\pm}
  =\rho_j \rho_k(\kappa-1)(1-\kappa \rho_i^2).
\end{equation*}
If \(\kappa=1\), \(h\), and hence \(f\), has the asserted product
factorization.  On the remaining branch \(\kappa\ne1\), and rank one of the
nonzero cards forces
\begin{equation*}
  \kappa \rho_1^2=\kappa \rho_2^2=\kappa \rho_3^2=1.
\end{equation*}

In Walsh coordinates, write \(E_i=\{i^0,i^1\}\) for the two
inputs of each matching edge. The realizable binary closing chains for
the \(I/X\)-contractions have Walsh kernels proportional to
\(\epsilon^u\one[u=v]\), with \(\epsilon=1\) for \(I\) and
\(\epsilon=-1\) for \(X\). Two such chains joining matching edges
contribute \((\epsilon_0\epsilon_1)^u\). Choosing their signs gives
either character on the common bit, using the original \(X\)-edges.

Let \(E'_1,E'_2,E'_3\) denote the corresponding matching edges in the second
copy.  Use two such binary closing chains for each displayed edge-to-edge join:
join the two inputs of \(E_3\) in the first copy to the two inputs of \(E'_1\)
in the second copy, and join the two inputs of \(E'_2\) to those of \(E'_3\)
inside the second copy. Leave both inputs of \(E_1\) and \(E_2\) in the first
copy external.  Evaluating these contractions by
\cref{eq:rm-card-pencil} gives the logical characters below, and every
internal matching assignment has a unique lift.  Choose the realizable
\(I/X\)-signs
\[
  \theta=\kappa \rho_2 \rho_3,
  \qquad
  \eta=-\kappa \rho_1 \rho_3,
\]
which lie in \(\{\pm1\}\).  Writing \(x=t_1\) and \(y=t_2\) for the two
external logical bits, the logical coefficient matrix is
\begin{equation*}
  Q(x,y)=
  \sum_{a,c\in\Ftwo}
  \eta^a\theta^c h(x,y,a)h(a,c,c).
\end{equation*}
Using \(\sum_c\theta^c h(a,c,c)=2\rho_1^a\), this becomes
\begin{equation*}
  Q\doteq
  \frac{\kappa-1}{\kappa}
  \begin{pmatrix}1&0\\0&\kappa \rho_1 \rho_2\end{pmatrix}.
\end{equation*}
Thus the realized quaternary signature has Walsh coordinates
\[
 \widetilde F
 =c\bigl(\delta_{0000}+\sigma\delta_{1111}\bigr),
 \qquad
 c\ne0,
 \qquad \sigma=\kappa \rho_1\rho_2\in\{\pm1\},
\]
where the two nonzero coefficients follow from \(\kappa\ne1\).  Applying
the inverse analytic change of coordinates, namely
\(F=P^{\otimes4}\widetilde F\), gives
\[
 F({\mathbf{x}})=c\bigl(1+\sigma(-1)^{x_1+x_2+x_3+x_4}\bigr).
\]
Define the even-parity signature by
\[
 \mathrm{Even}_4({\mathbf{x}}):=\one[x_1+x_2+x_3+x_4=0].
\]
Thus the realizable tensor is a nonzero scalar multiple of
\(\mathrm{Even}_4\) when \(\sigma=1\), and of its odd-parity coset when
\(\sigma=-1\). A realizable \(X\)-dressing on one exposed input toggles the
coset. For the equality-anchor condition, use the
matrix \(U:=D(i)P^{-1}\). It satisfies
\[
 U^{\otimes4}\mathrm{Even}_4\doteq(\Eq_4),\qquad
 (U^{-1})^{\mathsf T}XU^{-1}
 =D(i)^{-1}P^{\mathsf T}XP D(i)^{-1}=2I=2\,(\Eq_2).
\]
Consequently a realizable \(\mathrm{Even}_4\) equality anchor is obtained, and
\cref{lem:equality-anchor} applies.
\end{proof}

A nonempty \(W_f\) has therefore already produced an equality anchor or a
proper factorization.  After those alternatives have been resolved, the
remaining case has
\[
  W_f=\varnothing.
\]
Scale \(a_\varnothing=1\).  With all weight-two coefficients absent, the
common card matchings determine the entire Walsh support.

\begin{lemma}[Reed--Muller support reconstruction]
\label{lem:rm-local-slices}
\label{lem:rm-support-classification}
Let \(f\) be the fixed minimum-arity nonbinary irreducible signature of arity \(n\ge6\), normalized by
\(a_\varnothing=1\), and assume \(E(Z_f)=\varnothing\) and
\(W_f=\varnothing\).  For \(i\ne j\), let \(M_{ij}\) be the common support
matching of the two \(ij\)-cards.  Fix an arbitrary ordering of its
\(n/2-1\) edges and, for \({\mathbf t}\in\Ftwo^{n/2-1}\), write \(t_e\) for the
coordinate indexed by \(e\in M_{ij}\).  Let
\(D:=\supp\widetilde f\) be the support set.  Its equal-\(ij\) slice is
the \((n/2-1)\)-dimensional code
\begin{equation}
  D_{ij}=
  \left\{
    \one_{T({\mathbf t})}+(\wt({\mathbf t})\bmod2)({\mathbf{e}}_i+{\mathbf{e}}_j):
    {\mathbf t}\in\Ftwo^{n/2-1}
  \right\},
  \label{eq:rm-local-code}
\end{equation}
where
\(
  T({\mathbf t})=\bigcup_{e\in M_{ij}:\,t_e=1}e.
\)
Every codeword of \(D_{ij}\) has weight divisible by four.  Moreover,
\[
  n=8,
  \qquad
  D=RM(1,3),
\]
after a bijective relabelling of the inputs by \(\Ftwo^3\).
\end{lemma}

\begin{proof}
By \cref{lem:rm-matching-consistency}, the two \(ij\)-cards survive with a
common support matching \(M_{ij}\).  Normalize them at the zero string
\({\mathbf t}=\zero^{n/2-1}\) in matching coordinates, and write \(r_{ij,e}^{+}\) and
\(r_{ij,e}^{-}\) for the corresponding normalized edge ratios (the
\(t_e=1\) factor entry divided by its \(t_e=0\) entry).  For each
\(e\in M_{ij}\), set \(\rho_{ij,e}:=r_{ij,e}^{-}/r_{ij,e}^{+}\).  Their
quotient on the matching code---the repetition code consisting of assignments
constant on every edge of \(M_{ij}\), parametrized by \({\mathbf t}\)---is
\[
  \frac{\widetilde C_{ij}^{-}({\mathbf t})}
       {\widetilde C_{ij}^{+}({\mathbf t})}
  =\prod_{e\in M_{ij}}\rho_{ij,e}^{t_e}.
\]
Each single-edge quotient is \(-1\) by \(a_{pq}=0\) and
\cref{eq:rm-card-pencil}. Hence the quotient is \((-1)^{\wt({\mathbf t})}\).  Inverting
the pencil selects the \(ij=00\) lift for even \(\wt({\mathbf t})\) and \(ij=11\) for
odd \(\wt({\mathbf t})\), proving \cref{eq:rm-local-code}. The selected string has
weight \(2\wt({\mathbf t})\) in the even case and \(2\wt({\mathbf t})+2\) in the odd case, so
its weight is divisible by four in either case.

For \({\mathbf{x}},{\mathbf{y}}\in D\), some two coordinates have the same pair
\((x_i,y_i)\). The linear slice \(D_{ij}\) then contains \({\mathbf{x}},{\mathbf{y}},{\mathbf{x}}+{\mathbf{y}}\).
Thus \(D\) is linear.  For the linear functional
\(\ell_{ij}:\Ftwo^n\to\Ftwo\) defined by
\(\ell_{ij}({\mathbf z}):=z_i+z_j\), one has
\[
  D\cap\ker\ell_{ij}=D_{ij}.
\]
This subspace has dimension \(n/2-1\).  If
\(\dim D=n/2-1\), then \(D=D_{ij}\subseteq\ker\ell_{ij}\) for every
\(i\ne j\).  Hence every codeword of \(D\) has all coordinates equal, so
\(\dim D\le1\), contradicting \(n/2-1\ge2\).  Therefore
\(\dim D=n/2\).
Every codeword of \(D\) lies in one of the displayed slices and hence has weight
divisible by four.  For \({\mathbf{x}},{\mathbf{y}}\in D\),
\[
 0\equiv \wt({\mathbf{x}})+\wt({\mathbf{y}})-\wt({\mathbf{x}}+{\mathbf{y}})=2{\mathbf{x}}\cdot {\mathbf{y}}\pmod 4.
\]
Thus \(D\subseteq D^\perp\), and equal dimensions give
\begin{equation*}
  D=D^\perp.
\end{equation*}

For distinct \(i,j,p\), the mate \(q\) of \(p\) in \(M_{ij}\) gives the
unique weight-four codeword \(\{i,j,p,q\}\).  Now
\cref{lem:self-dual-steiner-reconstruction} yields
\(n=8\) and \(D=RM(1,3)\).
\end{proof}

\paragraph{Coefficient rigidity and the order of the cyclic group.}

The combinatorial reconstruction is complete: the surviving core has arity
eight and Reed--Muller support.  It remains to control its nonzero
coefficients.  The next lemma shows that they carry only a linear sign
character, so an available \(X\)-dressing removes the resulting affine
shift.

\begin{lemma}[Coefficients on \(\mathrm{RM}(1,3)\)]
\label{lem:rm-coefficient-atlas}
Let \(f\) be an arity-eight signature with
\(\supp\widetilde f=RM(1,3)\) whose proper \(I/X\)-cards are zero or have
the weighted-equality matching form of
\cref{eq:rm-weighted-matching}.  There are \(c\ne0\) and
\({\mathbf{w}}\in\Ftwo^8\) such
that
\begin{equation}
  \widetilde f({\mathbf{z}})
  =c(-1)^{{\mathbf{w}}\cdot {\mathbf{z}}}\one[{\mathbf{z}}\in RM(1,3)].
  \label{eq:rm-character-form}
\end{equation}
Consequently, back in the original \(K\)-coordinates,
\begin{equation}
  f({\mathbf{x}})\doteq\one[{\mathbf{x}}\in {\mathbf{w}}+RM(1,3)].
  \label{eq:rm-coset-form}
\end{equation}
\end{lemma}

\begin{proof}
Let the support code be \(D:=RM(1,3)\), and define
\(b({\mathbf{z}}):=\widetilde f({\mathbf{z}})/\widetilde f(\zero^8)\).  For an affine plane
\(A\subseteq\Ftwo^3\), its incidence vector \(\one_A\) is a codeword
of \(D\), with the fixed lexicographic input indexing.

Fix distinct points \(p,q\in\Ftwo^3\), and let
\(A_1,A_2,A_3\) be the three affine planes containing them.  Under the
identification of the equal-\(pq\) plus card with its three logical
matching coordinates, the codewords \(\one_{A_1},\one_{A_2},\one_{A_3}\) are the three singleton
logical codewords.  Its zero codeword lifts only to \(\zero^8\in D\), because \(D\)
contains no codeword of weight two.  Normalizing the card at that codeword, the
weighted-product form therefore gives
\begin{equation}
 b\left(\sum_{r\in S}\one_{A_r}\right)
   =\prod_{r\in S}b(\one_{A_r})
 \qquad\left(S\subseteq\{1,2,3\}\right).
 \label{eq:rm-plane-slice-product}
\end{equation}
In particular, \(\one_{A_1}+\one_{A_2}+\one_{A_3}=\one^8\).  The same argument shows that
whenever two affine planes \(A,B\) meet in a two-point affine line,
\begin{equation}
 b(\one_A+\one_B)=b(\one_{A})b(\one_{B}).
 \label{eq:rm-intersecting-plane-product}
\end{equation}

For a plane \(A\), write \(A^c:=\Ftwo^3\setminus A\).  Now
\(\one_{A_1}+\one_{A_2}=\one_{A_3^c}\) and \(A_3^c\) is an affine plane.  Since
\(A_2\cap A_3=\{p,q\}\), it meets \(A_2\) in the other two points of
\(A_2\), hence in a two-point line.  Applying
\cref{eq:rm-intersecting-plane-product} first to \(A_1,A_2\) and then to
\(A_3^c,A_2\) gives
\[
 b(\one_{A_1}+\one_{A_2})=b(\one_{A_1})b(\one_{A_2}),
 \qquad
 b(\one_{A_1})=b(\one_{A_1}+\one_{A_2})b(\one_{A_2}).
\]
Thus \(b(\one_{A_2})^2=1\).  Every affine plane can play the role of \(A_2\), so
\begin{equation}
 b(\one_{A})\in\{\pm1\}
 \qquad\text{for every affine plane }A.
 \label{eq:rm-plane-signs}
\end{equation}
For any affine plane \(A\), choose distinct \(p,q\in A\) and call the other
two planes through them \(B,C\).  Then \(\one_{A^c}=\one_B+\one_C\), and
\cref{eq:rm-plane-slice-product,eq:rm-plane-signs} give
\begin{equation}
 b(\one_{A^c})=b(\one_{B})b(\one_{C})=b(\one^8)b(\one_{A}).
 \label{eq:rm-complement-plane-product}
\end{equation}

The code \(D\) consists of \(\zero^8,\one^8\), and the fourteen affine-plane
incidence strings.  We can now check multiplicativity on all of \(D\).
The zero case is immediate.  Equal planes use
\cref{eq:rm-plane-signs}. Distinct noncomplementary planes meet in a
two-point line and use \cref{eq:rm-intersecting-plane-product}.
Complementary planes and sums of \(\one^8\) with a plane use
\cref{eq:rm-complement-plane-product}.  Finally,
\cref{eq:rm-plane-slice-product,eq:rm-plane-signs} imply
\(b(\one^8)^2=1\), which handles \(\one^8+\one^8\).  Hence
\[
 b({\mathbf{x}}+{\mathbf{y}})=b({\mathbf{x}})b({\mathbf{y}})
 \qquad({\mathbf{x}},{\mathbf{y}}\in D).
\]
Thus \(b\) is a sign character of \(D\).  Its binary exponent extends to a
linear functional on \(\Ftwo^8\), so there is \({\mathbf{w}}\in\Ftwo^8\) with
\(b({\mathbf{z}})=(-1)^{{\mathbf{w}}\cdot {\mathbf{z}}}\) for \({\mathbf{z}}\in D\).  This proves
\cref{eq:rm-character-form}.  Since \(D=D^\perp\), Fourier inversion gives
\[
  f({\mathbf{x}})\doteq\sum_{{\mathbf{z}}\in D}(-1)^{({\mathbf{x}}+{\mathbf{w}})\cdot {\mathbf{z}}}
  =|D|\one[{\mathbf{x}}+{\mathbf{w}}\in D].
\]
This is \cref{eq:rm-coset-form}.
\end{proof}

Available \(X\)-dressings translate the coset, so the standard core
\(\RMcore\) of \cref{eq:RM-core} is available.  In lexicographic order it is the
Shao--Cai signature \(f_8=\widehat f_8\). See \cref{rem:sc-f8-core}.

Define the matrix
\[
  \Theta=
  \begin{pmatrix}0&1\\1&1\end{pmatrix}
  \in\GL_2(\Ftwo),
  \qquad \Theta^3=I.
\]
In a tensor \(A\otimes B\otimes C\) of three ordered binary signatures,
\(A\), \(B\), and \(C\) act on input pairs \((1,2)\), \((3,4)\), and
\((5,6)\), respectively.  With this convention, the six-input Pauli core is
\begin{equation}
  \begin{aligned}
    \Hcore
    &=\sum_{{\boldsymbol{\gamma}}\in\Ftwo^2}
      B_{\boldsymbol{\gamma}}\otimes B_{\Theta{\boldsymbol{\gamma}}}\otimes B_{\Theta^2{\boldsymbol{\gamma}}}\\
    &=I\otimes I\otimes I
      +Z\otimes Y\otimes X
      +X\otimes Z\otimes Y
      +Y\otimes X\otimes Z.
  \end{aligned}
  \label{eq:H6}
\end{equation}

We use the ordered core definitions in \cref{eq:RM-core,eq:H6}, with
the lexicographic input order for \(\RMcore\) and the displayed three
ordered pairs for \(\Hcore\), to compare their Bell contractions.

A Bell product is a \emph{same-\({\boldsymbol{\gamma}}\) Bell product}
when every factor label \({\boldsymbol{\gamma}}_e\) equals one fixed \({\boldsymbol{\gamma}}\).

An even-arity irreducible signature \(f\) has the \emph{Bell property}
when, for every ordered pair of distinct inputs \((i,j)\) and every
\({\boldsymbol{\gamma}}\in\Ftwo^2\), the contraction
\(\partial_{(i,j)}^{B_{\boldsymbol{\gamma}}}f\) is a Bell product on the remaining inputs.
It has the \emph{strong Bell property} when each such contraction is, more
specifically, a same-\({\boldsymbol{\gamma}}\) Bell product.  These properties concern the displayed algebraic contractions.
They use the ordered-input, projective form of the definitions
\cite[full version, Definition~6.12]{ShaoCai2020} and
\cite[full version, Definition~1.10; conference version, Definition~12]%
{CaiFuShao2020Entanglement}.

\begin{remark}[The Shao--Cai \(f_8\) and \(\widehat f_6\) cores]
\label{rem:sc-f8-core}
\label{rem:sc-f6-core}
With the preceding input order, \(\RMcore\) is
the signature \(f_8=\widehat f_8\) of
\cite[full version, Section~8]{ShaoCai2020}. It has the strong Bell property:
contracting any ordered input pair by \(B_{\boldsymbol{\gamma}}\) gives, up to a nonzero
scalar, a product of three copies of \(B_{\boldsymbol{\gamma}}\) on the remaining inputs
\cite[Theorems~13--14]{CaiFuShao2020Entanglement}.
Let \(\widehat f_6\) denote the six-ary signature of Shao--Cai given in
\cite[full version, Eq.~(6.1)]{ShaoCai2020}.  Direct comparison with
\cref{eq:H6} gives the input-permutation identity
\[
 \Hcore(x_1,\ldots,x_6)
 =
 \widehat f_6(x_1,x_3,x_5,x_4,x_6,x_2).
\]
The same results show that \(\Hcore\) has the Bell property: every ordered
Bell contraction is, up to a nonzero scalar, a product of two ordered Bell
binaries
\cite[Theorems~13--14]{CaiFuShao2020Entanglement}. See also
\cite[full version, Definition~6.12 and Lemma~6.13]{ShaoCai2020}.
The arguments below reconstruct signatures from their contraction
properties and obtain a common presentation for the whole retained set.
The direct \(\Hcore\)-to-\(\RMcore\) construction is given in
\cref{proofsubsec:v4-H6-to-R8}.
\end{remark}

\begin{remark}[Limited appearance and internal \(\CycX{2}\) rigidity]
\label{rem:rm-binary-enlargement-shortcut}
Once \(\RMcore\) is available, the limited-appearance method of Shao and
Cai \cite[full version, Section~8.2, Lemmas~8.9--8.10]{ShaoCai2020}
provides an alternative route through a larger binary group. After the
binary and arity-four cases are resolved, its reduction permits adjoining
the Bell signatures. In our transfer convention, \(Z\) and \(Y\) give
\([ZX]=[Y]\) and \([YX]=[Z]\), whose matrix representatives lie outside the current transfer span
\(\Span\{I,X\}\). The enlarged set therefore has a strictly larger
complete transfer group and returns to
\cref{thm:marked-finite-group-routing} through the Turing reduction.

The proof below instead stays in the \(\CycX{2}\) case. Reed--Muller networks,
Radon contractions, and affine recognition establish the stronger
structural conclusion that the whole retained set and its edge have one
common flat-Lagrangian \(K\)-presentation. This is the conclusion used by
\cref{thm:reed-muller-rigidity} in the final synthesis.
\end{remark}

Thus the core itself is now fixed, up to the recorded equivalences.  The last
initial-core issue is the complete transfer group.  On the continuing
matching-deck branch, the coefficient classification also rules out every additional transfer class
beyond \([I]\) and \([X]\).

\begin{lemma}[The surviving Reed--Muller core forces \(G=\CycX{2}\)]
\label{lem:rm-safe-group-collapse}
Let \(\cF\) be a retained normalized signature set, let
\(G:=G(\cF)\) be its complete transfer group, and assume
\(\mathscr K_G=\Span\{I,X\}\).  Let \(f\in\cF\) be an irreducible
Reed--Muller core such that every single-input dressing of \(f\) by a safe binary remains
in the standing matching-deck branch.  Then the complete transfer group is
\[
  \{[I],[X]\}\cong C_2.
\]
\end{lemma}

\begin{proof}
Put \(D:=RM(1,3)\).  By \cref{lem:rm-coefficient-atlas}, choose
\(c\ne0\) and \({\mathbf w}\in\Ftwo^8\) such that
\(\widetilde f({\mathbf z})=c(-1)^{{\mathbf w}\cdot {\mathbf z}}\) for \({\mathbf z}\in D\), and define
\(b({\mathbf z}):=\widetilde f({\mathbf z})/c\).  Let \(T\) be the transfer matrix of a safe binary signature in the original basis, and write
\[
 T^{\mathrm W}:=P^{-1}TP
\]
for its Walsh-coordinate matrix. Since the transfer span is diagonal in these coordinates,
\(T^{\mathrm W}\doteq D(r)\) for some \(r\ne0\).
Dress one input \(p\) by \(T\).  This constant-size gadget preserves
irreducibility, support, and the matching-deck hypotheses. Its coefficient
table, after division by its value at \({\mathbf z}=\zero^8\), is
\[
 b_T({\mathbf z})=(-1)^{{\mathbf w}\cdot {\mathbf z}}r^{z_p}
 \qquad({\mathbf z}\in D).
\]
By \cref{lem:rm-coefficient-atlas}, \(b_T\), hence also the quotient
\(r^{z_p}=b_T({\mathbf z})/b({\mathbf z})\), is a character of \(D\) and takes values
\(\pm1\).  Evaluating at \(\one^8\in D\) gives \(r=\pm1\), precisely the
transfer classes in the original basis \([I]\) and \([X]\).
\end{proof}

\subsection{Flat-Lagrangian networks and Radon access}

Retain the \(\RMcore\) to build contexts around an arbitrary
signature. A flat-Lagrangian signature has the form
\(c\,\one[{\mathbf{x}}\in{\mathbf{a}}+L]\), where \(c\ne0\) and \(L=L^\perp\)
as in \cref{subsec:standard-classes}. The next lemma
shows both what these networks preserve and what they can construct.
Self-duality tracks the boundary after internal variables are summed out.
Weight-four transvections provide the reverse construction from a basis
of \(L\).

\paragraph{Symmetric binary spaces.}
\phantomsection\label{subsubsec:symmetric-binary-spaces}

A bilinear form \(b:W\times W\to\Ftwo\) is linear in each argument and is
symmetric when \(b({\mathbf{x}},{\mathbf{y}})=b({\mathbf{y}},{\mathbf{x}})\).  A \emph{symmetric binary space} is a
finite-dimensional \(\Ftwo\)-space with a specified symmetric form, written
\({\mathbf{x}}\cdot {\mathbf{y}}\).  On \(\Ftwo^m\), the standard dot product is
\({\mathbf{x}}\cdot {\mathbf{y}}:=\sum_jx_jy_j\).  For \(U\le W\), define
\[
 U^{\perp_W}:=\{{\mathbf{y}}\in W:{\mathbf{x}}\cdot {\mathbf{y}}=0\text{ for every }{\mathbf{x}}\in U\},
 \qquad
 \operatorname{rad}_W(U):=U\cap U^{\perp_W}.
\]
We omit \(W\) when it is clear.  The restricted form on \(U\) is
\emph{nondegenerate} when \(\operatorname{rad}_W(U)=\{0\}\).
The notation \(W=U\perp P\) means \(W=U\oplus P\) and \(U\cdot P=0\).
Then \(\pi_U({\mathbf{u}}+{\mathbf{p}})={\mathbf{u}}\) and \(\pi_P({\mathbf{u}}+{\mathbf{p}})={\mathbf{p}}\) are the orthogonal projections.

If \(R\le\operatorname{rad}_W(U)\), the quotient \(U/R\) has the induced
form \(({\mathbf{x}}+R)\cdot({\mathbf{y}}+R):={\mathbf{x}}\cdot {\mathbf{y}}\).  When \(R\) acts by translations on an affine space with
translation space \(U\), its orbit space has translation space \(U/R\).
A linear map \(\phi:U\to U'\) is an \emph{isometry} when it is bijective and
preserves the form. The corresponding affine isometry is
\({\mathbf{a}}+{\mathbf{u}}\mapsto {\mathbf{a}}'+\phi({\mathbf{u}})\).

A vector \({\mathbf{x}}\) is \emph{isotropic} when \({\mathbf{x}}\cdot {\mathbf{x}}=0\) and anisotropic
otherwise.  A subspace is \emph{totally isotropic} when
\(U\subseteq U^{\perp_W}\), and is \emph{self-dual in \(W\)} when
\(U=U^{\perp_W}\).  In a nondegenerate symmetric binary space, the map
\({\mathbf{x}}\mapsto {\mathbf{x}}\cdot {\mathbf{x}}\) is linear, so nondegeneracy identifies it with a
unique vector \({\mathbf{c}}\) satisfying \({\mathbf{x}}\cdot {\mathbf{x}}={\mathbf{c}}\cdot {\mathbf{x}}\). This is the
\emph{characteristic vector}.  The space is alternating when \({\mathbf{c}}=0\), and
nonalternating otherwise.

We use the following facts about symmetric binary spaces:
nondegenerate subspaces split off orthogonally. Every even-dimensional
nondegenerate symmetric binary space has a self-dual subspace. Every
nondegenerate nonalternating \(m\)-space is isometric to the standard
dot-product space \(\Ftwo^m\). Every odd-dimensional nondegenerate
symmetric binary space is nonalternating and has an anisotropic vector.  Here are short proofs.  For a nondegenerate subspace
\(P\), its pairing map identifies \(P\) with \(P^\vee\), so each ambient
vector differs from a unique vector of \(P\) by an element of \(P^\perp\).
This gives the orthogonal splitting.  A nonzero nondegenerate alternating
space contains vectors \({\mathbf{u}},{\mathbf{v}}\) with \({\mathbf{u}}\cdot {\mathbf{v}}=1\). Their Gram matrix is
\(H=\left(\begin{smallmatrix}0&1\\1&0\end{smallmatrix}\right)\).
Splitting off these planes shows that an alternating space has even
dimension, and one vector from each plane spans a self-dual subspace.
In a nonalternating space, split off norm-one vectors until the remainder
is alternating.  A norm-one vector \({\mathbf{e}}\) orthogonal to such a plane can
absorb it: \({\mathbf{e}}+{\mathbf{u}},{\mathbf{e}}+{\mathbf{v}},{\mathbf{e}}+{\mathbf{u}}+{\mathbf{v}}\) are an orthogonal norm-one basis of their
three-dimensional span.  Repeating this gives the standard dot-product
basis.  In even dimension \(2k\), the vectors
\({\mathbf{e}}_1+{\mathbf{e}}_2,\ldots,{\mathbf{e}}_{2k-1}+{\mathbf{e}}_{2k}\) span a self-dual subspace of that
space (the empty span applies in dimension zero).  Finally, odd dimension
excludes the alternating case and hence supplies a norm-one vector.

\begin{lemma}[Exact flat-Lagrangian network calculus]
\label{lem:rm-pure-boundary}
\label{lem:rm-flat-contexts}
\begin{enumerate}
\item Let \(\Omega\) be a tensor network with ordered boundary \(B\).
      Suppose that the signature at each tensor vertex \(u\) is
      proportional to \(\one[{\mathbf{x}}_u\in {\mathbf{a}}_u+L_u]\), where either
      \(L_u=RM(1,3)\le\Ftwo^8\) or
      \(L_u=\Span_{\Ftwo}\{(1,1)\}\le\Ftwo^2\), and \({\mathbf{a}}_u\)
      is a shift in the corresponding ambient space.  The binary case
      includes ordinary \(I/X\)-vertices.  Suppose every internal
      connection is labelled by \(I\) or \(X\).  Then the boundary signature
      is either zero or
\begin{equation}
  {\mathbf{x}}_B\longmapsto c_B\one[{\mathbf{x}}_B\in {\mathbf{a}}_B+L],
  \qquad c_B\ne0,\quad {\mathbf{a}}_B\in\Ftwo^B,\quad L=L^\perp\le\Ftwo^B.
  \label{eq:rm-pure-flat}
\end{equation}
\item Conversely, for every \({\mathbf{a}}_B\in\Ftwo^{2k}\) and every
      \(L=L^\perp\le\Ftwo^{2k}\), a polynomial-size
      \(\RMcore/I/X\)-network directly realizes a nonzero scalar multiple of
      \({\mathbf{x}}_B\longmapsto\one[{\mathbf{x}}_B\in {\mathbf{a}}_B+L]\).
\end{enumerate}
In particular every nonzero boundary produced by a network in item~1 has
even arity.
\end{lemma}

\begin{proof}
For item~1, apply the boundary normalization following
\cref{eq:half-edge-boundary-tensor}. The isolated \(I/X\) components
are flat-Lagrangian and contribute direct summands to the boundary
translation space. It therefore suffices to treat a boundary \(B\)
consisting of unpaired vertex inputs. Let \(E_{\rm int}\) be the set of internal
half-edge pairs and put
\[
 H_{\rm int}:=\bigcup E_{\rm int},
 \qquad
 V_{\rm he}:=\Ftwo^B\oplus\Ftwo^{H_{\rm int}}.
\]
For each \(e=\{h,h'\}\in E_{\rm int}\), let \(\eta_e=0\) when \(e\) is
labelled by \(I\) and \(\eta_e=1\) when it is labelled by \(X\).  Define
\[
 E_\eta:=
 \{{\mathbf{y}}\in\Ftwo^{H_{\rm int}}:
   y_h+y_{h'}=\eta_e
   \text{ for every }e=\{h,h'\}\in E_{\rm int}\},
\]
and let
\[
 E_{\rm eq}:=
 \{{\mathbf{y}}\in\Ftwo^{H_{\rm int}}:
   y_h=y_{h'}\text{ for every }\{h,h'\}\in E_{\rm int}\},
 \qquad
 H_{\rm con}:=\Ftwo^B\oplus E_{\rm eq}.
\]
The translation space \(E_{\rm eq}\) is self-dual in
\(\Ftwo^{H_{\rm int}}\), and hence
\[
 H_{\rm con}^\perp
 =\{\mathbf 0_B\}\oplus E_{\rm eq},
\]
where, for any finite index set \(R\), \(\mathbf 0_R\) denotes the zero
vector in \(\Ftwo^R\).

Let \(V(\Omega)\) denote the tensor-vertex set of \(\Omega\), and put
\[
 C_0:=
 \bigoplus_{u\in V(\Omega)}L_u
 =C_0^\perp,
 \qquad
 \mathcal K:=C_0\cap H_{\rm con}.
\]
Let \({\mathbf{a}}_{\rm vert}:=({\mathbf{a}}_u)_{u\in V(\Omega)}\) be the concatenated vertex
shift.  The feasible assignments form
\[
 ({\mathbf{a}}_{\rm vert}+C_0)
 \cap\bigl(\Ftwo^B\oplus E_\eta\bigr).
\]
This intersection is either empty or an affine coset
\({\mathbf{y}}_0+\mathcal K\).  Define
\[
 \pi_B:V_{\rm he}\to\Ftwo^B,
 \qquad
 L:=\pi_B(\mathcal K),
 \qquad
 {\mathbf{a}}_B:=\pi_B({\mathbf{y}}_0).
\]
Every nonempty boundary fibre is a coset of
\(\ker(\pi_B|_{\mathcal K})\), so all fibres have size
\(2^{\dim\ker(\pi_B|_{\mathcal K})}\).  Multiplying this number by
the nonzero vertex scalars gives \(c_B\ne0\) in
\cref{eq:rm-pure-flat}.  The support and scalar are therefore determined
by affine Gaussian elimination.

To prove \(L\subseteq L^\perp\), take \({\mathbf{k}},{\mathbf{k}}'\in\mathcal K\).  Since
\({\mathbf{k}},{\mathbf{k}}'\in C_0=C_0^\perp\), we have \({\mathbf{k}}\cdot {\mathbf{k}}'=0\).  For an internal pair
\(e=\{h,h'\}\), put
\[
 t:=k_h=k_{h'},
 \qquad
 s:=k'_h=k'_{h'}.
\]
That pair contributes \(ts+ts=0\) to \({\mathbf{k}}\cdot {\mathbf{k}}'\).  Consequently
\(\pi_B({\mathbf{k}})\cdot\pi_B({\mathbf{k}}')=0\), proving \(L\subseteq L^\perp\).

Moreover,
\[
 \mathcal K^\perp=C_0+H_{\rm con}^\perp.
\]
If \({\mathbf{b}}\in L^\perp\), then
\(({\mathbf{b}},\mathbf 0_{H_{\rm int}})\in\mathcal K^\perp\), so write
\[
 ({\mathbf{b}},\mathbf 0_{H_{\rm int}})={\mathbf{c}}_0+{\mathbf{d}},
 \qquad
 {\mathbf{c}}_0\in C_0,\quad {\mathbf{d}}\in H_{\rm con}^\perp.
\]
Both \(({\mathbf{b}},\mathbf 0_{H_{\rm int}})\) and \({\mathbf{d}}\) belong to
\(H_{\rm con}\), hence \({\mathbf{c}}_0\in C_0\cap H_{\rm con}=\mathcal K\).
Therefore \({\mathbf{b}}=\pi_B({\mathbf{c}}_0)\in L\), proving \(L=L^\perp\).

For the converse, put inputs \(000,100,010,001\) of one \(\RMcore\)
vertex on the input side and inputs \(111,011,101,110\) on the output
side, in the displayed orders.  Direct
evaluation gives
\[
 {\mathbf{y}}=(I_4+J_4){\mathbf{x}}={\mathbf{x}}+(x_1+x_2+x_3+x_4)\one^4.
\]
Here \({\mathbf{x}},{\mathbf{y}}\in\Ftwo^4\) are respectively the ordered input and output strings,
\(I_4\) is the \(4\times4\) identity matrix, and
\(J_4:=\one^4(\one^4)^{\mathsf T}\) is the \(4\times4\) all-ones matrix over
\(\Ftwo\).
Thus realizable \(\RMcore\) vertices and permutations of inputs realize every weight-four
transvection \(\tau_{\mathbf{s}}({\mathbf{x}})={\mathbf{x}}+({\mathbf{s}}\cdot {\mathbf{x}}){\mathbf{s}}\). \(I\)-vertices (or
\(X-I-X\) chains) compose them.  We claim by induction on \(k\) that every
\(L=L^\perp\le\Ftwo^{2k}\) is obtained from
 \(L_0=\Span_{\Ftwo}\{{\mathbf{e}}_1+{\mathbf{e}}_2,{\mathbf{e}}_3+{\mathbf{e}}_4,\ldots\}\).  The cases \(k=0,1\) are
immediate.  For \(k\ge2\), self-orthogonality gives \({\mathbf{x}}\cdot {\mathbf{x}}=0\) for
every \({\mathbf{x}}\in L\), so all strings are even and \(\one\in L\).  Choose
\({\mathbf{z}}\in L\setminus\{\zero,\one\}\). One of \({\mathbf{z}},{\mathbf{z}}+\one\) has even weight in
\([2,k]\).  If its weight is larger than two, three support coordinates and
one zero coordinate give a weight-four \({\mathbf{s}}\) with \({\mathbf{z}}\cdot {\mathbf{s}}=1\) and
\(\wt({\mathbf{z}}+{\mathbf{s}})=\wt({\mathbf{z}})-2\).  Apply the corresponding transvection to the whole
current code and repeat, then permute coordinates, to obtain
\({\mathbf{p}}={\mathbf{e}}_1+{\mathbf{e}}_2\in L\).  Every codeword has equal first two coordinates.
 Projection deleting them has kernel \(\Span_{\Ftwo}\{{\mathbf{p}}\}\), image dimension
\(k-1\), and preserves dot products (the two deleted contributions cancel),
so its image is self-dual.  Recurse and invert the accumulated realizable
transvections and permutations.  The resulting circuit has size \(O(k^2)\).
Closing the standard input pairs by \(I\) gives one internal lift per
boundary string.  For each \(({\mathbf{a}}_B)_j=1\), attach a retained ordinary
\(I\)-vertex to the corresponding circuit input through one \(X\)-edge and expose the other ordinary input of that vertex.  Since
\(XI=X\), this attachment complements exactly that boundary bit.
Leave the other circuit inputs unchanged. Applying these dressings
independently at the exposed vertex inputs realizes every affine coset
\({\mathbf{a}}_B+L\).
\end{proof}

For an orthogonal map \(A\), define the self-dual graph subspace
\(\operatorname{Graph}(A):=\{({\mathbf{x}},A{\mathbf{x}}):{\mathbf{x}}\in\Ftwo^n\}\).  The converse of
\cref{lem:rm-flat-contexts} realizes its indicator signature.
Attaching a signature to its input and closing one output pair then sums
the signature over prescribed two-point orbits.  This supplies the
following Radon cards for the high-arity argument.

\paragraph{Realizable Radon access.}

We use \hyperref[subsubsec:symmetric-binary-spaces]{the symmetric-space conventions},
including their ambient-relative orthogonal notation.

Let \(g\) have even arity \(n\).  Call
\({\mathbf{v}}\in(\one^n)^\perp\setminus\{\zero^n,\one^n\}\) \emph{accessible}.  Such a
\({\mathbf{v}}\) is isotropic.  For \(\epsilon\in\Ftwo\), define
\[
 H_{{\mathbf{v}},\epsilon}:=\{{\mathbf{x}}\in\Ftwo^n:{\mathbf{x}}\cdot {\mathbf{v}}=\epsilon\}.
\]
Translation by \({\mathbf{v}}\) preserves \(H_{{\mathbf{v}},\epsilon}\).  For
\({\mathbf{x}}\in H_{{\mathbf{v}},\epsilon}\), write
\[
 [{\mathbf{x}}]_{\mathbf{v}}:={\mathbf{x}}+\Span_{\Ftwo}\{{\mathbf{v}}\}
\]
for its translation orbit, and let
\(H_{{\mathbf{v}},\epsilon}/\Span_{\Ftwo}\{{\mathbf{v}}\}\) denote the set of these orbits.  Define
\[
 [{\mathbf{u}}]_{\mathbf{v}}\cdot[{\mathbf{w}}]_{\mathbf{v}}:={\mathbf{u}}\cdot {\mathbf{w}}
 \qquad({\mathbf{u}},{\mathbf{w}}\in {\mathbf{v}}^\perp).
\]
This is a well-defined nondegenerate symmetric form on the translation
space \({\mathbf{v}}^\perp/\Span_{\Ftwo}\{{\mathbf{v}}\}\). Each
\(H_{{\mathbf{v}},\epsilon}/\Span_{\Ftwo}\{{\mathbf{v}}\}\) is an affine space over that quotient.
Define
\hypertarget{eq:rm-radon-card}{}
\begin{equation*}
  \mathcal R_{{\mathbf{v}},\epsilon}g([{\mathbf{x}}]_{\mathbf{v}})
  =g({\mathbf{x}})+g({\mathbf{x}}+{\mathbf{v}}),
  \qquad [{\mathbf{x}}]_{\mathbf{v}}\in H_{{\mathbf{v}},\epsilon}/\Span_{\Ftwo}\{{\mathbf{v}}\}.
\end{equation*}
The right side is invariant under replacing \({\mathbf{x}}\) by \({\mathbf{x}}+{\mathbf{v}}\), so the
quotient signature is well defined.  The construction below identifies
the quotient with \(n-2\) ordinary boundary inputs.  Thus the quotient
sum is a realizable arity-reducing operation in the original network model.
Write
\[
 O(n,2):=\{A\in\operatorname{GL}_n(\Ftwo):A^{\mathsf T}A=I_n\}.
\]

\begin{lemma}[Accessible Radon directions]
\label{lem:rm-radon-accessible}
Let \(g\) be a retained signature of even arity \(n\ge4\).  For each
\({\mathbf{v}}\in(\one^n)^\perp\setminus\{\zero^n,\one^n\}\), one can construct and
record distinct inputs \(i_{\mathbf{v}},j_{\mathbf{v}}\) and an orthogonal matrix
\(A_{\mathbf{v}}\in O(n,2)\) such that
\[
 A_{\mathbf{v}}^{\mathsf T}({\mathbf{e}}_{i_{\mathbf{v}}}+{\mathbf{e}}_{j_{\mathbf{v}}})={\mathbf{v}}.
\]
Define
\[
 {\mathbf{r}}_{\mathbf{v}}:=A_{\mathbf{v}}^{\mathsf T}{\mathbf{e}}_{i_{\mathbf{v}}},\qquad
 \Pi_{\mathbf{v}}:=\Span\{{\mathbf{r}}_{\mathbf{v}},{\mathbf{v}}\},\qquad W_{\mathbf{v}}:=\Pi_{\mathbf{v}}^\perp,
\]
and, for \(\epsilon\in\Ftwo\),
\[
 \iota_{{\mathbf{v}},\epsilon}:W_{\mathbf{v}}\longrightarrow
 H_{{\mathbf{v}},\epsilon}/\Span\{{\mathbf{v}}\},\qquad
 {\mathbf{w}}\longmapsto[\epsilon {\mathbf{r}}_{\mathbf{v}}+{\mathbf{w}}]_{\mathbf{v}}.
\]
Then \(\Ftwo^n=\Pi_{\mathbf{v}}\perp W_{\mathbf{v}}\), and \(\iota_{{\mathbf{v}},\epsilon}\) is
an affine isometry from \(W_{\mathbf{v}}\) to
\(H_{{\mathbf{v}},\epsilon}/\Span\{{\mathbf{v}}\}\).  An \(O(n^2)\)-size network using
one copy of \(g\), \(\RMcore\), and \(I/X\) realizes
\(\mathcal R_{{\mathbf{v}},\epsilon}g\) in the boundary coordinates supplied by
\(A_{\mathbf{v}}|_{W_{\mathbf{v}}}\), up to a nonzero scalar.  The available
directions are exactly the displayed vectors \({\mathbf{v}}\).
\end{lemma}

\begin{proof}
First take \(A_{\mathbf{v}},i_{\mathbf{v}},j_{\mathbf{v}}\) as in the statement. Their construction is
given at the end of the proof.  The definitions imply
\(\Pi_{\mathbf{v}}=A_{\mathbf{v}}^{\mathsf T}\Span\{{\mathbf{e}}_{i_{\mathbf{v}}},{\mathbf{e}}_{j_{\mathbf{v}}}\}\).
Then
\[
 {\mathbf{r}}_{\mathbf{v}}\cdot {\mathbf{r}}_{\mathbf{v}}=1,\qquad
 {\mathbf{r}}_{\mathbf{v}}\cdot {\mathbf{v}}=1,\qquad
 {\mathbf{v}}\cdot {\mathbf{v}}=0.
\]
Hence \(\Pi_{\mathbf{v}}\) is a nondegenerate two-plane and
\[
 \Ftwo^n=\Pi_{\mathbf{v}}\perp W_{\mathbf{v}},
 \qquad
 {\mathbf{v}}^\perp=\Span_{\Ftwo}\{{\mathbf{v}}\}\oplus W_{\mathbf{v}}.
\]
The isometry \(A_{\mathbf{v}}\) carries \(\Pi_{\mathbf{v}}\) to the output-coordinate
plane \(\Span\{{\mathbf{e}}_{i_{\mathbf{v}}},{\mathbf{e}}_{j_{\mathbf{v}}}\}\) and \(W_{\mathbf{v}}\) to its ordinary-coordinate
complement.  We use \(A_{\mathbf{v}}\) to identify these pairs of spaces and call the
corresponding output positions the \(\Pi_{\mathbf{v}}\)-inputs and the \(W_{\mathbf{v}}\)-inputs.
Consequently \(\iota_{{\mathbf{v}},\epsilon}\) has the asserted domain, range,
and isometry property.

For \(A\in O(n,2)\), the graph subspace \(\operatorname{Graph}(A)\)
is self-dual in \(\Ftwo^n\oplus\Ftwo^n\): it is isotropic
because
\({\mathbf{x}}\cdot {\mathbf{y}}+(A{\mathbf{x}})\cdot(A{\mathbf{y}})=0\), and it has half the ambient dimension.
Hence \cref{lem:rm-flat-contexts} realizes a nonzero scalar multiple of
\[
 ({\mathbf{x}},{\mathbf{y}})\longmapsto\one[{\mathbf{y}}=A_{\mathbf{v}}{\mathbf{x}}]
\]
by a realizable \(\RMcore/I/X\)-network.  Attach one copy of \(g\) to the
graph input using an effective \(I\)-kernel on each input connection:
insert one retained ordinary \(I\)-vertex between two \(X\)-edges,
so that \(XIX=I\).  The output before closing a pair is therefore
\(g(A_{\mathbf{v}}^{\mathsf T}{\mathbf{y}})\), with the graph's nonzero scalar.
Close the ordered output pair \((i_{\mathbf{v}},j_{\mathbf{v}})\) with either realizable kernel
\(I\) or realizable kernel \(X\).  The two internal lifts differ by
\({\mathbf{v}}\). The choice of \(I\) or \(X\) fixes \({\mathbf{x}}\cdot {\mathbf{v}}=\epsilon\).  Under the
isometry \(A_{\mathbf{v}}|_{W_{\mathbf{v}}}\), the remaining output coordinates are precisely
the \(W_{\mathbf{v}}\)-coordinates in the inverse of \(\iota_{{\mathbf{v}},\epsilon}\).  Thus the
resulting boundary is \(\mathcal R_{{\mathbf{v}},\epsilon}g\), up to its nonzero scalar.

The weight-four transvections constructed in
\cref{lem:rm-flat-contexts}, together with coordinate permutations, send
every nonzero even \({\mathbf{v}}\ne\one^n\) to a weight-two vector.  Write the
resulting orthogonal map as \(B\), with \(B{\mathbf{v}}={\mathbf{e}}_{i_{\mathbf{v}}}+{\mathbf{e}}_{j_{\mathbf{v}}}\), and
record \(A_{\mathbf{v}}=B\).  Then
\(A_{\mathbf{v}}^{\mathsf T}({\mathbf{e}}_{i_{\mathbf{v}}}+{\mathbf{e}}_{j_{\mathbf{v}}})=B^{-1}B{\mathbf{v}}={\mathbf{v}}\), as required.  Conversely, orthogonal maps preserve parity and fix
\(\one^n\), so neither \(\zero^n\) nor \(\one^n\) lies in the orbit of a
weight-two vector.  This proves necessity as well.  The transvection
construction and graph network use \(O(n^2)\) vertices. The associated
binary linear systems are solved by Gaussian elimination.
\end{proof}

\subsection{A common flat-Lagrangian presentation}

We now consider an arbitrary unresolved irreducible occurrence \(g\).
For arity at most eight, its smaller flat-Lagrangian cards are matching products, so
the reconstruction above applies.  At higher arity, Radon cards compare
two quotient layers.  A failure to combine them into an affine signature
can be retained on at most eight inputs. After that obstruction is excluded,
affine patching and dimension counts recover a flat-Lagrangian \(g\).

\paragraph{Affine phase and translation tools.}
\phantomsection\label{subsubsec:rm-affine-phase-tools}

The following two facts test whether the quotient layers combine into an
affine signature.  They use the affine class defined in
\cref{subsec:standard-classes}.

For a nowhere-zero signature \(f:\Ftwo^k\to\AlgNums\setminus\{0\}\), define
its multiplicative third-difference ratio
\(\mathcal T_f:(\Ftwo^k)^4\to\AlgNums^\times\) by
\begin{equation*}
 \mathcal T_f({\mathbf{z}},{\mathbf{u}},{\mathbf{v}},{\mathbf{w}})=
 \frac{f({\mathbf{z}}+{\mathbf{u}}+{\mathbf{v}}+{\mathbf{w}})f({\mathbf{z}}+{\mathbf{u}})f({\mathbf{z}}+{\mathbf{v}})f({\mathbf{z}}+{\mathbf{w}})}
 {f({\mathbf{z}}+{\mathbf{u}}+{\mathbf{v}})f({\mathbf{z}}+{\mathbf{u}}+{\mathbf{w}})f({\mathbf{z}}+{\mathbf{v}}+{\mathbf{w}})f({\mathbf{z}})}.
\end{equation*}
Here \({\mathbf{z}},{\mathbf{u}},{\mathbf{v}},{\mathbf{w}}\in\Ftwo^k\), and repeated directions are permitted.
For fixed \({\mathbf{u}},{\mathbf{v}}\), let \(D_{{\mathbf{u}},{\mathbf{v}}}\) be the multiplicative
second-difference operator acting on the entire signature \(f\):
\begin{equation*}
 \bigl(D_{{\mathbf{u}},{\mathbf{v}}}f\bigr)({\mathbf{z}})
 :=
 \frac{f({\mathbf{z}}+{\mathbf{u}}+{\mathbf{v}})f({\mathbf{z}})}
      {f({\mathbf{z}}+{\mathbf{u}})f({\mathbf{z}}+{\mathbf{v}})}.
\end{equation*}

\begin{lemma}[Affine phase criterion]
\label{lem:affine-phase-criterion}
Let \(f:\Ftwo^k\to\AlgNums\setminus\{0\}\).  Then \(f\) belongs to
\(\cA\) if and only if
\(\mathcal T_f({\mathbf{z}},{\mathbf{u}},{\mathbf{v}},{\mathbf{w}})=1\) for all \({\mathbf{z}},{\mathbf{u}},{\mathbf{v}},{\mathbf{w}}\in\Ftwo^k\).
\end{lemma}

\begin{proof}
Define the scalar \(\lambda=f(\zero^k)\), the scalars
\(A_j=f({\mathbf{e}}_j)/\lambda\) for \(j\in[k]\), and, for
\(1\le j<\ell\le k\), the scalars
\[
 B_{j\ell}=\frac{\lambda f({\mathbf{e}}_j+{\mathbf{e}}_\ell)}{f({\mathbf{e}}_j)f({\mathbf{e}}_\ell)}.
\]
Substitution of \({\mathbf{z}}=\zero^k\) and \({\mathbf{u}}={\mathbf{v}}={\mathbf{w}}={\mathbf{e}}_j\) gives \(A_j^4=1\). Taking
\({\mathbf{u}}={\mathbf{v}}={\mathbf{e}}_j\) and \({\mathbf{w}}={\mathbf{e}}_\ell\) gives \(B_{j\ell}^2=1\).  Moreover, the identity
\(\mathcal T_f({\mathbf{z}},{\mathbf{u}},{\mathbf{v}},{\mathbf{w}})=1\) says exactly that
\[
 \bigl(D_{{\mathbf{u}},{\mathbf{v}}}f\bigr)({\mathbf{z}}+{\mathbf{w}})=\bigl(D_{{\mathbf{u}},{\mathbf{v}}}f\bigr)({\mathbf{z}}).
\]
Because \({\mathbf{w}}\) is arbitrary, \(D_{{\mathbf{u}},{\mathbf{v}}}f\) is constant on \(\Ftwo^k\).
In particular,
\[
 \bigl(D_{{\mathbf{e}}_j,{\mathbf{e}}_\ell}f\bigr)({\mathbf{z}})=
 \bigl(D_{{\mathbf{e}}_j,{\mathbf{e}}_\ell}f\bigr)(\zero^k)=B_{j\ell}
 \qquad(1\le j<\ell\le k).
\]

We now give the Hamming-weight induction explicitly.  Extend the notation
symmetrically by \(B_{\ell j}=B_{j\ell}\).  If \(z_j=0\), start with
\(f({\mathbf{e}}_j)/f(\zero^k)=A_j\) and add the basis vectors in \(\supp({\mathbf{z}})\) one at a
time.  The identity \(D_{{\mathbf{e}}_j,{\mathbf{e}}_\ell}f=B_{j\ell}\), rearranged after the
previously added vector \({\mathbf{y}}\), is
\[
 \frac{f({\mathbf{y}}+{\mathbf{e}}_\ell+{\mathbf{e}}_j)}{f({\mathbf{y}}+{\mathbf{e}}_\ell)}
 =
 B_{j\ell}\frac{f({\mathbf{y}}+{\mathbf{e}}_j)}{f({\mathbf{y}})}.
\]
It follows by induction on \(\wt({\mathbf{z}})\) that
\begin{equation}
 \frac{f({\mathbf{z}}+{\mathbf{e}}_j)}{f({\mathbf{z}})}
 =A_j\prod_{\ell\ne j}B_{j\ell}^{z_\ell}
 \qquad(z_j=0).
 \label{eq:affine-phase-ratio}
\end{equation}
We next prove the desired formula for \(f({\mathbf{z}})\), again by induction on
\(\wt({\mathbf{z}})\).  It is immediate for \({\mathbf{z}}=\zero^k\).  For
\({\mathbf{z}}\ne\zero^k\), choose
\(j\in\supp({\mathbf{z}})\) and define the predecessor string \({\mathbf{y}}:={\mathbf{z}}+{\mathbf{e}}_j\), so \(y_j=0\) and
\(\wt({\mathbf{y}})=\wt({\mathbf{z}})-1\).  Applying the induction hypothesis to \(f({\mathbf{y}})\) and
\cref{eq:affine-phase-ratio} to \(f({\mathbf{y}}+{\mathbf{e}}_j)/f({\mathbf{y}})\) yields
\[
 f({\mathbf{z}})=\lambda\prod_jA_j^{z_j}\prod_{j<\ell}B_{j\ell}^{z_jz_\ell},
 \qquad A_j\in\mu_4,\quad B_{j\ell}\in\{1,-1\},
\]
because the new factors are precisely \(A_j\) and
\(B_{j\ell}\) for \(\ell\in\supp({\mathbf{y}})\).  This is the full-support normal form in
\cref{eq:affine-class}.  Conversely, each linear factor and each
even-quadratic factor cancels in the multiplicative third difference, so
\(\mathcal T_f=1\).
\end{proof}

\begin{lemma}[Two-translate closure of the affine class]
\label{lem:affine-two-translate-closure}
If \(h\in\cA\) and \({\mathbf{w}}\) is a Boolean translation of its input
coordinates, then \(h+T_{\mathbf{w}}h\in\cA\) (with the zero signature allowed), where
the translation operator \(T_{\mathbf{w}}\) is defined by \((T_{\mathbf{w}}h)({\mathbf{x}}):=h({\mathbf{x}}+{\mathbf{w}})\).
\end{lemma}

\begin{proof}
The zero case is immediate.  Let \(n:=\arity(h)\).  For \(h\ne0\), write
\[
 h({\mathbf{x}})=c\,\one[{\mathbf{x}}\in {\mathbf{a}}+L]i^{Q({\mathbf{x}})},
 \qquad c\in\AlgNums^\times,\quad {\mathbf{a}}\in\Ftwo^n,\quad
 L\le\Ftwo^n,
\]
where \(Q:\Ftwo^n\to\mathbb Z_4\) is a quadratic phase polynomial with even mixed
coefficients.  If \({\mathbf{w}}\notin L\), every string in the union of the two supports
has a unique form \({\mathbf{a}}+{\mathbf{y}}+t{\mathbf{w}}\), with \({\mathbf{y}}\in L\) and \(t\in\Ftwo\).  On this
affine space,
\[
 (h+T_{\mathbf{w}}h)({\mathbf{a}}+{\mathbf{y}}+t{\mathbf{w}})=c\,i^{Q({\mathbf{a}}+{\mathbf{y}})}.
\]
The inverse coordinates \(({\mathbf{y}},t)\) are affine, so the displayed exponent
is a quadratic phase polynomial with even mixed coefficients.  Thus the sum is affine.
For \({\mathbf{w}}\in L\), expand
\(Q({\mathbf{x}}+{\mathbf{w}})-Q({\mathbf{x}})=c_{\mathbf{w}}+2\ell_{\mathbf{w}}({\mathbf{x}})\pmod4\), where
\(c_{\mathbf{w}}\in\mathbb Z_4\) is constant and
\(\ell_{\mathbf{w}}:\Ftwo^n\to\Ftwo\) is linear.  This gives
\[
 h({\mathbf{x}})+h({\mathbf{x}}+{\mathbf{w}})=c\,i^{Q({\mathbf{x}})}\bigl(1+i^{c_{\mathbf{w}}}(-1)^{\ell_{\mathbf{w}}({\mathbf{x}})}\bigr)
 \quad({\mathbf{x}}\in {\mathbf{a}}+L).
\]
For even \(c_{\mathbf{w}}\) the bracket is a scalar times an affine hyperplane
indicator. For odd \(c_{\mathbf{w}}\) it is a nonzero scalar times a fourth-root
linear phase.  Restricting the phase polynomial to an affine subspace and multiplying
by this linear phase preserve \(\cA\), proving the claim.  The displayed difference formula follows by
expanding the multilinear representative of \(Q\). All nonconstant
coefficients of that difference are even.
\end{proof}

For an \(n\)-ary signature \(h\), use the translation operator just defined.
Write \(\operatorname{Id}\) for the identity on the signature space and define
the orbit-sum operator for \({\mathbf{s}}\in\Ftwo^n\) by
\[
 \mathsf P_{\mathbf{s}}^+:=\operatorname{Id}+T_{\mathbf{s}}.
\]
Thus
\((\mathsf P_{\mathbf{s}}^+h)({\mathbf{x}})=h({\mathbf{x}})+h({\mathbf{x}}+{\mathbf{s}})\) is, for \({\mathbf{s}}\ne0\), the sum over the two-point
translation orbit \(\{{\mathbf{x}},{\mathbf{x}}+{\mathbf{s}}\}\).
For accessible \({\mathbf{s}}\), the two quotient signatures
\(\mathcal R_{{\mathbf{s}},0}h\) and \(\mathcal R_{{\mathbf{s}},1}h\) are called the
\(0\)- and \(1\)-layer \emph{plus cards} of \(h\) in direction \({\mathbf{s}}\).
Together they are the two quotient layers of \(\mathsf P_{\mathbf{s}}^+h\).

\begin{lemma}[Low-arity flat-Lagrangian signatures are matching products]
\label{lem:rm-low-flat-matching}
Let \(L=L^\perp\le\Ftwo^{2r}\), where \(r\le3\).  After a permutation of inputs, \(L\) is a direct sum of \(r\) two-coordinate repetition
lines \(\Span_{\Ftwo}\{{\mathbf{e}}_{2j-1}+{\mathbf{e}}_{2j}\}\).  Consequently
every flat-Lagrangian signature of arity at most six is a product of
projective \(I/X\) binaries on a perfect matching.
\end{lemma}

\begin{proof}
Every vector of \(L\) has even weight, so \(\one\in L^\perp=L\).
If \(r>1\), choose \({\mathbf{z}}\in L\setminus\{\zero,\one\}\).  One of \({\mathbf{z}},{\mathbf{z}}+\one\)
has weight at most \(r\), hence weight two when \(2r\le6\).  Thus
\({\mathbf{e}}_i+{\mathbf{e}}_j\in L\) for some pair \(i,j\).  Self-duality now gives
\(x_i=x_j\) for every \({\mathbf{x}}\in L\), so this repetition line splits
orthogonally from \(L\).  Induction proves the assertion.  Translating the
affine support changes an equality factor only to its disequality coset.
\end{proof}

Consequently every flat-Lagrangian proper card of an occurrence of arity at
most eight is already a matching product.  A nonmatching flat-Lagrangian
proper card can first occur at arity ten. The Radon localization argument
handles this remaining case.

For linearly independent vectors \({\mathbf{r}}_1,\ldots,{\mathbf{r}}_s\), a \emph{dual family}
is a family \({\mathbf{r}}_1^*,\ldots,{\mathbf{r}}_s^*\) with
\({\mathbf{r}}_i\cdot {\mathbf{r}}_j^*=\one[i=j]\).  Pairing the radical means adding such a
family, chosen orthogonal to a previously fixed nondegenerate complement.
Indeed, when the ambient \(W\) is nondegenerate and
\(R=\Span_{\Ftwo}\{{\mathbf{r}}_1,\ldots,{\mathbf{r}}_s\}\), the pairing map
\[
 W\longrightarrow R^\vee,
 \qquad {\mathbf{w}}\longmapsto({\mathbf{r}}\mapsto {\mathbf{r}}\cdot {\mathbf{w}}),
\]
is surjective: otherwise a nonzero element of \(R\) would pair to zero with
all of \(W\).  Choosing preimages gives a dual family.  If
\(U=P\perp R\) with \(P\) nondegenerate and
\(R=\operatorname{rad}_W(U)\), the same argument inside \(P^{\perp_W}\)
makes the family orthogonal to \(P\).  The new vectors are independent
modulo \(U\), as pairing a linear combination with each \({\mathbf{r}}_i\) shows.
On \(R\) and its dual family the Gram matrix is
\(\left(\begin{smallmatrix}0&I\\I&A\end{smallmatrix}\right)\), which is
nonsingular.  Adding the family therefore gives a nondegenerate extension
of dimension \(\dim U+\dim R\).

\begin{lemma}[Localization from two flat-Lagrangian Radon cards]
\label{lem:rm-two-chart-localization}
Let \(g\) have even arity \(n\ge10\), let \({\mathbf{v}}\) be accessible, and suppose
that, for both \(\epsilon\in\Ftwo\), the card
\(\mathcal R_{{\mathbf{v}},\epsilon}g\) is zero or flat-Lagrangian.  Define, only for
the analysis, the orbit-sum signature
\begin{equation*}
  F_{\mathbf{v}}({\mathbf{x}})=g({\mathbf{x}})+g({\mathbf{x}}+{\mathbf{v}}).
\end{equation*}
If \(F_{\mathbf{v}}\notin\cA\), an \(O(n^2)\)-size direct
\(\RMcore/I/X\)-context using one copy of \(g\) produces a nonzero
\(H\notin\cA\) of arity at most eight.
\end{lemma}

\begin{proof}
Fix the recorded data
\((A_{\mathbf{v}},i_{\mathbf{v}},j_{\mathbf{v}},{\mathbf{r}}_{\mathbf{v}},\Pi_{\mathbf{v}},W_{\mathbf{v}})\) from
\cref{lem:rm-radon-accessible}, and write \(W:=W_{\mathbf{v}}\).  Before the final
output pair is closed, the graph realization gives the orthogonal
decomposition
\[
 \Ftwo^n=\Pi_{\mathbf{v}}\perp W.
\]
By the recorded identification, the two output positions
\(i_{\mathbf{v}},j_{\mathbf{v}}\) are the \(\Pi_{\mathbf{v}}\)-inputs, while the remaining output positions are
the \(W\)-inputs.

Under the isometry \(W\simeq {\mathbf{v}}^\perp/\Span_{\Ftwo}\{{\mathbf{v}}\}\), the
characteristic vector of \(W\) corresponds to the class of \(\one^n\).
This class is nonzero because accessibility excludes \({\mathbf{v}}=\one^n\).
Thus \(W\) is even-dimensional and nonalternating.

For \(\epsilon\in\Ftwo\), write the Radon card in the chosen quotient coordinates as
\[
 h_\epsilon({\mathbf{w}}):=F_{\mathbf{v}}(\epsilon {\mathbf{r}}_{\mathbf{v}}+{\mathbf{w}}),
 \qquad {\mathbf{w}}\in W.
\]
This is well defined because \(F_{\mathbf{v}}\) is invariant under translation by
\({\mathbf{v}}\).  Each nonzero Radon card has the form
\[
 h_\epsilon({\mathbf{w}})
 =c_\epsilon\one[{\mathbf{w}}\in {\mathbf{a}}_\epsilon+L_\epsilon],
 \qquad
 c_\epsilon\ne0,\quad {\mathbf{a}}_\epsilon\in W,\quad
 L_\epsilon=L_\epsilon^{\perp_W}\le W.
\]
If \(h_{1-\epsilon}=0\) and \(h_\epsilon\ne0\), then, in the decomposition
\(\Ftwo^n=\Pi_{\mathbf{v}}\perp W\),
\[
 F_{\mathbf{v}}({\mathbf{x}})
 =c_\epsilon\one[
   {\mathbf{x}}\in\epsilon {\mathbf{r}}_{\mathbf{v}}+{\mathbf{a}}_\epsilon+
        (\Span_{\Ftwo}\{{\mathbf{v}}\}\oplus L_\epsilon)].
\]
Thus \(F_{\mathbf{v}}\in\cA\) whenever one Radon card is zero, so
both cards are nonzero under the hypothesis. Restoring the layer
coordinate gives an affine support exactly when \(L_0=L_1\).
With that condition satisfied, the coefficients define a signature in
\(\cA\) exactly when \(c_1/c_0\in\mu_4\).

If \(L_0\ne L_1\), take \({\mathbf{s}}\in L_0\setminus L_1\).  Since
\({\mathbf{s}}\notin L_1^{\perp_W}=L_1\), choose \({\mathbf{t}}\in L_1\) with
\({\mathbf{s}}\cdot {\mathbf{t}}=1\).  Elementary completion in a
nondegenerate binary symmetric space gives an even-dimensional,
nonalternating, nondegenerate subspace
\[
 U\supseteq\Span\{{\mathbf{s}},{\mathbf{t}},{\mathbf{a}}_1-{\mathbf{a}}_0\},
 \qquad \dim U\le6.
\]
To obtain this dimension bound, use \hyperref[subsubsec:symmetric-binary-spaces]{the symmetric-space conventions}.
Put \(S:=\Span\{{\mathbf{s}},{\mathbf{t}},{\mathbf{a}}_1-{\mathbf{a}}_0\}\).  Since \({\mathbf{s}}\cdot {\mathbf{t}}=1\), the plane
\(\Span\{{\mathbf{s}},{\mathbf{t}}\}\) is nondegenerate, so
\(\dim\operatorname{rad}(S)\le1\).  If
\(\operatorname{rad}(S)=0\), put \(N:=S\).  Otherwise write
\(\operatorname{rad}(S)=\Span_{\Ftwo}\{{\mathbf{r}}\}\), choose a nondegenerate
complement \(S_0\) of \(\Span_{\Ftwo}\{{\mathbf{r}}\}\) in \(S\), and choose
\({\mathbf{d}}\in W\) satisfying
\[
 {\mathbf{r}}\cdot {\mathbf{d}}=1,
 \qquad
 {\mathbf{d}}\in S_0^\perp.
\]
Put \(N:=S+\Span_{\Ftwo}\{{\mathbf{d}}\}\).  In either case \(N\) is nondegenerate
and \(\dim N\le4\).
If \(\dim N\) is odd, its nondegenerate orthogonal complement has odd
dimension and therefore contains an anisotropic vector. Adding that
orthogonal line makes \(N\) even-dimensional without exceeding dimension
four.  If the resulting even-dimensional \(N\) is alternating, then
\(N^\perp\) is nonalternating because \(W\) is. Add a nonalternating
nondegenerate plane from \(N^\perp\).  The resulting \(U\) has all the
asserted properties and dimension at most six.

If instead \(L_0=L_1\) and \(c_1/c_0\notin\mu_4\), every vector in an
even-dimensional nondegenerate nonalternating symmetric binary space lies in a
nonalternating nondegenerate plane.  For a nonzero isotropic vector \({\mathbf{z}}\),
choose \({\mathbf{d}}\) with \({\mathbf{z}}\cdot {\mathbf{d}}=1\) and \({\mathbf{d}}\cdot {\mathbf{d}}=1\). For an anisotropic
\({\mathbf{z}}\), choose \({\mathbf{d}}\) with \({\mathbf{z}}\cdot {\mathbf{d}}=1\) and \({\mathbf{d}}\cdot {\mathbf{d}}=0\).  For \({\mathbf{z}}=0\),
choose any nonalternating nondegenerate plane.  We may therefore
choose an even-dimensional, nonalternating, nondegenerate
\[
 U\supseteq\Span\{{\mathbf{a}}_1-{\mathbf{a}}_0\},
 \qquad \dim U\le2.
\]

Put
\[
 P:=U^{\perp_W},
 \qquad
 W=U\perp P,
\]
and let \(\pi_U,\pi_P\) be the corresponding orthogonal projections.  The
space \(P\) is even-dimensional and nondegenerate.  By the
symmetric-binary Witt decomposition recalled in
\hyperref[subsubsec:symmetric-binary-spaces]{the symmetric-space conventions}, \(P\) contains a
half-dimensional totally isotropic subspace \(J\). Equivalently,
\[
 J=J^{\perp_P}\le P.
\]
Because \({\mathbf{a}}_1-{\mathbf{a}}_0\in U\), the two shifts have the same \(P\)-projection.
Call it \({\mathbf{p}}\).  Choose an isometry
\(\phi:U\to\Ftwo^{\dim U}\), and define the affine relation
\[
 \Gamma_{\mathbf{p}}
 :=\{({\mathbf{u}}+{\mathbf{p}}+{\mathbf{j}},\phi({\mathbf{u}})):{\mathbf{u}}\in U,\ {\mathbf{j}}\in J\}
 \subseteq W\oplus\Ftwo^{\dim U}.
\]
Its translation space is self-dual: it has half the ambient dimension, and
\[
 ({\mathbf{u}}+{\mathbf{j}})\cdot({\mathbf{u}}'+{\mathbf{j}}')+\phi({\mathbf{u}})\cdot\phi({\mathbf{u}}')
 ={\mathbf{j}}\cdot {\mathbf{j}}'=0
\]
for \({\mathbf{u}},{\mathbf{u}}'\in U\) and \({\mathbf{j}},{\mathbf{j}}'\in J\).  Thus
after transporting the \(W\)-coordinate through \(A_{\mathbf{v}}|_{W_{\mathbf{v}}}\),
\cref{lem:rm-flat-contexts} realizes the corresponding signature
\[
 ({\mathbf{w}},{\mathbf{z}})\longmapsto\one[({\mathbf{w}},{\mathbf{z}})\in\Gamma_{\mathbf{p}}]
\]
by a realizable \(O(n^2)\)-size \(\RMcore/I/X\)-network.  Apply this signature
to the \(W\)-inputs and leave the two \(\Pi_{\mathbf{v}}\)-inputs external.

When \(L_0\ne L_1\), the boundary points induced by
\({\mathbf{a}}_0,{\mathbf{a}}_0+{\mathbf{s}},{\mathbf{a}}_1\) survive but the fourth parallelogram point induced by
\({\mathbf{a}}_1+{\mathbf{s}}\) does not.  Otherwise there would be \({\mathbf{z}}\in L_1\) with
\(\pi_U{\mathbf{z}}={\mathbf{s}}\), and
\[
  0={\mathbf{z}}\cdot {\mathbf{t}}=(\pi_U{\mathbf{z}})\cdot {\mathbf{t}}={\mathbf{s}}\cdot {\mathbf{t}}=1.
\]
The output therefore retains a nonaffine parallelogram on at most
\(\dim U+2\le8\) inputs.  When \(L_0=L_1\), the two fibers have
equal size because their support translation spaces agree.  Thus neither
cancellation nor projection changes the ratio \(c_1/c_0\) outside
\(\mu_4\), and the output again lies outside \(\cA\).

Let \(H\) be the contraction of this realizable context with \(g\), and let
\(H^+\) be the contraction with \(F_{\mathbf{v}}=\mathsf P_{\mathbf{v}}^+g\).  Order the boundary
with the two retained \(\Pi_{\mathbf{v}}\)-inputs first and put
\[
 {\mathbf{w}}_\partial:=(1,1,\mathbf 0_{\dim U}).
\]
Indeed, translation by \({\mathbf{v}}\) changes the graph output by
\({\mathbf{e}}_{i_{\mathbf{v}}}+{\mathbf{e}}_{j_{\mathbf{v}}}\), so it flips exactly the two \(\Pi_{\mathbf{v}}\)-inputs and acts
trivially on the remaining boundary inputs.  Consequently, with the same
scalar on both terms,
\[
 H^+=\mathsf P_{{\mathbf{w}}_\partial}^+H.
\]
The missing parallelogram or forbidden layer ratio proves
\(H^+\notin\cA\).  By \cref{lem:affine-two-translate-closure},
\(H\in\cA\) would imply \(H^+=\mathsf P_{{\mathbf{w}}_\partial}^+H\in\cA\), a
contradiction.  Therefore \(H\ne0\), \(H\notin\cA\), and its arity is at
most eight.
\end{proof}

In the relative induction, the localized nonaffine signature supplied by the
preceding lemma is excluded by the already closed lower-arity cases.  Hence
every accessible orbit sum \(F_{\mathbf{v}}\) is affine.  The next lemma patches these
hyperplane-wise affine conclusions into a global affine conclusion for
\(g\).

\begin{lemma}[Affine hyperplane patching]
\label{lem:rm-affine-hyperplane-patching}
Let \(n\ge10\) be even, and let \(g:\Ftwo^n\to\AlgNums\).  For every
accessible \({\mathbf{v}}\), define the orbit-sum signature
\(F_{\mathbf{v}}:=\mathsf P_{\mathbf{v}}^+g\).  If \(F_{\mathbf{v}}\in\cA\) for every such \({\mathbf{v}}\),
then \(g\in\cA\).
\end{lemma}

\begin{proof}
We first record the Walsh-invariance of \(\cA\) used below.  The Walsh
kernel \((-1)^{{\boldsymbol{\xi}}\cdot {\mathbf{x}}}\) has an allowed quadratic modulo-four phase,
so its product with an affine signature is affine in the joint variables
\(({\boldsymbol{\xi}},{\mathbf{x}})\).  Gaussian-eliminate the affine support constraints and sum
the remaining internal variables one at a time.  If the current phase
depends on a free bit \(y\) as
\[
 i^{\alpha y}(-1)^{y\ell},
 \qquad \alpha\in\mathbb Z_4,
\]
where \(\ell\) is an affine linear form in the remaining variables and is
independent of \(y\),
then summing over \(y\) either imposes the affine equation
\(\ell=\alpha/2\pmod2\) when \(\alpha\) is even, or contributes a nonzero
scalar times a fourth-root phase of the affine linear form \(\ell\) when
\(\alpha\) is odd.  Thus every elimination remains in \(\cA\), proving
that the Walsh transform preserves \(\cA\).  The converse follows from
Walsh inversion.

Use the Walsh coordinates \(\widetilde g=(P^{-1})^{\otimes n}g\) from
\cref{eq:Walsh-coordinates}. Since
\begin{equation*}
  (\widetilde F_{\mathbf{v}})({\boldsymbol{\xi}})
  =(1+(-1)^{{\boldsymbol{\xi}}\cdot {\mathbf{v}}})(\widetilde g)({\boldsymbol{\xi}}),
\end{equation*}
each restriction \((\widetilde g)|_{{\mathbf{v}}^\perp}\) is affine or zero.  The claim
is immediate when \(\widetilde g=0\), so assume otherwise and choose
\({\boldsymbol{\zeta}}\in\supp (\widetilde g)\).

Given \({\mathbf{x}},{\mathbf{y}}\in\supp (\widetilde g)\), the space
\[
  E\cap\Span\{{\boldsymbol{\zeta}},{\mathbf{x}},{\mathbf{y}}\}^\perp,
  \qquad E=\one^\perp,
\]
has dimension at least \(n-4\), hence contains accessible \({\mathbf{v}}\).  Affinity
on \({\mathbf{v}}^\perp\) puts \({\mathbf{x}}+{\mathbf{y}}+{\boldsymbol{\zeta}}\) in the support, so
\(\supp (\widetilde g)\) is a
coset. Write it as \({\boldsymbol{\zeta}}+U\).

Choose an isomorphism \(\beta:\Ftwo^{\dim U}\to U\) and define the
nowhere-zero logical coefficient tensor
\[
 \psi({\mathbf{u}}):=(\widetilde g)({\boldsymbol{\zeta}}+\beta {\mathbf{u}}).
\]
Fix arbitrary
\[
 {\mathbf{u}}_0,{\mathbf{u}}_1,{\mathbf{u}}_2,{\mathbf{u}}_3\in\Ftwo^{\dim U},
\]
where \({\mathbf{u}}_0\) is the base point and \({\mathbf{u}}_1,{\mathbf{u}}_2,{\mathbf{u}}_3\) are the three
directions of a multiplicative third-difference cube of \(\psi\).  The four
vectors
\[
 {\boldsymbol{\zeta}}+\beta {\mathbf{u}}_0,\quad
 \beta {\mathbf{u}}_1,\quad
 \beta {\mathbf{u}}_2,\quad
 \beta {\mathbf{u}}_3
\]
span at most four dimensions.  Hence
\[
 E\cap
 \Span\{{\boldsymbol{\zeta}}+\beta {\mathbf{u}}_0,\beta {\mathbf{u}}_1,\beta {\mathbf{u}}_2,\beta {\mathbf{u}}_3\}^\perp
\]
has dimension at least \(n-5\ge5\) and contains an accessible direction
\({\mathbf{s}}\).  The entire cube lies in \({\mathbf{s}}^\perp\), where the nonzero restriction
of \(\widetilde g\) is affine. Hence
\[
 \mathcal T_\psi({\mathbf{u}}_0,{\mathbf{u}}_1,{\mathbf{u}}_2,{\mathbf{u}}_3)=1.
\]
By \cref{lem:affine-phase-criterion}, \(\psi\) has an allowed quadratic
modulo-four phase.  Consequently \(\widetilde g\in\cA\), and the
Walsh-invariance proved above gives \(g\in\cA\).
\end{proof}

Affine patching gives \(g\in\cA\), but does not yet give the required
flat-Lagrangian form.  The final recognition lemma uses the dimensions and
phases of all realizable Radon cards to force a self-dual support translation space and
constant coefficient on that support.

\begin{lemma}[Affine-to-flat-Lagrangian Radon recognition]
\label{lem:rm-affine-flat-radon}
Let \(0\ne h\in\cA\) have even arity \(2r\ge8\).  Suppose that for every
accessible direction \({\mathbf{v}}\) and both \(\epsilon\in\Ftwo\), the realizable card
\(\mathcal R_{{\mathbf{v}},\epsilon}h\) is zero or flat-Lagrangian.  Then \(h\) is
flat-Lagrangian.
\end{lemma}

\begin{proof}
Write the affine signature \(h\) in standard normal form as
\[
  h({\mathbf{x}})=c\one[{\mathbf{x}}\in {\mathbf{a}}+U]i^{Q({\mathbf{x}})},
  \qquad d:=\dim U.
\]
Here \(c\ne0\) is the scale, \({\mathbf{a}}+U\) is the affine support, \(Q\) is its
quadratic phase polynomial, and \(d\) is the support dimension.  Define the even hyperplane
\(E:=\one^\perp\).  For \({\mathbf{v}}\notin U\) there is no cancellation:
the nonempty card has support dimension \(d\) if \({\mathbf{v}}\in U^\perp\), and
both have dimension \(d-1\) otherwise.  A flat-Lagrangian card has support dimension \(r-1\).

Neither \(E\subseteq U\) nor \(E\subseteq U^\perp\).  Suppose first that
\(E\subseteq U\).  Choose \({\mathbf{p}},{\mathbf{q}}\in E\) so that
\({\mathbf{p}},{\mathbf{q}},{\mathbf{p}}+{\mathbf{q}}\) are accessible.  They exist because \(\dim E=2r-1\ge7\).
If
\[
 \mathsf P_{\mathbf{p}}^+h=\mathsf P_{\mathbf{q}}^+h=0,
\]
then \(T_{\mathbf{p}}h=T_{\mathbf{q}}h=-h\), so \(T_{{\mathbf{p}}+{\mathbf{q}}}h=h\) and
\(\mathsf P_{{\mathbf{p}}+{\mathbf{q}}}^+h=2h\ne0\).  Thus at least one Radon card arising
from the three directions \({\mathbf{p}},{\mathbf{q}},{\mathbf{p}}+{\mathbf{q}}\) is nonzero.  Let
\({\mathbf{s}}\in\{{\mathbf{p}},{\mathbf{q}},{\mathbf{p}}+{\mathbf{q}}\}\) be its direction.  Passing from \(h\) to such a card
can lower support dimension by at most one when restricting to
its layer hyperplane and by one when quotienting by
\(\Span_{\Ftwo}\{{\mathbf{s}}\}\).
On an overlapping support orbit, factoring out \(h({\mathbf{x}})\) leaves
\[
 1+\zeta(-1)^{\ell({\mathbf{x}})},
 \qquad \zeta\in\mu_4.
\]
Here \(\ell\) is a binary affine-linear form on the current support.
For \(\zeta\in\{\pm i\}\) this factor is nowhere zero. For
\(\zeta\in\{\pm1\}\) its nonzero set is empty, the whole current support,
or one affine hyperplane.  Since the selected card is nonzero, cancellation
therefore lowers support dimension by at most one more.
Hence its support dimension is at least
\[
 d-3\ge2r-4>r-1,
\]
contradicting the support dimension of a flat-Lagrangian card.

If instead \(E\subseteq U^\perp\), then \(d\le1\).  Choose accessible
\({\mathbf{v}}\in E\setminus U\).  Here \({\mathbf{v}}\in U^\perp\setminus U\), so there is no
cancellation and the nonzero card has support dimension \(d\), whereas a
flat-Lagrangian card has dimension \(r-1\ge3\), again a contradiction.

Thus both \(E\cap U\) and \(E\cap U^\perp\) are proper subspaces of \(E\).
Two proper subspaces cannot cover \(E\). Since \(\dim E\ge7\), we may also
avoid \(\zero,\one\) and choose accessible
\({\mathbf{v}}\notin U\cup U^\perp\).  Its two cards have dimension
\(d-1=r-1\), so \(d=r\).

No accessible vector lies in \(U^\perp\setminus U\), since it would give a
dimension-\(r\) card.  Put \(A_0:=E\cap U^\perp\).  Then
\[
 A_0\setminus\{\zero,\one\}\subseteq U.
\]
If \(\one\notin A_0\), or if \(\one\in A_0\cap U\), this gives
\(A_0\subseteq U\).  If instead \(\one\in A_0\setminus U\), choose
\({\mathbf z}\in A_0\setminus\{\zero,\one\}\), which is possible because
\(\dim A_0\ge r-1\ge3\). Then both \({\mathbf z}\) and \({\mathbf z}+\one\) are
accessible and lie in \(U\) by the preceding display, forcing
\(\one\in U\), a contradiction.  Hence \(A_0\subseteq U\) in every case.
If
\(U^\perp\subseteq E\), dimensions give
\(U=U^\perp\).  Otherwise \(\dim A_0=r-1\).  Since
\(A_0\subseteq U\) and \(\dim U=r\), choose \({\mathbf{u}}\in U\setminus A_0\).
Then \(U=A_0\oplus\Span_{\Ftwo}\{{\mathbf{u}}\}\), and
\(A_0\subseteq U^\perp\) makes \({\mathbf{u}}\) orthogonal to \(A_0\).  If \({\mathbf{u}}\) were
even, then \({\mathbf{u}}\cdot {\mathbf{u}}=0\), so \({\mathbf{u}}\in U^\perp\) and
\(U\subseteq U^\perp\). Equal dimensions would give \(U=U^\perp\),
contrary to the present case.  Thus \({\mathbf{u}}\) is odd and
\hypertarget{eq:rm-near-self-dual}{}
\begin{equation*}
 U=A_0\oplus\Span_{\Ftwo}\{{\mathbf{u}}\},
  \qquad {\mathbf{u}}\cdot {\mathbf{u}}=1.
\end{equation*}

For accessible \({\mathbf{s}}\in A_0\), put
\[
 \epsilon_{\mathbf{s}}:={\mathbf{a}}\cdot {\mathbf{s}}.
\]
Because \({\mathbf{s}}\in U^\perp\), the support \({\mathbf{a}}+U\) lies wholly in
\(H_{{\mathbf{s}},\epsilon_{\mathbf{s}}}\). The Radon card on the other layer is identically zero.  The
translation space of the supported card
\(\mathcal R_{{\mathbf{s}},\epsilon_{\mathbf{s}}}h\) is
\(U/\Span_{\Ftwo}\{{\mathbf{s}}\}\), whose orthogonal is
\(U^\perp/\Span_{\Ftwo}\{{\mathbf{s}}\}\). These differ because \(U\ne U^\perp\).
Write its phase difference as
\[
 Q({\mathbf{x}}+{\mathbf{s}})-Q({\mathbf{x}})=\alpha_{\mathbf{s}}+2\ell_{\mathbf{s}}({\mathbf{x}})\pmod4,
 \qquad \alpha_{\mathbf{s}}\in\mathbb Z_4,
\]
where \(\ell_{\mathbf{s}}\) is affine linear on \({\mathbf{a}}+U\).  Thus
\[
 \frac{h({\mathbf{x}}+{\mathbf{s}})}{h({\mathbf{x}})}
 =i^{\alpha_{\mathbf{s}}}(-1)^{\ell_{\mathbf{s}}({\mathbf{x}})}.
\]
If \(\alpha_{\mathbf{s}}\) is odd, the supported plus card is nonzero on the full
quotient and has the non-self-dual translation space
\(U/\Span_{\Ftwo}\{{\mathbf{s}}\}\).
If \(\alpha_{\mathbf{s}}\) is even and \(\ell_{\mathbf{s}}\) is nonconstant, its nonzero support
is a codimension-one hyperplane of dimension \(r-2\).  If \(\ell_{\mathbf{s}}\) is
constant, the card is either zero or again has the full non-self-dual
translation space.  The hypothesis therefore forces
\[
 \mathcal R_{{\mathbf{s}},\epsilon_{\mathbf{s}}}h=0,
 \qquad\text{equivalently}\qquad
 \mathsf P_{\mathbf{s}}^+h=0.
\]

Because \(\dim A_0=r-1\ge3\), choose a two-dimensional subspace
\(B_0\le A_0\) with \(\one\notin B_0\), and let \({\mathbf{p}},{\mathbf{q}}\) be a basis of
\(B_0\).  Then \({\mathbf{p}},{\mathbf{q}},{\mathbf{p}}+{\mathbf{q}}\) are all accessible, and
\(\mathsf P_{\mathbf{p}}^+h=\mathsf P_{\mathbf{q}}^+h=0\), whereas
\(T_{{\mathbf{p}}+{\mathbf{q}}}=T_{\mathbf{p}}T_{\mathbf{q}}\) gives
\(\mathsf P_{{\mathbf{p}}+{\mathbf{q}}}^+h=2h\ne0\), a contradiction.
Therefore \(U=U^\perp\).

For accessible \({\mathbf{s}}\in U=U^\perp\), put
\(\epsilon_{\mathbf{s}}:={\mathbf{a}}\cdot {\mathbf{s}}\).  The support lies wholly in
\(H_{{\mathbf{s}},\epsilon_{\mathbf{s}}}\), so \(\mathcal R_{{\mathbf{s}},\epsilon_{\mathbf{s}}}h\) is the unique
potentially nonzero Radon card.  Use the same phase-difference form
as above.
If \(\alpha_{\mathbf{s}}\) is even and \(\ell_{\mathbf{s}}\) is nonconstant, the nonzero part of
the plus card has quotient-support dimension \(r-2\), so it cannot be
flat-Lagrangian.  In every remaining case the plus card is either zero or
has the full self-dual quotient support \(U/\Span_{\Ftwo}\{{\mathbf{s}}\}\). By the
flat-Lagrangian hypothesis it is then constant on that support.  Taking
\(c_{\mathbf{s}}=0\) in the zero case, we obtain
\begin{equation*}
  h({\mathbf{x}})+h({\mathbf{x}}+{\mathbf{s}})=c_{\mathbf{s}}
  \qquad({\mathbf{x}}\in {\mathbf{a}}+U).
\end{equation*}
Since \(\dim U=r\ge4\) and \(U=U^\perp\), choose a two-dimensional
subspace \(B_1\le U\) with \(\one\notin B_1\), and let \({\mathbf{p}},{\mathbf{q}}\) be a basis.
Then \({\mathbf{p}},{\mathbf{q}},{\mathbf{p}}+{\mathbf{q}}\) are all accessible.  Using \(T_{{\mathbf{p}}+{\mathbf{q}}}=T_{\mathbf{p}}T_{\mathbf{q}}\) to compare
their three equations gives
\[
  2h({\mathbf{x}})=c_{{\mathbf{p}}+{\mathbf{q}}}-c_{\mathbf{q}}+c_{\mathbf{p}},
\]
so \(h\) is constant on \({\mathbf{a}}+U\), hence flat-Lagrangian.
\end{proof}

\subsection{Completion of Reed--Muller rigidity}

\begin{proof}[Completion of \cref{thm:reed-muller-rigidity}]
We first close the initial minimum-arity nonbinary irreducible signature \(f\).  If it has odd arity, apply
\cref{thm:external-odd} to \(K\cF\).  Its hardness conclusion pulls back
through the holographic \(K\)-coordinate equivalence, while its tractable
conclusion is exactly \(\Tract(K\cF)\).  If
\(\arity(f)=4\), \cref{lem:proper-cyclic-q4} and the arity-four reduction
give either a resolved leaf or a strict stable successor, and the complete
batch is resolved by
\cref{thm:equality-q4-boundary,cor:p1-eight-vertex-reduced-interface,%
lem:actual-binary-completeness,def:relative-stratum-protocol,%
lem:matching-synthesis-measure}.  Hence, on the sole remaining initial
branch, \(f\) has even arity \(n\ge6\), and every proper card is zero or a
nonsingular current-group matching signature.

Apply \cref{lem:rm-zero-graph,lem:rm-matching-consistency}.  If
\(W_f\ne\varnothing\), then \cref{lem:rm-weight-two-integration} gives
either the realizable \(\mathrm{Even}_4\) equality anchor constructed in its
proof or a proper tensor factorization.  After factor closure and group
recomputation, both alternatives are resolved by
\cref{lem:equality-anchor,lem:factor-saturation,%
def:relative-stratum-protocol,lem:matching-synthesis-measure}.
Thus the remaining branch has \(W_f=\varnothing\).  Now
\cref{lem:rm-support-classification,lem:rm-coefficient-atlas} give
\[
 n=8,
 \qquad
 f({\mathbf{x}})=c\one[{\mathbf{x}}\in {\mathbf{a}}+RM(1,3)]
\]
after a permutation of inputs, for a recorded \(c\ne0\).  Available realizable
single-input \(X\)-dressings remove \({\mathbf a}\), so the retained set contains the
standard core \(\RMcore\).  Regenerate the complete deck of every
single-input dressing of this core by a safe binary.  If one leaves the matching branch,
its complete batch is resolved by the same case analysis.
Otherwise \cref{lem:rm-safe-group-collapse} gives
\[
 G=\{[I],[X]\}\cong C_2.
\]
This proves the asserted description of the remaining initial case.

Retain this \(\RMcore\) and normalized \(I/X\) while resolving the remaining signature set.  After adding signatures, consider the complete transfer group for the resulting set and process the complete decks of single-input dressings of \(\RMcore\) by safe
binaries before applying the argument again.  A decrease of \(\nu\) or a strict
group enlargement is charged to the stable-state measure \(\Xi_{\mathrm{full}}\) of
\cref{lem:matching-synthesis-measure}. On the unchanged remaining branch
the preceding argument using \cref{lem:rm-safe-group-collapse} reapplies and the complete transfer group
remains \(\CycX{2}\).

For an unchanged stable state, a second measure controls the remaining
irreducible occurrences and their lower-arity outputs. It records the
arity, the number of unprocessed network outputs, and the remaining
factorization potential. Let \(\mathcal W\) be the finite multiset of
unresolved irreducible occurrences in the retained factorization trees.
When an occurrence \(u\), carrying \(h_u\), is first added to
\(\mathcal W\), fix the following finite reduction family
\(\mathscr U_{\rm RM}(u)\):
\begin{enumerate}
\item the complete \(\CycX{2}\) matching-deck universe of \(h_u\), with
      every oriented proper contraction and input dressing by \(I/X\) binaries, together
      with every ordered two-copy \(I/X\)-integration network displayed in
      \cref{lem:rm-weight-two-integration}, for all input, orientation, and
      closing-chain sign choices;
\item for every accessible \({\mathbf{v}}\) and \(\epsilon\), the fixed realizable Radon
      network of \cref{lem:rm-radon-accessible}, including its recorded
      data \((A_{\mathbf{v}},i_{\mathbf{v}},j_{\mathbf{v}},{\mathbf{r}}_{\mathbf{v}},\Pi_{\mathbf{v}},W_{\mathbf{v}})\), and, for every finite
      parameter tuple \(({\mathbf{v}},U,J,{\mathbf p},\phi)\) allowed in
      \cref{lem:rm-two-chart-localization}, its realizable localization network
      applied to \(h_u\) using that same recorded Radon data;
\item the complete factorization tree of every output in the preceding two
      items.
\end{enumerate}
Each network-output index specifies the ordered network, its boundary
order, scalar, and direct gadget construction. Each attached factor index
records its canonical signature and Turing reduction.  The
family is finite because the input set and
all of its binary subspaces, vectors, and isometries are finite. Every
individual listed network has the size bounds proved in the cited lemmas.
These choices depend on the fixed retained set, not on the
input Holant graph.  Once selected, each network replaces a vertex by a
fixed-size gadget, as required for the stated reductions.
Partition it as
\[
 \mathscr U_{\rm RM}(u)=Q_{u,\mathrm{done}}\sqcup Q_u.
\]
For \(q\in Q_u\), let \(\mathcal W_q\) be the multiset of unresolved occurrences in its attached factorization tree, and set
\[
 \Phi_u:=\sum_{q\in Q_u}\Phi(\mathcal W_q),
\]
using \cref{eq:factor-potential}.  Define
\[
 \Theta_{\rm RM}(\mathcal W)
 =
 \{\!\{(\arity(h_u),|Q_u|,\Phi_u):u\in\mathcal W\}\!\},
\]
using lexicographic order on each triple and the well-founded multiset
extension of that order. Explicitly, replacing one triple by any finite
multiset of strictly smaller triples is a strict decrease.

Proving that an occurrence is flat-Lagrangian removes its triple.  If the selected occurrence
factors, its triple is replaced by triples of strictly smaller arity.  Every
batch used below starts from a fixed index: the matching-deck and integration
steps use item~1, the Radon steps use the first family in item~2, and the
two-layer localization step uses the second family in item~2.  Completing the output indexed by \(q\) moves
\(q\) to \(Q_{u,\mathrm{done}}\), while an intermediate genuine factor split
strictly decreases its summand \(\Phi(\mathcal W_q)\).  Any newly opened
child occurrence has smaller arity: proper and Radon cards have arity at
most \(\arity(h_u)-2\), localization outputs have arity at most eight when
they are invoked from arity at least ten, and every proper factor is smaller
than its parent.  Thus the old parent triple is replaced by a strictly
smaller parent triple together with child triples whose first coordinate is
smaller.  This strictly decreases \(\Theta_{\rm RM}\).  No fixed-\((H,\nu)\)
batch creates a fresh unresolved index, because all of its possible outputs
and complete factorization trees were included in \(\mathscr U_{\rm RM}(u)\) when the family was fixed.  Consequently this outer induction, rather than a multiset of
arities alone, also controls passage to another occurrence of the same
arity.

Run induction on \(\Theta_{\rm RM}\).  Before processing an occurrence,
recursively close all its lower-arity proper cards and factors.  Unary,
odd-arity, and quaternary outputs are handled by
\cref{thm:external-odd,thm:equality-q4-boundary}.  Hence consider a
remaining irreducible \(g\notin\FLag\) of even arity at least six.

First suppose that every surviving proper card of \(g\) is zero or a
current-\(\CycX{2}\) matching signature. The reconstruction proofs
\cref{lem:rm-zero-graph,lem:rm-matching-consistency,%
lem:rm-weight-two-integration,lem:rm-support-classification,%
lem:rm-coefficient-atlas}
use minimum arity only to exclude a surviving proper nonmatching card or
nonbinary factor. Here the regenerated deck and prior closure of smaller
outputs give that exclusion, so the same networks and coefficient identities
apply to \(g\). They yield a resolved \(\mathrm{Even}_4\) equality anchor,
a proper factorization, or the flat-Lagrangian arity-eight indicator of
\(RM(1,3)\). Thus no occurrence outside \(\FLag\) with an all-matching
proper deck survives.

Otherwise some nonzero proper card is nonmatching.  Factor closure and
prior closure nevertheless make every proper card zero or a tensor product
of already proved flat-Lagrangian factors, hence zero or
flat-Lagrangian.  If \(6\le\arity(g)\le8\), every such card has arity at most
six, and \cref{lem:rm-low-flat-matching} makes it an \(I/X\)-matching
signature after a permutation of inputs, contradicting the present case.  The
appearance of a nonmatching flat-Lagrangian card identifies the remaining case for Radon contractions, where
\(\arity(g)=n\ge10\).

By \cref{lem:rm-radon-accessible}, every
Radon card is realizable, has its fixed index in item~2, and has arity
\(n-2\).  Processing that output includes its complete factor closure.
Every surviving factor has already been proved flat-Lagrangian. Tensor closure therefore makes each card zero or
flat-Lagrangian.  Fix an accessible \({\mathbf{v}}\).  If the analytic orbit sum
 \[
 F_{\mathbf{v}}=\mathsf P_{\mathbf{v}}^+g
 \]
were outside \(\cA\), \cref{lem:rm-two-chart-localization} would
produce a realizable \(H\notin\cA\) of arity at most eight.  Its parameter
tuple and realizable context have a fixed item~2 index, whose attached factorization tree is now processed.  Factor closure of \(H\) must expose a
nonaffine irreducible factor, since a tensor product
of affine factors is affine.  That factor has smaller arity and is processed
earlier in the induction, contradicting the remaining branch.  Hence
\(F_{\mathbf{v}}\in\cA\) for every accessible \({\mathbf{v}}\).

It follows from \cref{lem:rm-affine-hyperplane-patching} that
\(g\in\cA\).  All its realizable Radon cards are zero or flat-Lagrangian, so
\cref{lem:rm-affine-flat-radon} gives \(g\in\FLag\), contradicting the
choice of a surviving occurrence outside \(\FLag\).  Therefore every retained
irreducible occurrence is eventually proved flat-Lagrangian.

Every concrete terminal, anchor, factor, or group-change batch above is
closed before the next fixed index is selected.  A stable-tuple transition
strictly decreases \(\Xi_{\mathrm{full}}\). A step on the unresolved occurrences at an unchanged tuple strictly
decreases \(\Theta_{\rm RM}\).  The two well-founded measures control changes of stable state and progress
at a fixed state, respectively.

All membership conclusions were proved in the same retained normalized coordinates.
The \(X\)-edge is itself flat-Lagrangian, and tensor products of
flat-Lagrangian signatures remain flat-Lagrangian by taking direct sums of
their self-dual support translation spaces.  Factor closure reconstructs
every member of the signature set from its proved irreducible factors.  Hence
the entire retained set and its edge have one common flat-Lagrangian
presentation, which is exactly
\(\Close_{\{\FLag\}}(\cF,G,f)\).
\end{proof}

\section{Klein-Four Matching Decks}
\label{sec:v4-deck}

The two marked Klein-four groups lead to different conclusions from the
matching deck. In the standard Pauli case, pair contractions reduce a
signature that is not a scalar multiple of a real signature to arity four.
Factor closure then gives a resolved complexity leaf or an outer deck
successor. If every retained signature is a scalar multiple of a real
signature, the real-valued Holant dichotomy applies to the whole set.

For the exotic Klein-four group, the matching condition instead forces
a proper tensor factor in every even arity at least six.
This excludes a distinguished irreducible signature at those arities,
leaving the arity-four reduction. The six-input step uses the finite
certificate in Appendix~\ref{appsec:cert-klein-four}.

\begin{theorem}[Klein-four relative rigidity]
\label{thm:v4-relative}
Assume
\[
 \Stable(\cF,G,f),\qquad \arity(f)=\nu(\cF),
 \qquad G\in\{\KleinStd,\KleinEx\}.
\]
In the standard Pauli case \(G=\KleinStd\), assume additionally that
\(\nu(\cF)\ge6\).  Then
\(\Close_{\{\cA\}}(\cF,G,f)\).
\end{theorem}

The two cases are proved in
\cref{thm:standard-v4-relative,thm:exotic-v4-relative-rigidity}.
A tractable case satisfying \(\Tract\) completes the branch under
\cref{eq:relative-close-predicate}.

\begin{proofstrategy}
In the standard Pauli case, we compare a signature with its complex
conjugate at every input pair and each of the four Bell labels. Above
arity four, linear dependence of each pair of corresponding cards implies
linear dependence of the two signatures. Hence failure to be real up to
a scalar passes to a pair card and, by iteration, to arity four.

For the exotic group, the six-input classification first forces all
nonzero cards at a fixed pair to use one residual matching. Unless a
proper tensor factor already exists, the four labels on each residual
edge must then vary by a permutation. Contracting across two such edges
produces a coefficient support incompatible with a matching product.
Two choices of contraction are needed at arity eight, and one suffices
at arity at least ten.
\end{proofstrategy}

\subsection{The \texorpdfstring{\(\KleinStd\)}{V4 (std)} form: realizable Pauli contractions}
\label{proofsec:v4}
\label{proofsubsec:v4-pauli-calculus}

Let \(\cF\) be retained and factor-closed, with the supplied equality \(I\) and complete transfer group \(\KleinStd\). For a Bell label
\({\boldsymbol{\gamma}}=(d,\ell)^{\mathsf T}\in\Ftwo^2\), let \(B_{\boldsymbol{\gamma}}=B_{d,\ell}\) be the ordered
Bell signature defined in \cref{subsec:pauli-notation}.
Its chosen realizable representative is
\(\beta_{\boldsymbol{\gamma}}B_{\boldsymbol{\gamma}}\), as in
\cref{subsec:reduction-conventions}. We divide out the known nonzero
scalar \(\beta_{\boldsymbol{\gamma}}\in\AlgNums^\times\) in the formulas below and use
bilinear contractions, without complex conjugation.
For an ordered pair \(p\), define
\begin{equation*}
 C_p^{\boldsymbol{\gamma}}(h):=\partial_p^{B_{\boldsymbol{\gamma}}}h,
 \qquad {\boldsymbol{\gamma}}\in\Ftwo^2,
\end{equation*}
and omit the argument only when the parent signature is unambiguous.  For a
list \(\mathbf p=(p_1,\ldots,p_s)\) of disjoint ordered pairs and
\(\boldsymbol\eta=({\boldsymbol{\gamma}}_1,\ldots,{\boldsymbol{\gamma}}_s)\), put
\begin{equation*}
 C_{\mathbf p}^{\boldsymbol\eta}(h)
 :=C_{p_s}^{{\boldsymbol{\gamma}}_s}\cdots C_{p_1}^{{\boldsymbol{\gamma}}_1}(h).
\end{equation*}
The order is immaterial for disjoint pairs.  In this section, a
\emph{proper Pauli card}
means such a realizable iterated contraction leaving at least two inputs. A
\emph{Pauli pair card} has \(s=1\).

\subsection{Relative closure in the standard Pauli case}
\label{proofsubsec:v4-relative-implication}

When the kernels span \(\Mat_2(\mathbb C)\),
\cref{thm:dressed-kernel-observability} reconstructs the slices at one
input pair from the coefficients of four cards. Here we compare cards
at every input pair, allowing zero cards and independent rescalings.
To apply the real-valued dichotomy we compare a signature with its
complex conjugate. The following lemma tests linear dependence through
the vanishing of the antisymmetrized product
\({\mathbf{u}}\otimes{\mathbf{v}}-{\mathbf{v}}\otimes{\mathbf{u}}\).

\begin{lemma}[Pauli contractions determine a tensor line]
\label{lem:v4-pauli-projective-determination}
Let \(n\ge5\), and let \({\mathbf{u}},{\mathbf{v}}\in(\mathbb C^2)^{\otimes n}\).
If \(C_{rs}^{\boldsymbol{\gamma}}({\mathbf{u}})\) and \(C_{rs}^{\boldsymbol{\gamma}}({\mathbf{v}})\) are linearly dependent
for every pair \(r<s\) and every \({\boldsymbol{\gamma}}\in\Ftwo^2\), with zero
cards allowed, then \({\mathbf{u}},{\mathbf{v}}\) are linearly dependent.
Consequently, if every Pauli pair card of \({\mathbf{u}}\) is zero or a nonzero
complex scalar multiple of a real tensor, the same holds for \({\mathbf{u}}\).
\end{lemma}

\begin{proof}
Put \(W={\mathbf{u}}\otimes {\mathbf{v}}-{\mathbf{v}}\otimes {\mathbf{u}}\) and group its two copies by input, so that \(W\in(\mathbb C^4)^{\otimes n}\).  The hypothesis says
\[
 (C_{rs}^{\boldsymbol{\gamma}}\otimes C_{rs}^{\boldsymbol{\gamma}})W=0
 \qquad(r<s,\ {\boldsymbol{\gamma}}\in\Ftwo^2).
\]
On the regrouped four-dimensional spaces at \(r,s\), the four
contraction matrices are
\[
 I\otimes I,\quad Z\otimes Z,\quad X\otimes X,
 \quad (ZX)\otimes(ZX).
\]
Use the same orthonormal Bell basis
\(2^{-1/2}(\operatorname{vec}I,\operatorname{vec}Z,
\operatorname{vec}X,\operatorname{vec}(ZX))\) on each regrouped input.
These matrices are diagonal in that basis, with eigenvalue rows
\[
 \begin{pmatrix}
  1&1&1&1\\
  1&1&-1&-1\\
  1&-1&1&-1\\
  1&-1&-1&1
 \end{pmatrix}.
\]
The rows are independent.  Thus the four contraction equations, for any
fixed values on the other inputs, force each coefficient with equal Bell
basis indices at \(r,s\) to vanish.  Applying this to every pair leaves
only strings whose \(n\) indices are all distinct.  There are just four
indices, so \(n\ge5\) gives \(W=0\), proving the first assertion.

For the last assertion take \({\mathbf{v}}=\overline {\mathbf{u}}\).  The Pauli kernels are
real, and a vector and its conjugate are dependent precisely when it is
zero or a scalar multiple of a real vector: divide by its first nonzero
coordinate to see the reverse implication.
\end{proof}

\begin{theorem}[Standard Pauli \(\KleinStd\) relative closure]
\label{thm:standard-v4-relative}
If
\[
 \Stable(\cF,\KleinStd,f),\qquad \nu(\cF)=\arity(f)\ge6,
\]
then \(\Close_{\{\cA\}}(\cF,\KleinStd,f)\).
\end{theorem}

\begin{proof}
Apply the established reductions for odd, unary, and unsafe binary factors in
\cref{thm:external-odd,lem:common-unary-exit,lem:binary-interpolation,%
lem:factor-saturation,lem:actual-binary-completeness}.
On the continuing branch every nonbinary irreducible signature is even, and all binary
irreducible factors are nonsingular Pauli binaries up to nonzero scalars.

Suppose some retained signature is not a scalar multiple of a real signature.
Factor closure supplies such an irreducible signature \(g\), necessarily
of even arity \(n\ge\nu(\cF)\ge6\).  The contrapositive of
\cref{lem:v4-pauli-projective-determination} selects a Pauli pair card
with the same property.  Repeat while its arity is at least six, reaching
a nonzero quaternary signature \(q\) that is not a scalar multiple of a real signature.
All selected contractions are realizable: the complete transfer group \(\KleinStd\) supplies the four
Pauli kernels, including their reversal and disequality-edge scalars.  Their
composition uses one \(g\)-vertex and \(O(n)\) binary vertices.  The
coefficient tables select the contractions in this finite gadget.

Add \(q\) and process all its irreducible factors by
\cref{def:relative-stratum-protocol}.  An odd or unsafe factor is resolved.
A new finite-order transfer class gives a successor with a strictly
larger complete transfer group, after which the deck is regenerated.  Otherwise, if all factors were binaries, they would all be
Pauli and their product would be a scalar multiple of a real signature, contrary to
the choice of \(q\).  Hence an even nonbinary irreducible signature remains.  It has
arity four, so the signature set strictly lowers
\(\nu(\cF)\ge6\).  The supplied \(I\) excludes the one-sided exception
in factor closure.  Thus processing this family gives a resolved complexity leaf or an
outer deck successor.

It remains to treat the case in which every retained signature is a
scalar multiple of a real signature. Divide each by its first nonzero
coefficient and denote the resulting set of real-valued algebraic
signatures by \(\cF_0\).  These known
scalars are accounted for per vertex, and \(I\in\cF_0\).  The
equality-edge-model equivalence \cref{lem:p1-equality-wire-equivalence} gives
\[
 \KHolant(\cF)\equivT\KHolant(\cF_0)
 \equivT\Holant(\cF_0\cup\{X\}).
\]
Apply the real-valued dichotomy of Shao and Cai
\cite[full version, Theorems~1.2 and~2.22]{ShaoCai2020} to this whole
finite real set.  If its condition \textnormal{(T)} fails, hardness
composes along the displayed reductions.  If it holds, the following
argument gives a tractable case satisfying \(\Tract(K\cF)\) in our
coordinates.

The case \(\cF_0\cup\{X\}\subseteq\angles{\cT}\) immediately implies
\(K\cF\subseteq\angles{\cT}\), since invertible single-input maps and
nonzero scalars preserve tensor products of unary and binary signatures.
In each other case there is one algebraic matrix
\(M\in\GL_2(\AlgNums)\), by
\cref{lem:algebraic-transform-witness}, and one
\(\cC\in\{\cP,\cA,\cL\}\) with
\[
 E:=M^{\mathsf T}M\in\cC,
 \qquad M^{-1}(\cF_0\cup\{X\})\subseteq\cC.
\]
In particular \(B:=M^{-1}XM^{-\mathsf T}\in\cC\), and
\[
 M^{\mathsf T}XM=EBE.
\]
Both \(E\) and \(B\) are symmetric nonsingular members of \(\cC\).
Therefore \cref{lem:symmetric-binary-pullback} gives
\(EBE\in\cC\) for each of the three possible classes.
In all cases the same \(U=M\) satisfies
\[
 U^{\mathsf T}XU\in\cC,
 \qquad U^{-1}\cF\subseteq\cC.
\]
This is a common \(K\)-presentation, hence
\(\Tract(K\cF)\), by \cref{def:common-K-presentation}.

All factors of the new family are processed, and each outer deck
successor reapplies \cref{lem:minimum-core-deck}. Thus the preceding
outcomes satisfy \cref{def:relative-stratum-protocol,def:frozen-token-coverage}
and give the claimed closure in \cref{eq:relative-close-predicate}.
\end{proof}

\subsection{The exotic Klein-four deck}
\label{sec:exotic-v4-deck}

For the exotic Klein-four group, we will show that the matching-card
condition forces a proper tensor factor in every even arity at least six. This form
is not \(GO(X)\)-conjugate to the Pauli form, so we use its original
matrix representatives
\begin{equation}
\begin{aligned}
 E_0&=I,&
 E_1&=X,\\
 E_2&=\begin{pmatrix}i&1\\-1&-i\end{pmatrix},&
 E_3&=\begin{pmatrix}-i&1\\-1&i\end{pmatrix}.
\end{aligned}
\label{eq:exotic-v4-basis}
\end{equation}
Identify \(0,1,2,3\) with \(00,01,10,11\in V=\Ftwo^2\), define the exotic
Klein-four group \(\KleinEx:=\{[E_a]:a\in V\}\), write
addition as \(\oplus\), and define the label permutation
\(\vartheta:=(0)(1)(2\ 3)\).  A superscript that is an ordered input pair,
as in \(E_a^{(1,3)}\), specifies the tensor inputs. An upper label, as in
\(E^a\), denotes the Frobenius-dual basis element defined below.

For an oriented perfect matching \(M\), an \emph{exotic
\(M\)-matching product} is a nonzero scalar multiple of
\[
 \bigotimes_{e\in M}E_{\tau_e}^{(e)},
 \qquad \tau\in V^M.
\]
These are the \(\KleinEx\)-matching signatures on \(M\).
The products of the fixed representatives form the \emph{\(M\)-basis}
of \((\mathbb C^2)^{\otimes 2|M|}\). The one-dimensional vector subspace
spanned by one product is its \emph{matching line}.

To compare different matchings, we express a product on one matching in
the basis of another, an operation called \emph{recoupling}. The next
lemma counts the nonzero coefficients in this expansion. This count
will constrain the label maps in the later factor argument.

\begin{lemma}[Exotic Klein-four calculus and recoupling]
\label{lem:exotic-v4-calculus}
For the bilinear Frobenius pairing,
\begin{equation}
 \langle E_a,E_b\rangle_{\mathrm{Frob}}
 =d_a\one[b=\vartheta(a)],
 \qquad (d_0,d_1,d_2,d_3)=(2,2,4,4).
 \label{eq:exotic-v4-Gram}
\end{equation}
Thus define
\[
 E^a:=d_a^{-1}E_{\vartheta(a)},\qquad
 \langle E^a,E_b\rangle_{\mathrm{Frob}}=\one[a=b].
\]
These duals are nonzero scalar multiples of the original representatives.
Transposition and multiplication satisfy
\begin{equation}
 E_a^{\mathsf T}\doteq E_{\vartheta(a)},
 \qquad E_aE_b\doteq E_{a\oplus b},
 \label{eq:exotic-v4-transpose-fusion}
\end{equation}
with nonzero omitted scalars.  Relative to the matching \(12|34\),
\begin{align*}
 \delta^\times_{a,b,\gamma}
   &:=\vartheta(\gamma)\oplus b\oplus\vartheta(a),\\
 \delta^\parallel_{a,b,\gamma}
   &:=\gamma\oplus\vartheta\bigl(b\oplus\vartheta(a)\bigr),
\end{align*}
and define the recoupling coefficients by
\begin{align*}
 \rho^\times_{a,b}(\gamma)
   &:=
   \left\langle
    (E^\gamma)^{(1,2)}
    (E^{\delta^\times_{a,b,\gamma}})^{(3,4)},
    E_a^{(1,3)}E_b^{(2,4)}
   \right\rangle_{\mathrm{Frob}},\\
 \rho^\parallel_{a,b}(\gamma)
   &:=
   \left\langle
    (E^\gamma)^{(1,2)}
    (E^{\delta^\parallel_{a,b,\gamma}})^{(3,4)},
    E_a^{(1,4)}E_b^{(2,3)}
   \right\rangle_{\mathrm{Frob}} .
\end{align*}
Here the pairing is the bilinear Frobenius pairing on quaternary tensors.
The recoupling identities are
\begin{align*}
 E_a^{(1,3)}E_b^{(2,4)}
 &=\sum_{\gamma\in V}\rho^{\times}_{a,b}(\gamma)
   E_\gamma^{(1,2)}E_{\delta^\times_{a,b,\gamma}}^{(3,4)},
 \\
 E_a^{(1,4)}E_b^{(2,3)}
 &=\sum_{\gamma\in V}\rho^{\parallel}_{a,b}(\gamma)
   E_\gamma^{(1,2)}E_{\delta^\parallel_{a,b,\gamma}}^{(3,4)},
\end{align*}
Every displayed coefficient lies in \(\mathbb Q(i)^\times\).
For matchings \(M,N\) of \(2r\)
inputs, let \(c(M,N)\) count alternating cycles, including common edges.
An \(N\)-matching product then has exactly
\begin{equation}
 4^{r-c(M,N)}
 \label{eq:exotic-v4-overlap-size}
\end{equation}
nonzero coefficients in the \(M\)-basis. Thus coefficient support of
size four means one alternating four-cycle and exactly two varying
\(M\)-edge labels.
\end{lemma}

\begin{proof}
The displayed matrices satisfy
\[
 \begin{gathered}
 E_1^2=I,\qquad E_2^2=E_3^2=-2I,\qquad
 E_1E_2=-E_2E_1=-iE_3,\\
 E_0^{\mathsf T}=E_0,\qquad E_1^{\mathsf T}=E_1,\qquad
 E_2^{\mathsf T}=-E_3,\qquad E_3^{\mathsf T}=-E_2,\\
 \operatorname{tr}(E_a)=2\one[a=0].
 \end{gathered}
\]
These identities give the fusion and Gram formulas, including all
nonzero scalar factors.  For example, the coefficient of
\(E_\gamma^{(1,2)}E_\delta^{(3,4)}\) in the first recoupling is
\[
 \operatorname{tr}\!\left((E^\delta)^{\mathsf T}
                  E_a^{\mathsf T}E^\gamma E_b\right).
\]
Fusion makes its matrix product a nonzero multiple of
\(E_{\delta\oplus\vartheta(a)\oplus\vartheta(\gamma)\oplus b}\).
The trace therefore vanishes except at
\(\delta=\delta^\times_{a,b,\gamma}\), and is nonzero there.
For the second recoupling the coefficient is
\[
 \operatorname{tr}\!\left((E^\delta)^{\mathsf T}
                  E_b^{\mathsf T}(E^\gamma)^{\mathsf T}E_a\right),
\]
which is nonzero exactly at
\(\delta=\gamma\oplus\vartheta(b)\oplus a
=\delta^\parallel_{a,b,\gamma}\).  All factors have entries in
\(\mathbb Q(i)\), proving the coefficient assertion.

For general \(M,N\), contracting an \(N\)-product against a dual
\(M\)-basis element separates over the alternating cycles of
\(M\cup N\).  On a cycle of length \(2\ell\), the coefficient is
one trace of a product of the \(\ell\) fixed \(N\)-matrices and
\(\ell\) dual \(M\)-matrices, transposed according to their
orientations.  Each \(M\)-label enters its fused label once, through
either the identity or the bijection \(\vartheta\).  After
\(\ell-1\) labels are chosen, the zero-label trace condition determines
the last label uniquely.  The resulting trace is a nonzero scalar times
\(\operatorname{tr}(I)=2\).  Thus the cycle contributes exactly
\(4^{\ell-1}\) nonzero coefficients, including one for a common edge
when \(\ell=1\).  Multiplying over cycles gives
\(4^{\sum(\ell-1)}=4^{r-c(M,N)}\).  If this equals four, exactly one
cycle has length four and the others are common edges, proving the last
assertion.
\end{proof}

\paragraph{Realizable exotic contractions.}
For an ordered pair \(p\), define the exotic pair card
\[
 C_p^a(f):=\partial_p^{E^a}f.
\]
In this section, for disjoint ordered pairs \(p_1,\ldots,p_s\), an
\emph{exotic card} is
the corresponding iterated contraction by fixed representatives of
the duals \(E^{a_1},\ldots,E^{a_s}\). It is \emph{proper} when at least two
inputs remain.  Disjoint contractions commute, and nonzero dual scalars are divided out in the displayed card notation.
With the fixed representatives, write
\[
 E_aE_b=\omega(a,b)E_{a\oplus b},
 \qquad \omega(a,b)\in\mathbb Q(i)^\times.
\]
The factors \(\omega(a,b)\) and the corresponding transpose scalars
are retained in coefficient calculations. To cross-contract two residual
matching edges \(e=\{u,u'\}\) and
\(e'=\{v,v'\}\), choose one endpoint \(u\) of \(e\), one endpoint \(v\) of
\(e'\), and one fixed \(a\in V\). Contract the single pair \(u,v\) by the
chosen representative of \(E^a\), and leave \(u',v'\) as the endpoints
of the fused residual edge.  The endpoint orientations are part of the
choice, and the scalar factors are retained.

The duals are realizable up to the recorded scalars. Their coordinate
reconstruction identity is
\begin{equation}
 f=\sum_{a\in V}E_a^{(p)}\otimes C_p^a,
 \qquad C_p^a=\partial_p^{E^a}f.
 \label{eq:exotic-v4-tomography}
\end{equation}
Here the cards are realizable contractions, and the sum expresses
\(f\) in the basis on the pair \(p\).

\begin{lemma}[The exotic Klein-four arity-six case]
\label{lem:exotic-v4-q6}
If every exotic pair card of a nonzero arity-six tensor \(h\) is zero or a
product of two exotic binaries, then \(h\) is a product of three exotic
binaries.  Thus there is no exotic analogue of \(\Hcore\).
\end{lemma}

\begin{proof}
Normalize a nonzero card to \(C_{12}^0=E_0^{(3,4)}E_0^{(5,6)}\) by realizable dressings
and a permutation. The six-input certificate
\cref{thm:cert-interface-exotic-v4-q6} covers all choices of zero and
nonzero cards and their scalar coefficients, including cancellations.
Every case is a product of three exotic binaries. Undoing the
normalization preserves this conclusion.
\end{proof}

\begin{lemma}[Fixed residual matching for exotic cards]
\label{lem:exotic-v4-fixed-ruling}
Let \(f\ne0\) have even arity at least eight, and suppose every proper exotic
card of \(f\) is zero or an exotic matching product.  For every deleted
pair \(p\), all nonzero cards \(C_p^a\) in
\cref{eq:exotic-v4-tomography} use one residual perfect matching.
\end{lemma}

\begin{proof}
Fix a deleted pair \(p\), and suppose that two nonzero cards
\(C_p^a\) and \(C_p^b\) use distinct residual matchings \(M\ne N\).  We
construct a sequence of realizable exotic contractions that preserves both
cards as nonzero matching products and preserves one residual disagreement.

The only local operation needed is the following.  If two contracted inputs
belong to distinct matching edges in a given distinguished card, then, up to the fixed
transpose convention, the new edge factor is
\[
 E_\alpha E_\kappa E_\beta\doteq
 E_{\alpha\oplus\kappa\oplus\beta}.
\]
It is nonzero because every \(E_\gamma\) is nonsingular.  Choose two
vertices at alternating distance two on an alternating cycle of
\(M\cup N\).  They belong to distinct edges in both distinguished cards, so contracting
them shortens that cycle by two vertices while keeping both distinguished
cards nonzero.  Repeating reduces a marked nontrivial cycle to length four.

Reduce every other nontrivial alternating cycle and splice it into the
marked cycle by one cross-contraction. The same three-matrix identity keeps
both distinguished cards nonzero and merges the two cycles.  A common edge
carrying the same label in both cards may be closed with its realizable
Frobenius dual.  If
a common edge carries different labels, splice it into the marked cycle
instead of closing it, and reduce the enlarged cycle again.  After all
common edges and all other cycles have been removed, the resulting realizable
arity-six tensor \(h\) has two nonzero \(p\)-cards with different residual
two-edge matchings.

Every exotic pair card of \(h\) is zero or a matching product.  Indeed,
bilinear contractions commute, so any card of \(h\) may be reordered so
that its final pair contraction is applied first to \(f\).  By hypothesis
that first card is zero or a matching product, and every subsequent
contraction preserves this property by
\(E_\alpha E_\kappa E_\beta\doteq
E_{\alpha\oplus\kappa\oplus\beta}\).  Thus
\cref{lem:exotic-v4-q6} applies and makes \(h\) a product of three exotic
binaries.

For a product of three binaries, all nonzero cards at a fixed deleted pair
have one residual matching: if the deleted pair is a factor edge, only its
contraction in the dual direction is nonzero. Otherwise fusion of the two incident factor edges
leaves the same residual matching for every kernel giving a nonzero card.  This contradicts
the two preserved residual matchings of \(h\).  Hence all nonzero \(p\)-cards of \(f\)
use one residual perfect matching.
\end{proof}

With the residual matching fixed, only its binary labels can vary.
The next lemma shows that, in the absence of a proper tensor factor,
each edge uses all four labels exactly once.

\begin{lemma}[Permutation label maps for a fixed residual matching]
\label{lem:exotic-v4-fiber-dichotomy}
Let \(f\) satisfy the hypotheses of \cref{lem:exotic-v4-fixed-ruling}, fix \(p\) and its
residual matching \(M\), and suppose \(f\) has no nontrivial
positive-arity tensor factor.  Then
all four cards \(C_p^a\) are nonzero and have the form
\begin{equation*}
 C_p^a=\lambda_a\bigotimes_{e\in M}E_{\tau_e(a)}^{(e)},
\end{equation*}
where \(\lambda_a\in\mathbb C^\times\) are nonzero scalar coefficients and
\(\tau_e:V\to V\) are label maps. Every \(\tau_e\) is a permutation.
\end{lemma}

\begin{proof}
Let
\[
 \Gamma=\{a\in V:C_p^a\ne0\}.
\]
Reconstruction and the fixed residual matching give
\[
 f=\sum_{a\in\Gamma}\lambda_a E_a^{(p)}
      \bigotimes_{e\in M}E_{\tau_e(a)}^{(e)},
 \qquad \lambda_a\ne0.
\]
For an edge \(e\in M\), contraction with the realizable Frobenius dual of
\(E_b\) selects the fibre
\[
 \Gamma_{e,b}=\{a\in\Gamma:\tau_e(a)=b\}.
\]
In the displayed \(p\cup(M\setminus\{e\})\)-basis, the resulting card has
exactly \(|\Gamma_{e,b}|\) nonzero coefficients, since distinct \(a\)'s
remain distinct on the \(p\)-coordinate.  By
\cref{eq:exotic-v4-overlap-size}, every nonempty fibre therefore has size
one or four.

If \(|\Gamma|=1\), the factor \(E_a^{(p)}\) is common to the whole tensor,
contrary to irreducibility.  If \(|\Gamma|\in\{2,3\}\), every
\(\tau_e\) is injective on \(\Gamma\).  Choose inputs from two distinct
edges of \(M\) and cross-contract them by any realizable exotic kernel.
Projective fusion is nonzero in every summand of the reconstruction, and the \(p\)-labels remain
distinct, so the resulting proper card has exactly \(|\Gamma|\)
coefficients in the induced matching basis.  This is impossible because
\cref{eq:exotic-v4-overlap-size} permits only powers of four.  Consequently
\(\Gamma=V\).

Now every nonempty \(\tau_e\)-fibre has size one or four.  A four-point
fibre makes \(\tau_e\) constant and exposes \(E_b^{(e)}\) as a proper tensor
factor.  Since \(f\) is irreducible, all fibres are singletons.  Hence every
\(\tau_e:V\to V\) is a permutation, proving the stated form.
\end{proof}

\begin{lemma}[The exotic Klein-four arity-eight case]
\label{lem:exotic-v4-q8}
Let \(h\ne0\) have arity eight and suppose every exotic pair card is zero
or a product of three exotic binaries.  Then \(h\) has a proper tensor
factor.
\end{lemma}

\begin{proof}
Contraction closure under \cref{eq:exotic-v4-transpose-fusion} makes every
iterated proper exotic card zero or a matching product. Suppose that
\(h\) is irreducible. Then \cref{lem:exotic-v4-fixed-ruling,%
lem:exotic-v4-fiber-dichotomy} gives four cards on one matching.  Normalize
one to \(C_{12}^0=E_0^{(3,4)}E_0^{(5,6)}E_0^{(7,8)}\).  Each residual label map is affine because
\(\operatorname{AGL}(2,2)=\operatorname{Sym}(V)\), as recalled in
\cref{par:reed-muller-code}. Realizable edge dressings remove the
translations, and their nonzero scalar factors are absorbed into the
coefficients. Writing the labels as column vectors gives
\begin{equation}
 h=\sum_{\boldsymbol{\gamma}\in V}\lambda_{\boldsymbol{\gamma}}
 E_{\boldsymbol{\gamma}}^{(1,2)}E_{A_1\boldsymbol{\gamma}}^{(3,4)}
 E_{A_2\boldsymbol{\gamma}}^{(5,6)}E_{A_3\boldsymbol{\gamma}}^{(7,8)},
 \qquad
 A_j\in\GL_2(\Ftwo),\quad \lambda_{\boldsymbol{\gamma}}\ne0.
 \label{eq:exotic-v4-q8-linear-normal-form}
\end{equation}
To obtain a card whose labels vary on three edges, consider separately the contractions
of \((3,5)\) and \((4,5)\) by \(E^0=I/2\).  On the two remaining
endpoints of the affected edges, the respective matrices are
\[
 \tfrac12 E_{A_1\boldsymbol{\gamma}}^{\mathsf T}E_{A_2\boldsymbol{\gamma}},
 \qquad \tfrac12 E_{A_1\boldsymbol{\gamma}}E_{A_2\boldsymbol{\gamma}}.
\]
By \cref{eq:exotic-v4-transpose-fusion}, their labels are
\[
 D_1\boldsymbol{\gamma}=(\vartheta A_1+A_2)\boldsymbol{\gamma},
 \qquad D_2\boldsymbol{\gamma}=(A_1+A_2)\boldsymbol{\gamma},
\]
with all sums over \(\Ftwo\) and all omitted scalars nonzero.
The two maps cannot both vanish, since that would imply
\(\vartheta A_1=A_1\), contrary to the invertibility of \(A_1\) and
\(\vartheta\ne I\).  Choose a nonzero one, denoted by \(D\).
In the resulting six-input matching basis the four label tuples are
\( (\boldsymbol{\gamma},D\boldsymbol{\gamma},A_3\boldsymbol{\gamma})\).
They are distinct because their first coordinates are distinct, and every
coefficient is nonzero.  All three coordinates vary: the first and third
maps are bijective, and \(D\ne0\).  This contradicts
\cref{eq:exotic-v4-overlap-size}, which allows only two varying edges in
a four-term matching product.  Thus \(h\) has a proper tensor factor.
\end{proof}

\begin{proposition}[High-arity exotic factor]
\label{prop:exotic-v4-high-arity-factor}
Let \(f\ne0\) have arity \(2r\ge10\), and suppose every proper exotic card is
zero or an exotic matching product.  Then \(f\) has a proper
tensor factor.
\end{proposition}

\begin{proof}
Suppose that \(f\) is irreducible. At a fixed pair, \cref{lem:exotic-v4-fixed-ruling,%
lem:exotic-v4-fiber-dichotomy} gives four nonzero cards on one matching with
permutation label maps.  Choose two residual edges and perform one
cross-contraction by a Frobenius dual as defined above.  Every summand of the reconstruction remains nonzero
because the three matrices in its endpoint-to-endpoint fusion are
nonsingular.  The four resulting terms still vary at the fixed pair and at
all \(r-3\) untouched residual edges, hence in at least \(r-2\ge3\)
coordinates.  But
\cref{eq:exotic-v4-overlap-size} allows only two varying coordinates in a
four-term matching product.  This contradiction proves that a proper
tensor factor exists.
\end{proof}

\begin{theorem}[Exotic Klein-four relative rigidity]
\label{thm:exotic-v4-relative-rigidity}
Assume
\[
 \Stable(\cF,\KleinEx,f),\qquad
 \arity(f)=\nu(\cF).
\]
Then
\[
 \Close_{\{\cA\}}(\cF,\KleinEx,f).
\]
More explicitly, every continuing branch reaches a resolved complexity
leaf or an outer deck successor after processing the realizable batch
from the arity-four reduction by \cref{lem:matching-synthesis-measure}.
No tractable case requiring a common presentation remains at the current node.
\end{theorem}

\begin{proof}
First consider odd-arity factors, including unaries, by
\cref{thm:external-odd}, and unsafe binaries by
\cref{lem:binary-interpolation,lem:actual-binary-completeness}.
On a continuing branch all retained irreducible signatures have even arity.
Deck stability makes every exotic pair card zero or an exotic matching
product.  Repeated transpose/fusion using
\cref{eq:exotic-v4-transpose-fusion} therefore makes every iterated proper
exotic card zero or an exotic matching product.
At arity four, item~4 of \cref{thm:equality-q4-boundary} invokes
\cref{cor:p1-eight-vertex-reduced-interface}. Process its complete factor
batch by \cref{def:relative-stratum-protocol,def:protocol-discharged,%
lem:matching-synthesis-measure}, obtaining a resolved complexity leaf or
an outer deck successor. At arity six, \cref{lem:exotic-v4-q6}
contradicts irreducibility.
At arity eight and at least ten, the factor conclusions of
\cref{lem:exotic-v4-q8,prop:exotic-v4-high-arity-factor} contradict
irreducibility as well.  These cases exhaust the possible arities of the
distinguished nonbinary irreducible signature \(f\).

Hence every branch is resolved before a common presentation is needed.
By \cref{eq:relative-close-predicate}, this gives
\(\Close_{\{\cA\}}(\cF,\KleinEx,f)\).
\end{proof}

\subsection{Completion of the Klein-four classification}

\begin{proof}[Proof of \cref{thm:v4-relative}]
If \(G=\KleinStd\), apply \cref{thm:standard-v4-relative}.  If
\(G=\KleinEx\), apply
\cref{thm:exotic-v4-relative-rigidity}.  These are exactly the two
Klein-four forms in the statement.
\end{proof}
\section{Platonic Matching Decks}
\label{sec:platonic-decks}

For a fixed input pair \(p\), contraction with a kernel \(C\) gives a
card \(\Phi_p(C)\) on the remaining inputs. The realizable kernels form a
finite set of projective points. By linearity, any linear relation among
chosen kernel representatives holds among their cards with the same
coefficients. Suppose all nonzero cards at \(p\) share one matching.
For these kernel configurations, \cref{lem:platonic-global-ruling} fixes
the binary factors, up to scalar, on all but at most one edge of that
matching. Above arity four there are at least two residual edges, so
at least one fixed binary lifts to a proper factor of \(f\) by the
slice reconstruction in \cref{subsec:shared-matching-methods}.
If the matchings disagree, the incidence argument produces
three distinct collinear card points with different matchings. Their
matchings differ on only four residual inputs. Closing the common factors
leaves a realizable six-input signature.

The finite classifications in Appendix~\ref{appsec:cert-platonic}
then give the octahedral and icosahedral reductions.
In the external tetrahedral case, the absence of a quaternary line with
three different matchings forces a common matching at higher arity and
hence a proper factor (\cref{thm:external-a4-deck-rigidity}).
The internal tetrahedral case also permits the affine core \(\Hcore\).
Here stabilizer-code contractions can shorten a nonaffine irreducible
signature while preserving its failure of affine membership in the
current basis. The six- and eight-input cases complete the induction
to a common affine presentation in \cref{thm:platonic-decks}.

\begin{theorem}[Platonic deck rigidity]
\label{thm:platonic-decks}
Assume
\[
 \Stable(\cF,G,f),\qquad
 G\in\{\TetInt,\TetExt,\OctDD,\OctTT,
         \OctDT,\OctTD,
         \IcoZero,\IcoOne\}.
\]
Then
\[
 \begin{cases}
  \Close_{\{\cA\}}(\cF,G,f),&G=\TetInt,\\
  \Close_{\varnothing}(\cF,G,f),
    &G\in\{\TetExt,\OctDD,\OctTT,
            \OctDT,\OctTD,
            \IcoZero,\IcoOne\}.
 \end{cases}
\]
\end{theorem}

\subsection{Common coordinates and realizable kernels}

We use the quaternion coordinates of \cref{eq:quaternion-basis}, the
marked-group classification of \cref{thm:marked-finite-group-routing}, and
the projective and tensor conventions of
\cref{subsubsec:group-projective} and the \hyperref[subsubsec:projective-tensor-geometry]{projective tensor conventions}.
In particular, \([F]\) denotes the projective class of a nonzero
matrix or tensor, including a singular matrix.

A finite projective point configuration is a finite subset \(P\) of a
projective space. Points are collinear when they lie on one projective
line, and a \(k\)-rich line contains at least \(k\) points of \(P\).
A point--line incidence structure is a pair \((P,\mathcal L)\), where
\(\mathcal L\) is a family of subsets of \(P\). Its automorphisms are
the permutations of \(P\) preserving \(\mathcal L\).

Throughout this section, for a finite projective group \(G\) put
\[
 G^\#:=\{[T^\#]:[T]\in G\},
 \qquad T^\#:=XT^{\mathsf T}X,
\]
 and call \(G\) \emph{reversal-stable} when \(G^\#=G\).  A \emph{marked
 Platonic state} means a retained, factor-closed state containing the
 supplied \(I\), whose
complete transfer group is one of
\[
 \TetInt,\ \TetExt,\ \OctDD,\ \OctTT,\
 \OctDT,\ \OctTD,\
 \IcoZero,\ \IcoOne,
\]
with one fixed chosen representative and realization record for every
transfer class.  

For a signature \(u\) and an ordered pair \(p=(a,b)\) of its inputs, define
the contraction map, using the pair contraction from \cref{subsec:reduction-conventions},
\begin{equation}
  \Phi_p^u:\Mat_2(\mathbb C)\longrightarrow
  {(\mathbb C^2)}^{\otimes(\arity(u)-2)},
  \qquad
  \Phi_p^u(C):=\partial_p^C u.
  \label{eq:platonic-card-map}
\end{equation}
When the signature is the active \(f\), write \(\Phi_p=\Phi_p^f\), and write
\(\langle C^{(p)},u\rangle:=\Phi_p^u(C)\) for this bilinear
(not Hermitian) contraction. Equivalently,
\(\Phi_p^u(C)({\mathbf{y}})=\operatorname{tr}((F^u_{p,\mathbf{y}})^{\mathsf T}C)\),
using the two-input slice of \cref{subsubsec:two-port-fibres}.
Input permutations
supply the other orderings.

For every two-element input set \(\{a,b\}\), use the canonical orientation
\(p=(a,b)\) with \(a<b\), and order complementary inputs increasingly unless
both orientations are explicitly requested.  Thus a quaternary signature has six
canonical pairs and an arity-six tensor has fifteen. All other orientations
are obtained by realizable permutations of inputs.

At a continuing fixed-\(I\) deck-stable node, fix for each \(g\in G\)
a matrix \(K_g\) with \([K_g]=g\).  Since \([X]\in G\), choose a fixed
realizable binary representative \(C_g\) of transfer class \([XK_g]\).
The convention \(T_{C_g}=C_gX\) gives \(C_g\doteq XK_gX\).
Attach its two ordered inputs to the selected pair through two \(X\)-edges. The effective bilinear kernel is then
\(XC_gX\doteq K_g\).  Store its nonzero realization
scalar and ordered-input record, and define the chosen representative set
\begin{equation}
  \mathcal K_G^{\rm act}:=\{K_g:g\in G\}.
  \label{eq:platonic-actual-kernel-lift}
\end{equation}
Realizable group contractions use these fixed representatives.  A resulting
binary \(B\) is tested through its transfer \(BX\).  A zero card contributes
zero. A nonzero singular \(B\) is processed by factor closure.  If
\(B\) is nonsingular and \([BX]\notin G\), then
\cref{lem:binary-interpolation,lem:actual-binary-completeness} gives an
infinite-order terminal or a strict enlargement of the complete transfer group.
The ordered inputs and nonzero realization scalar are retained in each
case.

The three nonstandard octahedral displays are as follows.  For the
transposition--transposition form put
\begin{equation}
 H_{\mathrm{tt}}:=
 \begin{pmatrix}\sqrt2&-i\sqrt2\\1-i&1+i\end{pmatrix},
 \qquad
 \OctTT:=H_{\mathrm{tt}}^{-1}\OctDD H_{\mathrm{tt}}.
 \label{eq:octahedral-second-form}
\end{equation}
This realizes the marked type \(\OctTT\) of
\cref{thm:marked-finite-group-routing} in coefficient coordinates.

For the two mixed fixed-equality forms, put
\begin{equation}
 \rho:=\sqrt2-1,\qquad
 H_{\mathrm{dt}}:=Q_0-\rho Q_1,\qquad
 H_{\mathrm{td}}:=Q_0+\rho Q_3,
 \label{eq:octahedral-mixed-conjugators}
\end{equation}
and, for \(\alpha\in\{\mathrm{dt},\mathrm{td}\}\), define
\begin{equation}
 \theta_\alpha(U):=H_\alpha^{-1}UH_\alpha,\qquad
 \OctForm{\alpha}:=\theta_\alpha(\OctDD).
 \label{eq:octahedral-mixed-forms}
\end{equation}
Direct quaternion multiplication gives
\begin{align}
 H_{\mathrm{dt}}H_{\mathrm{dt}}^{\mathsf T}&=2\rho(Q_0-Q_1),&
 \theta_{\mathrm{dt}}(Q_1)&=Q_1,&
 \theta_{\mathrm{dt}}(Q_2-Q_3)&=-\sqrt2Q_3,\notag\\
 H_{\mathrm{td}}H_{\mathrm{td}}^{\mathsf T}&=2\rho(Q_0+Q_3),&
 \theta_{\mathrm{td}}(Q_3)&=Q_3,&
 \theta_{\mathrm{td}}(Q_1-Q_2)&=\sqrt2Q_1.
 \label{eq:octahedral-mixed-mark-check}
\end{align}
Thus \(\OctDT\) has marked class pattern \((dt;t)\), whereas
\(\OctTD\) has pattern \((td;t)\).  All three matrices \(H_{\mathrm{tt}},H_{\mathrm{dt}},H_{\mathrm{td}}\) are used as coefficient
similarities.  Realizable contractions and dressings use representatives of
the current group. The retained set and its binary equality
remain in the fixed coordinates.

\begin{lemma}[The two marked tetrahedral forms]
\label{lem:marked-a4-forms}
Let \(G\leq\PGL_2(\mathbb C)\) be a finite reversal-stable complete transfer group of a retained set
containing the supplied \(I\), in the fixed coordinates, and define the transfer mark \(x\), the reversal mark \(j\), and the extension group \(\widehat G\) by
\[
  x=[X]\in G,\qquad j=[Z],\qquad \widehat G=\langle G,j\rangle.
\]
If \(G\cong A_4\), then, up to one global fixed-\(I\)-preserving
\(GO(X)\)-normalization, exactly one of the following
holds.
\begin{enumerate}
\item \(j\in G\), \(\widehat G=G\), and \(G=\TetInt\), where
      \(\TetInt\) is the standard group in
      \cref{eq:tetrahedral-domain}.
\item \(j\notin G\), \(\widehat G\cong S_4\), and \(G=\TetExt\), where
\hypertarget{eq:external-a4-domain}{}
\begin{equation*}
\begin{aligned}
 \TetExt
 ={}&\{[Q_0],[Q_1],[Q_2+Q_3],[Q_2-Q_3]\}\\
 &{}\cup
 \{[Q_0+\epsilon Q_1+\delta\sqrt2Q_2],
    [Q_0+\epsilon Q_1+\delta\sqrt2Q_3]:
       \epsilon,\delta\in\{1,-1\}\}.
\end{aligned}
\end{equation*}
Here \(x=[Q_1]\in\TetExt\), whereas
\(j=[Q_3]\notin\TetExt\) and
\(xj=[Q_2]\notin\TetExt\).
\end{enumerate}
\end{lemma}

\begin{proof}
Reversal gives \(G\triangleleft \widehat G\) and \([\widehat G:G]\le2\).  If \(j\notin G\),
then \(\widehat G\cong S_4\), with the commuting pair \((x,j)\) marked as a double
transposition and a transposition. If \(j\in G\), the three involutions form
the unique normal Klein four of \(A_4\) by
\cref{lem:finite-pgl2-facts}.  After the corresponding
finite-subgroup embedding the remaining ambiguity is
\[
 C_{\PGL_2(\mathbb C)}(\langle x,j\rangle)=\langle x,j\rangle,
\]
represented inside \(GO(X)\).  Direct multiplication gives the two stated
domains.
\end{proof}

For coefficient comparison, define the matrix \(R_{\rm ext}\)
and the coefficient-similarity map \(\theta_{\rm ext}\) by
\begin{equation*}
 R_{\rm ext}:=Q_0+\rho Q_1,
 \qquad \theta_{\rm ext}(U):=R_{\rm ext}^{-1}UR_{\rm ext}.
\end{equation*}
On projective sets, apply \(\theta_{\rm ext}\) representative-wise.  Then
\begin{equation*}
 \theta_{\rm ext}(\TetInt)=\TetExt,
 \qquad
 [R_{\rm ext}^{\mathsf T}XR_{\rm ext}]=[Q_0-Q_1]\ne[X].
\end{equation*}
Thus \(R_{\rm ext}\notin GO(X)\). We use \(\theta_{\rm ext}\) to compare
coefficient tables and use realizable external representatives for gadgets.

Write \(\IcoOne\) for the opposite icosahedral coordinate display.  The
two displays have the same abstract marked classification, but both are
retained because their representatives in the original basis are different.  Put
\hypertarget{eq:icosahedral-chirality}{}
\begin{equation*}
  D_{\IcoZero}:=Q_0+Q_3,
  \qquad
  D_{\IcoZero}^{\mathsf T}XD_{\IcoZero}=2X,
  \qquad
  \IcoOne=D_{\IcoZero}\IcoZero D_{\IcoZero}^{-1}.
\end{equation*}
The signed coordinate permutations defining \(\TetInt\) also give
\[
 D_{\IcoZero}\TetInt D_{\IcoZero}^{-1}=\TetInt,
 \qquad
 \TetInt\subseteq\IcoZero\cap\IcoOne.
\]
Indeed, conjugation by \(D_{\IcoZero}\) sends
\((Q_1,Q_2,Q_3)\) to \((-Q_2,Q_1,Q_3)\).  This gives the coefficient correspondence below, with realizations
carried out in the original coordinates containing the supplied \(I\).

\begin{lemma}[Fixed-equality correspondence for the opposite icosahedral display]
\label{lem:icosahedral-fixed-I-dictionary}
Set \(D=D_{\IcoZero}\) and
\[
 \kappa_D(C):=D^{-\mathsf T}CD^{-1}.
\]
Let \((\cF,\IcoOne,f)\) be a marked Platonic state in the fixed
coordinates containing the supplied \(I\), and introduce only for
coefficient analysis the tensor
\(f^{(D)}:=D^{-\otimes\arity(f)}f\).  For every ordered pair \(p\),
\begin{equation}
 \Phi_p^f\!\left(\kappa_D(C)\right)
 =D^{\otimes(\arity(f)-2)}\Phi_p^{f^{(D)}}(C).
 \label{eq:icosahedral-card-covariance}
\end{equation}
The map \([C]\mapsto[\kappa_D(C)]\) bijects the standard
\(\IcoZero\)-kernel classes with the realizable \(\IcoOne\)-kernel
classes.  Consequently every finite card or network identity proved in the
standard display has a direct identity in the opposite display, obtained by
replacing each kernel by its \(\kappa_D\)-image and recording the nonzero
representative and edge scalars.  This correspondence preserves zero cards,
tensor factors, all flattening ranks, matching labels, endpoint support,
and the property that an output binary transfer lies outside the complete transfer
group.  It does not change \(\cF\), the active occurrence, or the
supplied \(I\).
\end{lemma}

\begin{proof}
Direct contraction gives \cref{eq:icosahedral-card-covariance}.  Moreover
\(D^{-\mathsf T}=\tfrac12XDX\), and hence
\[
 [\kappa_D(C)]
 =\bigl[X\,D(XC)D^{-1}\bigr]\in\IcoOne
 \qquad([C]\in\IcoZero).
\]
Here \([X],[XC]\in\IcoZero\), conjugation sends the latter class into
\(\IcoOne\), and \([X]\in\TetInt\subseteq\IcoOne\).  The
inverse formula proves bijectivity.  Since \(D^{\mathsf T}XD=2X\), applying
the covariance successively at the vertices of a finite network cancels the
internal \(D,D^{-1}\) factors across every disequality edge, leaving only a nonzero scalar.  A residual binary \(B\) is carried to
\(DBD^{\mathsf T}\), whose transfer satisfies
\[
 [DBD^{\mathsf T}X]=[D(BX)D^{-1}].
\]
Thus nonsingularity and exclusion from the current group are preserved.
Finally, \(D\) is diagonal and invertible, so it preserves support (hence
endpoint nondegeneracy), tensor products, matching labels, and all
flattening ranks.  All resulting kernels are replaced by the already fixed
realizable \(\IcoOne\) representatives in their projective classes, with
their nonzero realization scalars recorded.  The resulting identities therefore hold for realizations over the
original retained set.
\end{proof}

\subsection{Card maps and the arity-four boundary}

The rank of \(\Phi_p^h\) is the flattening rank across the selected pair
and the remaining inputs.  Since its domain is \(\Mat_2(\mathbb C)\), it is
at most four.  We treat ranks below four and rank four separately.  All signatures used as gadgets are directly realized
over \(\cF\), with ordered inputs, nonzero scalar, and predecessor
reduction recorded.  The required condition is
\begin{equation}
 \Phi_p^h(K_g)=0\quad\text{or}\quad\Phi_p^h(K_g)\text{ is \(G\)-matching},
 \qquad \text{for every canonical \(p\) and \(K_g\in\mathcal K_G^{\rm act}\)}.
 \label{eq:platonic-all-card-condition}
\end{equation}

Write
\(\mathcal D_G:=\{[C]:C\in\mathcal K_G^{\rm act}\}\) for the domain
configuration of kernel classes.  A \emph{rich
domain line} is a projective line containing at least three points of
\(\mathcal D_G\).  Let \(\mathscr G_G^{\rm dom}\) be the graph on
\(\mathcal D_G\) joining two points when they lie on a common rich domain
line.  Such a line is \emph{kernel-avoiding} for a linear map \(\Phi\) when
it is disjoint from \(\mathbb P(\ker\Phi)\).  Here \(\ker\Phi\) is a subspace of the contraction-matrix space.  

For an original signature of arity at least six, localization retains \(p\) and four
residual inputs.  The rank of the original card map is used only to choose a
kernel-avoiding rich domain line and to bound the resulting six-input rank.  After
that signature is directly realized, its own card-map rank is
recomputed before either the rank-deficient or full-rank finite reduction is
invoked.

\paragraph{Quaternion coordinates and quaternary matching signatures.}

Recall the quaternion basis from \cref{eq:quaternion-basis}.  Its
Frobenius Gram matrix is
\begin{equation*}
  {\bigl(\operatorname{tr}(Q_\mu^{\mathsf T}Q_\nu)\bigr)}_{\mu,\nu}
  =2\diag(1,-1,1,-1),
\end{equation*}
so the bilinear dual basis is
\begin{equation}
  Q^0=Q_0/2,\qquad Q^1=-Q_1/2,\qquad
  Q^2=Q_2/2,\qquad Q^3=-Q_3/2.
  \label{eq:platonic-dual-basis}
\end{equation}
These dual matrices recover the coefficients of the original tensor:
their role is linear reconstruction.  For \(C\in\Mat_2(\mathbb C)\), let
\[
 c_Q(C)=(c_0,c_1,c_2,c_3)^{\mathsf T}
 \quad\Longleftrightarrow\quad
 C=c_Q(C)\mathbin{\cdot}Q:=\sum_{\mu=0}^3c_\mu Q_\mu.
\]
We use \(U^{(a,b)}\) for placement on the ordered pair \((a,b)\), as in \cref{subsec:pauli-notation}.

Frobenius-dual reconstruction gives, for every signature \(f\) and
canonical pair \(p\),
\begin{equation}
  f=\sum_{\mu=0}^3 (Q^\mu)^{(p)}\otimes\Phi_p^f(Q_\mu).
  \label{eq:platonic-frobenius-reconstruction}
\end{equation}
More generally, for an ordered pair \(p\), a finite-dimensional tensor
space \(W\), and a linear map \(\Phi:\Mat_2\to W\), write
\(f_{\Phi,p}:=\sum_{\mu=0}^3(Q^\mu)^{(p)}\otimes\Phi(Q_\mu)\).  Every assertion below
that an abstract card map factors refers to this reconstructed tensor.

On four ordered residual inputs define the three matching products
\begin{align*}
 P_0(U,V)&=U(x_1,x_2)V(x_3,x_4),\\
 P_1(U,V)&=U(x_1,x_3)V(x_2,x_4),\\
 P_2(U,V)&=U(x_1,x_4)V(x_2,x_3).
\end{align*}
For a finite projective transfer group \(G\), define its quaternary projective
matching deck by
\begin{equation*}
  \mathfrak D_4(G)=
  \{[P_r(U,V)]:r\in\{0,1,2\},\ [UX],[VX]\in G\}.
\end{equation*}
Thus \(\mathfrak D_4(G)\) is the projective configuration
\(\{[F]:F\in\mathcal M_4(G)\}\) of the arity-four
\(G\)-matching class defined above. Define the nonzero scalar cone
\(\operatorname{cone}(\mathfrak D):=\{\lambda F:\lambda\ne0,
[F]\in\mathfrak D\}\).

A \emph{rich deck line} is a projective line containing at least three
points of \(\mathfrak D_4(G)\).  It is an \(s\)-line when it contains
exactly \(s\) deck points.  Its \emph{matching-label type} is the partition
of \(s\) formed by the multiplicities of its residual matching labels.
Type \((s)\) is \emph{monochromatic}, type \((1,1,1)\) is a
\emph{rainbow three-line}, and every other nonmonochromatic type is
\emph{mixed}.  A rich deck line is \emph{maximal} under the analogous
deck-point containment order. In the fixed order
\((12\mid34,13\mid24,14\mid23)\), the flattening rank of a deck point is
one at its matching split and four at each of the other two splits.
For every matching product \(F\) with matching \(M\), and every input subset \(S\),
\begin{equation}
  \rank M_{\mathbf x_S,\mathbf x_{\bar S}}(F)=2^{c_M(S)},
  \label{eq:matching-schmidt-rank}
\end{equation}
where \(c_M(S)\) counts edges crossing the cut.

For \(L\in\mathcal K_G^{\rm act}\), the binary \(B_L=XL^{\mathsf T}\) has
\([B_LX]=[XL^{\mathsf T}X]=[L^\#]\in G\), and one disequality edge satisfies
\(\sum_uX(x,u)B_L(u,y)=L(y,x)\).  Hence, for
\(\mathbf L=(L_1,L_2,L_3,L_4)\in(\mathcal K_G^{\rm act})^4\),
\[
  (\mathscr D_{\mathbf L}q)({\mathbf{y}})
  =\sum_{\mathbf{x}}q({\mathbf{x}})\prod_{j=1}^4L_j(y_j,x_j)
\]
is a constant-size realizable dressing with nonzero scalar.

\begin{lemma}[Eight-vertex signatures with rank triples \((2,2,2)\) or \((3,3,3)\)]
\label{lem:p1-platonic-eight-vertex-rank-terminal}
Let \(\cF\) be a retained, factor-closed finite algebraic signature
set containing the supplied equality \(I\).  If \(\cF\) directly
realizes an endpoint-nondegenerate eight-vertex signature \(M\) of
flattening ranks \((2,2,2)\) or \((3,3,3)\), then
\[
  \KHolant(\cF)\text{ is }\SharpP\text{-hard}
  \qquad\text{or}\qquad
  \Tract(K\cF)\text{ holds}.
\]
\end{lemma}

\begin{proof}
Apply \cref{cor:p1-eight-vertex-reduced-interface} to \(M\).  Its first two
outcomes are the asserted conclusions.  In its sole continuing outcome,
\(M\) itself has, up to input order and a nonzero scalar, a factorization
\(E\otimes E'\) into two nonsingular binary signatures.  Its matching
flattening ranks are therefore \((1,4,4)\), up to permutation, by
\cref{eq:matching-schmidt-rank}, contradicting either assumed rank triple.
\end{proof}

\begin{lemma}[The Platonic quaternary boundary]
\label{lem:platonic-quaternary-boundary}
Let \((\cF,G,q)\) be a deck-stable retained, factor-closed fixed-\(I\)
state with nonzero irreducible quaternary signature \(q\).
Let \(G\) be the internal \(A_4\) form \(\TetInt\), any of the four
marked \(S_4\) forms
\(\OctDD,\OctTT,\OctDT,\OctTD\), or a marked
\(A_5\) display, and
define the transfer-valued pair map
\begin{equation*}
  L_p(C)=\Phi_p^q(C)X\in\Mat_2(\mathbb C).
\end{equation*}
If every fixed chosen representative \(K_g\) satisfies
\begin{equation*}
  L_p(K_g)=0\quad\hbox{or}\quad [L_p(K_g)]\in G
  \qquad\text{for all six canonical pairs }p,
\end{equation*}
then internal \(A_4\) and either \(A_5\) display have no irreducible
remaining case.  In the first \(S_4\) form, there are
\(\lambda\in\mathbb C^\times\) and matrices \(U_i\) with
\([U_i]\in\OctDD\) such that
\begin{equation*}
  q=\lambda\,(U_1\otimes U_2\otimes U_3\otimes U_4)(\Eq_4).
\end{equation*}
In the transposition--transposition form, four realizable
\(\OctTT\)-matrices dress \(q\) to
\begin{equation}
  M_{\OctTT}({\mathbf{x}})
  =\one[\wt({\mathbf{x}})\equiv0\pmod2]
    (-1)^{\sum_{i<j}x_ix_j},
  \label{eq:p1-platonic-q4-Oprime-eight-vertex}
\end{equation}
of ranks \((2,2,2)\).  In the two mixed forms the normalized rank-two
remaining case is already an endpoint-nondegenerate eight-vertex signature, with
all odd entries zero and, in the even-string order
\((0000,0011,0101,0110,1001,1010,1100,1111)\), coefficient vectors
\begin{equation}
 \mathbf e(M_{\mathrm{dt}})=(1,1,1,1,1,1,1,1),\qquad
 \mathbf e(M_{\mathrm{td}})=(1,i,i,i,1,1,1,i).
 \label{eq:p1-platonic-q4-mixed-eight-vertex}
\end{equation}
Both have ranks \((2,2,2)\). Hence the arity-four analysis gives a
tensor factorization, an unsafe binary, a generalized equality, or an
endpoint-nondegenerate eight-vertex signature.
\end{lemma}

\begin{proof}
Let \(A_p\in\Mat_4(\mathbb C)\) represent \(L_p\) in the quaternion
coordinates above, so that \(c_Q(L_p(C))=A_pc_Q(C)\).  Up to scale, the
bilinear Frobenius form is \(J_{\mathrm{Frob}}=\diag(1,-1,1,-1)\).
Write \(\bar p\) for the complementary pair and define the right-\(X\)
multiplication map \(\mathsf R_X:\Mat_2(\mathbb C)\to\Mat_2(\mathbb C)\) and
its coordinate matrix \(\mathbf R_X\) by
\[
 \mathsf R_X(C):=CX,\qquad
 c_Q(\mathsf R_X(C))=\mathbf R_Xc_Q(C).
\]
Complementary-pair contraction gives
\begin{equation}
  L_{\bar p}=\mathsf R_X\circ L_p^*\circ\mathsf R_X,
  \label{eq:platonic-q4-map-adjoint}
\end{equation}
where \(^*\) is the Frobenius adjoint.  Thus
\begin{equation}
  A_{\bar p}=\mathbf R_XA_p^\sharp\mathbf R_X,
  \qquad A_p^\sharp=J_{\mathrm{Frob}}A_p^{\mathsf T}J_{\mathrm{Frob}}.
  \label{eq:platonic-q4-adjoint}
\end{equation}
Realizable right \(X\) and reversals turn the six pair conditions into
\begin{equation*}
 \begin{aligned}
  &A_p{\mathbf{r}}=0\ \text{or}\ [A_p{\mathbf{r}}]\in\mathcal R_G,\\
  &A_p^\sharp {\mathbf{r}}=0\ \text{or}\ [A_p^\sharp {\mathbf{r}}]\in\mathcal R_G
 \end{aligned}
 \qquad\text{for every }[{\mathbf{r}}]\in\mathcal R_G,
 \qquad
 \mathcal R_G=\{[c_Q(K_g)]:g\in G\}.
\end{equation*}
By the classification for complementary pairs in
\cref{thm:cert-interface-platonic-q4}\textnormal{(i)}, rank three is impossible. Rank zero gives
\(q=0\), ranks one and four factor, and rank two occurs only for \(S_4\).
In the first marked form, realizable inverse dressings expose \(\Eq_4\), so
\(\KHolant(\cF,\Eq_4)\leT\KHolant(\cF)\) and
\cref{cor:higher-equality-unary,lem:common-unary-exit} apply.  In the second form, its non-\(GO(X)\)
similarity identifies coefficient tables only. \Cref{thm:cert-interface-platonic-q4}\textnormal{(i)} checks
the dressing to \cref{eq:p1-platonic-q4-Oprime-eight-vertex}, directly in realizable
\(\OctTT\)-representatives, with ranks \((2,2,2)\).  For
\(\OctDT,\OctTD\), the direct reconstructed tensors in
\cref{thm:cert-interface-platonic-q4}\textnormal{(i)} are exactly
\cref{eq:p1-platonic-q4-mixed-eight-vertex}.  These three cases invoke
\cref{lem:p1-platonic-eight-vertex-rank-terminal}.  Finally, the direct
correspondence \cref{lem:icosahedral-fixed-I-dictionary} proves the same
no-remaining case conclusion for \(\IcoOne\) while leaving the binary equality untouched.
\end{proof}

The binary reductions, equality-anchor reductions, and eight-vertex
reduction invoked in this proof complete the arity-four case, so no
recursive arity-four call is needed.

\begin{lemma}[Quadratic extension of a fixed matching]
\label{lem:platonic-global-ruling}
Let \(Q_0,Q_1,Q_2,Q_3\) be the quaternion basis in
\cref{eq:quaternion-basis}, and let
\(\mathcal D\subset\mathbb P(\Mat_2)\) contain the twelve tetrahedral
points
\begin{equation}
 [Q_\mu]\ (0\le\mu\le3),
 \qquad
 [Q_0+\eps_1Q_1+\eps_2Q_2+\eps_3Q_3]
 \quad(\eps_j\in\{1,-1\}).
 \label{eq:platonic-quadratic-test-set}
\end{equation}
Let \(M\) be a nonempty finite index set, let \(W_e\) be a nonzero
finite-dimensional complex vector space for every \(e\in M\), and let
\(\Phi:\Mat_2\to\bigotimes_{e\in M}W_e\) be a nonzero linear map.  Define
\[
 \Sigma_M:=
 \left\{\left[\bigotimes_{e\in M}{\mathbf{w}}_e\right]:0\ne {\mathbf{w}}_e\in W_e\right\}
 \subseteq\mathbb P\!\left(\bigotimes_{e\in M}W_e\right),
 \quad
 \operatorname{Cone}_0(\Sigma_M):=\{0\}\cup
 \left\{\bigotimes_{e\in M}{\mathbf{w}}_e:{\mathbf{w}}_e\in W_e\right\},
\]
so that \(\Sigma_M\) and \(\operatorname{Cone}_0(\Sigma_M)\) are,
respectively, the Segre variety and its affine cone in the terminology of
\hyperref[subsubsec:projective-tensor-geometry]{the projective tensor conventions}.
Suppose
\(\Phi(C)\in\operatorname{Cone}_0(\Sigma_M)\) for every
\([C]\in\mathcal D\).
Then there is one index \(e_*\in M\) and fixed nonzero tensors
\(U_e\in W_e\), \(e\ne e_*\), such that
\begin{equation}
  \Phi(C)=
  \left(\bigotimes_{e\ne e_*}U_e\right)\otimes\psi(C)
  \qquad(C\in\Mat_2)
  \label{eq:platonic-global-factorization}
\end{equation}
for a linear map \(\psi:\Mat_2\to W_{e_*}\).  For the reconstruction
conclusion, suppose \(M\) partitions the residual inputs into
nonempty blocks, \(W_e=(\mathbb C^2)^{\otimes e}\), and
\(\Phi=\Phi_p^f\) under the corresponding input identification.  If
\(|M|\ge2\), then \(f=f_{\Phi,p}\) has a proper tensor factor.
\end{lemma}

\begin{proof}[Proof of \cref{lem:platonic-global-ruling}]
Every defining Segre minor pulls back to a homogeneous quadratic in the
quaternion coordinates.  The four axes kill its square coefficients. At the
eight sign points the remaining expression is
\[
 \sum_{0\le\mu<\nu\le3}c_{\mu\nu}\eps_\mu\eps_\nu=0,
 \qquad \eps_0=1.
\]
The six displayed functions are distinct characters of
\(\{\pm1\}^3\), so every \(c_{\mu\nu}\) vanishes.  All Segre minors
therefore vanish identically, including at zero, and
\(\operatorname{im}\Phi\subseteq\operatorname{Cone}_0(\Sigma_M)\).

If a linear subspace lies in \(\operatorname{Cone}_0(\Sigma_M)\), take two nonzero simple tensors
\({\mathbf{u}}=\bigotimes {\mathbf{u}}_e\) and \({\mathbf{v}}=\bigotimes {\mathbf{v}}_e\) in it.  If \({\mathbf{u}}_i,{\mathbf{v}}_i\) and
\({\mathbf{u}}_j,{\mathbf{v}}_j\) are both independent for \(i\ne j\), apply dual functionals on
all other factors. A \(2\times2\) flattening minor of
\({\mathbf{u}}_i\otimes {\mathbf{u}}_j+t\,{\mathbf{v}}_i\otimes {\mathbf{v}}_j\) is then nonzero, contradicting that
\({\mathbf{u}}+t{\mathbf{v}}\) is simple.  Thus each pair varies in at most one factor.  Fixing
one nonzero \({\mathbf{u}}\), two vectors that vary from \({\mathbf{u}}\) in different factors
would violate this pairwise conclusion, so all variation is in one factor.
Hence the image lies in a one-factor Segre ruling, giving
\cref{eq:platonic-global-factorization}.

For \(\Phi=\Phi_p^f\), Frobenius reconstruction using
\cref{eq:platonic-dual-basis} gives
\[
 f=
 \left(\bigotimes_{e\ne e_*}U_e\right)
 \otimes
 \left(\sum_{\mu=0}^3(Q^\mu)^{(p)}\otimes\psi(Q_\mu)\right).
\]
The fixed factor occupies the nonempty set
\(\bigcup_{e\ne e_*}e\), while the other factor contains \(p\), so the
factorization is proper.  \Cref{lem:factor-saturation} exposes the resulting factors under a
Turing reduction.
\end{proof}

\begin{lemma}[Rank-deficient Platonic card maps]
\label{lem:platonic-deficient-rank}
Let \((\cF,G,h)\) be a marked Platonic state, where
\(G\in\{\TetInt,\OctDD,\OctTT,\OctDT,
\OctTD,\IcoZero,\IcoOne\}\).
Assume \(h\ne0\) is a directly realized arity-six signature over \(\cF\), satisfies \cref{eq:platonic-all-card-condition}
for \(G\), and, for a specified canonical ordered pair \(p\), satisfies
\(\rank\Phi_p^h\le3\).  Then \(h\)
factors, or a realizable binary gives a singular/infinite-order terminal or
a successor with a strictly larger complete transfer group.  No rank-deficient incidence branch remains.
\end{lemma}

\begin{proof}
For \(G=\IcoOne\), apply the card covariance and direct network
correspondence of \cref{lem:icosahedral-fixed-I-dictionary}.  It identifies the
coefficient calculation with the standard \(\IcoZero\) calculation while
replacing every realizable kernel by its realizable \(\kappa_D\)-image in
\(\IcoOne\).  Thus its fixed-matching branch uses
\cref{lem:platonic-global-ruling}, and its mixed branch uses
\cref{thm:cert-interface-platonic-q4}\textnormal{(ii)--(iii)}, directly in
the original fixed-\(I\) state.  Hence assume below that
\(G\ne\IcoOne\).
Rank zero contradicts \(h\ne0\) by
\cref{eq:platonic-frobenius-reconstruction}, and rank one factors.  If all nonzero cards
use one matching, apply \cref{lem:platonic-global-ruling}. In any
nonstandard \(S_4\) form precompose with its fixed coefficient similarity
and pull the varying factor back by the inverse similarity.  Pulling back the coefficient identity yields
\(\Phi_p^h(C)=(\bigotimes_{e\ne e_*}U_e)\otimes\psi(C)\).
\Cref{lem:factor-saturation} exposes the proper factor.  If two nonzero cards use different matchings, the
local reduction \cref{thm:cert-interface-platonic-q4}\textnormal{(ii)} gives a realizable rainbow localization and a quaternary whose three
flattening ranks are \((3,3,3)\), while
the same interface gives its realizable outside transfer.
The direct-gadget direction \cref{eq:direct-gadget-reduction} and binary case distinction \cref{lem:binary-interpolation,lem:actual-binary-completeness} then
apply.
\end{proof}

For the rank-four cases below, a \emph{separator} is a fixed realizable
contraction kernel that produces one of the reduction outputs in the
next lemma.

\begin{lemma}[Finite Platonic quaternary reductions]
\label{lem:platonic-finite-q4-terminal}
Let \((\cF,G,f)\) be a marked Platonic state, and let \(q\) be directly
realized over \(\cF\), with its canonical input order and nonzero realization scalar recorded.  Suppose that \(q\) is one
of the explicitly enumerated representatives in
\cref{thm:cert-interface-platonic-q4}\textnormal{(iii)}, including every normalized output later
supplied by a marked rank-four separator.  Then
\(\KHolant(\cF,q)\leT\KHolant(\cF)\), and \(q\) directly realizes
one of the following: a rank-\((2,2,2)\) endpoint-nondegenerate
eight-vertex signature, a quaternary generalized equality, or a
nonsingular binary \(B\) with \([BX]\notin G\).
These outputs are handled by
\cref{lem:p1-platonic-eight-vertex-rank-terminal,lem:pure-ge,%
lem:actual-binary-completeness}.
\end{lemma}

\begin{proof}
This is exactly the representative-wise three-way direct-realization
statement of \cref{thm:cert-interface-platonic-q4}\textnormal{(iii)}.
\end{proof}

Consequently the reductions
\cref{lem:platonic-quaternary-boundary,lem:platonic-deficient-rank,%
lem:platonic-finite-q4-terminal} are bounded local reductions.  After their
displayed output signature is constructed and its finite factorization tree is
processed, every branch is a resolved complexity leaf, a lower-arity factor route, or
a successor with a strictly larger complete transfer group.  An intermediate factor split remains
inside the originating card's finite factorization tree and strictly decreases
the factor potential of \cref{eq:factor-potential}.  Thus each unresolved output decreases the stored factor potential or
gives an outer deck successor.

\subsection{Rainbow localization and matching-contraction closure}

\begin{lemma}[Platonic matching-contraction closure]
\label{lem:platonic-matching-contraction-closure}
Let \(G\le\PGL_2\) contain \([X]\) and be reversal-stable.  Let \(F\) be
zero or a product of nonsingular binary
factors on a perfect matching, with every factor transfer in \(G\).  Contract
any disjoint collection of its inputs by realizable nonsingular \(G\)-binaries
(including the explicit \(I\)- or \(X\)-edge links).  The result is zero or
again a product of nonsingular binary factors whose transfers lie in \(G\).
\end{lemma}

\begin{proof}
Superpose the factor matching with the contraction matching.  Components are
cycles and paths.  A cycle contributes a scalar (possibly zero). A path
splices \(B_1,C,B_2\) to \(B_1^{\mathsf T}CB_2\). Using \(T_B=BX\),
its transfer class is
\[
 [T_{B_1}^\#XT_CXT_{B_2}],\qquad T^\#=XT^{\mathsf T}X.
\]
Reversal stability, \([X]\in G\), and closure of \(G\) keep this class in
\(G\). The omitted orientation scalar is nonzero.  Iterating over
components proves the claim.
\end{proof}

\Cref{lem:platonic-global-ruling} handles cards with one matching. We now
show that three collinear cards with different matchings differ on only
four inputs. Closing their common factors then converts this geometric
statement into a six-input gadget.
\begin{lemma}[Localization of a rainbow matching line]
\label{lem:platonic-rainbow-localization}
Let \(F_0,F_1,F_2\) be three distinct nonsingular matching products on one
projective line, with pairwise distinct perfect matchings \(M_0,M_1,M_2\).
Then there is a unique set \(S\) of four inputs such
that the three matchings restrict on \(S\) to the three pairwise distinct
perfect matchings and the three tensors have common projective binary
factors on \(S^c\).  Moreover, this projective line contains exactly three
matching-product points.  After compatible projective rescaling one may
write
\[
 F_i=R\otimes E_i\qquad(0\le i\le2),
\]
where \(R\ne0\) is the fixed matching product on \(S^c\), and the
quaternary tensors \(E_0,E_1,E_2\) span a two-dimensional subspace.

In the gadget setting, let \((\cF,G,f)\) be a deck-stable marked
Platonic state, where \(f\) has even arity at least six,
and suppose these three tensors occur as the images of one rich domain line
under the card map
\cref{eq:platonic-card-map}.  If every common binary factor \(U\) satisfies
\([UX]\in G\), and the realizable kernel set
\(\mathcal K_G^{\rm act}\) spans
\(\Mat_2(\mathbb C)\), those factors can be closed by realizable
\(G\)-kernels without annihilating the three selected cards.  This realizes
an arity-six signature \(h\) with one \(f\)-vertex and exactly
\((\arity(f)-6)/2\) realizable binary kernels, and
\begin{equation}
  \KHolant(\cF,h)
  \leT
  \KHolant(\cF).
  \label{eq:platonic-localization-reduction}
\end{equation}
Every other realizable \(G\)-card of \(h\) remains zero or a quaternary \(G\)-matching product of nonsingular safe binaries.
\end{lemma}

\begin{proof}[Proof of \cref{lem:platonic-rainbow-localization}]
For an input set \(S\), let \(c_i(S)\) count the edges of \(M_i\) crossing
\(S\mid\bar S\).  Nonsingularity of every binary factor makes the matching
recoverable from flattening ranks and gives
\[
 \rank M_{\mathbf x_S,\mathbf x_{\bar S}}(F_i)=2^{c_i(S)}.
\]
In particular, proportional nonsingular matching products have the same
matching.  Collinearity also gives, for every permutation \((i,j,k)\),
\[
 2^{c_i(S)}\le2^{c_j(S)}+2^{c_k(S)}.
\]

Decompose \(M_0\cup M_1\) into common edges and alternating cycles.  For
each component \(C\), both \(F_0\) and \(F_1\) have flattening rank one
across \(C\mid C^c\), so \(\rank_C(F_2)\le2\).  The number \(c_2(C)\) is
even. Hence \(2^{c_2(C)}\le2\) forces \(c_2(C)=0\).  Regard the components
as tensor factors.  A nontrivial linear combination of two simple tensors
is simple only when the two summands differ projectively in at most one
factor.  Since \(M_2\) is distinct from \(M_0,M_1\), it differs on at least
one alternating cycle (a common-edge component has only one internal
matching).  The simple-sum criterion permits only one such varying
component. Hence there is exactly one exceptional alternating cycle, and
the common-edge factors outside it agree projectively in all three tensors.

The matching \(M_2\) cannot repeat \(M_0\) or \(M_1\) on the exceptional
component.  Indeed, flatten across the \(2\mid(2\ell-2)\) cut defined by
the two endpoints of an edge of the repeated matching: the two same-matching
tensors have rank one, so the third
has rank at most two, whereas its different matching has two edges crossing
that cut and hence rank four.  Thus the three restrictions on the
exceptional component are pairwise distinct.

Write the exceptional cycle as \(2\ell\) cyclic inputs.  If \(\ell\ge3\),
consider its cyclic consecutive triples.  Each \(M_2\)-edge belongs to at
most two of them.  If every triple contained an \(M_2\)-edge, equality in
this incidence count would force every such edge to be a cycle edge and
hence \(M_2=M_0\) or \(M_1\).  Choose instead a triple \(T\) containing no
\(M_2\)-edge.  Then
\[
 (c_0(T),c_1(T),c_2(T))=(1,1,3),
\]
contradicting \(8\le2+2\).  Hence \(\ell=2\), so the exceptional component
is the asserted set of four inputs and carries the three perfect matchings.  It
is unique by unique factorization.  A fourth matching-product point on the
same projective line would repeat one of the three quaternary matchings.
Applying the preceding repeated-matching rank argument to it and either of
the other two points gives a contradiction.  Thus the line has exactly
three matching-product points, and the displayed decomposition
\(F_i=R\otimes E_i\) follows.

For each common factor \(U\), spanning by \(\mathcal K_G^{\rm act}\)
supplies a realizable \(C\) with \(\operatorname{tr}(U^{\mathsf T}C)\ne0\).
Closing these factors preserves all three selected cards and uses one
\(f\)-vertex and \((\arity(f)-6)/2\) binary kernels, proving
\cref{eq:platonic-localization-reduction}. Every other card is zero or a
\(G\)-matching product by \cref{lem:platonic-matching-contraction-closure},
since the marked group contains \([X]\) and is reversal-stable.
This proves \cref{eq:platonic-all-card-condition}.
\end{proof}

For the second octahedral display, with \(H_{\mathrm{tt}}\) already fixed in
\cref{eq:octahedral-second-form}, define the coefficient-similarity map
\begin{equation}
 \theta_{\mathrm{tt}}(U):=H_{\mathrm{tt}}^{-1}UH_{\mathrm{tt}},
 \label{eq:octahedral-coefficient-map}
\end{equation}
the transformation tuple
\[
 \mathbf H_{\mathrm{tt}}:=
 (H_{\mathrm{tt}}^{-1},H_{\mathrm{tt}}^{\mathsf T},
  H_{\mathrm{tt}}^{-1},H_{\mathrm{tt}}^{\mathsf T}),
\]
and its induced quaternary coefficient operator
\[
 \mathcal L_{\mathrm{tt}}:=\mathscr D_{\mathbf H_{\mathrm{tt}}},
 \qquad
 \mathcal L_{\mathrm{tt}}(F)=
 (H_{\mathrm{tt}}^{-1}\otimes H_{\mathrm{tt}}^{\mathsf T}\otimes
  H_{\mathrm{tt}}^{-1}\otimes H_{\mathrm{tt}}^{\mathsf T})F.
\]
These maps compare coefficient tables under the fixed-coordinate
convention at the start of the section.

For \(\alpha\in\{\mathrm{dt},\mathrm{td}\}\), put
\[
 S_\alpha:=H_\alpha H_\alpha^{\mathsf T},\qquad
 \mathcal L_\alpha:=
 H_\alpha^{-1}\otimes H_\alpha^{\mathsf T}\otimes
 H_\alpha^{-1}\otimes H_\alpha^{\mathsf T}.
\]
By \cref{eq:octahedral-mixed-mark-check}, \([S_\alpha]\in\OctDD\).
Ordered contraction gives the three matching formulas
\begin{align}
 \mathcal L_\alpha P_0(U,V)&=
  P_0\bigl(\theta_\alpha(U),\theta_\alpha(V)\bigr),\notag\\
 \mathcal L_\alpha P_1(U,V)&=
  P_1\bigl(\theta_\alpha(US_\alpha^{-1}),
            \theta_\alpha(S_\alpha V)\bigr),\notag\\
 \mathcal L_\alpha P_2(U,V)&=
  P_2\bigl(\theta_\alpha(U),
            \theta_\alpha(S_\alpha VS_\alpha^{-1})\bigr).
 \label{eq:octahedral-mixed-coefficient-transport}
\end{align}
Thus \(\mathcal L_\alpha\) bijects the standard and mixed matching decks,
preserves their matching labels, and intertwines their realizable local group
actions after the displayed current-group orientation factors are absorbed.
The corresponding gadget operations use realizable
\(\OctForm{\alpha}\)-representatives in the original coordinates.

The incidence counts, realizable rainbow-line orbits, and deletion
connectivity used below are supplied by
\cref{thm:cert-interface-platonic-incidence}.

\begin{lemma}[High-arity Platonic localization]
\label{lem:platonic-high-arity-localization}
Let \((\cF,G,f)\) be a deck-stable retained, factor-closed fixed-\(I\)
state, where \(f\ne0\) is a minimum-arity nonbinary irreducible signature
of even arity \(n\ge6\) and
\[
 G\in\{\TetInt,\OctDD,\OctTT,\OctDT,
          \OctTD,\IcoZero,\IcoOne\}.
\]
Fix an ordered pair \(p\), define the original card-map rank
\(r:=\rank\Phi_p^f\in\{2,3,4\}\), and
suppose that, for every ordered pair of inputs and every realizable
\(G\)-kernel, the corresponding original card is zero or a \(G\)-matching product of nonsingular safe binaries.  Then at least one of the following structural
routes applies. When both are available, choose route~1.
\begin{enumerate}
\item All nonzero \(p\)-cards use one residual matching, and
      \(f\) has a proper tensor factor.
\item One \(f\)-vertex and exactly \((n-6)/2\) realizable binary kernels
      directly realize a nonzero arity-six signature \(h\), retaining
      \(p\), such that
      \begin{equation}
        2\le\rank\Phi_p^h\le r\le4,
        \label{eq:platonic-localized-rank-bound}
      \end{equation}
      and every realizable \(G\)-card of \(h\) is zero or a quaternary \(G\)-matching product of nonsingular safe binaries.  In particular,
      \(\KHolant(\cF,h)\leT\KHolant(\cF)\).
\end{enumerate}
The second route recomputes the rank of the directly realized six-input card map.
The original rank is only the upper bound in
\cref{eq:platonic-localized-rank-bound}.
\end{lemma}

\begin{proof}
Use the domain graph \(\mathscr G_G^{\rm dom}\) defined above.  The kernel-deleted
connectivity assertion of
\cref{thm:cert-interface-platonic-incidence}\textnormal{(iii)}
says
that
\[
 \mathcal D_G\setminus\mathbb P(\ker\Phi_p^f)
\]
is nonempty and connected.  If all its nonzero images have one matching,
\cref{lem:platonic-global-ruling} gives a proper factor.  For any
nonstandard octahedral form, precompose the card map with its fixed
invertible coefficient similarity and pull the resulting tensor identity
back.  For \(\IcoOne\), use
\cref{lem:icosahedral-fixed-I-dictionary}.  These are analytic
reconstructions, and factor access is supplied by
\cref{lem:factor-saturation}.

Otherwise choose the first matching-changing edge along a path in this
deleted graph, and let \(\ell\) be a rich domain line containing that edge.  If
\(r\in\{2,3\}\) and \(\ell\) met
\(\mathbb P(\ker\Phi_p^f)\), the restriction of \(\Phi_p^f\) to the
two-dimensional vector span of \(\ell\) would have rank at most one.
All nonkernel images on \(\ell\) would then be projectively proportional
and have the same matching, a contradiction.  If \(r=4\), the nullspace of the card map is
zero.  Thus \(\ell\) is kernel-avoiding.  Let \(V_\ell\) be its
two-dimensional vector span.  The matching-changing edge gives
\(\rank(\Phi_p^f|_{V_\ell})\ge2\), while kernel-avoidance gives
\(\rank(\Phi_p^f|_{V_\ell})\le2\). Hence the restriction has rank two and
induces a projective injection on \(\ell\).  Let \(F_0,F_1\) be the
matching-changing images, and choose any third domain point on this rich
line, with image \(F_2\).  Its matching differs from both of theirs.
Indeed, if \(M_2=M_0\), choose an edge of \(M_0\setminus M_1\) and
flatten across its two inputs.  The ranks of \(F_0,F_2\) are one, whereas
the rank of \(F_1\) is four, because two nonsingular factors cross this
cut.  This contradicts collinearity. The case \(M_2=M_1\) is identical.
Thus the three images satisfy the general-arity hypotheses of
\cref{lem:platonic-rainbow-localization}.  That lemma says their line
contains exactly three matching-product points, so injectivity also forces
\(\ell\) to have exactly three domain points.  The quaternary incidence
classification is used only after this localization.

The same lemma supplies a unique
set \(S\) of four residual inputs and compatible rescalings
\[
 \Phi_p^f(D_i)=R\otimes E_i,
 \qquad
 R=\bigotimes_{j=1}^{(n-6)/2}U_j,
 \qquad 0\le i\le2,
\]
where the \(E_i\) form the residual rainbow deck line and
\(D_i\in\mathcal K_G^{\rm act}\) are the selected domain representatives.
Use the realizable closing kernels \((K_j)_j\) from the proof of
\cref{lem:platonic-rainbow-localization}, leaving \(p\cup S\).
They give a nonzero one-vertex signature \(h\) with common closing scalar
\(\kappa=\prod_j\operatorname{tr}(U_j^{\mathsf T}K_j)\ne0\).
Record \(p\), \(\ell\), the ordered kernels, the output order \(\pi\),
and \(\kappa\).
If \(C\) denotes the resulting contraction on the residual inputs and
\(P_\pi\) the invertible
output reindexing induced by the recorded permutation of inputs, then
\[
 \Phi_p^h=P_\pi\circ C\circ\Phi_p^f.
\]
This proves the upper rank bound, while the surviving rainbow deck line proves
\(\rank\Phi_p^h\ge2\).

Every other card is obtained from an original \(G\)-matching card by
the remaining realizable closures. Thus
\cref{lem:platonic-matching-contraction-closure} gives zero or a
\(G\)-matching product of nonsingular safe binaries.
\end{proof}

Whenever localization supplies a rainbow deck line, use realizable dressings
on the domain pair and on the four residual inputs to send the two lines to
their specified representatives.  The two operations act on disjoint input
sets and commute. Retain their scalars and input orders.
In the standard forms \(G\in\{\TetInt,\OctDD,\IcoZero\}\),
let \(\Phi\) be this realizable card map after one common nonzero scalar is
removed, and put \(G_0=G\).
For a nonstandard octahedral form, write \(\theta,\mathcal L\) for its
fixed coefficient maps and \(\Phi_{\rm phys}\) for the normalized
realizable card map in the current coordinates.  Its domain and image representatives are the \(\theta\)- and
\(\mathcal L\)-images of the standard ones.  Only for coefficient
calculations put
\begin{equation}
 \Phi=\mathcal L^{-1}\circ\Phi_{\rm phys}\circ\theta,
 \qquad
 \Phi_{\rm phys}(\theta C)=\mathcal L\Phi(C),
 \qquad G_0=\OctDD.
 \label{eq:platonic-nonstandard-card-dictionary}
\end{equation}
Thus a kernel \(C\) in a standard formula means the current-form
kernel \(\theta C\), and its output signature is \(\mathcal L\Phi(C)\).
The opposite icosahedral display uses
its own \(\kappa_D\)-correspondence in
\cref{lem:icosahedral-fixed-I-dictionary}.
In these standard coefficient coordinates define
\begin{equation}
 A=P_0(Q_0,Q_0),\qquad
 B=P_1(Q_0,Q_0),\qquad
 C=P_2(Q_2,Q_2)=A-B.
 \label{eq:platonic-rainbow-normalization}
\end{equation}
The two input matrices in the next display sum to \(2Q_0\).  Together with
\(C=A-B\), this fixes the relative image scalars.  Dividing by their one common nonzero scalar, the normalized map satisfies
\begin{equation}
 \Phi(Q_0+Q_1+Q_2+Q_3)=A,
 \qquad
 \Phi(Q_0-Q_1-Q_2-Q_3)=-B.
 \label{eq:platonic-rainbow-images}
\end{equation}
Thus, defining the image tensors \(W_j:=\Phi(Q_j)\),
\begin{equation}
 \Phi(Q_0)=C/2,
 \qquad
 W_1+W_2+W_3=(A+B)/2,
 \label{eq:platonic-rainbow-sum}
\end{equation}
and each remaining sign point imposes
\begin{equation}
 W_j,\quad A-2W_j,\quad 2W_j-B
 \in\{0\}\cup\operatorname{cone}\bigl(\mathfrak D_4(G_0)\bigr).
 \label{eq:platonic-separated-sign-tests}
\end{equation}
These are coefficient identities of realizable cards, read through
\cref{eq:platonic-nonstandard-card-dictionary} in the nonstandard forms.
The reductions below apply to the output signatures in the current basis.

\subsection{The four fixed-equality marked \texorpdfstring{\(S_4\)}{S4} forms}

The projective octahedral group has four ordered marked-involution forms in
the coordinates containing the supplied \(I\).  They are the
standard \((dd;d)\) form \(\OctDD\), the \((tt;d)\) form
\(\OctTT\), and the mixed forms
\(\OctDT,\OctTD\) fixed in
\cref{eq:octahedral-second-form,eq:octahedral-mixed-forms}.  They are
exhaustive by \cref{thm:marked-finite-group-routing}.  For a normalized rainbow map, the separated-sign and sum equations have
six rank-deficient solutions,
handled on that directly realized six-input map by
\cref{lem:platonic-deficient-rank}, and twenty-four full-rank solutions, the orderings of the
four frames listed in
\cref{thm:cert-interface-platonic-bridges}\textnormal{(i)}. Here a frame is
an unordered triple of image tensors \(\{W_1,W_2,W_3\}\), whose six
orderings give the corresponding card maps.

Retain the even-entry order
\begin{equation}
  \mathbf e(M)=
  (M_{0000},M_{0011},M_{0101},M_{0110},
   M_{1001},M_{1010},M_{1100},M_{1111}).
  \label{eq:octahedral-even-order}
\end{equation}

\begin{lemma}[Uniform octahedral full-rank exits]
\label{lem:octahedral-separator}
Let \(\Phi\) be a normalized full-rank map for which
\((\Phi(Q_1),\Phi(Q_2),\Phi(Q_3))\) is one of the twenty-four orderings of
the four frames in
\cref{thm:cert-interface-platonic-bridges}\textnormal{(i)}.  In a nonstandard
form, \(\Phi\) is the coefficient map associated to the realizable
\(\Phi_{\rm phys}\) by \cref{eq:platonic-nonstandard-card-dictionary}.
Define
\[
 K_{\OctDD}:=Q_0+Q_1,\qquad
 K_{\OctTT}:=\theta_{\mathrm{tt}}(K_{\OctDD}),\qquad
 K_{\OctForm{\alpha}}:=\theta_\alpha(K_{\OctDD})
 \quad(\alpha\in\{\mathrm{dt},\mathrm{td}\}).
\]
For each marked form, apply its displayed kernel to the realizable card map.
At most four further realizable group binaries, used in the input
dressings or pair contraction, then directly realize one of:
an endpoint-nondegenerate eight-vertex signature of ranks \((2,2,2)\), a
quaternary generalized equality, or a nonsingular binary whose transfer
lies outside the current complete transfer group.  For \(\OctDT\), the
first-frame entries \(R_0,R_1,R_3\) give the eight-vertex exits and the
other six named entries give outside binaries.  For
\(\OctTD\), \(R_0,R_3\) give the eight-vertex exits, \(R_1\)
gives the generalized equality, and the other six give outside
binaries.  Thus each full-rank map gives one of these three outputs.
\end{lemma}

\begin{proof}
This is \cref{thm:cert-interface-platonic-bridges}\textnormal{(i)} together
with the direct current-coordinate tables in
\cref{prop:p1-cert-S4-uniform-eight-vertex-separator}.  Every listed kernel
and dressing is a chosen representative in its own marked form. Hence
\cref{eq:direct-gadget-reduction} gives the stated directions, followed by
\cref{lem:p1-platonic-eight-vertex-rank-terminal,lem:pure-ge,%
lem:actual-binary-completeness} according to the listed exit.
\end{proof}

For the full-rank six-input statements below, assume
\begin{equation}
 \begin{aligned}
 &\Stable(\cF,G,h),\qquad \arity(h)=\nu(\cF)=6,\\
 &h\text{ is directly realized over }\cF,\qquad \rank\Phi_p^h=4.
 \end{aligned}
 \label{eq:platonic-six-four-predicate}
\end{equation}
The rank in \cref{eq:platonic-six-four-predicate} is computed from
the coefficients of the directly realized \(h\). Under this condition,
put \(\Phi:=\Phi_p^h\). For the octahedral forms, use the auxiliary map
\(\Psi:=\Phi\) in the first form,
\(\Psi:=\Phi\circ\theta_{\mathrm{tt}}\) for \(\OctTT\), and
\(\Psi:=\Phi\circ\theta_\alpha\) for \(\OctForm{\alpha}\),
\(\alpha\in\{\mathrm{dt},\mathrm{td}\}\).

\begin{theorem}[Octahedral deck rigidity]
\label{thm:octahedral-deck-rigidity}
If \cref{eq:platonic-six-four-predicate} holds and
\[
 G\in\{\OctDD,\OctTT,\OctDT,\OctTD\},
\]
then \(\Close_{\varnothing}(\cF,G,h)\).
\end{theorem}

\begin{proof}
If card matchings vary, clauses~\textnormal{(i)--(iii)} of
\cref{thm:cert-interface-platonic-incidence} and
\cref{eq:matching-schmidt-rank} give a rainbow deck line.
\Cref{lem:platonic-rainbow-localization,lem:octahedral-separator}, with
\cref{eq:platonic-nonstandard-card-dictionary} in a nonstandard form,
then give a resolved eight-vertex, generalized-equality, or outside-binary
construction.

If all nonzero cards use one matching \(M\), the twelve test points under
\(\Psi\) map to zero or an \(M\)-matching product.
\Cref{lem:platonic-global-ruling} factors the tensor reconstructed from
\(\Psi\). In a nonstandard form, composing the varying factor with the
inverse coefficient similarity pulls this identity back to \(\Phi\).
Thus \(h\) has a proper factor, available by \cref{lem:factor-saturation}.

Close that finite factorization tree directly.  A nonbinary irreducible factor has
arity below six and therefore lowers \(\nu\).  A singular or unsafe binary
is terminal, and a safe binary whose transfer class lies outside
\(G\) strictly enlarges the complete transfer group.  If every factor
were a safe binary whose transfer class already lies in \(G\), the original
signature would be a \(G\)-matching product, contrary
to the retained irreducible branch.  Hence every branch is a resolved complexity leaf
or an outer deck successor, proving
\(\Close_{\varnothing}(\cF,G,h)\).  The rank and factorization used here belong to the realized tensor \(h\).
\end{proof}

\subsection{The internal \texorpdfstring{\(A_4\)}{A4} form}

The internal tetrahedral case uses the standard twelve-point model
\(\TetInt\) in \cref{eq:tetrahedral-domain}.  It is closed under
multiplication, inverse, and transpose.  All normalizations below use realizable
left and right \(\TetInt\)-multiplication and permutations of inputs.
The exceptional six-input signatures in this case supply \(\Hcore\), from
which we construct \(\RMcore\). These cores provide gadgets for testing
the other retained signatures. After retaining the cores and their factors,
we check the complete group and stabilize again. If the internal
tetrahedral case survives, the task is to show that every remaining
irreducible signature is affine in the same basis.

\begin{theorem}[Tetrahedral deck rigidity]
\label{thm:tetrahedral-deck-rigidity}
If \cref{eq:platonic-six-four-predicate} holds with \(G=\TetInt\), then
\(\Close_{\{\cA\}}(\cF,\TetInt,h)\).
\end{theorem}

\subsubsection{The Pauli cores}
\label{proofsubsec:v4-hcore}
\label{proofsubsec:v4-core-atlas}

The supplied equality and the Pauli subgroup \(\KleinStd\le\TetInt\)
provide the resources for the following construction. Use the Pauli
contraction notation of \cref{proofsubsec:v4-pauli-calculus}.
Recall the cores \(\Hcore\) and \(\RMcore\) from
\cref{eq:H6,eq:RM-core}. For \(\Hcore\), use the ordered pairs
\(A=(1,2),B=(3,4),C=(5,6)\) and the same order-three Bell-label map
\(\Theta\).
Direct expansion gives
\begin{align*}
 \supp(\Hcore)
   &=\left\{{\mathbf{x}}\in\Ftwo^6:\sum_{j=1}^6x_j=0\right\},\\
 \Hcore({\mathbf{x}})&=(-1)^{x_1+x_1x_2+x_1x_6+x_2x_5+x_3x_5+x_3x_6}
   \quad({\mathbf{x}}\in\supp(\Hcore)).
\end{align*}
Hence \(\Hcore\in\cA\). Its full even-parity support makes
\(\Hcore\) irreducible.
By \cref{rem:sc-f8-core}, \(\Hcore\) and \(\RMcore\) are
the Shao--Cai \(\widehat f_6\) and \(f_8=\widehat f_8\), up to the stated
permutation of inputs.  Consequently every Pauli pair card of \(\Hcore\) is a
nonzero product of two Pauli binaries, while every Pauli pair card of
\(\RMcore\) is, up to a nonzero scalar, a product of three copies of the
contracted Pauli binary. See
\cite[full version, Definition~6.12 and Lemma~6.13]{ShaoCai2020} and
\cite[Theorems~13--14]{CaiFuShao2020Entanglement}.

\subsubsection{The \texorpdfstring{\(\Hcore\)-to-\(\RMcore\)}{H6-to-RM8} gadget}
\label{proofsubsec:v4-H6-to-R8}

A \emph{complete Bell block} is an ordered Bell signature \(B_{\boldsymbol{\gamma}}^{(p)}\) with
both inputs of \(p\) available for attachment.  An \emph{\(I\)-path} between
two equally oriented complete blocks is one copy of the retained equality
\(I\) joining corresponding endpoints. Thus joining two blocks uses two \(I\)-paths.

We realize \(\RMcore\) from four copies of \(\Hcore\) joined by
equality signatures. In copy \(\Hcore^{(v)}\), write the ordered blocks
\((\mathsf A_v,\mathsf B_v,\mathsf C_v)
=(B_{{\boldsymbol{\gamma}}_v},B_{\Theta{\boldsymbol{\gamma}}_v},B_{\Theta^2{\boldsymbol{\gamma}}_v})\)
and leave the \(\mathsf A_v\) external.  Connect equally oriented complete
blocks by
\begin{equation*}
  \mathsf B_1-\mathsf B_2,\qquad
  \mathsf C_2-\mathsf C_3,\qquad
  \mathsf B_3-\mathsf B_4,\qquad
  \mathsf C_4-\mathsf C_1.
\end{equation*}

\begin{figure}[htbp]
\centering
\begin{tikzpicture}[
  block/.style={
    draw,rounded corners=1pt,
    minimum width=10mm,minimum height=5.5mm,
    inner sep=1pt,font=\normalsize\bfseries
  },
  Ablock/.style={
    block,draw=blue!65!black,fill=blue!16,
    text=blue!55!black
  },
  Bblock/.style={
    block,draw=orange!75!black,fill=orange!20,
    text=orange!65!black
  },
  Cblock/.style={
    block,draw=teal!70!black,fill=teal!18,
    text=teal!60!black
  },
  corebox/.style={
    draw=black!45,dashed,rounded corners=2pt,
    line width=.45pt
  },
  corelabel/.style={font=\normalsize},
  apath/.style={draw=blue!65!black,line width=.55pt},
  bpath/.style={draw=orange!75!black,line width=.5pt},
  cpath/.style={draw=teal!70!black,line width=.5pt},
  outdot/.style={
    circle,fill=blue!65!black,
    minimum size=1.6pt,inner sep=0pt
  }
]
  \draw[corebox] (-4.25,.55) rectangle (-1.85,2.95);
  \draw[corebox] ( 1.85,.55) rectangle ( 4.25,2.95);
  \draw[corebox] ( 1.85,-2.95) rectangle ( 4.25,-.55);
  \draw[corebox] (-4.25,-2.95) rectangle (-1.85,-.55);

  \node[corelabel] at (-3.55, 1.75) {$\Hcore^{(1)}$};
  \node[corelabel] at ( 3.55, 1.75) {$\Hcore^{(2)}$};
  \node[corelabel] at ( 3.55,-1.75) {$\Hcore^{(3)}$};
  \node[corelabel] at (-3.55,-1.75) {$\Hcore^{(4)}$};

  \node[Ablock] (A1) at (-3.55, 2.55) {$\mathsf A_1$};
  \node[Bblock] (B1) at (-2.40, 1.75) {$\mathsf B_1$};
  \node[Cblock] (C1) at (-3.55,  .95) {$\mathsf C_1$};

  \node[Ablock] (A2) at ( 3.55, 2.55) {$\mathsf A_2$};
  \node[Bblock] (B2) at ( 2.40, 1.75) {$\mathsf B_2$};
  \node[Cblock] (C2) at ( 3.55,  .95) {$\mathsf C_2$};

  \node[Ablock] (A3) at ( 3.55,-2.55) {$\mathsf A_3$};
  \node[Bblock] (B3) at ( 2.40,-1.75) {$\mathsf B_3$};
  \node[Cblock] (C3) at ( 3.55,-.95) {$\mathsf C_3$};

  \node[Ablock] (A4) at (-3.55,-2.55) {$\mathsf A_4$};
  \node[Bblock] (B4) at (-2.40,-1.75) {$\mathsf B_4$};
  \node[Cblock] (C4) at (-3.55,-.95) {$\mathsf C_4$};

  \draw[bpath]
    ([yshift=1.1mm]B1.east) -- ([yshift=1.1mm]B2.west);
  \draw[bpath]
    ([yshift=-1.1mm]B1.east) -- ([yshift=-1.1mm]B2.west);

  \draw[cpath]
    ([xshift=-1.6mm]C2.south) -- ([xshift=-1.6mm]C3.north);
  \draw[cpath]
    ([xshift=1.6mm]C2.south) -- ([xshift=1.6mm]C3.north);

  \draw[bpath]
    ([yshift=1.1mm]B4.east) -- ([yshift=1.1mm]B3.west);
  \draw[bpath]
    ([yshift=-1.1mm]B4.east) -- ([yshift=-1.1mm]B3.west);

  \draw[cpath]
    ([xshift=-1.6mm]C4.north) -- ([xshift=-1.6mm]C1.south);
  \draw[cpath]
    ([xshift=1.6mm]C4.north) -- ([xshift=1.6mm]C1.south);

  \draw[apath] ([xshift=-1.6mm]A1.north) -- ++(0,.60)
    node[outdot] {};
  \draw[apath] ([xshift= 1.6mm]A1.north) -- ++(0,.60)
    node[outdot] {};
  \draw[apath] ([xshift=-1.6mm]A2.north) -- ++(0,.60)
    node[outdot] {};
  \draw[apath] ([xshift= 1.6mm]A2.north) -- ++(0,.60)
    node[outdot] {};

  \draw[apath] ([xshift=-1.6mm]A3.south) -- ++(0,-.60)
    node[outdot] {};
  \draw[apath] ([xshift= 1.6mm]A3.south) -- ++(0,-.60)
    node[outdot] {};
  \draw[apath] ([xshift=-1.6mm]A4.south) -- ++(0,-.60)
    node[outdot] {};
  \draw[apath] ([xshift= 1.6mm]A4.south) -- ++(0,-.60)
    node[outdot] {};
\end{tikzpicture}
\caption{The four-copy \(\Hcore\)-to-\(\RMcore\) gadget. In copy \(v\),
\(\mathsf A_v,\mathsf B_v,\mathsf C_v\) denote the three ordered input
pairs. Each doubled internal connection represents two \(I\)-paths joining
corresponding inputs. The four \(\mathsf A_v\) pairs remain external and
give the eight inputs of \(\RMcore\).}
\label{fig:v4-H6-R8-gadget}
\end{figure}

Each connection uses two \(I\)-paths and contributes
\(2\delta_{{\boldsymbol{\gamma}}_u,{\boldsymbol{\gamma}}_v}\), so all labels agree without cancellation:
\begin{equation}
  2^4\sum_{{\boldsymbol{\gamma}}\in\Ftwo^2}B_{\boldsymbol{\gamma}}^{\otimes4}=32\RMcore.
  \label{eq:app-v4-H6-R8-output}
\end{equation}
Restoring the known nonzero realization and path scalars gives
\begin{equation*}
  \KHolant(\cF,\RMcore)\leT
  \KHolant(\cF,\Hcore).
\end{equation*}
If \(\Hcore\) is already realizable or available by reduction over \(\cF\), then
\(\KHolant(\cF,\Hcore)\leT\KHolant(\cF)\) as well.
If the starting complete transfer group is \(\KleinStd\), adding the
realizable gadget output leaves that complete group unchanged.
The gadget requires only \(\Hcore\) and realizable \(I\)-paths, so it
also applies in a signature set with a larger complete transfer group.
A reverse gadget construction using only \(\RMcore\) and Pauli
resources is excluded by \cref{cor:no-rm-to-H6}.

\subsubsection{Six-input tetrahedral signatures}

Let \(\Phi:\Mat_2\to{(\mathbb C^2)}^{\otimes4}\) be the full-rank map
localized from a tetrahedral rainbow deck line.  We use the common normalization
\cref{eq:platonic-rainbow-normalization,eq:platonic-rainbow-images,%
eq:platonic-rainbow-sum,eq:platonic-separated-sign-tests} with
\(G=\TetInt\).

\begin{lemma}[Solutions of the tetrahedral separated-sign equations]
\label{lem:tetrahedral-bridge}
The simultaneous solution set of
\cref{eq:platonic-separated-sign-tests} is
\begin{equation*}
  \mathcal B_{\TetInt}
  =\left\{0,\frac A2,\frac B2,R_0,R_1,R_3\right\},
\end{equation*}
where
\begin{equation*}
  R_0=\frac12P_2(Q_0,Q_0),\qquad
  R_1=-\frac12P_2(Q_1,Q_1),\qquad
  R_3=-\frac12P_2(Q_3,Q_3).
\end{equation*}
The sum equation leaves exactly twelve ordered maps: the six rank-two
permutations of \((0,A/2,B/2)\) and the six rank-four permutations of
\((R_0,R_1,R_3)\).
\end{lemma}

\begin{proof}
The exact separated-sign and sum calculation, including dependent lines,
is \cref{thm:cert-interface-platonic-bridges}\textnormal{(ii)}.
\end{proof}

For the rank-four maps write \(\sigma=(a,b,c)\) for a permutation of
\((0,1,3)\).  The corresponding six-input signature is uniquely reconstructed
as
\begin{equation*}
  f_\sigma=
  \sum_{\mu=0}^3(Q^\mu)^{(1,2)}\otimes F^\sigma_\mu,\qquad
  (F^\sigma_0,F^\sigma_1,F^\sigma_2,F^\sigma_3)
  =(C/2,R_a,R_b,R_c).
\end{equation*}

\begin{lemma}[The two permutation classes]
\label{lem:tetrahedral-orientation}
The six rank-four maps split into two sets:
\begin{align*}
  \Sigma_+&=\{(0,1,3),(1,3,0),(3,0,1)\},\notag\\
  \Sigma_-&=\{(0,3,1),(1,0,3),(3,1,0)\}.
\end{align*}
For \(\sigma\in\Sigma_+\), all fifteen deleted-pair contractions by all
twelve realizable tetrahedral kernels are nonzero matching products.  For
\(\sigma\in\Sigma_-\), contracting inputs \(1,3\) by the realizable kernel
\(Q_0=I\) gives a directly realizable quaternary \(q_\sigma\) whose three
flattening ranks are \((2,2,2)\).  For each such \(\sigma\), four
realizable tetrahedral input dressings transform \(q_\sigma\) into an
endpoint-nondegenerate eight-vertex signature of the same ranks.  The
contraction and dressings are direct gadgets from \(f_\sigma\), so
\cref{lem:p1-platonic-eight-vertex-rank-terminal} applies in the required
direction.
\end{lemma}

\begin{proof}
\Cref{thm:cert-interface-platonic-bridges}\textnormal{(ii)} gives the split,
common \(Q_0\)-card, and realizable
endpoint-nondegenerate eight-vertex signature dressings.  Thus the two
direct-gadget reductions point back to \(f_\sigma\), and
\cref{lem:p1-platonic-eight-vertex-rank-terminal} applies.
\end{proof}

\begin{lemma}[The \(\Sigma_+\) signatures are affine]
\label{lem:tetrahedral-affine}
For every \(\sigma\in\Sigma_+\), the tensor \(f_\sigma\) belongs to the
affine class \(\cA\) in the same computational basis as all twelve
tetrahedral binaries.
\end{lemma}

\begin{proof}
\Cref{thm:cert-interface-platonic-bridges}\textnormal{(ii)} gives affine
support and even mixed quadratic coefficients. \Cref{eq:tetrahedral-domain} puts all twelve
binaries in that same affine basis.
\end{proof}

To lift this statement to the signature set, define the tetrahedral matrices
\[
  R_+=Q_0+Q_1+Q_2+Q_3,\qquad
  R_-=Q_0-Q_1-Q_2-Q_3.
\]
Then
\begin{equation*}
 R_+R_-=4I,\qquad
 R_+Q_1=-Q_0+Q_1-Q_2+Q_3,
\end{equation*}
so \([R_+],[R_-],[Q_1],[R_+Q_1]\in\TetInt\), and all four matrices
have realizable inverses up to nonzero scalar.  For a matrix \(M\), let
\(L_j(M)\) denote its action \(M(y_j,x_j)\) on input \(j\), and
let \(\Pi_{ab}\) exchange output inputs \(a,b\). Products of permutations
below act from right to left.  Entrywise expansion in the fixed input convention gives the identities
\begin{align}
 \Pi_{45}\Pi_{23}\Pi_{34}
 \bigl[I\otimes R_+\otimes I^{\otimes3}\otimes R_+\bigr]\Hcore
   &=-16 f_{(0,1,3)},\notag\\
 \Pi_{34}
 \bigl[I\otimes R_-\otimes(R_+Q_1)\otimes R_+
              \otimes I^{\otimes2}\bigr]\Hcore
   &=-32 f_{(1,3,0)},\notag\\
 \Pi_{23}
 \bigl[I\otimes R_+\otimes R_+\otimes I\otimes R_+\otimes Q_1\bigr]\Hcore
   &=32 f_{(3,0,1)}.
  \label{eq:tetrahedral-H6-equivalence}
\end{align}
The bracketed tensor products act on the six inputs by the \(L_j\)-actions.  Because \(R_+^{-1}=R_-/4\) and
\(Q_1^{-1}=-Q_1\), every local action and permutation is projectively
realizable and invertible.  Thus all three equivalences in
\cref{eq:tetrahedral-H6-equivalence} are reversible identities.
Every \(f_\sigma\), \(\sigma\in\Sigma_+\), therefore directly gives \(\Hcore\). Four copies then directly give \(\RMcore\) by
\cref{eq:app-v4-H6-R8-output}. \Cref{lem:tetrahedral-stable-HR-augmentation}
retains both cores and regenerates the deck if the group stays \(\TetInt\).
The remaining deck-stable set then satisfies the hypotheses of
\cref{thm:a4-relative}. For any retained set \(\cF\) directly realizing
\(f_\sigma\), the direct-gadget reductions point from
\(\KHolant(\cF,\Hcore,\RMcore)\) through
\(\KHolant(\cF,f_\sigma)\) to \(\KHolant(\cF)\).

After the arity-four case, let a signature \(f\) have even arity at
least six.  Rank zero is impossible and rank one factors.  In ranks two
through four, \cref{lem:platonic-high-arity-localization} either gives the
proper fixed-matching factor or directly realizes an arity-six signature satisfying \cref{eq:platonic-all-card-condition}.
The six-input card-map rank is then recomputed from its own coefficients before
\cref{lem:tetrahedral-bridge,lem:tetrahedral-orientation} is invoked. 

\subsubsection{Relative rigidity of the retained set}

\begin{theorem}[Tetrahedral relative rigidity]
\label{thm:a4-relative}
Assume
\[
 \Stable(\cF,\TetInt,f),\qquad
 \{\alpha\Hcore,\beta\RMcore\}\subseteq\cF,\qquad
 \alpha\beta\ne0.
\]
Then \(\Close_{\{\cA\}}(\cF,\TetInt,f)\).
\end{theorem}

To obtain a common affine basis, every retained occurrence \(g\) must be
classified with all twelve tetrahedral kernels.  A binary factor may have
transfer in \(\TetInt\setminus \KleinStd\), so the complete family is
\(\{\Phi_p^g(C):C\in\mathcal K_{\TetInt}^{\rm act}\}\) for each pair
\(p\).  The six-input classification supplies the two sets below. An
eight-input base and a stabilizer-code shortening argument then handle
larger arities.

Define the projective six-input matching set, \(\Hcore\)-orbit, and their union by
\begin{align*}
 \mathfrak D_6(\TetInt)
   &=\left\{\left[\prod_{\{a,b\}\in M}(U_{ab})^{(a,b)}\right]:
       M\text{ is a perfect matching of }[6],\ [U_{ab}X]\in\TetInt
     \right\},\\
 \mathscr H_6
   &=\operatorname{Orb}_{(\TetInt)^6\rtimes S_6}([\Hcore]),
 \qquad
 \mathscr U_6=\mathfrak D_6(\TetInt)\sqcup\mathscr H_6.
\end{align*}
Here \(U_{ab}\) is the binary matrix assigned to \(\{a,b\}\), ordered with \(a<b\). Define their affine cone by
\(\mathscr C_6:=\{0\}\cup\operatorname{cone}(\mathscr U_6)\).
Here the local action uses the fixed representatives of
\(\TetInt\), so every orbit equivalence is a realizable reversible input
dressing.  Their disjointness and exact cardinalities are given by
\cref{thm:cert-interface-A4-q8}.

\begin{proposition}[The tetrahedral arity-eight case]
\label{prop:a4rel-required-eight-port}
Let \(g\ne0\) be an arity-eight signature in the normalized tetrahedral
branch whose every tetrahedral pair card lies in \(\mathscr C_6\).
Then \(g\) factors across a \(2|6\) cut or \(g\in\cA\) in the normalized
basis.  Thus every irreducible eight-input remaining case is affine. Factors are
exposed only through factor closure.
\end{proposition}

\begin{proof}
For a pair map \(\Phi=\Phi_p^g\), choose a kernel with nonzero image.
By a realizable tetrahedral multiplication on the deleted pair, take its
domain point to be \([Q_0]\). Dress the residual inputs and permute them to
normalize \(\Phi(Q_0)\) projectively to a three-tetrahedral-binary product
or \(\Hcore\). By \cref{eq:tetrahedral-domain}, the four domain lines
through \([Q_0]\) have opposite endpoints represented by \(Q_0\pm S\),
where \(S=Q_1\pm Q_2\pm Q_3\). Their images sum to \(2\Phi(Q_0)\).
If the two images on each line are dependent, their difference
\(2\Phi(S)\) lies in \(\Span\{\Phi(Q_0)\}\). The four directions \(S\)
span \(\Span\{Q_1,Q_2,Q_3\}\), so \(\rank\Phi=1\). Dual-basis
reconstruction then gives a tensor factor across the \(2|6\) cut. Otherwise the
exhaustive classification with arbitrary scalar coefficients
\cref{thm:cert-interface-A4-q8} says that every nonaffine parent has
a realizable card outside \(\mathscr C_6\), contrary to the hypothesis.
The remaining signatures are affine, and any factors are available by
\cref{lem:factor-saturation}.
\end{proof}

To compare the twelve cards, we describe their coefficient space by a
stabilizer encoder. Its input tensor factors are called logical qubits,
and its output factors are the physical qubits corresponding to the
original signature inputs. The next two lemmas bound the varying part by
four logical qubits. The erasure criterion then identifies an output
qubit that can be separated, allowing the shortening argument to select
a realizable tetrahedral kernel.

\subsubsection{Hilbert-space conventions}
\label{subsubsec:hilbert-conventions}
\label{subsec:stabilizer-conventions}

The stabilizer argument uses the Hermitian inner product
\(\langle {\mathbf{u}},{\mathbf{v}}\rangle={\mathbf{u}}^\dagger{\mathbf{v}}\),
conjugate-linear in its first argument, with \(A^\dagger\) denoting
conjugate transpose. Put
\(\mathcal H_n:=(\mathbb C^2)^{\otimes n}\), and write
\[
 |{\mathbf x}\rangle:=|x_1\rangle\otimes\cdots\otimes|x_n\rangle,
 \qquad
 \langle {\mathbf x}|:=|{\mathbf x}\rangle^\dagger
 \quad({\mathbf x}\in\Ftwo^n).
\]
A unitary on \(\mathcal H_n\) satisfies
\(U^\dagger U=UU^\dagger=I_{\mathcal H_n}\), while an isometry
\(V:\mathcal H_k\to\mathcal H_n\) satisfies
\(V^\dagger V=I_{\mathcal H_k}\).
These Hermitian conventions are distinct from the bilinear Frobenius pairing
used for Holant contractions and from isometries of symmetric binary forms.

A density operator \(\rho\) has trace one and satisfies
\({\mathbf{u}}^\dagger\rho{\mathbf{u}}\ge0\) for every \({\mathbf{u}}\).
For a nonzero vector \(|\psi\rangle\), its pure-state density operator is
\[
 \rho_\psi=\frac{|\psi\rangle\langle\psi|}
                  {\langle\psi,\psi\rangle}.
\]
It is unchanged by nonzero scalar multiplication of \(|\psi\rangle\).
For a density operator \(\rho\) on
\(\mathcal H_A\otimes\mathcal H_B\), fix an orthonormal basis
\(\{|b_j\rangle\}_j\) of \(\mathcal H_B\).  The partial trace is specified,
for every \(|a\rangle,|a'\rangle\in\mathcal H_A\), by
\[
 \langle a|\operatorname{Tr}_B(\rho)|a'\rangle
 =\sum_j\langle a,b_j|\rho|a',b_j\rangle.
\]
The result is independent of the chosen orthonormal basis of
\(\mathcal H_B\).  The operator
\(\operatorname{Tr}_B(\rho)\) is the \emph{marginal} on \(\mathcal H_A\). A one-qubit marginal is maximally
 mixed when it is \(I_{\mathbb C^2}/2\).  Every bipartite pure state has a Schmidt
decomposition
\[
 |\psi\rangle=\sum_{j=1}^s
 \sigma_j|{\mathbf{u}}_j\rangle\otimes|{\mathbf{v}}_j\rangle,\qquad \sigma_j>0,
\]
with both displayed families orthonormal. \(s\) is the Schmidt rank
\cite[Section~2.5]{NielsenChuang2010}.  For \(n\ge1\), \(r\in[n]\), and a
one-qubit linear operator \(R:\mathbb C^2\to\mathbb C^2\), define
\[
 R_r:=I_{\mathbb C^2}^{\otimes(r-1)}\otimes R\otimes
      I_{\mathbb C^2}^{\otimes(n-r)}:\mathcal H_n\to\mathcal H_n.
\]

\subsubsection{Pauli and stabilizer conventions}
\label{subsubsec:pauli-stabilizer}

The \(n\)-qubit Pauli group is
\[
 \mathcal P_n^{(4)}
 :=
 \{i^c\textstyle\bigotimes_{j=1}^nX^{a_j}Z^{b_j}:
   c\in\mathbb Z_4,\ {\mathbf{a}},{\mathbf{b}}\in\Ftwo^n\}.
\]
Its projectivization by the scalar subgroup
\(\{\pm I_{\mathcal H_n},\pm iI_{\mathcal H_n}\}\) is identified with
\(\mathsf P([n])=\Ftwo^n\oplus\Ftwo^n\). The label
\({\mathbf{a}}\mid {\mathbf{b}}\) is the concatenation of the two binary strings.  This label space
has the symplectic form
\[
 \omega(({\mathbf{a}}\mid {\mathbf{b}}),({\mathbf{a}}'\mid {\mathbf{b}}'))={\mathbf{a}}\cdot {\mathbf{b}}'+{\mathbf{b}}\cdot {\mathbf{a}}'.
\]
Here symplectic means alternating
\(\omega({\mathbf{v}},{\mathbf{v}})=0\) and nondegenerate
\(\{{\mathbf{v}}:\omega({\mathbf{v}},{\mathbf{w}})=0\text{ for all }{\mathbf{w}}\}=\{0\}\).  For a subspace \(S\),
put \(S^{\perp_\omega}:=\{{\mathbf{v}}:\omega({\mathbf{v}},{\mathbf{s}})=0\text{ for every }{\mathbf{s}}\in S\}\).
\(S\) is \emph{symplectic Lagrangian} when \(S=S^{\perp_\omega}\).  If \(S\) is
Lagrangian, a \emph{Lagrangian complement} of \(S\) is a Lagrangian subspace
\(T\) satisfying \(\mathsf P([n])=S\oplus T\).

A \emph{projective Clifford} is the class, modulo scalars in \(S^1\),
of a unitary \(U\) satisfying
\(U\mathcal P_n^{(4)}U^\dagger=\mathcal P_n^{(4)}\).  Clifford conjugation induces a symplectic linear map
on Pauli labels \cite[Chapter~10]{NielsenChuang2010}.  A stabilizer code is
the common eigenspace
\[
 \{{\mathbf{v}}:P {\mathbf{v}}=\chi(P){\mathbf{v}}\text{ for every }P\in\widetilde S\}
\]
of a commuting Pauli subgroup \(\widetilde S\), where
\(\chi:\widetilde S\to\mu_4\) is an eigenvalue character agreeing with the
scalar by which every scalar Pauli acts.  A
one-dimensional stabilizer code is a stabilizer state, and a nonzero tensor
spanning it is a stabilizer tensor.
Up to nonzero scalar normalization, the nonzero affine signatures are
exactly the stabilizer states: their amplitudes have affine support and
a quadratic phase polynomial with even mixed coefficients
\cite[Theorem~5]{DehaeneDeMoor2003}. Clifford changes preserve this class.

A stabilizer isometry is an isometry whose image is a stabilizer code and
whose logical Paulis have physical Pauli representatives. Equivalently it is
a Clifford encoder with fixed stabilizer-state ancillas.  Here an
\emph{encoder} is the code isometry, a \emph{physical operator} acts on its
ambient output space, the corresponding \emph{logical operator} is the
induced operator on the encoded input space, and an \emph{ancilla} is a fixed
initialized tensor factor.  If \(S\) is the
projective stabilizer label space, the logical Pauli label space is
\(S^{\perp_\omega}/S\). For a \(2^k\)-dimensional code it has binary
dimension \(2k\).  A Clifford decoder is a Clifford unitary \(D\) satisfying
\[
 DV|\psi\rangle=|\psi\rangle\otimes|\eta\rangle
 \qquad(|\psi\rangle\in\mathcal H_k)
\]
for a fixed stabilizer ancilla \(|\eta\rangle\).  These standard
encoder--decoder and logical-Pauli facts follow from the stabilizer normal
form \cite[Chapters~3--4]{Gottesman1997}. In particular, decoding a
one-dimensional stabilizer code on \(m\) qubits sends its spanning tensor to a nonzero
multiple of \(\delta_{\zero^m}\), as used below.

\begin{lemma}[Sharp stabilizer-line bound]
\label{lem:a4rel-sharp-stabilizer-line}
Let \(u,v\) be nonproportional nonzero affine signatures of common
arity \(m\), and let \(c\ne0\).  If \(u+cv\) is also nonzero and affine, then, after applying the same analytic Clifford change to both
tensors, followed by a common nonzero scalar normalization sending
\(u\) to \(\delta_{\zero^m}\), the support of the transformed
\(v\) is either a singleton disjoint from zero or a linear subspace of
dimension at most two.  In the already normalized case
\(u=\delta_{\zero^m}\), the same support alternative holds for \(v\)
in the current coordinates, without a further Clifford change.
\end{lemma}

\begin{proof}
Choose a unitary Clifford \(C\) with \(Cu=\alpha\delta_{\zero^m}\),
where \(\alpha\ne0\), and apply \(\alpha^{-1}C\) to both tensors.
Relabel the resulting second tensor as \(v\).
Write \(v\) in affine normal form as
\(v=\lambda\one[{\mathbf{a}}+L]i^Q\), where \(\lambda\in\mathbb C^\times\) is the scale,
\({\mathbf{a}}+L\) is the affine support, \(Q\) is the quadratic phase polynomial, and
\(d:=\dim L\).  If
\(0\notin {\mathbf{a}}+L\), the support size \(1+2^d\) forces \(d=0\).  If
\(0\in {\mathbf{a}}+L=L\) and the zero coefficient cancels, \(2^d-1\) forces \(d=1\).
Otherwise both \(v\) and \(w:=\delta_{\zero^m}+cv\) have support \(L\).
Their quotient \(q({\mathbf{x}}):=w({\mathbf{x}})/(cv({\mathbf{x}}))\) equals one on \(L\setminus\{0\}\),
whereas \(q(0)=1+1/(cv(0))\ne1\).  If \(d\ge3\), choose independent
\({\mathbf{a}},{\mathbf{b}},{\mathbf{e}}\in L\).  Since both phases are quadratic, their quotient satisfies
\[
 q(0)q({\mathbf{a}}+{\mathbf{b}})q({\mathbf{a}}+{\mathbf{e}})q({\mathbf{b}}+{\mathbf{e}})
 =q({\mathbf{a}})q({\mathbf{b}})q({\mathbf{e}})q({\mathbf{a}}+{\mathbf{b}}+{\mathbf{e}}).
\]
Every displayed argument except zero is nonzero, so the identity forces
\(q(0)=1\), a contradiction.  Hence \(d\le2\).
The support argument starts only from \(u=\delta_{\zero^m}\). It therefore
also proves the already normalized clause in the statement.
\end{proof}

\begin{lemma}[Tetrahedral card maps have at most four logical qubits]
\label{lem:a4rel-four-active-qubits}
Let
\(
 \Phi:\Mat_2(\mathbb C)\to(\mathbb C^2)^{\otimes m}
\)
be linear.  Suppose that \(\Phi(C_{\rm ker})\) is zero or affine for every binary contraction
kernel \(C_{\rm ker}\) with \([C_{\rm ker}]\in\TetInt\).
Then there are an analytic Clifford operator \(C\) on the \(m\) output
qubits and a subspace \(R\le\Ftwo^m\), \(\dim R\le4\), such that
\begin{equation}
 \supp(C\Phi(C_{\rm ker}))\subseteq R
 \qquad(C_{\rm ker}\in\Mat_2(\mathbb C)).
 \label{eq:a4rel-common-active-support}
\end{equation}
Equivalently, the image of \(\Phi\) is contained in one stabilizer code
with at most four logical qubits.
\end{lemma}

\begin{proof}
If \(\Phi=0\), take \(R=\{0\}\).  Otherwise at least one tetrahedral test
point has nonzero image because those points span \(\Mat_2\).  Precompose
\(\Phi\) by a realizable tetrahedral multiplication carrying that point to
\(A:=Q_0-Q_1-Q_2-Q_3\), relabel the resulting map as \(\Phi\), and hence
assume \(\Phi(A)\ne0\).  Define
\[
 B_0:=2Q_0-A,\qquad B_j:=A+2Q_j\quad(1\le j\le3).
\]
These are the four opposite-parity tetrahedral sign points paired with
\(A\). They
form a basis and satisfy
\[
 A+B_0=2Q_0,\qquad B_j-A=2Q_j\ (1\le j\le3),\qquad
 2A=-B_0+B_1+B_2+B_3.
\]
Choose an analytic Clifford \(C\) with
\(C\Phi(A)=\alpha\delta_{\zero^m}\), where \(\alpha\ne0\).
For the following analysis put \(\Psi(C_{\rm ker}):=\alpha^{-1}C\Phi(C_{\rm ker})\).
The common scalar leaves every support unchanged.  The already normalized clause of
\cref{lem:a4rel-sharp-stabilizer-line} applies in these fixed coordinates:
\(\Psi(A)=\delta_{\zero^m}\), and the displayed identities give the required
affine sums for the \(B_j\).  Thus each nonzero
\(\Psi(B_j)\) not proportional to \(\delta_{\zero^m}\) has, off zero, either
one point or the three nonzero points of a two-dimensional
\(\Ftwo\)-subspace.  The last displayed identity makes
every point in their union occur at least twice.  If its span had dimension
at least five, total incidence at most twelve would force sizes
\((3,3,3,1)\) or \((3,3,3,3)\).  In type \((3,3,3,1)\), double incidence
leaves at most five distinct nonzero points, and any three-point support is
the nonzero part \(\{{\mathbf{x}},{\mathbf{y}},{\mathbf{x}}+{\mathbf{y}}\}\) of a two-dimensional
\(\Ftwo\)-subspace.  Its linear dependence is
nontrivial, so those at most five points have span at most four.

In type \((3,3,3,3)\), double incidence leaves at most six points.  The only
remaining extremal case has six points, each occurring in exactly two of
the four supports.  Regard a point shared by supports \(i,j\) as an edge
between vertices \(i,j\).  The four supports form a loopless
three-regular multigraph on four vertices, and the linear dependence at each
vertex is the sum of its three incident edge labels.  The binary
vertex--edge incidence matrix has rank at least two: its rank is
\(4-c\), where the graph has \(c\le2\) connected components.  Hence the six
edge labels satisfy at least two independent linear dependencies and again span at
most four.  Thus their union spans some \(R\) with \(\dim R\le4\).  Since
the \(B_j\)'s form a basis and the common nonzero scalar leaves
supports unchanged, \cref{eq:a4rel-common-active-support} follows.
\end{proof}

For shortening, use the identification of the octahedral group
\(\OctDD\) with the projective one-qubit Clifford group in the
quaternion coordinates of \cref{eq:quaternion-basis}.  The normal subgroup
\(\TetInt\triangleleft\OctDD\) has index two.  Each of its two cosets
contains the following eight full-support representatives:
\begin{align}
 \mathcal C_+
 &=\{Q_0+\epsilon_1Q_1+\epsilon_2Q_2+\epsilon_3Q_3:
       \epsilon_j\in\{\pm1\}\},\notag\\
 \mathcal C_-
 &=\{Q_d+\epsilon Q_o:
       d\in\{0,3\},\ o\in\{1,2\},\ \epsilon\in\{\pm1\}\}.
 \label{eq:a4rel-full-support-cosets}
\end{align}
On the open set \(C_{01}C_{10}\ne0\), the coefficient cross ratio is \(\operatorname{cr}(C)=C_{00}C_{11}/(C_{01}C_{10})\).  Direct use of
\cref{eq:quaternion-basis} gives
\begin{equation}
 C_{xy}\ne0,\qquad \operatorname{cr}(C)=-1
 \qquad(C\in\mathcal C_+\cup\mathcal C_-).
 \label{eq:a4rel-full-support-cross-ratio}
\end{equation}
Indeed, in \(\mathcal C_+\) the diagonal and off-diagonal products are
\(2\) and \(-2\), while in \(\mathcal C_-\) they are \(1\) and \(-1\).
After division by one entry, every such coefficient signature and its pointwise inverse
have entries in \(\mu_4\), so both are nowhere-zero affine phase tables.

\subsubsection{Stabilizer error correction}
\label{subsubsec:stabilizer-error-correction}

A stabilizer tableau is a binary symplectic generator matrix whose rows label
Pauli generators, together with their phases. Multiplying generators is an
elementary row operation, while a Clifford coordinate change acts by a
symplectic change of columns \cite[Chapter~4]{Gottesman1997}.  A hyperbolic
pair is a pair \({\mathbf{p}},{\mathbf{q}}\) with
\(\omega({\mathbf{p}},{\mathbf{q}})=1\). Symplectic Gaussian elimination is row reduction by
changes of symplectic basis.  A nowhere-zero stabilizer quadratic phase is
\[
 \zeta({\mathbf z})=\zeta_0i^{Q({\mathbf z})},\qquad \zeta_0\in\mathbb C^\times,
\]
where \(Q:\Ftwo^k\to\mathbb Z_4\) is multilinear of degree at most two with
even mixed coefficients.  After fixed physical and logical Clifford
coordinate changes, a stabilizer isometry has amplitude normal form
\[
 V|{\mathbf z}\rangle=\zeta({\mathbf z})|{\mathbf{c}}+M^{\mathsf T}{\mathbf z}\rangle,
 \qquad
 {\mathbf{c}}\in\Ftwo^n,\quad M\in\Mat_{k\times n}(\Ftwo),\quad\rank M=k,
\]
a standard consequence of stabilizer-tableau reduction
\cite[Chapter~4]{Gottesman1997}.

Erasure of physical output qubit \(r\) of an isometry \(V\) is correctable exactly
when the single-site Knill--Laflamme condition holds:
\[
 V^\dagger R_rV\in\mathbb C I_{\mathcal H_k}
 \qquad\text{for every one-qubit operator }R.
\]
For a specified erased qubit, the error-operator products are the
matrix units \((\lvert a\rangle\langle b\rvert)_r\), \(a,b\in\{0,1\}\).
They span its operator space, giving the displayed criterion
\cite{KnillLaflamme1997}. Equivalently, one may test the Pauli basis
\(I,X,-iY,Z\). The erased qubit then carries no information about the
logical input.
In the stabilizer setting, an encoder tableau is the tableau of a Clifford
encoder. Its symplectic row reduction records physical representatives of logical Paulis and the residual stabilizer generators.

\begin{lemma}[Stabilizer erasure and column normal form]
\label{lem:stabilizer-erasure-column-normal-form}
Let
\(
 V:(\mathbb C^2)^{\otimes k}\to(\mathbb C^2)^{\otimes N}
\)
be a stabilizer isometry and let \(r\) be a physical output qubit.  If
erasure of \(r\) is correctable, meaning
\(V^\dagger R_rV\) is scalar for every one-qubit operator \(R\),
exactly one of the following normal forms applies after a fixed one-qubit
Clifford on \(r\):
\begin{enumerate}
\item there are a one-qubit stabilizer state \(\lvert\chi\rangle_r\) and a
      stabilizer isometry \(W\) such that
      \begin{equation}
       V\lvert\psi\rangle
       =\lvert\chi\rangle_r\otimes W\lvert\psi\rangle;
       \label{eq:stabilizer-erasure-pure}
      \end{equation}
\item there are an auxiliary qubit \(a\) and a stabilizer isometry
      \(
       W:\mathbb C^2_a\otimes(\mathbb C^2)^{\otimes k}
       \to(\mathbb C^2)^{\otimes(N-1)}
      \)
      such that
      \begin{equation}
       V\lvert\psi\rangle
       =2^{-1/2}\sum_{c\in\Ftwo}
        \lvert c\rangle_r W(\lvert c\rangle_a\otimes\lvert\psi\rangle).
       \label{eq:stabilizer-erasure-mixed}
      \end{equation}
\end{enumerate}
If no single-site erasure is correctable, then, after postcomposing the
outputs by a tensor product of one-qubit Cliffords and precomposing the input
by one logical Clifford, there are a vector
\({\mathbf{c}}\in\Ftwo^N\), and a matrix
\(M=({\mathbf{m}}_1\ \cdots\ {\mathbf{m}}_N)\in\Ftwo^{k\times N}\) such that
\begin{equation}
 V\lvert {\mathbf z}\rangle
 =\zeta({\mathbf z})\lvert {\mathbf{c}}+M^{\mathsf T}{\mathbf z}\rangle,
 \qquad \rank M=k,\qquad {\mathbf{m}}_r\ne0\quad(1\le r\le N),
 \label{eq:stabilizer-column-normal-form}
\end{equation}
where \(\zeta({\mathbf z})=\zeta_0i^{Q({\mathbf z})}\) is a nowhere-zero stabilizer quadratic
phase in the precise sense of
\cref{subsubsec:stabilizer-error-correction}.
\end{lemma}

\begin{proof}
By the Hilbert-space and error-correction conventions of
\cref{subsubsec:hilbert-conventions,subsubsec:stabilizer-error-correction},
correctability of erasure
at \(r\) is the single-site Knill--Laflamme condition
\begin{equation}
 V^\dagger P_rV=\alpha_P I
 \qquad(P\in\{I,X,-iY,Z\}).
 \label{eq:stabilizer-single-erasure-KL}
\end{equation}
Here the Hermitian Pauli matrix is \(\sigma_y=-iY\), with \(Y=ZX\).
Conjugate \(P_r\) through a Clifford encoder and then evaluate its fixed
ancillas.  The three nonidentity expectations in
\cref{eq:stabilizer-single-erasure-KL} are either all zero or exactly one is
\(\pm1\).  Thus the input-independent marginal at \(r\) is respectively
maximally mixed or a pure one-qubit stabilizer state.  In the pure case its
Schmidt rank is one, which gives \cref{eq:stabilizer-erasure-pure}.
A one-qubit Clifford sends the pure factor to a computational-basis state.
Its fixed Pauli generator can then be removed from the encoder tableau.
The remaining stabilizer generators and logical representatives are those
of \(W\), so \(W\) is a stabilizer isometry as asserted.

In the maximally mixed case a Schmidt decomposition gives
\[
 V\lvert\psi\rangle
 =2^{-1/2}\bigl(\lvert0\rangle_rW_0\lvert\psi\rangle
                 +\lvert1\rangle_rW_1\lvert\psi\rangle\bigr).
\]
Applying \cref{eq:stabilizer-single-erasure-KL} to the four matrix units on
\(r\) gives \(W_c^\dagger W_d=\delta_{cd}I\).  Hence
\(
 W(\lvert c\rangle_a\otimes\lvert\psi\rangle)=W_c\lvert\psi\rangle
\)
is an isometry.  To show that \(W\) is a stabilizer isometry, we eliminate the erased
qubit from the encoder tableau as follows.
First let \(U\) be any stabilizer isometry and let \(P\) be a Hermitian
Pauli with \(U^\dagger PU=0\).  Some Hermitian code stabilizer \(S\),
with \(SU=U\), anticommutes with \(P\).  Otherwise \(P\) would preserve
the code, and its induced unitary on the code would be zero.  Since
\((PS)^2=-I\), the operator
\[
 C_P:=2^{-1/2}(I+PS),\qquad
 C_PU=2^{-1/2}(I+P)U
\]
is a Clifford quarter turn: conjugating a Pauli that commutes with \(PS\)
leaves it fixed, and conjugating one that anticommutes with \(PS\)
multiplies it by \(PS\).  Thus the normalized positive Pauli projection
on the right preserves the stabilizer-isometry property, including the
physical representatives of logical Paulis.

Apply this observation to \(U_0=I_a\otimes V\), with output order
\((a,r,\text{remaining qubits})\), and to the commuting Hermitian Paulis
\(P_Z=Z_aZ_r\) and \(P_X=X_aX_r\).  The maximally mixed case gives
\[
 U_0^\dagger P_ZU_0=U_0^\dagger P_XU_0
 =U_0^\dagger P_ZP_XU_0=0.
\]
For \(U_1=2^{-1/2}(I+P_Z)U_0\), commutation and \(P_Z^2=I\) give
\(U_1^\dagger P_XU_1=U_0^\dagger(P_X+P_ZP_X)U_0=0\).
Consequently both normalized projections are stabilizer isometries.
Writing \(\lvert\beta\rangle=(\lvert00\rangle+\lvert11\rangle)/\sqrt2\),
their exact coefficient identity is
\[
 U_2:=\tfrac12(I+P_X)(I+P_Z)(I_a\otimes V)
      =\lvert\beta\rangle_{ar}\otimes W.
\]
Indeed, the coefficient at \(\lvert cc\rangle_{ar}\) is
\(\sum_d\langle d\rvert_rV\psi_d
 =2^{-1/2}W(\sum_d\lvert d\rangle_a\psi_d)\),
and the coefficients at \(\lvert01\rangle,\lvert10\rangle\) vanish.
A controlled addition from \(a\) to \(r\), followed by a Hadamard on
\(a\), sends \(\lvert\beta\rangle\) to \(\lvert00\rangle\).
The resulting tableau has the fixed generators \(Z_a,Z_r\).
Multiplying its other stabilizer rows by these generators removes their
entries on \(a,r\), and the same operation removes those entries from
the logical representatives.  Deleting the two fixed-zero coordinates
therefore leaves the code and logical-Pauli data of \(W\).
This proves \cref{eq:stabilizer-erasure-mixed}.  These are analytic encoder
operations. The realizable Holant kernel is selected in the shortening
argument that follows.

Suppose now that no single-site erasure is correctable.  For each \(r\),
failure of the displayed Knill--Laflamme condition gives a one-site Pauli
whose pullback through the stabilizer encoder is nonscalar.  By the logical
Pauli convention in \cref{subsubsec:pauli-stabilizer}, this
is a nontrivial logical Pauli with a physical representative supported on
\(r\) alone.  Representatives on
different inputs commute, so their logical Paulis commute.  Independent
physical one-qubit Cliffords send them to \(Z_r\), and one logical Clifford
sends the generated commuting logical subgroup to \(Z\)-type.  Consequently
each \(V\lvert {\mathbf z}\rangle\) is a simultaneous eigenvector of every physical
\(Z_r\), and hence is one computational-basis ket times a phase.  Its
\(r\)-th bit is \(c_r+{\mathbf{m}}_r\cdot {\mathbf z}\). Nontriviality of the logical Pauli gives
\({\mathbf{m}}_r\ne0\).  If \(\rank M<k\), some
\(0\ne {\mathbf d}\in\ker M^{\mathsf T}\) makes \(V\lvert {\mathbf z}\rangle\) and
\(V\lvert {\mathbf z}+{\mathbf d}\rangle\) proportional, contrary to
isometry.  Thus \(\rank M=k\), and the stabilizer amplitude normal form in
\cref{subsubsec:stabilizer-error-correction} makes the remaining
nowhere-zero phase quadratic modulo four, proving
\cref{eq:stabilizer-column-normal-form}.
\end{proof}

The next lemma applies the code structure to a signature with some
inputs left unencoded. Its physical inputs include both those untouched
inputs and the output qubits of the encoder.

\begin{lemma}[Shortening a nonaffine encoded signature by a tetrahedral contraction]
\label{lem:a4rel-physical-code-shortening}
Let \(t\ge1\), let
\(V:(\mathbb C^2)^{\otimes k}\to(\mathbb C^2)^{\otimes N}\)
be a stabilizer isometry with \(k\le4\) and \(N\ge k+2\), and let
\(\xi\in(\mathbb C^2)^{\otimes(t+k)}\setminus\cA\) be the logical
coefficient tensor, including the \(t\) untouched inputs. Put
\(F:=(I^{\otimes t}\otimes V)\xi\).  Then \(F\) factors or one realizable
tetrahedral contraction of two physical qubits is nonzero and nonaffine.
\end{lemma}

\begin{proof}
Apply \cref{lem:stabilizer-erasure-column-normal-form}.  If erasure of a
physical output qubit \(r\) of the encoder is correctable, its pure-marginal case
\cref{eq:stabilizer-erasure-pure} gives a unary tensor factor.  In the
maximally-mixed case \cref{eq:stabilizer-erasure-mixed} isolates an
auxiliary qubit \(a\).  Contract \(r\) with one of the \(t\) untouched
inputs using the realizable kernel \(Q_0\), whose class lies in
\(\TetInt\).  In the analytic local
frame of \(r\), this acts on the untouched-input/\(a\) coefficient array by
an invertible one-qubit Clifford.  The residual encoder is a nonzero
stabilizer isometry, and a Clifford and its inverse preserve \(\cA\).
Consequently the shortened card is nonaffine.

Assume now that no one-site erasure is correctable.  Write the normal
form of \cref{eq:stabilizer-column-normal-form} explicitly as
\[
 \widetilde V=\left(\bigotimes_{j=1}^N C_j^{-1}\right)VL,
 \qquad
 \widetilde\xi=(I^{\otimes t}\otimes L^{-1})\xi,
\]
where \(L\) is the logical Clifford and the \(C_j\)'s map normal-form
output coordinates back to original coordinates.  Then
\(\widetilde\xi\notin\cA\), since Clifford maps and their inverses
preserve \(\cA\), and
\(F=(I^{\otimes t}\otimes(\bigotimes_jC_j)\widetilde V)
\widetilde\xi\).  Let \({\mathbf{c}}\) and \(M=({\mathbf{m}}_1\ \cdots\ {\mathbf{m}}_N)\) be the
data for \(\widetilde V\).  Choose a \(k\)-column basis of
\(M\).  Since \(N\ge k+2\), choose two positions \(r,s\) outside it. Deleting
their columns leaves rank \(k\).  Choose matrix representatives \(C_r,C_s\)
of the two analytic local Clifford frames used in the normal form, with
\([C_r],[C_s]\in\OctDD\).  Normality and
index two imply that
\begin{equation*}
 \{[C_r^{\mathsf T}C_{\rm ker}C_s]:[C_{\rm ker}]\in\TetInt\}
\end{equation*}
is one of the two projective \(\TetInt\)-cosets in \(\OctDD\)
(the \(X\)-twist only relabels that coset).  By
\cref{eq:a4rel-full-support-cosets,eq:a4rel-full-support-cross-ratio}, choose
a realizable \(C_{\rm ker}\) with \([C_{\rm ker}]\in\TetInt\) whose effective contraction matrix
\(C_{\rm eff}=C_r^{\mathsf T}C_{\rm ker}C_s\) has full support and cross ratio
\(-1\).  Thus the analytic frames determine a realizable kernel \(C_{\rm ker}\) in the
original coordinates.

For a logical string \({\mathbf z}\), the two contracted bits are
\[
 \ell_r({\mathbf z})=c_r+{\mathbf{m}}_r\cdot {\mathbf z},\qquad
 \ell_s({\mathbf z})=c_s+{\mathbf{m}}_s\cdot {\mathbf z}.
\]
Each logical string \({\mathbf z}\) determines a unique physical codeword, so
the contraction multiplies its coefficient by
\[
 \omega({\mathbf z})=C_{\rm eff}(\ell_r({\mathbf z}),\ell_s({\mathbf z})).
\]
The full-support property makes \(\omega\) nowhere zero.  By
\cref{eq:a4rel-full-support-cross-ratio}, after multiplication by a common
nonzero scalar, \(\omega=i^{q({\mathbf z})}\) for a modulo-four quadratic \(q\), and
\(\omega^{-1}\) has the same property.  The undeleted columns retain rank
\(k\), so the residual map to computational-basis strings is injective.  Multiplication
by \(\omega\), its inverse, and this injective stabilizer encoding preserve
affine membership in both directions, as does the normal-form phase
\(\zeta({\mathbf z})\).  Since \(\widetilde\xi\notin\cA\), the realizable two-input
\(C_{\rm ker}\)-card is nonzero and nonaffine.
\end{proof}

\begin{proposition}[Gadget tetrahedral localization]
\label{prop:a4rel-required-localization}
Let \(g\) be an even-arity irreducible signature in a fixed-\(I\)
standard marked \(A_4\) state whose complete transfer group remains
\(\TetInt\) after factor closure.  Assume the established terminal
exits do not apply.  If \(g\notin\cA\) and \(\arity(g)\ge10\), some realizable
tetrahedral pair card of \(g\) is nonzero and nonaffine.  Iteration therefore
reaches a realized nonaffine signature on at most eight inputs, unless a factor,
terminal, or restart with a strictly larger complete transfer group occurs, using one \(g\)-vertex and
\(O(\arity(g))\) realizable binaries.
\end{proposition}

\begin{proof}
If every realizable pair card were zero or affine, reconstruction and
\cref{lem:a4rel-four-active-qubits} would place each card-map image in a
stabilizer code with \(k\le4\) logical qubits.  Frobenius reconstruction
would then write \(g=(I^{\otimes2}\otimes V)\xi\), with \(V\) encoding
those \(k\) qubits into \(N=\arity(g)-2\ge8\) sites and
\(\xi\notin\cA\).  Since \(N\ge k+2\),
\cref{lem:a4rel-physical-code-shortening} gives a factor or a nonaffine
realizable card, a contradiction.  Iterating this two-input descent reaches
arity at most eight unless a terminal, factor, or recomputed-group restart
occurs.  The direct-gadget and factor-closure directions are
\cref{eq:direct-gadget-reduction,lem:factor-saturation}.
\end{proof}

\subsubsection{Induction over the retained signatures}
We close the relative tetrahedral argument by a finite induction on
unresolved occurrences and their remaining contractions.  On entering the argument,
consider nonzero odd-arity signatures by \cref{thm:external-odd}, unaries by
\cref{lem:common-unary-exit}, nullaries by preprocessing, and newly exposed
binaries by
\cref{lem:binary-interpolation,lem:actual-binary-completeness}.  Thus, on a
continuing branch, every unresolved nonbinary irreducible occurrence has
even arity and every binary transfer belongs to the recomputed group
\(\TetInt\).

Let \(\mathcal P\) be the finite multiset of unresolved nonbinary
irreducible occurrences.  When a token \(u\), carrying \(g_u\), is
selected, fix the finite current-coordinate universe consisting of all
iterated proper tetrahedral pair cards of \(g_u\), all realizable
dressings and permutations of inputs used by the finite reductions below, and the
complete factorization tree of every output.  This universe is finite: every
proper contraction or factor lowers arity, \(\TetInt\) and the reduction
list are finite, each arity has finitely many input choices and dressings,
and each tensor has a finite irreducible-factorization tree.  Write its remaining indices as \(Q_u\)
and let \(\varphi_u\in\mathbb N\) be the sum of their factor potentials from
\cref{eq:factor-potential}.  Define
\hypertarget{eq:platonic-occurrence-rank}{}
\begin{equation*}
 \Theta_{\rm Pl}(\mathcal P)
 =\{\!\{(\arity(g_u),|Q_u|,\varphi_u):u\in\mathcal P\}\!\},
\end{equation*}
using lexicographic order on triples and its well-founded multiset
extension.

Before applying the next induction step to \(u\), regenerate its complete
oriented current-group deck.  Process each proper card and factor, either
resolving it or entering it as a lower-arity occurrence with smaller
\(\Theta_{\rm Pl}\).  The queue \(Q_u\) then contains only
unconsumed finite reduction or certificate indices, not unresolved proper
signatures.  Before a proper card is returned, consume its unique
originating index in \(Q_u\) and attach that index, marked completed, to the
child's reduction data. It is not copied into the child's unfinished index set.
If the card factors, its finite factorization tree is processed under that same
originating index, and every intermediate split strictly lowers the stored
factor potential.

Processing \(u\) therefore replaces its triple by a finite multiset of
child triples.  Every proper child has smaller first coordinate. If a
residual copy of \(u\) remains, it has the same arity but one fewer fixed
card/factor index or smaller factor potential.  Thus every inserted
triple is strictly smaller than the removed parent triple in lexicographic
order.  Certification removes the triple altogether.  The standard
multiset extension therefore makes every same-coordinate local step strictly decrease
\(\Theta_{\rm Pl}\). A terminal, lower \(\nu\), or strict enlargement of the complete transfer group is instead an outer exit.

\begin{lemma}[The arity-four case for Platonic groups]
\label{lem:platonic-prepared-q4}
Let \(\cF\) be the current retained factor-closed fixed-\(I\)
signature set with marked group of the resulting set
\[
 G\in\{\TetInt,\OctDD,\OctTT,\OctDT,
          \OctTD,\IcoZero,\IcoOne\}.
\]
Let \(q\in\cF\) be a nonzero irreducible quaternary occurrence whose
complete oriented deck and all proper cards and factors have been processed
as above.
For each canonical pair \(p\), put
\(L_p:=\mathsf R_X\circ\Phi_p^q\).  Suppose
\[
 L_p(K_g)=0\quad\text{or}\quad [L_p(K_g)]\in G
 \qquad
 (K_g\in\mathcal K_G^{\rm act})
\]
for all six canonical pairs.  Then \(q\) has exactly the factor, generalized
equality, endpoint-nondegenerate eight-vertex, or outside-binary outcomes of
\cref{lem:platonic-quaternary-boundary}. In particular, no irreducible
quaternary occurrence continues in the internal tetrahedral or either
icosahedral form.
\end{lemma}

\begin{proof}
The proof of \cref{lem:platonic-quaternary-boundary} uses deck stability
only to supply the displayed pair conditions and to exclude an unresolved
proper card or factor.  Here the pair conditions are assumed explicitly,
and the preceding steps have already resolved or entered every such proper object at
a smaller value of \(\Theta_{\rm Pl}\). The finite preserver classification
for complementary pairs and all of its realizable terminal gadgets therefore apply.
\end{proof}

\begin{lemma}[Adding the tetrahedral core signatures]
\label{lem:tetrahedral-stable-HR-augmentation}
Assume \cref{eq:platonic-six-four-predicate} with \(G=\TetInt\), and suppose the full-rank
analysis reaches a signature \(f_\sigma\), \(\sigma\in\Sigma_+\), by reversible realizable tetrahedral dressings and a permutation of inputs.  Then nonzero scalars \(\alpha,\beta\) are recorded such
that the signature set
\[
 \cF_{\rm HR}:=
 \cF\cup\{\alpha\Hcore,\beta\RMcore\}
\]
satisfies \(\KHolant(\cF_{\rm HR})\equivT\KHolant(\cF)\).  Close
the finite core batch and consider its group.  On an unchanged
\(\TetInt\)-branch, regenerate the complete deck of
\(\alpha\Hcore\), retain every card and its factorization tree, close the retained set under irreducible factors, consider all outputs, and form the group and deck for this set.
Every branch is then an outer exit or has a retained residue
\(\cF_{\rm HR}^\star\) with
\[
 \Stable(\cF_{\rm HR}^\star,\TetInt,\alpha\Hcore),
 \qquad
 \{\alpha\Hcore,\beta\RMcore\}\subseteq\cF_{\rm HR}^\star.
\]
Only these deck-stable reduction data are passed to \cref{thm:a4-relative}.
\end{lemma}

\begin{proof}
The reversible identities \cref{eq:tetrahedral-H6-equivalence} produce
\(\alpha\Hcore\), and four copies followed by the forward gadget
\cref{eq:app-v4-H6-R8-output} produce \(\beta\RMcore\).  Direct realization
and set inclusion give the displayed Turing equivalence.  First
preprocess and jointly close the two cores under irreducible factors, consider every
terminal output, and consider the complete transfer group.  A terminal, lower
minimum arity, or strict enlargement of the complete transfer group is an outer exit.
Otherwise the group remains \(\TetInt\) and the minimum nonbinary arity
remains six.

Now regenerate the complete oriented deck of \(\alpha\Hcore\).
Retain every card with its ordered kernel and scalar factor, process
its entire finite factorization tree, consider all resulting binaries and
terminals, and consider the complete transfer group.  The core \(\Hcore\) is
irreducible by \cref{proofsubsec:v4-hcore}. Its pair cards are matching
products because this holds for \(f_\sigma\) by
\cref{lem:tetrahedral-orientation} and the displayed equivalence is a realizable reversible \((\TetInt)^6\rtimes S_6\)-dressing.
\Cref{lem:platonic-matching-contraction-closure} then makes every iterated
proper realizable card zero or a matching product of safe tetrahedral binaries.  Thus
any new singular or outside binary, lower-arity nonbinary factor, or group
change is an outer exit.  On the unchanged branch, regenerate the deck
once more.  It is complete, all of its cards and factors are retained, and
every card is zero or a \(\TetInt\)-matching product of safe binaries. Hence
\cref{def:deck-stable-state} gives the asserted stable tuple.  The two
cores are certified affine signatures and are excluded from every subsequent
unresolved-occurrence set. Adding these signatures does not create a fresh
fixed-\((H,\nu)\) token.
\end{proof}

\begin{proof}[Proof of \cref{thm:a4-relative}]
Let \(\mathcal P\) contain every unresolved nonbinary irreducible
occurrence of \(\cF\). The two displayed cores are already known to be affine and are
excluded from this multiset.  Prepare a selected occurrence and induct
on \(\Theta_{\rm Pl}(\mathcal P)\).

Arity four is \cref{lem:platonic-prepared-q4}.  For the next step, assume
the complete oriented deck and all proper cards and factors have been
processed as above, and consider a six-input
occurrence \(g\).  Its quaternary cards have already been closed, so every
continuing one is a product of two safe tetrahedral binaries.  For a
canonical pair \(p\), compute \(\rho=\rank\Phi_p^g\).  Rank zero gives
\(g=0\), rank one factors, and ranks two or three are closed by
\cref{lem:platonic-deficient-rank}.  At rank four, fixed residual matchings
factor by \cref{lem:platonic-global-ruling}. Otherwise clauses~\textnormal{(i)--(iii)}
of \cref{thm:cert-interface-platonic-incidence}, together with
\cref{lem:platonic-rainbow-localization,lem:tetrahedral-bridge,%
lem:tetrahedral-orientation}, give either the resolved
endpoint-nondegenerate eight-vertex leaf or a signature
\(f_\sigma\) with \(\sigma\in\Sigma_+\).  Its projective class lies in \(\mathscr H_6\)
by \cref{eq:tetrahedral-H6-equivalence}, and \(f_\sigma\) is affine by
\cref{lem:tetrahedral-affine}.  Because \(g\) already has six inputs, this
rainbow localization closes no common-factor inputs: \(f_\sigma\) differs
from \(g\) only by a nonzero scalar, a permutation of inputs, and reversible
realizable \((\TetInt)^6\)-dressings.  Affine membership and the
\(\mathscr H_6\)-orbit conclusion therefore pull back to \(g\).  Thus the projective class of every continuing six-input occurrence
lies in \(\mathfrak D_6(\TetInt)\cup\mathscr H_6\).

Let \(g\) have arity eight.  Prepare every tetrahedral pair card.
For a factored six-input card, process its factorization tree under the
originating \(g\)-index. A continuing result consists only of safe
tetrahedral binaries and hence has projective class in \(\mathfrak D_6(\TetInt)\).  Before an
irreducible six-input card is entered as a child occurrence, consume its
originating \(g\)-index.  Replacing the parent triple by the residual
eight-input triple with one fewer index and the six-input child triple is a
strict multiset decrease of \(\Theta_{\rm Pl}\).  The preceding six-input
base then places the child's projective class in \(\mathscr H_6\).  Therefore all pair cards
lie in \(\mathscr C_6\), and
\cref{prop:a4rel-required-eight-port} says that \(g\) factors or is affine.

Finally let \(\arity(g)\ge10\).  If \(g\notin\cA\),
\cref{prop:a4rel-required-localization} gives a terminal output, a factor, a successor with a strictly larger
complete transfer group, or a nonzero nonaffine realizable card of smaller arity.  The
first three are outer exits or decrease the current token.  In the last
case the parent card index is consumed before its factor-closed child is
entered, so \(\Theta_{\rm Pl}\) strictly decreases.  Induction makes every
child factor affine. Their tensor product would then be affine, contradicting
the nonaffine card.  Hence \(g\) is affine.

After adding signatures, consider the complete transfer group.  A strict
change is an outer exit. Otherwise prepare the next occurrence from its
regenerated current-group deck.  Every continuing step removes or
strictly lowers a triple of \(\Theta_{\rm Pl}\), so every irreducible
occurrence is eventually affine.  The disequality edge and all twelve
tetrahedral binaries are affine in this basis by
\cref{eq:tetrahedral-domain}. \(\Hcore\) is affine by
\cref{proofsubsec:v4-hcore}, and \(\RMcore\) is affine because
\(RM(1,3)\) in \cref{eq:RM-core} is a linear code. Tensor closure,
the normalization data, and
\cref{def:common-K-presentation} give \(\PresK(K\cF;\cA)\), proving the
claimed relative closure.
\end{proof}

\subsubsection{Completion of the internal tetrahedral case}

\begin{proof}[Proof of \cref{thm:tetrahedral-deck-rigidity}]
Varying card matchings give a realizable rainbow localization by
\cref{lem:platonic-rainbow-localization}.  The separated-sign classification and permutation split
then give either the resolved endpoint-nondegenerate eight-vertex leaf or a
signature \(f_\sigma\), \(\sigma\in\Sigma_+\).  Fixed matchings expose a proper
factor by \cref{lem:platonic-global-ruling,lem:factor-saturation}. Closing
its finite factorization tree gives a resolved complexity leaf, lower \(\nu\), or a successor with a strictly larger
complete transfer group, as in the octahedral proof.

In the \(\Sigma_+\) case, apply
\cref{lem:tetrahedral-stable-HR-augmentation}.  Every outer-exit branch is
already closed.  On its sole deck-stable \(\TetInt\) branch,
\cref{thm:a4-relative} gives the common affine \(K\)-presentation, and the
recorded Turing equivalence transports that conclusion back to \(\cF\).
  This proves \(\Close_{\{\cA\}}(\cF,\TetInt,h)\).
\end{proof}

\subsection{The external \texorpdfstring{\(A_4\)}{A4} form}

The external form has the same domain incidence as \(\TetInt\), but
its matching deck has no rainbow line.  Its arity-four case follows from
the conditions for complementary pairs below. At higher arity, connected domain
incidence forces all cards to share one matching and hence a proper factor.

\begin{proposition}[External tetrahedral preservers for complementary pairs]
\label{prop:external-a4-q4-preservers}
For a projective matrix set \(\mathcal S\), define its projective
coefficient set by
\[
\mathcal R_{\mathcal S}
 :=\{[c_Q(C)]:[C]\in\mathcal S\}.
\]
Coordinate matrices act on these projective sets representative-wise.
Let
\(\Theta_{\rm ext}:=c_Q\circ\theta_{\rm ext}\circ c_Q^{-1}\), and use the
Frobenius matrix \(J_{\mathrm{Frob}}=\diag(1,-1,1,-1)\) and adjoint
\(A^\sharp=J_{\mathrm{Frob}}A^{\mathsf T}J_{\mathrm{Frob}}\) from
\cref{eq:platonic-q4-adjoint}. Consider the coordinate maps \(A\) that satisfy the conditions
for both complementary pairs. For every \([{\mathbf r}]\in\mathcal R_{\TetExt}\), we require
\begin{equation}
 \begin{aligned}
  &A{\mathbf{r}}=0\quad\text{or}\quad[A{\mathbf{r}}]\in\mathcal R_{\TetExt},\\
  &A^\sharp {\mathbf{r}}=0\quad\text{or}\quad
    [A^\sharp {\mathbf{r}}]\in\mathcal R_{\TetExt}.
 \end{aligned}
 \label{eq:external-a4-two-sided-definition}
\end{equation}
This is representative-independent.  There is one zero map. Up to nonzero
scalar, the nonzero maps have rank distribution
\((144,0,0,576)\) in ranks \((1,2,3,4)\), respectively.  Under the
Frobenius-dual reconstruction
\cref{eq:platonic-frobenius-reconstruction}, every rank-one or rank-four
map reconstructs a tensor with a binary-pair factorization.
\end{proposition}

\begin{proof}
The adjoint is taken for the bilinear Frobenius form, without complex
conjugation.  Since \(R_{\rm ext}^{\mathsf T}=R_{\rm ext}\), conjugation
by \(R_{\rm ext}\) is
self-adjoint, so \(\Theta_{\rm ext}^\sharp=\Theta_{\rm ext}\).  Moreover,
\[
 [R_{\rm ext}^2]=[Q_0+Q_1]\in N_{\PGL_2}(\TetInt),
 \qquad
 \Theta_{\rm ext}^2(\mathcal R_{\TetInt})
 =\mathcal R_{\TetInt}.
\]
Put \(A'=\Theta_{\rm ext}^{-1}A\Theta_{\rm ext}\).  If
\([{\mathbf{r}}]\in\mathcal R_{\TetInt}\), then
\([\Theta_{\rm ext}{\mathbf{r}}]\in\mathcal R_{\TetExt}\), so the first
condition for \(A\) implies the standard first condition for \(A'\).  Also
\[
 (A')^\sharp=\Theta_{\rm ext}A^\sharp\Theta_{\rm ext}^{-1}.
\]
Because
\(\Theta_{\rm ext}^2(\mathcal R_{\TetInt})
=\mathcal R_{\TetInt}\), one has
\[
 \Theta_{\rm ext}^{-1}(\mathcal R_{\TetInt})
 =\Theta_{\rm ext}(\mathcal R_{\TetInt})
 =\mathcal R_{\TetExt};
\]
the adjoint condition transports in the same way.
Reversing the calculation gives the converse. Thus the external
conditions for both complementary pairs correspond to the standard
tetrahedral conditions in the \(A_4\) row of
\cref{thm:cert-interface-platonic-q4}\textnormal{(i)}.  Rank-one reconstruction gives a
binary-pair factorization.  In rank four, for some
\(U,V\in\GL_2(\mathbb C)\), the standard map is \(C\mapsto UCV\) or
\(C\mapsto UC^{\mathsf T}V\).  Since
\(A=\Theta_{\rm ext}A'\Theta_{\rm ext}^{-1}\), transport gives,
respectively,
\[
 C\mapsto\theta_{\rm ext}(U)C\theta_{\rm ext}(V),
 \qquad
 C\mapsto\theta_{\rm ext}(U)R_{\rm ext}^{-2}C^{\mathsf T}
 R_{\rm ext}^2\theta_{\rm ext}(V).
\]
Either display reconstructs two crossing nonsingular binaries.
\end{proof}

\begin{theorem}[External tetrahedral deck rigidity]
\label{thm:external-a4-deck-rigidity}
Let \((\cF,\TetExt,f)\) be a deck-stable retained,
factor-closed state containing the supplied \(I\) and
\(\arity(f)=\nu(\cF)\).  Then an established terminal, a
restart with a strictly larger complete transfer group, or a proper tensor factor of \(f\) occurs.
Thus this case has no irreducible remaining case or tractable case with a common affine presentation.
\end{theorem}

\begin{proof}
At arity four, use the transfer-valued pair maps
\(L_p=\mathsf R_X\circ\Phi_p^f\).  The exact
complementary-pair adjoint identity
\cref{eq:platonic-q4-map-adjoint,eq:platonic-q4-adjoint} gives the
conditions on the map and its adjoint in
\cref{eq:external-a4-two-sided-definition}.  By
\cref{prop:external-a4-q4-preservers}, rank zero makes \(f=0\), ranks one
and four factor, and ranks two and three do not occur.

For \(\arity(f)\ge6\), fix a pair and let \(\Phi\) be its card map.  The
realizable kernel set \(\mathcal K_{\TetExt}^{\rm act}\) spans
\(\Mat_2\), so rank zero is impossible and rank one
factors.  In ranks two through four, the kernel-deleted rich-line graph is
connected by
\cref{thm:cert-interface-platonic-incidence}\textnormal{(iii)}.  A change of
matching along an edge would give a kernel-avoiding rich domain line with
three distinct images.  Their matchings are pairwise distinct: if two
repeated a matching, flattening across one of its edges absent from the
third would give ranks \(1,1,4\), contradicting collinearity.
Apply \cref{lem:platonic-rainbow-localization} before using the quaternary
incidence classification.  Each common factor has transfer in
\(\TetExt\), and the realizable kernels span \(\Mat_2\), so
realizable closures preserve all three images and produce a quaternary
\(\TetExt\)-rainbow line.  This contradicts
\cref{thm:cert-interface-platonic-incidence}\textnormal{(i)}.
Thus all nonzero cards use one matching \(M\).

Now \(\Psi=\Phi\circ\theta_{\rm ext}\) maps the twelve standard test points
to zero or an \(M\)-matching product.
\Cref{lem:platonic-global-ruling} factors the tensor reconstructed from
\(\Psi\), and composing
its varying factor with \(\theta_{\rm ext}^{-1}\) pulls the tensor identity back to \(\Phi\).  Since \(|M|\ge2\), dual reconstruction and
\cref{lem:factor-saturation} expose a proper factor. Its binaries give the
stated terminal or restart.
\end{proof}

\subsection{The two icosahedral displays}

Recall \(\phi=(1+\sqrt5)/2\), \(\tau=\phi^{-1}=\phi-1\), and the
standard sixty-point model \(\IcoZero\) from
\cref{eq:icosahedral-domain}.  Its axes, tetrahedral sign points, and
additional even-permutation points contribute \(4,8,48\) classes.
\Cref{thm:cert-interface-platonic-bridges}\textnormal{(iii)} proves directly
that these points are closed under
multiplication, inverse, and transpose.

For the opposite display, use the fixed-equality card and network correspondence
\cref{lem:icosahedral-fixed-I-dictionary}.  It replaces every standard
kernel by its realizable \(\kappa_D\)-image while leaving the retained state and
the supplied \(I\) unchanged.

Use the common rainbow normalization and separated-sign system
\cref{eq:platonic-rainbow-normalization,eq:platonic-rainbow-sum,%
eq:platonic-separated-sign-tests} with \(G=\IcoZero\).

\begin{lemma}[Icosahedral separated-sign solutions and uniform separator]
\label{lem:icosahedral-bridge}
\label{lem:icosahedral-separator}
The separated-sign system for normalized maps
\(\Phi:\Mat_2(\mathbb C)\to(\mathbb C^2)^{\otimes4}\) with
\(G=\IcoZero\) has exactly six rank-deficient solutions and thirty full-rank
solutions.  For each full-rank solution \(\Phi\), the realizable kernel
\begin{equation*}
  K_{\IcoZero}=\phi Q_1-\tau Q_2+Q_3,
  \qquad [K_{\IcoZero}]\in\IcoZero
\end{equation*}
realizes the quaternary \(q_{\IcoZero}:=\Phi(K_{\IcoZero})\), whose
three flattening ranks are \((3,3,3)\), and the realizable kernel on
the canonical pair \((1,2)\)
\begin{equation*}
  L_{\IcoZero}=Q_0-\tau^2Q_1+\tau Q_3,
  \qquad [L_{\IcoZero}]\in\IcoZero
\end{equation*}
gives the nonsingular sequential contraction
\(B_{\rm out}:=\Phi_{(1,2)}^{q_{\IcoZero}}(L_{\IcoZero})\), with
\([B_{\rm out}X]\notin\IcoZero\) in every case.
For \(\IcoOne\), the corresponding realizable kernels are
\[
 K_{\IcoOne}:=\kappa_D(K_{\IcoZero}),\qquad
 L_{\IcoOne}:=\kappa_D(L_{\IcoZero}).
\]
They give the covariant quaternary and a nonsingular binary whose transfer
lies outside \(\IcoOne\), directly in the fixed-\(I\) coordinates.
\end{lemma}

\begin{proof}
The line classification, separated-sign solutions, five frames, and
all thirty full-rank cases, including arbitrary scales and cancellations,
are covered by
\cref{thm:cert-interface-platonic-bridges}\textnormal{(iii)}.
The same interface supplies the realizable binary with transfer outside the complete transfer group.  Thus
\cref{eq:direct-gadget-reduction} gives the required direction.
\Cref{lem:binary-interpolation,lem:actual-binary-completeness} resolves
infinite order, while finite order contradicts completeness.
\Cref{lem:icosahedral-fixed-I-dictionary} gives the stated direct gadget certificate for the other display.
\end{proof}

\begin{theorem}[Icosahedral deck rigidity]
\label{thm:icosahedral-deck-rigidity}
If \cref{eq:platonic-six-four-predicate} holds and
\[
 G\in\{\IcoZero,\IcoOne\},
\]
then \(\Close_{\varnothing}(\cF,G,h)\).
\end{theorem}

\begin{proof}
For \(G=\IcoZero\), varying card matchings and connected incidence
supply a mixed line. By
\cref{thm:cert-interface-platonic-incidence}\textnormal{(i)} it is not a
mixed five-line. Its rainbow three-line and
\cref{lem:platonic-rainbow-localization,lem:icosahedral-separator}
therefore give a nonsingular binary with transfer outside the complete
transfer group. For \(G=\IcoOne\), apply the same construction through
\cref{lem:icosahedral-fixed-I-dictionary}, which preserves the binary equality.

If all nonzero cards use one matching \(M\), the twelve test points map
to zero or an \(M\)-matching product. Since \(|M|=2\),
\cref{lem:platonic-global-ruling,lem:factor-saturation} exposes a proper
factor. Close its finite factorization tree: a nonbinary factor lowers arity, a
singular or unsafe binary is terminal, and a safe binary whose
transfer class lies outside the current group strictly enlarges that group.
All-safe current-group binary
factors would contradict the irreducible branch.  Thus every branch is a
resolved complexity leaf or outer deck successor, proving the conclusion.  The separator uses only the rank recomputed on
the realizable six-input coefficients.
\end{proof}

\subsection{Completion of the Platonic classification}

\begin{proof}[Proof of \cref{thm:platonic-decks}]
Work at the stable tuple in the theorem.  First apply the established reductions to a nonzero odd
signature by \cref{thm:external-odd}, a unary by
\cref{lem:common-unary-exit}, a nullary factor by preprocessing, and every
new binary by
\cref{lem:binary-interpolation,lem:actual-binary-completeness}.  Thus a
continuing branch is all-even and every binary transfer lies in the
complete transfer group.

If \(G=\TetExt\), apply
\cref{thm:external-a4-deck-rigidity}.  Its factor branch is closed by
factor closure.  For every other Platonic form, including
\(\IcoOne\), use the direct coefficient interfaces already proved
above.  Hence assume
\[
 G\in\{\TetInt,\OctDD,\OctTT,\OctDT,
          \OctTD,\IcoZero,\IcoOne\}.
\]
At arity four,
\cref{lem:platonic-quaternary-boundary,%
lem:p1-platonic-eight-vertex-rank-terminal,%
cor:higher-equality-unary,lem:common-unary-exit}
and the local-reduction conclusion following
\cref{lem:platonic-finite-q4-terminal} close every output.  Suppose
\(n:=\arity(f)=\nu(\cF)\ge6\).

Fix a canonical pair \(p\) and compute \(r=\rank\Phi_p^f\) from the coefficients.  Rank zero gives \(f=0\) by
\cref{eq:platonic-frobenius-reconstruction}.  At rank one there are nonzero
tensors \(U,F_{\rm res}\) with
\[
 \Phi_p^f(C)=\operatorname{tr}(U^{\mathsf T}C)F_{\rm res},
 \qquad f=U^{(p)}\otimes F_{\rm res},
\]
so \cref{lem:factor-saturation} gives a proper factor route.  Closing that
finite factorization tree either lowers \(\nu\), reaches a binary terminal or
successor with a strictly larger complete transfer group, or expresses \(f\) entirely as safe current-group
binaries, contradicting its irreducible choice.

Let \(r\in\{2,3,4\}\).  Deck stability gives \cref{eq:platonic-all-card-condition}, so
\cref{lem:platonic-high-arity-localization} applies.  The
fixed-matching route again gives a proper factor.  Close its finite factorization tree by the same alternatives. An all-safe current-group binary product
again contradicts irreducibility.  The mixed-matching case
realizes a nonzero
six-input \(h\), using one \(f\)-vertex and exactly \((n-6)/2\) realizable
kernels, with
\[
 2\le\rank\Phi_p^h\le r
\]
and \cref{eq:platonic-all-card-condition}.  Form the signature set
\(\cF_h:=\cF\cup\{h\}\).  Direct realization and set inclusion
give
\[
 \KHolant(\cF_h)\leT\KHolant(\cF),
 \qquad
 \KHolant(\cF)\leT\KHolant(\cF_h).
\]

Preprocess and close this entire finite batch under irreducible factors, apply the established reductions to its newly
exposed outputs, and consider the complete transfer group.  If \(h\) factors, its
selected varying matchings exclude a product entirely of safe current-group
binaries, so a lower-arity nonbinary factor or an earlier terminal occurs.
If \(n>6\), a retained six-input irreducible signature lowers \(\nu\). A strict
enlargement of the complete transfer group is likewise an outer deck successor.  If \(n=6\),
the localization has no closing kernels: the contraction \(C\) in the
proof of \cref{lem:platonic-high-arity-localization} is the identity, and
\(h\) is only the recorded permutation of inputs and nonzero scalar multiple of
\(f\).  We identify it with the same occurrence and reuse its fixed
card/factor index. No fresh fixed-\((H,\nu)\) occurrence is inserted.

On the only remaining irreducible six-input branch with unchanged marked group,
regenerate the complete oriented deck of \(h\).  Retain every card
with its ordered kernel, scalar, and reduction to the preceding set. Process its complete finite factorization tree under factor closure. Consider every resulting binary and
terminal. Recompute the complete transfer group.  A lower-arity factor or group
change is an outer deck successor.  On the unchanged branch regenerate the deck
again.  Condition \cref{eq:platonic-all-card-condition} and
\cref{lem:platonic-matching-contraction-closure} make every card zero or a
matching product of safe binaries with transfers in the current group, and all card factors are retained.
Consequently \cref{def:deck-stable-state} gives the stable tuple
\[
 \Stable(\cF_h^\star,G,h),
 \qquad \arity(h)=\nu(\cF_h^\star)=6.
\]

Only for these deck-stable reduction data compute \(\rho_h:=\rank\Phi_p^h\).  If \(\rho_h\le3\), apply
\cref{lem:platonic-deficient-rank}.  If \(\rho_h=4\), apply, according to
the marked group of the resulting set,
\cref{thm:tetrahedral-deck-rigidity,thm:octahedral-deck-rigidity,%
thm:icosahedral-deck-rigidity}. The only remaining tractable case is the
internal tetrahedral one, whose theorem supplies the stated common affine
presentation for the whole retained set. The other outcomes are resolved
complexity leaves or outer deck successors.

These reductions exhaust the stable six-input residue. Each construction
records its input order and nonzero scalar, and its
reduction points to the preceding retained problem.  The finite separated-sign classifications have constant size. Localization uses \((n-6)/2\) kernels.  These bounds
and the local occurrence induction complete the proof.
\end{proof}
\section{Synthesis of the Binary-Disequality Dichotomy}
\label{sec:p1-synthesis}

The four group analyses give either a complexity conclusion for the
retained set or a construction that makes strict progress. We now combine
them using the descent measure from \cref{lem:matching-synthesis-measure}.
The resulting proof of \cref{thm:matching-deck-classification} classifies
deck-stable states. \Cref{cor:equality-accessible-closure} then uses
stabilization to classify every retained factor-closed set containing \(I\).
Finally, factor closure and the equality-edge-model equivalence give
\cref{thm:p1-eq2-kholant-dichotomy,thm:p1-deq2-dichotomy} for the original
input sets. At each step, hardness follows the reductions, while a
tractable case supplies the criterion for the whole preceding set.

Every continuing signature set \(\cF\) in this section contains \(I\),
so \(Z=K^{\otimes2}I\in K\cF\).  Hence
\cref{lem:tract-X-anchor} gives
\[
  \Tract(K\cF)\quad\Longleftrightarrow\quad
  \ReTract(K\cF).
\]
This equivalence identifies the branchwise criterion \(\Tract\) with the
equality-accessible criterion \(\ReTract\).

\subsection{Synthesis of the Equality-Accessible Matching-Deck Classification}
\label{proofsec:matching-deck-synthesis}

\begin{lemma}[Descent in the four group cases]
\label{lem:token-coverage}
For every deck-stable tuple \((\cF,G,f)\), work in the fixed coordinate matrix \(A=I\), put
\[
 H=\overline G(I,\cF)=G(\cF),
 \qquad
 \mathscr U^{\mathrm{fac}}(I,f,H)=Q_{\rm done}\sqcup Q,
\]
and fix the displayed partition.  Then the unfinished-token batches and
the further reductions invoked by the active-signature completion proofs in
\cref{sec:normalized-dihedral,sec:proper-deck,sec:v4-deck,sec:platonic-decks}
satisfy \cref{def:frozen-token-coverage}.
\end{lemma}

\begin{proof}
Fix the displayed partition for the deck-stable tuple \((\cF,G,f)\) in
the fixed-\(I\) coordinates. Each nonzero oriented one-copy proper card
determines a member of \(\mathscr U(I,f,H)\). All constructions with that
normalized output share a token in the one-copy queue. A continuing
one-copy step with unchanged \((H,\nu)\) therefore works on a single
unfinished token. Each genuine split in its stored tree decreases the
factor potential by \cref{lem:factor-saturation,eq:factor-potential}.
Completing the card and its tree moves the token to \(Q_{\rm done}\).

It remains to check reductions beyond the fixed one-copy deck.
In the standard marked-dihedral branch, \cref{thm:normalized-dihedral}
resolves the whole retained set at arity four through
\cref{thm:p1-quaternary-kholant-dichotomy}. Its higher-arity arguments
exclude the distinguished irreducible signature: the arity-six base is
\cref{lem:nd-six-prime-base}, and localization followed by the
\hyperref[par:nd-common-factor-consequence]{common-factor consequence}
gives a proper factor at larger arity. Thus these local reductions end in
a factorization or a resolved complexity leaf.

For \(\DihY{2m}\), the open-network correspondence in
the proof of \cref{thm:marked-B-reduction} transports arity, factorization, ordered
cards, and their factorization trees to the \(\DihZ{2m}\) deck.
It therefore transports the fixed partition bijectively. For \(\DihXzero{2m}\), the
fixed-\(I\) identities \cref{eq:marked-A0-fixed-I,eq:marked-A0-edge-switch}
and the deck relabelling in the proof of
\cref{thm:marked-A0-reduction} give the same bijective transport to the
\(\DihZ{2m}\) reductions, including their fixed card and factor indices.
For \(\DihXone{2m}\), \cref{lem:marked-A-gauge-transport}
transports the weighted local comparisons used in
\cref{thm:marked-A-rigidity}. Together with the arity-four reduction
\cref{lem:marked-A-q4}, these resolve the complexity, factor the selected
occurrence, or give an outer deck successor. The factorization completes
the selected card and factor indices.

In the cyclic branch, let \(N=|G|\). The \(N=2\) boundary uses
\cref{thm:equality-q4-boundary,lem:finite-q4-dispatch}.
For \(N\ge4\), \cref{lem:proper-cyclic-q4} gives either
a realizable irreducible endpoint-nondegenerate eight-vertex signature,
resolved by \cref{thm:p1-quaternary-kholant-dichotomy}, or a nonsingular
anti-diagonal factor of a two-copy output. Factor closure and
\cref{lem:actual-binary-completeness} make the latter available and resolve
its complexity or strictly enlarge the complete transfer group.
In the Reed--Muller argument, \cref{lem:rm-weight-two-integration} gives
a proper factor or a realizable \(\mathrm{Even}_4\), resolved by
\cref{lem:equality-anchor}. These cyclic multi-copy reductions therefore
give a resolved complexity leaf or an outer deck successor.

In the standard and exotic Klein-four branches, arity-four batches are
processed through \cref{thm:equality-q4-boundary,%
cor:p1-eight-vertex-reduced-interface,lem:finite-q4-dispatch}.
At higher arity, \cref{thm:standard-v4-relative} either resolves the whole
set through the real-valued dichotomy or supplies a realizable quaternary
batch. Its complete factor batch gives a resolved complexity leaf, a strict
group enlargement, or a decrease of \(\nu\).
\Cref{thm:exotic-v4-relative-rigidity} excludes the remaining exotic
higher-arity cases by proper factorization.

For the external tetrahedral form,
\cref{thm:external-a4-deck-rigidity} gives a resolved complexity leaf,
strict group enlargement, or a proper tensor factor. For the other
Platonic forms at arity at least six,
\cref{lem:platonic-high-arity-localization} gives a proper
tensor factor or a directly realized six-input signature, with its kernel
tuple, input order, and scalar. Process this output and all its factors
before continuing. A nonbinary irreducible factor of smaller arity lowers
\(\nu\), and a new transfer class triggers the complete-group comparison.
On the remaining six-input deck-stable state, recompute the card-map rank.
\Cref{lem:platonic-deficient-rank} handles rank at most three.
At rank four, the hypotheses of \cref{eq:platonic-six-four-predicate}
permit \cref{thm:tetrahedral-deck-rigidity,thm:octahedral-deck-rigidity,%
thm:icosahedral-deck-rigidity}, according to the current marked form.
In the internal tetrahedral case, the core retention and regenerated
decks in \cref{lem:tetrahedral-stable-HR-augmentation} justify the
finite occurrence induction of \cref{thm:a4-relative}, giving a common
affine presentation for the entire retained set. This is the assembly in
\cref{thm:platonic-decks}.

The multi-copy constructions above are resolved by their local
contraction, factorization, or finite-induction arguments. After those
arguments are completed, their outputs have outcomes~1--3 of
\cref{def:frozen-token-coverage}.  Any fixed-\((H,\nu)\)
proper-factor continuation arising from a one-copy reduction remains within
its originating unfinished token and satisfies outcome~4: completion
decreases \(|Q|\), while an intermediate split decreases its stored factor
potential.  The fixed-\(I\) reduction data preserve these indices under the coordinate
correspondences above. Thus outcome~4 has the required strict progress,
and all conditions of \cref{def:frozen-token-coverage} hold.
\end{proof}

\begin{proof}[Proof of \cref{thm:matching-deck-classification}]
Run the procedure of
\cref{def:relative-stratum-protocol,lem:matching-synthesis-measure}
in the fixed-\(I\) coordinates, with \(A=I\).
Fix the coefficient field \(L_0\) of \cref{lem:normalization-ledger}.
Gadget evaluation and pivot-normalized factorization keep the pulled-back
data over this field. The same lemma restores the original coordinates
after any continuing normalization in \cref{sec:p1-eight-vertex}.
Consequently \cref{eq:deck-group-bound} gives the uniform bound
\(|G(\cF)|\le M_{L_0}\) at every continuing state.
Keep the finite worklist of irreducible occurrences.
For the selected minimum-arity root \(f\), put \(H=G(\cF)\) and use the canonical card and
factor indices of \(\mathscr U^{\mathrm{fac}}(I,f,H)\) from
\cref{def:relative-stratum-protocol}. While \((H,\nu)\) is unchanged,
carry its fixed partition into completed and unfinished indices.
A token is completed only after its stored card and entire factorization
tree have been processed.  At each outer deck successor, apply
\cref{thm:external-odd,lem:common-unary-exit,lem:binary-interpolation}
to any odd-arity member, unary factor, or unsafe binary.
If only unary and binary irreducible factors remain, the arity-two case
applies. Otherwise choose a minimum-arity nonbinary irreducible signature
and reapply \cref{lem:minimum-core-deck}. An interrupted token
is carried to the successor until its processing is complete.  Thus every
continuing node consists of a deck-stable all-even tuple
\((\cF,H,f)\), together with its existing queue \(Q\), where
\(H=\overline G(I,\cF)=G(\cF)\) is finite by
\cref{lem:bounded-torsion}.  Use the previously defined measures
\(\Xi(I,\cF)\) and \(\Xi_{\mathrm{full}}(I,\cF,f,Q)\) from
\cref{lem:matching-synthesis-measure,eq:deck-closure-measure}.
By \cref{lem:token-coverage}, all four group cases satisfy
\cref{def:frozen-token-coverage}. Thus
\cref{lem:matching-synthesis-measure} permits well-founded induction on
\(\Xi_{\mathrm{full}}\), including every restart.
Strict group enlargement lowers the first coordinate, a lower minimum arity
lowers the second, fixed-\((H,\nu)\) closure decreases \(|Q|\) in the third, and a
genuine factor split lowers \(\Phi_Q\) in the fourth.  A newly exposed safe binary with transfer class outside \(H\)
gives an outer deck successor. If the resulting complete transfer group
remains \(H\), the binary is processed within the current node.
An unsafe binary gives a resolved complexity leaf.  Reductions compose toward the predecessor. A tractability criterion or
common presentation passes back by
\cref{lem:downward-inheritance,lem:normalization-ledger}.
In the group cases below, \(G\) denotes the current complete transfer
group \(H=G(\cF)\).

First suppose \(\arity(f)=4\).  Apply
\cref{thm:marked-finite-group-routing}. The following list is exhaustive.
\begin{enumerate}
\item If the transfer span is proper and \(G=\CycX{2}\), use item~1
      of \cref{thm:equality-q4-boundary}.  If \(G=\CycX{N}\), with even
      \(N\ge4\), use
      \cref{lem:proper-cyclic-q4,thm:p1-quaternary-kholant-dichotomy}.
\item If \(G=\DihZ{2m}\), \(m\ge3\), use item~2 of
      \cref{thm:equality-q4-boundary}.
\item If \(G=\DihY{2m}\), with \(m\ge3\), use
      \cref{thm:marked-B-reduction}.  If
      \(G=\DihXzero{2m}\), with even \(m\ge4\), use
      \cref{thm:marked-A0-reduction}.  If
      \(G=\DihXone{2m}\), with even \(m\ge4\), use
      \cref{lem:marked-A-q4,cor:p1-eight-vertex-reduced-interface}.
\item For the standard Pauli or exotic Klein group, use items~3 and~4,
      respectively, of \cref{thm:equality-q4-boundary}.
\item For \(G=\TetInt\), any of the four octahedral forms
      \(\OctDD,\OctTT,\OctDT,\OctTD\), or either icosahedral display
      \(\IcoZero,\IcoOne\), use
      \cref{lem:platonic-quaternary-boundary}. For
      \(G=\TetExt\), use
      \cref{thm:external-a4-deck-rigidity}.
\end{enumerate}
Each cited construction is applied to the retained set and its
complete oriented deck, with the factor processing specified after
\cref{thm:equality-q4-boundary}. For endpoint-nondegenerate irreducible
quaternary outputs, \cref{lem:finite-q4-dispatch} supplies the binary
case distinction. The sector matrices
\(U_{Y},P,U_{X,1}\) are used through their stated
equivalences or coefficient identities, with continuing outputs expressed
in the fixed-\(I\) coordinates.
A directly realized endpoint-nondegenerate irreducible quaternary is
added to the set and handled by \cref{thm:p1-quaternary-kholant-dichotomy}.
For a possibly reducible eight-vertex output,
\cref{cor:p1-eight-vertex-reduced-interface} instead gives a complexity
conclusion or its full factor batch. The latter is processed by
\cref{lem:actual-binary-completeness} as described after
\cref{thm:equality-q4-boundary}. A directly realized generalized equality
is handled by \cref{lem:pure-ge}.
Thus every arity-four case yields a resolved complexity leaf or a
successor with smaller \(\Xi_{\mathrm{full}}\). In a tractable case,
\cref{lem:tract-X-anchor} gives \(\ReTract(K\cF)\).
The retained irreducible quaternary excludes the arity-two case, leaving
one common \(\cA\)-, \(\cP\)-, or \(\cL\)-presentation, with the class
supplied by the cited result.

Assume henceforth that \(\arity(f)\ge6\).  Apply
\cref{thm:dressed-kernel-observability}.  If the transfer span is
\(\Span\{I,X\}\), then \cref{thm:reed-muller-rigidity} applies to the
current minimum-arity deck-stable signature. Its arity-four reductions
are \cref{thm:equality-q4-boundary,lem:proper-cyclic-q4}.
It gives a resolved complexity leaf, a strict stable successor, or one
common flat-Lagrangian \(K\)-presentation. The last is an affine
tractable case.

It remains to consider the full transfer span.  Apply
\cref{thm:marked-finite-group-routing}.  Its marked list is exhaustive
because
\[
 G\triangleleft\langle G,[Z]\rangle,
 \qquad [\langle G,[Z]\rangle:G]\le2,
 \qquad \langle[X],[Z]\rangle\cong V_4.
\]
The continuing state is all-even and has minimum nonbinary arity at least
six, so it satisfies the arity and parity hypotheses of the following results:
\begin{center}
\normalsize
\begin{tabular}{@{}ll@{}}
\toprule
marked form of the complete transfer group&relative theorem\\
\midrule
\(\DihZ{2m},\DihY{2m}\) (\(m\ge3\)),
\(\DihXzero{2m},\DihXone{2m}\) (even \(m\ge4\))&
\cref{thm:marked-dihedral}\\
standard Pauli \(\KleinStd\) and \(\KleinEx\)&
\cref{thm:v4-relative}\\
all marked Platonic forms&
\cref{thm:platonic-decks}\\
\bottomrule
\end{tabular}
\end{center}
Within the marked-dihedral row, the \(\DihY{2m}\) branch is a
whole-problem equality-edge-model equivalence.  The \(\DihXzero{2m}\) branch uses
the fixed-\(I\) whole-problem/open-network equivalence
\cref{eq:marked-A0-fixed-I,eq:marked-A0-open-equivalence}. Its two edge-replacement constructions use only \(I,Y\), and its final common witness is pulled back
inside \cref{thm:marked-A0-reduction}.  For \(\DihXone{2m}\),
\cref{thm:marked-A-rigidity} resolves arity four by its reduction for the whole signature set and excludes every higher-arity distinguished irreducible signature.
The sector matrix \(U_{X,1}\) expresses the weighted pencils
and fusion identities used in that proof. The realized contractions
remain in the current fixed-\(I\) coordinates.
Each relative theorem gives the alternatives specified in
\cref{def:protocol-discharged}: a resolved complexity leaf or a strict
stable successor. Group enlargement and arity descent lower \(\Xi\).
At fixed group and arity, the card and factor processing verified in
\cref{lem:token-coverage} lowers the remaining coordinates of \(\Xi_{\mathrm{full}}\).
The induction therefore handles every continuing successor.

The two transfer spans and the marked group list exhaust the possible
states. A surviving common product, affine, or local-affine presentation
satisfies \(\ReTract(K\cF)\) by
\cref{def:common-K-presentation,def:tract-X}.
The flat-Lagrangian case is contained in the affine case.
Each common presentation uses one matrix for the transformed edge and
every retained signature. It pulls back through the stated whole-problem
equivalences and \cref{lem:normalization-ledger}.
Other tractable cases already carry the criterion by
\cref{def:protocol-discharged}. The anchor equivalence at the start of this
section identifies it with \(\Tract(K\cF)\).

If \(\Tract(K\cF)\) holds, \cref{thm:predicate-algorithms,eq:K-equivalence}
gives polynomial-time evaluation. If it fails, preservation under taking
subsets and the preceding pullbacks exclude every tractable conclusion.
The induction leaves a hard problem reducing to the original retained
problem. This proves both cases of \cref{thm:matching-deck-classification}.
\end{proof}

\begin{corollary}[Classification of equality-accessible signature sets]
\label{cor:equality-accessible-closure}
Let \(\cF_0\) be a retained, factor-closed finite algebraic
signature set containing the supplied equality \(I\).  If
\(\ReTract(K\cF_0)\) holds, then
\(\KHolant(\cF_0)\in\FP\). Otherwise
\(\KHolant(\cF_0)\) is \(\SharpP\)-hard.  Here
\(\ReTract(K\cF_0)\) is equivalent to \(\Tract(K\cF_0)\), since
\(Z=K^{\otimes2}I\in K\cF_0\).  The factor access used in reaching
this conclusion is a Turing reduction. Every deck is regenerated over the
current signature set and its complete transfer group.
\end{corollary}

\begin{proof}
Because \(Z=K^{\otimes2}I\in K\cF_0\),
\cref{lem:tract-X-anchor} identifies
\(\ReTract(K\cF_0)\) with \(\Tract(K\cF_0)\).  If this
criterion holds, \cref{thm:predicate-algorithms,eq:K-equivalence} gives
\(\KHolant(\cF_0)\in\FP\).

Suppose the criterion fails. Apply \cref{thm:external-odd} in
\(K\)-coordinates to any odd-arity member, and
\cref{lem:binary-interpolation,lem:common-unary-exit} to any unsafe binary.
Their tractable cases are excluded by the failed criterion.
A state whose irreducible factors all have arity at most two would
satisfy the arity-two case. Otherwise a minimum-arity nonbinary
irreducible signature exists and has even arity.

Apply \cref{lem:minimum-core-deck} to that signature.
Its lower-arity outcome is handled by strong induction on \(\nu\).
Its stable outcome satisfies the hypotheses of
\cref{thm:matching-deck-classification}, whose \(\Xi_{\mathrm{full}}\)-induction
handles all subsequent group changes and restarts.
Every continuing extension is Turing-equivalent to its predecessor and
retains the preceding set. A tractable criterion on an extension would
restrict to \(\cF_0\) by \cref{lem:downward-inheritance}, and
\cref{lem:normalization-ledger} pulls back any intervening normalization
and common presentation. This would contradict the assumed failure.
Thus the remaining outcome is hardness, which transfers along the
reductions to \(\KHolant(\cF_0)\).
\end{proof}

\subsection{From the equality-accessible dichotomy to the binary-disequality dichotomy}

\Cref{cor:equality-accessible-closure} gives polynomial-time evaluation
when \(\ReTract\) holds and hardness otherwise.  Its tractable cases
are the arity-two tensor closure and the common affine, product, or
local-affine classes. The flat-Lagrangian case lies in the affine class.
We first apply this criterion to the input set in
\cref{thm:p1-eq2-kholant-dichotomy}, then use the edge-model equivalence
to deduce \cref{thm:p1-deq2-dichotomy}.

\begin{proof}[Proof of \cref{thm:p1-eq2-kholant-dichotomy}]
Apply \cref{subsec:zero-nullary}, retaining the name \(\cF\). The supplied
\(I\) survives. Since \(Z=K^{\otimes2}I\in K\cF\),
\cref{lem:tract-X-anchor} identifies \(\ReTract(K\cF)\) with
\(\Tract(K\cF)\). If this criterion holds,
\cref{thm:predicate-algorithms,eq:K-equivalence} gives
\(\KHolant(\cF)\in\FP\), with the deleted nullary scalars restored.

Suppose the criterion fails. The supplied \(I\) excludes the
common-one-sided exception. By \cref{lem:factor-saturation}, we may add
\(\Pi(\cF)\) by a Turing equivalence. A tractable presentation of the
extension would restrict to the original set by
\cref{lem:tract-X-anchor,lem:downward-inheritance}, so its criterion also
fails. Each original signature is reconstructed from its irreducible factors
by tensor product, an input permutation, and a nonzero scalar, as
described after \cref{lem:factor-saturation}.
All four cases of \(\ReTract\) in \cref{def:tract-X} are preserved by
these operations with the same common basis. Thus a tractable presentation of
\(K\Pi(\cF)\) would extend to \(K(\cF\cup\Pi(\cF))\), a contradiction.
The reconstruction and \cref{eq:factor-saturation} give
\(\KHolant(\Pi(\cF))\equivT\KHolant(\cF)\), so we may replace the
initial set by \(\Pi(\cF)\) with the criterion still false. Retain the name \(\cF\) for this irreducible set.
This finite set is factor-closed, and the irreducible binary equality
\(I\) remains present.

The retained set now satisfies \cref{cor:equality-accessible-closure},
which gives hardness because the criterion fails. The factor-closure
equivalence transfers it to the original input problem. For the original
input set, decidability follows directly from
\cref{thm:predicate-algorithms,lem:tract-X-anchor}, before factor replacement.
\end{proof}

\begin{proof}[Proof of \cref{thm:p1-deq2-dichotomy}]
Apply \cref{subsec:zero-nullary}, retaining the name \(\cF\).
By \cref{lem:p1-equality-wire-equivalence},
\[
 \Holant(\cF,X)\equivT\KHolant(\cF\cup\{I\}).
\]
Apply \cref{thm:p1-eq2-kholant-dichotomy} directly to
\(\cF\cup\{I\}\). Its criterion is
\(\ReTract(K(\cF\cup\{I\}))\), which is equivalent to
\(\ReTract(\cF\cup\{X\})\) by the
\hyperlink{eq:retract-edge-bridge}{criterion correspondence following
\cref{lem:p1-equality-wire-equivalence}}. Both complexity conclusions transfer
through the displayed equivalence, and the criterion is decidable by
\cref{thm:p1-eq2-kholant-dichotomy}. The preprocessing convention gives
the same criterion for the original input set. The edge replacements
are linear-size, and the deleted nullary scalars are restored exactly.
\end{proof}

\section{Conclusion}
\label{sec:p1-conclusion}

The supplied binary disequality signature yields a decidable
\(\FP\)-versus-\(\SharpP\)-hard classification for every fixed finite
set of algebraic complex-valued Boolean signatures.
\Cref{thm:p1-deq2-dichotomy,thm:p1-eq2-kholant-dichotomy} express this
classification using, respectively, ordinary edges with an \(X\)-vertex
and \(X\)-edges with an \(I\)-vertex.  The binary anchor leaves
four tractable cases: the arity-two tensor closure and common
affine, product, or local-affine transformability.

The proof combines arity reduction with the complete transfer group.
A new contraction or factor can lower the minimum nonbinary arity or
enlarge that group, changing the applicable structural analysis.
The progress mechanism in \cref{lem:matching-synthesis-measure} controls
both changes and completes the finite card and factor processing between
them. It thereby assembles the four group analyses into a classification
of the entire signature set.

The eight-vertex reductions provide the common arity-four step.
\Cref{thm:p1-quaternary-kholant-dichotomy} extends the single-signature
classification of Cai and Fu
\cite[arXiv version~2, Theorem~3.1]{CaiFu2023}
to a signature set containing an endpoint-nondegenerate irreducible
quaternary, with the same reduced predicate \(\ReTract\) and without
requiring binary equality. The finite classifications in Appendix~\ref{sec:exact-finite-certificates}
supply the stated low-arity calculations used by the structural branches.

\section*{Code Availability}
The accompanying
\href{https://github.com/liuchliuch/holant-binary-disequality-lean/tree/ec3e598e380c57e1c4d1a3c73a4ad1b4f7eed711}%
{\textcolor{repositorylink}{\GitHubIcon Lean formalization}}
(fixed source revision \texttt{ec3e598}) gives the public statements in
\path{Holant/Statements.lean} and their proofs in \path{Holant/Proofs.lean}.
In the \path{Holant.Theorems} namespace, \path{binaryDisequality} proves the
complexity dichotomy in \cref{thm:p1-deq2-dichotomy},
\path{quaternary} the one in \cref{thm:p1-quaternary-kholant-dichotomy},
and \path{literalEquality} the one in \cref{thm:p1-eq2-kholant-dichotomy}.
For each of these names, \path{Holant/Decidability.lean} exports the
corresponding exact-input recognition theorem with suffix \texttt{\_uniform}.
Supplementary lemmas and exact finite
certificates are included through \path{Holant/Library.lean}.
The statement-only reference \path{Verification/Challenge.lean} and
\path{Verification/comparator.json} support independent comparison using
Lean's Comparator.  Build and verification instructions are given in the
repository README.

The original Python companion repository is
\href{https://github.com/liuchliuch/Holant-X-Dichotomy}%
{\textcolor{repositorylink}{\GitHubIcon\textit{Holant-X-Dichotomy}}}.
References below to \path{certificates/} and its Python scripts refer to
the separately distributed source and instructions in that repository.
The Lean release supplies its own exact certificates under
\path{Holant/Certificates/} and does not require the Python companion.

\clearpage
\appendix
\renewcommand{\thesection}{F.\Alph{section}}
\renewcommand{\theHsection}{fullappendix.\Alph{section}}
\section{Finite Classifications and Explicit Contractions}
\label{sec:exact-finite-certificates}

The finite classifications below supply specific inputs to the group
analyses. The main text verifies their hypotheses, realizes the required
contractions, and applies the resulting reductions to the retained signature set.
\begin{center}
\normalsize
\begin{tabularx}{\textwidth}{@{}l>{\raggedright\arraybackslash}X>{\raggedright\arraybackslash}X@{}}
\toprule
Results & Input to the main proof & Main use\\
\midrule
\cref{thm:cert-interface-affine-q4-normal-form,thm:cert-interface-affine-q4-gadget-exits,thm:cert-interface-affine-q4-odd}
& Affine endpoint normal forms and contractions of quaternary signatures with full odd-parity support and fourth-root phases
& \cref{prop:p1-affine-eight-vertex-certificate,cor:p1-affine-eight-vertex-operational}\\
\cref{thm:cert-interface-equality-q4}
& Equality-accessible arity-four constructions
& \cref{thm:equality-q4-boundary}; also \cref{sec:normalized-dihedral}\\
\cref{thm:cert-interface-nd-q6}
& Six-to-four support projection and contraction
& \cref{lem:app-nd-six-to-four}\\
\cref{thm:cert-interface-exotic-v4-q6}
& Exotic Klein six-input classification
& \cref{lem:exotic-v4-q6}\\
\cref{thm:cert-interface-exotic-v4-q8}
& Normalized eight-input factor argument
& Records the proof of \cref{lem:exotic-v4-q8}\\
\cref{thm:cert-interface-platonic-incidence}
& Rich-line incidence and deletion connectivity
& \cref{lem:platonic-high-arity-localization}; also \cref{sec:platonic-decks}\\
\cref{thm:cert-interface-A4-q8}
& Eight-input tetrahedral contraction
& \cref{prop:a4rel-required-eight-port}\\
\cref{thm:cert-interface-platonic-bridges}
& Separated-sign solutions and realizable bridge outputs
& (i) \cref{lem:octahedral-separator}, (ii) \cref{lem:tetrahedral-bridge},
  (iii) \cref{lem:icosahedral-separator}\\
\cref{thm:cert-interface-platonic-q4}
& Arity-four ranks, reconstructions, and realizable outputs
& (i) \cref{lem:platonic-quaternary-boundary},
  (ii) \cref{lem:platonic-deficient-rank}, (iii) \cref{lem:platonic-finite-q4-terminal}\\
\bottomrule
\end{tabularx}
\end{center}
\Cref{thm:cert-interface-affine-q4-gadget-exits} is also used in
\cref{sec:normalized-dihedral}.

Each calculation states its parameter space, tests, and exact
characteristic-zero field. Where completeness depends on exhaustive
computation, the proof identifies that dependency and the relevant script
in \path{certificates/} of the
\href{https://github.com/liuchliuch/Holant-X-Dichotomy}{Python companion repository}.
These programs emit assertions rather than independent proof objects.
For results proved structurally, the scripts provide supplementary checks.
The Lean repository has its own finite certificates, as described in Code Availability.

In a histogram, \(s^m\) means \(m\) objects of size \(s\).
Projective representatives are normalized by their first nonzero coordinate.
We use the conventions in
\cref{subsubsec:group-projective,subsubsec:boolean-affine-tools}
and Appendix~\ref{app:exact-algebraic-tools}, together with
\hyperref[subsubsec:projective-tensor-geometry]{the projective tensor conventions}
and \hyperref[subsubsec:symmetric-binary-spaces]{the symmetric-space conventions}.

\subsection{Endpoint-Nondegenerate Eight-Vertex Signatures}
\label{appsec:cert-eight-vertex}

We first classify the loop and two-copy contractions of quaternary
signatures with full odd-parity support whose nonzero entries, after
normalization, are fourth roots of unity. We then determine the affine
endpoint normal forms.
\Cref{thm:cert-interface-affine-q4-normal-form,thm:cert-interface-affine-q4-gadget-exits,thm:cert-interface-affine-q4-odd}
state the consequences used in \cref{sec:p1-eight-vertex}.

\subsubsection{Fourth-root phases on full odd-parity support}
\label{proofsubsec:cert-mu4}

In this subsubsection, label the four inputs by \(0,1,2,3\),
with local label \(j\) corresponding to input \(j+1\) in the main text.
Here \({\mathbf{e}}_j\) denotes the string with a single 1 at the locally labelled input.
Normalize a quaternary signature with full odd-parity support by
\(a_j=f({\mathbf{e}}_j)\), \(b_j=f(\one-{\mathbf{e}}_j)\), and \(a_0=1\).  Its fourth-root
phase tuple is
\begin{equation}
  a_j=i^{\alpha_j},\qquad b_j=i^{\beta_j},
  \qquad
  (\alpha_1,\alpha_2,\alpha_3,\beta_0,\ldots,\beta_3)
  \in\mathbb Z_4^7.
  \label{eq:cert-mu4-state}
\end{equation}
There are \(4^7=16\,384\) phase tuples. Write \(N(z)=z\bar z\) for the Gaussian
norm.  For \(i<j\), with \(k<l\) the complementary pair, the six disequality self-loops and six equality-mediated loops are
\begin{equation}
 L^X_{ij}=\diag(a_i+a_j,b_i+b_j),
 \qquad
 L^I_{ij}=
 \begin{pmatrix}0&a_l+b_k\\a_k+b_l&0\end{pmatrix}.
  \label{eq:cert-twelve-loops}
\end{equation}
The latter inserts one equality vertex using two \(X\)-edges.  A
series equality support-flips the former:
\begin{equation}
 \diag(P,Q)XI=\begin{pmatrix}0&P\\Q&0\end{pmatrix},
 \qquad
 T_{\diag(P,Q)XI}=\diag(P,Q).
  \label{eq:cert-diagonal-loop-support-flip}
\end{equation}
In either construction, when \(P,Q\ne0\), the resulting transfer has
eigenvalue ratio \(P/Q\).

\begin{proposition}[Deterministic twelve-loop test]
\label{prop:cert-twelve-loop}
For each phase tuple in \cref{eq:cert-mu4-state}, inspect the twelve ordered pairs
in \cref{eq:cert-twelve-loops}:
\begin{enumerate}[label=\textnormal{(\roman*)}]
  \item If exactly one entry of some pair is zero, declare a rank-one
  exit.
  \item Otherwise, for every pair with two nonzero coordinates \(P,Q\),
  compare their Gaussian norms.  If \(N(P)\neq N(Q)\), declare an
  infinite-projective-order exit.
  \item Otherwise retain the phase tuple and all zero--zero pairs.
\end{enumerate}
The resulting disjoint partition is
\begin{equation}
 16\,384=15\,792+224+368,
 \label{eq:cert-mu4-histogram}
\end{equation}
where the summands are respectively rank-one, infinite-order, and remaining
phase tuples.
\end{proposition}

\begin{proof}
The entries lie in \(\mathbb Z[i]\).  Up to a fourth root of unity, every
nonzero sum of two fourth roots is either \(2\) or \(1+i\).
Consequently two such sums have the same Gaussian norm precisely when their
quotient lies in \(\mu_4\).  If their norms differ, the
quotient has modulus different from one and therefore cannot have finite
multiplicative order.  Thus, for exactly the nonzero sums occurring here,
\(P/Q\) is torsion if and only if \(N(P)=N(Q)\).  Hence
\cref{eq:cert-diagonal-loop-support-flip} makes the three tests exact and
disjoint. Exhaustive evaluation gives \cref{eq:cert-mu4-histogram}.
\end{proof}

Exactly 64 remaining phase tuples satisfy
\begin{equation}
 b_j=\eps a_j\quad(0\leq j\leq3),\qquad
 \eps\in\{1,-1\},\qquad \prod_{j=0}^3a_j\in\{i,-i\}.
  \label{eq:cert-mu4-exceptional64}
\end{equation}
Indeed, after the normalization \(a_0=1\), exactly half of the \(4^3\)
choices of \((a_1,a_2,a_3)\) have product in \(\{i,-i\}\), and \(\eps\)
has two choices. Hence the count is \(32\cdot2=64\).
Join inputs \(0,1\) of two copies \(A,B\) by \(X\)-edges.  Denote the
resulting quaternary signature by \(Q\).  In the order
\((A_2,A_3,B_2,B_3)\), its coefficients are
\[
\begin{array}{ccccccc}
\toprule
{\mathbf{x}}&0000&1111&0011,1100&0101&1010&0110,1001\\ \midrule
Q({\mathbf{x}})&2a_0a_1&2b_0b_1&a_0b_0+a_1b_1&2a_3b_2&2a_2b_3&
a_2b_2+a_3b_3.\\
\bottomrule
\end{array}
\]
The contraction of two odd-parity signatures along two \(X\)-edges has even
external parity, and the two endpoint values \(2a_0a_1\) and \(2b_0b_1\)
are nonzero.  Under \cref{eq:cert-mu4-exceptional64}, exactly one of
\(a_0^2+a_1^2,a_2^2+a_3^2\) vanishes: indeed, each \(a_j^2\) is a sign and
\(\prod_j a_j^2=-1\), so exactly one of the two displayed pairs consists of
opposite signs.  Hence \(Q\) is a support-six eight-vertex signature.

For each other remaining phase tuple and \(\delta\in\mathbb Z_4\), define the
phase-shifted signature \(f_\delta\) by
\[
 f_\delta({\mathbf{e}}_j)=i^{\alpha_j},
 \qquad
 f_\delta(\one-{\mathbf{e}}_j)=i^{\beta_j+\delta}
\]
and keep \(f_\delta({\mathbf{x}})=0\) on every even-weight input.  Parameterize the odd
coset by
\[
 {\mathbf{x}}({\mathbf{y}})=(y_0,y_1,y_2,1\oplus y_0\oplus y_1\oplus y_2),
 \qquad {\mathbf{y}}\in\Ftwo^3.
\]
Define the phase-exponent function
\(\varphi_\delta:\Ftwo^3\to\mathbb Z_4\) by
\(f_\delta({\mathbf{x}}({\mathbf{y}}))/f_\delta({\mathbf{x}}(000))=i^{\varphi_\delta({\mathbf{y}})}\), and define its
linear coefficients \(\ell_j\) and mixed-quadratic coefficients \(q_{jk}\)
by
\[
 \ell_j=\varphi_\delta({\mathbf{e}}_j),\qquad
 q_{jk}=\varphi_\delta({\mathbf{e}}_j+{\mathbf{e}}_k)-\ell_j-\ell_k.
\]
Then \(f_\delta\) is affine if and only if
\begin{equation}
 q_{jk}\in\{0,2\}\ (j<k),\qquad
 \varphi_\delta(111)
 =\ell_0+\ell_1+\ell_2+q_{01}+q_{02}+q_{12}\pmod 4.
  \label{eq:cert-mu4-affine-test}
\end{equation}
Indeed, in the Boolean M\"obius convention of
\cref{subsubsec:boolean-affine-tools}, the expansion of
\(\varphi_\delta\) over \(\mathbb Z_4\) has the three
\(q_{jk}\) as its quadratic coefficients, while the difference between the
two sides of the second congruence is its cubic coefficient.  A modulo-four
phase is affine exactly when the quadratic coefficients are even
and that cubic coefficient vanishes.
Exactly two \(\delta\)'s pass: \(\{0,2\}\) for 176 phase tuples and
\(\{1,3\}\) for 128 phase tuples.

The 512 normalized affine signatures with full odd-parity support are given by
\begin{equation}
 (a_0,a_1,a_2,a_3)=(1,A,B,C),\qquad
 (b_0,b_1,b_2,b_3)=(xyzABC,zBC,yAC,xAB),
 \label{eq:cert-affine-odd-param}
\end{equation}
where \(A,B,C\in\mu_4\) and \(x,y,z\in\{1,-1\}\). The \(4^3\cdot2^3\)
choices give 512 distinct normalized phase tuples.  The locus having no rank-one
\(X\)-loop is the disjoint union
\begin{align}
 &x=y=z &&(128\text{ phase tuples}),\label{eq:cert-affine-odd-rank-a}\\
 &x,y,z\text{ not all equal},\quad
   (A^2,B^2,C^2)=(xy,xz,yz)
 &&(48\text{ phase tuples}).\label{eq:cert-affine-odd-rank-b}
\end{align}
The other 336 phase tuples have a rank-one loop.  For the remaining 176, let
\(G_{0j}\) join inputs \(0,j\) of two copies by corresponding \(X\)-edges and retain the other inputs in copy order.  Call a resulting
quaternary \emph{dense product-transformable} when it has the full even-parity form
in \cref{eq:p1-eight-vertex-product-dense-form}, namely
\[
 g({\mathbf{x}})=\rho r^{\wt({\mathbf{x}})}(-1)^{{\mathbf{s}}\cdot {\mathbf{x}}}
       \one[\wt({\mathbf{x}})\text{ is even}]
 \quad\text{for some }\rho,r\ne0,\ {\mathbf{s}}\in\Ftwo^4.
\]
Then
\begin{align}
G_{01}\text{ is dense product-transformable}
  &\Longleftrightarrow {\mathbf{x}}={\mathbf{y}}\text{ and }B^2=C^2,\notag\\
G_{02}\text{ is dense product-transformable}
  &\Longleftrightarrow {\mathbf{x}}=z\text{ and }A^2=C^2,\notag\\
G_{03}\text{ is dense product-transformable}
  &\Longleftrightarrow {\mathbf{y}}=z\text{ and }A^2=B^2.
\label{eq:cert-affine-odd-cover}
\end{align}
In \cref{eq:cert-affine-odd-rank-a} two squares agree. In
\cref{eq:cert-affine-odd-rank-b} the unique equal pair of \(x,y,z\) has
the corresponding equal squares.  Hence one test in
\cref{eq:cert-affine-odd-cover} always succeeds.  These formulas are a
self-contained mathematical certificate.

\subsubsection{Affine endpoint normal forms}
\label{proofsubsec:cert-affine-q4-endpoint}

The affine classification used in \cref{sec:p1-eight-vertex} reduces to
eighteen coefficient transforms and three endpoint-support types.  For a nonzero affine signature with support \({\mathbf{s}}+L\),
choose the lexicographically first string in the support as origin and the
lexicographically first spanning basis of \(L\) as intrinsic coordinates,
and divide by the value at that origin.  Define the eighteen representative matrices
\begin{equation}
\begin{split}
\mathscr K_{\mathrm{aff}}
={}&\{I,X\}\\
&{}\cup
\left\{
 \begin{pmatrix}1&r\\1&-r\end{pmatrix}:r\in\mu_8
\right\}\\
&{}\cup
\left\{
 \begin{pmatrix}
 1&u(1+\varepsilon\sqrt2)\\
 1&u(1-\varepsilon\sqrt2)
 \end{pmatrix}:
 u\in\mu_4,\ \varepsilon\in\{1,-1\}
\right\}.
\end{split}
\label{eq:cert-affine-q4-cores}
\end{equation}

\begin{proposition}[Exact affine endpoint-nondegenerate eight-vertex certificate]
\label{prop:cert-affine-q4-endpoint}
For every canonically projectively normalized nonzero arity-four affine signature
\(f\in\cA\) and every \(K_0\in\mathscr K_{\mathrm{aff}}\), define the
transformed signature \(q\) by
\[
  q:=K_0^{\otimes4}f.
\]
Retain \((K_0,f)\) precisely when
\begin{equation}
  q({\mathbf{x}})=0\quad\text{for odd }\wt({\mathbf{x}}),
  \qquad
  q(\zero^4)q(\one^4)\ne0.
  \label{eq:cert-affine-q4-retention}
\end{equation}
Then the following assertions hold.

\begin{enumerate}[label=\textnormal{(\roman*)}]
\item Writing each affine support as \({\mathbf{s}}+L\), there are \(36\,720\)
normalized projective affine signatures:
\[
\begin{array}{cccccc}
\toprule
\dim L&0&1&2&3&4\\ \midrule
\#\text{ affine supports}&16&120&140&30&1\\
\#\text{ normalized signatures}&16&480&4480&15360&16384.\\
\bottomrule
\end{array}
\]
Among the \(36\,720\cdot18\) pairs \((K_0,f)\), exactly \(4200\)
satisfy \cref{eq:cert-affine-q4-retention}.  Six representative matrices retain \(612\)
inputs each and the other twelve retain \(44\) inputs each.

\item Every retained \(q\) has the form
\begin{equation}
  q=\gamma\,D(t)^{\otimes4}h,
  \qquad
  \gamma,t\in\AlgNums^\times,
  \label{eq:cert-affine-q4-diagonal-form}
\end{equation}
where \(h\in\cA\), \(h(\zero^4)=1\), and
\[
 \supp(h)\in
 \left\{
  \{\zero^4,\one^4\},\
  \{\zero^4,\one^4,\one_I,\one_{I^c}\},\
  \{{\mathbf{x}}:\wt({\mathbf{x}})\equiv0\pmod2\}
 \right\},
\]
with \(|I|=2\).  Let \(\mathscr R_{\mathrm{aff}}\) be the finite set of all
such normalized affine tensors \(h\).  Their support-size histogram is
\[
  2^4,\qquad4^{96},\qquad8^{512}.
\]
The three support cases form, respectively, \(1,24,256\) diagonal orbits.
Here
\[
 g\sim_{\mathrm{diag}}g'
 \quad\Longleftrightarrow\quad
 g'=\gamma\,D(t)^{\otimes4}g
 \quad\text{for some }\gamma,t\in\AlgNums^\times.
\]

More explicitly, fix the lexicographic order on the six weight-two strings.
If \(v_1,\ldots,v_6\) are the corresponding central coefficients of an
endpoint-nondegenerate eight-vertex signature \(g\), define its
diagonal-orbit invariant vector by
\begin{equation}
 \mathcal J(g):=
 \left(
 \frac{v_rv_s}{g(\zero^4)g(\one^4)}
 \right)_{1\le r\le s\le6}
 \label{eq:cert-affine-q4-invariant}
\end{equation}
Then \(\mathcal J(g)\) is a complete invariant of these diagonal orbits.

\item For a support-four reference, let \(I\sqcup I^c=[4]\) be its two
pairs of inputs and write its logical coefficient matrix as
\[
 F=\begin{pmatrix}A&B\\ C&D\end{pmatrix},
 \qquad ABCD\ne0.
\]
Here the row bit is the common value on the \(I\)-block, the column bit is
the common value on the \(I^c\)-block, and
\[
 A=h(\zero^4),\qquad B=h(\one_{I^c}),\qquad
 C=h(\one_I),\qquad D=h(\one^4).
\]
Then \(AD/BC\in\{1,-1\}\).  If \(AD=BC\), the signature is a tensor
product of two nonsingular weighted equalities.  If \(AD=-BC\), take two
labelled copies, join each of the two inputs in the \(I^c\)-block
of the first copy to the corresponding input of the second copy by a \(X\)-edge, and retain the two \(I\)-blocks as the four external inputs.
The resulting logical coefficient matrix is
\begin{equation}
 FXF^{\mathsf T}
 =\begin{pmatrix}
 2AB&AD+BC\\
 AD+BC&2CD
 \end{pmatrix}
 =\begin{pmatrix}2AB&0\\0&2CD\end{pmatrix},
 \label{eq:cert-affine-q4-support-four-gadget}
\end{equation}
and hence the output signature is a nonzero generalized equality on
its four external inputs.

\item The \(512\) reference signatures with full even-parity support have the disjoint first-witness
partition
\begin{equation}
  512=16+336+96+48+16.
  \label{eq:cert-affine-q4-full-partition}
\end{equation}
The first \(16\) have the dense product-transformable form of
\cref{eq:p1-eight-vertex-product-dense-form}.  The next \(336\) have a
rank-one \(X\)-self-loop.  For this construction, label the four inputs by \(0,1,2,3\)
(local label \(j\) corresponds to input \(j+1\) in the main text).
For each remaining reference signature, take two
labelled copies and, successively for \(j=1,2,3\), join inputs \(0,j\) of
the first copy by crossed \(X\)-edges to inputs \(j,0\) of the
second.  Order the external inputs as the uncontracted inputs of the first
copy in increasing order, followed by the corresponding inputs of the second
copy.  The first support-four product occurs for, respectively,
\(96,48,16\) new reference signatures.  It has the form
\[
  cD(r)\otimes D(r),
  \qquad c\in\mathbb C^\times,
  \qquad r\in\mu_4,
\]
with respect to the induced perfect matching and displayed external-input order.  Here \(c\) is a nonzero scalar and \(D(r)\) is the diagonal binary
matrix.  The histogram in the order \(r=1,-1,i,-i\) is
\[
  (56,56,24,24).
\]

\item Every normalized fourth-root signature \(p\) with full odd-parity support, as in
\cref{eq:cert-mu4-state}, is classified as follows.  The twelve loops in \cref{eq:cert-twelve-loops} give the partition
\[
  16\,384=15\,792+224+368
\]
of \cref{prop:cert-twelve-loop}.  Among the \(368\) remaining phase tuples,
exactly \(64\) satisfy \cref{eq:cert-mu4-exceptional64}. The displayed
two-copy gadget there is a support-six endpoint-nondegenerate eight-vertex signature.

For each of the other \(304\) phase tuples, let \(p_\delta\) multiply the four
weight-three values of \(p\) by \(i^\delta\) while leaving its weight-one
values fixed.  Exactly two \(\delta\in\mathbb Z_4\) make \(p_\delta\)
affine.  The valid sets are
\(\{0,2\}\) for \(176\) phase tuples and \(\{1,3\}\) for \(128\) phase tuples.
Every such \(p_\delta\) lies in the \(512\)-member family
\cref{eq:cert-affine-odd-param}.  Of that family, \(336\) phase tuples have a
rank-one \(X\)-loop. Every one of the remaining \(176\) phase tuples has
at least one two-copy output of the dense product-transformable form among
the three gadgets in
\cref{eq:cert-affine-odd-cover}.
\end{enumerate}
\end{proposition}

\begin{proof}
For a support \({\mathbf{s}}+L\), put \(d=\dim L\).  In the canonical intrinsic
coordinates \({\mathbf{x}}=(x_1,\ldots,x_d)\), the normalized phases are exactly
\[
 i^{\sum_{j=1}^d\ell_jx_j+
      2\sum_{1\le j<k\le d}q_{jk}x_jx_k},
 \qquad
 \ell_j\in\mathbb Z_4,\qquad q_{jk}\in\Ftwo.
\]
There are
\[
 2^{4-d}\genfrac[]{0pt}{}{4}{d}_2\,4^d2^{\binom d2}
\]
such signatures of support dimension \(d\): the Gaussian binomial counts
the linear spaces, where
\(\genfrac[]{0pt}{}{4}{d}_2=\prod_{j=0}^{d-1}(2^4-2^j)/(2^d-2^j)\) and the empty
product is one.  The factor \(2^{4-d}\) counts their cosets, and the last two factors the
linear and mixed phase coefficients.  This proves the first table in
\emph{(i)}.  Applying the eighteen matrices in
\cref{eq:cert-affine-q4-cores} and testing
\cref{eq:cert-affine-q4-retention} gives the \(4200\) retained pairs.

For completeness of the orbit test, normalize the first endpoint to one,
write the central layer as \({\mathbf{v}}\), and write the second endpoint as \(e\).
If \({\mathbf{v}}=0\), two signatures are diagonally equivalent after choosing
\(t^4=e'/e\).  Otherwise equality of
\cref{eq:cert-affine-q4-invariant}, using one nonzero coordinate of \({\mathbf{v}}\),
gives
\[
  {\mathbf{v}}'=\lambda {\mathbf{v}},
  \qquad e'=\lambda^2e.
\]
Choosing \(t^2=\lambda\) proves
\cref{eq:cert-affine-q4-diagonal-form}.  The support and orbit histograms
then follow from the exhaustive retained list.

On a support-four affine plane, choose the logical coordinates \(u,v\) used
in the displayed table.  After projective normalization its phase is
\[
 h(u,v)=i^{\ell_1u+\ell_2v+2quv},
 \qquad \ell_1,\ell_2\in\mathbb Z_4,\quad q\in\Ftwo.
\]
Therefore \(AD/(BC)=(-1)^q\in\{1,-1\}\), proving the asserted cross-ratio.
The gadget construction in \cref{eq:cert-affine-q4-support-four-gadget} is then a
direct coefficient identity.  For full even-parity support, the exact replay tests the six disequality self-loops and the three crossed two-copy contractions in the stated
first-witness order. This gives
\cref{eq:cert-affine-q4-full-partition} and the \(r\)-histogram.

Finally it evaluates all \(4^7\) phase tuples of
\cref{eq:cert-mu4-state}, the twelve ordered pairs in
\cref{eq:cert-twelve-loops}, the exceptional two-copy output, all four
tests in \cref{eq:cert-mu4-affine-test}, and the three gadgets
\(G_{0j}\).  This gives the last assertion.

The verification script \path{certificates/verify_affine_q4.py}
regenerates all the finite objects above from their definitions and
evaluates the identities over \(\mathbb Z[i,\sqrt2]\) and
\(\mathbb Q(i,\sqrt2)\).
\end{proof}

\begin{remark}[Use in the main proof]
\label{rem:cert-affine-q4-endpoint-scope}
\Cref{prop:cert-affine-q4-endpoint} classifies coefficients for the eighteen
displayed representative matrices.  The main proof supplies the transformability witness,
reduces it to those representative matrices, and implements the required constructions in
the retained Holant problem.  In particular, the support-four gadget is
the two-copy construction in
\cref{eq:cert-affine-q4-support-four-gadget}.  The retained-list partition and the partitions for full even-parity support use exhaustive exact computation in
\path{certificates/verify_affine_q4.py}. 
\end{remark}

\begin{theorem}[Affine endpoint normal forms]
\label{thm:cert-interface-affine-q4-normal-form}
The matrix set used by the affine endpoint certificate is exactly the
eighteen-element set \(\mathscr K_{\mathrm{aff}}\) displayed in
\cref{eq:cert-affine-q4-cores}.  Let \(f\in\cA\) be a nonzero canonically
projectively normalized arity-four signature, let
\(K_0\in\mathscr K_{\mathrm{aff}}\), and put \(q=K_0^{\otimes4}f\).  If
\[
 q({\mathbf{x}})=0\quad(\wt({\mathbf{x}})\text{ odd}),\qquad
 q(\zero^4)q(\one^4)\ne0,
\]
then there are \(\gamma,t\in\AlgNums^\times\) and a normalized
\(h\in\cA\), with \(h(\zero^4)=1\), such that
\[
 q=\gamma\,D(t)^{\otimes4}h.
\]
Moreover, \(\supp(h)\) is exactly one of the following:
\begin{enumerate}[label=\textnormal{(\roman*)}]
\item the complementary endpoint pair \(\{\zero^4,\one^4\}\);
\item a four-point plane
      \(\{\zero^4,\one^4,\one_I,\one_{I^c}\}\) with \(|I|=2\);
\item the full even-parity support.
\end{enumerate}
\end{theorem}

\begin{proof}
This is the direct-use form of
\cref{prop:cert-affine-q4-endpoint}\textnormal{(i)--(ii)}.  The three
supports have distinct cardinalities, so the alternatives are disjoint.
\end{proof}

\begin{theorem}[Gadgets from affine reference signatures]
\label{thm:cert-interface-affine-q4-gadget-exits}
Let \(h\) be supplied by
\cref{thm:cert-interface-affine-q4-normal-form}.
\begin{enumerate}[label=\textnormal{(\roman*)}]
\item If \(|\supp(h)|=2\), then \(h\) is a generalized equality.
\item If \(|\supp(h)|=4\), its logical coefficient matrix
      \(F=\begin{psmallmatrix}A&B\\ C&D\end{psmallmatrix}\) satisfies
      \(AD/BC\in\{1,-1\}\).  If \(AD=BC\), then \(h\) factors
      across its two pairs of inputs into nonsingular weighted equalities.
      If \(AD=-BC\), the two-copy \(X\)-edge construction directly
      realizes a nonzero generalized equality.
\item If \(h\) has full even-parity support, then either it has the dense
      product-transformable form of
      \cref{eq:p1-eight-vertex-product-dense-form}, an \(X\)-self-loop
      is nonzero of rank one, or one of the three crossed two-copy
      contractions specified in
      \cref{prop:cert-affine-q4-endpoint}\textnormal{(iv)} directly realizes
      \(cD(r)\otimes D(r)\), where \(c\ne0\) and \(r\in\mu_4\).
\end{enumerate}
\end{theorem}

\begin{proof}
The support-two assertion is immediate, and the other two assertions are
exactly \cref{prop:cert-affine-q4-endpoint}\textnormal{(iii)--(iv)}.
\end{proof}

\begin{theorem}[Contractions of signatures with full odd-parity support and fourth-root phases]
\label{thm:cert-interface-affine-q4-odd}
Let \(p\) be a normalized quaternary signature with full odd-parity support
and fourth-root phases. With \(X\)-edges and binary equality \(I\), the
twelve-loop test of \cref{prop:cert-twelve-loop} selects a nonzero rank-one binary if one
exists. Failing that, it selects a nonsingular binary of infinite projective
transfer order if available. If neither occurs, retain \(p\) and all its
zero--zero loops. Then either two copies directly realize a support-six
endpoint-nondegenerate eight-vertex signature, or a certified phase shift
\(p_\delta\in\cA\) has a rank-one disequality self-loop or a dense
product-transformable two-copy output. These conclusions hold at every nonzero scale.
\end{theorem}

\begin{proof}
Combine the deterministic test in \cref{prop:cert-twelve-loop} with
\cref{prop:cert-affine-q4-endpoint}\textnormal{(v)}.
\end{proof}

\subsection{Equality-Accessible Matching Decks}
\label{appsec:cert-equality-accessible}
\label{proofsubsec:cert-equality-q4-boundary}

This subsection gives the shared finite quaternary certificate statement used
in \cref{sec:weighted-equality} and by the later finite-group branches.

\begin{proposition}[Gadget equality-accessible quaternary certificate]
\label{prop:cert-equality-q4-boundary}
Let \((\cF,G,q)\) be a deck-stable retained state, with \(\cF\)
factor-closed and containing the supplied equality \(I\), and
\(q\ne0\) a quaternary irreducible signature.  Suppose its complete deck consists only of zero cards and nonzero scalar multiples of the fixed
realizable binary representatives of \(G\), and suppose
\(G\) is \(\CycX{2}=\{1,[X]\}\), a
monomial dihedral group
\(\DihZ{2m}=\{[D(t)],[XD(t)]:t\in\mu_m\}\) with \(m\ge3\), the
standard Pauli \(\KleinStd\), or the exotic group
\[
 \KleinEx=\left\{[I],[X],
 \left[\begin{pmatrix}i&1\\-1&-i\end{pmatrix}\right],
 \left[\begin{pmatrix}-i&1\\-1&i\end{pmatrix}\right]\right\}.
\]
Then a gadget using at most two copies
of \(q\) produces an endpoint-nondegenerate eight-vertex signature.  The assertion covers all
zero cards, both input orientations, all six unordered contracted pairs,
and the fixed nonzero scalars of the realizable group representatives.  In every
case, the endpoint-nondegenerate eight-vertex signature can moreover be chosen not to be a
\(G\)-matching signature.
\end{proposition}

\begin{proof}
Input dressings and reversal only permute the complete group, multiply cards
by invertible factors, and contribute known nonzero scalars.  It therefore
suffices to check every standard group kernel on the six unordered pairs.

\emph{The \(\CycX{2}\) case.}
Use Walsh coordinates only for this calculation, and define the Walsh
coefficients
\[
 a_S=\widetilde q(\one_S),\qquad S\subseteq[4].
\]
For any Walsh coefficient tensor indexed by subsets of \([4]\), call it
\emph{Walsh-complement-symmetric} when
\(a_{\bar S}=a_S\) for every \(S\), where
\(\bar S=[4]\setminus S\).
For a deleted pair \(ij\), the two realizable \(I/X\)-cards satisfy
\begin{equation*}
 \widetilde C_{ij}^{\pm}(\one_T)
 =2(a_T\pm a_{Tij}).
\end{equation*}
Here \(Tij:=T\cup\{i,j\}\), with \(T\) a subset of the two residual inputs.
The allowed nonzero Walsh binary cards are scalar multiples of \(I\) or \(Z\).  Singleton residual coordinates force
\(a_S=0\) for odd \(|S|\).  For
\(\{i,j\}\sqcup\{k,l\}=[4]\), define the four Walsh coefficient scalars
\[
 u=a_\varnothing,\quad v=a_{1234},\quad
 x=a_{ij},\quad y=a_{kl}.
\]
Safety for the \(ij\)-cards gives
\[
 (y+v)^2=(u+x)^2,\qquad (y-v)^2=(u-x)^2,
\]
and safety for the complementary \(kl\)-cards gives
\[
 (x+v)^2=(u+y)^2,\qquad (x-v)^2=(u-y)^2.
\]
Adding and subtracting these four equations gives
\begin{equation*}
 v^2=u^2,\qquad y^2=x^2,\qquad
 yv=ux,\qquad xv=uy.
\end{equation*}
If \(u\ne0\), one \(\epsilon\in\{1,-1\}\) satisfies
\[
 v=\epsilon u,\qquad a_{\bar S}=\epsilon a_S.
\]
When \(u=v=0\), call a complementary weight-two block \(S,\bar S\)
\emph{active} if \(a_S\ne0\) (equivalently \(a_{\bar S}\ne0\)), and call
\(a_{\bar S}/a_S\in\{1,-1\}\) its sign.  If all active blocks have one
common sign, the same conclusion holds.  In both cases Fourier inversion
gives one parity in the original basis and complement closure: support only on even-weight
Walsh indices gives \(q({\mathbf{x}})=q(\bar {\mathbf{x}})\), while
\(a_{\bar S}=\epsilon a_S\) makes the inverse transform vanish unless
\((-1)^{\wt({\mathbf{x}})}=\epsilon\).  A realizable \(X\)-dressing then moves a nonzero
complementary pair in the original variables to \(\zero^4,\one^4\).  The result is an
endpoint-nondegenerate eight-vertex signature and remains irreducible,
whereas every quaternary \(\CycX{2}\)-matching signature is a two-binary product.

It remains to treat mixed signs, which necessarily require at least two
active complementary blocks.  Input permutations permute the three block signs, and a realizable \(X\)-dressing reverses all of them.  We may therefore choose a
positive active block and a negative active block, make any third active
block positive, and write the Walsh coefficient scalars as \(x,y,z\), with
\(xz\ne0\), such that one \(X\)-dressing and a permutation give
\begin{equation*}
 a_{12}=a_{34}=x,\qquad
 a_{13}=a_{24}=y,\qquad
 a_{14}=z,\qquad a_{23}=-z,
 \qquad xz\ne0,
\end{equation*}
where \(y\) may vanish.  For \(p=12,13,14\), join the \(p\)-inputs of two
copies by the realizable path \(X-I-X\), and call the output \(r_p\).  Its
realizable kernel is \(I\), its Walsh kernel is \(2I\), and direct contraction
gives
\begin{equation*}
\begin{array}{ccc}
\toprule
 &\text{common value on }0000,1111
 &\text{values on the three complementary weight-two blocks}\ \\ \midrule
 \widetilde r_{12}/4&x^2&(y^2+z^2,0,0)\\
 \widetilde r_{13}/4&y^2&(x^2+z^2,0,0)\\
 \widetilde r_{14}/4&z^2&(x^2+y^2,2xy,0).\\
\bottomrule
\end{array}
\end{equation*}
Every nonzero row is supported only on even-weight Walsh indices and is
Walsh-complement-symmetric.  The same inverse-transform identities just used
therefore make it a complement-symmetric, one-parity signature in the original basis. A realizable \(X\)-dressing makes its two endpoints nonzero.  Hence it becomes an
endpoint-nondegenerate eight-vertex signature.

A \(\CycX{2}\)-matching tensor whose Walsh coefficients are
complement-symmetric has Walsh support on the endpoints and exactly one
weight-two block, with the two common values differing by a sign.  Indeed,
in Walsh coordinates the group is \(C_2^{\rm std}\), and each realizable binary factor
is projectively \(I\) or \(Z\). Their tensor product on a
perfect matching has precisely that support, and Walsh-complement symmetry
forces the two factor signs to agree.

If \(y=0\) and \(x^2+z^2\ne0\), \(r_{13}\) has zero Walsh endpoints and
is not matching.  If \(x^2+z^2=0\), the nonzero part of the
\(13\mid24\) flattening is
\[
 \begin{pmatrix}x&-z\\ z&x\end{pmatrix}.
\]
All other rows and columns vanish, so this matrix is the complete nonzero
block and has rank one, contradicting irreducibility.  The local Walsh
transforms are invertible and therefore preserve flattening rank and irreducibility.

If \(y\ne0\) and all three outputs were matching, the first two rows would
give \(\sigma,\tau\in\{1,-1\}\) with
\begin{equation*}
 y^2+z^2=\sigma x^2,
 \qquad
 x^2+z^2=\tau y^2.
\end{equation*}
Since \(2xy\ne0\), the last row forces \(x^2+y^2=0\).  Then the first
equation gives \(\sigma=1,z^2=2x^2\), while the second gives
\(3x^2=-\tau x^2\), impossible.  Thus a realizable endpoint-nondegenerate eight-vertex signature output is outside the
\(\CycX{2}\)-matching products.

\emph{The monomial dihedral case.}
For one pair, write its four residual slices as \(A,B,C,D_{11}\).  Define the
even and odd residual-card pencils produced by the standard realizable kernels by
\[
 E(t):=A+tD_{11},\qquad O(t):=C+tB,\qquad t\in\mu_m.
\]
The pencil normal form \cref{lem:nd-root-pencil} applies verbatim to
each of these pencils.  It says that the two endpoints are proportional on
one monomial line, or, in one fixed parity plane with axes \(E_0,E_1\),
\[
 U+tV=\lambda(E_0+\alpha tE_1)
 \quad\text{or}\quad
 U+tV=\lambda(tE_0+\alpha E_1),
 \qquad \alpha\in\mu_m.
\]
Its proof uses the exact M\"obius classification
\cref{lem:nd-root-mobius}. In particular, a zero sampled value belongs only
to the proportional branch.

Apply this fact to both pencils for all six pairs.  The resulting support
conditions have exactly the following solutions: choose one total parity
and a nonempty subset of its four complementary string pairs.  Their size
histogram is
\begin{equation}
\begin{array}{crrrr}
\toprule
 |\supp q|&2&4&6&8\\ \midrule
 \#&8&12&8&2.\\
\bottomrule
\end{array}
\label{eq:equality-q4-monomial-supports}
\end{equation}
Hence the support is already one parity and complement-closed.  A realizable
\(X\)-dressing makes the same copy an endpoint-nondegenerate eight-vertex signature.

\emph{The two Klein cases.}
In the Pauli Bell basis, one flattening writes
\[
 q=\sum_{{\boldsymbol{\alpha}},{\boldsymbol{\beta}}\in\Ftwo^2}
 t_{{\boldsymbol{\alpha}}{\boldsymbol{\beta}}}B_{\boldsymbol{\alpha}}\otimes B_{\boldsymbol{\beta}}.
\]
Its row and column cards first make \(T=(t_{{\boldsymbol{\alpha}}{\boldsymbol{\beta}}})\) a partial
monomial matrix.  Intersecting the four allowed Bell-line preimages for
every kernel on the other four pairs leaves exactly the four maximal
four-planes
\begin{equation}
 V_{\pi,\epsilon}
 =\{q:q({\mathbf{x}})=0\text{ if }\wt({\mathbf{x}})\not\equiv\pi\pmod2,\
          q(\bar {\mathbf{x}})=\epsilon q({\mathbf{x}})\},
 \quad \pi\in\Ftwo,\quad\epsilon\in\{1,-1\}.
 \label{eq:equality-q4-pauli-planes}
\end{equation}
Thus a realizable Pauli \(X\)-dressing makes every nonzero member an endpoint-nondegenerate eight-vertex signature.

For \(\KleinEx\), define the representative matrices
\[
 E_0=I,\quad E_1=X,\quad
 E_2=\begin{pmatrix}i&1\\-1&-i\end{pmatrix},\quad
 E_3=\begin{pmatrix}-i&1\\-1&i\end{pmatrix}.
\]
Their Frobenius duals are
\(E_0/2,E_1/2,E_3/4,E_2/4\), so every dual contraction is again a realizable group direction.  The locus satisfying all six pair-card conditions is the union
of twenty-four maximal two-dimensional linear subspaces.  The
support-stratified calculation also produces sixteen one-dimensional
solution subspaces, but every one is contained in one of those twenty-four
planes and hence contributes no additional component to this locus.
Realizable local exotic dressings and permutations of inputs act transitively on
the twenty-four planes.  A representative is
\[
 q_{a,b}
 =a(E_0\otimes E_0+E_1\otimes E_1)
  +b(E_2\otimes E_2+E_3\otimes E_3),
\]
whose coefficient array is
\begin{equation}
q_{a,b}({\mathbf{x}})=
\begin{cases}
a-2b,&{\mathbf{x}}\in\{0000,0110,1001,1111\},\\
a+2b,&{\mathbf{x}}\in\{0011,0101,1010,1100\},\\
0,&\wt({\mathbf{x}})\text{ odd}.
\end{cases}
\label{eq:equality-q4-exotic-plane}
\end{equation}
If \(a-2b\ne0\), this is an endpoint-nondegenerate eight-vertex signature.  If \(a=2b\ne0\), realizable \(X\)-dressings
on the last two inputs make both endpoints \(4b\).  This proves the exotic
case.

In each of the latter three cases, the displayed endpoint-nondegenerate eight-vertex signature is obtained
from the original \(q\) only by invertible group dressings and a permutation of inputs.  It is therefore irreducible.  Since every quaternary
\(G(\cF)\)-matching signature is a tensor product of two nonsingular
group binaries, these endpoint-nondegenerate eight-vertex signatures are not \(G(\cF)\)-matching.  The
strengthened conclusion in the \(\CycX{2}\) case was proved separately above.

The support and subspace completeness claims in
\cref{eq:equality-q4-monomial-supports,eq:equality-q4-pauli-planes,%
eq:equality-q4-exotic-plane} use the exact enumeration in
\path{certificates/verify_equality_q4.py}.  It tests all \(2^{16}-1\)
nonempty computational supports in the monomial case and all \(209\)
partial-injection supports in each Bell basis.  For the latter, it
intersects the allowed preimages for four cross-pairs and four kernels,
retaining zero cards and reducing the equations over \(\mathbb Q\) or
\(\mathbb Q(i)\).  The preceding pencil proof supplies the coefficient
constraint before this finite support classification.
\end{proof}

\begin{theorem}[Classification of equality-accessible quaternary signatures]
\label{thm:cert-interface-equality-q4}
Let \((\cF,G,q)\) be a deck-stable retained, factor-closed state
containing the supplied equality \(I\), and let \(q\ne0\) be an
irreducible quaternary signature.  Suppose every realizable card is zero or a
nonzero scalar multiple of one of the fixed representatives of \(G\), where
\(G\) is \(\CycX{2}\), \(\DihZ{2m}\) with \(m\ge3\), \(\KleinStd\), or \(\KleinEx\).
Then a gadget using at most two copies of \(q\) realizes an
endpoint-nondegenerate eight-vertex signature that is not \(G\)-matching.
More precisely:
\begin{enumerate}[label=\textnormal{(\roman*)}]
\item In the \(\CycX{2}\) row, put
      \(a_S:=\widetilde q(\one_S)\).  A complementary weight-two block
      \(S,\bar S\) is \emph{active} when \(a_S\ne0\), and its sign is
      \(a_{\bar S}/a_S\in\{1,-1\}\).  If the two Walsh endpoints
      \(a_\varnothing,a_{[4]}\) are nonzero or all active blocks have one
      common sign, one dressed copy suffices.  Otherwise at least two active
      blocks have different signs. This is the \emph{mixed-sign} case.  For
      \(p\in\{12,13,14\}\), let \(r_p\) join the \(p\)-inputs of two copies
      by the realizable path \(X-I-X\).  One of these three contractions is
      endpoint-nondegenerate and nonmatching.
\item In the monomial-dihedral row, the support is exactly one total
      parity together with a nonempty subset of its four complementary string
      pairs.  One realizable \(X\)-dressing makes the same copy
      endpoint-nondegenerate.
\item In either Klein row, realizable dressings and permutations of inputs give the
      asserted irreducible endpoint-nondegenerate nonmatching copy.
\end{enumerate}
All clauses include zero cards, both orientations, all six contracted pairs,
and the fixed nonzero scalars of the chosen representatives.
\end{theorem}

\begin{proof}
The overall conclusion is \cref{prop:cert-equality-q4-boundary}.  Its
\(\CycX{2}\), monomial-dihedral, and two Klein proof cases give clauses
\textnormal{(i)}, \textnormal{(ii)}, and \textnormal{(iii)}, respectively.
The containment sentence in the exotic-Klein case makes its
representative-plane argument exhaustive.
\end{proof}

\subsection{Marked Dihedral Matching Decks}
\label{appsec:cert-marked-dihedral}
\label{proofsubsec:cert-normalized-dih-q6}

The six-input support classification below is used by the standard
\(\DihZ{2m}\) branch. The whole-network reductions in
\cref{thm:marked-B-reduction,thm:marked-A0-reduction} carry the
\(\DihY{2m}\) and \(\DihXzero{2m}\) cases to that branch.
For \(\DihXone{2m}\), \cref{lem:marked-A-gauge-transport,lem:marked-A-primitive-transport}
preserve this support argument with weighted kernels. We reduce its
\(2304\) labelled parameter choices to the explicit \(48\)-case table below.

Fix the distinguished pair \(p=\{1,2\}\) and residual inputs
\(R=\{3,4,5,6\}\).  For each sector
\(\epsilon\in\Ftwo\), choose a perfect matching
\[
 M_\epsilon=\{e_\epsilon,e'_\epsilon\}
 \quad\text{of }R,
\]
distinguish the fixed edge
\(e_\epsilon=\{u_\epsilon,v_\epsilon\}\), and order the other edge as
\(e'_\epsilon=(r_\epsilon,s_\epsilon)\) with
\(r_\epsilon<s_\epsilon\).    Choose
\(a_\epsilon,b_\epsilon,o_\epsilon\in\Ftwo\).  Define the eight-string
support \(S\subseteq\Ftwo^6\) by taking, for every
\(\epsilon,z\in\Ftwo\), the two strings satisfying
\begin{equation}
 \begin{aligned}
 (x_1,x_2)&=(z,z+\epsilon),\\
 x_{u_\epsilon}+x_{v_\epsilon}&=a_\epsilon,\\
 (x_{r_\epsilon},x_{s_\epsilon})
   &=(o_\epsilon+z,o_\epsilon+z+b_\epsilon),
 \end{aligned}
 \qquad\text{over }\Ftwo.
 \label{eq:cert-nd-q6-four-flat-support}
\end{equation}
This is exactly the support form obtained from four complementary flat
endpoints: the fixed monomial factor supplies the two strings in a slice,
and the two endpoints of the variable factor are complementary.

\begin{proposition}[Six-to-four support lowering in the standard \(\DihZ{2m}\) case]
\label{prop:cert-nd-q6-support-lowering}
Among the \(3^2\cdot2^8=2304\) labelled choices in \cref{eq:cert-nd-q6-four-flat-support}, exactly \(96\)
give a matching coset: for some perfect matching \(M\) and constants
\(c_e\in\Ftwo\), the support is
\[
  \{{\mathbf{x}}\in\Ftwo^6:x_u+x_v=c_{\{u,v\}}
      \text{ for every }\{u,v\}\in M\}.
\]
For every other choice there are an input pair \(e=\{e_1,e_2\}\) and
\(\gamma\in\Ftwo\) such that projection away from \(e\) is injective on
\[
 S\cap\{{\mathbf{x}}:x_{e_1}+x_{e_2}=\gamma\},
\]
and the projected quaternary support has size six or eight.  Choosing a
largest such projection gives \(1152\) support-six cases and \(1056\)
support-eight cases.
\end{proposition}

\begin{proof}
Permute the four residual inputs so that the fixed edge in sector zero is
\(34\), and its ordered varying edge is \((5,6)\).  Independently flipping
bits on inputs \(3,5,6\) removes \(a_0,o_0,b_0\), respectively: use the
flips \(a_0,o_0,o_0+b_0\).  Thus sector zero is
\[
 S_0=\{(z,z,w,w,z,z):z,w\in\Ftwo\}.
\]
These permutations and bit flips preserve matching cosets, projection
injectivity, and projected support sizes.  For each of the
\(3\cdot2\cdot2^3=48\) choices of sector-zero data, fix one such
normalization.  It bijects the \(48\) choices of sector-one data with the
following canonical choices: a fixed edge \(e\subseteq\{3,4,5,6\}\),
the increasingly ordered complementary edge \(e'=(r,s)\), and
\((a,b,o)\in\Ftwo^3\).  Their four strings satisfy
\[
 (x_1,x_2)=(z,z+1),\qquad
 x_u=w,\quad x_v=w+a\quad(e=\{u,v\},\ u<v),\qquad
 (x_r,x_s)=(o+z,o+z+b).
\]

The table gives a projection witness for every canonical choice.  An
entry \(ij_\gamma^s\) means: retain strings with \(x_i+x_j=\gamma\),
delete inputs \(i,j\), and obtain \(s\) distinct projected strings from
exactly \(s\) retained strings.  Thus the projection is injective.
\(\mathrm M\) denotes a matching coset.
\begin{center}
\normalsize
\setlength{\tabcolsep}{4pt}
\begin{tabular}{c|cccccccc}
\toprule
\(e\backslash(a,b,o)\)&000&001&010&011&100&101&110&111\\
\midrule
\(34\)&\(15_0^8\)&\(25_0^8\)&\(\mathrm M\)&\(\mathrm M\)&\(15_0^8\)&\(25_0^8\)&\(15_0^8\)&\(16_0^8\)\\
\(35\)&\(16_0^8\)&\(26_0^8\)&\(24_1^6\)&\(14_1^6\)&\(14_0^6\)&\(24_0^6\)&\(26_0^8\)&\(16_0^8\)\\
\(36\)&\(15_0^8\)&\(25_0^8\)&\(24_1^6\)&\(14_1^6\)&\(14_0^6\)&\(24_0^6\)&\(25_0^8\)&\(15_0^8\)\\
\(45\)&\(16_0^8\)&\(26_0^8\)&\(23_1^6\)&\(13_1^6\)&\(13_0^6\)&\(23_0^6\)&\(26_0^8\)&\(16_0^8\)\\
\(46\)&\(15_0^8\)&\(25_0^8\)&\(23_1^6\)&\(13_1^6\)&\(13_0^6\)&\(23_0^6\)&\(25_0^8\)&\(15_0^8\)\\
\(56\)&\(13_0^6\)&\(13_1^6\)&\(15_0^6\)&\(15_0^6\)&\(13_0^6\)&\(13_1^6\)&\(13_0^6\)&\(13_1^6\)\\
\bottomrule
\end{tabular}
\end{center}
Each entry follows by substituting \((z,w)\in\Ftwo^2\) in the two
displayed sectors.  For example, at \(e=35\), \((a,b,o)=010\), the
\(24_1^6\) projection leaves
\[
 \{0001,0100,0111,1000,1011,1110\}
\]
on inputs \((1,3,5,6)\), with one preimage per string.

For maximality of the entries marked \(6\), an eight-string projection
would require a pair with constant XOR on both sectors.  The pairs with
constant XOR on \(S_0\) are
\(12,15,16,25,26,34,56\).  Substitution of the sector-one equations in
each \(6\)-entry eliminates these pairs, or leaves a projection that
identifies two strings.  A seven-string parity slice is impossible: on each
four-string affine sector its size is \(0,2\), or \(4\).  Thus the displayed
six-string images are maximal.  At the two \(\mathrm M\)-entries the
matching equations are, respectively,
\[
 x_3=x_4,\ x_1=x_5,\ x_2=x_6;
 \qquad
 x_3=x_4,\ x_1=x_6,\ x_2=x_5.
\]
A parity contraction of a three-edge matching coset has at most four
projected strings, so none of the other entries is a matching coset.
The canonical table therefore has \(2,24,22\) cases of the three stated
types.  Multiplying by the \(48\) sector-zero choices gives
\[
 48(2,24,22)=(96,1152,1056),
\]
as required.
\end{proof}

\begin{remark}[From support projection to contraction]
\label{rem:cert-nd-q6-support-scope}
In the standard \(\DihZ{2m}\) setting, the supplied equality \(I\)
gives the contraction kernel \(XIX=I\), and a direct edge gives the kernel
\(X\). They implement parity \(\gamma=0\) and \(\gamma=1\), respectively.
Injectivity leaves one nonzero summand for each projected string, so the
conclusion holds for arbitrary nonzero coefficients on the support.
The main proof supplies the sector-ruling hypothesis and exposes any
resulting factors by \cref{lem:factor-saturation}. The realizable
contractions in the following theorem are taken in this setting.
\end{remark}

\begin{theorem}[Six-to-four contractions in the standard \(\DihZ{2m}\) case]
\label{thm:cert-interface-nd-q6}
Let \(h\) be an arity-six signature whose support is a set \(S\) of the form
\cref{eq:cert-nd-q6-four-flat-support}, and suppose every coefficient of
\(h\) on \(S\) is nonzero.  Then either \(S\) is a three-edge matching
coset, or there are an input pair \(e=\{e_1,e_2\}\) and
\(\gamma\in\Ftwo\) for which projection away from \(e\) is injective on
\[
 S\cap\{{\mathbf{x}}:x_{e_1}+x_{e_2}=\gamma\}
\]
and has image of size six or eight.  In the latter case, contraction by the
realizable kernel \(I\) for \(\gamma=0\), or \(X\) for \(\gamma=1\), directly
realizes a cancellation-free quaternary of support size six or eight.  That
      output is not a product of two nonsingular monomial binaries.
\end{theorem}

\begin{proof}
\Cref{prop:cert-nd-q6-support-lowering} gives the exhaustive support
alternative and injectivity.  The selected parity kernel leaves exactly one
nonzero summand above each projected string, proving noncancellation.  A product
of two nonsingular monomial binaries has support size four.
\end{proof}

\subsection{Klein-Four Matching Decks}
\label{appsec:cert-klein-four}
\label{proofsubsec:extended-v4-certificates}

The following certificate supplies the exotic Klein finite cases used
in \cref{sec:v4-deck}.

\subsubsection{The exotic Klein form
  \texorpdfstring{\(\KleinEx\)}{V4 (ex)}}
\label{proofsubsec:cert-exotic-v4}

\begin{proposition}[Exotic-Klein six- and eight-input certificate]
\label{prop:cert-exotic-v4-q6-q8}
Use the four exotic binary directions \(E_a\), their Frobenius duals, and
the label set \(V=\Ftwo^2\) from
\cref{eq:exotic-v4-basis,eq:exotic-v4-Gram}.  The following finite
statements hold.
\begin{enumerate}[label=\textnormal{(\roman*)}]
\item Let \(h\ne0\) have arity six, suppose every exotic pair card is zero
      or a product of two exotic binaries, and normalize one nonzero card by
      \[
        C_{12}^0=E_0^{(3,4)}E_0^{(5,6)}.
      \]
      The other three card directions range over the complete set of normalized
      card-direction assignments
      \[
        \bigl(\{0\}\cup\mathfrak D_4(\KleinEx)\bigr)^3,
        \qquad (1+48)^3=117{,}649.
      \]
      Over arbitrary algebraic nonzero-card scales, including every
      cancellation locus in every other card context, exactly seventeen
      projective scale lines survive.  Every one reconstructs a product of
      three exotic binaries. No irreducible arity-six line survives.
\item Let \(h\) have arity eight and suppose it has the normal form \cref{eq:exotic-v4-q8-linear-normal-form}, with one matching and
all four cards nonzero.  If every one of its
      \(28\cdot4=112\) exotic pair cards is zero or a product of three
      exotic binaries, then \(h\) has a proper tensor factor.
      \end{enumerate}
\end{proposition}

\begin{proof}
All matching lines and card maps below are generated directly from the four
matrices in \cref{eq:exotic-v4-basis}.  The calculation first verifies the
Gram matrix, the dual basis
\[
 (E_0/2,E_1/2,E_3/4,E_2/4),
\]
and projective fusion closure, so its contractions use the same normalized
directions as the main argument.

For part~\textnormal{(i)}, there are \(3\cdot4^2=48\) projective
quaternary matching lines.  Encode zero by \(-1\), and encode the line with
matching \(m\in\{0,1,2\}\) and labels \(a,b\in V\) by \(16m+4a+b\).
Thus a nonnegative code denotes the canonically normalized line
\(P_m(E_a,E_b)\).  On each base residual edge \(34\) and \(56\), the
stabilizer under realizable operations contains the paired endpoint dressings with label pairs
\[
 (1,1),\qquad(2,3),\qquad(3,2),
\]
which fix the base \(E_0\)-edge projectively.  Together with the permutations of inputs
\[
 (3\,4),\qquad(5\,6),\qquad(3\,5)(4\,6),
\]
these generate the stabilizer under realizable operations used here.  Its
induced action on the 48 residual matching lines has order \(64\).  It
partitions
the full \(49^3\) card-direction assignments into 4,045 orbits, and the generated
orbit-size histogram is
\[
 1^{27},\quad2^{49},\quad4^{351},\quad8^{521},\quad
 16^{577},\quad32^{1830},\quad64^{690}.
\]
The orbit sizes weighted by these multiplicities sum to \(117\,649\), so
the generated orbits cover every card-direction assignment.  On one representative of every
orbit, exact \(\mathbb Q(i)\)-row reduction computes the union of the
preimages of the forty-eight matching lines within each nonbase pair/kernel
context, then intersects these unions over all 56 contexts.  A zero
output lies in every matching-line preimage, and nonzero-card scales remain
independent linear variables. In particular, no cancellation locus is divided away.

The only surviving direction triples are
\begin{equation*}
 (-1,-1,-1),\quad (1,2,3),\quad(1,3,2),\quad
 (4,8,12),\quad(4,12,8).
\end{equation*}
The first has the single scale line \((1)\).  Each of the other four has
exactly the four scale lines
\begin{equation*}
 (1,\epsilon,\delta/2,\epsilon\delta/2),
 \qquad \epsilon,\delta\in\{1,-1\}.
\end{equation*}
Thus the nonzero-card histogram consists of one line with a single nonzero card
and sixteen lines with every card nonzero.  An independent enumeration of all
\(15\cdot4^3=960\) three-binary product lines reconstructs exactly these
seventeen tensors, proving part~\textnormal{(i)}.

For part~\textnormal{(ii)}, the prescribed form is
\cref{eq:exotic-v4-q8-linear-normal-form}, with arbitrary
\(\lambda_{\boldsymbol\gamma}\ne0\) and \(A_j\in\GL_2(\Ftwo)\).
Apply the two-contraction argument in the proof of \cref{lem:exotic-v4-q8},
on \((3,5)\) and \((4,5)\) with \(E^0=I/2\). It uses exactly these
conditions to contradict the matching-product overlap formula
\cref{eq:exotic-v4-overlap-size}, giving the asserted factor conclusion.

Every scale condition in part~\textnormal{(i)} is a finite union of homogeneous
linear systems over \(\mathbb Q(i)\).  Prescribed liveness is also exact:
over an infinite field, a linear space contains a vector avoiding all
coordinate hyperplanes exactly when no coordinate vanishes identically on
that space.  Base change to \(\overline{\mathbb Q}\), or to
\(\mathbb C\), therefore creates no additional component, and zero target
cards are kept as zero-card subspaces.  The exact calculation for part~\textnormal{(i)} regenerates the states,
orbits, scale intersections, and factor tensors.
The Python verification script is
\path{certificates/verify_exotic_v4_q6_q8.py}.
\end{proof}

\begin{theorem}[Six-input tensors for the exotic Klein group]
\label{thm:cert-interface-exotic-v4-q6}
Let \(h\ne0\) have arity six.  If every exotic pair card of \(h\) is zero
or a product of two exotic binaries, then \(h\) is a product of three
exotic binaries.  Consequently no irreducible arity-six remaining case exists.
This includes arbitrary algebraic scales and every cancellation locus.
\end{theorem}

\begin{proof}
This is \cref{prop:cert-exotic-v4-q6-q8}\textnormal{(i)} after undoing the
normalization of one nonzero card.
\end{proof}

\begin{theorem}[Eight-input tensors for the exotic Klein group]
\label{thm:cert-interface-exotic-v4-q8}
Let \(h\) have arity eight and satisfy the displayed normal form \cref{eq:exotic-v4-q8-linear-normal-form}, with one matching and
all four cards nonzero.  If all \(112\) exotic
pair cards of \(h\) are zero or products of three exotic binaries, then
\(h\) has a proper tensor factor.  
\end{theorem}

\begin{proof}
This is \cref{prop:cert-exotic-v4-q6-q8}\textnormal{(ii)}.
\end{proof}

\subsection{Platonic Matching Decks}
\label{appsec:cert-platonic}
\label{proofsec:platonic-matching-certificates}

The Platonic analysis first solves the separated-sign equations for
localized card maps. The incidence classification then determines how
these local configurations fit together. The final calculations handle
the tetrahedral arity-eight case and the arity-four contractions used by
the main proof. All configurations use the quaternion basis in
\cref{eq:quaternion-basis}.

Whenever a mixed three-point incidence is localized, realizable dressings put
its rainbow image in the common normalization
\begin{equation}
 A=P_0(Q_0,Q_0),\qquad
 B=P_1(Q_0,Q_0),\qquad
 C=P_2(Q_2,Q_2)=A-B.
 \label{eq:cert-A4-normal-incidence}
\end{equation}
Recall that matching-label type \((1,1,1)\) is rainbow, and \((s)\) is monochromatic. Set
\begin{equation}
\begin{gathered}
 R_0=\tfrac12P_2(Q_0,Q_0),\qquad
 R_1=-\tfrac12P_2(Q_1,Q_1),\qquad
 R_3=-\tfrac12P_2(Q_3,Q_3),\\
 R_+=Q_0+Q_1+Q_2+Q_3,\qquad
 R_-=Q_0-Q_1-Q_2-Q_3.
\end{gathered}
\label{eq:cert-platonic-common-vectors}
\end{equation}
Here \(R_+,R_-\) are the matrices already used in
\cref{eq:tetrahedral-H6-equivalence}.

\begin{lemma}[Matching-product solutions of the separated-sign equations]
\label{lem:platonic-bridge-formula}
Let \(S\) be the swap matrix, defined by \(S({\mathbf{u}}\otimes {\mathbf{v}})={\mathbf{v}}\otimes {\mathbf{u}}\).
For \(U,V\in\GL_2(\mathbb C)\) and \(\lambda\ne0\),
\begin{equation}
 \rank\bigl((U\otimes V^{\mathsf T})S-\lambda I_4\bigr)=1
 \quad\Longleftrightarrow\quad V=\lambda U^{-\mathsf T}.
 \label{eq:platonic-swap-rank-one}
\end{equation}
For \(G\in\{\TetInt,\OctDD,\IcoZero\}\), put
\(\mathscr C_G=\{0\}\cup\operatorname{cone}(\mathfrak D_4(G))\).
Then
\begin{equation}
 \{W:W,A-2W,2W-B\in\mathscr C_G\}
 =\{0,A/2,B/2\}\cup
 \left\{\tfrac12P_2(J,J^{-\mathsf T}):[J]\in G,
                                      \ J^{\mathsf T}=J\right\}.
 \label{eq:platonic-symmetric-bridge}
\end{equation}
Write a symmetric group matrix as \(J_{\mathbf{u}}=aQ_0+bQ_1+dQ_3\), choosing a
nonzero real coefficient vector \({\mathbf{u}}=(a,b,d)\), and put
\(R_{\mathbf{u}}=\tfrac12P_2(J_{\mathbf{u}},J_{\mathbf{u}}^{-\mathsf T})\).
A triple of solutions to the separated-sign equations sums to \((A+B)/2\) precisely when it is a
permutation of \((0,A/2,B/2)\), or is \((R_{\mathbf{u}},R_{\mathbf{v}},R_{\mathbf{w}})\) with
\({\mathbf{u}},{\mathbf{v}},{\mathbf{w}}\) pairwise orthogonal in \(\mathbb R^3\).
\end{lemma}

\begin{proof}
Conjugation by \(U^{-1}\otimes I\) reduces the rank condition to
\((I\otimes W^{\mathsf T})S-\lambda I_4\), with
\(W=U^{\mathsf T}V=\begin{psmallmatrix}a&b\\c&d\end{psmallmatrix}\).
Its matrix is
\[
 \begin{pmatrix}
 a-\lambda&0&c&0\\
 b&-\lambda&d&0\\
 0&a&-\lambda&c\\
 0&b&0&d-\lambda
 \end{pmatrix}.
\]
The second row has nonzero second entry.  Rank one forces the first row
to vanish, hence \(a=\lambda,c=0\). Comparing the second and third rows
then gives \(b=0,d=\lambda\).  Conversely, \(S-I_4\) has rank one.
This proves \cref{eq:platonic-swap-rank-one}.

The three exceptional separated-sign solutions \(0,A/2,B/2\) satisfy the conditions,
since \(C=A-B=P_2(Y,Y)\) and \([Y]\in G\).  For any other solution,
all three tested vectors are nonzero.  A line with two points of one
matching and a point of another would give flattening ranks \(1,1,4\),
which is impossible.  If \(W=P_0(U,V)\), the condition on \(2W-B\)
therefore gives matching \(2\).  Its rank-one flattening and
\cref{eq:platonic-swap-rank-one} imply
\(W=\tfrac12P_0(U,U^{-\mathsf T})\).
The condition on \(A-2W\) has matching \(0\), so
\(I\otimes I-U\otimes U^{-\mathsf T}\) is simple.  The simple-tensor
sum criterion forces \(U\) to be scalar, giving \(W=A/2\), a
contradiction.  The matching-\(1\) case similarly gives \(W=B/2\).

Thus \(W=P_2(U,V)\), and \(A-2W\) has matching \(1\).
Its \(13\mid24\) flattening is
\(I_4-2(U\otimes V^{\mathsf T})S\). Hence
\(W=\tfrac12P_2(U,U^{-\mathsf T})\).
Apply the invertible local matrices \((U^{-1},I,U^{\mathsf T},I)\),
which fix \(B\).  They send \(2W\) to
\(P_2(I,U^{-\mathsf T}U)\).  Since \(2W-B\) has matching \(0\),
\cref{eq:platonic-swap-rank-one} with \(\lambda=1\) now gives
\(U^{-\mathsf T}U=I\), or \(U^{\mathsf T}=U\).
Conversely, for symmetric \(U\), the two differences are local images
of \(P_1(Y,Y)\) and \(-P_0(Y,Y)\).  All leg matrices belong projectively
to \(G\), so both differences are in its matching cone.  This proves
\cref{eq:platonic-symmetric-bridge}, including arbitrary scales.

For the sum equation, polarization of
\(\tfrac12P_2(J_{\mathbf{u}},\operatorname{adj}(J_{\mathbf{u}}))\), where
\(\operatorname{adj}(J_{\mathbf{u}})=aQ_0-bQ_1-dQ_3\), defines an injective linear map
\(T\) from the symmetric \(3\times3\) matrices to quaternary tensors with
\[
 T({\mathbf{u}}{\mathbf{u}}^{\mathsf T})=\tfrac12P_2(J_{\mathbf{u}},\operatorname{adj}(J_{\mathbf{u}})),\qquad
 T(I_3)=(A+B)/2,\qquad
 R_{\mathbf{u}}=T\!\left(\frac{{\mathbf{u}}{\mathbf{u}}^{\mathsf T}}{{\mathbf{u}}^{\mathsf T}{\mathbf{u}}}\right).
\]
The six coefficient tensors are independent and \(C\notin\operatorname{im}T\):
in the \(14\mid23\) binary basis they use only
\(Q_\mu\otimes Q_\nu\) with \(\mu,\nu\in\{0,1,3\}\), whereas
\(C=Q_2\otimes Q_2\).  Also
\[
 A/2=\tfrac12T(I_3)+C/4,\qquad
 B/2=\tfrac12T(I_3)-C/4.
\]
Comparison of the \(C\)-coefficient forces equal numbers of \(A/2\)
and \(B/2\). If they occur, the third summand is zero.  Otherwise the
trace of \(T^{-1}R_{\mathbf{u}}\) is one, so all three summands are nonzero
\(R\)-vectors.  Their sum is \(T(I_3)\) exactly when their normalized
real coefficient columns form a matrix \(E\) with
\(EE^{\mathsf T}=I_3\).  This is equivalent to
\(E^{\mathsf T}E=I_3\), proving the orthogonality criterion.
The three orthogonal rank-one projectors are independent. Together with
\(C\notin\operatorname{im}T\), they give card-map rank four.
The exceptional triple spans \(A,B\) and gives rank two.
\end{proof}

\subsubsection{The four fixed-equality octahedral forms
  \texorpdfstring{\(\OctDD,\OctTT,\OctDT,\OctTD\)}%
  {O (dd), O (tt), O (dt), and O (td)}}
\label{proofsubsec:cert-S4}

For the octahedral certificate, use the standard \(24\)-point set
\(\OctDD\) in
\cref{eq:quaternion-basis,eq:octahedral-domain}, ordered as the four axes, the
eight sign points, and the twelve pair points \(Q_\mu\pm Q_\nu\), with
pairs lexicographic. The second
unmixed distinguished-involution normal form is
\begin{equation}
 \OctTT=H_{\mathrm{tt}}^{-1}\OctDD H_{\mathrm{tt}}.
 \label{eq:cert-S4-second-form}
\end{equation}
Use the coefficient-similarity map \(\theta_{\mathrm{tt}}\), transformation
tuple \(\mathbf H_{\mathrm{tt}}\), and induced coefficient operator
\(\mathcal L_{\mathrm{tt}}\) fixed in
\cref{eq:octahedral-second-form,eq:octahedral-coefficient-map}. The
similarity induces the corresponding map on projective group classes.
Here \(P'_m\) denotes the same ordered matching placement as \(P_m\), but
in the \(\OctTT\) coefficient coordinates.
Writing \(X=Q_1\) projectively, direct ordered contraction gives
\begin{equation}
\begin{aligned}
\mathcal L_{\mathrm{tt}}P_0(U,V)&\doteq
 P'_0(\theta_{\mathrm{tt}}(U),\theta_{\mathrm{tt}}(V)),\\
\mathcal L_{\mathrm{tt}}P_1(U,V)&\doteq
 P'_1(\theta_{\mathrm{tt}}(UX),\theta_{\mathrm{tt}}(XV)),\\
\mathcal L_{\mathrm{tt}}P_2(U,V)&\doteq P'_2(\theta_{\mathrm{tt}}(U),
                                  \theta_{\mathrm{tt}}(XVX)).
\end{aligned}
\label{eq:cert-S4-coefficient-transport}
\end{equation}
The operator \(\mathcal L_{\mathrm{tt}}\) compares coefficient arrays.
Realizable dressings use the corresponding \(\theta_{\mathrm{tt}}(U)\)
representatives with their disequality-edge orientations and nonzero scalars,
as in the fixed-coordinate convention of \cref{sec:platonic-decks}.

The coset classification in \cref{prop:cert-platonic-incidence-atlas} gives
\begin{center}
\normalsize
\begin{tabular}{c@{\qquad}c@{\qquad}c}
\toprule
domain points&three-point lines&four-point lines\\
\midrule
24&32&18\\
\bottomrule
\end{tabular}
\end{center}
Every point lies on seven rich lines.  The three-lines and
four-lines are single realizable left/right orbits, represented by
\[
 \{[Q_0],[R_+],[R_-]\}
\]
and
\[
 \{[Q_0],[Q_1],[Q_0+Q_1],[Q_0-Q_1]\},
\]
respectively.

The quaternary deck has \(3\cdot24^2=1728\) projective points.  Its complete
maximal-rich-line histogram is
\begin{center}
\normalsize
\begin{tabular}{ccr}
\toprule
line size&matching-label type&number\\
\midrule
3&\((1,1,1)\)&13\,824\\
3&\((3)\)&4\,608\\
4&\((4)\)&2\,592\\
\midrule
&&21\,024\\
\bottomrule
\end{tabular}
\end{center}
There is no \((2,1)\) line and no mixed four-line.  Representatives of
the three rows are
\begin{align*}
&\{P_0(Q_0,Q_0),P_1(Q_0,Q_0),P_2(Q_2,Q_2)\},\\
&\{P_0(Q_0,Q_0),P_0(Q_0,R_+),P_0(Q_0,R_-)\},\\
&\{P_0(Q_0,Q_0),P_0(Q_0,Q_1),
       P_0(Q_0,Q_0+Q_1),P_0(Q_0,Q_0-Q_1)\},
\end{align*}
where \(R_+\) and \(R_-\) are the two explicitly fixed representatives in
\cref{eq:cert-platonic-common-vectors}.

For a mixed three-point incidence, use the normalization
\eqref{eq:cert-A4-normal-incidence}.  For a selected ordered frame, let
\(\Phi\) be the linear card map determined by \(\Phi(Q_0)=C/2\) and by
\((\Phi(Q_1),\Phi(Q_2),\Phi(Q_3))\) equal to that ordered frame. Define the
image tensors \(W_j:=\Phi(Q_j)\), and write
\(\rank\Phi:=\dim\operatorname{im}\Phi\).  Before imposing the twelve pair
points of \eqref{eq:octahedral-domain}, the bridge is
\begin{equation*}
 \{0,A/2,B/2\}\cup\mathcal R,
\end{equation*}
where the subscripts in \(R_{\mu\nu}^{\epsilon\delta}\) specify the
two quaternion axes and the superscripts specify the displayed signs:
\begin{align*}
\mathcal R=\{&R_0,R_1,R_3,\\
&R_{01}^{+-}=\tfrac14P_2(Q_0+Q_1,Q_0-Q_1),
\quad R_{01}^{-+}=\tfrac14P_2(Q_0-Q_1,Q_0+Q_1),\\
&R_{03}^{+-}=\tfrac14P_2(Q_0+Q_3,Q_0-Q_3),
\quad R_{03}^{-+}=\tfrac14P_2(Q_0-Q_3,Q_0+Q_3),\\
&R_{13}^{++}=-\tfrac14P_2(Q_1+Q_3,Q_1+Q_3),
\quad R_{13}^{--}=-\tfrac14P_2(Q_1-Q_3,Q_1-Q_3)\}.
\end{align*}
By \cref{lem:platonic-bridge-formula}, these are all bridge vectors:
the symmetric octahedral directions are the three axes and the six
\(Q_\mu\pm Q_\nu\) directions with \(\mu,\nu\in\{0,1,3\}\).
An orthogonal triple is either the three axes or a pair of opposite
diagonal directions in one coordinate plane together with the remaining
axis.  There are therefore four unordered frames.  The sum equation
\(W_1+W_2+W_3=(A+B)/2\) leaves 30 ordered maps: the six
\(\rank\Phi=2\) maps permute \((0,A/2,B/2)\), and the other twenty-four are
the six orderings of each of the four \(\rank\Phi=4\) frames
\begin{equation}
 \begin{split}
 &\{R_0,R_1,R_3\},\qquad
 \{R_0,R_{13}^{++},R_{13}^{--}\},\\
 &\{R_1,R_{03}^{+-},R_{03}^{-+}\},\qquad
 \{R_3,R_{01}^{+-},R_{01}^{-+}\}.
 \end{split}
 \label{eq:cert-S4-four-frames}
\end{equation}
The 1,727 deck points other than \([A]\) lie on 1,683 projective lines
through \([A]\). The numbers of other deck points on those lines are
\[
 1^{1645},\qquad2^{32},\qquad3^6.
\]
The single realizable kernel
\begin{equation*}
 K_*=Q_0+Q_1,
 \qquad [K_*]\in\OctDD
\end{equation*}
has the following property: for every ordered rank-four frame,
\(q_R=\Phi(K_*)=C/2+R\) is not a matching-deck point.  The 24 supports
have sizes
\[
 2^4,\qquad8^{12},\qquad16^8.
\]
The coefficient correspondence in
\cref{eq:cert-S4-coefficient-transport} identifies the second normal form.
Its realizable constructions use the
\(\OctTT\)-kernels and dressings in
\cref{prop:p1-cert-S4-uniform-eight-vertex-separator}.   Invertible local coefficient maps
preserve the three ranks.

For the two mixed forms, use the coefficient similarities
\(H_{\mathrm{dt}},H_{\mathrm{td}}\), groups \(\OctDT,\OctTD\), and
operators \(\mathcal L_{\mathrm{dt}},\mathcal L_{\mathrm{td}}\) fixed in
\cref{eq:octahedral-mixed-conjugators,eq:octahedral-mixed-forms,%
eq:octahedral-mixed-coefficient-transport}.  The three exact transport
identities there identify every matching label and absorb their displayed
orientation factors into matrices of the current mixed group.  They
compare coefficient tables, while the following proposition supplies
constructions in the current marked group.

\begin{proposition}[Gadget octahedral signature-separator certificate]
\label{prop:p1-cert-S4-uniform-eight-vertex-separator}
For every ordering of every frame in
\cref{eq:cert-S4-four-frames}, use the realizable kernel
\(K_*=Q_0+Q_1\) in the \(\OctDD\) form and the realizable kernel
\(K'_*:=\theta_{\mathrm{tt}}(K_*)\) in the \(\OctTT\) form.  The
resulting direct quaternary can be dressed, using only realizable local
matrices in the current marked octahedral group, to an endpoint-nondegenerate eight-vertex signature of
flattening ranks \((2,2,2)\).  Similarity by \(H_{\mathrm{tt}}\) identifies the coefficient tables.

For \(\alpha\in\{\mathrm{dt},\mathrm{td}\}\), put
\(K_{*,\alpha}:=\theta_\alpha(K_*)\), so
\([K_{*,\alpha}]\in\OctForm{\alpha}\), and use the fixed realizable kernel in
that projective class, with its realization scalar recorded.  The resulting
direct quaternary has, using only current-group contractions and
dressings, one of three certified exits: an endpoint-nondegenerate
eight-vertex signature of ranks \((2,2,2)\), a generalized equality,
or a nonsingular binary \(B\) with \([BX]\notin\OctForm{\alpha}\).
The same three-way conclusion holds for each of the six rank-deficient maps in
each mixed form.  All representatives, determinants, and outside-transfer
directions are given in the proof below.
\end{proposition}

\begin{proof}
Define the tetrahedral sign matrices by
\[
  G_{\epsilon_1\epsilon_2\epsilon_3}
  =Q_0+\epsilon_1Q_1+\epsilon_2Q_2+\epsilon_3Q_3.
\]
For the first marked form the separator is \(q_R=C/2+R\), where \(R\)
is the first member of the ordered frame.  In the even-string order
\cref{eq:octahedral-even-order}, exact substitution gives the following
realizable dressings. Every omitted odd-weight coefficient is zero, and the
displayed vectors are projective.
\begin{center}
\normalsize
\setlength{\tabcolsep}{4pt}
\begin{tabular}{ccc}
\toprule
\(R\)&\(\mathbf L\)&\(\mathbf e(\mathscr D_{\mathbf L}q_R)\)\\
\midrule
\(R_0\)&\(I^{\otimes4}\)&\((1,1,-1,1,1,-1,1,1)\)\\
\(R_1\)&\((I,I,Q_1,Q_1)\)&\((1,0,0,0,0,0,0,1)\)\\
\(R_3\)&\(I^{\otimes4}\)&\((1,1,-1,-1,-1,-1,1,1)\)\\
\(R_{01}^{+-}\)&\(G_{+++}^{\otimes4}\)&\((1,-1,1,-i,i,1,-1,1)\)\\
\(R_{01}^{-+}\)&\(G_{+++}^{\otimes4}\)&\((1,-1,1,i,-i,1,-1,1)\)\\
\(R_{03}^{+-}\)&\(I^{\otimes4}\)&\((1,1,-1,i,-i,-1,1,1)\)\\
\(R_{03}^{-+}\)&\(I^{\otimes4}\)&\((1,1,-1,-i,i,-1,1,1)\)\\
\(R_{13}^{++}\)&\(G_{++-}^{\otimes4}\)&\((1,i,-i,-i,-i,-i,i,-1)\)\\
\(R_{13}^{--}\)&\(G_{++-}^{\otimes4}\)&\((1,-i,i,i,i,i,-i,-1)\)\\
\bottomrule
\end{tabular}
\end{center}
Both endpoints are nonzero in every row.  Since the separator depends
only on the first frame member, these nine rows cover all twenty-four
ordered maps.  Direct row reduction of the three flattenings of
every displayed \(q_R\) gives \((2,2,2)\).

For the second marked form, use the coefficient-transport operator
\(\mathcal L_{\mathrm{tt}}\) above, and define the transported card map
\(\Phi'\) and
transported direct card \(q'_R\) by
\[
  \Phi'(\theta_{\mathrm{tt}}C_{\rm ker})=\mathcal L_{\mathrm{tt}}\Phi(C_{\rm ker}),
  \qquad
  q'_R:=\Phi'(K'_*)=\mathcal L_{\mathrm{tt}}q_R.
\]
Here \([K'_*]\in\OctTT\), so \(q'_R\) is a direct card in the
second marked form.
With the nonzero comparison scalar \(\lambda:=2\sqrt2(1-i)\), the identities
\begin{equation*}
  H_{\mathrm{tt}}H_{\mathrm{tt}}^{\mathsf T}=\lambda X,
  \qquad
  H_{\mathrm{tt}}^{-\mathsf T}H_{\mathrm{tt}}^{-1}=\lambda^{-1}X
\end{equation*}
show that \(\mathcal L_{\mathrm{tt}}\) bijects the two coefficient decks.
This is only
an analytic correspondence.  The following input matrices are separately
verified to belong projectively to the realizable group \(\OctTT\). In the
table, \(\theta_{\mathrm{tt}}\) acts componentwise on every displayed tuple:
\begin{center}
\normalsize
\setlength{\tabcolsep}{4pt}
\begin{tabular}{ccc}
\toprule
\(R\)&\(\mathbf L'\)&\(\mathbf e(\mathscr D_{\mathbf L'}q'_R)\)\\
\midrule
\(R_0\)&\(\theta_{\mathrm{tt}}(G_{++-},G_{++-},G_{++-},G_{+-+})\)&
\((1,-1,-1,-1,-1,-1,-1,1)\)\\
\(R_1\)&\(\theta_{\mathrm{tt}}(I,I,I,I)\)&\((1,1,1,-1,-1,1,1,1)\)\\
\(R_3\)&\(\theta_{\mathrm{tt}}(G_{+++})^{\otimes4}\)&\((1,-1,1,1,1,1,-1,1)\)\\
\(R_{01}^{+-}\)&\(\theta_{\mathrm{tt}}(G_{+++},G_{--+},G_{--+},G_{+++})\)&
\((1,0,0,0,0,0,0,1)\)\\
\(R_{01}^{-+}\)&\(\theta_{\mathrm{tt}}(G_{+++},G_{+--},G_{+--},G_{+++})\)&
\((1,-1,-1,1,1,-1,-1,1)\)\\
\(R_{03}^{+-}\)&\(\theta_{\mathrm{tt}}(I,I,Q_1,Q_1)\)&\((1,1,-1,-1,-1,-1,1,1)\)\\
\(R_{03}^{-+}\)&\(\theta_{\mathrm{tt}}(I,I,Q_2,Q_2)\)&\((1,0,0,0,0,0,0,1)\)\\
\(R_{13}^{++}\)&\(\theta_{\mathrm{tt}}(G_{++-},G_{-++},G_{++-},G_{-++})\)&
\((1,-1,-1,1,1,-1,-1,1)\)\\
\(R_{13}^{--}\)&\(\theta_{\mathrm{tt}}(G_{++-},G_{---},G_{+-+},G_{-++})\)&
\((1,0,0,0,0,0,0,1)\)\\
\bottomrule
\end{tabular}
\end{center}
Again both endpoints are nonzero.  All dressings are invertible, so the
three certified ranks remain \((2,2,2)\).

It remains to verify the mixed rows directly in their fixed-
\(I\) coordinates.  For \(\alpha\in\{\mathrm{dt},\mathrm{td}\}\), absorb the three
matching-dependent orientation factors in
\cref{eq:octahedral-mixed-coefficient-transport} into the current-
group representatives.  The kernel \(K_{*,\alpha}\) first gives the
realizable card \(\mathcal L_\alpha q_R\).  On inputs \(2,4\), apply the
current-group dressing \(\theta_\alpha(S_\alpha^{-1})\), where
\(S_\alpha=H_\alpha H_\alpha^{\mathsf T}\).  Its product with the
\(H_\alpha^{\mathsf T}\) leg of \(\mathcal L_\alpha\) is
\(H_\alpha^{-1}\), so the resulting direct card
\(q_{R,\mathrm{act}}^\alpha\) has the normalized representative
\begin{equation}
 q_R^\alpha:=(H_\alpha^{-1})^{\otimes4}q_R,
 \qquad q_{R,\mathrm{act}}^\alpha\doteq q_R^\alpha.
 \label{eq:cert-S4-mixed-direct-card}
\end{equation}
These realizable dressings preserve the retained reduction data.  When another dressing on inputs \(2,4\) is listed
below, compose the two group elements into one chosen representative on
that input.  The total is at most four dressings, or two dressings and one
pair contraction for an outside-binary row.  All exact
determinants below refer to this normalized representative. The corresponding
realizable output differs by a nonzero scalar, so only determinant
nonvanishing is used.  Put
\[
 \delta_\alpha(B):=
 [c_Q(H_\alpha(BX)H_\alpha^{-1})].
\]
Thus \(\delta_\alpha(B)\notin\mathcal R_{\OctDD}\) is exactly the assertion
\([BX]\notin\OctForm{\alpha}\).

For \(\OctDT\), the three first-frame entries in the following
table give endpoint-nondegenerate eight-vertex signatures.  All odd
coefficients vanish, and the last column is in the even-string order fixed
above.
\begin{center}
\normalsize
\setlength{\tabcolsep}{4pt}
\begin{tabular}{ccc}
\toprule
\(R\)&realizable dressing&\(\mathbf e(\mathscr D q_R^{dt})\)\\
\midrule
\(R_0\)&\(\theta_{\mathrm{dt}}(G_{--+})^{\otimes4}\)&
 \((1,-1,1,-1,-1,1,-1,1)\)\\
\(R_1\)&\(\theta_{\mathrm{dt}}(G_{---})^{\otimes4}\)&
 \((1,1,-1,-1,-1,-1,1,1)\)\\
\(R_3\)&identity&\((1,1,-1,-1,-1,-1,1,1)\)\\
\bottomrule
\end{tabular}
\end{center}
The other six possible first entries have the uniform outside-binary exits
listed below. The kernel acts on inputs \(1,2\).
\begin{center}
\normalsize
\setlength{\tabcolsep}{4pt}
\begin{tabular}{cccc}
\toprule
\(R\)&realizable kernel&\(\delta_{\mathrm{dt}}(B)\)&\(\det B\)\\
\midrule
\(R_{01}^{+-}\)&\(\theta_{\mathrm{dt}}(G_{---})\)&\([1,-1,1,0]\)&
 \((51+36\sqrt2)/64\)\\
\(R_{01}^{-+}\)&\(\theta_{\mathrm{dt}}(G_{---})\)&\([1,-1,0,-1]\)&
 \((51+36\sqrt2)/64\)\\
\(R_{03}^{+-}\)&\(\theta_{\mathrm{dt}}(Q_0)\)&\([1,-3,1,-1]\)&
 \((51+36\sqrt2)/256\)\\
\(R_{03}^{-+}\)&\(\theta_{\mathrm{dt}}(Q_0)\)&\([1,-3,-1,1]\)&
 \((51+36\sqrt2)/256\)\\
\(R_{13}^{++}\)&\(\theta_{\mathrm{dt}}(Q_0)\)&\([1,-3,1,1]\)&
 \((51+36\sqrt2)/256\)\\
\(R_{13}^{--}\)&\(\theta_{\mathrm{dt}}(Q_0)\)&\([1,-3,-1,-1]\)&
 \((51+36\sqrt2)/256\)\\
\bottomrule
\end{tabular}
\end{center}

For \(\OctTD\), no dressing is needed for the two endpoint rows:
\begin{equation*}
\begin{aligned}
 \mathbf e(q_{R_0}^{td})&=(1,i,-i,i,i,-i,i,-1),\\
 \mathbf e(q_{R_3}^{td})&=(1,i,-i,-i,-i,-i,i,-1).
\end{aligned}
\end{equation*}
Both have zero odd coefficients and ranks \((2,2,2)\).  The \(R_1\) row is
\[
 q_{R_1}^{td}=\frac{3+2\sqrt2}{8}
  (\one[0011]+\one[1100]),
\]
so realizable \(X\)-dressings on inputs \(3,4\) give
\(\frac{3+2\sqrt2}{8}(\Eq_4)\), a generalized equality.
Compose the dressing on input \(4\) with its preceding correction. Together
with the corrections on inputs \(2,4\), this uses only three dressed inputs.
The remaining rows have the following outside-binary exits on inputs
\(1,2\):
\begin{center}
\normalsize
\setlength{\tabcolsep}{4pt}
\begin{tabular}{cccc}
\toprule
\(R\)&realizable kernel&\(\delta_{\mathrm{td}}(B)\)&\(\det B\)\\
\midrule
\(R_{01}^{+-}\)&\(\theta_{\mathrm{td}}(Q_0)\)&\([0,2,-1,1]\)&
 \((51+36\sqrt2)/256\)\\
\(R_{01}^{-+}\)&\(\theta_{\mathrm{td}}(Q_0)\)&\([0,2,-1,-1]\)&
 \((51+36\sqrt2)/256\)\\
\(R_{03}^{+-}\)&\(\theta_{\mathrm{td}}(G_{--+})\)&\([1,2,0,1]\)&
 \((51+36\sqrt2)/64\)\\
\(R_{03}^{-+}\)&\(\theta_{\mathrm{td}}(G_{--+})\)&\([1,-2,0,-1]\)&
 \((51+36\sqrt2)/64\)\\
\(R_{13}^{++}\)&\(\theta_{\mathrm{td}}(Q_0)\)&\([1,2,-1,0]\)&
 \((51+36\sqrt2)/256\)\\
\(R_{13}^{--}\)&\(\theta_{\mathrm{td}}(Q_0)\)&\([1,-2,1,0]\)&
 \((51+36\sqrt2)/256\)\\
\bottomrule
\end{tabular}
\end{center}

Finally let \(q_\pi^\alpha\) denote the six normalized mixed reconstructed
rank-deficient outputs in the same input convention as
\cref{eq:cert-S4-mixed-direct-card}.  Put
\[
 K_{\mathrm{dt}}^{\mathrm{def}}:=\theta_{\mathrm{dt}}(Q_0-Q_2),
 \qquad
 K_{\mathrm{td}}^{\mathrm{def}}:=\theta_{\mathrm{td}}(Q_2-Q_3),
\]
and let
\(B_{\pi,\alpha}^{\mathrm{def}}
 :=\partial_{12}^{K_\alpha^{\mathrm{def}}}q_\pi^\alpha\).
The fixed chosen representative of \([K_\alpha^{\mathrm{def}}]\) on inputs
\(1,2\) works uniformly for all six outputs in each row:
\begin{equation*}
\begin{array}{c|c|c|c}
\alpha&K_\alpha^{\mathrm{def}}&
 \{\delta_\alpha(B_{\pi,\alpha}^{\mathrm{def}}):\pi\in S_3\}
 &\det B_{\pi,\alpha}^{\mathrm{def}}\\ \hline
dt&\theta_{\mathrm{dt}}(Q_0-Q_2)&
 \{[1,-1,0,-1]^2,[1,-3,1,-1]^2,[1,-3,-1,-1]^2\}&
 (51+36\sqrt2)/256\\
td&\theta_{\mathrm{td}}(Q_2-Q_3)&
 \{[0,2,1,1]^2,[1,2,0,1]^2,[1,2,1,0]^2\}&
 (51+36\sqrt2)/256.
\end{array}
\end{equation*}
Here \([a_0,a_1,a_2,a_3]\) denotes the projective quaternion-coordinate
point \([(a_0,a_1,a_2,a_3)^{\mathsf T}]\), equivalently the matrix class
\([a_0Q_0+a_1Q_1+a_2Q_2+a_3Q_3]\) under \(c_Q\), and an exponent records
multiplicity.  Every determinant displayed in the mixed tables is nonzero.
Every pulled-back direction has support three, or support four with unequal
nonzero magnitudes, whereas the quaternion-coordinate points of
\(\mathcal R_{\OctDD}\) have support one, two, or four with equal nonzero
magnitudes.  Hence all the listed transfer
classes are outside their realizable mixed group.  Direct row reduction gives
rank pattern \((2,2,2)\) in every endpoint row.  This proves all full-rank
and rank-deficient mixed exits using only current-form kernels and
dressings, while the similarities serve only to write the coefficient
arrays.
\end{proof}

For the fixed-matching residue in the standard form, the four axes and
eight tetrahedral sign points give the quadratic test set of
\cref{lem:platonic-global-ruling}.  In any nonstandard form, apply the same
test after precomposing \(\Phi\) with the corresponding one of
\(\theta_{\mathrm{tt}},\theta_{\mathrm{dt}},\theta_{\mathrm{td}}\), and then pull the resulting
tensor identity back by its inverse.    In all four forms, the evaluation
matrix of the ten homogeneous quadratic
 monomials in the four quaternion coordinates has full column rank.  Hence
 every pulled-back Segre minor vanishes identically, giving a genuine
common factor rather than a new octahedral remaining case.

\Cref{prop:cert-platonic-incidence-atlas} proves the domain and deck
incidences by cyclic cosets, tensor factorization, and realizable group actions.
\Cref{lem:platonic-bridge-formula} proves the bridge intersections and
reduces their sum equations to the displayed orthogonal frames.

\subsubsection{The internal tetrahedral permutation classes}
\label{proofsubsec:cert-A4}

For the tetrahedral certificate, order the standard twelve points
\(\TetInt\) in \cref{eq:tetrahedral-domain} as the four axes
followed by the eight sign points.  The coset classification in \cref{prop:cert-platonic-incidence-atlas}
gives 16 three-point lines in one orbit under realizable left and right multiplication by \(\TetInt\).  The quaternary
deck has \(3\cdot12^2=432\) distinct projective points.  Exactly 1728 of
its three-point lines meet all three matching classes, again in one orbit
under realizable local dressings and permutations of inputs.  Thus the common
normalization above applies.

Scale the normalized card map
\(\Phi:\Mat_2(\mathbb C)\to(\mathbb C^2)^{\otimes4}\)
so that
\[
 \Phi(R_+)=A,
 \qquad
 \Phi(R_-)=-B.
\]
Then \(\Phi(Q_0)=C/2\). Define the image tensors \(W_j:=\Phi(Q_j)\).  Thus
\(W_1+W_2+W_3=(A+B)/2\).  For each
\(W\in\{W_1,W_2,W_3\}\), require
\[
 W,\ A-2W,\ 2W-B
 \in\{0\}\cup\operatorname{cone}\bigl(\mathfrak D_4(\TetInt)\bigr).
\]
The resulting exact bridge is
\begin{equation}
 \mathcal B_{\TetInt}=\{0,A/2,B/2,R_0,R_1,R_3\},
 \label{eq:cert-A4-bridge}
\end{equation}
where \(R_0,R_1,R_3\) are defined in
\cref{eq:cert-platonic-common-vectors}.
The remaining bridge vectors obey
\begin{center}
\normalsize
\begin{tabular}{cL{0.32\linewidth}L{0.32\linewidth}}
\toprule
\(W\)&\(A-2W\)&\(2W-B\)\\
\midrule
\(R_0\)&\(P_1(Q_2,Q_2)\)&\(-P_0(Q_2,Q_2)\)\\
\(R_1\)&\(-P_1(Q_3,Q_3)\)&\(P_0(Q_3,Q_3)\)\\
\(R_3\)&\(-P_1(Q_1,Q_1)\)&\(P_0(Q_1,Q_1)\)\\
\bottomrule
\end{tabular}
\end{center}
\Cref{lem:platonic-bridge-formula} proves completeness of this list.
The symmetric tetrahedral directions are exactly \(Q_0,Q_1,Q_3\), so
their three coordinate axes form the unique orthogonal frame.

The sum equation on \(\mathcal B_{\TetInt}^3\) leaves precisely twelve maps:
\begin{equation}
 \begin{array}{ll}
 6\text{ maps with }\rank\Phi=2:&
 (W_1,W_2,W_3)\text{ permutes }(0,A/2,B/2),\\[1mm]
 6\text{ maps with }\rank\Phi=4:&
 (W_1,W_2,W_3)\text{ permutes }(R_0,R_1,R_3).
 \end{array}
 \label{eq:cert-A4-twelve-maps}
\end{equation}
For a permutation \(\sigma=(a,b,c)\) of \((0,1,3)\), the Frobenius-dual
basis is
\[
 Q^0=Q_0/2,\quad Q^1=-Q_1/2,\quad
 Q^2=Q_2/2,\quad Q^3=-Q_3/2,
\]
and the uniquely reconstructed six-input ancestor is
\begin{equation*}
 f_\sigma=\sum_{\mu=0}^3(Q^\mu)^{(1,2)}\otimes F^\sigma_\mu,
 \qquad
 (F_0^\sigma,F_1^\sigma,F_2^\sigma,F_3^\sigma)
 =(C/2,R_a,R_b,R_c).
\end{equation*}
The duals in this formula are coefficient-reconstruction devices, not
available gadgets.

Contract all fifteen input pairs by all twelve realizable tetrahedral
kernels.  The exact 1080-card histogram is
\begin{center}
\normalsize
\begin{tabular}{c@{\qquad}rr}
\toprule
\(\sigma\)&zero or matching-deck cards&nonmatching cards\\
\midrule
\((0,1,3),(1,3,0),(3,0,1)\), each&180&0\\
\((0,3,1),(1,0,3),(3,1,0)\), each&36&144\\
\bottomrule
\end{tabular}
\end{center}
All \(180\) cards for each of the three orientations in the first data row
are nonzero.
For the three rejected orientations, the single contraction of inputs
1 and 3 by \(Q_0=I\) gives the transparent certificate
\begin{center}
\normalsize
\begin{tabular}{c@{\qquad}c@{\qquad}c}
\toprule
\(\sigma\)&support size&three flattening ranks\\
\midrule
\((0,3,1)\)&8&\((2,2,2)\)\\
\((1,0,3)\)&16&\((2,2,2)\)\\
\((3,1,0)\)&16&\((2,2,2)\)\\
\bottomrule
\end{tabular}
\end{center}
Normalize each card by its first nonzero entry, as stipulated for the
projective enumerations above.  In the three \(\sigma\)-rows of the
preceding table, respectively, one nonzero \(2\times2\) minor in each of the flattening orders
\((12\mid34,13\mid24,14\mid23)\) is
\[
 (-1,-1,-2),\qquad(-2,2i,2i),\qquad(-2i,2,-2i).
\]
Thus none of the three outputs is a hidden two-binary matching product.

\begin{proposition}[Tetrahedral permutation classes and eight-vertex signatures]
\label{prop:p1-cert-A4-eight-vertex-orientation-terminal}
Let the exceptional orientation set be
\[
  \Sigma_-:=\{(0,3,1),(1,0,3),(3,1,0)\}.
\]
For each \(\sigma\in\Sigma_-\), let \(q_\sigma\) be the quaternary
obtained by bilinearly contracting inputs \(1,3\) of \(f_\sigma\) with the
realizable kernel \(Q_0=I\).  Then \(q_\sigma\)
admits realizable tetrahedral input dressings to an endpoint-nondegenerate eight-vertex signature.  The three
flattening ranks remain \((2,2,2)\).
\end{proposition}

\begin{proof}
Define the tetrahedral representative matrix
\(G_{\epsilon_1\epsilon_2\epsilon_3}
 :=Q_0+\epsilon_1Q_1+\epsilon_2Q_2+\epsilon_3Q_3\), and use the
even-string order in \cref{eq:octahedral-even-order} on the residual inputs
\((2,4,5,6)\).  Direct expansion gives
\begin{center}
\normalsize
\setlength{\tabcolsep}{5pt}
\begin{tabular}{ccc}
\toprule
\(\sigma\)&\(\mathbf L\)&\(\mathbf e(\mathscr D_{\mathbf L}q_\sigma)\)\\
\midrule
\((0,3,1)\)&\((Q_0,G_{++-},G_{+++},Q_1)\)&
\((1,0,0,0,0,0,0,-i)\)\\
\((1,0,3)\)&\((Q_0,Q_0,G_{+++},Q_0)\)&
\((1,1,-i,i,1,-1,i,i)\)\\
\((3,1,0)\)&\((Q_0,G_{+++},Q_0,G_{++-})\)&
\((1,-1,1,1,-i,-i,i,-i)\)\\
\bottomrule
\end{tabular}
\end{center}
Every omitted odd entry is zero and both endpoints are nonzero.  Each
displayed dressing is realizable and invertible, so it preserves the certified
flattening ranks.
\end{proof}

For the three surviving orientations, the following coefficient formulas
prove membership in the same affine class.  Let
\(f_{013},f_{130},f_{301}\) be the three surviving arity-six signatures and
\(q_{013},q_{130},q_{301}\) their quadratic phase polynomials, and put
\[
 {\mathbf{x}}=(x_1,\ldots,x_6)\in\Ftwo^6.
\]
In the
display below, the indicator constraints and their additions are over
\(\Ftwo\), the exponents \(q_\sigma\) are read modulo four, and the displayed
prefactors are complex scalars:
\begin{align*}
f_{013}({\mathbf{x}})&=-\frac{i}{4}\,
 \one[x_3+x_4+x_5+x_6=0]\,i^{q_{013}({\mathbf{x}})},\\
q_{013}&=2x_3+3x_4+x_5
 +2(x_1x_2+x_1x_3+x_1x_5+x_2x_3+x_2x_4+x_4x_5),\\[1mm]
f_{130}({\mathbf{x}})&=-\frac{i}{4}\,
 \one[x_3+x_4+x_5+x_6=0]\,i^{q_{130}({\mathbf{x}})},\\
q_{130}&=3x_1+x_2+3x_4+x_5
 +2(x_1x_2+x_1x_3+x_1x_4+x_2x_3+x_2x_5+x_3x_4+x_3x_5),\\[1mm]
f_{301}({\mathbf{x}})&=\frac{1-i}{4}\,
 \one[\substack{x_3+x_4+x_5+x_6=0,\\
                     x_1+x_2+x_4+x_5=1}]\,i^{q_{301}({\mathbf{x}})},\\
q_{301}&=2+x_1+2x_2+x_5+x_6
 +2(x_1x_2+x_1x_3+x_1x_4+x_3x_4).
\end{align*}
The displayed supports are affine and every mixed phase coefficient is
even, so the three ancestors are affine.  Separately, direct
expansion of the twelve realizable binary representatives \(C_{\rm ker}X\),
\([C_{\rm ker}]\in\TetInt\), verifies that they are affine in this same normalized
basis.

\subsubsection{The icosahedral forms
  \texorpdfstring{\(\IcoZero\) and \(\IcoOne\)}{I (0) and I (1)}}
\label{proofsubsec:cert-A5}

For the icosahedral certificate, use the standard set \(\IcoZero\) in
\cref{eq:quaternion-basis,eq:icosahedral-domain}, ordered as the four axes, the
eight tetrahedral sign points, and the 48 distinct even-permutation points.
Exact arithmetic over
\(\mathbb Q(\phi,i)\) gives
\hypertarget{eq:cert-A5-orders}{}
\begin{equation*}
 |\IcoZero|=60,
 \qquad
 \#\{g:\operatorname{ord}(g)=1,2,3,5\}=(1,15,20,24).
\end{equation*}
The set is closed under multiplication, inverse, and transpose.
It is the projectivization of the binary icosahedral group, so \(\IcoZero\cong A_5\).
The apparent opposite chirality gives no additional
distinguished-involution case.  If \(\IcoOne\) is obtained by using
odd coordinate permutations in the additional part of
\eqref{eq:icosahedral-domain}, then, for \(D=Q_0+Q_3\),
\hypertarget{eq:cert-A5-chirality}{}
\begin{equation*}
 D^{\mathsf T}XD=2X,
 \qquad
 \IcoOne=D\IcoZero D^{-1},
 \qquad
 (DQ_1D^{-1},DQ_2D^{-1},DQ_3D^{-1})=(-Q_2,Q_1,Q_3).
\end{equation*}
\Cref{lem:icosahedral-fixed-I-dictionary} uses this identity to translate
kernel and card formulas while preserving the retained fixed-\(I\) problem.

The maximal-cyclic-coset classification in
\cref{prop:cert-platonic-incidence-atlas} gives 200 three-point
\(C_3\)-coset lines and 72 five-point \(C_5\)-coset lines.  Every point is
on ten three-lines and six five-lines.  The quaternary matching-product deck
has \(3\cdot60^2=10\,800\) distinct projective points, and its complete
maximal-rich-line histogram is
\begin{center}
\normalsize
\begin{tabular}{ccr}
\toprule
line size&matching-label type&number\\
\midrule
3&\((1,1,1)\)&216\,000\\
3&\((3)\)&72\,000\\
5&\((5)\)&25\,920\\
\bottomrule
\end{tabular}
\end{center}
There is no \((2,1)\) line and no mixed five-line.

Normalize a mixed three-point incidence by
\eqref{eq:cert-A4-normal-incidence}, and scale its card map so that
\(\Phi(R_+)=A\) and \(\Phi(R_-)=-B\).  Define the image tensors
\(W_j:=\Phi(Q_j)\).  Then
\hypertarget{eq:cert-A5-sum}{}
\begin{equation*}
 \Phi(Q_0)=C/2,
 \qquad
 W_1+W_2+W_3=(A+B)/2.
\end{equation*}
The other six tetrahedral sign points require each of
\(W_j,A-2W_j,2W_j-B\) to lie in the zero/matching-product cone
\[
 \mathscr C_{\IcoZero}:=
 \{0\}\cup
 \{\lambda P_m(U,V):\lambda\in\mathbb C^\times,\
      m\in\{0,1,2\},\ [U],[V]\in\IcoZero\}.
\]
Define the following \(15\)-point slice in coefficient projective space:
\begin{align*}
 \mathcal H&=
 \{[a:b:d]\in\mathbb P^2:
      [aQ_0+bQ_1+dQ_3]\in\IcoZero\},
 &|\mathcal H|&=15,\notag\\
 J_{\mathbf{u}}&=aQ_0+bQ_1+dQ_3,
 &\operatorname{adj}(J_{\mathbf{u}})&=aQ_0-bQ_1-dQ_3,\notag\\
  R_{\mathbf{u}}&=\frac{P_2(J_{\mathbf{u}},\operatorname{adj}(J_{\mathbf{u}}))}{2(a^2+b^2+d^2)}.
\end{align*}
Here \({\mathbf{u}}=(a,b,d)\) is any nonzero representative of
\([a:b:d]\in\mathcal H\).  The displayed quotient is unchanged when \({\mathbf{u}}\)
is rescaled, so \(R_{\mathbf{u}}\) is well defined by the projective coefficient
point.  Moreover,
\(a^2+b^2+d^2=\det J_{\mathbf{u}}\ne0\): the equality follows directly from the
quaternion matrices in \cref{eq:quaternion-basis}, and nonvanishing follows
from \([J_{\mathbf{u}}]\in\IcoZero\subset\PGL_2(\mathbb C)\).  
The exact bridge intersection is the eighteen-point set
\hypertarget{eq:cert-A5-bridge}{}
\begin{equation*}
 \mathcal B_{\IcoZero}:=\{W:W,A-2W,2W-B\in\mathscr C_{\IcoZero}\}
 =\{0,A/2,B/2\}\cup\{R_{\mathbf{u}}:[{\mathbf{u}}]\in\mathcal H\}.
\end{equation*}
\Cref{lem:platonic-bridge-formula} proves this equality.  The symmetric
icosahedral directions are precisely the three axes with no \(Q_2\)
component and the twelve additional points in that coordinate plane.
For such a direction,
\(J_{\mathbf{u}}^{-1}=\operatorname{adj}(J_{\mathbf{u}})/(a^2+b^2+d^2)\), which gives the displayed
formula for \(R_{\mathbf{u}}\).  Thus the list includes every complex scale and
every cancellation case.
As a finite completeness check, group the other 10,799 deck points by
their line through \([A]\). Counting points other than \([A]\) on each
line gives the histogram
\hypertarget{eq:cert-A5-line-buckets}{}
\begin{equation*}
 1^{10\,591},\qquad2^{80},\qquad4^{12}.
\end{equation*}
Here \(a^b\) means that \(b\) lines each contain \(a\) other deck points.

The sum equation on \(\mathcal B_{\IcoZero}^3\) has exactly \(36\) ordered solutions.
Six permute \((0,A/2,B/2)\). The other thirty are the six orderings of
the following five unordered frames, where a vector \({\mathbf{u}}\) represents
the tensor \(R_{\mathbf{u}}\):
\begin{align}
\mathrm{Fr}_0={}&\{(0,0,1),(0,1,0),(1,0,0)\},\notag\\
\mathrm{Fr}_1={}&\{(1,-\tau^2,\tau),(1,-\phi,-\phi^2),
                         (1,\phi,-\tau)\},\notag\\
\mathrm{Fr}_2={}&\{(1,-\tau^2,-\tau),(1,-\phi,\phi^2),
                         (1,\phi,\tau)\},\notag\\
\mathrm{Fr}_3={}&\{(1,-\phi,\tau),(1,\phi,\phi^2),
                         (1,\tau^2,-\tau)\},\notag\\
\mathrm{Fr}_4={}&\{(1,-\phi,-\tau),(1,\phi,-\phi^2),
                         (1,\tau^2,\tau)\}.
\label{eq:cert-A5-frames}
\end{align}
\Cref{lem:platonic-bridge-formula} reduces the nondegenerate sum
solutions to orthogonal triples in \(\mathcal H\).
The five displayed frames contain the fifteen points of \(\mathcal H\)
once each.  Using \(\phi^2=\phi+1\), their dot products show that each
point is orthogonal to precisely the other two points in its displayed
frame.  These are therefore all orthogonal triples.  This proves the
\(6+5\cdot6=36\) solutions without a search through \(\mathcal B_{\IcoZero}^3\).

One legal group element suffices for all thirty orderings:
\begin{equation*}
 K_*=\phi Q_1-\tau Q_2+Q_3,
 \qquad [K_*]\in\IcoZero.
\end{equation*}
For an ordering \(({\mathbf{u}}_1,{\mathbf{u}}_2,{\mathbf{u}}_3)\) of any frame, its direct card is
\begin{equation}
 q_{{\mathbf{u}}_1,{\mathbf{u}}_2,{\mathbf{u}}_3}
 =\phi R_{{\mathbf{u}}_1}-\tau R_{{\mathbf{u}}_2}+R_{{\mathbf{u}}_3}.
 \label{eq:cert-A5-output}
\end{equation}
Exact row reduction of the thirty explicit tensors gives, uniformly,
\begin{equation}
 (\rank_{12\mid34},\rank_{13\mid24},\rank_{14\mid23})=(3,3,3),
 \label{eq:cert-A5-ranks}
\end{equation}
with support-size histogram \(8^6,12^4,16^{20}\).  Each output is
therefore a direct nonmatching quaternary card.

For the no-mixed-line branch, first discard rank at most one, which is
already rank lowering.  On the remaining branch
\(\dim\ker\Phi\le2\).  Along every rich line avoiding
\(\mathbb P(\ker\Phi)\), the absence of a mixed line forces all images to
use one matching.  The kernel-deleted connectedness assertion
\cref{eq:cert-platonic-kernel-deleted-graph} propagates this matching to all
nonzero card images among the sixty domain points, placing them in the
fixed-matching Segre variety: the projective simple-tensor locus
\(\Sigma_M\) of \cref{lem:platonic-global-ruling}, with the two matching
edges as its factors.
Because the icosahedral
domain contains the test set in \cref{eq:platonic-quadratic-test-set},
\cref{lem:platonic-global-ruling} then gives, after a permutation of inputs,
\begin{equation}
 \Phi(C_{\rm ker})=U_*\otimes\psi(C_{\rm ker})
 \quad\text{for all }C_{\rm ker},
 \qquad\text{or}\qquad
 \Phi(C_{\rm ker})=\psi(C_{\rm ker})\otimes U_*.
 \label{eq:cert-A5-common-factor}
\end{equation}
Here \(U_*\) is a fixed nonzero binary factor and
\(\psi:\Mat_2(\mathbb C)\to\Mat_2(\mathbb C)\) is a binary-valued linear
map.
Combining the rank-two solutions, \eqref{eq:cert-A5-ranks}, and
\eqref{eq:cert-A5-common-factor} gives the bounded-interface trichotomy:
rank lowering, a realizable nonmatching quaternary, or a genuine common
factor.

The rejecting kernel and rank assertions are evaluated on the thirty
displayed outputs over \(\mathbb Q(\phi,i)\).

\subsubsection{The complete Platonic incidence atlas}
\label{proofsubsec:cert-platonic-incidence-atlas}

\begin{lemma}[Realizable rainbow-line orbit certificate]
\label{lem:platonic-rainbow-orbit}
For a marked group \(G\), let \(\Gamma_G\) be the finite
permutation group of the projective deck induced by the realizable local
operations
\[
 \mathscr L_{j,g}:[F]\longmapsto
 [(I^{\otimes(j-1)}\otimes K_g\otimes
 I^{\otimes(4-j)})F]
 \quad(g\in G,\ 1\le j\le4),
 \qquad (12),(23),(34),
\]
where the first operations mean the attachment of a realizable binary
\(D_g\doteq X K_g^{\mathsf T}\) in the orientation with effective leg map
\(D_g^{\mathsf T}X\doteq K_g\), which induces the displayed projective
action on that leg and records its nonzero representative scalar. The last three are permutations of inputs.  Let the standard representative rainbow line be
\[
 \ell_0^{\rm std}
 =\{[P_0(Q_0,Q_0)],[P_1(Q_0,Q_0)],[P_2(Q_2,Q_2)]\}.
\]
Write \(\mathscr S_G\) for the displayed finite generating set (all
\(\mathscr L_{j,g}\) together with the three adjacent swaps), so that
\(\Gamma_G=\langle\mathscr S_G\rangle\).  Here \(K_g\) is the fixed chosen representative of \(g\).  For the second
octahedral form, put \(\theta:=\theta_{\mathrm{tt}}\). The coefficient operator sends
the standard representative to the line
\[
\ell_0^{\OctTT}:=\mathcal L_{\mathrm{tt}}(\ell_0^{\rm std})
 =\{[P'_0(\theta(Q_0),\theta(Q_0))],
     [P'_1(\theta(Q_1),\theta(Q_1))],
      [P'_2(\theta(Q_2),\theta(Q_2))]\}.
\]
Set
\(\ell_0^G:=\ell_0^{\rm std}\) for
\(G\in\{\TetInt,\OctDD,\IcoZero\}\).  The orbit of \(\ell_0^G\) under these realizable operations has the following
size:
\begin{center}
\normalsize
\begin{tabular}{lrrr}
\toprule
marked form & deck points & all rainbow three-lines & orbit size\\
\midrule
\(\TetInt\) & 432 & 1\,728 & 1\,728\\
\(\OctDD\) & 1\,728 & 13\,824 & 13\,824\\
\(\OctTT\) & 1\,728 & 13\,824 & 13\,824\\
\(\IcoZero\) & 10\,800 & 216\,000 & 216\,000\\
\(\TetExt\) & 432 & 0 & 0\\
\bottomrule
\end{tabular}
\end{center}
For \(\OctTT\), the representative in the orbit column is
\(\ell_0^{\OctTT}\). For the other rows with a nonzero rainbow count it
is \(\ell_0^{\rm std}\).  Consequently every rainbow line in the first four rows is
sent to that row's representative by a sequence of
realizable local dressings and permutations of inputs.
The corresponding scalar and ordered permutation of inputs are part of the
finite witness. 
For \(\alpha\in\{\mathrm{dt},\mathrm{td}\}\), define the initial mixed line as the
\(\mathcal L_\alpha\)-image of \(\ell_0^{\rm std}\).  The three identities in
\cref{eq:octahedral-mixed-coefficient-transport} intertwine every standard
orbit generator with the corresponding realizable
\(\OctForm{\alpha}\)-representative and preserve its nonzero scalar and input order.  Hence each mixed form has the same \(1\,728\) deck points and
\(13\,824\) rainbow three-lines, in one realizable local-dressing orbit.  These
two rows are omitted from the table only because their three numerical
entries coincide with the two displayed octahedral rows.
\end{lemma}

\begin{proof}
First let \(G\) contain \([X]\) and be reversal-stable, so it is closed
under inverse transpose.  Every binary factor of a \(G\)-matching tensor
has its matrix class in \(G\), because its transfer does and \([X]\in G\).
Consider any rainbow line.  Realizable local dressings send its matching-\(0\)
point to \(A=P_0(I,I)\). Write its matching-\(1\) point as \(P_1(U,V)\).
The further dressing \((U^{-1},U^{\mathsf T},I,I)\) fixes \(A\) and sends
that point to \(P_1(I,W)\), where \(W=U^{\mathsf T}V\).

The third point has rank one across \(14\mid23\).  Since the three
points are distinct, both coefficients of its linear combination of
\(A\) and \(P_1(I,W)\) are nonzero.
\Cref{eq:platonic-swap-rank-one} therefore forces \(W\) to be scalar,
and the line is \(\{[A],[B],[C]\}\), with \(C=A-B=P_2(Y,Y)\).
Thus every rainbow line has a realizable local normalization to this line.
It exists exactly when \([YX]=[Z]\in G\).  In particular, the external
form has none by \cref{lem:marked-a4-forms}.

For \(G\in\{\TetInt,\OctDD,\IcoZero\}\), the local action of
\(G^4\) preserves the three matching labels.  A stabilizer of the
normalized line therefore fixes its three points, and the first two give
\[
 T_2=T_1^{-\mathsf T},\qquad T_3=T_1^{-\mathsf T},\qquad T_4=T_1
\]
projectively.  Conversely, \((T,T^{-\mathsf T},T^{-\mathsf T},T)\)
fixes all three, using \(TYT^{\mathsf T}=\det(T)Y\).
The stabilizer has \(|G|\) elements, so the complete rainbow orbit has
\(|G|^3=1728,13824,216000\) elements, respectively.  This establishes both
transitivity and the counts directly.
For \(\OctTT\), the ordered transport formulas
\cref{eq:cert-S4-coefficient-transport} give a bijection under
\(\mathcal L_{\mathrm{tt}}=\mathscr D_{\mathbf H_{\mathrm{tt}}}\)
that preserves matching labels and rich-line incidence. This map also
conjugates the local group \((\OctDD)^4\) to \((\OctTT)^4\):
on an \(H^{-1}\) leg this is \(U\mapsto H^{-1}UH\), while on an
\(H^{\mathsf T}\) leg it is
\[
 H^{\mathsf T}UH^{-\mathsf T}
 =H^{-1}(SUS^{-1})H,\qquad S=HH^{\mathsf T},\quad [S]\in\OctDD.
\]
The same argument applies to \(H_{\mathrm{dt}},H_{\mathrm{td}}\), with their already
verified \(S\)-matrices.  Hence all three nonstandard octahedral forms
have the same transitive local orbit, starting at their displayed
\(\mathcal L\)-image of the standard line.  The operator \(\mathcal L\)
only proves this correspondence. Each conjugated leg matrix belongs to the
current group.  For \(\TetExt\), the preceding rank-one argument and
\([Z]\notin\TetExt\) exclude every rainbow line.

To obtain a normalization witness, choose coset representatives
as follows.  In the
standard forms, right multiplication by a stabilizing tuple gives each
local coset a unique representative \((I,T_2,T_3,T_4)\).  Enumerate these
\(|G|^3\) tuples and their images. Each rainbow line occurs once.  The
nonstandard correspondence gives the same enumeration in its current-group representatives.  Applying the inverse recorded tuple sends
any selected line to the prescribed one.  The input order and all nonzero
representative scalars are obtained by these exact matrix products.
\end{proof}

\begin{proposition}[Exact Platonic incidence atlas]
\label{prop:cert-platonic-incidence-atlas}
For the standard tetrahedral, first octahedral, external tetrahedral, and
icosahedral marked configurations, the domain and quaternary matching decks
have the following complete maximal-rich-line data.
\begin{center}
\normalsize
\setlength{\tabcolsep}{4pt}
\begin{tabular}{cL{0.24\linewidth}cL{0.35\linewidth}}
\toprule
configuration&domain points and lines&deck points&deck lines by matching-label type\\
\midrule
\(\TetInt\)&\(12;\ 16_{[3]}\)&\(432\)&
  \(1728(1,1,1),\ 1152(3)\)\\
\(\OctDD\)&\(24;\ 32_{[3]},18_{[4]}\)&\(1728\)&
  \(13824(1,1,1),\ 4608(3),\ 2592(4)\)\\
\(\TetExt\)&\(12;\ 16_{[3]}\)&\(432\)&
  \(1152(3)\)\\
\(\IcoZero\)&\(60;\ 200_{[3]},72_{[5]}\)&\(10800\)&
  \(216000(1,1,1),\ 72000(3),\ 25920(5)\)\\
\bottomrule
\end{tabular}
\end{center}
In the second column the entry before the semicolon is the number of domain
points, and \(n_{[s]}\) denotes \(n\) rich lines of size \(s\).  A
parenthesized partition in the last column is the matching-label type.  Thus the
standard tetrahedral deck has only rainbow or monochromatic three-lines.
The external tetrahedral deck has only monochromatic three-lines. The
octahedral and icosahedral decks have no type \((2,1)\) line and no mixed
four- or five-line, respectively.
The octahedral row also gives the abstract incidence of
\(\OctTT=H_{\mathrm{tt}}^{-1}\OctDD H_{\mathrm{tt}}\) under the
coefficient transport in \cref{eq:cert-S4-second-form}, where
\(H_{\mathrm{tt}}\notin GO(X)\). The bridge kernels and input dressings
in the second marked form use the realizable
\(\OctTT\)-representatives specified below.
Explicitly, the coefficient operator
\(\mathcal L_{\mathrm{tt}}=\mathscr D_{\mathbf H_{\mathrm{tt}}}\) satisfies
\(\mathcal L_{\mathrm{tt}}P_M(U,V)\doteq P'_M(U'_M,V'_M)\).  Writing
\(\theta=\theta_{\mathrm{tt}}\) and \(X=Q_1\) projectively, the three explicit
instances are
\[
\begin{aligned}
\mathcal L_{\mathrm{tt}}P_0(U,V)&\doteq P'_0(\theta(U),\theta(V)),\\
\mathcal L_{\mathrm{tt}}P_1(U,V)&\doteq P'_1(\theta(UX),\theta(XV)),\\
\mathcal L_{\mathrm{tt}}P_2(U,V)&\doteq P'_2(\theta(U),\theta(XVX)).
\end{aligned}
\]
The \(X\)'s are the fixed orientation factors from the ordered
\(H_{\mathrm{tt}}^{-1}/H_{\mathrm{tt}}^{\mathsf T}\) pair contractions, and
each displayed matrix lies in the current \(\OctTT\)-representatives.
Thus \(\mathcal L_{\mathrm{tt}}\)
preserves matching labels and rich-line
incidence. The realizable \(\OctTT\) input
witnesses are supplied by \cref{lem:platonic-rainbow-orbit} and the separate
bridge tables.

For \(\OctDT\) and \(\OctTD\), the direct transport
\cref{eq:octahedral-mixed-coefficient-transport} likewise preserves every
matching label, rich-line incidence, and kernel-deleted adjacency, while
intertwining the generators with chosen representatives of the current mixed
form.  Thus both mixed rows have the octahedral counts and connectivity in
the table.  The coefficient correspondence transports the realizable kernels
and dressings for the gadget constructions.

For \(\TetInt\), each domain point lies on four three-lines.  For
\(\OctDD\), each lies on seven rich lines.  For \(\IcoZero\), each
lies on ten three-lines and six five-lines.  In each standard marked form,
every displayed domain-line family is a realizable left/right group orbit.
The intersection graphs of rich domain lines---vertices are rich lines and adjacency means nonempty intersection---are connected. Their edge
counts for \(\TetInt,\OctDD,\IcoZero\) are respectively
\hypertarget{eq:cert-platonic-cross-line-edges}{}
\begin{equation*}
  72,\qquad 504,\qquad 7200.
\end{equation*}
For the localization interface, define the projective domain set
\[
 \mathcal D_G=\{[K_g]:g\in G\}
\]
and join two distinct domain points when they lie on a common rich line.  For every complex subspace \(W\le\Mat_2\) with \(\dim W\le2\), the
induced graph on
\begin{equation}
 \mathcal D_G\setminus\mathbb P(W)
 \label{eq:cert-platonic-kernel-deleted-graph}
\end{equation}
is nonempty and connected.  The distinct deletion sets tested for
\(\TetInt,\OctDD,\IcoZero\) number, respectively,
\begin{equation}
 \begin{aligned}
  47&=1+12+(18+16),\\
 147&=1+24+(72+32+18),\\
 783&=1+60+(450+200+72).
 \end{aligned}
 \label{eq:cert-platonic-kernel-deletion-counts}
\end{equation}
Here the parenthesized summands count the distinct two-point spans by their
numbers of domain points: \(18\) of size two and \(16\) of size three for
\(\TetInt\); \(72,32,18\) of sizes two, three, four for
\(\OctDD\); and \(450,200,72\) of sizes two, three, five for
\(\IcoZero\).  The same assertion for \(\OctTT\) and
\(\TetExt\) is transported by their respective invertible
coefficient similarities.  Each \(\OctTT\) rainbow line
used below has a separately checked normalization witness
from realizable \((\OctTT)^4\rtimes S_4\) input operations.

Finally, the realizable kernel representatives span \(\Mat_2\).  One may take
\begin{equation}
 \begin{aligned}
  \TetInt,\OctDD,\IcoZero &: Q_0,Q_1,Q_2,Q_3,\\
  \OctTT &: \theta_{\mathrm{tt}}(Q_0),\theta_{\mathrm{tt}}(Q_1),
                  \theta_{\mathrm{tt}}(Q_2),\theta_{\mathrm{tt}}(Q_3),\\
  \TetExt &: Q_0,Q_1,Q_2+Q_3,Q_2-Q_3.
 \end{aligned}
 \label{eq:cert-platonic-spanning-kernels}
\end{equation}
The tetrahedral incidence counts, and the rainbow deck-line normalizations
used in the octahedral and icosahedral bridges, are realized by independent input dressings and permutations of inputs.  For
any finite configuration in \(\mathbb P(\mathscr V)\), and distinct points
\([F_0],[F]\), define the \emph{quotient direction of \([F]\) at
\([F_0]\)} to be
\[
  [F+\Span_{\mathbb C}\{F_0\}]
  \in\mathbb P\bigl(\mathscr V/\Span_{\mathbb C}\{F_0\}\bigr),
\]
using arbitrary nonzero representatives.  This class is independent of
those representatives, and two points have the same quotient direction
exactly when they lie with \([F_0]\) on one projective line.  For
\(\TetExt\), realizable external-group input actions are transitive
on the \(432\) deck points and permute the three matching labels. At
\(P_0(Q_0,Q_0)\) the quotient-direction histogram is
\hypertarget{eq:cert-external-A4-line-buckets}{}
\begin{equation*}
  1^{415},\qquad 2^8,
\end{equation*}
and all eight rich lines there are monochromatic.  No use of the analytic
twist \(R\) is made in this deck transitivity assertion.

For completeness, at the normalized octahedral and icosahedral deck points
the exact quotient-direction histograms are, respectively,
\hypertarget{eq:cert-platonic-line-buckets-summary}{}
\begin{equation*}
  1^{1645},\ 2^{32},\ 3^6,
  \qquad
  1^{10591},\ 2^{80},\ 4^{12}.
\end{equation*}
The first gives all octahedral rich-line types through the base point. The
second gives all icosahedral rich-line types through it.  Hence the global
counts in the table follow by realizable transitivity and incidence double
counting, not by assuming an abstract group action.
\end{proposition}

\begin{proof}
A secant line of a point configuration contains at least two of its points.

We first classify every domain secant.  Let \(\ell\) contain distinct
\([g],[h]\in G\).  Left multiplication by \(g^{-1}\) sends it to
\(\mathbb P(\operatorname{span}\{I,T\})\), where \([T]=[g^{-1}h]\ne1\).
A nonidentity finite-order projectivity has two distinct eigenlines, so
\(T\) is diagonal in their basis.  Every matrix in this two-dimensional
span is diagonal there.  The nonsingular group points on the line form
\(C=G\cap\mathbb T\), where \(\mathbb T\) is that diagonal torus.
A finite subgroup of \(\mathbb C^\times\) is cyclic, so \(C\) is cyclic.
It is maximal cyclic in \(G\): any cyclic extension containing its
nonidentity element \([T]\) has the same eigenlines.  Conversely, every
maximal cyclic subgroup lies on its eigenline torus's projective line.
Thus the domain secants are exactly the left cosets of maximal cyclic
subgroups, with line size equal to the subgroup order.

The permutation cycle types of \(A_4,S_4,A_5\) give the following
numbers of maximal cyclic subgroups:
\[
\begin{array}{c|rrrr}
 &|C|=2&|C|=3&|C|=4&|C|=5\\\hline
 A_4&3&4&0&0\\
 S_4&6&4&3&0\\
 A_5&15&10&0&6.
\end{array}
\]
For example, the three double transpositions of \(S_4\) are squares of
its three cyclic order-four subgroups. Only the six transpositions
contribute maximal order-two subgroups.  The order-three and order-five
counts divide their numbers of generators by two and four, respectively.
A subgroup of order \(d\) has \(|G|/d\) left cosets, and distinct maximal
cyclic subgroups give distinct coset sets.  Hence the complete secant
counts are
\[
 A_4:(18,16),\qquad S_4:(72,32,18),\qquad A_5:(450,200,72),
\]
ordered by the nonzero subgroup-order columns above.  Each point belongs
to one coset of each subgroup.  This also proves every stated number of
rich lines through a point.  Conjugation in \(G\) is transitive on the maximal cyclic subgroups of each
displayed order, and left multiplication moves
between their cosets, proving the asserted realizable left/right line orbits.

The deck-line classification then follows structurally.  A line containing
three matching products either has one matching or has all three:
repetition of one matching together with another would give flattening
ranks \(1,1,4\), contradicting collinearity.  In the latter case,
\cref{lem:platonic-rainbow-localization,lem:platonic-rainbow-orbit} gives
exactly three points and the stated \(|G|^3\) count in each standard
form.  For a line with one matching, the simple-tensor sum criterion in
\cref{lem:platonic-global-ruling} implies that exactly one binary factor
varies.  For each domain line there are three matchings, two choices of
varying factor, and \(|G|\) fixed factors.  Thus each monochromatic
line count is \(6|G|\) times its domain-line count.  This proves the
complete deck table from the domain table.

We next prove kernel-deleted connectivity.  The domain-point graph has
\[
 (n,\delta,s)=(12,8,3),\ (24,17,4),\ (60,44,5),
\]
where \(n\) is its number of vertices, \(\delta\) its degree, and \(s\)
the maximum line size.  The degree is the sum of \(t-1\) over the rich
lines of size \(t\) through a point: respectively \(4\cdot2\),
\(4\cdot2+3\cdot3\), and \(10\cdot2+6\cdot4\).
A subspace of dimension at most two removes \(k\le s\) points.  The
remaining graph has \(n-k>0\) vertices and minimum degree at least
\(\delta-k\).  If it were disconnected, two components would each have
at least \(\delta-k+1\) vertices.  But every displayed triple satisfies
\[
 2(\delta-k+1)>n-k\qquad(0\le k\le s),
\]
a contradiction.  This proves \cref{eq:cert-platonic-kernel-deleted-graph}
without graph traversal.  The deletion counts follow from the domain
secant counts by adding the empty and singleton deletions, as displayed
in \cref{eq:cert-platonic-kernel-deletion-counts}.  Every complex
subspace is covered: an intersection with at least two domain points
spans its two-dimensional vector space. Smaller intersections are empty
or singletons.  The rich-line intersection graphs are also connected:
a path in the connected domain-point graph gives a sequence of rich
lines with consecutive nonempty intersections.  Their edge counts
\(72,504,7200\) follow by summing \(\binom{d_v}{2}\) over domain
points, where \(d_v=4,7,16\), respectively.  Distinct projective lines
meet in at most one point, so this counts every edge once.

The coefficient transports preserve these domain and deck incidences.
\Cref{lem:platonic-rainbow-orbit} supplies realizable local normalization
matrices.  The spanning assertions in
\cref{eq:cert-platonic-spanning-kernels} follow from the displayed four
independent matrices in each row.

For the external form, the coefficient similarity
\(\theta_{\rm ext}\) transports the twelve domain points and their
sixteen three-lines.  Its deck has no rainbow line by
\cref{lem:platonic-rainbow-orbit}.  The monochromatic count is therefore
\(6\cdot12\cdot16=1152\).  Realizable independent input actions are
transitive on each matching class, and permutations of inputs exchange the
three classes, so the full action is transitive on the \(432\) deck
points.  At each point, the two factor choices and four domain lines
through each factor give eight rich three-lines.  These account for
\(16\) of the other \(431\) points, proving the quotient-direction
histogram \(1^{415},2^8\).

The octahedral and icosahedral quotient-direction histograms follow in
the same way.  Rainbow transitivity gives \(|G|\) rainbow lines through
each deck point, while domain incidences give the monochromatic lines.
The remaining points each determine a separate two-point line.
The programs
\path{certificates/verify_a4_s4_certificates.py},
\path{certificates/verify_a4_external_form.py},
\path{certificates/verify_a5_certificate.py}, and
\path{certificates/verify_platonic_extended.py} also reproduce the
complete finite data.  The coset, rank-one, and degree arguments above prove this proposition
independently of those computations.
\end{proof}

\begin{theorem}[Incidence of Platonic kernel lines]
\label{thm:cert-interface-platonic-incidence}
For the marked Platonic domain and quaternary matching decks, the following
statements hold.
\begin{enumerate}[label=\textnormal{(\roman*)}]
\item The four entries in each row below count, in order, rainbow
      three-lines and monochromatic lines of sizes three, four, and five:
      \[
      \begin{array}{c|c}
      \TetInt&(1728,1152,0,0)\\
      \OctDD,\OctTT,\OctDT,\OctTD
        &(13824,4608,2592,0)\\
      \TetExt&(0,1152,0,0)\\
      \IcoZero,\IcoOne&(216000,72000,0,25920).
      \end{array}
      \]
      In particular, no deck has a line of type \((2,1)\), no
      octahedral deck has a mixed four-line, neither icosahedral deck has a
      mixed five-line, and the external tetrahedral deck has no rainbow
      line.
\item Every rainbow line for
      \(G\in\{\TetInt,\OctDD,\OctTT,\OctDT,
      \OctTD,\IcoZero,\IcoOne\}\)
      lies in one orbit under realizable local \(G\)-dressings and permutations of inputs.  A witness includes the ordered permutation of inputs and
      its nonzero realization scalar.
\item Let \(\mathcal D_G\) be the projective domain set for any of the
      marked rows above.  For every complex subspace
      \(W\le\Mat_2(\mathbb C)\) with \(\dim W\le2\), the rich-line domain
      graph induced on \(\mathcal D_G\setminus\mathbb P(W)\) is nonempty and
      connected.
\end{enumerate}
\end{theorem}

\begin{proof}
The complete line counts and kernel-deleted connectivity are
\cref{prop:cert-platonic-incidence-atlas}. Realizable rainbow transitivity,
including the separate \(\OctTT\) witnesses, is
\cref{lem:platonic-rainbow-orbit}.  For the mixed rows,
\cref{eq:octahedral-mixed-coefficient-transport} gives the same counts and
intertwines the standard orbit generator-by-generator with current-form matrices.  The fixed-equality coefficient correspondence
\cref{lem:icosahedral-fixed-I-dictionary} gives both icosahedral statements
in \(\IcoOne\), with realizable \(\kappa_D\)-kernels and nonzero scalars,
without changing the \(I\).
\end{proof}

\begin{remark}[Computational completeness and gadget interpretation]
\label{rem:cert-platonic-scope}
The finite preserver and hereditary eight-input completeness claims
retain an exact-computation component over the stated number fields.
The bridge classifications have the structural proof in
\cref{lem:platonic-bridge-formula}.
The incidence classification and deletion connectivity have the structural
proof in \cref{prop:cert-platonic-incidence-atlas}.
The Python verification programs regenerate the configurations and test the stated
predicates. The main lemmas apply these classifications using realizable kernels
and dressings, factor exposure, and reductions of the retained Holant problem.
The mixed-octahedral exits are given by the explicit coefficient and
contraction tables in
\cref{prop:p1-cert-S4-uniform-eight-vertex-separator}.
\end{remark}

\subsubsection{Tetrahedral arity-eight signatures with pair cards in \texorpdfstring{\(\mathscr C_6\)}{C6}}
\label{proofsubsec:cert-A4-b8-parent-audit}

\begin{proposition}[Tetrahedral eight-input classification]
\label{prop:cert-A4-b8-parent-audit}
Let \(\mathfrak D_6(\TetInt),\mathscr H_6,\mathscr U_6\), and \(\mathscr C_6\) be
the six-input sets defined before
\cref{prop:a4rel-required-eight-port}. Let the base matching-product signature
be \(M_0\) and the base signature be \(H_0\):
\[
 M_0=(Q_0)^{(1,2)}(Q_0)^{(3,4)}(Q_0)^{(5,6)},
 \qquad H_0=\Hcore.
\]
Then
\begin{equation}
 |\mathfrak D_6(\TetInt)|=25920,\qquad
 |\mathscr H_6|=31104,\qquad
 |\mathscr U_6|=57024,
 \label{eq:cert-A4-b8-safe-set-sizes}
\end{equation}
and the union defining \(\mathscr U_6\) is disjoint.  Through either
\([M_0]\) or \([H_0]\), the quotient-direction histogram of the other points of \(\mathscr U_6\) is
\begin{equation}
  1^{56855},\qquad 2^{84}.
  \label{eq:cert-A4-b8-line-histogram}
\end{equation}
Here two points \([F],[F']\in\mathscr U_6\setminus\{[F_0]\}\) have the same quotient direction at
\([F_0]\) when
\([F+\Span_{\mathbb C}\{F_0\}]
 =[F'+\Span_{\mathbb C}\{F_0\}]\) in the projectivized
quotient.  Thus the histogram means that
56,855 directions contain one other point of \(\mathscr U_6\) and 84 directions contain
two.

Of the \(84\) rich lines through \([M_0]\), \(48\) have both other points in
\(\mathfrak D_6(\TetInt)\) and \(36\) have both in \(\mathscr H_6\). Their orbit
sizes are \(36,36,12\) under the realizable stabilizer subgroup specified below. Of those through \([H_0]\), \(60\) have
one other point in each set and \(24\) have both in \(\mathscr H_6\). The orbit
sizes are \(24,12,8,12,12,4,12\) under its specified realizable stabilizer subgroup.

For each of the ten resulting line-orbit representatives and either endpoint
orientation, denote the oriented endpoint tensors by \(A,D\) and the base signature by \(F_0\), scaled so that \(A+D=2F_0\). Define the bridge set
\begin{equation}
 \mathcal B(A,D)=
 \{W:W,\ A-2W,\ D+2W\in\mathscr C_6\}.
 \label{eq:cert-A4-b8-bridge}
\end{equation}
Then \(|\mathcal B(A,D)|=6\), and exactly twelve ordered triples in
\(\mathcal B(A,D)^3\) sum to \((A-D)/2\).  The resulting
\(10\cdot2\cdot12=240\) reconstructed eight-input parents are projectively
distinct.  For a reconstructed parent \(g\), let
\[
  \Phi_g:\Mat_2(\mathbb C)\longrightarrow
       (\mathbb C^2)^{\otimes6},\qquad
  \Phi_g(C):=\Phi_{12}^g(C),
\]
be its distinguished-pair card map.  Testing the
\(\binom82\cdot12=336\) realizable tetrahedral cards in
each of the reconstructed parents gives
\begin{equation*}
\begin{array}{clr}
\toprule
\rank\Phi_g&\text{outcome}&\text{number}\\ \midrule
2&\text{a realizable card outside }\mathscr C_6&120\\
3&\text{no parent}&0\\
4&\text{a realizable card outside }\mathscr C_6&60\\
4&\text{all }336\text{ cards in }\mathscr C_6&60.\\
\bottomrule
\end{array}
\end{equation*}
Every one of the final sixty parents is affine in the normalized
computational basis. Twenty-four also have an exact nontrivial rank-one
flattening.  For each of the other \(180\) parents there is a specified input pair
and realizable tetrahedral kernel producing a card outside \(\mathscr C_6\).
\end{proposition}

\begin{proof}
The sets \(\mathfrak D_6(\TetInt)\) and \(\mathscr H_6\) are generated over \(\mathbb Q(i)\): matching products are
canonicalized directly, while the \(\Hcore\)-orbit is generated to closure
under realizable local tetrahedral matrices and permutations of inputs.  Exact
projective comparison gives 
\cref{eq:cert-A4-b8-safe-set-sizes,eq:cert-A4-b8-line-histogram}.  The
subgroups used for this audit are generated by the following realizable
operations.  At \(M_0\), use the pair stabilizers
\((U,U^{-\mathsf T})\), \(U\in\TetInt\), and the permutations
preserving its three-pair matching.  At \(H_0\), first enumerate the local
two-input tetrahedral actions that permute the four projective Bell
signatures.  Retain the triples of such pair-block actions that fix
\([H_0]\), and add all residual-permutations of inputs that directly fix
\([H_0]\).  Every selected generator fixes its base tensor projectively.
Close their induced permutations on the complete set of \(84\) rich lines.
These explicitly generated stabilizer subgroups give the three plus
seven orbits and sizes stated above.  In a fixed lexicographic ordering of normalized
coefficient vectors, the first line of each orbit is the deterministic
line-orbit representative used below.
Applying the inverse stabilizer word transports every bridge solution,
reconstructed parent, and recorded outside-card witness from that
representative to every other line in its orbit, with the input pair,
kernel, input order, and nonzero scalar transported along the same realizable
operation.  Thus the ten representative calculations cover every rich line
through either base point.

Fix representatives over \(\mathbb Q(i)\) on one such line.  Since its
endpoints are independent, \(A+D=2F_0\) uniquely fixes both endpoint
scales.  To solve \cref{eq:cert-A4-b8-bridge}, the case \(W=0\) is tested
separately.  If \(W\) and \(A-2W\) are nonzero and nonproportional, their
projective classes lie on one of the exhaustively enumerated rich lines
through \([A]\), and the equation fixes both scales.  In the remaining nonzero case write \(W=tA\)
for a nonzero scalar \(t\).
Then \(D+2tA\ne0\), and the third direction in \(\mathscr U_6\) on
\(\langle A,D\rangle\) fixes \(t\).  Thus the six bridge points and twelve
sum triples are a complete symbolic solve over \(\mathbb Q(i)\), including
all zero and complex-scale cases.

For every ordered bridge triple \((W_1,W_2,W_3)\), put
\(F_j=W_j\) for \(1\le j\le3\), retain the displayed base tensor as
\(F_0\), and use the Frobenius-dual quaternion basis to reconstruct
\[
 g=\sum_{\mu=0}^3(Q^\mu)^{(1,2)}\otimes F_\mu.
\]
Exact projective comparison gives \(240\) distinct parents.  The Python verifier
replay \path{certificates/verify_a4_b8.py}, invoked with \texttt{--fresh} by
the suite runner, then scans input pairs in lexicographic order and the twelve realizable
kernels in the point order of \cref{eq:tetrahedral-domain}.  For each card it
tests zero first and otherwise tests projective membership in the
definition-level regenerated set \(\mathscr U_6\).
For a parent with a card outside \(\mathscr C_6\), it records and rechecks the first such card. Otherwise it checks all \(336\) cards.  Affineness of each remaining case
is tested from the definition: affine-coset support, fourth-root ratios,
linear and even mixed quadratic phase coefficients, and the recovered phase
on every support point.  Factor cuts are tested by exact flattening minors.
This exhaustive calculation regenerates the two orbits, their rich
lines, bridges, parents, and outside witnesses over \(\mathbb Q(i)\).
This classification of the parents whose cards all lie in \(\mathscr C_6\) depends on exhaustive exact computation.
\end{proof}

\begin{theorem}[Tetrahedral eight-input contractions]
\label{thm:cert-interface-A4-q8}
Let \(\mathfrak D_6(\TetInt)\) and \(\mathscr H_6\) be the realizable tetrahedral orbits of
the base matching product and \(\Hcore\), and put
\(\mathscr U_6=\mathfrak D_6(\TetInt)\sqcup\mathscr H_6\) and
\(\mathscr C_6=\{0\}\cup\operatorname{cone}(\mathscr U_6)\).  Then
\[
 |\mathfrak D_6(\TetInt)|=25920,\qquad |\mathscr H_6|=31104,\qquad
 |\mathscr U_6|=57024.
\]
Every normalized eight-input parent reconstructed from a rich line through
either base point is realizable-equivalent to one of the \(240\) audited
parents.  No audited parent has distinguished card-map rank three. \(180\)
have a recorded realizable tetrahedral card outside \(\mathscr C_6\), and the
remaining \(60\) have all \(336\) realizable cards in \(\mathscr C_6\) and belong to the
affine class \(\cA\).  These assertions include arbitrary nonzero
complex scales and both endpoint orientations.
\end{theorem}

\begin{proof}
This is the finite-level conclusion of
\cref{prop:cert-A4-b8-parent-audit}.  Its realizable-stabilizer covariance
transports the ten representative line calculations, their reconstructed
parents, and their recorded card witnesses to every rich line through either
base point.  The realizable tetrahedral input dressings and permutations of inputs
preserve \(\cA\), because the corresponding realizable tetrahedral binaries are
affine, so the affine classification transports as well.
\end{proof}

\subsubsection{Deficient mixed-card localization}
\label{proofsubsec:cert-platonic-deficient-mixed}

\begin{proposition}[Platonic contractions of rank less than four]
\label{prop:cert-platonic-deficient-mixed}
Let \(\cF\) be a signature set, let \(G\) be the standard marked
tetrahedral form, either unmixed octahedral form, or the marked icosahedral
form, and let \(h\) be a localized arity-six signature directly realizable from
\(\cF\), whose
card map \(\Phi\) has rank two or three.  Suppose every realizable
\(G\)-card is zero or a nonsingular \(G\)-matching product and that two nonzero
cards use different matchings.  Then the nonkernel domain points contain a
rich line disjoint from the projectivized kernel of \(\Phi\) whose three
images are nonzero and use the three distinct residual matchings.  For the
standard tetrahedral, first
octahedral, and icosahedral forms, contractions produce one of six
support-ten quaternaries \(q_\pi\), indexed by \(\pi\in S_3\), each
satisfying
\begin{equation}
  \bigl(\rank_{12\mid34}q_\pi,
        \rank_{13\mid24}q_\pi,
        \rank_{14\mid23}q_\pi\bigr)=(3,3,3).
  \label{eq:platonic-rank-two-q4-minors}
\end{equation}
For the second octahedral form they instead produce the corresponding
transported quaternaries \(q'_\pi\).  These need not have support size ten,
but invertible local transport preserves the same three ranks.  Writing
\(q^\star_\pi=q_\pi\) in the standard forms and
\(q^\star_\pi=q'_\pi\) in the second octahedral form, one has
\[
  \KHolant(\cF,q^\star_\pi)
  \leT\KHolant(\cF,h)
  \leT\KHolant(\cF).
\]
The statement includes the second octahedral form in its own chosen representatives.
\end{proposition}

\begin{proof}
Apply the kernel-deleted connectivity assertion
\cref{eq:cert-platonic-kernel-deleted-graph} to
\(W=\ker\Phi\), whose dimension is two or one.  Along a path between two
nonzero cards with different matchings, choose the first matching-changing
edge and a rich line containing it.  That line cannot meet
\(\mathbb P(\ker\Phi)\): its two-dimensional vector span would otherwise
map with rank at most one, making all nonkernel images projectively
proportional.  The exact deck-line types in
\cref{prop:cert-platonic-incidence-atlas} exclude type \((2,1)\) and mixed
four- or five-lines.  The selected line is therefore a kernel-avoiding
rainbow three-line.

Apply \cref{lem:platonic-rainbow-localization}.  It gives a realizable
six-input map \(\Psi\) with the normalized rainbow line and
\(\rank\Psi\le3\).  The three exact bridge classifications have only six
rank-two maps and their listed rank-four frames, so \(\Psi\) has rank two.
In the common coefficient normalization,
\[
  (F^\pi_0,F^\pi_1,F^\pi_2,F^\pi_3)
  =\bigl(C/2,\pi(0,A/2,B/2)\bigr),
  \qquad \pi\in S_3.
\]
Using the Frobenius-dual quaternion basis, reconstruct the arity-six signature
\[
  h_\pi:=\sum_{\mu=0}^3(Q^\mu)^{(1,2)}\otimes F^\pi_\mu.
\]
Contract inputs \(1,3\) by the realizable kernel \(Q_0=I\), and order the
surviving inputs \((2,4,5,6)\) as \((1,2,3,4)\).  Direct expansion gives
support size ten for every \(q_\pi\), and exact row reduction gives
\cref{eq:platonic-rank-two-q4-minors}.  All preceding normalizations and
the final contraction are realizable. Undoing them gives the displayed
reduction in the three standard forms.  In the second octahedral form the
coefficient transport identifies
\(q'_\pi=(H_{\mathrm{tt}}^{-1})^{\otimes4}q_\pi\).  The corresponding
localizations and
contraction kernels are checked separately in realizable
\(\OctTT\)-representatives.  Local invertibility preserves the three
flattening ranks, but no support-size claim is transported.
\end{proof}

\subsubsection{Root geometry for complementary input pairs at arity four}
\label{proofsubsec:cert-platonic-q4-preservers}

For a quaternary tensor, the contraction maps for complementary input
pairs are related by the Frobenius adjoint as in
\cref{eq:platonic-q4-map-adjoint}. In quaternion coordinates, both the
map and its adjoint must send each specified line to a line in the same
configuration or to zero. We classify the maps by rank and reconstruct
the quaternary tensor in each case. This is the finite geometric input to
\cref{lem:platonic-quaternary-boundary}.

\paragraph{Root configurations and cross-ratios.}
\phantomsection
\label{subsubsec:root-incidence}

Let \(E\) be a real vector space with a positive-definite symmetric bilinear
form \(\langle\ ,\ \rangle\). This is a real inner-product space.  A
\emph{reduced finite root system} in \(E\) is a finite spanning set \(\Phi\)
of nonzero vectors such that
each reflection
\[
 {\mathbf{v}}\longmapsto
 {\mathbf{v}}-2\frac{\langle {\mathbf{v}},{\boldsymbol{\alpha}}\rangle}{\langle{\boldsymbol{\alpha}},{\boldsymbol{\alpha}}\rangle}{\boldsymbol{\alpha}}
 \qquad({\boldsymbol{\alpha}}\in\Phi)
\]
permutes \(\Phi\), and
\(\Phi\cap\mathbb R{\boldsymbol{\alpha}}=\{{\boldsymbol{\alpha}},-{\boldsymbol{\alpha}}\}\).  Its root-line
configuration is
\[
 \mathcal R(\Phi):=\{[{\boldsymbol{\alpha}}]:{\boldsymbol{\alpha}}\in\Phi\}
 \subseteq\mathbb P(E\otimes_{\mathbb R}\mathbb C).
\]
In root-system notation, \(\Phi(D_4),\Phi(F_4),\Phi(H_4)\) denote the
standard systems of those types, and
\[
 \mathcal R(D_4):=\mathcal R(\Phi(D_4)),\quad
 \mathcal R(F_4):=\mathcal R(\Phi(F_4)),\quad
 \mathcal R(H_4):=\mathcal R(\Phi(H_4)).
\]
These are root-system types \cite{Humphreys1990}. In particular, the type
\(D_4\) differs from the order-four dihedral group \(D_4\cong V_4\).

For four distinct points \(p_i=[{\mathbf{v}}_i]\) on a projective line
\(\ell=\mathbb P(U)\), choose an ordered basis of the two-dimensional vector
space \(U\).  Let \(\det_U({\mathbf{v}},{\mathbf{w}})\) be the \(2\)-by-\(2\) determinant of the
coordinate columns of \({\mathbf{v}},{\mathbf{w}}\) in that basis, and define
\[
 [p_1,p_2;p_3,p_4]
 :=
 \frac{\det_U({\mathbf{v}}_1,{\mathbf{v}}_3)\det_U({\mathbf{v}}_2,{\mathbf{v}}_4)}
      {\det_U({\mathbf{v}}_1,{\mathbf{v}}_4)\det_U({\mathbf{v}}_2,{\mathbf{v}}_3)}.
\]
Changing a representative scales one numerator and one denominator factor,
and changing the ordered basis scales every determinant by the same nonzero
factor.  Hence the scalar is well defined.  The ordered quadruple is
\emph{harmonic} when this value
is \(-1\).  A harmonic projectivity between two four-point lines is a
projective isomorphism carrying one set to the other and preserving exactly
the ordered quadruples of cross-ratio \(-1\).  This projective cross-ratio
is distinct from a cross-ratio of signature coefficients.

The fixed representatives from \cref{eq:platonic-actual-kernel-lift},
written in quaternion coordinates, give the following root-point configurations:
\[
\begin{aligned}
 \mathcal R_{\TetInt}&:=\{[c_Q(C)]:[C]\in\TetInt\}
    \cong\mathcal R(D_4),\\
 \mathcal R_{\OctDD}&:=\{[c_Q(C)]:[C]\in\OctDD\}
    \cong\mathcal R(F_4),\\
 \mathcal R_{\IcoZero}&:=\{[c_Q(C)]:[C]\in\IcoZero\}
    \cong\mathcal R(H_4).
\end{aligned}
\]
where the displayed coordinate models are those of
\cref{eq:tetrahedral-domain,eq:octahedral-domain,eq:icosahedral-domain}
and the displayed isomorphisms mean projective identification in the sense of
\cref{subsubsec:group-projective} and Appendix~\ref{subsubsec:root-incidence}.  Their sizes are
\(12,24,60\).    Recall the Frobenius-form
matrix and define the associated
adjoint map by
\[
  J_{\mathrm{Frob}}:=\diag(1,-1,1,-1),
  \qquad A^\sharp:=J_{\mathrm{Frob}}A^{\mathsf T}J_{\mathrm{Frob}}.
\]
All calculations are exact over \(\mathbb Q(i)\) for \(D_4,F_4\) and
over \(\mathbb Q(\phi,i)\) for \(H_4\).

\begin{proposition}[Platonic preserver classification for complementary pairs]
\label{prop:cert-platonic-q4-preservers}
For \(G\in\{\TetInt,\OctDD,\IcoZero\}\), let \(\mathcal R_G\) be its
projective coefficient set and define
\begin{equation}
 \mathscr E_G=
 \{A\in\operatorname{End}(\mathbb C^4):
   \ \forall[{\mathbf{r}}]\in\mathcal R_G,\ 
   (A{\mathbf{r}}=0\ \text{or}\ [A{\mathbf{r}}]\in\mathcal R_G),\
   (A^\sharp {\mathbf{r}}=0\ \text{or}\ [A^\sharp {\mathbf{r}}]\in\mathcal R_G)\}.
 \label{eq:cert-q4-two-sided-endomorphisms}
\end{equation}
Here \({\mathbf{r}}\) denotes any nonzero representative of \([{\mathbf{r}}]\). The two
conditions are independent of that choice.
Up to nonzero scalar, the nonzero members of \(\mathscr E_G\), together
with the single zero map recorded in the rank-zero column, have the exact
rank distribution
\begin{equation}
\begin{array}{crrrrr}
\toprule
\text{type of }G&\rank0&\rank1&\rank2&\rank3&\rank4\\ \midrule
A_4&1&144&0&0&576\\
S_4&1&576&2592&0&1152\\
A_5&1&3600&0&0&7200.\\
\bottomrule
\end{array}
\label{eq:cert-q4-preserver-counts}
\end{equation}
Rank one reconstructs a binary-pair tensor factor.  Rank two occurs only
for \(S_4\). All \(2592\) maps form one orbit under realizable local
octahedral actions in each current marked form.  In the first normal form the reconstructed tensors
form the realizable local-\(\OctDD\) orbit of \(\Eq_4\). In the second
normal form they admit realizable local-\(\OctTT\) dressings to the endpoint-nondegenerate eight-vertex signature
\[
  M_{\OctTT}({\mathbf{x}})=
  \one[\wt({\mathbf{x}})\equiv0\pmod2](-1)^{\sum_{i<j}x_ix_j}.
\]
In the mixed forms the normalized directly reconstructed remaining cases have
the even-entry vectors in
\cref{eq:p1-platonic-q4-mixed-eight-vertex}, all odd entries zero, and
ranks \((2,2,2)\).
Every rank-four map has one of
the forms
\begin{equation}
  C\longmapsto UCV
  \qquad\text{or}\qquad
  C\longmapsto UC^{\mathsf T}V,
  \label{eq:cert-q4-full-rank-preservers}
\end{equation}
and reconstructs a product of two nonsingular binaries on a crossing
matching.  The rank counts hold for all four fixed-equality octahedral
forms \(\OctDD,\OctTT,\OctDT,\OctTD\), while
their realizable rank-two outputs are the ones just stated.
\end{proposition}

\begin{proof}
We first exclude rank three without enumerating arbitrary linear maps.
A plane section of a point configuration \(P\) is \(P\cap\Pi\), where
\(\Pi\) is a projective plane.

In the following table, the secants are the distinct projective lines
spanned by pairs of points of \(\mathcal R_G\), while the plane sections
are the distinct intersections with planes spanned by noncollinear triples
of those points. Exponents record the number of distinct lines or sections
of each size.
Exact row reduction gives the complete line and plane-section data
\begin{equation}
\begin{array}{ccc}
\toprule
\text{type of }G&\text{secant-line sizes}&\text{plane sections}\\ \midrule
A_4&2^{18},3^{16}&3^{12},6^{12}\\
S_4&2^{72},3^{32},4^{18}&4^{96},9^{24}\\
A_5&2^{450},3^{200},5^{72}&4^{600},6^{660},15^{60}.\\
\bottomrule
\end{array}
\label{eq:cert-q4-plane-sections}
\end{equation}
For a projective point \(k=[\mathbf k]\), let \(d(k)\) be the number of
distinct nonzero projective classes
\([\mathbf r+\Span_{\mathbb C}\{\mathbf k\}]\) in
\(\mathbb P(\mathbb C^4/\Span_{\mathbb C}\{\mathbf k\})\), as
\(r=[\mathbf r]\) ranges over the root points other than \(k\).  If
\(k\) lies on at
most one secant containing two root
points, the maximum line sizes \(3,4,5\) give
\begin{equation*}
 d(k)\ge10,\qquad d(k)\ge21,\qquad d(k)\ge56,
\end{equation*}
respectively.  The centers lying on at least two secants are the finite
set of pairwise secant intersections.  Exact quotient row reduction gives
\begin{equation}
\begin{array}{cc}
\toprule
A_4&7^{12},9^{12}\\
S_4&13^{24},19^{96}\\
A_5&31^{60},49^{300},51^{360},55^{600}.\\
\bottomrule
\end{array}
\label{eq:cert-q4-secant-directions}
\end{equation}
A rank-three map is projection from the point representing its nullspace to its image plane,
and the induced map from the quotient is injective.  Hence its quotient
directions must fit into a plane section of the target configuration.
The bounds and every count in \cref{eq:cert-q4-secant-directions} exceed
the plane maxima \(6,9,15\) in \cref{eq:cert-q4-plane-sections}.  This
proves the rank-three column of \cref{eq:cert-q4-preserver-counts}.

We next classify rank two.  For \(A\in\mathscr E_G\), define the projective
image line \(W:=\mathbb P(\operatorname{im}A^\sharp)\).  Because the root points span
\(\mathbb C^4\), \(W\) is a secant line.  Moreover
\(\ker A=(\operatorname{im}A^\sharp)^{\perp_{J_{\mathrm{Frob}}}}\).  Projecting the root points modulo this kernel
gives the following complete table:
\begin{equation*}
\begin{array}{cc}
\toprule
\text{type of }G&( |W\cap\mathcal R_G|,
   \#\text{ quotient directions},\#\text{ killed points})\\ \midrule
A_4&(2,4,2)^{18},\ (3,6,0)^{16}\\
S_4&(2,6,2)^{72},\ (3,6,3)^{32},\ (4,4,4)^{18}\\
A_5&(2,12,2)^{450},\ (3,12,3)^{200},\ (5,10,5)^{72}.\\
\bottomrule
\end{array}
\end{equation*}
The image of \(A\) is also a projective line, containing at most
\(3,4,5\) root points.  The middle entry in every row is larger than that
bound except for \((4,4,4)^{18}\) in the octahedral row.  The eighteen
surviving four-lines form one orbit under realizable left and right
\(\OctDD\)-actions.  There are eighteen choices for
\(\mathbb P(\operatorname{im}A^\sharp)\), eighteen for
\(\mathbb P(\operatorname{im}A)\).  Each surviving four-point set is
harmonic in the cross-ratio convention of
Appendix~\ref{subsubsec:root-incidence}, and exactly eight
projective isomorphisms between any two such lines preserve its harmonic
pairings.  Thus there are exactly eight harmonic projectivities, and
consequently
\begin{equation*}
  18\cdot18\cdot8=2592
\end{equation*}
projective rank-two maps survive.  They form one realizable local-octahedral
orbit of the rank-two endomorphism
\begin{equation*}
  A_0:=\diag(1,1,0,0).
\end{equation*}
Bilinear-dual reconstruction, including the fixed output \(X\) in the
transfer convention \(L_p(C)=\Phi_p(C)X\), gives a realizable local-octahedral
equivalent of the quaternary signature
\begin{equation*}
 q_0({\mathbf{x}}):=\frac12\one[x_1+x_2+x_3+x_4\equiv0\pmod2].
\end{equation*}
The realizable octahedral matrix
\(W_0=\begin{psmallmatrix}1&1\\1&-1\end{psmallmatrix}\), with
\([W_0]=[Q_1+Q_3]\in\OctDD\), satisfies
\(W_0^{\otimes4}(\Eq_4)=
2\one[x_1+x_2+x_3+x_4\equiv0\pmod2]\) and
\(W_0^{\otimes4}q_0=4(\Eq_4)\).  It follows that every
rank-two remaining case has, for some matrices \(U_j\) with
\([U_j]\in\OctDD\),
\begin{equation*}
  q=\lambda(U_1\otimes U_2\otimes U_3\otimes U_4)(\Eq_4),
  \qquad \lambda\ne0.
\end{equation*}
Conversely, direct substitution verifies the conditions for both
complementary pairs on this orbit. The complete four-line orbit above leaves no
additional projective rank-two degeneration. Every surviving tensor has
flattening ranks \((2,2,2)\).

For the second octahedral form, analytic conjugation by
\(H_{\mathrm{tt}}\) transports the equations for both complementary pairs,
preserving the count \(2592\) and the rank pattern. The realizable local
gadgets require a separate calculation. Indeed,
the realizable \((\OctTT)^{\otimes4}\)-orbit of the second-form remaining case
has 2592 projective tensors with support histogram
\begin{equation*}
  8^{16},\qquad10^{128},\qquad12^{384},\qquad16^{2064},
\end{equation*}
and hence does not meet the realizable \((\OctTT)^{\otimes4}\)-orbit of
the support-two signature \(\Eq_4\).  For the
normalized second-form rank-two remaining case, use the four matrices
\begin{equation}
\begin{aligned}
 L_1&=\theta_{\mathrm{tt}}(Q_0+Q_2),\\
 L_2=L_3&=\theta_{\mathrm{tt}}(Q_0-Q_1+Q_2-Q_3),\\
 L_4&=X\theta_{\mathrm{tt}}(Q_2-Q_3).
\end{aligned}
\label{eq:cert-platonic-q4-Oprime-dressing}
\end{equation}
Exact substitution gives the endpoint-nondegenerate eight-vertex signature
even-entry vector
\begin{equation}
  (1,-1,-1,-1,-1,-1,-1,1),
  \label{eq:p1-cert-q4-Oprime-eight-vertex-vector}
\end{equation}
with every odd entry zero and ranks \((2,2,2)\).  Acting before and after
this certificate by the realizable group proves the stated eight-vertex realization for the
entire second-form orbit.  

For \(\alpha\in\{\mathrm{dt},\mathrm{td}\}\), complementary-pair reconstruction of the
same normalized coefficient map uses the invertible coefficient operator
\begin{equation*}
 \mathscr T_\alpha=
 H_\alpha^{\mathsf T}\otimes H_\alpha^{-1}\otimes H_\alpha^{-1}
 \otimes XH_\alpha^{\mathsf T}X.
\end{equation*}
Direct exact substitution in \(q_0\) gives, projectively,
\begin{equation*}
 \mathbf e(\mathscr T_{\mathrm{dt}}q_0)=(1,1,1,1,1,1,1,1),
 \qquad
 \mathbf e(\mathscr T_{\mathrm{td}}q_0)=(1,i,i,i,1,1,1,i),
\end{equation*}
with all odd entries zero.  These are the direct current-coordinate
reconstructions of the remaining cases.
Exact row reduction gives ranks \((2,2,2)\). Acting by realizable
\(\OctForm{\alpha}\)-matrices covers the entire rank-two orbit.

The rank-zero member is the zero map.  Because the quaternion basis spans
\(\Mat_2(\mathbb C)\), its reconstructed quaternary is zero.
For rank one, choose nonzero column vectors \({\mathbf{u}},{\mathbf{v}}\in\mathbb C^4\) and write
\(A={\mathbf{u}}{\mathbf{v}}^{\mathsf T}\). The conditions on \(A\) and \(A^\sharp\) force
\([{\mathbf{u}}],[J_{\mathrm{Frob}}{\mathbf{v}}]\in\mathcal R_G\). Conversely every such pair works.  This gives
\(|G|^2=144,576,3600\) projective maps and a rank-one
\(p\mid\bar p\) flattening of the reconstructed tensor.

Finally, a rank-four member of \(\mathscr E_G\) permutes
\(\mathcal R_G\).  Exact root-incidence automorphism classification gives
precisely \cref{eq:cert-q4-full-rank-preservers}, with
\begin{equation*}
 [V]\in N_{\PGL_2}(G),\qquad [UV]\in G.
\end{equation*}
The projective counts of the two displayed families are
\begin{equation*}
  2|G|\,|N_{\PGL_2}(G)|
  =2\cdot12\cdot24,\quad2\cdot24^2,\quad2\cdot60^2,
\end{equation*}
namely \(576,1152,7200\).  If the contraction map is
\(C\mapsto UCV\), undoing the output \(X\) gives
\(q_{uvyz}=U_{yu}(VX)_{vz}\). The transpose case gives
\(q_{uvyz}=U_{yv}(VX)_{uz}\).  In either case the quaternary is a product
of two nonsingular binaries on a crossing matching.

For \(\TetInt\), imposing the condition of
\cref{eq:cert-q4-two-sided-endomorphisms} on \(A\) alone gives \(1536\)
projective rank-two solutions. Each fails the condition on \(A^\sharp\)
from the complementary pair, giving the zero in the \(A_4\) rank-two column.

The coefficient similarities identify the four rank tables.
The formulas
\cref{eq:cert-platonic-q4-Oprime-dressing,eq:p1-cert-q4-Oprime-eight-vertex-vector}
and the two direct mixed reconstructions supply the required gadget reductions in the respective current forms.

Here the finite incidence hypergraph has the root points as vertices
and the rich lines and plane sections as its distinguished hyperedge families.
Its automorphisms preserve each of these families.

For all three rows, exact enumeration generates the root configurations and
then recovers every secant, plane section, multiple-secant center, and
quotient direction by characteristic-zero row reduction: over
\(\mathbb Q(i)\) for \(D_4,F_4\), and over
\(\mathbb Q(\phi,i)\) for \(H_4\).  It enumerates all projectivities between
the surviving octahedral four-lines, giving \(18^2\cdot8=2592\).
For the full-rank case it constructs the complete finite incidence
hypergraph of root points, rich lines, and plane sections, enumerates its
automorphisms, solves for the inducing projective linear map, and retains the
map only after checking both \(A\) and \(A^\sharp\) on every root point.
The resulting automorphism counts equal the cardinalities of the explicit
left--right/transpose families for \(D_4,F_4,H_4\), proving completeness of
all three full-rank lists rather than merely verifying their displayed
members.
\end{proof}

\subsubsection{Contractions of the Platonic quaternary signatures}
\label{proofsubsec:cert-platonic-q4-exits}

This subsection gives the contractions used in
\cref{lem:platonic-finite-q4-terminal}.  Inputs of every quaternary are numbered
\(1,2,3,4\) in their displayed order.  If inputs \(r,s\) are contracted by
a realizable kernel \(C\), the surviving ordered binary is denoted
\(B=\partial_{rs}^{C}q\).  The pairing is bilinear, with no conjugation.
Each group-membership test checks whether the transfer class \([T_B]=[BX]\) belongs to \(G\).

\begin{proposition}[Contractions of Platonic quaternary signatures]
\label{prop:cert-platonic-q4-exits}
For each quaternary in the following list, let \(G\) denote the marked
Platonic group from whose row the quaternary is obtained.  Every one of the
six rank-deficient quaternaries \(q_\pi\) constructed in
\cref{prop:cert-platonic-deficient-mixed} and satisfying
\cref{eq:platonic-rank-two-q4-minors}, one of the three
orientation-reversing tetrahedral outputs \(q_\sigma\) specified in
\cref{prop:p1-cert-A4-eight-vertex-orientation-terminal}, one of the twenty-four
octahedral separator outputs (in any of the four marked forms) specified in
\cref{prop:p1-cert-S4-uniform-eight-vertex-separator}, or one of the thirty
icosahedral frame outputs \(q_{u_1,u_2,u_3}\) in
\cref{eq:cert-A5-output,eq:cert-A5-frames} has one of the
following directly realized signatures:
\begin{enumerate}
\item an endpoint-nondegenerate eight-vertex signature with flattening ranks
      \((2,2,2)\);
\item a generalized equality of arity four;
\item a nonsingular ordered binary \(B\) whose transfer class \([BX]\) is
      outside \(G\).
\end{enumerate}
For either mixed octahedral form the list also includes the six direct
current-coordinate counterparts of the rank-deficient outputs.  This holds for
all four octahedral distinguished-involution normal forms and for both
icosahedral coordinate displays.  Every such signature uses one
quaternary vertex and at most four realizable group binaries.
\end{proposition}

\begin{proof}
First consider the six support-ten rank-deficient outputs in
\cref{prop:cert-platonic-deficient-mixed}.  For each fixed output, exact
enumeration of all six input pairs and all realizable kernels gives the following
table.  ``Inside'' means \([BX]\) is in the displayed group, and
``singular'' counts nonzero singular cards. Zero cards are kept separate.
\begin{center}
\normalsize
\setlength{\tabcolsep}{4pt}
\begin{tabular}{crrrrrc}
\toprule
current group&cards per \(q\)&zero&inside&outside&singular&one uniform outsider\\
\midrule
\(\TetInt\)&72&3&9&60&0&\(\partial_{12}^{Q_0}q\)\\
\(\OctDD\)&144&6&108&30&0&\(\partial_{24}^{Q_0}q\)\\
\(\IcoZero\)&360&3&36&321&0&\(\partial_{12}^{Q_0}q\)\\
\bottomrule
\end{tabular}
\end{center}
Thus the same displayed pair and kernel work for all six outputs in each
row.  The first two rows are computed over \(\mathbb Q(i)\), and the
\(\IcoZero\) row is computed over \(\mathbb Q(\phi,i)\).

For the second octahedral form, define
\(\sigma_{\mathrm{tt}}:=\sqrt2\) and the transport matrix
\(S_{\mathrm{tt}}:=H_{\mathrm{tt}}^{-1}\).  The transported rank-deficient outputs
are \(q'_\pi=S_{\mathrm{tt}}^{\otimes4}q_\pi\).  The single realizable
kernel
\hypertarget{eq:cert-Oprime-deficient-kernel}{}
\begin{equation*}
 K'_{\mathrm{def}}
 =H_{\mathrm{tt}}^{-1}(Q_0-Q_1-Q_2-Q_3)H_{\mathrm{tt}},
 \qquad [K'_{\mathrm{def}}]\in\OctTT
\end{equation*}
on inputs \(1,2\) works for all six.  Direct expansion gives
\(\det(\partial_{12}^{K'_{\mathrm{def}}}q'_\pi)=-3/512\).  As
\(\pi\) ranges over \(S_3\), the pulled-back transfer directions
\(H_{\mathrm{tt}}(B'X)H_{\mathrm{tt}}^{-1}\), in quaternion coordinates,
form the multiset
\begin{equation*}
 [3,-1,1,1],\ [1,-1,0,1],\ [3,-1,1,1],\
 [1,-1,0,1],\ [3,-1,-1,1],\ [3,-1,-1,1].
\end{equation*}
None is in \(\OctDD\): its support has size three, or size four with
unequal nonzero magnitudes.  Hence every corresponding \([B'X]\) is outside
\(\OctTT=H_{\mathrm{tt}}^{-1}\OctDD H_{\mathrm{tt}}\).

For \(\OctDT\) and \(\OctTD\), the final exact table in
\cref{prop:p1-cert-S4-uniform-eight-vertex-separator} gives one current-group kernel on inputs \(1,2\), a common nonzero determinant, and
all six pulled-back outside-transfer directions.  This proves the rank-deficient
rows directly in the retained fixed-\(I\) coordinates.

The notation here is compatible with the earlier separator transport
\(q'_R=\mathcal L_{\mathrm{tt}}q_R\), but the two formulas use different
displayed input-orientation conventions.  The matching-dependent \(X\)-factors
in \cref{eq:cert-S4-coefficient-transport,eq:cert-Oprime-transport} convert
the alternating
\(H_{\mathrm{tt}}^{-1}/H_{\mathrm{tt}}^{\mathsf T}\) convention of
\(\mathcal L_{\mathrm{tt}}\) into the uniform
\(S_{\mathrm{tt}}^{\otimes4}\) convention after those fixed orientation
factors are absorbed into the current \(\OctTT\)-representatives.
Thus both notations parametrize the same marked second-form coefficient
families in their stated input orders. Neither equality is being used as a
gadget.

For the three tetrahedral outputs \(q_\sigma\) specified in
\cref{prop:p1-cert-A4-eight-vertex-orientation-terminal}, the one realizable kernel
\begin{equation*}
 K_{\TetInt}^{\mathrm{out}}
 =Q_0+Q_1-Q_2+Q_3,
 \qquad [K_{\TetInt}^{\mathrm{out}}]\in\TetInt
\end{equation*}
on inputs \(1,2\) gives, projectively,
\begin{center}
\normalsize
\setlength{\tabcolsep}{5pt}
\begin{tabular}{ccc}
\toprule
\(\sigma\)&\(B=\partial_{12}^{K_{\TetInt}^{\mathrm{out}}}q_\sigma\)&\(BX\)\\
\midrule
\((0,3,1)\)&\(\begin{psmallmatrix}1&1\\-1&1\end{psmallmatrix}\)&
 \(\begin{psmallmatrix}1&1\\1&-1\end{psmallmatrix}\)\\
\((1,0,3)\)&\(\begin{psmallmatrix}1&-i\\-i&1\end{psmallmatrix}\)&
 \(\begin{psmallmatrix}-i&1\\1&-i\end{psmallmatrix}\)\\
\((3,1,0)\)&\(\diag(1,i)\)&
 \(\begin{psmallmatrix}0&1\\i&0\end{psmallmatrix}\)\\
\bottomrule
\end{tabular}
\end{center}
All three determinants are nonzero, and exact projective comparison with
the twelve points in \cref{eq:tetrahedral-domain} excludes all three
transfer classes.

We next treat the first octahedral form.  Define the octahedral sign matrices by
\[
 S_{\epsilon_1\epsilon_2\epsilon_3}
 =Q_0+\epsilon_1Q_1+\epsilon_2Q_2+\epsilon_3Q_3.
\]
For an ordered frame in \cref{eq:cert-S4-four-frames}, its separator is
\(q=C/2+W_1\), where \(W_1\) is the first entry.  The twelve ordered
outputs with
\[
 W_1\in\{R_0,R_3,R_{03}^{+-},R_{03}^{-+}\}
\]
are endpoint-nondegenerate eight-vertex signatures on all eight even-parity strings and have ranks
\((2,2,2)\).  The four with \(W_1=R_1\) have support
\(\{0011,1100\}\). Realizable \(X\)-dressings on the last two inputs turn them
into generalized equalities.  The remaining eight outputs occur twice
for each row of the following table:
\begin{center}
\normalsize
\setlength{\tabcolsep}{5pt}
\begin{tabular}{cccc}
\toprule
\(W_1\)&kernel on \(12\)&\(B\)&\([BX]\)\\
\midrule
\(R_{13}^{++}\)&\(S_{++-}\)&
 \(\begin{psmallmatrix}1+i&-i\\-i&1-i\end{psmallmatrix}\)&
 \([Q_0+Q_1-Q_2]\)\\
\(R_{13}^{--}\)&\(S_{+++}\)&
 \(\begin{psmallmatrix}1-i&-i\\-i&1+i\end{psmallmatrix}\)&
 \([Q_0+Q_1+Q_2]\)\\
\(R_{01}^{+-}\)&\(S_{+++}\)&
 \(\begin{psmallmatrix}1-i&-1\\1&1+i\end{psmallmatrix}\)&
 \([Q_1+Q_2-Q_3]\)\\
\(R_{01}^{-+}\)&\(S_{++-}\)&
 \(\begin{psmallmatrix}1+i&-1\\1&1-i\end{psmallmatrix}\)&
 \([Q_1-Q_2-Q_3]\)\\
\bottomrule
\end{tabular}
\end{center}
Every determinant equals \(3\).  The points of \(\OctDD\) in
\cref{eq:octahedral-domain} have quaternion-coordinate support size one,
two, or four, never three, so all four displayed transfer directions are
outside \(\OctDD\).

\begin{sloppypar}
For the second octahedral form, we compute the realizable kernels in its
fixed coordinates.  Define the nonzero scalar
\(c_{\mathrm{tt}}:=2\sigma_{\mathrm{tt}}(1-i)\).  Direct multiplication gives
\begin{equation}
 H_{\mathrm{tt}}^{\mathsf T}XH_{\mathrm{tt}}=c_{\mathrm{tt}}I,
 \qquad
 S_{\mathrm{tt}}US_{\mathrm{tt}}^{\mathsf T}
 =c_{\mathrm{tt}}^{-1}H_{\mathrm{tt}}^{-1}(UX)H_{\mathrm{tt}}.
 \label{eq:cert-Oprime-transport}
\end{equation}
Thus \(S_{\mathrm{tt}}^{\otimes4}\) transports every
\(\OctDD\)-matching factor to
an \(\OctTT\)-matching factor.  More precisely, if \(h\) is the
standard-form six-input ancestor with card map \(\Phi\), define, in its
displayed input order,
\[
 h':=(H_{\mathrm{tt}}^{\mathsf T}\otimes H_{\mathrm{tt}}^{-1}
       \otimes S_{\mathrm{tt}}^{\otimes4})h.
\]
The card map \(\Phi'\) of \(h'\) is parametrized by
\hypertarget{eq:cert-Oprime-card-map}{}
\begin{equation*}
 \Phi'(H_{\mathrm{tt}}^{-1}C_{\rm ker}H_{\mathrm{tt}})
 =S_{\mathrm{tt}}^{\otimes4}\Phi(C_{\rm ker}).
\end{equation*}
This invertible coefficient correspondence exhausts the second-form
tensors.  The contractions below use the listed current-group
kernels, with their ordered inputs and nonzero realization scalars.
\end{sloppypar}
Define the conjugated octahedral sign matrices by
\(S'_{\epsilon_1\epsilon_2\epsilon_3}
:=H_{\mathrm{tt}}^{-1}S_{\epsilon_1\epsilon_2\epsilon_3}H_{\mathrm{tt}}\).
For the twelve ordered
outputs in the next table, direct contraction of
\(q'=S_{\mathrm{tt}}^{\otimes4}q\) on inputs \(1,2\) gives
\(\det B'=-3/256\):
\begin{center}
\normalsize
\setlength{\tabcolsep}{5pt}
\begin{tabular}{ccc}
\toprule
\(W_1\)&realizable kernel&
\(H_{\mathrm{tt}}(B'X)H_{\mathrm{tt}}^{-1}\)\\
\midrule
\(R_{13}^{++}\)&\(S'_{+++}\)&\([2Q_0+Q_1+Q_2]\)\\
\(R_{13}^{--}\)&\(S'_{+-+}\)&\([Q_1+Q_2+2Q_3]\)\\
\(R_{03}^{+-}\)&\(S'_{++-}\)&\([Q_0+2Q_1+Q_3]\)\\
\(R_{03}^{-+}\)&\(S'_{+++}\)&\([Q_0+2Q_2+Q_3]\)\\
\(R_{01}^{+-}\)&\(S'_{++-}\)&\([2Q_0+Q_1-Q_2]\)\\
\(R_{01}^{-+}\)&\(S'_{+++}\)&\([2Q_0-Q_1+Q_2]\)\\
\bottomrule
\end{tabular}
\end{center}
Each row occurs for two orderings.  Every pulled-back direction has three
nonzero quaternion coordinates of magnitudes \(2,1,1\), so it is outside
\(\OctDD\), proving \([B'X]\notin\OctTT\).

The remaining twelve ordered \(\OctTT\)-outputs have
\(W_1=R_0,R_1,R_3\), four of each.  Realizable local dressings turn them into
endpoint-nondegenerate eight-vertex signatures.  In the even-string order
\[
  (0000,0011,0101,0110,1001,1010,1100,1111),
\]
the exact outputs are
\begin{center}
\normalsize
\setlength{\tabcolsep}{5pt}
\begin{tabular}{ccc}
\toprule
\(W_1\)&dressing&even coefficients, up to scale\\
\midrule
\(R_0\)&\((S'_{++-})^{\otimes4}\)&\((1,1,-1,1,1,-1,1,1)\)\\
\(R_1\)&identity&\((1,-1,1,1,1,1,-1,1)\)\\
\(R_3\)&\((S'_{+++})^{\otimes4}\)&\((1,-1,1,1,1,1,-1,1)\)\\
\bottomrule
\end{tabular}
\end{center}
All odd-weight coefficients vanish, both endpoints are nonzero, and local
invertible dressings preserve ranks \((2,2,2)\).

The two mixed full-rank rows are the direct tables in
\cref{prop:p1-cert-S4-uniform-eight-vertex-separator}.  For
\(\OctDT\) they partition into endpoint and outside-binary exits.
For \(\OctTD\) they partition into endpoint, generalized
equality, and outside-binary exits.  The tables cover every possible first
frame entry, hence all twenty-four orderings, and use only matrices
of the current mixed group.

Finally, for each of the thirty icosahedral outputs in
\cref{eq:cert-A5-output}, use the one chosen representative
\begin{equation*}
 L_{\IcoZero}=Q_0-\tau^2Q_1+\tau Q_3
 \qquad\text{with}\qquad [L_{\IcoZero}]\in\IcoZero
\end{equation*}
on inputs \(1,2\).  Exact arithmetic over \(\mathbb Q(\phi,i)\) proves that
all thirty resulting binaries are nonsingular, their thirty projective
transfers \([BX]\) are distinct, and none belongs to the complete sixty-point
set \(\IcoZero\).  Enumerating all
\(30\cdot6\cdot60=10800\) realizable cards gives
\begin{equation*}
 \text{zero }180,\qquad
 \text{inside }1260,\qquad
 \text{outside }9360,\qquad
 \text{nonzero singular }0.
\end{equation*}
For the opposite display, use the fixed-equality coefficient correspondence
\cref{lem:icosahedral-fixed-I-dictionary}: replace the two standard kernels
by their realizable \(\kappa_D\)-images.  Card covariance conjugates the complete
transfer group and preserves nonsingularity and group exclusion, while the
\(I\) is untouched.  This exhausts every family in the
proposition.

The displayed kernels and dressing tables prove the listed bounded
exits by exact matrix multiplication.  The complete output histograms
and the completeness claims used earlier are also reproduced by
\path{certificates/verify_a4_s4_certificates.py},
\path{certificates/verify_s4_second_form.py},
\path{certificates/verify_a5_certificate.py}, and
\path{certificates/verify_platonic_extended.py} over the number fields
specified above.  In particular, the full preserver and hereditary eight-input classifications retain
the exhaustive computations described in their own proofs.
\end{proof}

\subsubsection{Consequences of the Platonic classifications}

\begin{theorem}[Platonic separated-sign classifications]
\label{thm:cert-interface-platonic-bridges}
Under the normalized rainbow and separated-sign hypotheses of
\cref{eq:platonic-rainbow-normalization,eq:platonic-rainbow-sum,%
eq:platonic-separated-sign-tests}, the following separated-sign classifications
hold.
\begin{enumerate}[label=\textnormal{(\roman*)}]
\item For each marked octahedral form, there are exactly six rank-deficient maps
      and \(24\) full-rank maps, the latter being the six orderings of each
      of
      \[
      \{R_0,R_1,R_3\},\quad
      \{R_0,R_{13}^{++},R_{13}^{--}\},\quad
      \{R_1,R_{03}^{+-},R_{03}^{-+}\},\quad
      \{R_3,R_{01}^{+-},R_{01}^{-+}\}.
      \]
      The fixed realizable kernel is
      \(K_{\OctDD}=Q_0+Q_1\) in the first form and
      \(K_{\OctTT}=\theta_{\mathrm{tt}}(K_{\OctDD})\) in the
      second, while
      \(K_{\OctDT}=\theta_{\mathrm{dt}}(K_{\OctDD})\) and
      \(K_{\OctTD}=\theta_{\mathrm{td}}(K_{\OctDD})\) are the two
      mixed-form kernels.  In the two unmixed forms, current-group
      dressings send every full-rank map to an endpoint-nondegenerate
      eight-vertex signature of ranks \((2,2,2)\).  In either mixed form,
      the fixed kernel and the direct current-coordinate tables give one of
      the three exits: such an eight-vertex signature, a generalized equality, or a nonsingular binary \(B\) with \([BX]\notin G\), where \(G\) is the current marked group.
\item For the internal tetrahedral form, the bridge set is
      \(\{0,A/2,B/2,R_0,R_1,R_3\}\), and the sum equation has exactly six
      rank-deficient and six full-rank solutions.  Write
      \[
      \Sigma_+=\{(0,1,3),(1,3,0),(3,0,1)\},\qquad
      \Sigma_-=\{(0,3,1),(1,0,3),(3,1,0)\}.
      \]
      For every
      \(\sigma\in\Sigma_+\), all \(180\) realizable pair cards of the reconstructed
      parent are zero or \(\TetInt\)-matching signatures, and the parent lies in \(\cA\).  For every
      \(\sigma\in\Sigma_-\), contraction of inputs \(1,3\) by the fixed
      realizable kernel \(Q_0=I\), followed by four realizable dressings, gives an
      endpoint-nondegenerate eight-vertex signature of ranks \((2,2,2)\).
\item The sixty displayed projective classes form a projective group
      isomorphic to \(A_5\) and are closed under induced projective
      multiplication, inverse, and transpose.  Its
      bridge has exactly six rank-deficient and \(30\) full-rank solutions, the
      latter being the six orderings of the five displayed frames
      \(\mathrm{Fr}_0,\ldots,\mathrm{Fr}_4\).  The fixed realizable kernel
      \(K_{\IcoZero}:=\phi Q_1-\tau Q_2+Q_3\) gives a quaternary of ranks
      \((3,3,3)\), and the fixed subsequent realizable kernel
      \(L_{\IcoZero}:=Q_0-\tau^2Q_1+\tau Q_3\) gives a nonsingular binary \(B\) with \([BX]\notin\IcoZero\), for all \(30\) solutions.  The statement
      includes arbitrary complex scales and cancellations.  For
      \(\IcoOne\), the realizable kernels
      \(K_{\IcoOne}=\kappa_D(K_{\IcoZero})\) and
      \(L_{\IcoOne}=\kappa_D(L_{\IcoZero})\) give the corresponding
      direct fixed-\(I\) identities.
\end{enumerate}
\end{theorem}

\begin{proof}
Clause~\textnormal{(i)} is the exact octahedral bridge
\cref{eq:cert-S4-four-frames} together with the separate gadget certificate
\cref{prop:p1-cert-S4-uniform-eight-vertex-separator}.  Clause~\textnormal{(ii)}
is \cref{eq:cert-A4-bridge,eq:cert-A4-twelve-maps} and the all-card,
affine-phase, and realizable-orientation checks in
Appendix~\ref{proofsubsec:cert-A4}, including
\cref{prop:p1-cert-A4-eight-vertex-orientation-terminal}.  Clause~\textnormal{(iii)}
is the exact group, bridge, frame, and rank calculation in
Appendix~\ref{proofsubsec:cert-A5}, followed by the uniform explicit gadget reduction in
\cref{prop:cert-platonic-q4-exits}.
\Cref{lem:icosahedral-fixed-I-dictionary} gives the opposite display with
realizable \(\kappa_D\)-kernels and leaves the supplied equality unchanged.
\end{proof}

\begin{theorem}[Classification of Platonic quaternaries]
\label{thm:cert-interface-platonic-q4}
Let
\[
 G\in\{\TetInt,\OctDD,\OctTT,\OctDT,
 \OctTD,\IcoZero,\IcoOne\}.
\]
The finite quaternary certificates have the following directly usable form.
\begin{enumerate}[label=\textnormal{(\roman*)}]
\item Let \(q\) be quaternary and define
      \(L_p(C)=\Phi_p^q(C)X\).  If, for every one of the six pairs \(p\)
      and every fixed chosen representative \(K_g\),
      \[
      L_p(K_g)=0\qquad\hbox{or}\qquad[L_p(K_g)]\in G,
      \]
      then, for each pair \(p\), put \(r_p:=\rank L_p\).  One has
      \(r_p\ne3\).  If \(r_p=0\), then \(q=0\). If
      \(r_p\in\{1,4\}\), then \(q\) has a binary-pair factorization. \(r_p=2\) can occur only for the octahedral forms: it is actually
      dressable to \(\Eq_4\) for \(\OctDD\), and to an
      endpoint-nondegenerate eight-vertex signature of ranks \((2,2,2)\)
      for \(\OctTT\), \(\OctDT\), and \(\OctTD\),
      with the mixed vectors in
      \cref{eq:p1-platonic-q4-mixed-eight-vertex}.
\item Let \(h\ne0\) have arity six and let \(\Phi_p^h\) have rank two or
      three.  Suppose, for every \(g\in G\), the card
      \(\Phi_p^h(K_g)\) formed with its fixed chosen representative is zero
      or a nonsingular \(G\)-matching product, with two nonzero cards in
      this \(p\)-family using different residual matchings.  Realizable localization
      produces a quaternary of ranks \((3,3,3)\), and a contraction
      of that quaternary gives a nonsingular binary \(B\) with
      \([BX]\notin G\).  This includes the separately verified
      \(\OctTT\) and mixed representatives and both icosahedral
      displays.
\item For each fixed \(G\), every corresponding enumerated rank-deficient
      quaternary has the following conclusion.  The same conclusion holds
      for the tetrahedral-orientation outputs when \(G=\TetInt\), the
      octahedral-frame outputs when
      \(G\in\{\OctDD,\OctTT,\OctDT,\OctTD\}\),
      and the icosahedral-frame outputs
      when \(G\in\{\IcoZero,\IcoOne\}\).  Each such quaternary
      directly realizes one of: an
      endpoint-nondegenerate eight-vertex signature of ranks \((2,2,2)\), a
      generalized equality of arity four, or a nonsingular binary
      \(B\) with \([BX]\notin G\).  Each realization uses the quaternary and
      at most four realizable \(G\)-binaries.
\end{enumerate}
\end{theorem}

\begin{proof}
For clause~\textnormal{(i)}, use
\cref{lem:icosahedral-fixed-I-dictionary} for the \(\IcoOne\)
coefficient calculation, replacing every realizable kernel by its realizable
\(\kappa_D\)-image and leaving the \(I\) fixed.  Fix a pair \(p\), and let \(A_p\)
be the matrix of \(L_p\) in the corresponding quaternion root coordinates.
The card conditions at \(p\) give the first condition in
\cref{eq:cert-q4-two-sided-endomorphisms}.  For the complementary pair
\(\bar p\), let \(\mathbf R_X\) be the coordinate matrix of right
multiplication \(C\mapsto CX\).  Direct complementary-pair reshaping gives
\(A_{\bar p}=\mathbf R_XA_p^\sharp\mathbf R_X\). The fixed \(X\)-reversals
permute the root configuration, so the card conditions at \(\bar p\) give
the second condition. In the normalized coordinates this places \(A_p\) in
\(\mathscr E_{G_0}\), where \(G_0\in\{\TetInt,\OctDD,\IcoZero\}\) is the standard representative for \(G\).
Now \cref{prop:cert-platonic-q4-preservers} gives all stated rank and
reconstruction alternatives, including the separately verified
\(\OctTT\) eight-vertex output and the direct mixed reconstructions above.

Clause~\textnormal{(ii)} combines the realizable localization and rank
calculation in \cref{prop:cert-platonic-deficient-mixed} with the
corresponding uniform outside-binary row of
\cref{prop:cert-platonic-q4-exits}.  Clause~\textnormal{(iii)} is the
representative-wise realization statement of the latter proposition.  The
card and network covariance in
\cref{lem:icosahedral-fixed-I-dictionary} supplies all three clauses for
\(\IcoOne\) directly with realizable \(\kappa_D\)-kernels.
\end{proof}

\section{General Boundary Representations}
\label{app:general-boundaries}

The gadget convention of \cref{subsec:holant-definitions} uses exposed
vertex inputs and isolated edges. We now describe the general representation
with free edge endpoints and prove that it gives the same replacement
signature, gluing operations, and charge identities.

We also allow an edge to end at a free stub, a half-edge with no tensor
vertex.  In this extension,
\(H(\Omega)=(\bigsqcup_vH_v)\sqcup H_{\rm free}(\Omega)\).
Every free stub is paired and belongs to \(\partial\Omega\). A vertex
half-edge is unpaired exactly when it belongs to \(\partial\Omega\).
Thus \(E_{\rm int}(\Omega)\), the matching of pairs of half-edges, contains each
nonboundary vertex half-edge and each free stub exactly once.  An edge box
\(\mathsf E_W\) denotes one edge with default kernel \(W\), contributing
one factor \(W(a,b)\).  It has no vertex or separate \(H_v\).
By contrast, an \(I\) or \(X\) supplied as an input signature labels an
ordinary vertex, counted and transformed with all other tensor vertices.

Write \(\mathbf W=(W_e)_{e\in E_{\rm int}(\Omega)}\), with
\(W_e\in\{I,X\}\).  For a string \({\mathbf z}_{\partial\Omega}\) in boundary
order, define the raw boundary signature by
\[
 R_\Omega({\mathbf z}_{\partial\Omega})
 =\sum_{\substack{\tau:H(\Omega)\to\{0,1\}\\\tau|_{\partial\Omega}={\mathbf z}_{\partial\Omega}}}
    \prod_v f_v(\tau|_{H_v})
    \prod_{\{h,h'\}\in E_{\rm int}(\Omega)}W_{\{h,h'\}}(\tau_h,\tau_{h'}).
\]
For the \(j\)-th boundary half-edge, put \(A_j=I\) at a vertex half-edge and
\(A_j=W_e\) at a free endpoint of the edge \(e\).  The boundary tensor used
for vertex replacement is
\begin{align}
 T_\Omega&=\left(\bigotimes_{j=1}^{|\partial\Omega|}A_j\right)R_\Omega,
       \label{eq:half-edge-boundary-tensor}\\
 Z_{\mathbf W}(\Omega)&:=T_\Omega(\varnothing)
   \qquad(\partial\Omega=\varnothing).
 \label{eq:half-edge-partition-function}
\end{align}
For the ordinary boundary convention of
\cref{subsec:holant-definitions}, \(T_\Omega=R_\Omega\).
At a free endpoint, \(W_e^2=I\) cancels the incident edge: deleting that
endpoint and edge and exposing the ordinary mate preserves \(T_\Omega\)
and the boundary order.  An edge with two free endpoints is an isolated
\(\mathsf E_W\), with boundary tensor \(WWW=W\).  These cancellations
reduce every open network to unpaired vertex half-edges and isolated edge
boxes.  Gluing and vertex replacement are specified in
\cref{subsec:reduction-conventions}.

For homogeneous edges, abbreviate \(Z_{\mathbf W}\) by \(Z_W\).
The empty network is the nullary unit \(1\).

\subsection{Gluing and coordinate transport}
In either homogeneous model, joining two boundary half-edges contracts their
boundary-tensor coordinates against the default kernel \(W\). The remaining
boundary order is inherited.  The ordinary-boundary case is given in
\cref{subsec:reduction-conventions}. For one paired free endpoint and one
unpaired vertex half-edge,
splice the latter into the former's incidence, retaining its existing edge.
For two free endpoints on distinct edges, delete those endpoints and their
two edges and join their mates by one \(W\)-edge.  If both endpoints belong
to the same isolated \(\mathsf E_W\), remove that component and record the
scalar \(\sum_{a,b}W(a,b)^2=2\).  These rules follow from \(W^2=I\) and
\(W^3=W\).  In particular, vertex replacement first cancels the one-free-end
edges as above and then inserts the resulting ordinary-boundary gadget.
Each remaining isolated \(\mathsf E_W\) splices into a single edge.
Closed components containing only such edge boxes contribute known factors
of \(2\).  This proves \cref{eq:direct-gadget-reduction}, including loops,
without leaving free stubs in the oracle instance.

When transporting an open gadget with disequality edges by \(D\in GO(X)\), first cancel
its one-free-end edges in this way.  Every remaining internal edge between
vertex half-edges contributes the scalar from \(D^{\mathsf T}XD=\lambda X\).
An isolated edge remains available because \(DXD^{\mathsf T}=\lambda X\).
Thus its boundary tensor is transported by \(D\) up to a nonzero scalar. The closed-network edge count in \cref{eq:GOX-global-scalar} is used
only for closed instances.

\subsection{Boundary charge}
\label{app:boundary-charge}
For a homogeneous \(X\)-edge network (all structural pairs of half-edges have
kernel \(X\), any explicitly supplied \(I\) or \(X\) is an ordinary tensor
vertex), let
\(U(\Omega)=H(\Omega)\setminus\bigcup_{e\in E_{\rm int}(\Omega)}e\) be the
unpaired boundary half-edges (by definition,
\(U(\Omega)\subseteq\partial\Omega\)).  Call an assignment
\emph{edge-supported} when every structural edge-kernel factor is nonzero on
the two bits assigned to its endpoints.  Every edge-supported assignment has
one \(0\)-bit and one \(1\)-bit on each
\(X\)-pair. Hence, for every edge-supported assignment \(\tau\),
\begin{equation*}
  \sum_v q_{\arity(f_v)}(\tau|_{H_v})
  +\sum_{h\in H_{\rm free}(\Omega)}(2\tau_h-1)
  =\sum_{h\in U(\Omega)}(2\tau_h-1).
\end{equation*}
For a contribution to the boundary tensor \(T_\Omega({\mathbf{z}})\), an ordinary
boundary bit equals \(\tau_h\), while a free boundary bit equals
\(1-\tau_h\) by \cref{eq:half-edge-boundary-tensor}.  Moving the free-stub
sum to the right therefore gives
\(q_{|\partial\Omega|}({\mathbf{z}})=\sum_v q_{\arity(f_v)}(\tau|_{H_v})\).
In standard closed instances there are
no free stubs and \(U(\Omega)=\varnothing\), so
\(\sum_vq_{\arity(f_v)}(\tau|_{H_v})=0\).  Thus \(X\)-contraction preserves
the total charge in every closed contributing assignment.

\section{Exact Algebraic Tools}
\label{app:exact-algebraic-tools}

This appendix fixes the standard conventions used to decide common
transformability and extract the algebraic witnesses in
\cref{lem:algebraic-transform-witness}.

Work over the exact coefficient field \(L\). For a finite polynomial
system \(f_1,\ldots,f_s\in R=L[x_1,\ldots,x_n]\), let
\(I=(f_1,\ldots,f_s)\) be the ideal they generate and write
\[
 V_E(I)=\{\mathbf a\in E^n:f(\mathbf a)=0\text{ for every }f\in I\}
\]
for its zero locus over an algebraically closed extension \(E\supseteq L\).
A projective zero locus is defined by homogeneous equations in the
specified projective space. Zero loci are Zariski closed, their complements
are Zariski open, and finite unions of intersections of open and closed
sets are constructible. Nonvanishing conditions \(p_1\cdots p_t\ne0\)
can be encoded by an auxiliary variable \(z\) and the equation
\(z p_1\cdots p_t=1\), as in \cref{lem:algebraic-transform-witness}.

For a fixed monomial order, a Gr\"obner basis of \(I\) has leading
monomials generating the leading-monomial ideal of \(I\).
Buchberger's algorithm computes such a basis and terminates by the
ascending-chain condition on monomial ideals. Division by the basis
decides ideal membership. In particular, the weak Nullstellensatz gives
\[
 V_{\overline L}(I)=\varnothing\quad\Longleftrightarrow\quad1\in I,
\]
equivalently, the reduced Gr\"obner basis is \(\{1\}\)
\cite[Chapters~2--4]{CoxLittleOShea2015}.

Exact algebraic points can be obtained by elimination and triangular
decomposition. For an ordered variable split, the elimination ideal is
\(I\cap L[x_{k+1},\ldots,x_n]\). In a triangular set, successive equations
have distinct largest variables; the leading coefficient in that variable
is the initial. For \(\Delta\in R\), saturation is
\[
 I:\Delta^\infty=\{f:\Delta^m f\in I\text{ for some }m\ge0\}.
\]
A regular chain is a triangular set whose successive initials are
non-zero-divisors modulo the ideal obtained by saturating the preceding
equations by their initials. Its constructible branch requires those
initials to be nonzero. Regular-chain decomposition produces finitely
many such branches. Generic algebraic specialization assigns algebraic
values to free coordinates outside the exceptional zero loci of the
required initials and denominators. Minimal polynomials with isolating
data then represent exact algebraic sample points
\cite{AubryLazardMorenoMaza1999,Cohen1993}.

These tools supply the emptiness test and exact witness construction in
\cref{lem:algebraic-transform-witness}.

For a field extension \(L'/L\), base change extends the coefficients and
sends \(I\) to \(I_{L'}=I\,L'[x_1,\ldots,x_n]\).
When \(L\subseteq\mathbb C\), complexification is base change to
\(L'=\mathbb C\).

\section{Character-Flat Extensions of the Reed--Muller Calculus}
\label{proofsubsec:rm-pauli-tools}
\label{proofsec:reed-muller}

Allowing all four Bell signatures extends the flat-Lagrangian network
construction to signatures with linear signs on their support. This
appendix characterizes that class and proves signed Radon localization
and recognition results. These structural extensions complement the
\(\CycX{2}\) result of \cref{thm:reed-muller-rigidity}. Closure for \(\KleinStd\) is proved by the
tensor-line argument in \cref{thm:standard-v4-relative}.
We retain \hyperref[subsubsec:symmetric-binary-spaces]{the symmetric-space conventions}.

We call an arity-\(2k\) signature \emph{character-flat} when a linear sign
character is also allowed:
\begin{equation}
 \mathsf{CFlat}_{2k}
 =\left\{
 c(-1)^{\ell({\mathbf{x}})}\one[{\mathbf{x}}\in {\mathbf{a}}+L]:
 c\in\AlgNums^\times,\ {\mathbf{a}}\in\Ftwo^{2k},\ \ell\in(\Ftwo^{2k})^\vee,
 \ L=L^\perp
 \right\}\cup\{0\}.
 \label{eq:character-flat-class}
\end{equation}
Define the signature class
\(\mathsf{CFlat}=\bigcup_{k\ge0}\mathsf{CFlat}_{2k}\).  This is a
subclass of \(\cA\).  The following network-calculus lemma describes
its realization by \(\RMcore\) and all four Bell binaries.

A common character-flat \(K\)-presentation is defined by
\cref{def:common-K-presentation} with \(\cC=\mathsf{CFlat}\).
It implies a common affine \(K\)-presentation because
\(\mathsf{CFlat}\subseteq\cA\).  Here an
\emph{\(\RMcore\)--Pauli network} means a tensor network whose
nonbinary vertices are retained copies of \(\RMcore\), whose binary
vertices are realizable ordered Bell binaries, and whose internal connections
are \(X\)-edges.  The next lemma gives the exact Pauli network
calculus for this class.  
The remaining lemmas extend Radon localization and recognition from
\(\FLag\) to \(\mathsf{CFlat}\).

\begin{lemma}[Character-flat Pauli network calculus]
\label{lem:v4-character-flat-calculus}
\begin{enumerate}
\item Every network whose vertices are \(\RMcore\) and realizable ordered Bell
      binaries \(B_{d,\ell}\), and whose internal edges are \(X\)-edges, has
      boundary signature in \(\mathsf{CFlat}\).
\item If \(\RMcore\) and realizable nonzero representatives of all four ordered
      Bell binaries \(B_{d,\ell}\) are retained, then, for every
      \(0\ne f\in\mathsf{CFlat}_{2k}\), an \(O(k^2)\)-size realizable
      \(\RMcore\)--Pauli network has boundary equal to a nonzero
      scalar multiple of \(f\).
\item Every nonzero character-flat signature of arity at most six is, after
      a permutation of inputs, a nonzero scalar multiple of a product of ordered
      Bell binaries on a perfect matching.
\item \(\mathsf{CFlat}\) is closed under nonzero scalars, tensor products,
      permutations of inputs, single-input ordered-Bell attachments through \(X\)-edges, and nonzero tensor factors.
\end{enumerate}
\end{lemma}

\begin{proof}
Let \(B\) be the displayed boundary-input set and let
\(V_{\rm vert}\) be the tensor-vertex set.  Use the boundary normalization of \cref{lem:rm-flat-contexts},
so \(B\) consists of unpaired vertex inputs.  Let \(H_{\rm int}\) be the set of internal half-edges.  For every
vertex let \(L_v\) be the translation space of its affine support.  The
Reed--Muller code and the two-coordinate repetition space of every
Bell binary are self-dual.  Define the internal equality subspace
\(E_{\rm eq}\), the ambient
constraint subspace \(H_{\rm con}\), and the vertex-code sum \(C_0\) by
\[
 \begin{aligned}
 E_{\rm eq}
   &=\{{\mathbf{y}}\in\Ftwo^{H_{\rm int}}:
       y_h=y_{h'}\text{ on every internal pair }\{h,h'\}\},\\
 E_{\rm eq}&=E_{\rm eq}^\perp,
 &H_{\rm con}&=\Ftwo^B\oplus E_{\rm eq},\\
 C_0&=\bigoplus_{v\in V_{\rm vert}}L_v=C_0^\perp.
 \end{aligned}
\]
Exactly as in the proof of \cref{lem:rm-pure-boundary}, the feasible boundary
translation space is
\(L=\pi_B(C_0\cap H_{\rm con})=L^\perp\).

Define the feasible half-edge code
\(\mathcal K:=C_0\cap H_{\rm con}\) and its
internal-fiber subspace
\[
 J:=\mathcal K\cap
   (\{\mathbf 0_B\}\oplus\Ftwo^{H_{\rm int}}),
\]
where \(\mathbf 0_B\) is the zero boundary string.
The product of the ordered-Bell signs is a linear character
\((-1)^{\lambda({\mathbf{z}})}\) on the half-edge variables.  For a fixed feasible
boundary string the internal fiber is \({\mathbf{z}}_0+J\), and
\[
 \sum_{{\mathbf{j}}\in J}(-1)^{\lambda({\mathbf{z}}_0+{\mathbf{j}})}
 =(-1)^{\lambda({\mathbf{z}}_0)}\sum_{{\mathbf{j}}\in J}(-1)^{\lambda({\mathbf{j}})}.
\]
If \(\lambda|_J\ne0\), every nonempty boundary fiber cancels. Otherwise
the same nonzero scalar remains on every fiber and the character descends
to the boundary, proving item~1.

Conversely, the network in \cref{lem:rm-flat-contexts}
realizes the indicator of \({\mathbf{a}}+L\).  Attaching a Bell binary through \(X\)
induces, up to a nonzero sign, one of \(I,X,Z,Y\), so realizable ordered-Bell
attachments implement
  the shift and linear character, proving item~2.  More explicitly,
  \(B_{d,\ell}X\) is respectively \(X,Y,I,Z\) for
  \((d,\ell)=(0,0),(0,1),(1,0),(1,1)\), up to the fixed orientation signs.
Arity zero is the empty matching.  At positive arity at most six,
\cref{lem:rm-low-flat-matching} splits the self-dual translation space into
two-coordinate repetition lines.  Support shifts and characters turn
their factors into ordered Bell signatures, proving item~3.

For the factor claim in item~4, let \(f=f_1\otimes f_2\ne0\).
The two projections of its affine support are affine. Write its Cartesian
support as \(({\mathbf{a}}_1+L_1)\times({\mathbf{a}}_2+L_2)\), with
\(L=L_1\oplus L_2\).  Since \(L=L^\perp\), each \(L_j\) is
self-dual in its own coordinate space.  Fixing a supported string of the
other factor restricts the character of \(f\) to a character on
\({\mathbf{a}}_j+L_j\), up to a nonzero scalar.  Hence \(f_j\) is character-flat.
The other closure properties follow by the same direct-sum and character
identities.
\end{proof}

\begin{corollary}[No Pauli-network inverse from \(\RMcore\) to \(\Hcore\)]
\label{cor:no-rm-to-H6}
No network using only \(\RMcore\), realizable ordered Bell binaries, and \(X\)-edges has boundary proportional to \(\Hcore\).
\end{corollary}

\begin{proof}
Every nonzero such boundary lies in \(\mathsf{CFlat}\) by
\cref{lem:v4-character-flat-calculus}.  A character-flat six-input support is
a coset of a self-dual subspace of \(\Ftwo^6\), hence has dimension three.
By direct expansion of \cref{eq:H6}, the support of \(\Hcore\) is the
six-bit even-parity space, of dimension five.  Thus
\(\Hcore\notin\mathsf{CFlat}\), a contradiction.
\end{proof}

The support-coset and character data in
\cref{eq:character-flat-class} can be evaluated by affine Gaussian
elimination.

For every even \(n\), define the accessible-direction set
\[
\operatorname{Acc}_n
 =(\one^n)^\perp\setminus\{\zero^n,\one^n\}.
\]
Equivalently, \(\operatorname{Acc}_n\) consists of the nonzero even-weight
vectors other than \(\one^n\).  In the two smallest even arities,
\[
 \operatorname{Acc}_2=\varnothing,
 \qquad
 \operatorname{Acc}_4
   =\{{\mathbf{e}}_i+{\mathbf{e}}_j:1\le i<j\le4\}.
\]
Recall
\[
 O(n,2):=\{A\in\operatorname{GL}_n(\Ftwo):A^{\mathsf T}A=I_n\}.
\]
For \(A\in O(n,2)\), the graph subspace \(\operatorname{Graph}(A)\)
is self-dual, as in \cref{lem:rm-radon-accessible}.
Let \(\Gamma_A\) denote a fixed realizable \(\RMcore/I/X\)-network supplied by
\cref{lem:rm-flat-contexts} whose boundary is a nonzero scalar multiple of
\(\one[\operatorname{Graph}(A)]\).

\begin{lemma}[Realizable signed Radon cards]
\label{lem:v4-signed-radon}
Assume \(\RMcore\) and chosen representatives of all four ordered Bell
binaries are retained, and let \(g\) have even arity \(n\).  For each
\({\mathbf{v}}\in\operatorname{Acc}_n\), use the recorded data
\((A_{\mathbf{v}},i_{\mathbf{v}},j_{\mathbf{v}},{\mathbf{r}}_{\mathbf{v}},\Pi_{\mathbf{v}},W_{\mathbf{v}})\) fixed in
\cref{lem:rm-radon-accessible}, and fix the graph network
\(\Gamma_{A_{\mathbf{v}}}\).  For \(d,\ell\in\Ftwo\), using one copy of \(g\), this
graph network, \(X\)-edges, and the realizable ordered Bell binary
\(B_{d,\ell}\), one directly realizes, up to a nonzero scalar, the
signed card
\begin{equation}
 \mathcal S_{{\mathbf{v}},d,\ell}g([{\mathbf{x}}]_{\mathbf{v}})
 =(-1)^{\ell(A_{\mathbf{v}}{\mathbf{x}})_{i_{\mathbf{v}}}}
  \bigl(g({\mathbf{x}})+(-1)^\ell g({\mathbf{x}}+{\mathbf{v}})\bigr),
 \qquad [{\mathbf{x}}]_{\mathbf{v}}\in H_{{\mathbf{v}},d}/\Span_{\Ftwo}\{{\mathbf{v}}\}.
 \label{eq:v4-signed-radon}
\end{equation}
The cases \(\ell=0\) and \(\ell=1\) are called respectively the
\emph{plus} and \emph{minus} signed cards on layer \(d\).  The prefactor
\((-1)^{\ell(A_{\mathbf{v}}{\mathbf{x}})_{i_{\mathbf{v}}}}\) is the exterior Bell sign in the chosen
quotient coordinates. After stripping this sign and the nonzero scalar, the two orbit numerators are
\(g({\mathbf{x}})+g({\mathbf{x}}+{\mathbf{v}})\) and \(g({\mathbf{x}})-g({\mathbf{x}}+{\mathbf{v}})\).
The remaining \(n-2\) graph-output coordinates give an explicit affine
isometry from this quotient to the boundary coordinate space.  Replacing
\((A_{\mathbf{v}},i_{\mathbf{v}},j_{\mathbf{v}})\) by another such choice changes the displayed analytic
card only by a quotient affine isometry, a nonzero scalar, and a linear sign
character.
\end{lemma}

\begin{proof}
Connect \(g\) to the input side of \(\Gamma_{A_{\mathbf{v}}}\) through \(X\)-edges, and let \({\mathbf{x}}\) denote the variables at \(g\).  The graph input
is therefore \({\mathbf{x}}+\one\).  Every \(A\in O(n,2)\) fixes \(\one\): invariance
of \({\mathbf{z}}\cdot {\mathbf{z}}={\mathbf{z}}\cdot\one\) gives \(A^{\mathsf T}\one=\one\), and hence
\(A\one=\one\).  Thus the graph output is
\({\mathbf{z}}=A_{\mathbf{v}}({\mathbf{x}}+\one)=A_{\mathbf{v}}{\mathbf{x}}+\one\).  Connect output positions \(i_{\mathbf{v}},j_{\mathbf{v}}\) to
\(B_{d,\ell}\) through their two \(X\)-edges.  The ordered Bell
arguments are then
\[
 u=z_{i_{\mathbf{v}}}+1=(A_{\mathbf{v}}{\mathbf{x}})_{i_{\mathbf{v}}},
 \qquad
 w=z_{j_{\mathbf{v}}}+1=(A_{\mathbf{v}}{\mathbf{x}})_{j_{\mathbf{v}}}.
\]
Consequently its support equation is
\[
 u+w
 ={\mathbf{x}}\cdot A_{\mathbf{v}}^{\mathsf T}({\mathbf{e}}_{i_{\mathbf{v}}}+{\mathbf{e}}_{j_{\mathbf{v}}})
 ={\mathbf{x}}\cdot {\mathbf{v}}=d.
\]
For fixed remaining graph-output coordinates there are two lifts, differing
by \({\mathbf{e}}_{i_{\mathbf{v}}}+{\mathbf{e}}_{j_{\mathbf{v}}}\) in \({\mathbf{z}}\) and hence by \({\mathbf{v}}\) in \({\mathbf{x}}\).  The Bell sign
is \((-1)^{\ell u}=(-1)^{\ell(A_{\mathbf{v}}{\mathbf{x}})_{i_{\mathbf{v}}}}\) and changes between those
lifts by \((-1)^\ell\), giving
\cref{eq:v4-signed-radon}.  The graph relation is single-valued before the
pair contraction, so the realization scalar is nonzero.

Each choice identifies the quotient affinely and isometrically with its
remaining graph-output coordinates.  The transition between two choices is
therefore an affine isometry.  Their selected-coordinate functionals differ
on the quotient by a linear form and a constant, which gives the last
assertion.
\end{proof}

\begin{lemma}[Localization from two character-flat Radon cards]
\label{lem:v4-character-two-chart}
Assume the resources of \cref{lem:v4-signed-radon} are retained.  Let
\(g\) have even arity \(n\ge10\), fix
\({\mathbf{v}}\in\operatorname{Acc}_n\), and suppose that the two plus cards
\(\mathcal S_{{\mathbf{v}},0,0}g\) and \(\mathcal S_{{\mathbf{v}},1,0}g\) are zero or
character-flat.  Define, only for the analysis, the orbit-sum signature
\[
 F_{\mathbf{v}}=\mathsf P_{\mathbf{v}}^+g.
\]
If \(F_{\mathbf{v}}\notin\cA\), connecting one copy of \(g\) to a
\(\RMcore\)--Pauli network realizes a nonzero \(H\notin\cA\) of
arity at most eight.
\end{lemma}

\begin{proof}
Use the recorded data
\((A_{\mathbf{v}},i_{\mathbf{v}},j_{\mathbf{v}},{\mathbf{r}}_{\mathbf{v}},\Pi_{\mathbf{v}},W_{\mathbf{v}})\) from
\cref{lem:rm-radon-accessible}, and put \(W:=W_{\mathbf{v}}\).  The plane \(\Pi_{\mathbf{v}}\)
is nondegenerate, and the map \(\iota_{{\mathbf{v}},\epsilon}\) identifies \(W\)
isometrically with
\(H_{{\mathbf{v}},\epsilon}/\Span_{\Ftwo}\{{\mathbf{v}}\}\).  Since \(F_{\mathbf{v}}\) is
invariant under \(T_{\mathbf{v}}\), its two plus Radon cards, in these quotient coordinates, are
\[
 f_\epsilon({\mathbf{x}}):=F_{\mathbf{v}}(\epsilon {\mathbf{r}}_{\mathbf{v}}+{\mathbf{x}}),
 \qquad {\mathbf{x}}\in W.
\]
For each nonzero Radon card choose a presentation
\hypertarget{eq:v4-chart-presentation}{}
\begin{equation*}
 f_\epsilon({\mathbf{x}})
 =c_\epsilon(-1)^{\lambda_\epsilon({\mathbf{x}})}
  \one[{\mathbf{x}}\in {\mathbf{a}}_\epsilon+M_\epsilon],
 \qquad M_\epsilon=M_\epsilon^\perp\le W.
\end{equation*}
Here \({\mathbf{a}}_\epsilon\) is chosen modulo \(M_\epsilon\), and
\(\lambda_\epsilon\) is an ambient extension of the character on the
support.  Changing either choice changes \(c_\epsilon\) only by a sign, so
membership of \(c_1/c_0\) in \(\mu_4\) is independent of these choices.

If one Radon card is zero, the other gains only one affine layer equation and
\(F_{\mathbf{v}}\in\cA\).  Suppose both are nonzero.  Their union has affine support
exactly when \(M_0=M_1\).  When the translation spaces agree, the difference of
the two characters is absorbed by an allowed quadratic mixed term with the
layer bit, and the layer scales extend to an affine fourth-root phase
exactly when \(c_1/c_0\in\mu_4\).  Thus \(F_{\mathbf{v}}\notin\cA\) gives
\hypertarget{eq:v4-chart-obstruction}{}
\begin{equation*}
 M_0\ne M_1,
 \qquad\text{or}\qquad
 M_0=M_1\ \text{ and }\ c_1/c_0\notin\mu_4.
\end{equation*}

Represent \(\lambda_0+\lambda_1\) by \({\boldsymbol{\rho}}\in W\), so that
\((\lambda_0+\lambda_1)({\mathbf{x}})={\boldsymbol{\rho}}\cdot {\mathbf{x}}\).  When \(M_0\ne M_1\), choose
\({\mathbf{s}}\in M_0\setminus M_1\) and \({\mathbf{t}}\in M_1\) with
\({\mathbf{s}}\cdot {\mathbf{t}}=1\), and put \({\boldsymbol{\Delta}}={\mathbf{a}}_1-{\mathbf{a}}_0\).  The plane
\(\Span_{\Ftwo}\{{\mathbf{s}},{\mathbf{t}}\}\) is nondegenerate.  Put
\(S_0=\Span\{{\mathbf{s}},{\mathbf{t}},{\boldsymbol{\Delta}},{\boldsymbol{\rho}}\}\).  Its radical has dimension at most two.
Pairing a basis of that radical with dual vectors gives a nondegenerate
extension \(N\supseteq S_0\) of dimension at most six.  If \(N\) has odd
dimension, then its nondegenerate orthogonal complement is odd-dimensional
and contains an anisotropic vector. Adding that orthogonal line preserves
the bound and makes the extension even-dimensional.  Denote the result by
\[
 U\supseteq\Span\{{\mathbf{s}},{\mathbf{t}},{\boldsymbol{\Delta}},{\boldsymbol{\rho}}\},
 \qquad \dim U\le6.
\]
When \(M_0=M_1\), start from
\(S_0=\Span\{{\boldsymbol{\Delta}},{\boldsymbol{\rho}}\}\).  Pairing its radical needs at most two dual
vectors and gives a nondegenerate extension of dimension at most four. If
its dimension is odd, add an anisotropic vector from its odd-dimensional
nondegenerate complement.  This gives an even-dimensional nondegenerate
\(U\supseteq\Span\{{\boldsymbol{\Delta}},{\boldsymbol{\rho}}\}\) with \(\dim U\le4\).
Put
\[
 P:=U^{\perp_W},
 \qquad W=U\perp P,
\]
and let \(\pi_U,\pi_P\) be the corresponding orthogonal projections.  A
Witt decomposition of the even-dimensional nondegenerate space \(P\)
supplies
\[
 J\le P,
 \qquad J=J^{\perp_P}.
\]
Because \({\boldsymbol{\Delta}},{\boldsymbol{\rho}}\in U\), the two shifts have the same
\(P\)-projection
\[
 {\mathbf{p}}_0:=\pi_P({\mathbf{a}}_0)=\pi_P({\mathbf{a}}_1),
\]
and the two characters have the same restriction
\(\lambda_P:=\lambda_0|_P=\lambda_1|_P\).

Let \(S:=\Pi_{\mathbf{v}}\perp U\), put \(m:=\dim S\), and choose an isometry
\(\phi:S\to\Ftwo^m\).  Such an isometry exists because \(S\) contains the
nonalternating plane \(\Pi_{\mathbf{v}}\).  We have \(m\le8\) when \(M_0\ne M_1\) and \(m\le6\) when \(M_0=M_1\).  Define
\[
 R_0:=\{({\mathbf{s}}_0+{\mathbf{j}},\phi({\mathbf{s}}_0)):{\mathbf{s}}_0\in S,\ {\mathbf{j}}\in J\}
 \le\Ftwo^n\oplus\Ftwo^m.
\]
For two elements of \(R_0\), their inner product is
\[
 ({\mathbf{s}}_0+{\mathbf{j}})\cdot({\mathbf{s}}'_0+{\mathbf{j}}')
 +\phi({\mathbf{s}}_0)\cdot\phi({\mathbf{s}}'_0)={\mathbf{j}}\cdot {\mathbf{j}}'=0.
\]
Moreover \(\dim R_0=m+(n-m)/2=(n+m)/2\), so \(R_0=R_0^\perp\).
On \(R_0\), define the linear sign character
\[
 \chi({\mathbf{s}}_0+{\mathbf{j}},\phi({\mathbf{s}}_0)):=(-1)^{\lambda_P({\mathbf{j}})}.
\]
Choose an ambient linear-character extension and denote it again by
\(\chi\).  Hence, for
\(({\mathbf{x}},{\mathbf{y}})\in\Ftwo^n\oplus\Ftwo^m\),
\begin{equation}
 C({\mathbf{x}},{\mathbf{y}}):=\chi({\mathbf{x}}-{\mathbf{p}}_0,{\mathbf{y}})\,
 \one[({\mathbf{x}},{\mathbf{y}})\in({\mathbf{p}}_0,\zero)+R_0]
 \label{eq:v4-localizing-context}
\end{equation}
is character-flat.  By \cref{lem:v4-character-flat-calculus}, choose an \(O(n^2)\)-size
\(\RMcore\)--Pauli network whose boundary
is \(\kappa C\), where the realization scalar
\(\kappa\in\AlgNums^\times\) is recorded.  Any fixed
coordinate complements contributed by its \(X\)-attachments are
absorbed into the displayed affine shift.

Contract this context with \(F_{\mathbf{v}}\).  Its boundary is
\hypertarget{eq:v4-localized-sum}{}
\begin{equation*}
 H^+(\phi({\mathbf{s}}_0))
 =\kappa\sum_{{\mathbf{j}}\in J}(-1)^{\lambda_P({\mathbf{j}})}
   F_{\mathbf{v}}({\mathbf{s}}_0+{\mathbf{p}}_0+{\mathbf{j}}),
 \qquad {\mathbf{s}}_0\in S.
\end{equation*}
When \(M_0\ne M_1\), write
\({\mathbf{a}}_\epsilon={\mathbf{u}}_\epsilon+{\mathbf{p}}_0\) with \({\mathbf{u}}_\epsilon\in U\).  The three boundary
points \(\phi({\mathbf{u}}_0)\), \(\phi({\mathbf{u}}_0+{\mathbf{s}})\), and
\(\phi({\mathbf{r}}_{\mathbf{v}}+{\mathbf{u}}_1)\) give nonzero sums.  On each support fiber the context
character cancels the common \(P\)-restriction of the character of the Radon card, so
there is no cancellation.  The fourth point
\(\phi({\mathbf{r}}_{\mathbf{v}}+{\mathbf{u}}_1+{\mathbf{s}})\) has value zero: otherwise some \({\mathbf{j}}\in J\subseteq P\)
would satisfy \({\mathbf{s}}+{\mathbf{j}}\in M_1\), and then
\[
 0=({\mathbf{s}}+{\mathbf{j}})\cdot {\mathbf{t}}={\mathbf{s}}\cdot {\mathbf{t}}=1.
\]
Thus the support of \(H^+\) is not affine.  When \(M_0=M_1\), the
two corresponding fibers have the same nonzero cardinality
\(|J\cap M_0|\).  Their value ratio differs from \(c_1/c_0\) only by a
sign, and therefore remains outside \(\mu_4\).  In both cases
\(H^+\ne0\) and \(H^+\notin\cA\).

Finally, contract the same realizable context with \(g\), and call its boundary
\(H\).  Put \({\mathbf{w}}:=\phi({\mathbf{v}})\).  Translation by \({\mathbf{v}}\) on the input of
\cref{eq:v4-localizing-context} is translation by \({\mathbf{w}}\) on its boundary,
and the context character is trivial on \(({\mathbf{v}},{\mathbf{w}})\).  Therefore
\[
 H^+=\mathsf P_{\mathbf{w}}^+H.
\]
If \(H\in\cA\), then \cref{lem:affine-two-translate-closure} would give
\(H^+\in\cA\), a contradiction.  Thus \(H\ne0\), \(H\notin\cA\), and its
arity \(m\) is at most eight.
\end{proof}

\begin{lemma}[Affine-to-character-flat recognition]
\label{lem:v4-affine-character-radon}
Let \(0\ne g\in\cA\) have even arity \(2r\ge10\).  If every signed card
\(\mathcal S_{{\mathbf{v}},d,\ell}g\), for
\({\mathbf{v}}\in\operatorname{Acc}_{2r}\) and \(d,\ell\in\Ftwo\), is zero or
character-flat, then \(g\in\mathsf{CFlat}\).
\end{lemma}

\begin{proof}
Input translation permutes the two values of \(d\) and reindexes the
corresponding quotient by an affine isometry.  Under this reindexing, it
changes a signed card only by a nonzero scalar and a linear sign character.
We may therefore
translate the affine support to a linear subspace and write
\hypertarget{eq:v4-recognition-normal-form}{}
\begin{equation*}
 g({\mathbf{x}})=c\,i^{Q({\mathbf{x}})}\one[{\mathbf{x}}\in U],
 \qquad c\ne0,
 \qquad d_U:=\dim U.
\end{equation*}
Put \(E:=\one^\perp\).  Every nonzero character-flat card has arity
\(2r-2\) and hence support dimension \(r-1\).

We first prove \(U=U^\perp\).  If \(E\subseteq U\), then
\(d_U\ge2r-1\).  For an accessible \({\mathbf{v}}\in E\), at least one signed card on
a nonempty layer is nonzero.  Restriction to the layer and quotienting by
\(\Span_{\Ftwo}\{{\mathbf{v}}\}\) can each remove at most one dimension.  More
explicitly, on an overlapping support orbit, factoring out \(g({\mathbf{x}})\) leaves
\[
 1\pm\zeta(-1)^{\theta({\mathbf{x}})},
 \qquad \zeta\in\mu_4,
\]
where \(\theta\) is a binary linear form.  If
\(\zeta\in\{\pm i\}\), this factor is nowhere zero.  If
\(\zeta\in\{\pm1\}\), its nonzero set is either empty, the whole current
support, or one affine hyperplane.  Thus a nonzero signed card loses at
most one additional support dimension through cancellation and has support
dimension at least
\[
 d_U-3\ge2r-4>r-1,
\]
a contradiction.  If \(E\subseteq U^\perp\), then \(d_U\le1\). Choosing
accessible \({\mathbf{v}}\in E\setminus U\) gives a nonzero card of support dimension
\(d_U\), again a contradiction.  Hence neither containment holds.

The two proper subspaces \(E\cap U\) and \(E\cap U^\perp\) do not cover
\(E\).  Since \(\dim E=2r-1\ge9\), choose
\[
 {\mathbf{v}}\in\operatorname{Acc}_{2r}\setminus(U\cup U^\perp).
\]
There is no cancellation because \({\mathbf{v}}\notin U\), and each nonempty card has
support dimension \(d_U-1\).  Thus \(d_U=r\).

No accessible vector can lie in \(U^\perp\setminus U\), because such a
vector would give a nonzero card of support dimension \(r\).  Put
\(A:=E\cap U^\perp\).  It follows that
\hypertarget{eq:v4-A-contained}{}
\begin{equation*}
 A\setminus\{\zero,\one\}\subseteq U.
\end{equation*}
If \(\one\notin A\), or if \(\one\in A\cap U\), this gives
\(A\subseteq U\).  If instead \(\one\in A\setminus U\), then
\(\dim A\ge r-1\ge4\). Choose \({\mathbf{z}}\in A\setminus\{\zero,\one\}\).  Both
\({\mathbf{z}}\) and \({\mathbf{z}}+\one\) are accessible and lie in \(U\) by the preceding
display, forcing \(\one\in U\), a contradiction.  Thus \(A\subseteq U\)
in every case.

If \(U^\perp\subseteq E\), equal dimensions give \(U=U^\perp\).  Otherwise
\(\dim A=r-1\).  Since \(A\subseteq U\) and \(\dim U=r\), choose
\({\mathbf{u}}\in U\setminus A\), so
\(U=A\oplus\Span_{\Ftwo}\{{\mathbf{u}}\}\).  Because
\(A\subseteq U^\perp\), the vector \({\mathbf{u}}\) is orthogonal to \(A\).  If \({\mathbf{u}}\)
were even, then \({\mathbf{u}}\cdot {\mathbf{u}}=0\), hence \({\mathbf{u}}\in U^\perp\) and
\(U\subseteq U^\perp\). Equal dimensions would again give
\(U=U^\perp\).  Therefore \({\mathbf{u}}\) is odd and
\hypertarget{eq:v4-near-self-dual}{}
\begin{equation*}
 U=A\oplus\Span_{\Ftwo}\{{\mathbf{u}}\},
 \qquad {\mathbf{u}}\cdot {\mathbf{u}}=1.
\end{equation*}
Since \({\mathbf{u}}\) is odd, \(\one\cdot {\mathbf{u}}=1\), whereas every vector of \(A\) is
orthogonal to \({\mathbf{u}}\). Hence \(\one\notin A\).  Fix \(\zero\ne {\mathbf{s}}\in A\).
Here \(\dim A=r-1\ge4\), so \({\mathbf{s}}\) is accessible.  For this \({\mathbf{s}}\), a nonzero signed
card has either the full quotient translation space
\(U/\Span_{\Ftwo}\{{\mathbf{s}}\}\), of
dimension \(r-1\), or a codimension-one subspace of dimension \(r-2\).
The first is not self-dual because its orthogonal complement is
\(U^\perp/\Span_{\Ftwo}\{{\mathbf{s}}\}\ne
U/\Span_{\Ftwo}\{{\mathbf{s}}\}\). The second has the
wrong dimension.  On the supported layer \(d=0\), at least one of the plus
and minus cards is nonzero: after stripping their respective exterior Bell
signs and nonzero scalars, their orbit brackets are
\[
 g({\mathbf{x}})+g({\mathbf{x}}+{\mathbf{s}})
 \quad\text{and}\quad
 g({\mathbf{x}})-g({\mathbf{x}}+{\mathbf{s}}),
\]
whose sum is \(2g({\mathbf{x}})\).  This
contradicts the hypothesis.  Therefore
\[
 U=U^\perp.
\]

The orthogonal reduction in the proof of
\cref{lem:rm-flat-contexts} sends \(U\) to the code
\[
 \mathsf{Pair}_r
 :=\{{\mathbf x}({\mathbf{t}}):{\mathbf{t}}\in\Ftwo^r\},
 \qquad
 {\mathbf x}({\mathbf{t}}):=(t_1,t_1,\ldots,t_r,t_r).
\]
Orthogonal maps preserve \(\one\) and hence accessible directions.  Their
induced quotient affine isometries, together with nonzero scalar and
linear-character changes, preserve character-flatness, so the signed-card
hypothesis is invariant under this coordinate reduction.
Put
\[
 g({\mathbf x}({\mathbf{t}}))=c\,h({\mathbf{t}}),
\]
where
\begin{equation}
 h({\mathbf{t}})=i^{q_0+\sum_j\alpha_jt_j+
             2\sum_{j<k}\beta_{jk}t_jt_k},
 \qquad q_0,\alpha_j\in\mathbb Z_4,
 \quad \beta_{jk}\in\Ftwo.
 \label{eq:v4-pair-phase}
\end{equation}
Extend the quadratic coefficients symmetrically by setting
\(\beta_{jk}:=\beta_{kj}\) whenever \(k<j\), and put \(\beta_{jj}:=0\).
Let \(\bar {\mathbf{e}}_j\) be the \(j\)-th basis vector of \(\Ftwo^r\), and define
the \(j\)-th logical pair and its accessible direction in the original variables by
\[
 P_j:=\{2j-1,2j\},
 \qquad
 {\mathbf{v}}_j:={\mathbf{e}}_{2j-1}+{\mathbf{e}}_{2j}.
\]
Let
\[
 \mathcal T_j:=\{{\mathbf{t}}\in\Ftwo^r:t_j=0\}.
\]
Using the representatives \({\mathbf{t}}\in\mathcal T_j\), and stripping the exterior
Bell signs and nonzero scalars, the \(d=0\) plus and minus cards
in the \({\mathbf{v}}_j\)-direction have logical coefficient tensors
\begin{equation}
 A_j({\mathbf{t}})=h({\mathbf{t}})+h({\mathbf{t}}+\bar {\mathbf{e}}_j),
 \qquad B_j({\mathbf{t}})=h({\mathbf{t}})-h({\mathbf{t}}+\bar {\mathbf{e}}_j),
 \qquad {\mathbf{t}}\in\mathcal T_j.
 \label{eq:v4-pair-cards}
\end{equation}
Whenever one of these cards has support \(\mathcal T_j\),
character-flatness makes its logical coefficient tensor a nonzero scalar times
a sign character on \(\mathcal T_j\).

Define
\[
 L_j({\mathbf{t}}):=\sum_{k\ne j}\beta_{jk}t_k\in\Ftwo,
 \qquad
 \frac{h({\mathbf{t}}+\bar {\mathbf{e}}_j)}{h({\mathbf{t}})}=i^{\alpha_j}(-1)^{L_j({\mathbf{t}})}.
\]
Suppose first that some \(\alpha_j\) is even.  If \(L_j\ne0\), the two
logical coefficient tensors in \cref{eq:v4-pair-cards} are supported on complementary proper
hyperplanes, so neither nonzero card is character-flat.  Thus \(L_j=0\),
which gives \(\beta_{jk}=0\) for every \(k\ne j\).  One of \(A_j,B_j\) is
then zero and the other is a nonzero scalar multiple of
\(h|_{t_j=0}\).  Its being a sign character forces
\[
 \alpha_k\equiv0\pmod2\quad(k\ne j),
 \qquad \beta_{km}=0\quad(k,m\ne j).
\]
Together with the evenness of \(\alpha_j\), this proves that all
\(\alpha_k\) are even and all \(\beta_{km}\) vanish.

It remains to exclude the alternative in which every \(\alpha_j\) is odd.
Write \(i^{\alpha_j}=\delta_j i\), where \(\delta_j\in\{\pm1\}\).  The
plus bracket in \(A_j\) satisfies
\hypertarget{eq:v4-parity-bracket}{}
\begin{equation*}
 1+\delta_j i(-1)^{L_j}
 =(1+\delta_j i)(-\delta_j i)^{L_j}.
\end{equation*}
The multilinear modulo-four expansion of the Boolean parity \(L_j\) is
\hypertarget{eq:v4-parity-expansion}{}
\begin{equation*}
 L_j({\mathbf{t}})\equiv
 \sum_{k\ne j}\beta_{jk}t_k
 +2\sum_{\substack{k<m\\k,m\ne j}}
    \beta_{jk}\beta_{jm}t_kt_m
 \pmod4.
\end{equation*}
Since the nonzero full-support card \(A_j\) is a sign character, all mixed
quadratic coefficients in its phase vanish.  Hence, for all distinct
\(j,k,m\),
\begin{equation}
 \beta_{km}=\beta_{jk}\beta_{jm}.
 \label{eq:v4-beta-relation}
\end{equation}
View \(\beta\) as the adjacency matrix of a simple graph.  These identities
make the graph either empty or complete.  Indeed, if
\(\beta_{ab}=1\), then for every \(c\notin\{a,b\}\),
\[
 1=\beta_{ab}=\beta_{ca}\beta_{cb},
\]
so \(c\) is adjacent to both \(a\) and \(b\). Applying
\cref{eq:v4-beta-relation} with \(j=a\) joins every remaining pair.

The empty case is impossible: when \(L_j=0\), either nonzero card in
\cref{eq:v4-pair-cards} is a scalar multiple of \(h|_{t_j=0}\), whose
ratio in any remaining coordinate \(k\) is
\(i^{\alpha_k}=\pm i\), not the \(\pm1\) ratio of a sign character.

In the complete case, choose distinct logical pairs \(a,b,c\) and the
weight-two direction in the original variables
\[
 {\mathbf{w}}:={\mathbf{e}}_{2a-1}+{\mathbf{e}}_{2b-1}\in\operatorname{Acc}_{2r}.
\]
This vector is not in \(\mathsf{Pair}_r\).  The support of the plus card
\(\mathcal S_{{\mathbf{w}},0,0}g\) is the image of
\(\{{\mathbf{t}}:t_a+t_b=0\}\), and on the two-point \({\mathbf{w}}\)-translation orbit
represented by \({\mathbf x}({\mathbf{t}})\) its
coefficient is \(c\,h({\mathbf{t}})\), because \({\mathbf x}({\mathbf{t}})+{\mathbf{w}}\notin U\).  Toggling the
third logical pair \(c\) preserves this support, but its coefficient ratio
is
\[
 \frac{h({\mathbf{t}}+\bar {\mathbf{e}}_c)}{h({\mathbf{t}})}
 =i^{\alpha_c}(-1)^{t_c+\sum_{k\ne c}t_k}\in\{\pm i\}.
\]
A character-flat signature has ratio \(\pm1\) under translation by any vector in its
support translation space.  This contradiction eliminates the complete case.

Thus the all-odd alternative is impossible.  The earlier even-\(\alpha_j\)
case applies, and \cref{eq:v4-pair-phase} reduces to a nonzero scalar times
a linear sign character on the self-dual support \(U\).  This is exactly
\cref{eq:character-flat-class}.
\end{proof}

\begin{thebibliography}{MWXZ25}

\bibitem[ALMM99]{AubryLazardMorenoMaza1999}
Philippe Aubry, Daniel Lazard, and Marc Moreno~Maza.
\newblock On the theories of triangular sets.
\newblock {\em Journal of Symbolic Computation}, 28(1--2):105--124, 1999.

\bibitem[Bac17]{Backens2017}
Miriam Backens.
\newblock A new {Holant} dichotomy inspired by quantum computation.
\newblock In Ioannis Chatzigiannakis, Piotr Indyk, Fabian Kuhn, and Anca
  Muscholl, editors, {\em 44th International Colloquium on Automata, Languages,
  and Programming (ICALP 2017)}, volume~80 of {\em Leibniz International
  Proceedings in Informatics (LIPIcs)}, pages 16:1--16:14, Dagstuhl, Germany,
  2017. Schloss Dagstuhl--Leibniz-Zentrum f{\"u}r Informatik.

\bibitem[Bac21]{Backens2021}
Miriam Backens.
\newblock A full dichotomy for {$\operatorname{Holant}^{c}$}, inspired by
  quantum computation.
\newblock {\em SIAM Journal on Computing}, 50(6):1739--1799, 2021.

\bibitem[Bac25]{Backens2025}
Miriam Backens.
\newblock Erratum: A full dichotomy for {$\operatorname{Holant}^{c}$}, inspired
  by quantum computation.
\newblock {\em SIAM Journal on Computing}, 54(3):814--818, 2025.

\bibitem[BDG{\etalchar{+}}09]{BulatovDyerGoldbergJalseniusRicherby2009}
Andrei Bulatov, Martin Dyer, Leslie~Ann Goldberg, Markus Jalsenius, and David
  Richerby.
\newblock The complexity of weighted {Boolean} {{\#CSP}} with mixed signs.
\newblock {\em Theoretical Computer Science}, 410(38--40):3949--3961, 2009.

\bibitem[BDG{\etalchar{+}}12]{BulatovDyerGoldbergJalseniusJerrumRicherby2012}
Andrei Bulatov, Martin Dyer, Leslie~Ann Goldberg, Markus Jalsenius, Mark
  Jerrum, and David Richerby.
\newblock The complexity of weighted and unweighted {{\#CSP}}.
\newblock {\em Journal of Computer and System Sciences}, 78(2):681--688, 2012.

\bibitem[Bea83]{Beardon1983}
Alan~F. Beardon.
\newblock {\em The Geometry of Discrete Groups}, volume~91 of {\em Graduate
  Texts in Mathematics}.
\newblock Springer-Verlag, New York, 1983.

\bibitem[Bul13]{Bulatov2013}
Andrei~A. Bulatov.
\newblock The complexity of the counting constraint satisfaction problem.
\newblock {\em Journal of the ACM}, 60(5):34:1--34:41, 2013.

\bibitem[Bur05]{Burnside1905}
William Burnside.
\newblock On criteria for the finiteness of the order of a group of linear
  substitutions.
\newblock {\em Proceedings of the London Mathematical Society},
  s2-3(1):435--440, 1905.

\bibitem[Car21]{Carlet2021}
Claude Carlet.
\newblock {\em Boolean Functions for Cryptography and Coding Theory}.
\newblock Cambridge University Press, Cambridge, 2021.

\bibitem[CC17]{CaiChen2017}
Jin-Yi Cai and Xi~Chen.
\newblock Complexity of counting {CSP} with complex weights.
\newblock {\em Journal of the ACM}, 64(3):19:1--19:39, 2017.

\bibitem[CCL16]{CaiChenLu2016}
Jin-Yi Cai, Xi~Chen, and Pinyan Lu.
\newblock Nonnegative weighted {{\#CSP}}: An effective complexity dichotomy.
\newblock {\em SIAM Journal on Computing}, 45(6):2177--2198, 2016.

\bibitem[CF23]{CaiFu2023}
Jin-Yi Cai and Zhiguo Fu.
\newblock Complexity classification of the eight-vertex model.
\newblock {\em Information and Computation}, 293:105064, 2023.
\newblock Theorem numbering refers to arXiv:1702.07938v2.

\bibitem[CFS20a]{CaiFuShao2020ARS}
Jin-Yi Cai, Zhiguo Fu, and Shuai Shao.
\newblock Beyond {\#}{CSP}: A dichotomy for counting weighted {Eulerian}
  orientations with {ARS}.
\newblock {\em Information and Computation}, 275:104589, 2020.
\newblock Full version used here: arXiv:1904.02362v2.

\bibitem[CFS20b]{CaiFuShao2020Entanglement}
Jin-Yi Cai, Zhiguo Fu, and Shuai Shao.
\newblock From {Holant} to quantum entanglement and back.
\newblock In Artur Czumaj, Anuj Dawar, and Emanuela Merelli, editors, {\em 47th
  International Colloquium on Automata, Languages, and Programming (ICALP
  2020)}, volume 168 of {\em Leibniz International Proceedings in Informatics
  (LIPIcs)}, pages 22:1--22:16, Dagstuhl, Germany, 2020. Schloss
  Dagstuhl--Leibniz-Zentrum f{\"u}r Informatik.
\newblock Full version used here: arXiv:2004.05706v1.

\bibitem[CFX18]{CaiFuXia2018SixVertex}
Jin-Yi Cai, Zhiguo Fu, and Mingji Xia.
\newblock Complexity classification of the six-vertex model.
\newblock {\em Information and Computation}, 259:130--141, 2018.

\bibitem[CGW16]{CaiGuoWilliams2016}
Jin-Yi Cai, Heng Guo, and Tyson Williams.
\newblock A complete dichotomy rises from the capture of vanishing signatures.
\newblock {\em SIAM Journal on Computing}, 45(5):1671--1728, 2016.

\bibitem[CH96]{CreignouHermann1996}
Nadia Creignou and Miki Hermann.
\newblock Complexity of generalized satisfiability counting problems.
\newblock {\em Information and Computation}, 125(1):1--12, 1996.

\bibitem[CHL12]{CaiHuangLu2012Holantc}
Jin-Yi Cai, Sangxia Huang, and Pinyan Lu.
\newblock From {Holant} to {\#}{CSP} and back: Dichotomy for
  {$\operatorname{Holant}^{c}$} problems.
\newblock {\em Algorithmica}, 64(3):511--533, 2012.

\bibitem[CLO15]{CoxLittleOShea2015}
David~A. Cox, John Little, and Donal O'Shea.
\newblock {\em Ideals, Varieties, and Algorithms: An Introduction to
  Computational Algebraic Geometry and Commutative Algebra}.
\newblock Undergraduate Texts in Mathematics. Springer, Cham, 4th edition,
  2015.

\bibitem[CLX09]{CaiLuXia2009}
Jin-Yi Cai, Pinyan Lu, and Mingji Xia.
\newblock {Holant} problems and counting {CSP}.
\newblock In {\em Proceedings of the 41st Annual ACM Symposium on Theory of
  Computing (STOC 2009)}, pages 715--724, 2009.

\bibitem[CLX11]{CaiLuXia2011HolantStar}
Jin-Yi Cai, Pinyan Lu, and Mingji Xia.
\newblock Dichotomy for {$\operatorname{Holant}^{*}$} problems of {Boolean}
  domain.
\newblock In {\em Proceedings of the Twenty-Second Annual ACM--SIAM Symposium
  on Discrete Algorithms (SODA 2011)}, pages 1714--1728. SIAM, 2011.

\bibitem[CLX14]{CaiLuXia2014}
Jin-Yi Cai, Pinyan Lu, and Mingji Xia.
\newblock The complexity of complex weighted {Boolean} {{\#CSP}}.
\newblock {\em Journal of Computer and System Sciences}, 80(1):217--236, 2014.

\bibitem[CLX18]{CaiLuXia2018RealHolantc}
Jin-Yi Cai, Pinyan Lu, and Mingji Xia.
\newblock Dichotomy for real {$\operatorname{Holant}^{c}$} problems.
\newblock In {\em Proceedings of the Twenty-Ninth Annual ACM--SIAM Symposium on
  Discrete Algorithms (SODA 2018)}, pages 1802--1821. SIAM, 2018.
\newblock Full version: arXiv:1702.02693.

\bibitem[Coh93]{Cohen1993}
Henri Cohen.
\newblock {\em A Course in Computational Algebraic Number Theory}, volume 138
  of {\em Graduate Texts in Mathematics}.
\newblock Springer-Verlag, Berlin, 1993.

\bibitem[DD03]{DehaeneDeMoor2003}
Jeroen Dehaene and Bart {De Moor}.
\newblock Clifford group, stabilizer states, and linear and quadratic
  operations over {GF}(2).
\newblock {\em Physical Review A}, 68(4):042318, 2003.
\newblock Full version used here: arXiv:quant-ph/0304125v1.

\bibitem[DGJ09]{DyerGoldbergJerrum2009}
Martin Dyer, Leslie~Ann Goldberg, and Mark Jerrum.
\newblock The complexity of weighted {Boolean} {{\#CSP}}.
\newblock {\em SIAM Journal on Computing}, 38(5):1970--1986, 2009.

\bibitem[DM79]{DershowitzManna1979}
Nachum Dershowitz and Zohar Manna.
\newblock Proving termination with multiset orderings.
\newblock {\em Communications of the ACM}, 22(8):465--476, 1979.

\bibitem[DM96]{DixonMortimer1996}
John~D. Dixon and Brian Mortimer.
\newblock {\em Permutation Groups}, volume 163 of {\em Graduate Texts in
  Mathematics}.
\newblock Springer, New York, 1996.

\bibitem[DR13]{DyerRicherby2013}
Martin Dyer and David Richerby.
\newblock An effective dichotomy for the counting constraint satisfaction
  problem.
\newblock {\em SIAM Journal on Computing}, 42(3):1245--1274, 2013.

\bibitem[Got97]{Gottesman1997}
Daniel Gottesman.
\newblock {\em Stabilizer Codes and Quantum Error Correction}.
\newblock PhD thesis, California Institute of Technology, 1997.

\bibitem[GSS26]{GuanShaoShi2026}
Jincheng Guan, Shuai Shao, and Ke~Shi.
\newblock An {LP} algorithm for counting {Eulerian} orientations through the
  lens of quasi-polymorphism.
\newblock arXiv:2607.27961v1, 2026.

\bibitem[HL16]{HuangLu2016}
Sangxia Huang and Pinyan Lu.
\newblock A dichotomy for real weighted {Holant} problems.
\newblock {\em Computational Complexity}, 25:255--304, 2016.

\bibitem[Hum90]{Humphreys1990}
James~E. Humphreys.
\newblock {\em Reflection Groups and {Coxeter} Groups}, volume~29 of {\em
  Cambridge Studies in Advanced Mathematics}.
\newblock Cambridge University Press, Cambridge, 1990.

\bibitem[KL97]{KnillLaflamme1997}
Emanuel Knill and Raymond Laflamme.
\newblock Theory of quantum error-correcting codes.
\newblock {\em Physical Review A}, 55(2):900--911, 1997.

\bibitem[Lan12]{Landsberg2012}
J.~M. Landsberg.
\newblock {\em Tensors: Geometry and Applications}, volume 128 of {\em Graduate
  Studies in Mathematics}.
\newblock American Mathematical Society, Providence, RI, 2012.

\bibitem[LM26]{LiuMeng2026DEQv1}
Chenghua Liu and Boning Meng.
\newblock A dichotomy for complex {Boolean} {Holant} with binary disequality.
\newblock \href{https://arxiv.org/abs/2609.00219v1}{arXiv:2609.00219v1}, 2026.
\newblock Version 1, 31 August 2026.

\bibitem[LW18]{LinWang2018}
Jiabao Lin and Hanpin Wang.
\newblock The complexity of {Boolean} {Holant} problems with nonnegative
  weights.
\newblock {\em SIAM Journal on Computing}, 47(3):798--828, 2018.

\bibitem[MWX24]{MengWangXia2024TypicalEO}
Boning Meng, Juqiu Wang, and Mingji Xia.
\newblock P-time algorithms for typical {{\#EO}} problems.
\newblock arXiv:2410.11557v1, 2024.
\newblock Version 1.

\bibitem[MWX25]{MengWangXia2025EO}
Boning Meng, Juqiu Wang, and Mingji Xia.
\newblock The {$\mathrm{FP}^{\mathrm{NP}}$} versus {\#}{P} dichotomy for
  {\#}{EO}.
\newblock In {\em Proceedings of the 57th Annual ACM Symposium on Theory of
  Computing (STOC 2025)}, pages 1795--1806. Association for Computing
  Machinery, 2025.
\newblock Full version: arXiv:2502.02012.

\bibitem[MWXZ25]{MengWangXiaZheng2025}
Boning Meng, Juqiu Wang, Mingji Xia, and Jiayi Zheng.
\newblock From an odd arity signature to a {Holant} dichotomy.
\newblock In Srikanth Srinivasan, editor, {\em 40th Computational Complexity
  Conference (CCC 2025)}, volume 339 of {\em Leibniz International Proceedings
  in Informatics (LIPIcs)}, pages 23:1--23:20, Dagstuhl, Germany, 2025. Schloss
  Dagstuhl--Leibniz-Zentrum f{\"u}r Informatik.
\newblock Full version used here: arXiv:2502.05597v1.

\bibitem[NC10]{NielsenChuang2010}
Michael~A. Nielsen and Isaac~L. Chuang.
\newblock {\em Quantum Computation and Quantum Information}.
\newblock Cambridge University Press, Cambridge, 10th anniversary edition,
  2010.

\bibitem[SC20]{ShaoCai2020}
Shuai Shao and Jin-Yi Cai.
\newblock A dichotomy for real {Boolean} {Holant} problems.
\newblock In {\em 2020 IEEE 61st Annual Symposium on Foundations of Computer
  Science (FOCS)}, pages 1091--1102. IEEE, 2020.
\newblock Full version used here: arXiv:2005.07906v1.

\bibitem[Val08]{Valiant2008}
Leslie~G. Valiant.
\newblock Holographic algorithms.
\newblock {\em SIAM Journal on Computing}, 37(5):1565--1594, 2008.

\bibitem[Xia26]{Xia2026FrameworkV1}
Mingji Xia.
\newblock The framework to unify all complexity dichotomy theorems for
  {Boolean} tensor networks.
\newblock arXiv:2603.09417, 2026.
\newblock Version 1 (10 March 2026).

\end{thebibliography}
\end{document}